\documentclass[]{pasj01}
\draft

\usepackage[]{natbib}
\usepackage{lscape}
\usepackage{longtable}
\usepackage{subfigure}
\usepackage{hangcaption}
\usepackage{multirow}
\usepackage{color}
\usepackage{bm}

\newcommand{\mic}{${\mu}{\rm m}$ }
\newcommand{\emic}{${\mu}{\rm m}$}
\newcommand{\degreee}{$^{\circ}$}

\begin{document} 
\Received{}
\Accepted{}

\title{Mid- and far-infrared properties of Spitzer Galactic bubbles revealed by the AKARI all-sky surveys}

\author{Yasuki \textsc{Hattori}\altaffilmark{1,*}}
\altaffiltext{1}{Graduate School of Science, Nagoya University, Furo-cho, Chikusa-ku, Nagoya 464-8602, Japan}
\email{hattori@u.phys.nagoya-u.ac.jp}
\author{Hidehiro \textsc{Kaneda}\altaffilmark{1,*}}
\email{kaneda@u.phys.nagoya-u.ac.jp}
\author{Daisuke \textsc{Ishihara}\altaffilmark{1}}
\author{Yasuo \textsc{Fukui}\altaffilmark{1}}
\author{Kazufumi \textsc{Torii}\altaffilmark{1}}
\author{Misaki \textsc{Hanaoka}\altaffilmark{1}}
\author{Takuma \textsc{Kokusho}\altaffilmark{1}}
\author{Akino \textsc{Kondo}\altaffilmark{1}}
\author{Kazuyuki \textsc{Shichi}\altaffilmark{1}}
\author{Sota \textsc{Ukai}\altaffilmark{1}}
\author{Mitsuyoshi \textsc{Yamagishi}\altaffilmark{1}}
\author{Yuta \textsc{Yamaguchi}\altaffilmark{1}}


\maketitle

\begin{abstract}

We have carried out a statistical study on the mid- and far-infrared (IR) properties of Galactic IR bubbles observed by Spitzer. Using the Spitzer 8 \mic images, we estimated the radii and covering fractions of their shells, and categorized them into closed, broken and unclassified bubbles with our data analysis method. Then, using the AKARI all-sky images at wavelengths of 9, 18, 65, 90, 140 and 160 \micron, we obtained the spatial distributions and the luminosities of polycyclic aromatic hydrocarbon (PAH), warm and cold dust components by decomposing 6-band spectral energy distributions with model fitting. As a result, 180 sample bubbles show a wide range of the total IR luminosities corresponding to the bolometric luminosities of a single B-type star to many O-type stars. For all the bubbles, we investigated relationships between the radius, luminosities and luminosity ratios, and found that there are overall similarities in the IR properties among the bubbles regardless of their morphological types. In particular, they follow a power-law relation with an index of $\sim$3 between the total IR luminosity and radius, as expected from the conventional picture of the Str$\rm{\ddot{o}}$mgren sphere. The exceptions are large broken bubbles; they indicate higher total IR luminosities, lower fractional luminosities of the PAH emission, and dust heating sources located nearer to the shells. We discuss the implications of those differences for a massive star-formation scenario. 

\end{abstract}

\section{Introduction}
Galactic infrared (IR) bubbles were first detected and cataloged in \citet{Churchwell2006} and (2007), using the 8 \mic band images of the Galactic Legacy Infrared Mid-plane Survey Extraordinaire (GLIMPSE; \citealt{Benjamin2003}) program with the Spitzer satellite (\citealt{Werner2004}). The IR bubbles are observed toward the Galactic plane, which show shell or broken shell structures in the 8 \mic images. In this catalogue, difference in the morphology of the bubbles was distinguished by visual classification. Closed bubbles have well-defined shell structures in the 8 \mic images, while broken bubbles have incomplete shell structures, which do not cover the whole direction. Figure \ref{fig:Introfig1} shows examples of the cataloged bubbles. The 8 \mic band intensity is dominated by the emission of polycyclic aromatic hydrocarbons (PAHs), which are mainly distributed in photodissociation regions (PDRs).

\citet{Deharveng2010} investigated 102 Galactic IR bubbles, using radio-continuum observational data by the Multi-Array Galactic Plane Imaging Survey (MAGPIS; \citealt{Helfand2006}) at 20 cm and the VLA Galactic Plane Survey (VGPS; \citealt{Stil2006}) at 21 cm. They examined spatial correlation between the Spitzer 8 \mic and radio continuum data, which trace PAH shells and HII regions, respectively, and found that 86\% of the bubbles enclose HII regions. They also found that 98\% of the sample show the 24 \mic emission associated with the shells, which traces warm dust. These facts strongly suggest that most of the Galactic IR bubbles contain massive young stars which heat the interstellar medium (ISM) such as PAH, dust grains and gas by strong ultraviolet (UV) radiation.

Massive stars severely affect the ambient ISM. Intense UV radiation changes ISM conditions through photo-ionization of gas, photo-dissociation of molecules, and heating and photo-evaporation of dust grains. Although a lot of studies on interaction between massive stars and ambient ISM have been carried out, the formation process of a massive star itself is not well understood. One of the biggest problems in massive star formation is radiation feedback of a central star to accreting matters. \citet{Wolfire1987} calculated the relation between the mass and the mass accretion rate of a massive star and concluded that a massive star is difficult to be formed with the accretion rate of $\sim \rm{10^{-5}}$ $M_{\rm{\solar}}$ $\rm{yr^{-1}}$, which is a typical value expected from the isothermal collapse of a gaseous sphere (\citealt{Stahler1980a}), because of intense radiation pressure. Therefore very efficient mass accretion is indispensable to form massive stars.

\citet{Elmegreen1998} suggested three major processes for massive star formation, which are $``$globule squeezing$"$, $``$collect and collapse$"$ and $``$cloud-cloud collision$"$. All the processes create dense molecular cores through gas compression and trigger massive star formation as a result of the self-gravitational collapses of the cores. \citet{Deharveng2010} suggested the observational evidence for the $``$globule squeezing$"$ process by investigating spatial correlation between the pre-existing dust condensations observed by the APEX Telescope Large Area Survey of the Galactic plane (ATLASGAL; \citealt{Schuller2009}) at 870 \mic and the ionization front of HII regions traced by the Spitzer 8 and 24 \mic images of Galactic IR bubbles. They found that 28\% of 65 spatially resolved bubbles include dust condensations associated with bright rims, which indicates the compression of pre-existing dense gas by the expanding HII regions. They also showed that other 40\% of the spatially resolved bubbles are surrounded by cold dust emission, which implies that these bubbles are candidates of massive star-forming regions triggered by the $``$collect and collapse$"$ process. \citet{Ishihara2007} studied IC4954 and IC4955 using the AKARI mid- and far-IR data, and found that young stellar objects are located on the 9 \mic emission arc created by the central source indicating star formation in the ambient ISM compressed by the central source. This is also observational evidence for $``$collect and collapse$"$.

Observational evidence for $``$cloud-cloud collision$"$ was first reported by \citet{Loren1976} using the velocity information of two clouds in NGC1333. \citet{Habe1992} carried out the simulation of head-on collisions and suggested that collision between two molecular clouds generates strong bow shocks which compress parental molecular clouds and create gravitationally unstable cores. Massive stars can be formed by self-gravitational collapse of the compressed unstable cores (e.g., \citealt{Takahira2014}). Recent observations of Galactic massive star-forming regions such as RCW49, Westerlund2, M20, NGC3603 and RCW120 with the NANTEN/NANTEN2 radio telescopes indicated several pieces of observational evidence for the $``$cloud-cloud collision$"$ scenario (\citealt{Furukawa2009}; \citealt{Ohama2010}; \citealt{Torii2011}; \citealt{Fukui2014}; \citealt{Torii2015}). They found that two molecular clouds possessing different radial velocities are associated with the star-forming regions; large velocity difference (about 15 km $\rm{s^{-1}}$) between two clouds indicates that the clouds cannot be bound gravitationally by the observed cloud masses. These facts suggest a possibility that the massive star formation in these regions may have been triggered by cloud-cloud collisions. Thus a lot of studies for the $``$cloud-cloud collision$"$ process have been carried out, emphasizing the importance of the process in massive star formation. It is notable that the resultant structure of the colliding clouds is similar to the structure of the Galactic IR bubbles, especially for the broken bubbles.

Mid- and far-IR observations of massive star-forming regions are essential to obtain information on PAHs and dust grains, which surround exciting sources. PAH emission traces the morphology of the surfaces of PDRs which are located on the borders of HII regions and molecular clouds. Warm dust emission with a typical temperature of about 60 $\rm{K}$ traces the morphology of HII regions, while cold dust emission with a typical temperature of about 20 $\rm{K}$ traces that of cold gas regions. Therefore, using these IR dust emissions, we can investigate the complicated geometry of HII regions and PDRs in star-forming regions with respect to the positions of massive stars, revealing the interaction of each component. In addition, and more importantly, we can estimate the total energy of exciting sources from the total IR dust emission because these embedded dust components receive a dominant fraction of UV photons from young massive stars and re-radiate the absorbed energy at mid- and far-IR wavelengths. Hence IR observations provide many pieces of information on the Galactic IR bubbles; nevertheless, a systematic study using both mid- and far-IR data has not been carried out yet.

In this paper, we investigate the mid- and far-IR properties of the Spitzer Galactic IR bubbles using the AKARI all-sky survey data. The AKARI all-sky surveys cover a wide IR wavelength range with the 6 photometric bands at wavelengths of 9, 18, 65, 90, 140 and 160 \mic so that we can estimate the total IR flux of each bubble. We obtained the central position and the radius of each bubble by fitting a circle to its image. We also estimated the covering fractions of bubbles on the basis of our quantitative criteria to classify them into closed and broken bubbles. We created a local spectral energy distribution (SED) from the 6 photometric bands with a \timeform{15''} spatial grid and fitted the SED by the model consisting of PAHs, warm dust and cold dust components. Then for each dust component, we created a distribution map with a \timeform{15''} grid.

\begin{figure*}[ht]
\centering
\subfigure{
\makebox[180mm][l]{\raisebox{0mm}[0mm][0mm]{ \hspace{12.5mm} \small{8 \mic}} \hspace{29.5mm} \small{9 \mic} \hspace{27mm} \small{18 \mic} \hspace{26.5mm} \small{90 \mic}}%
}
\subfigure{
\makebox[180mm][l]{\raisebox{0mm}[0mm][0mm]{ \hspace{9mm} \small{RA (J2000)}} \hspace{19.5mm} \small{RA (J2000)} \hspace{19.5mm} \small{RA (J2000)} \hspace{19.5mm} \small{RA (J2000)}}%
}
\subfigure{
\mbox{\raisebox{6mm}{\rotatebox{90}{\small{DEC (J2000)}}}}%
\mbox{\raisebox{0mm}{\includegraphics[width=40mm]{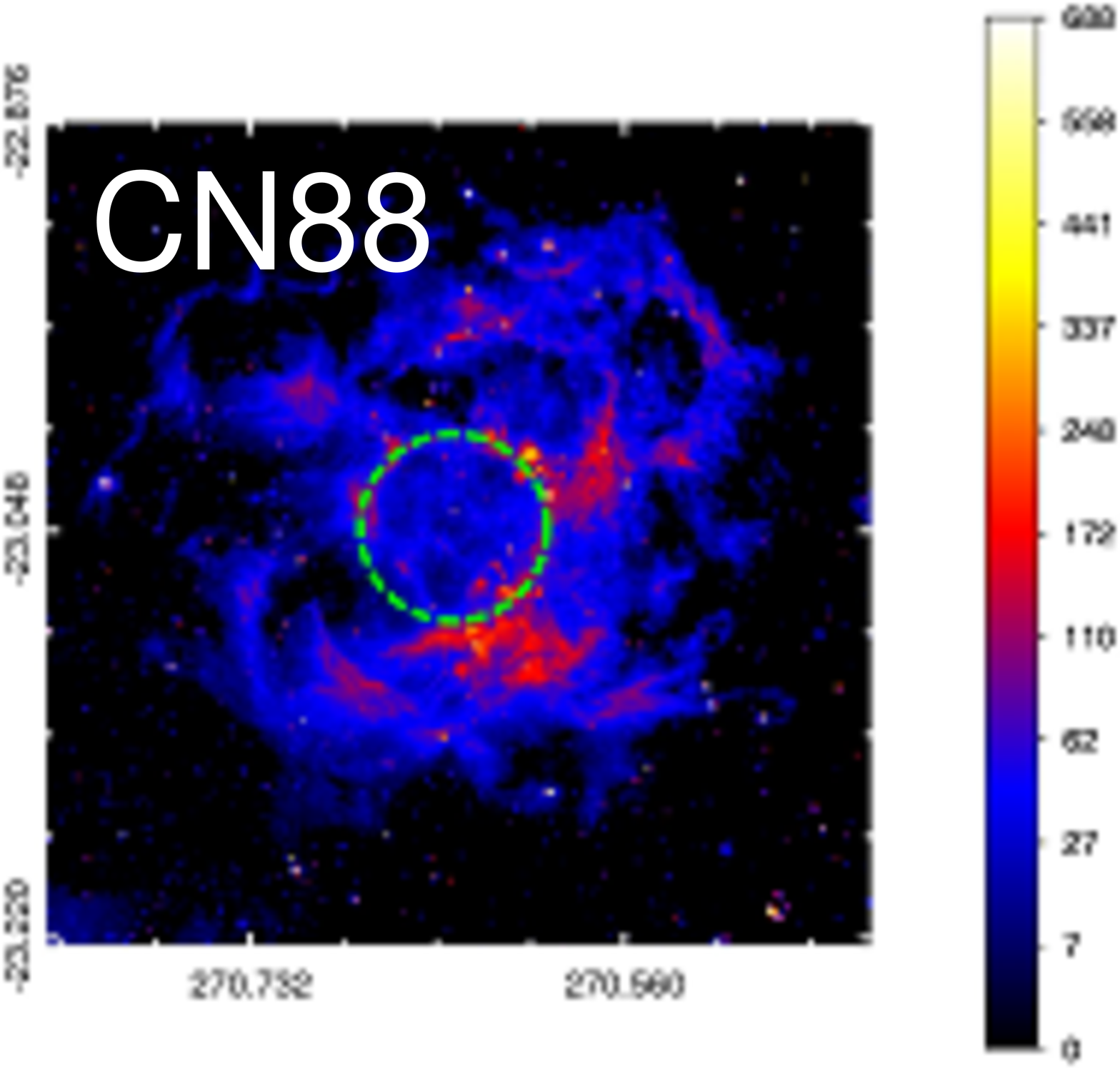}}}%
\mbox{\raisebox{0mm}{\includegraphics[width=40mm]{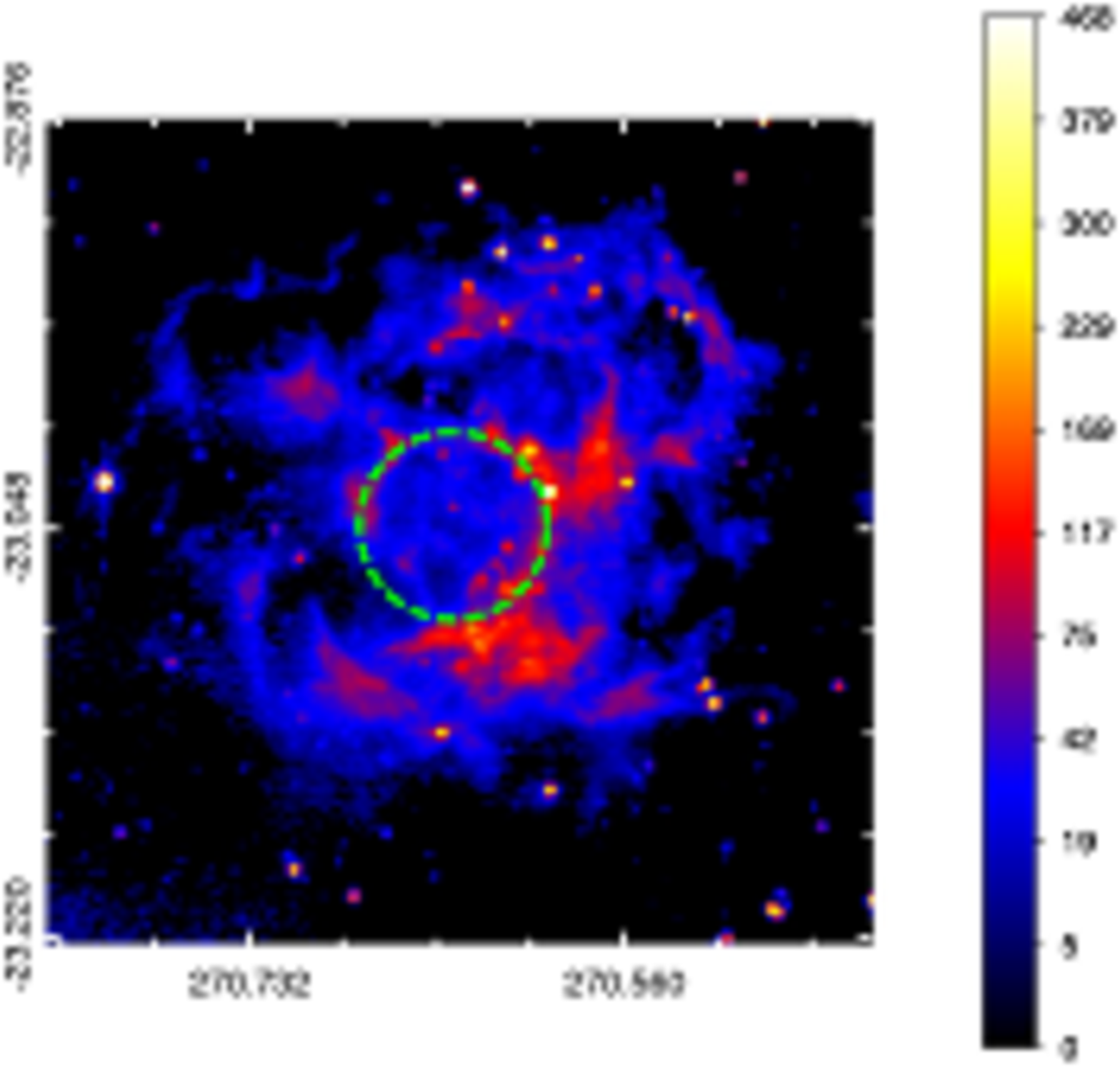}}}%
\mbox{\raisebox{0mm}{\includegraphics[width=40mm]{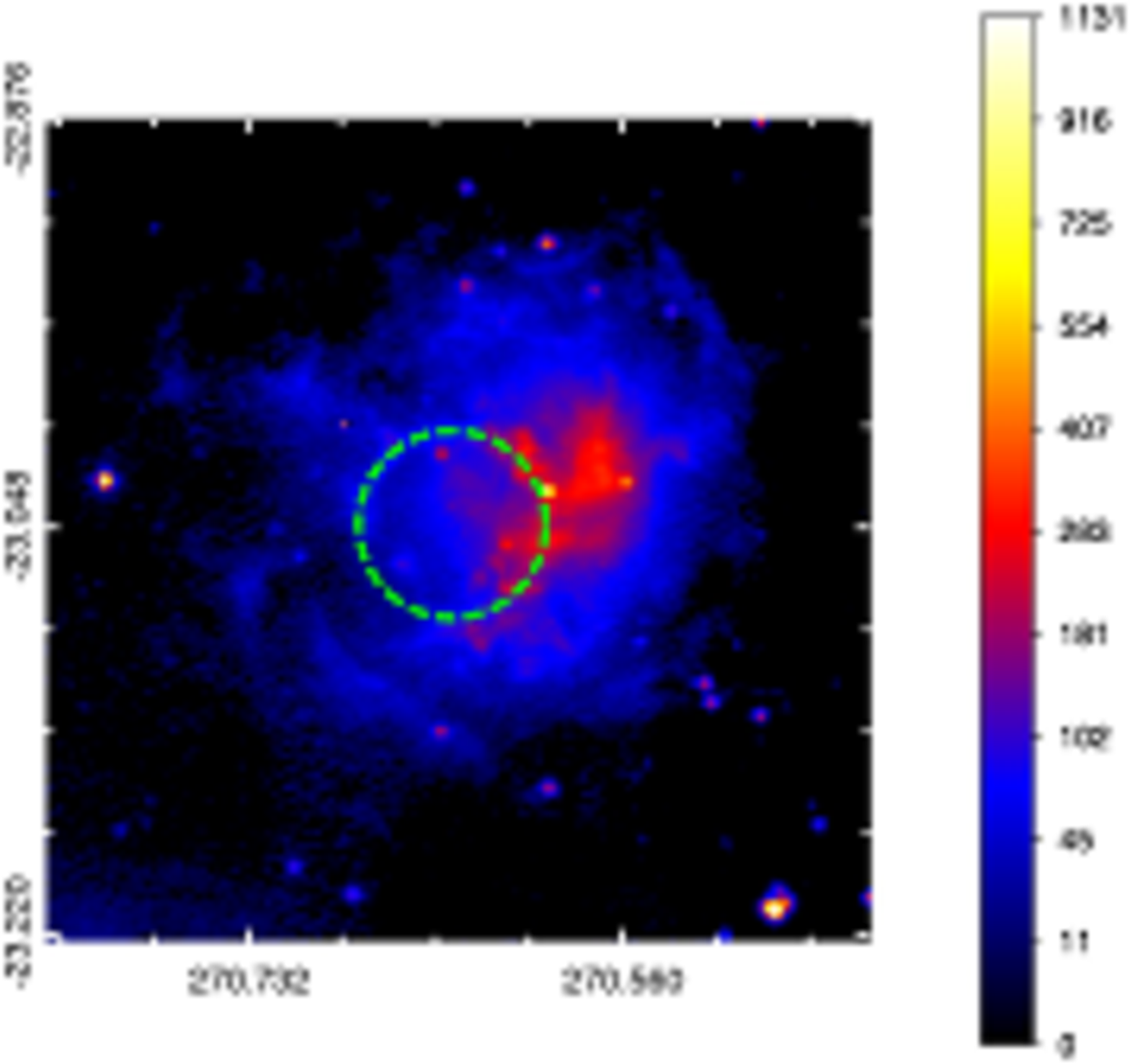}}}%
\mbox{\raisebox{0mm}{\includegraphics[width=40mm]{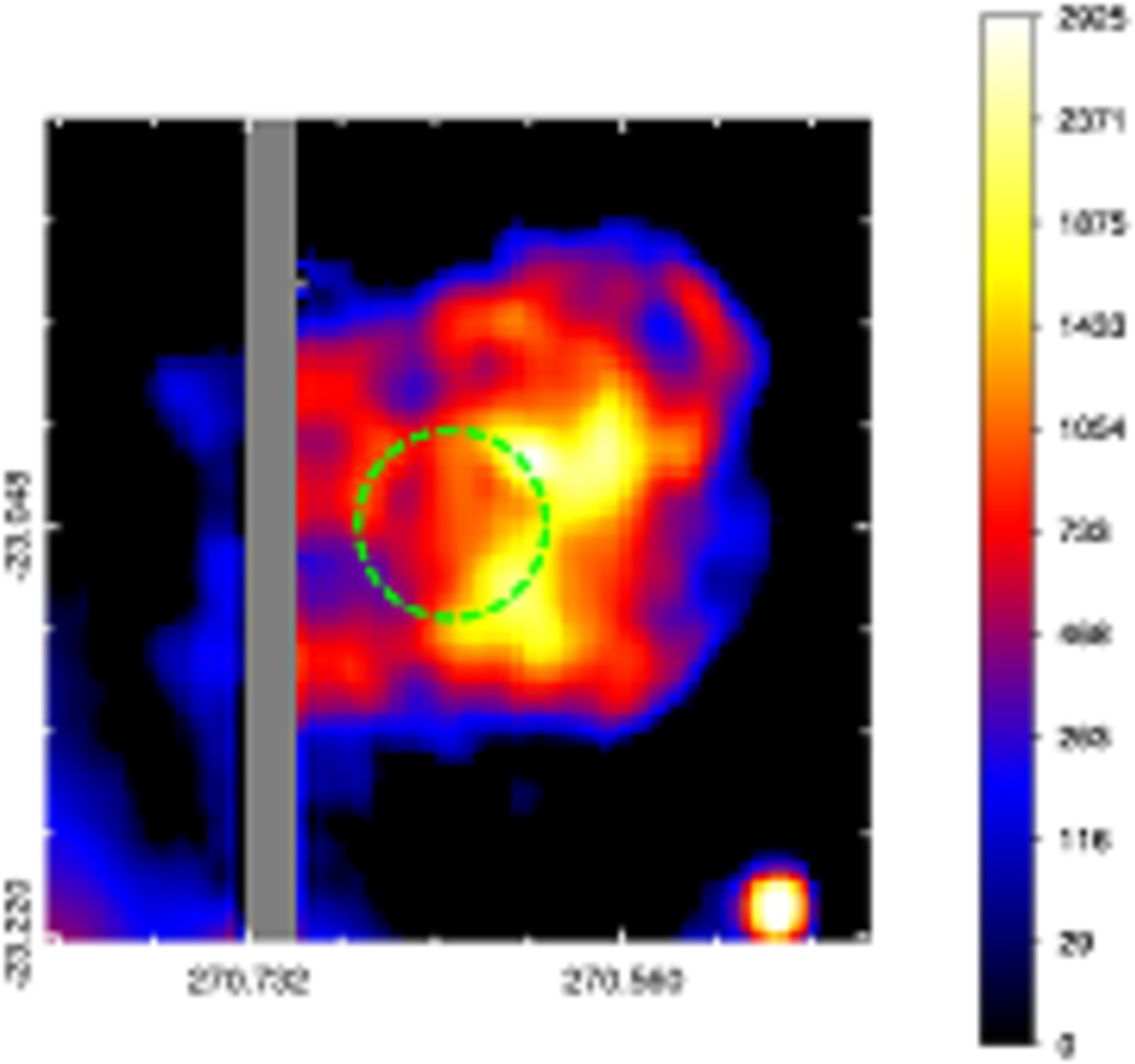}}}%
}
\subfigure{
\mbox{\raisebox{6mm}{\rotatebox{90}{\small{DEC (J2000)}}}}%
\mbox{\raisebox{0mm}{\includegraphics[width=40mm]{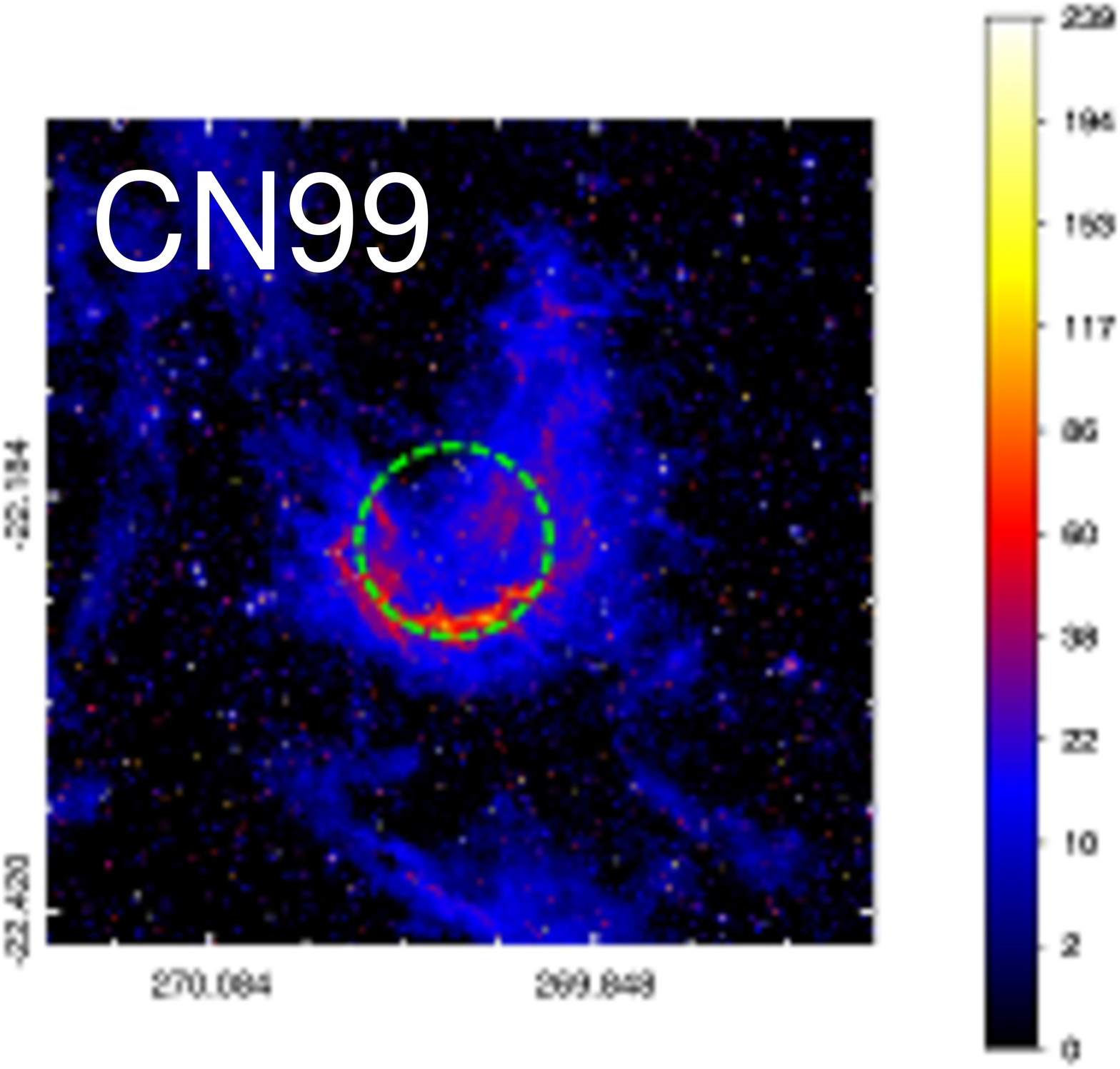}}}%
\mbox{\raisebox{0mm}{\includegraphics[width=40mm]{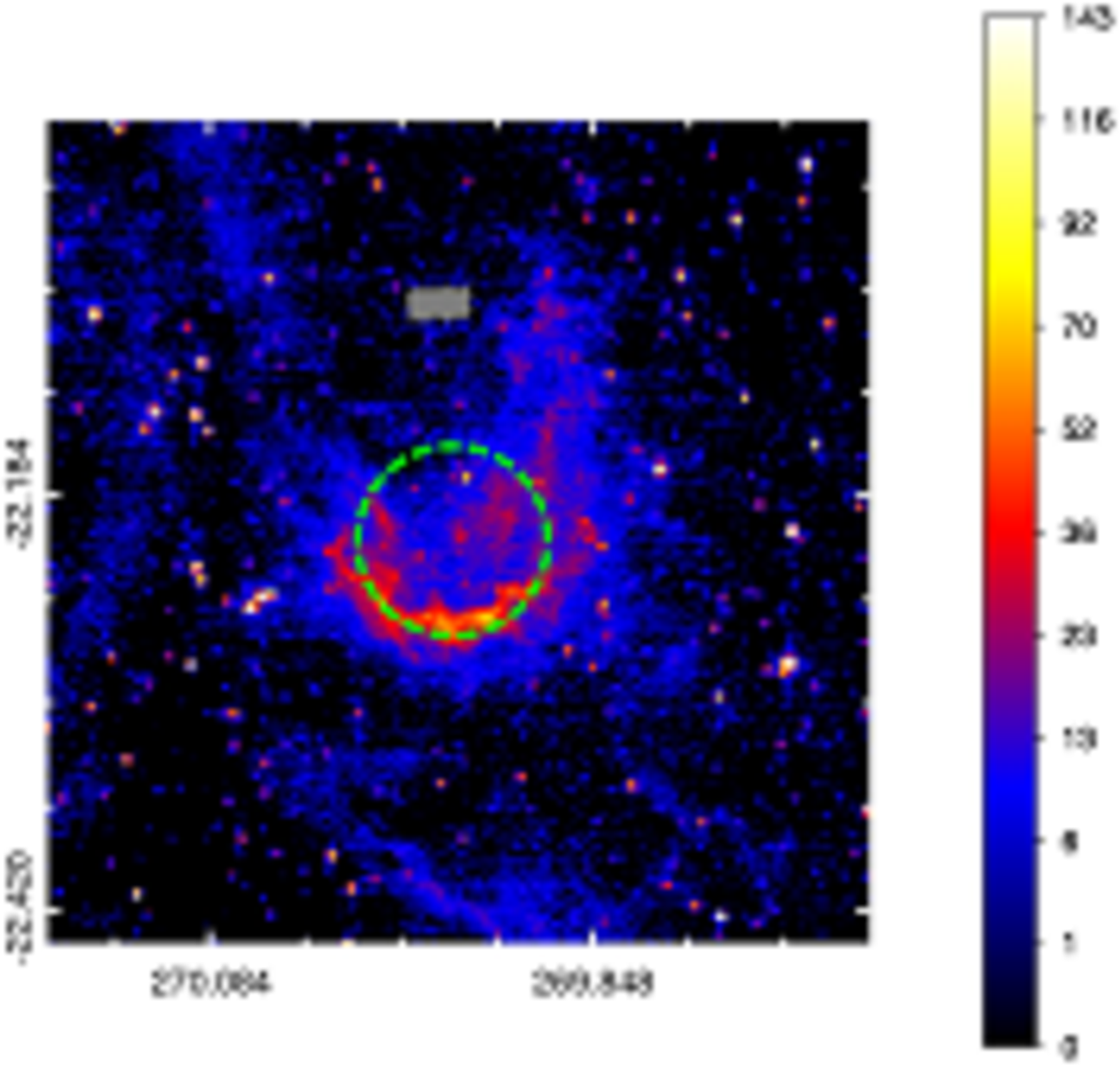}}}%
\mbox{\raisebox{0mm}{\includegraphics[width=40mm]{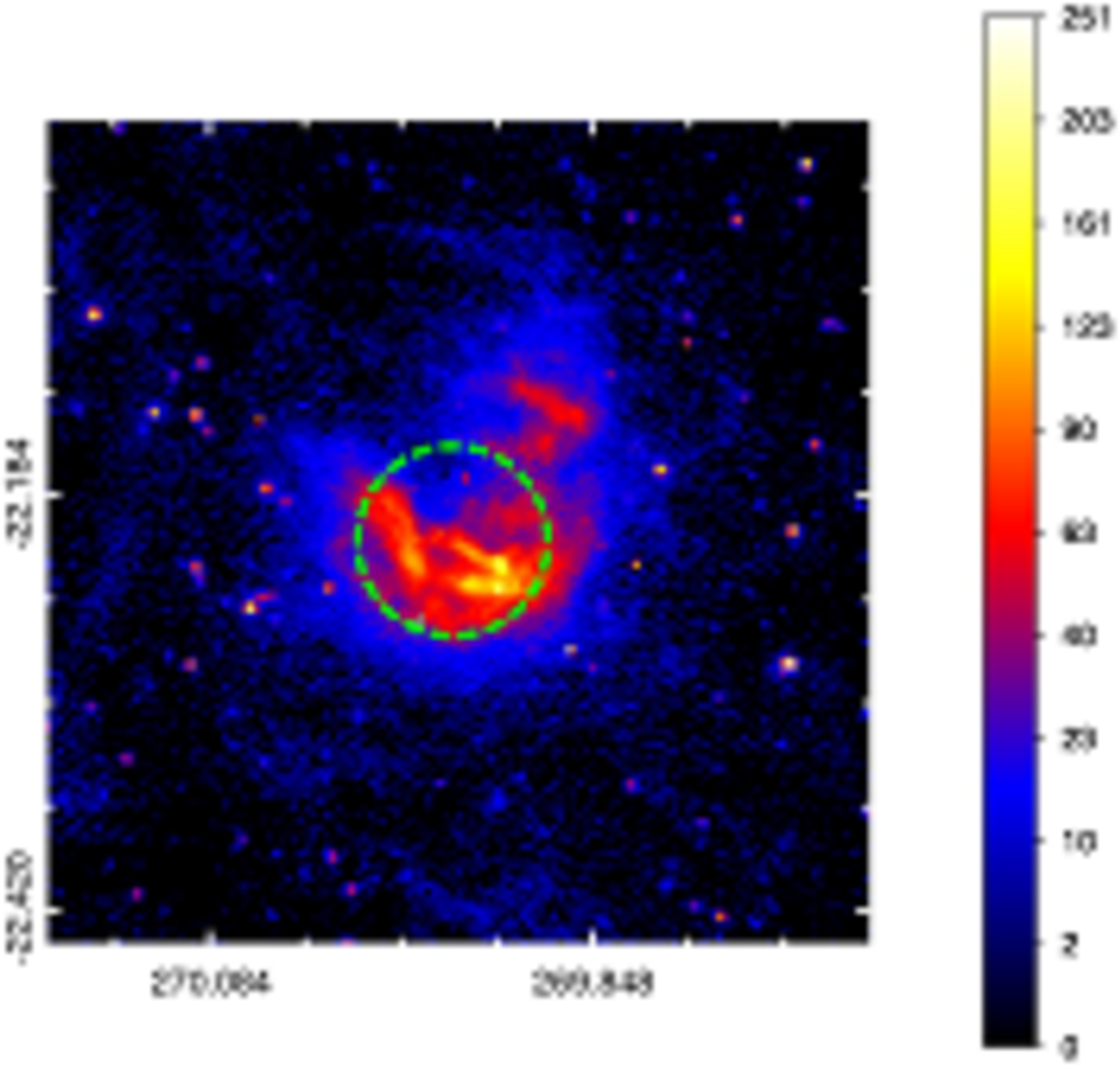}}}%
\mbox{\raisebox{0mm}{\includegraphics[width=40mm]{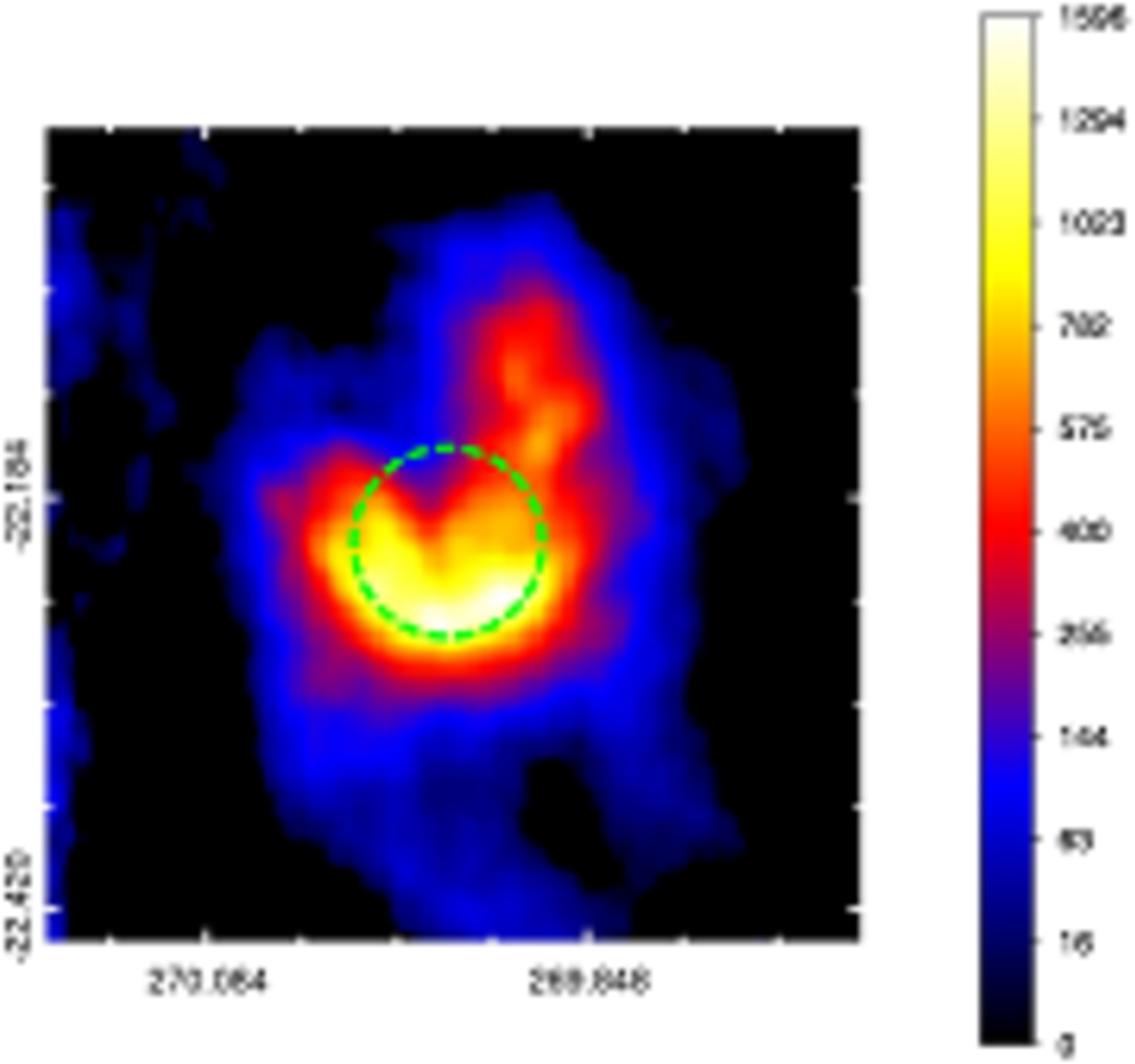}}}%
}
\subfigure{
\mbox{\raisebox{6mm}{\rotatebox{90}{\small{DEC (J2000)}}}}%
\mbox{\raisebox{0mm}{\includegraphics[width=40mm]{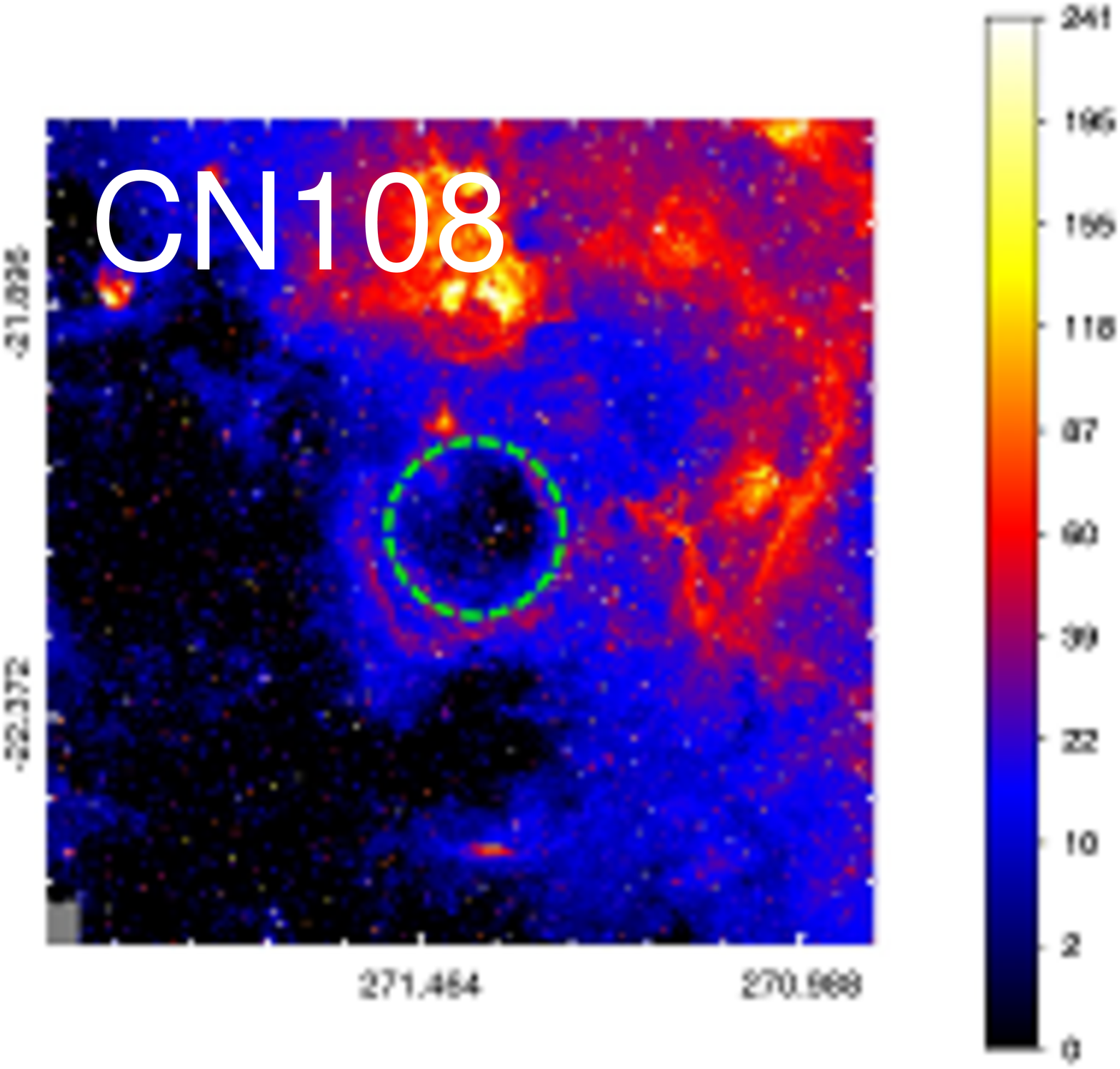}}}%
\mbox{\raisebox{0mm}{\includegraphics[width=40mm]{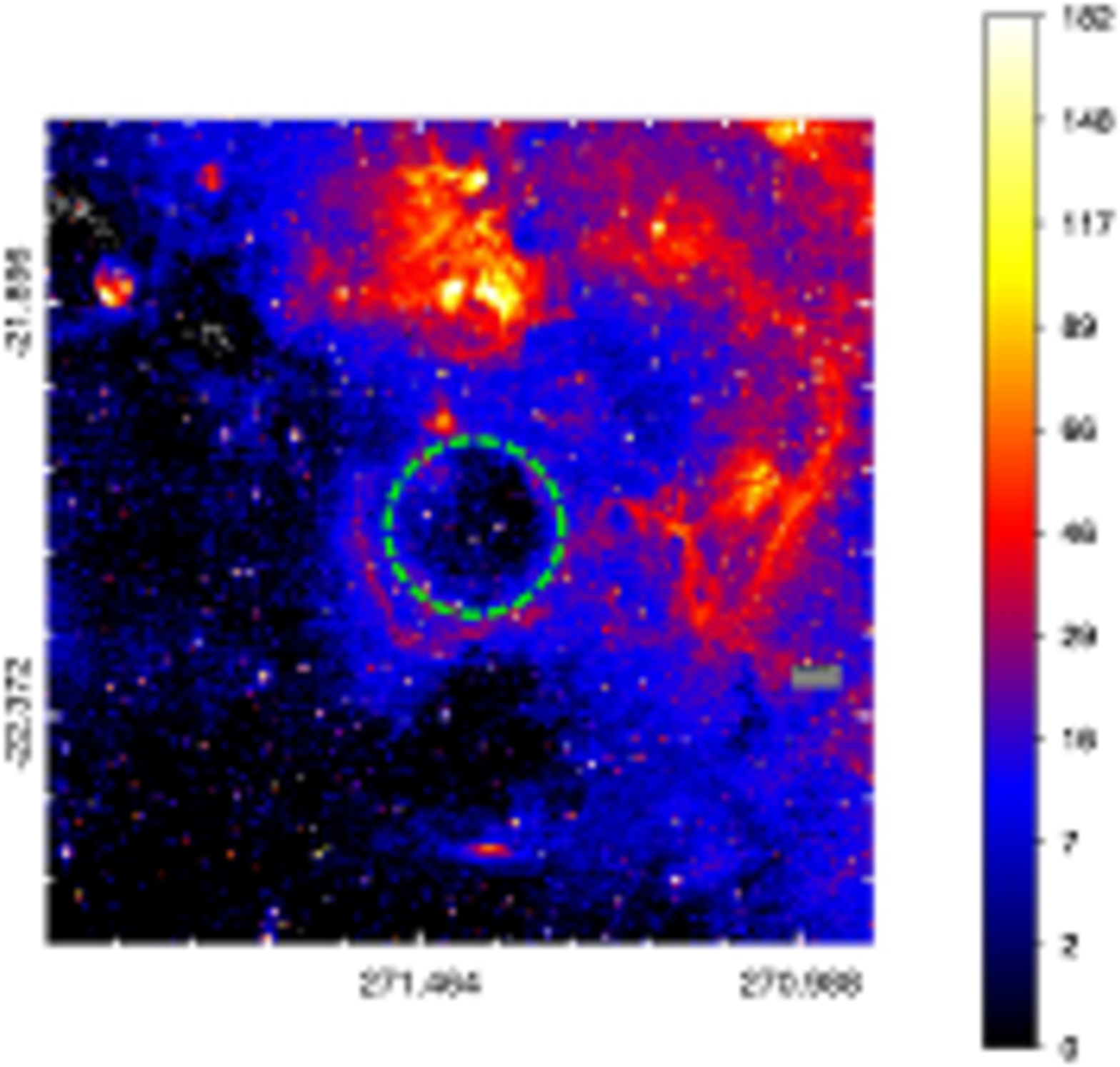}}}%
\mbox{\raisebox{0mm}{\includegraphics[width=40mm]{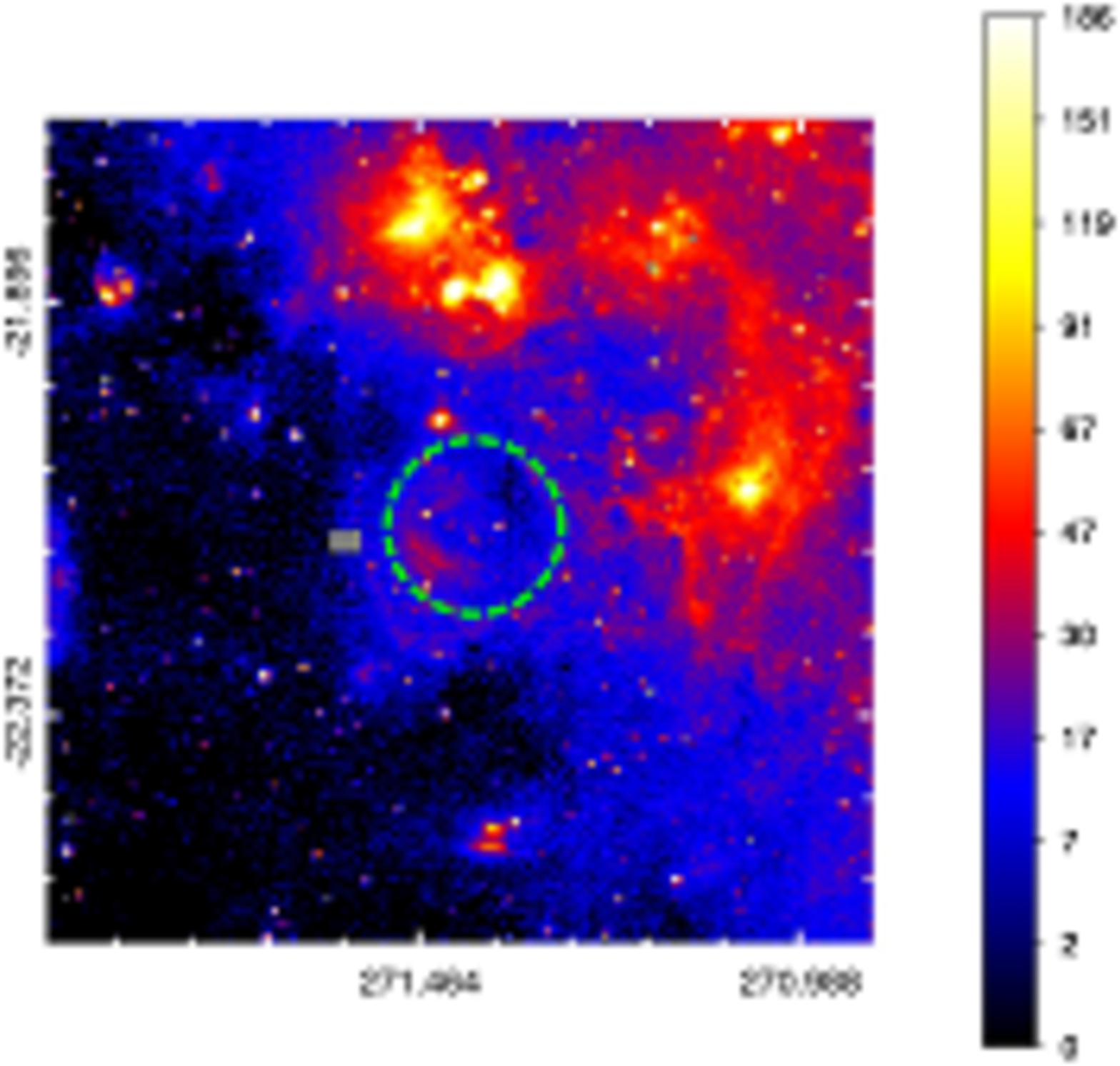}}}%
\mbox{\raisebox{0mm}{\includegraphics[width=40mm]{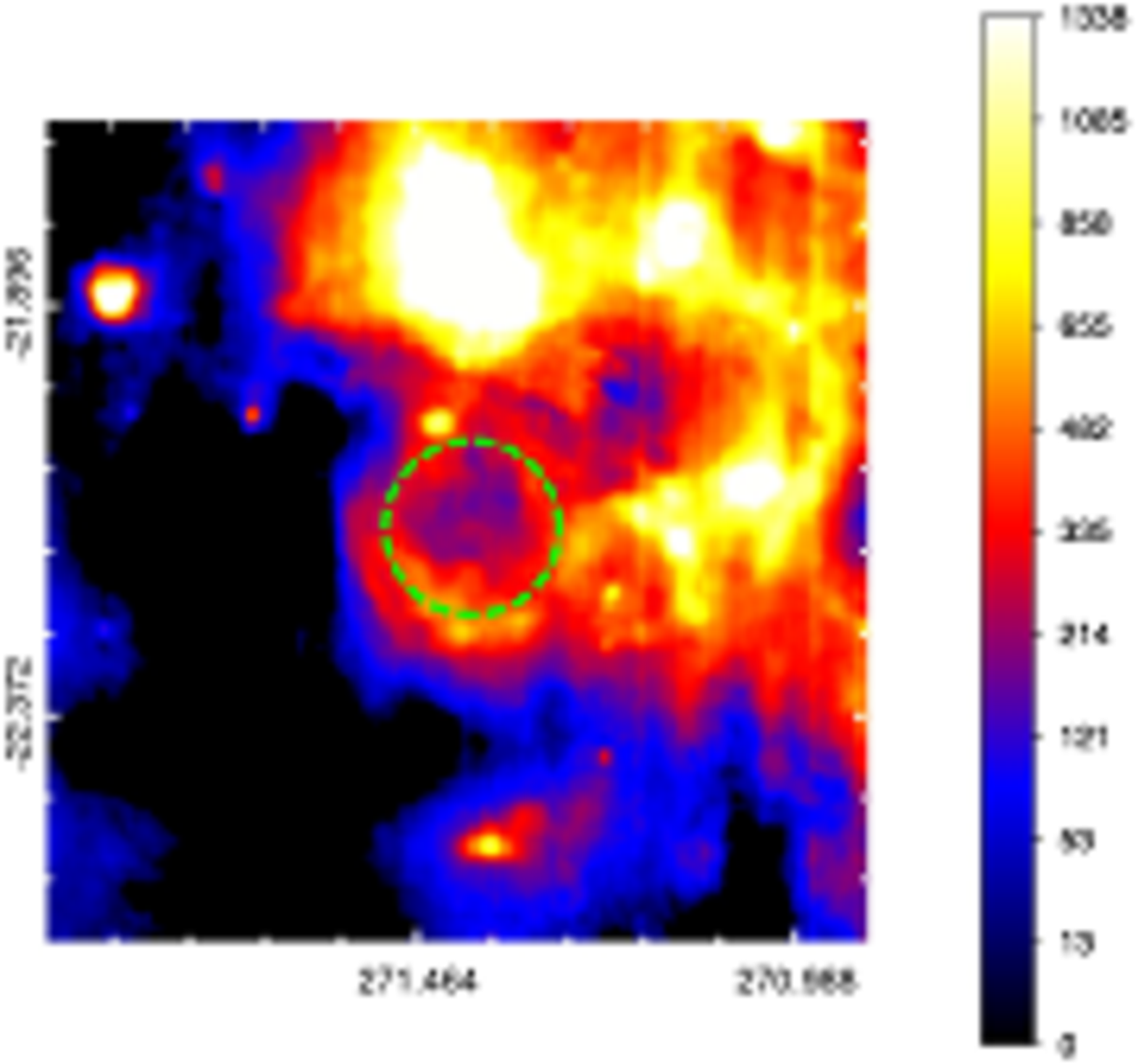}}}%
}
\subfigure{
\mbox{\raisebox{6mm}{\rotatebox{90}{\small{DEC (J2000)}}}}%
\mbox{\raisebox{0mm}{\includegraphics[width=40mm]{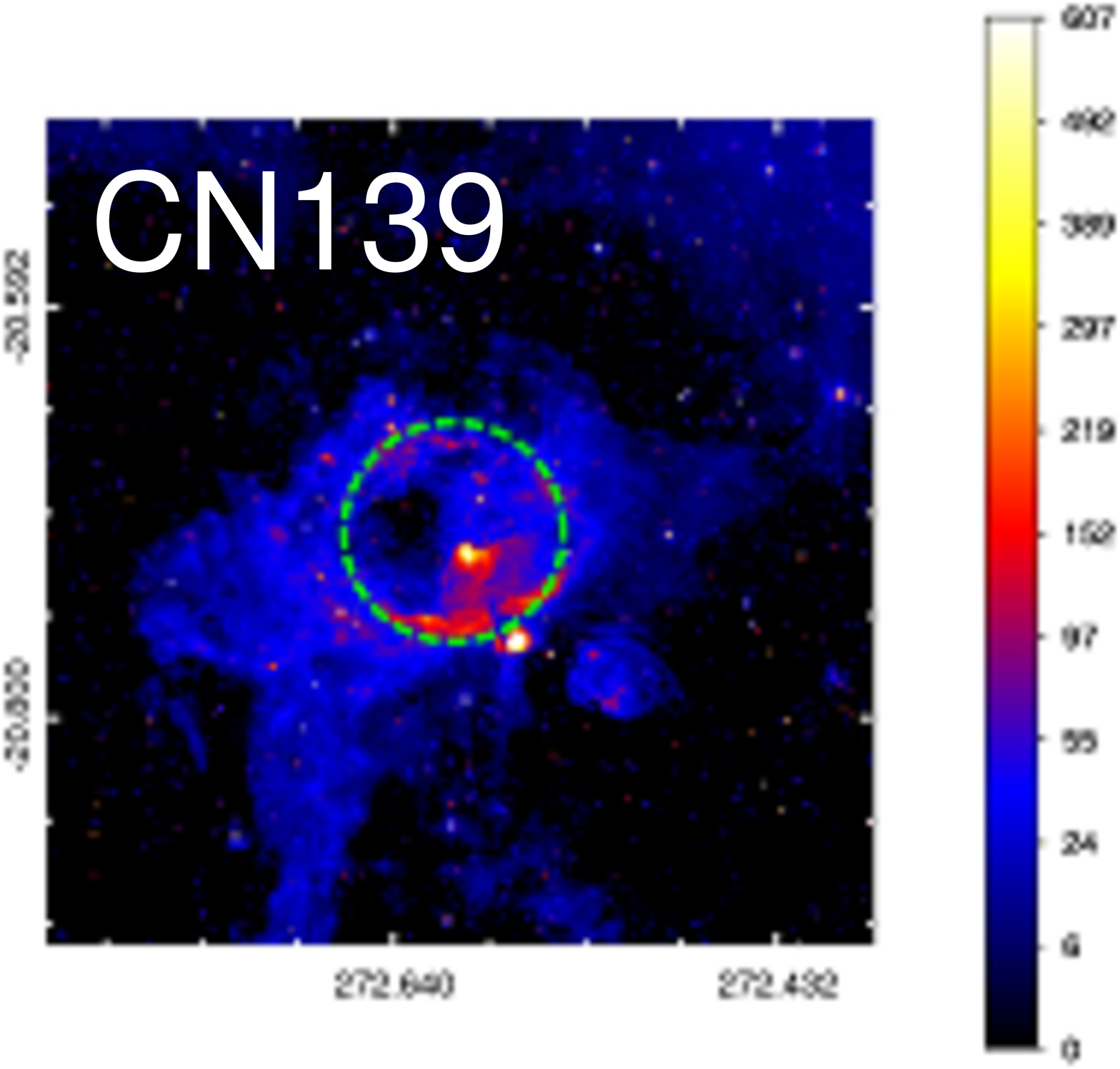}}}%
\mbox{\raisebox{0mm}{\includegraphics[width=40mm]{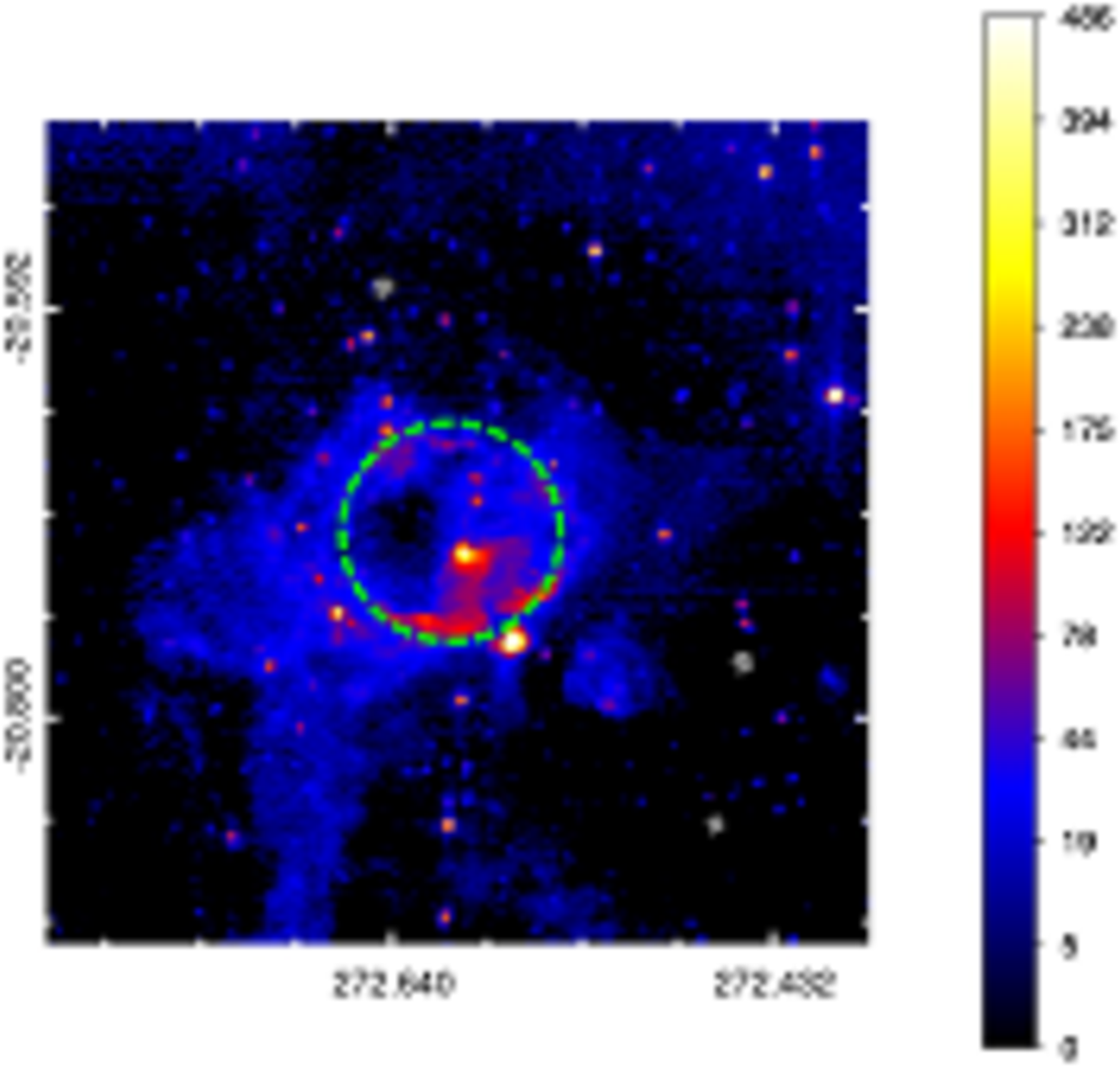}}}%
\mbox{\raisebox{0mm}{\includegraphics[width=40mm]{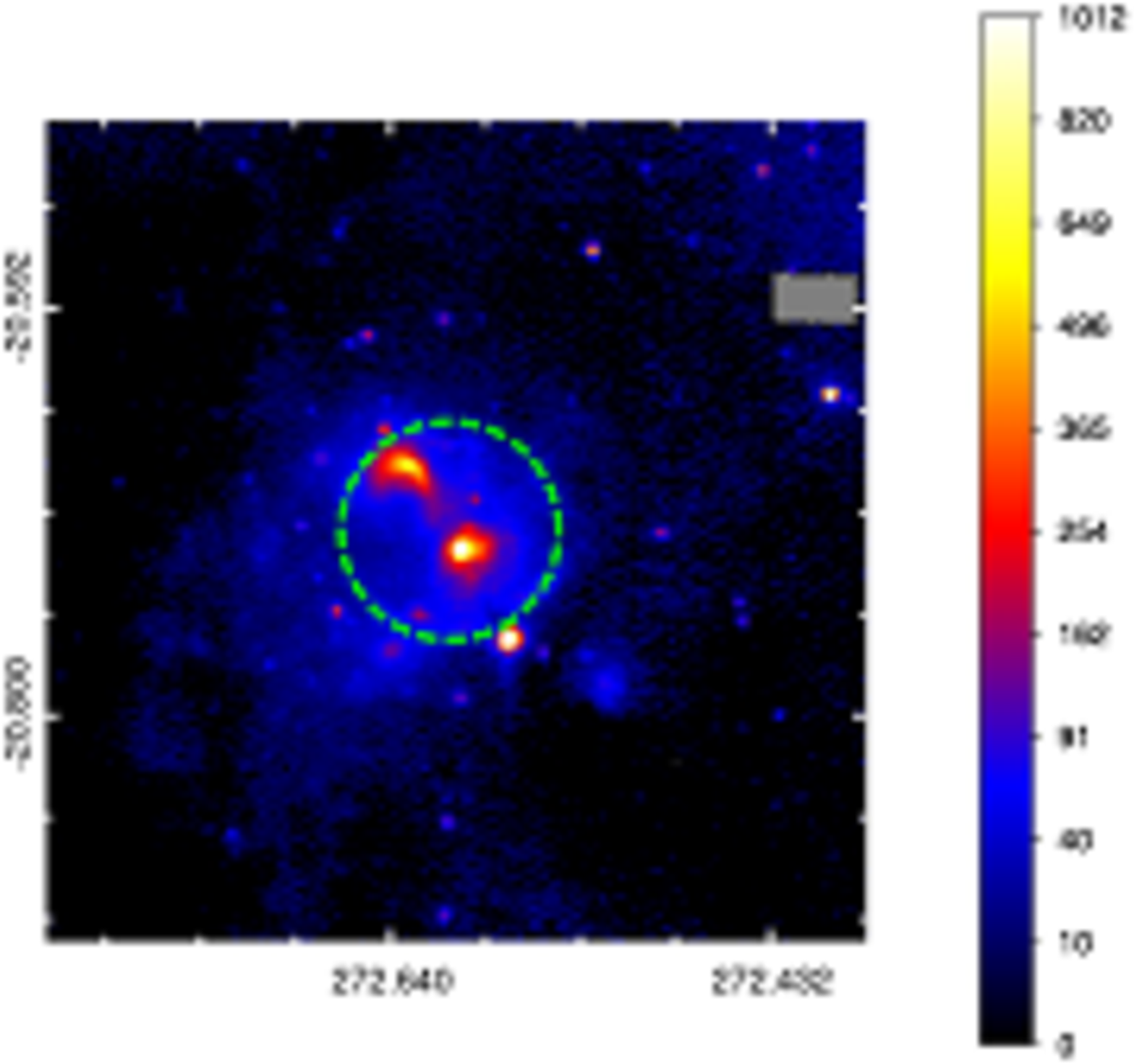}}}%
\mbox{\raisebox{0mm}{\includegraphics[width=40mm]{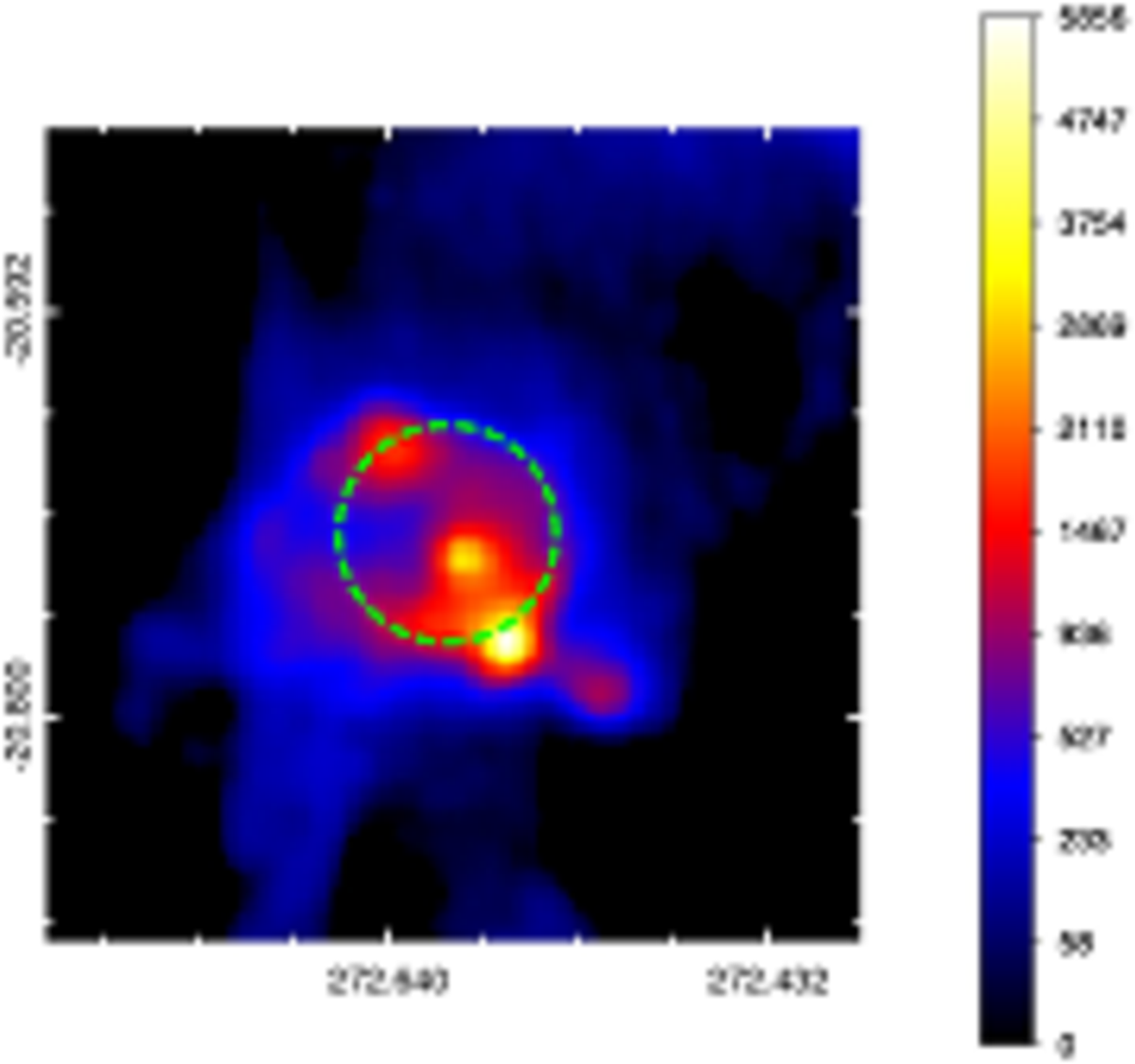}}}%
}
\subfigure{
\mbox{\raisebox{6mm}{\rotatebox{90}{\small{DEC (J2000)}}}}%
\mbox{\raisebox{0mm}{\includegraphics[width=40mm]{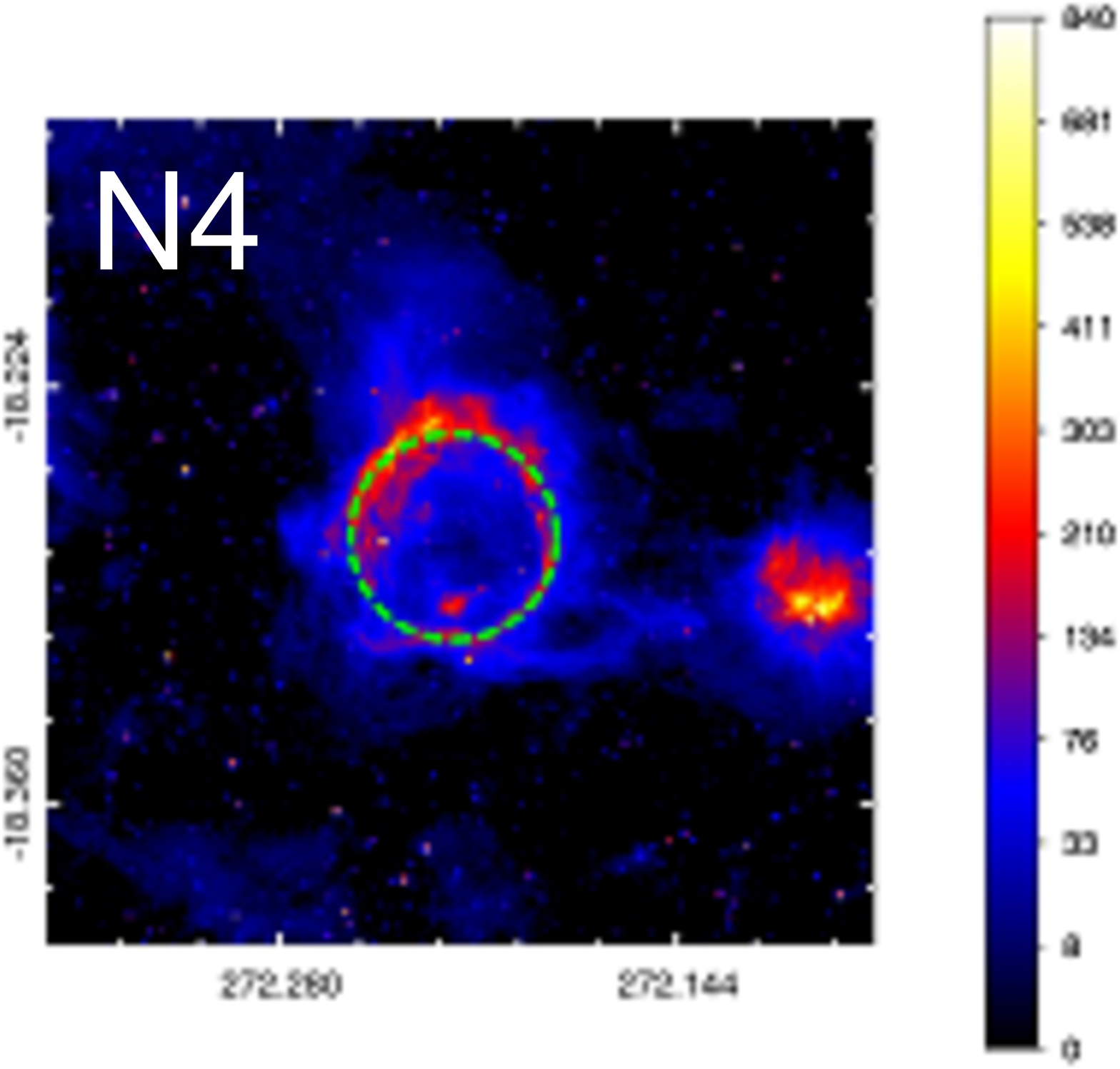}}}%
\mbox{\raisebox{0mm}{\includegraphics[width=40mm]{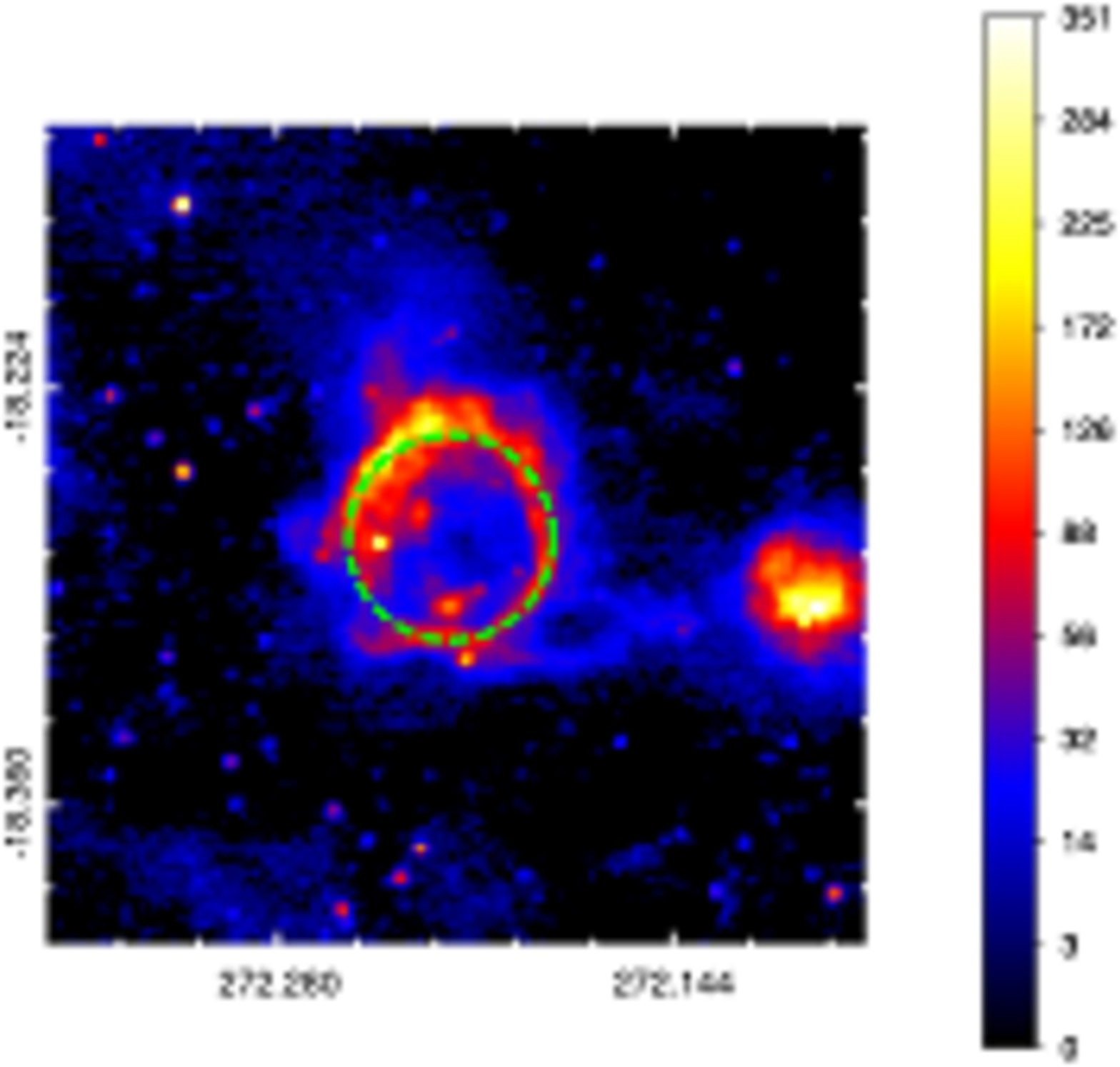}}}%
\mbox{\raisebox{0mm}{\includegraphics[width=40mm]{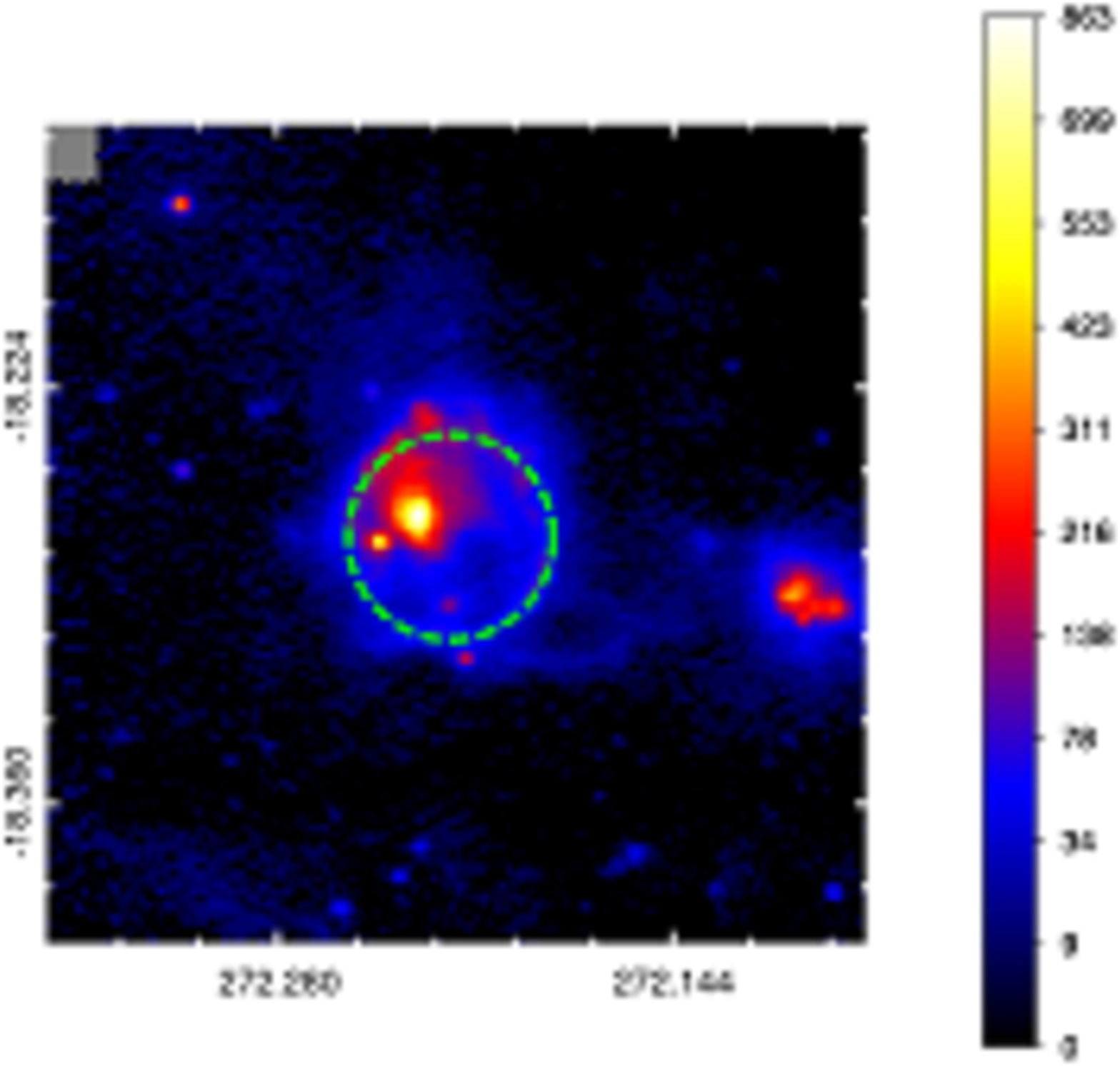}}}%
\mbox{\raisebox{0mm}{\includegraphics[width=40mm]{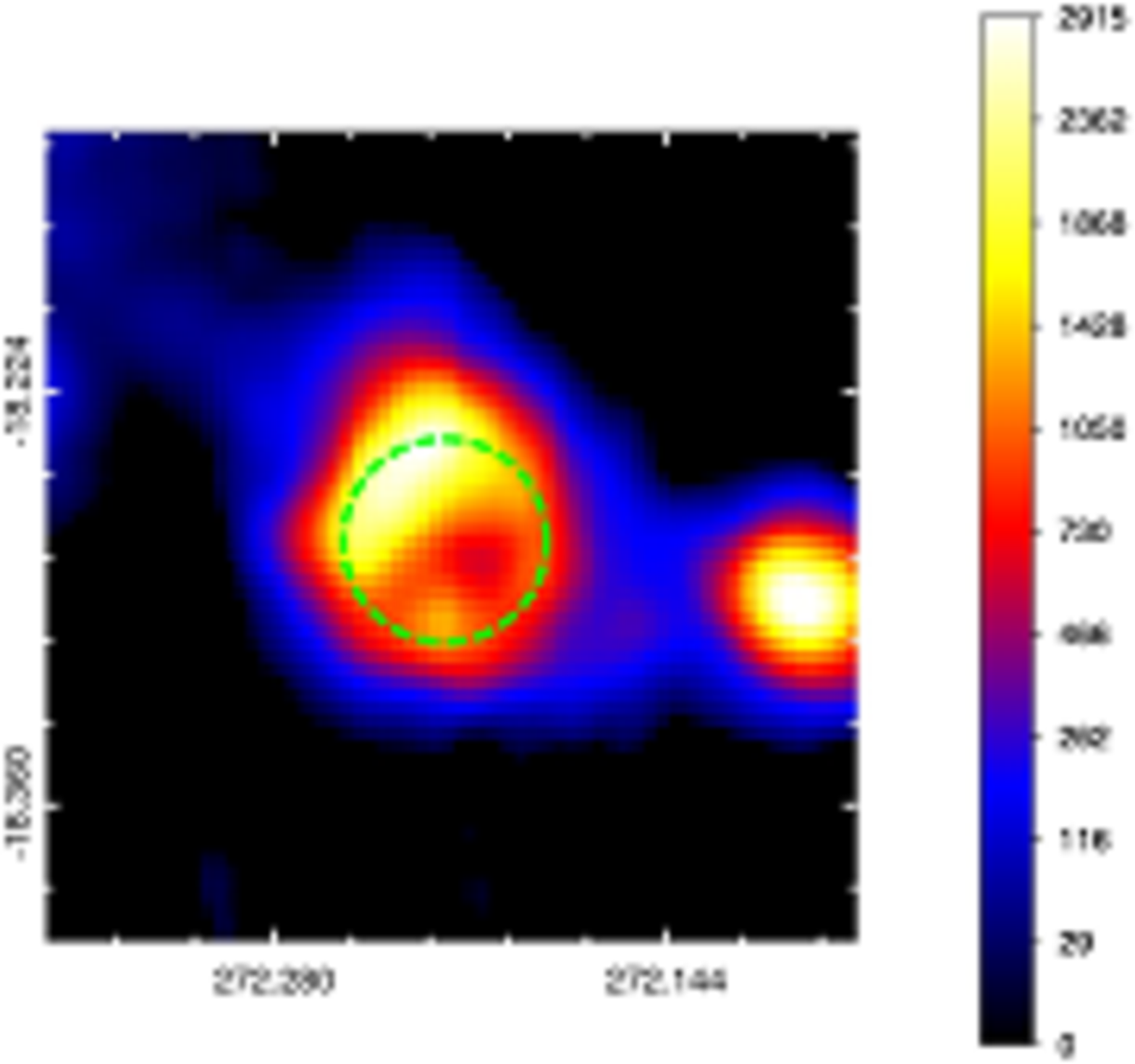}}}%
}
\raggedright
\caption{Examples of the Spitzer 8 \mic band and the AKARI 9, 18 and 90 \mic band images of the Spitzer Galactic IR bubbles cataloged in \citet{Churchwell2006} and (2007). For each row, the name of the bubble is given in the left panel, and the 8, 9, 18 and 90 \mic images are arrayed from left to right. The green dashed circles are drawn with the central positions and radii of the bubbles given in \citet{Churchwell2006} and (2007). The color levels are given in units of MJy $\rm{sr^{-1}}$. The angular size of each panel is $4R{\times}4R$, where $R$ is the bubble radius estimated with our circular fitting (see section 2). \label{fig:Introfig1}} %
\end{figure*}

\addtocounter{figure}{-1}
\begin{figure*}[ht]
\addtocounter{subfigure}{1}
\centering
\subfigure{
\makebox[180mm][l]{\raisebox{0mm}[0mm][0mm]{ \hspace{15mm} \small{8 \mic}} \hspace{29.5mm} \small{9 \mic} \hspace{27mm} \small{18 \mic} \hspace{26.5mm} \small{90 \mic}}%
}
\subfigure{
\makebox[180mm][l]{\raisebox{0mm}[0mm][0mm]{ \hspace{11mm} \small{RA (J2000)}} \hspace{19.5mm} \small{RA (J2000)} \hspace{20mm} \small{RA (J2000)} \hspace{20mm} \small{RA (J2000)}}%
}
\subfigure{
\mbox{\raisebox{6mm}{\rotatebox{90}{\small{DEC (J2000)}}}}%
\mbox{\raisebox{0mm}{\includegraphics[width=40mm]{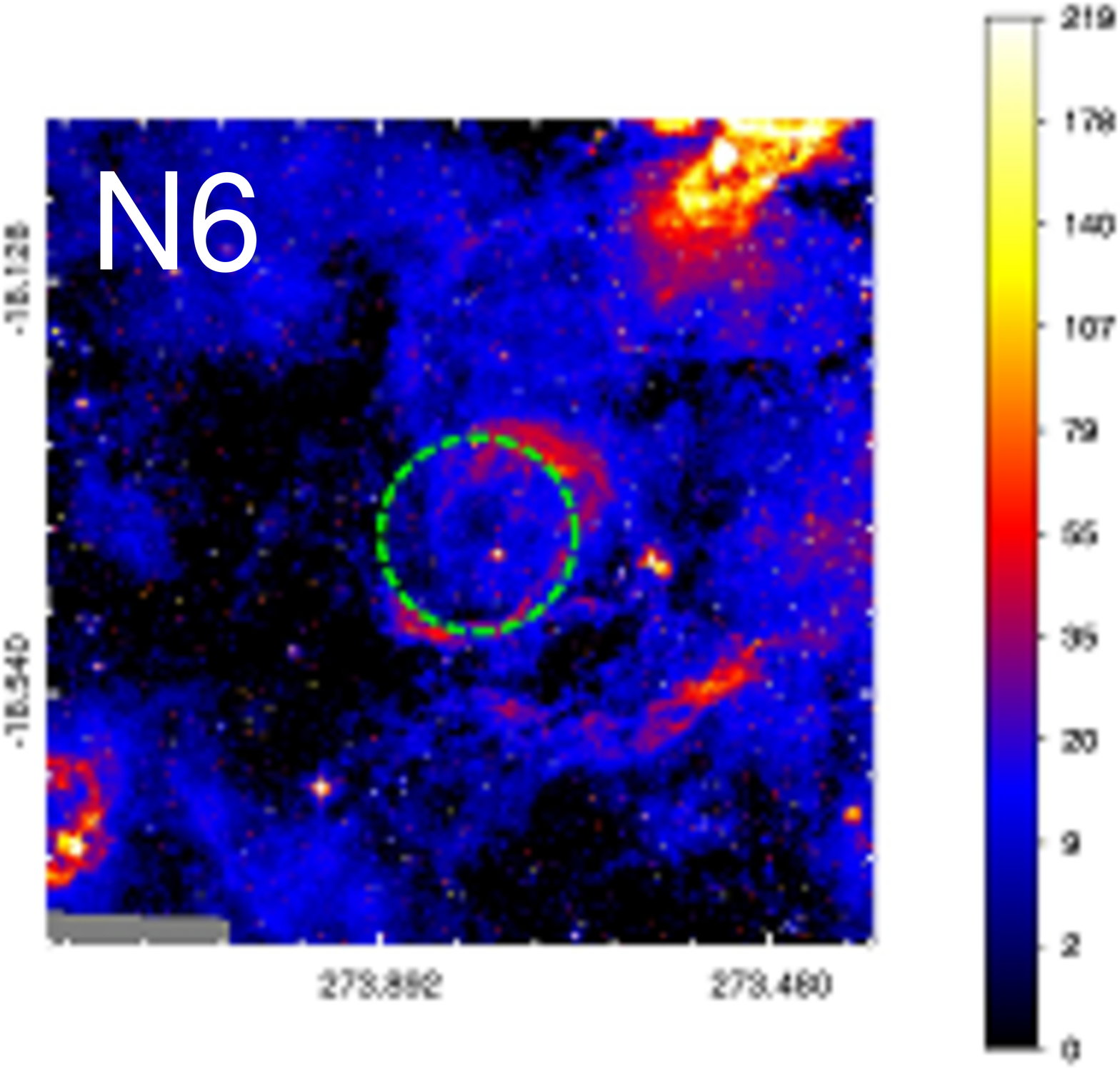}}}%
\mbox{\raisebox{0mm}{\includegraphics[width=40mm]{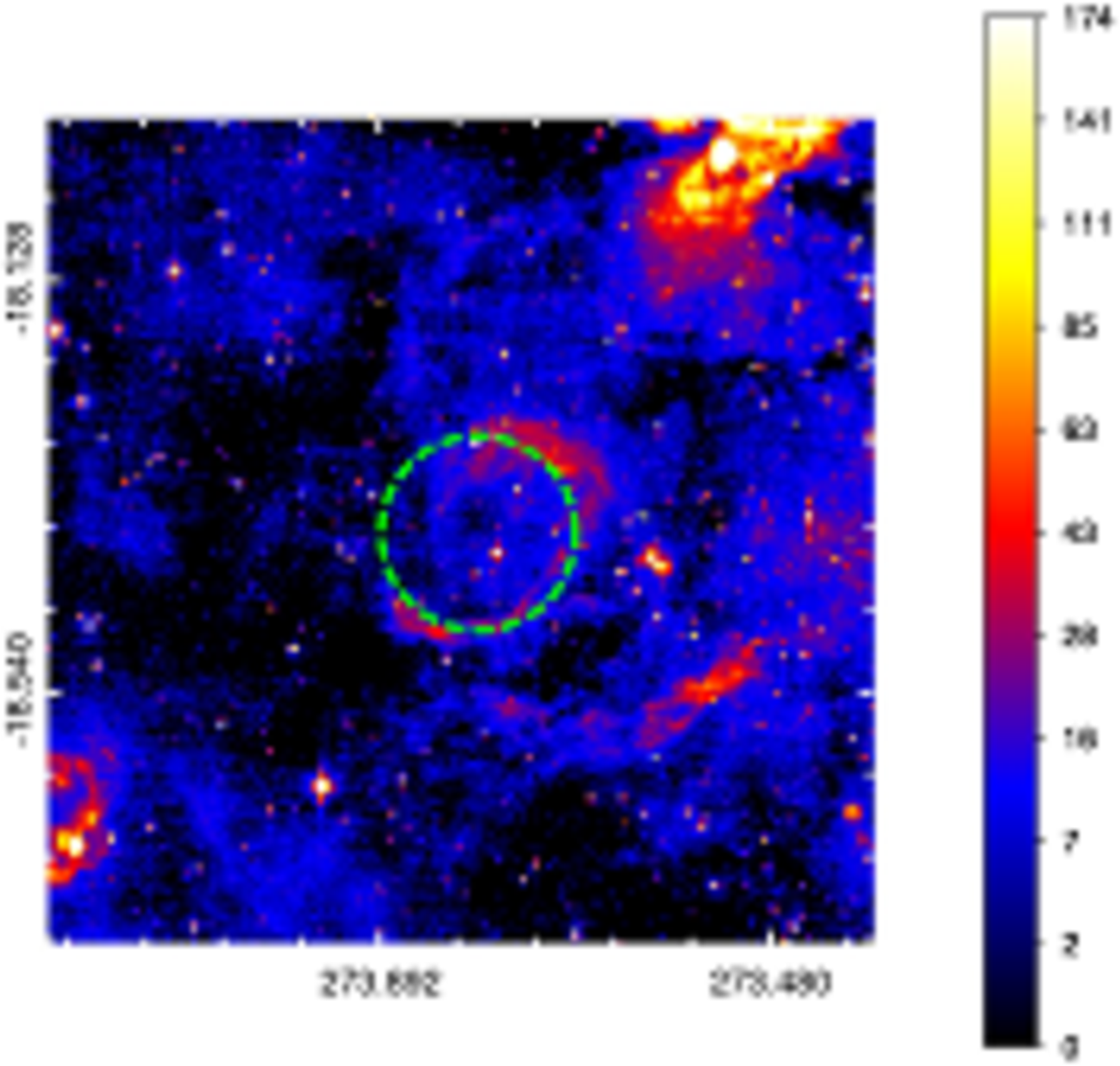}}}%
\mbox{\raisebox{0mm}{\includegraphics[width=40mm]{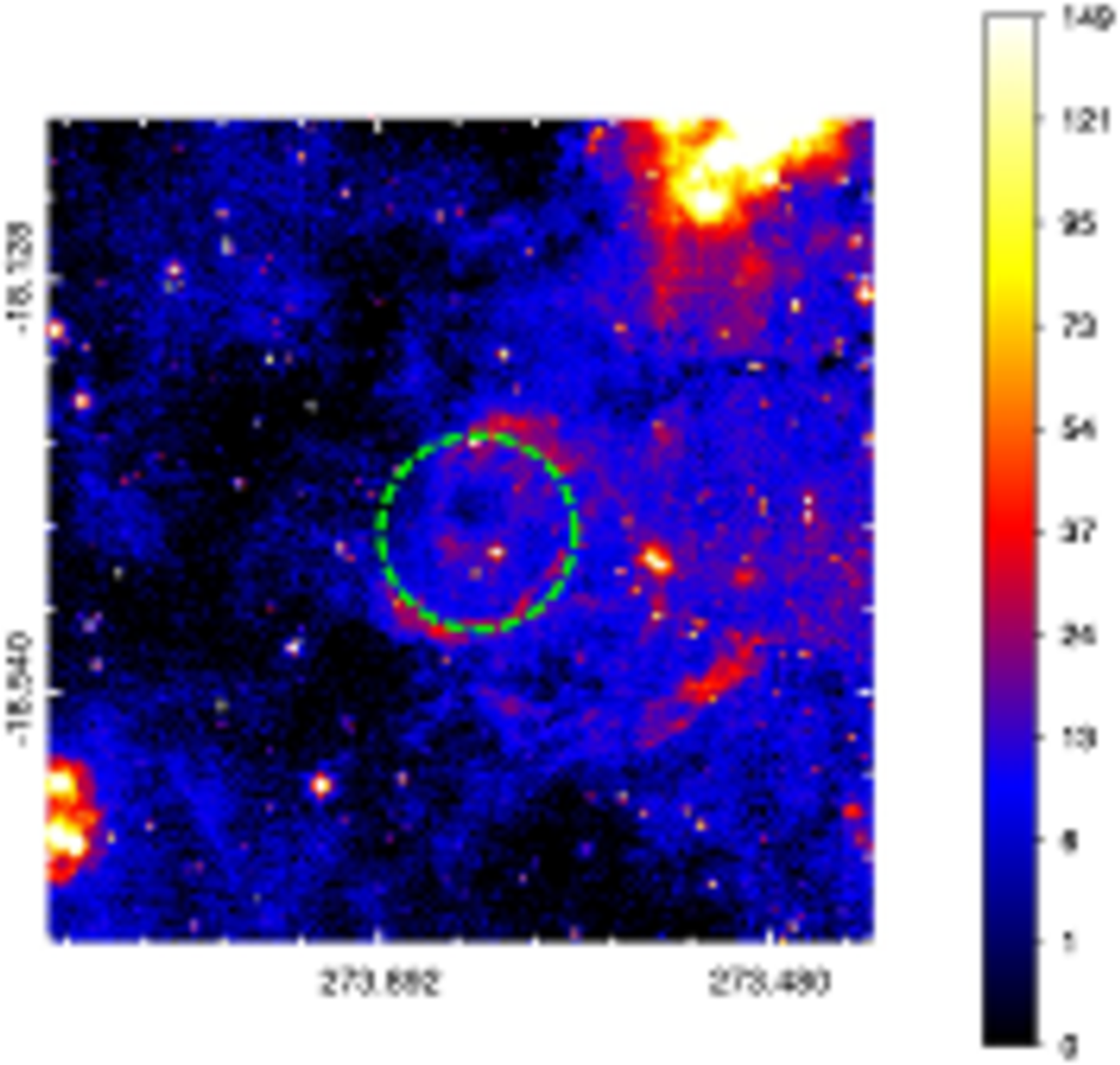}}}%
\mbox{\raisebox{0mm}{\includegraphics[width=40mm]{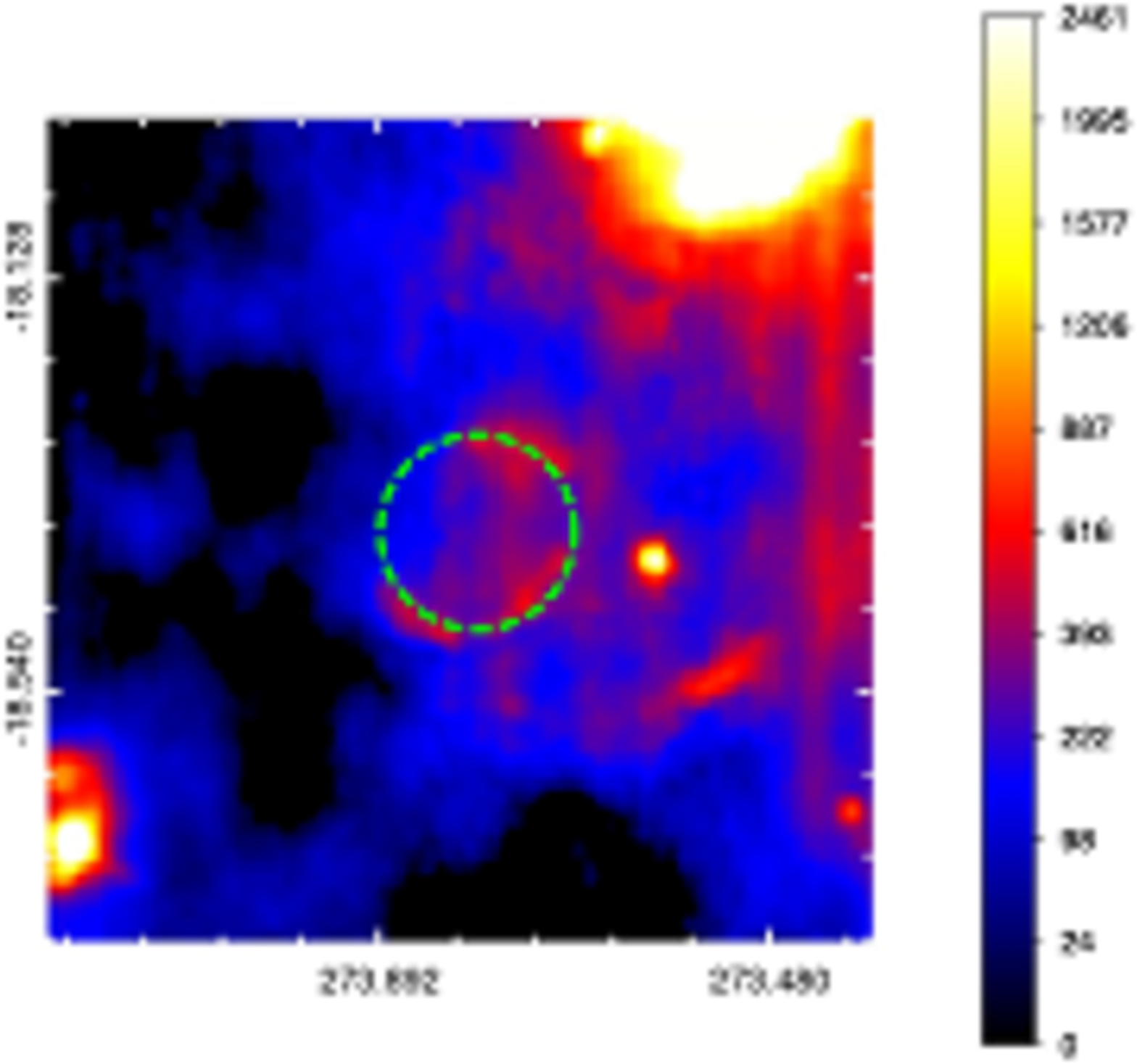}}}%
}
\subfigure{
\mbox{\raisebox{6mm}{\rotatebox{90}{\small{DEC (J2000)}}}}%
\mbox{\raisebox{0mm}{\includegraphics[width=40mm]{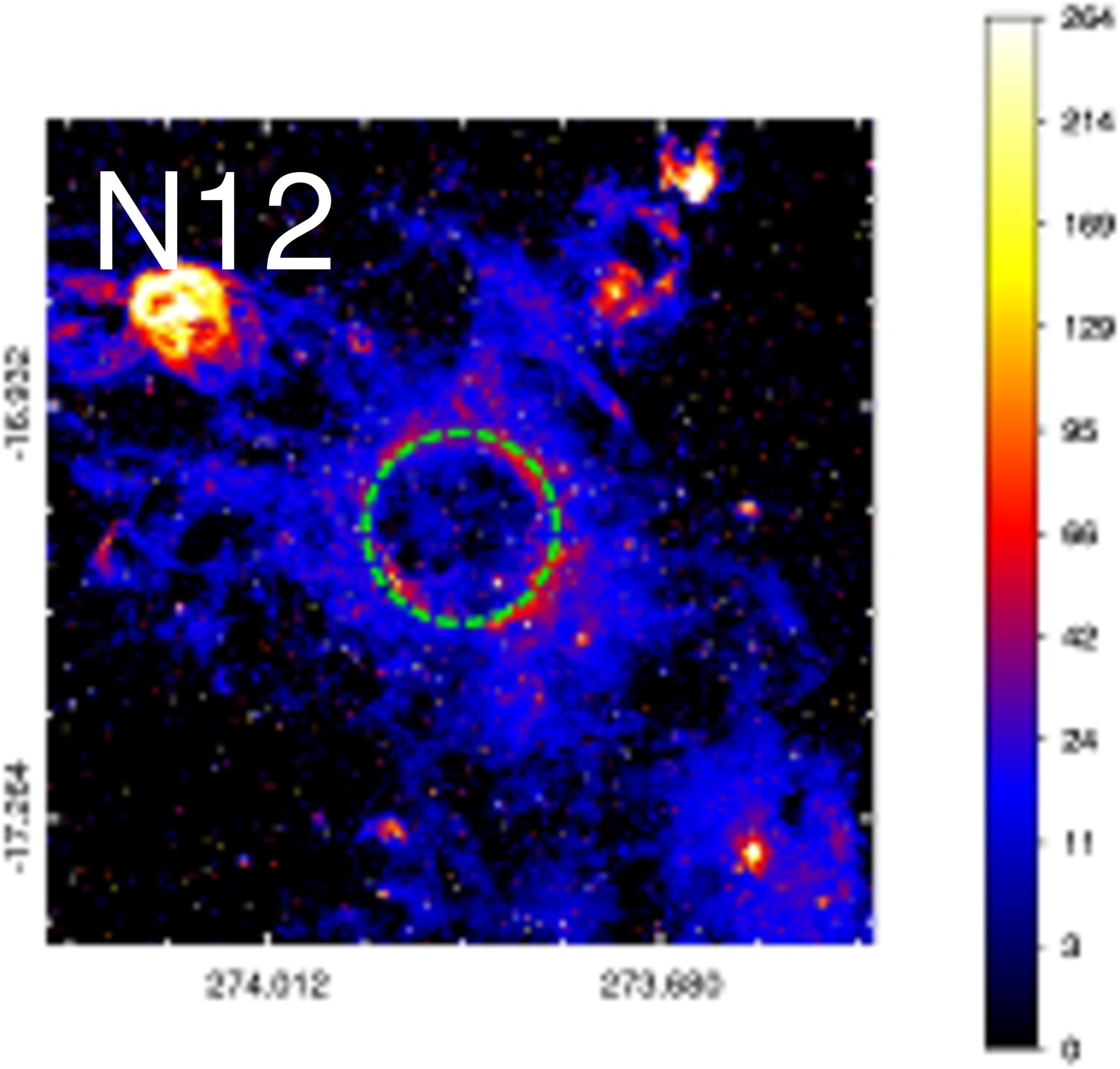}}}%
\mbox{\raisebox{0mm}{\includegraphics[width=40mm]{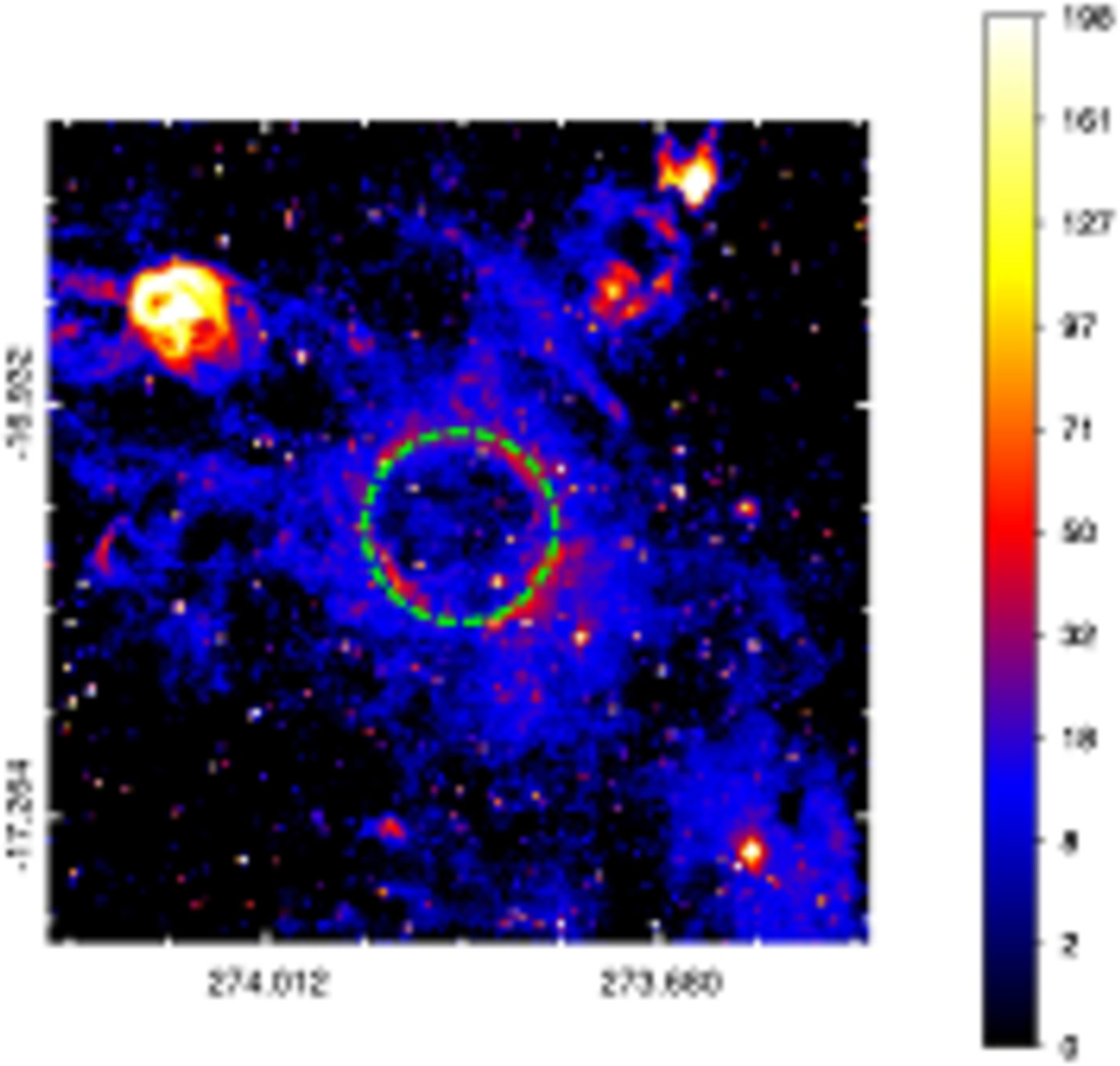}}}%
\mbox{\raisebox{0mm}{\includegraphics[width=40mm]{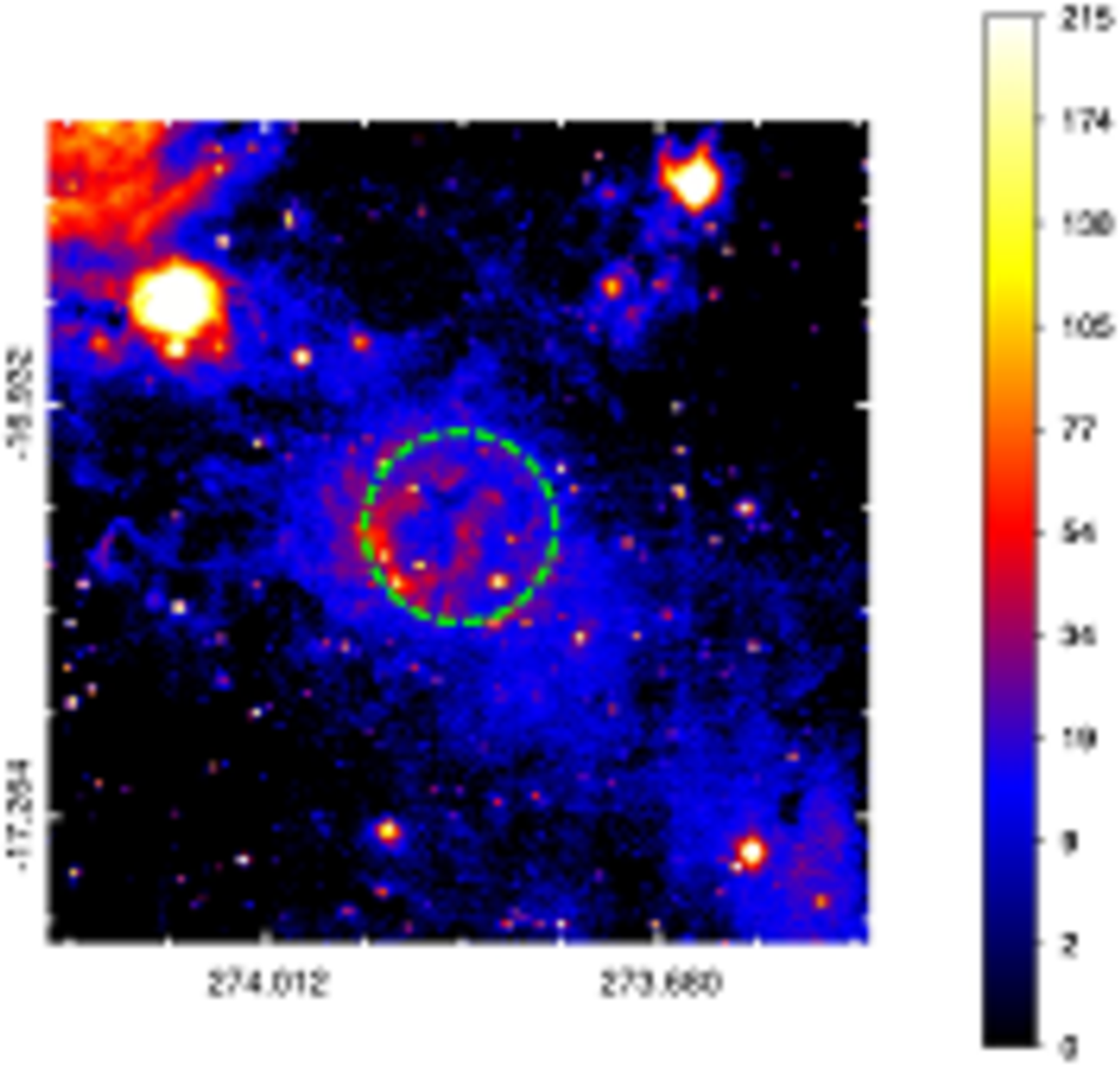}}}%
\mbox{\raisebox{0mm}{\includegraphics[width=40mm]{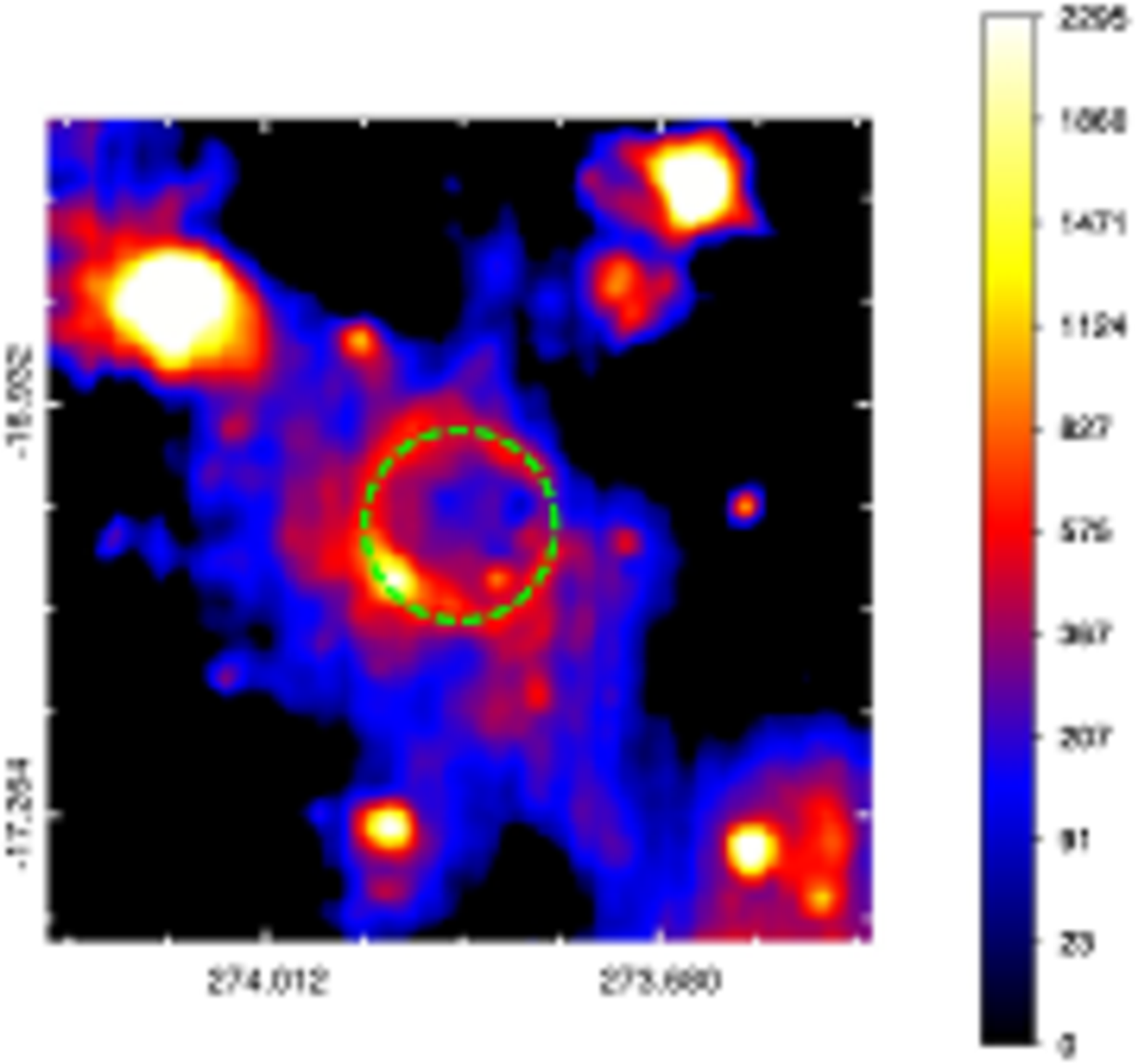}}}%
}
\subfigure{
\mbox{\raisebox{6mm}{\rotatebox{90}{\small{DEC (J2000)}}}}%
\mbox{\raisebox{0mm}{\includegraphics[width=40mm]{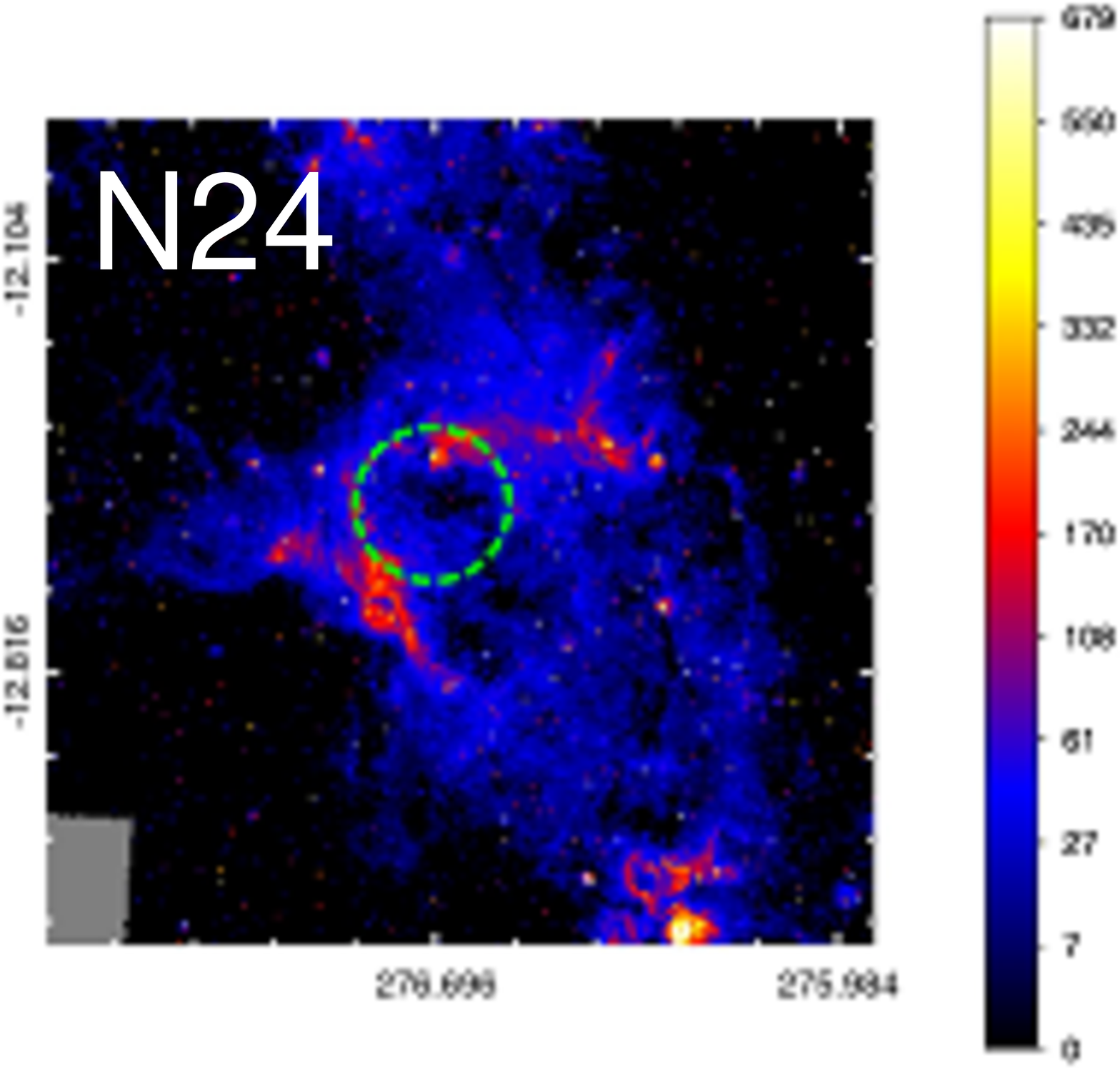}}}%
\mbox{\raisebox{0mm}{\includegraphics[width=40mm]{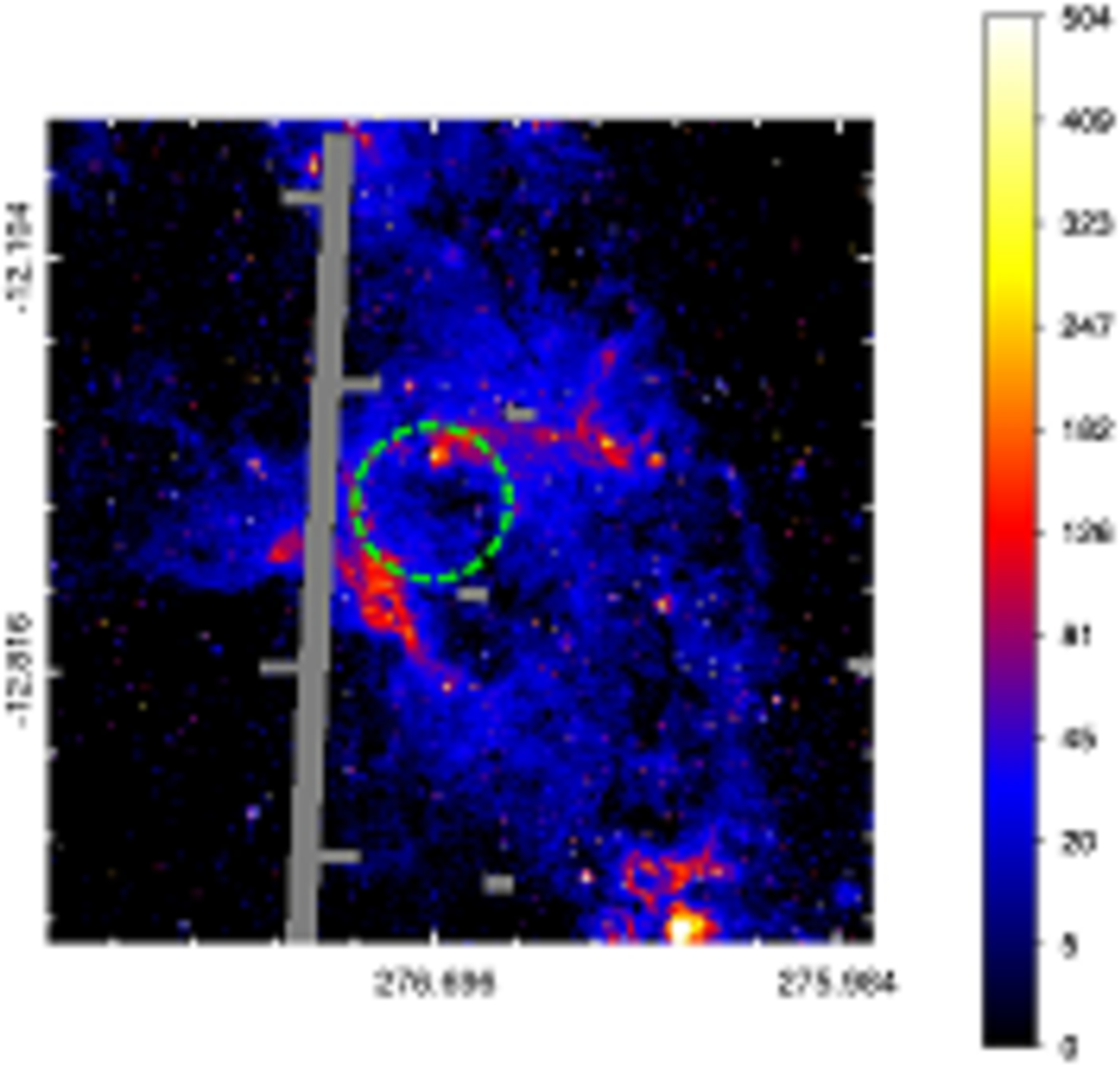}}}%
\mbox{\raisebox{0mm}{\includegraphics[width=40mm]{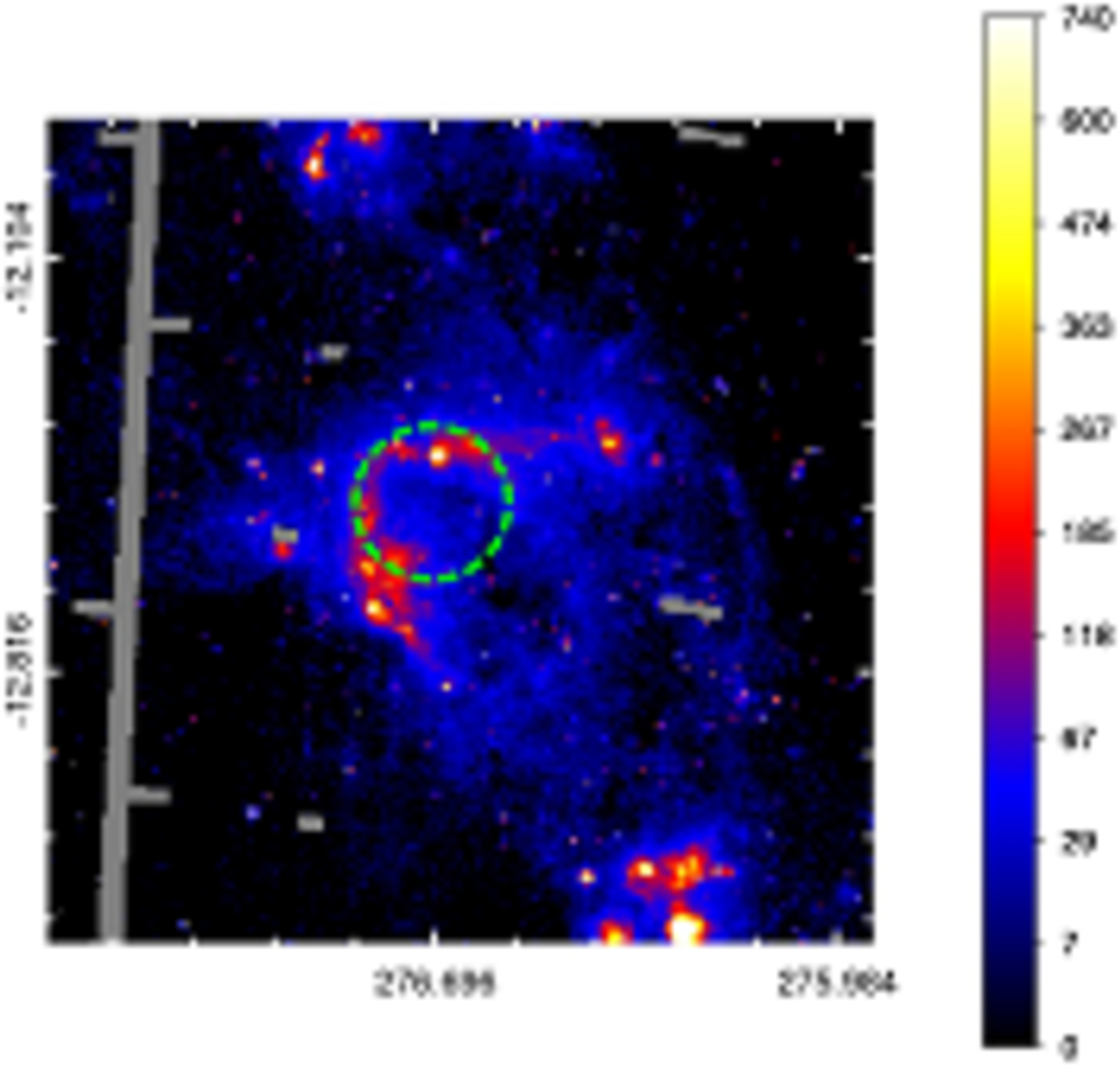}}}%
\mbox{\raisebox{0mm}{\includegraphics[width=40mm]{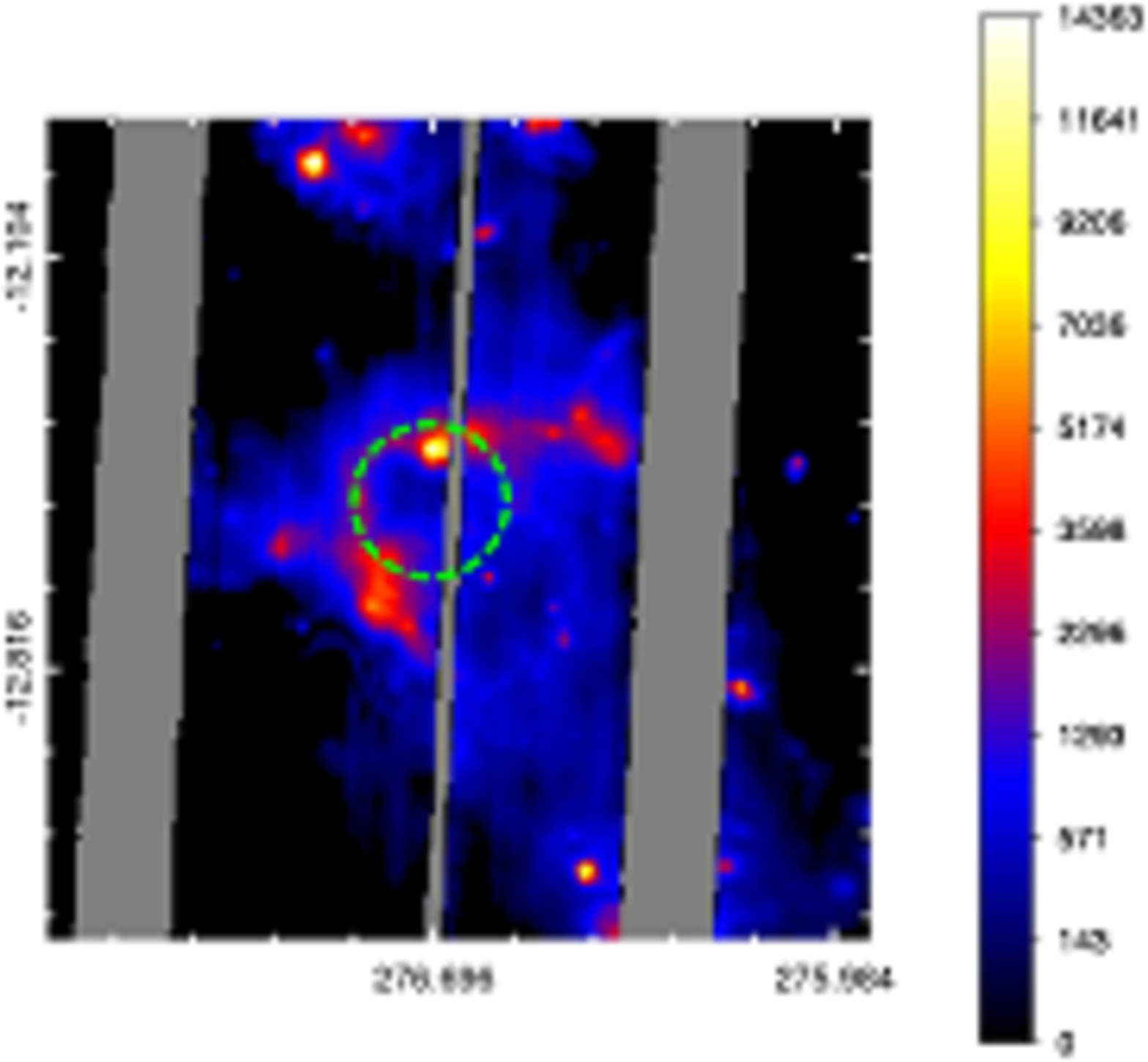}}}%
}
\subfigure{
\mbox{\raisebox{6mm}{\rotatebox{90}{\small{DEC (J2000)}}}}%
\mbox{\raisebox{0mm}{\includegraphics[width=40mm]{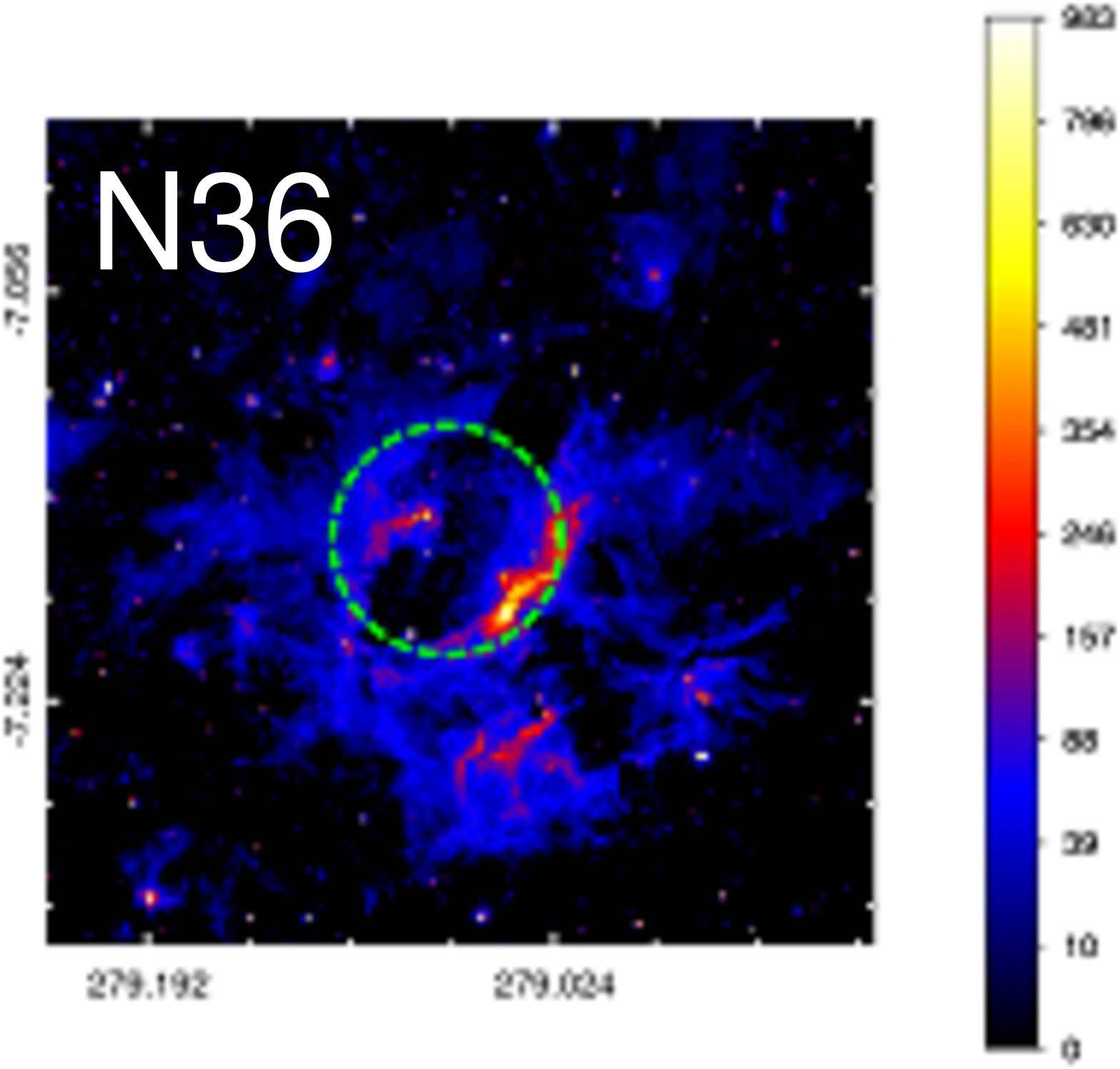}}}%
\mbox{\raisebox{0mm}{\includegraphics[width=40mm]{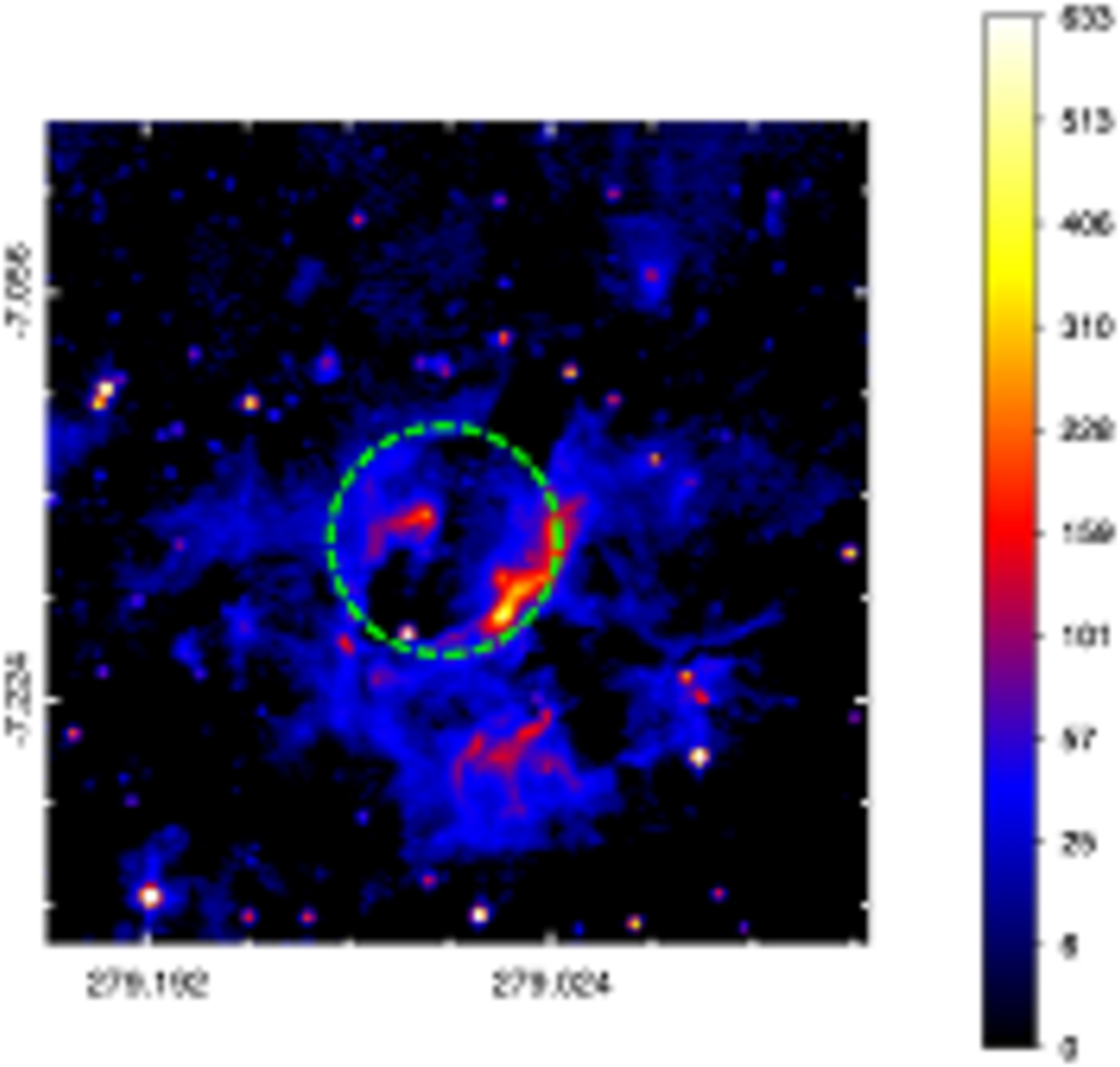}}}%
\mbox{\raisebox{0mm}{\includegraphics[width=40mm]{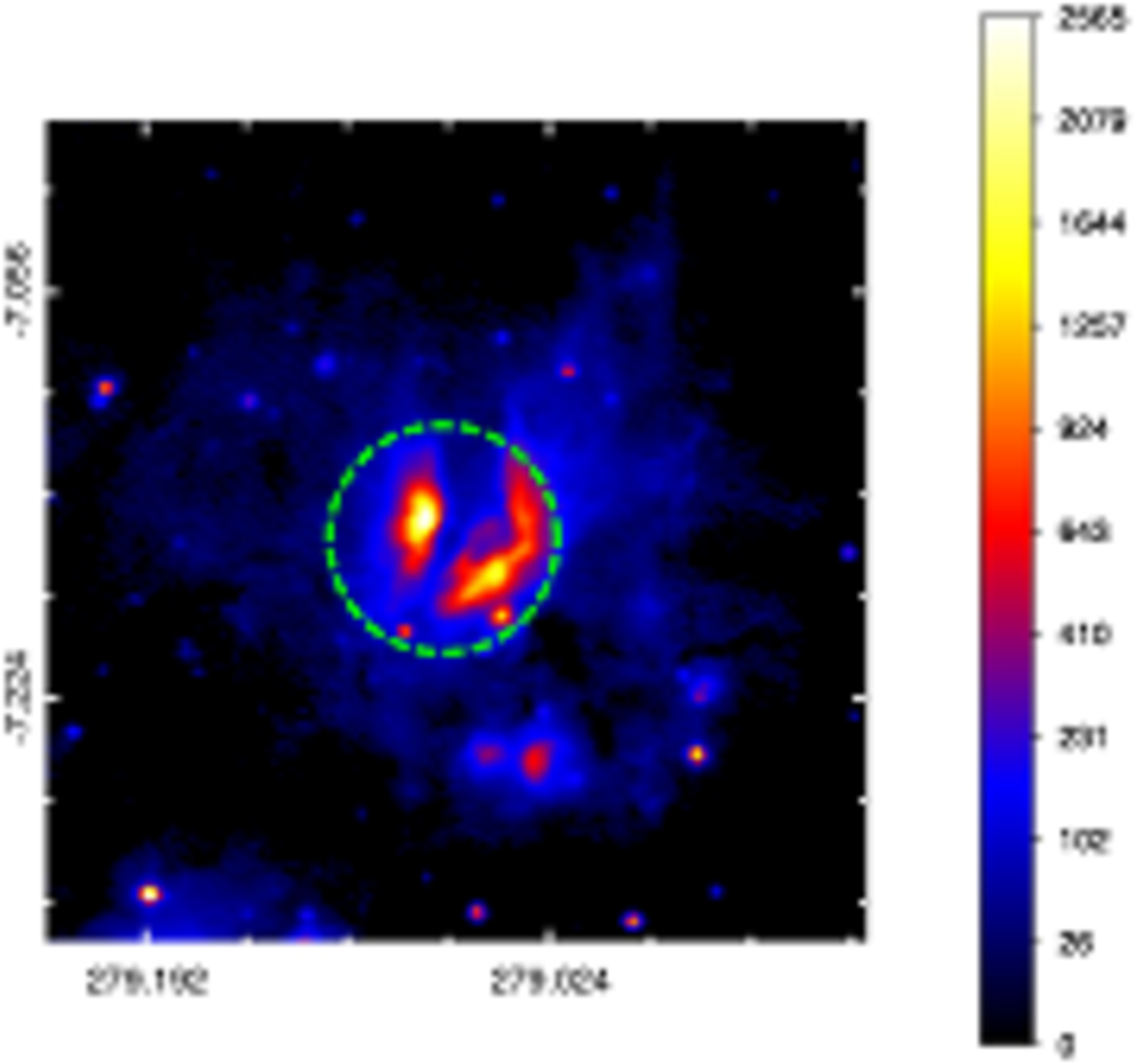}}}%
\mbox{\raisebox{0mm}{\includegraphics[width=40mm]{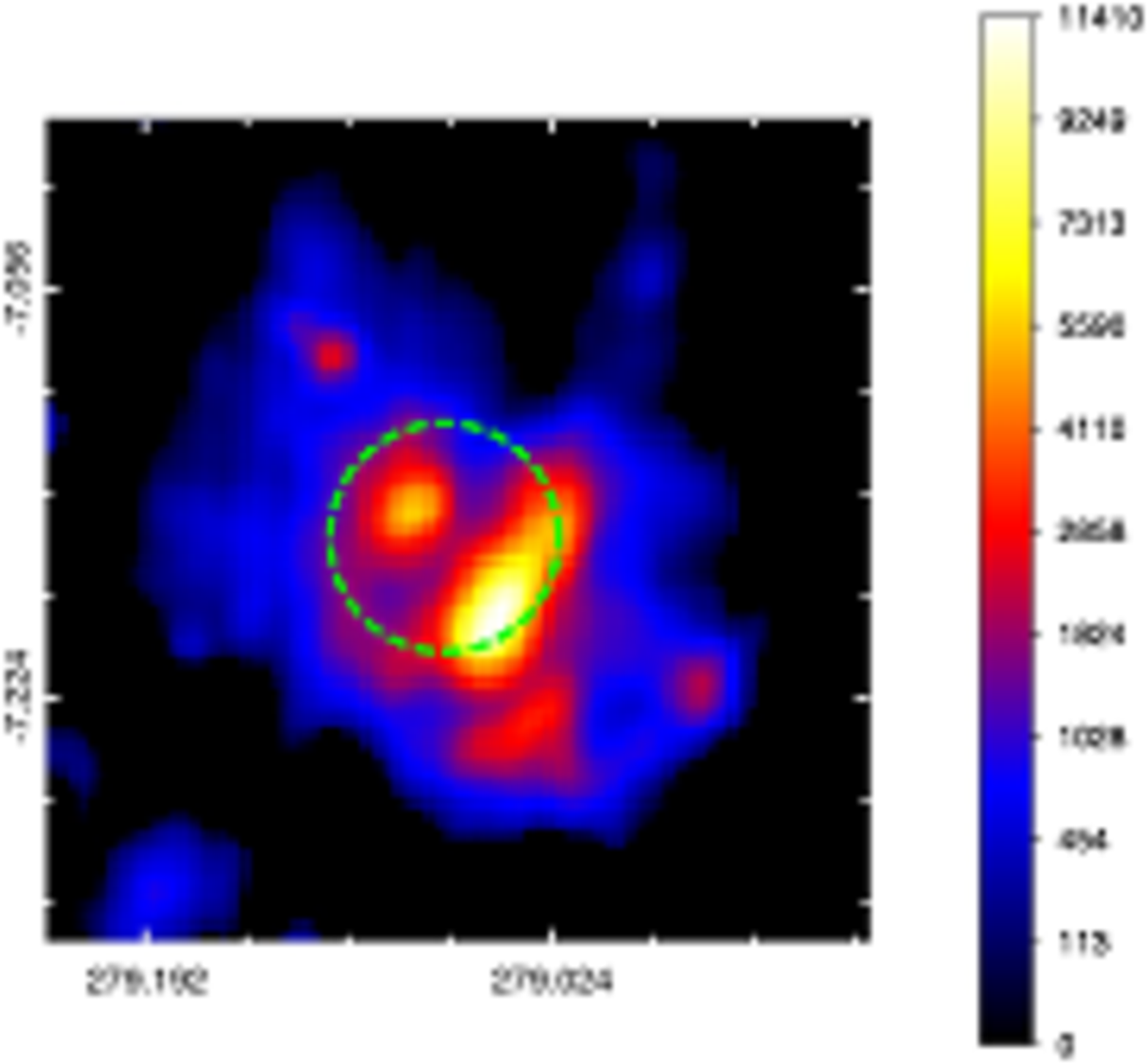}}}%
}
\subfigure{
\mbox{\raisebox{6mm}{\rotatebox{90}{\small{DEC (J2000)}}}}%
\mbox{\raisebox{0mm}{\includegraphics[width=40mm]{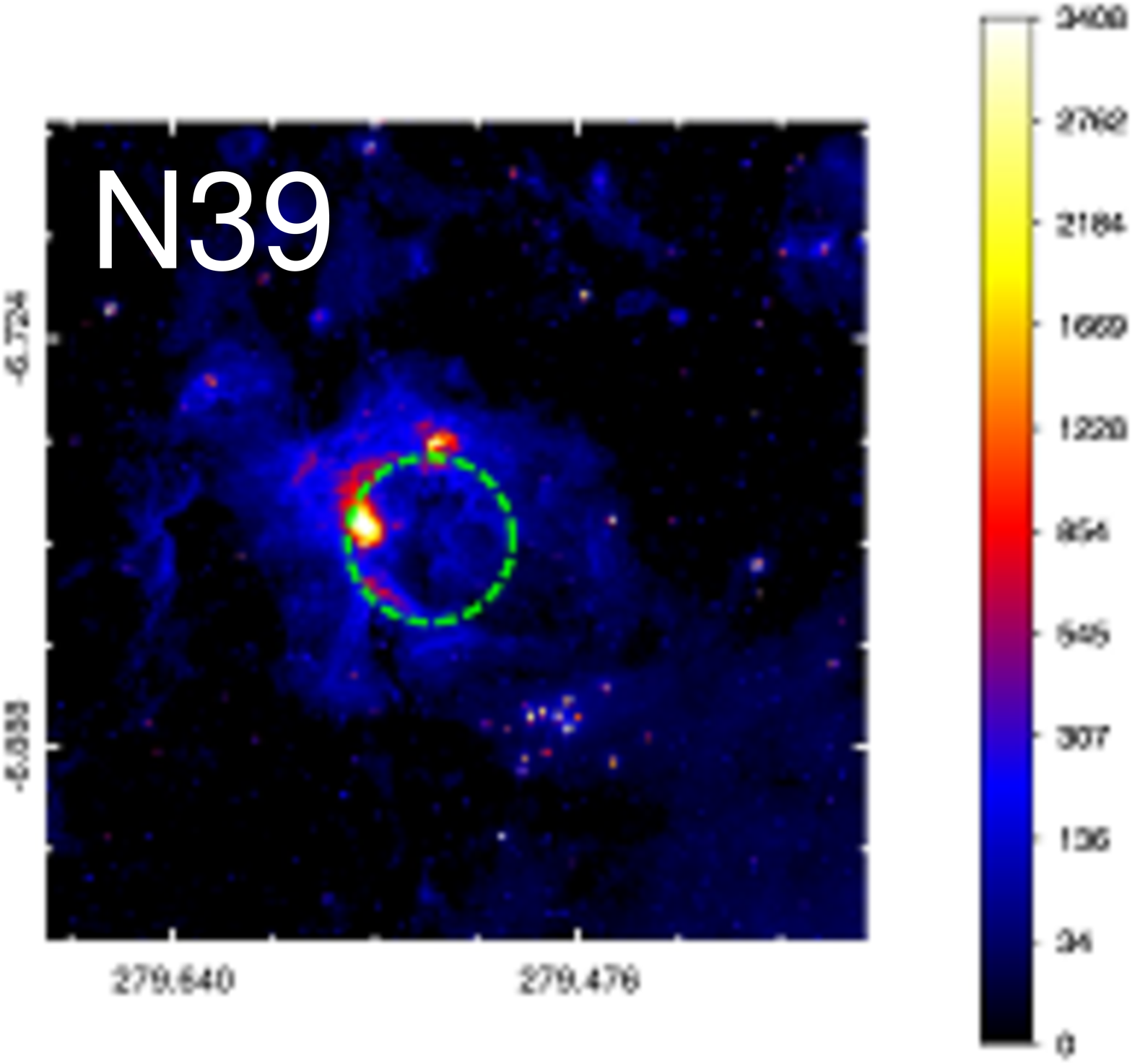}}}%
\mbox{\raisebox{0mm}{\includegraphics[width=40mm]{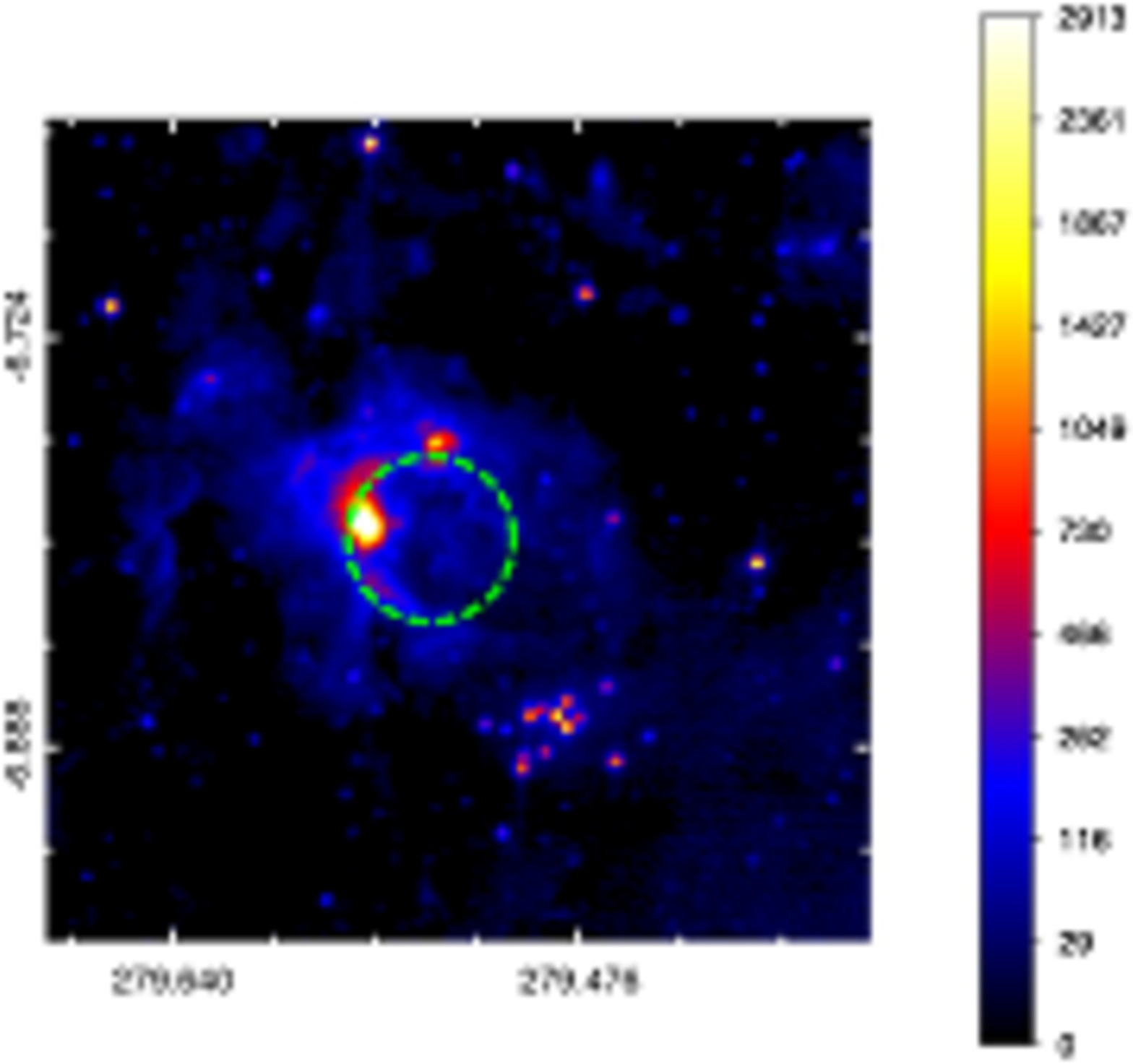}}}%
\mbox{\raisebox{0mm}{\includegraphics[width=40mm]{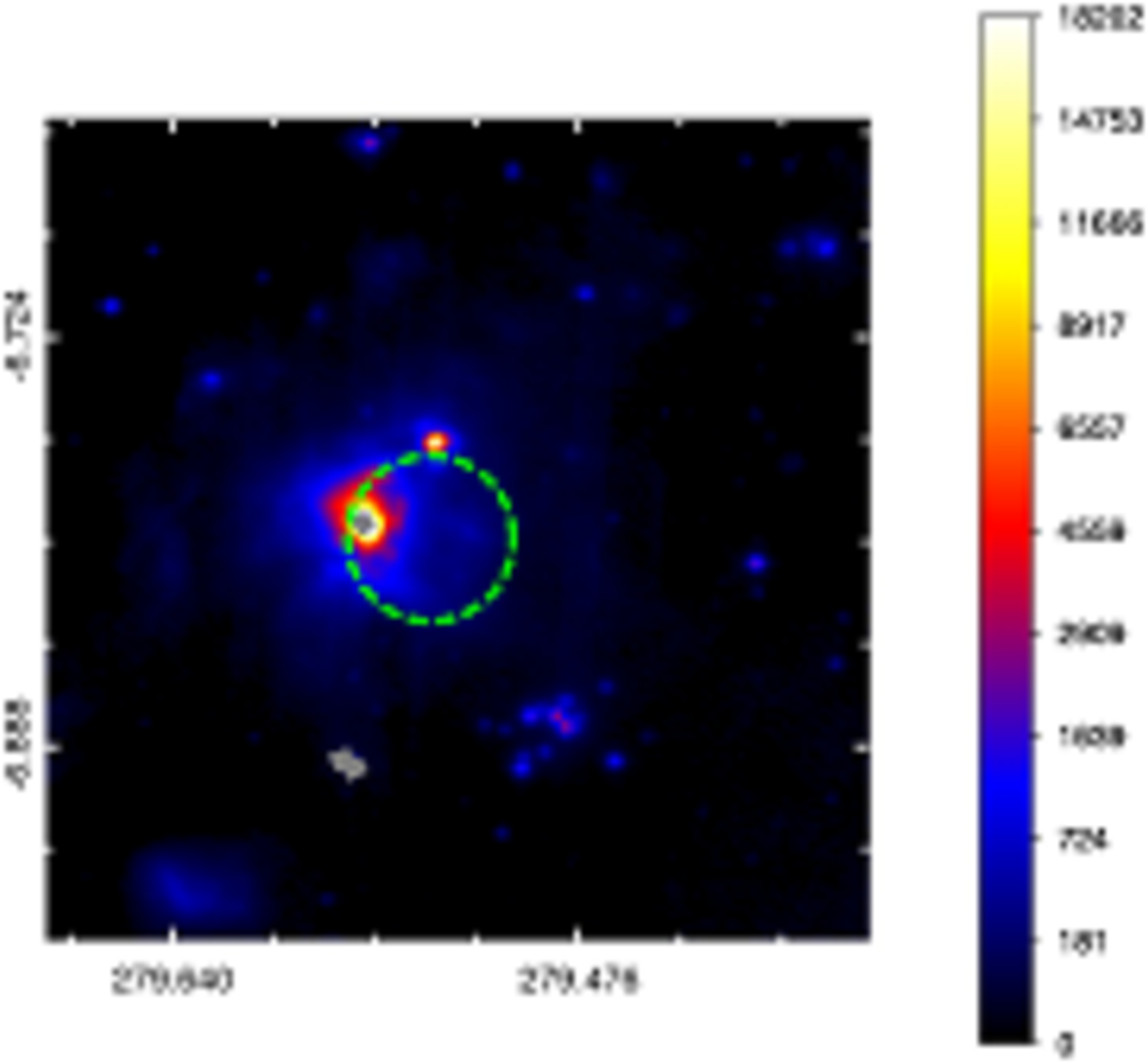}}}%
\mbox{\raisebox{0mm}{\includegraphics[width=40mm]{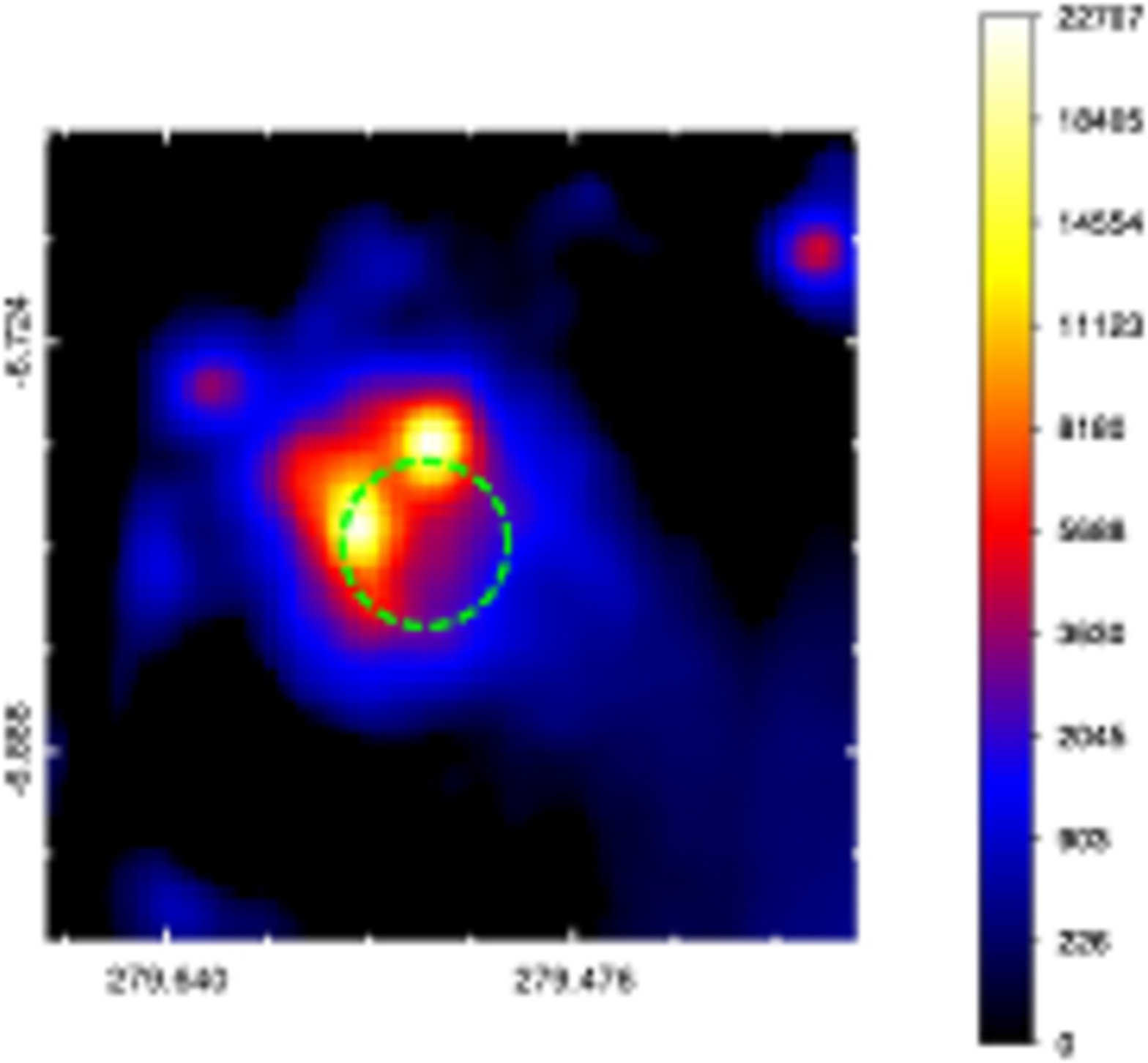}}}%
}
\caption{Continued.} \label{fig:Introfig1:b}
\end{figure*}

\addtocounter{figure}{-1}
\begin{figure*}[ht]
\addtocounter{subfigure}{1}
\centering
\subfigure{
\makebox[180mm][l]{\raisebox{0mm}[0mm][0mm]{ \hspace{15mm} \small{8 \mic}} \hspace{29.5mm} \small{9 \mic} \hspace{27mm} \small{18 \mic} \hspace{26.5mm} \small{90 \mic}}%
}
\subfigure{
\makebox[180mm][l]{\raisebox{0mm}[0mm][0mm]{ \hspace{11mm} \small{RA (J2000)}} \hspace{19.5mm} \small{RA (J2000)} \hspace{20mm} \small{RA (J2000)} \hspace{20mm} \small{RA (J2000)}}%
}
\subfigure{
\mbox{\raisebox{6mm}{\rotatebox{90}{\small{DEC (J2000)}}}}%
\mbox{\raisebox{0mm}{\includegraphics[width=40mm]{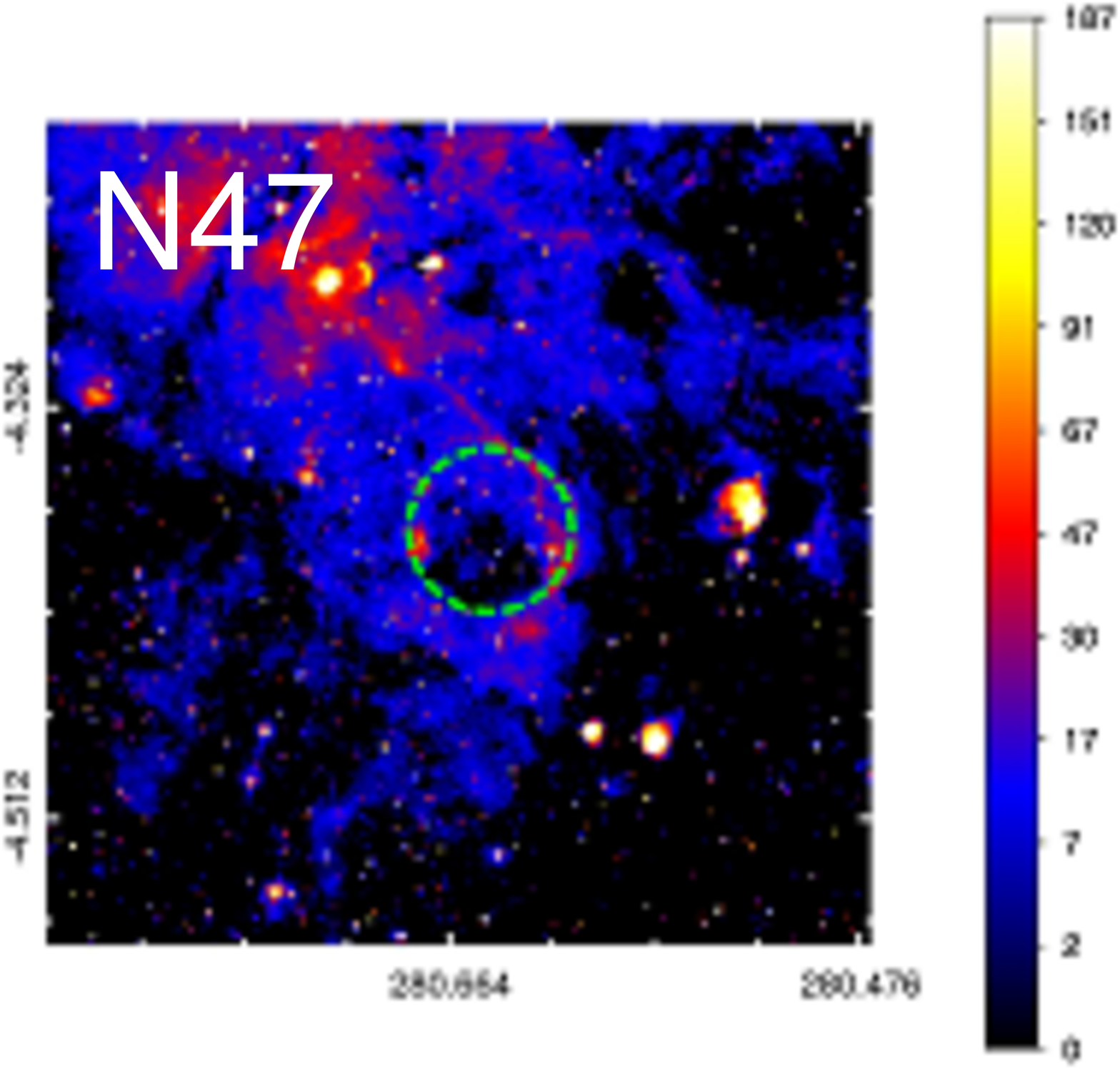}}}%
\mbox{\raisebox{0mm}{\includegraphics[width=40mm]{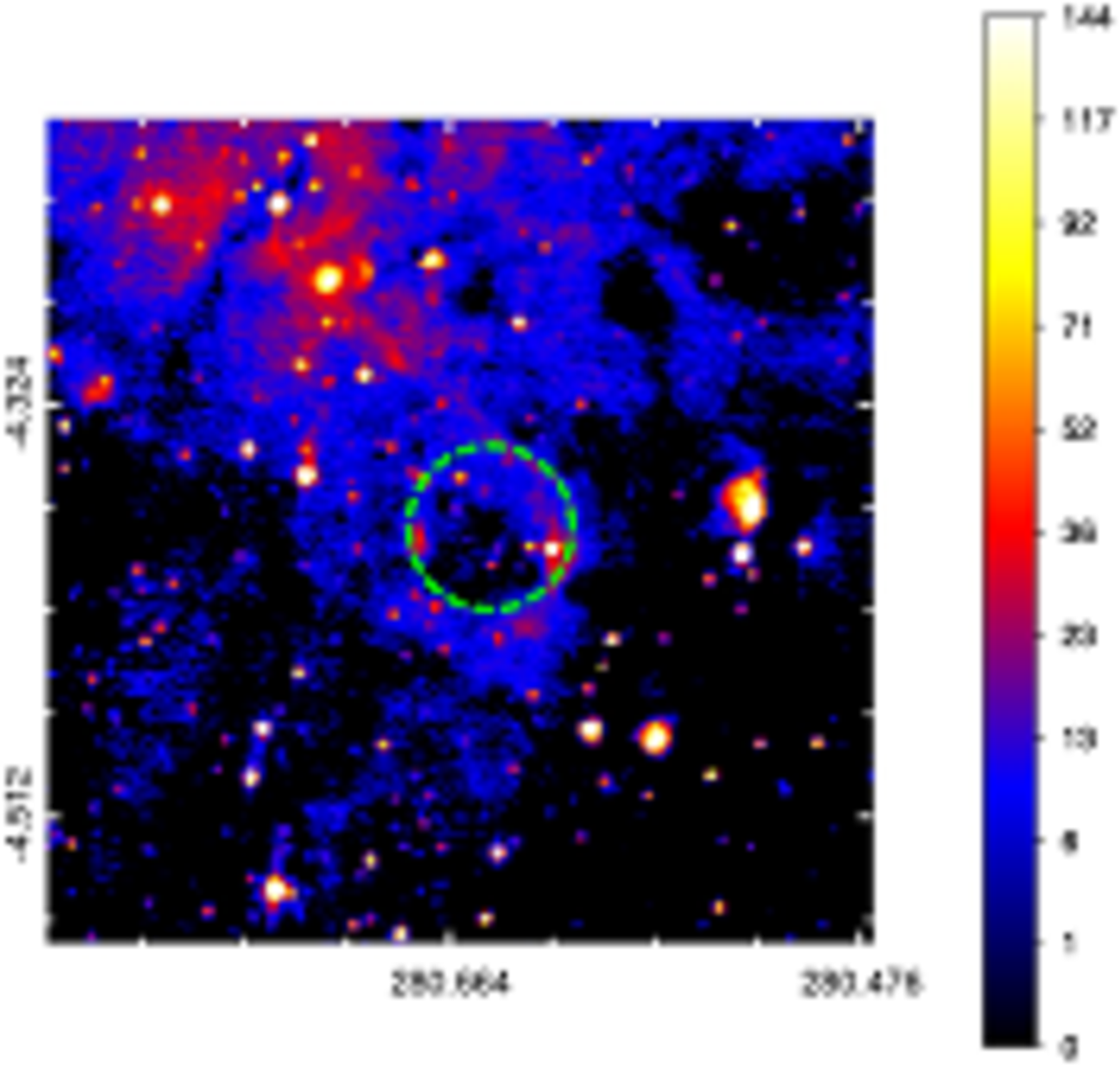}}}%
\mbox{\raisebox{0mm}{\includegraphics[width=40mm]{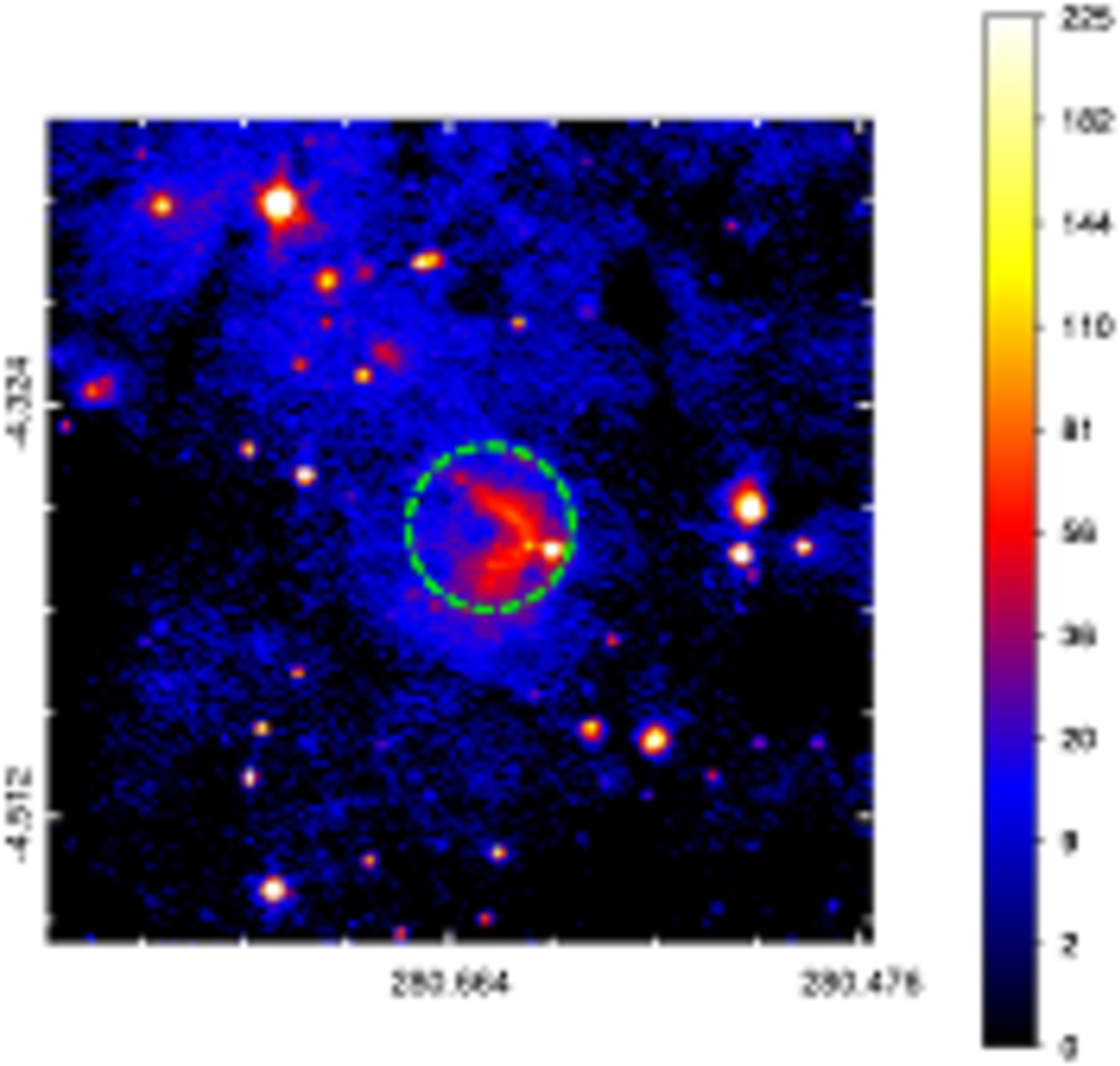}}}%
\mbox{\raisebox{0mm}{\includegraphics[width=40mm]{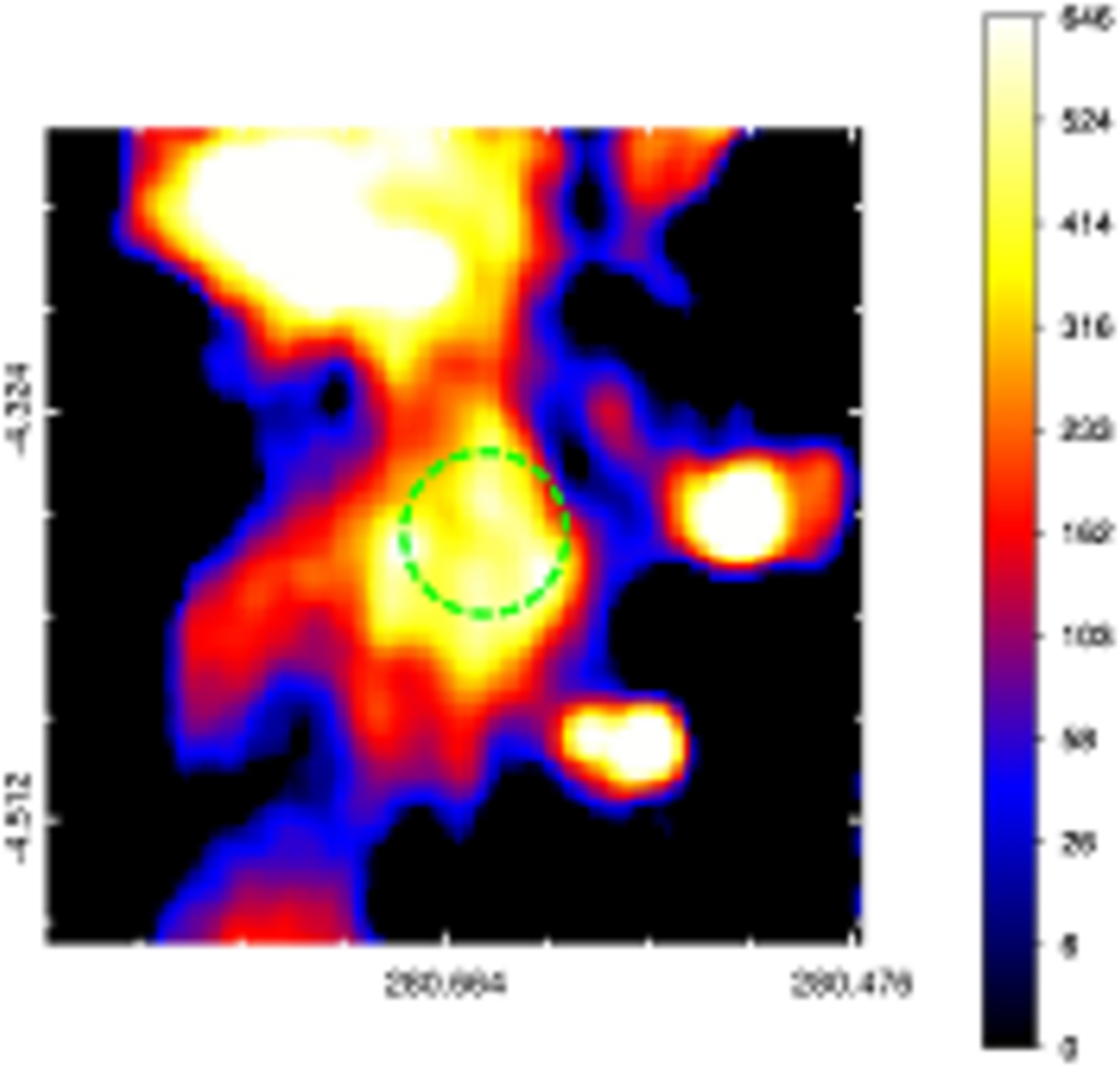}}}%
}
\subfigure{
\mbox{\raisebox{6mm}{\rotatebox{90}{\small{DEC (J2000)}}}}%
\mbox{\raisebox{0mm}{\includegraphics[width=40mm]{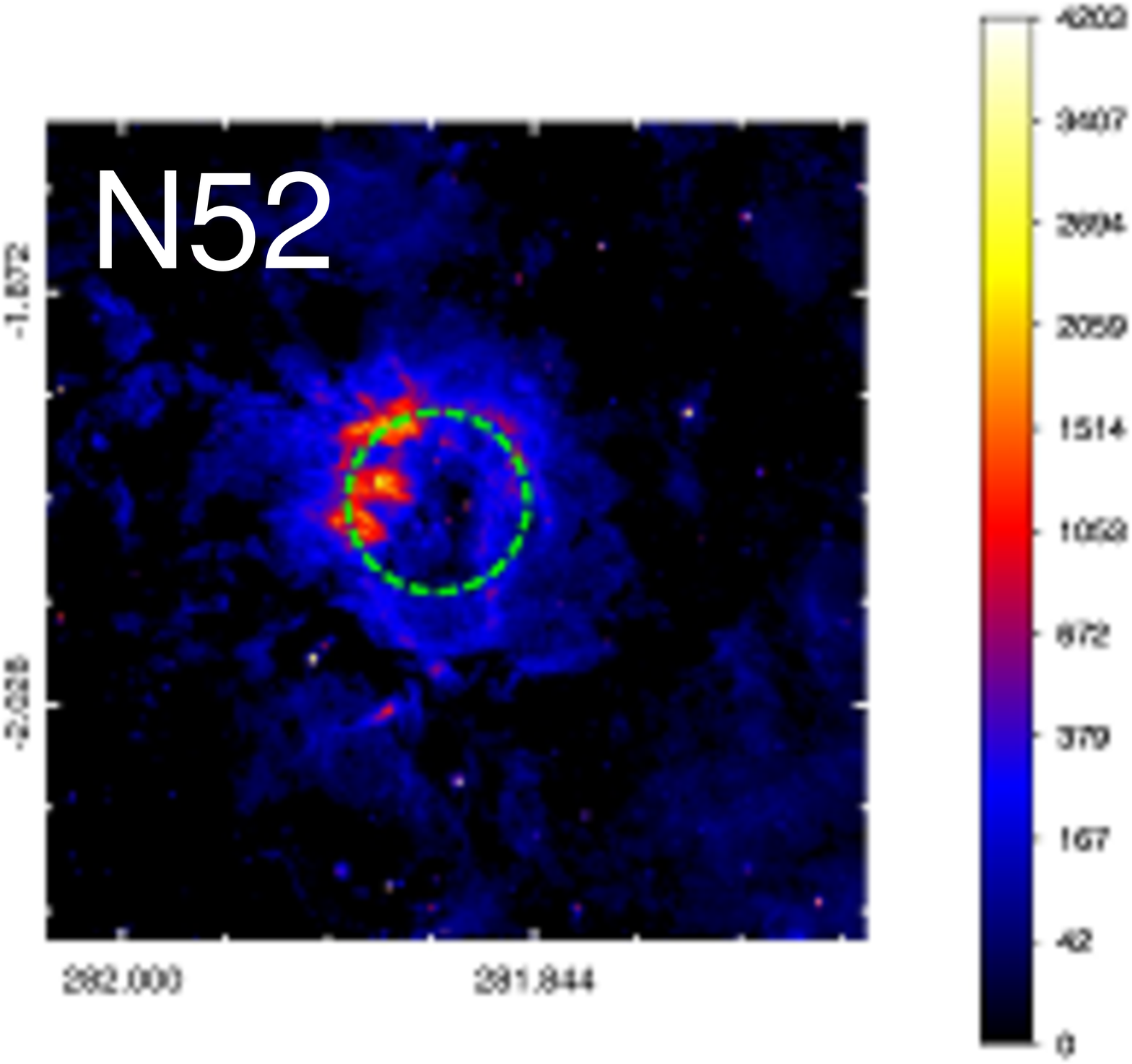}}}%
\mbox{\raisebox{0mm}{\includegraphics[width=40mm]{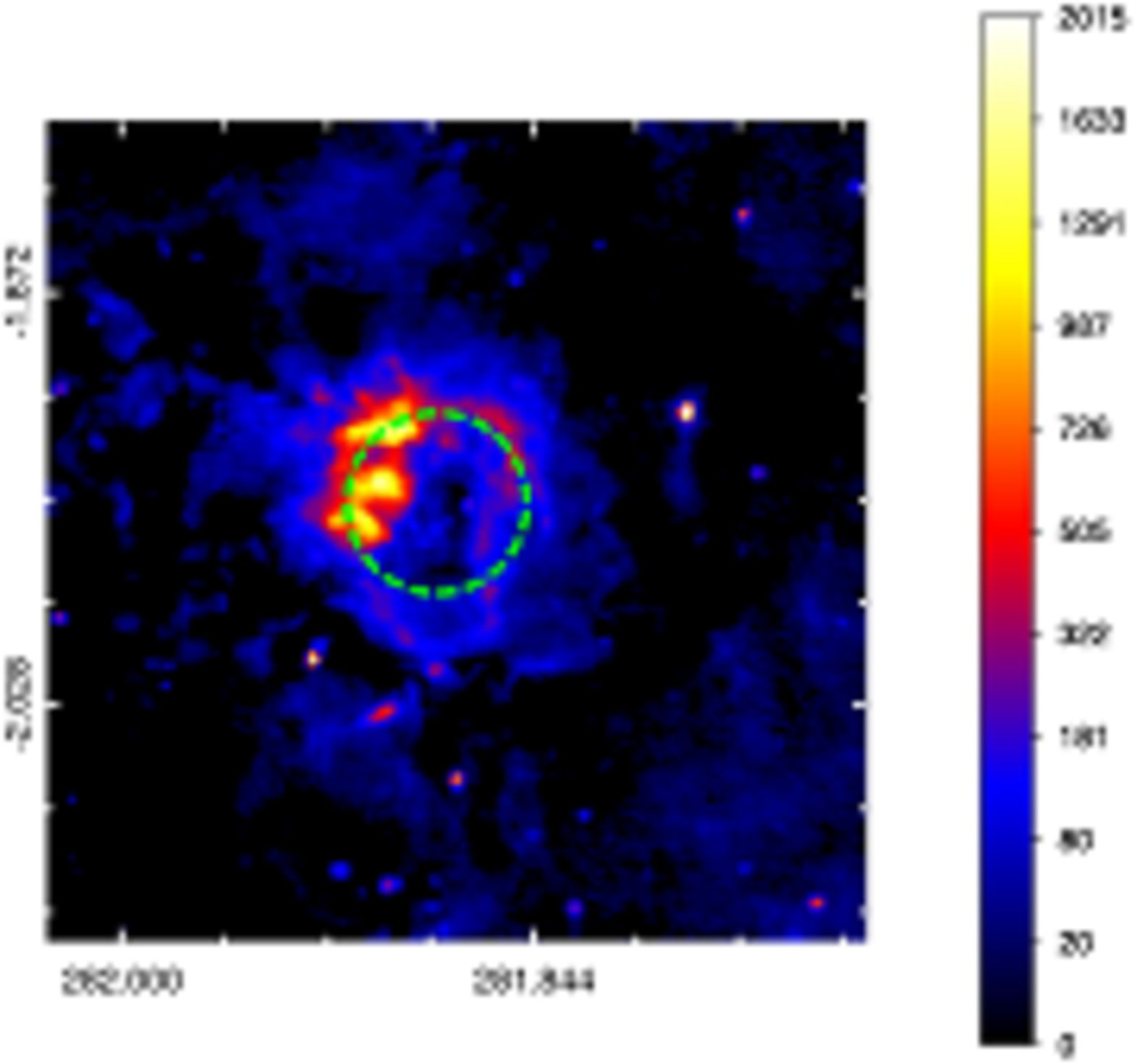}}}%
\mbox{\raisebox{0mm}{\includegraphics[width=40mm]{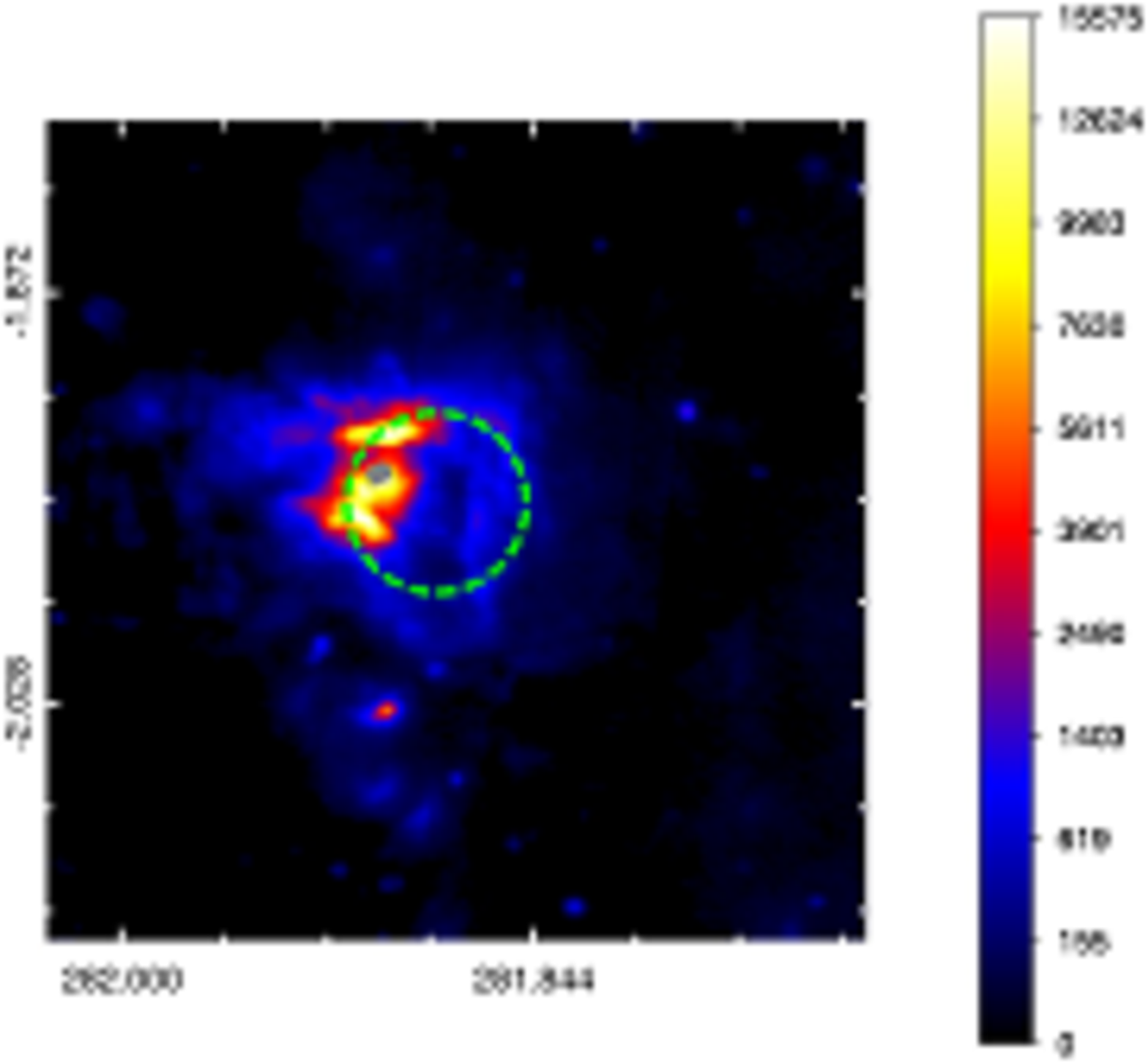}}}%
\mbox{\raisebox{0mm}{\includegraphics[width=40mm]{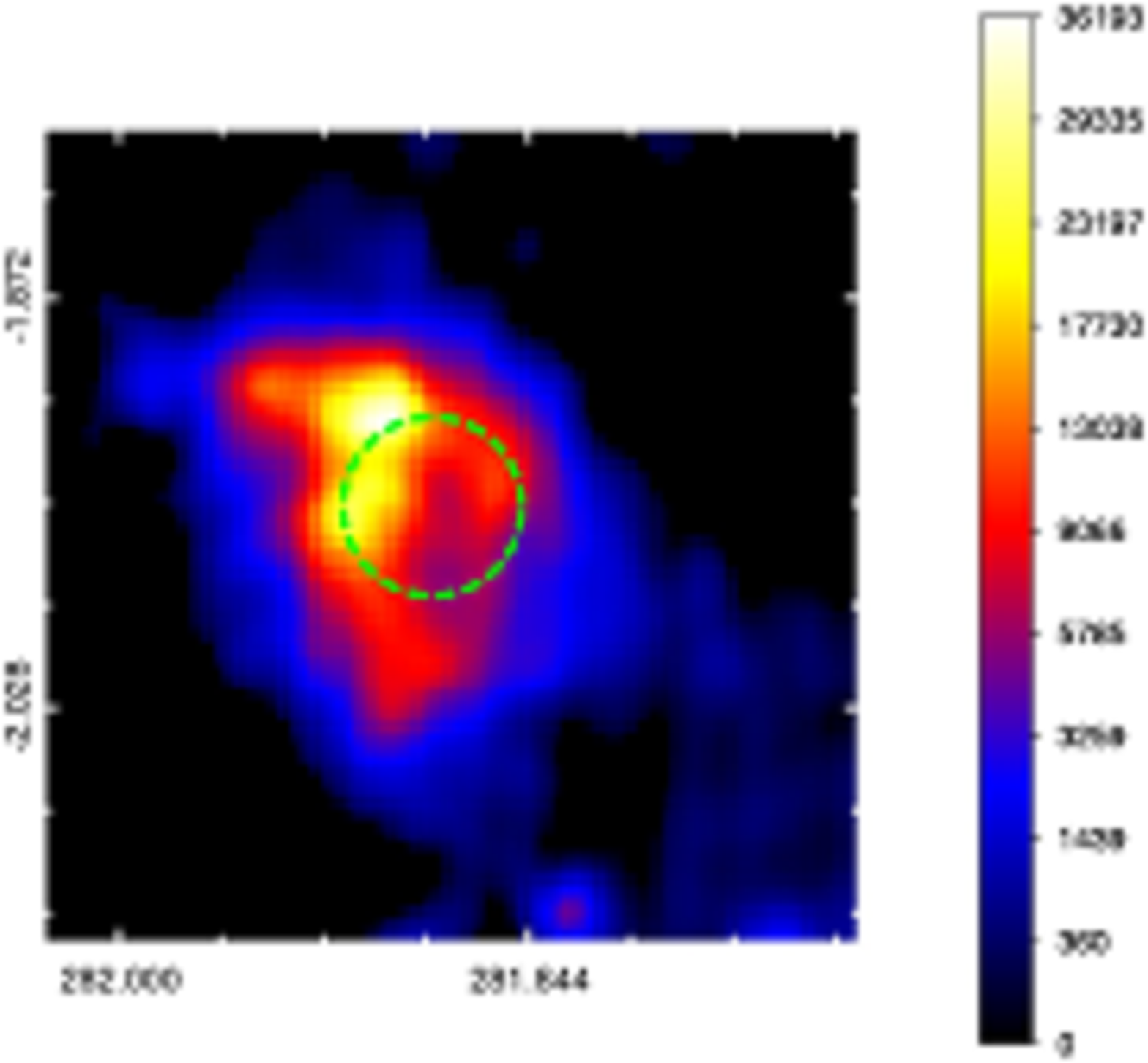}}}%
}
\subfigure{
\mbox{\raisebox{6mm}{\rotatebox{90}{\small{DEC (J2000)}}}}%
\mbox{\raisebox{0mm}{\includegraphics[width=40mm]{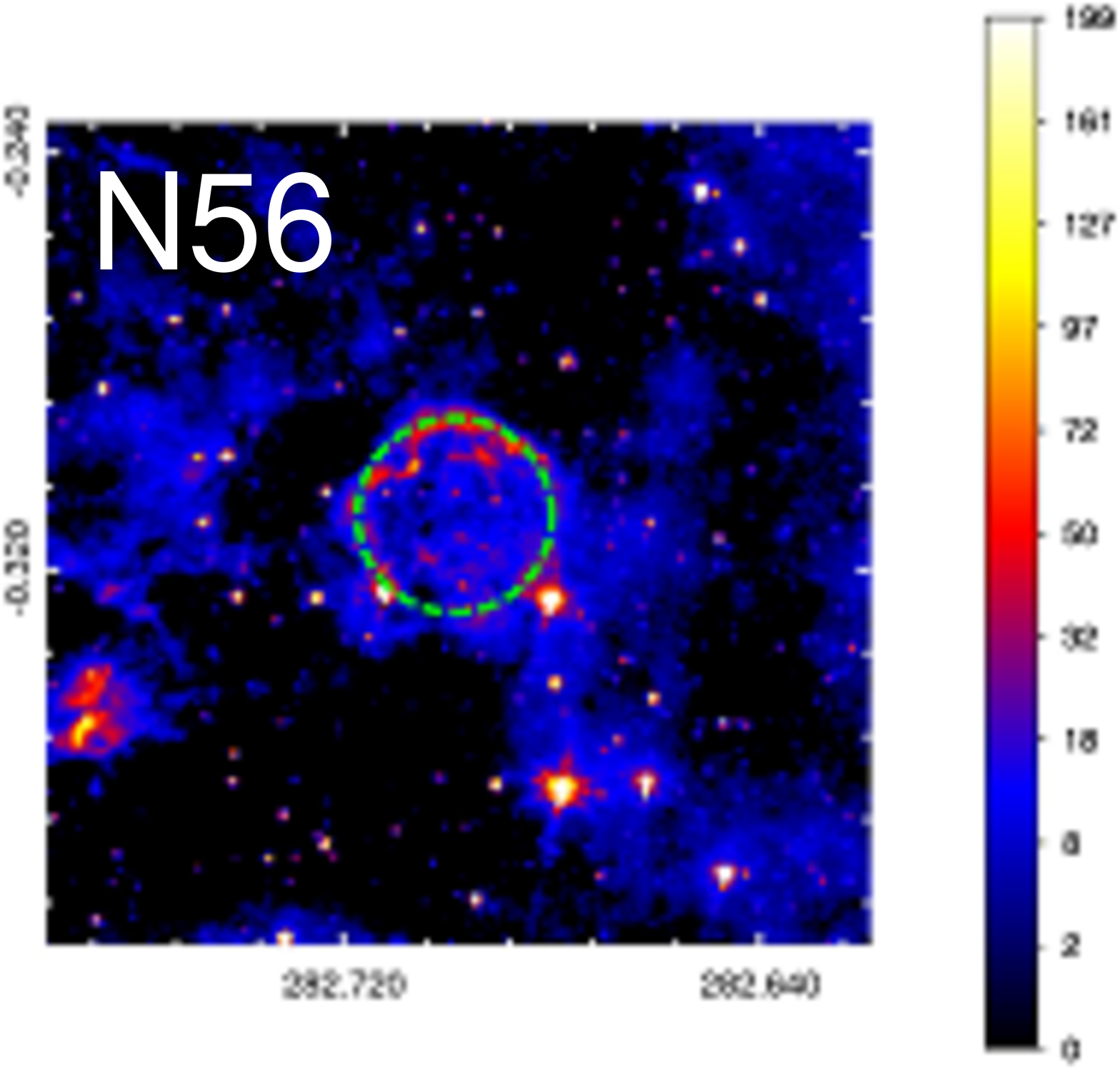}}}%
\mbox{\raisebox{0mm}{\includegraphics[width=40mm]{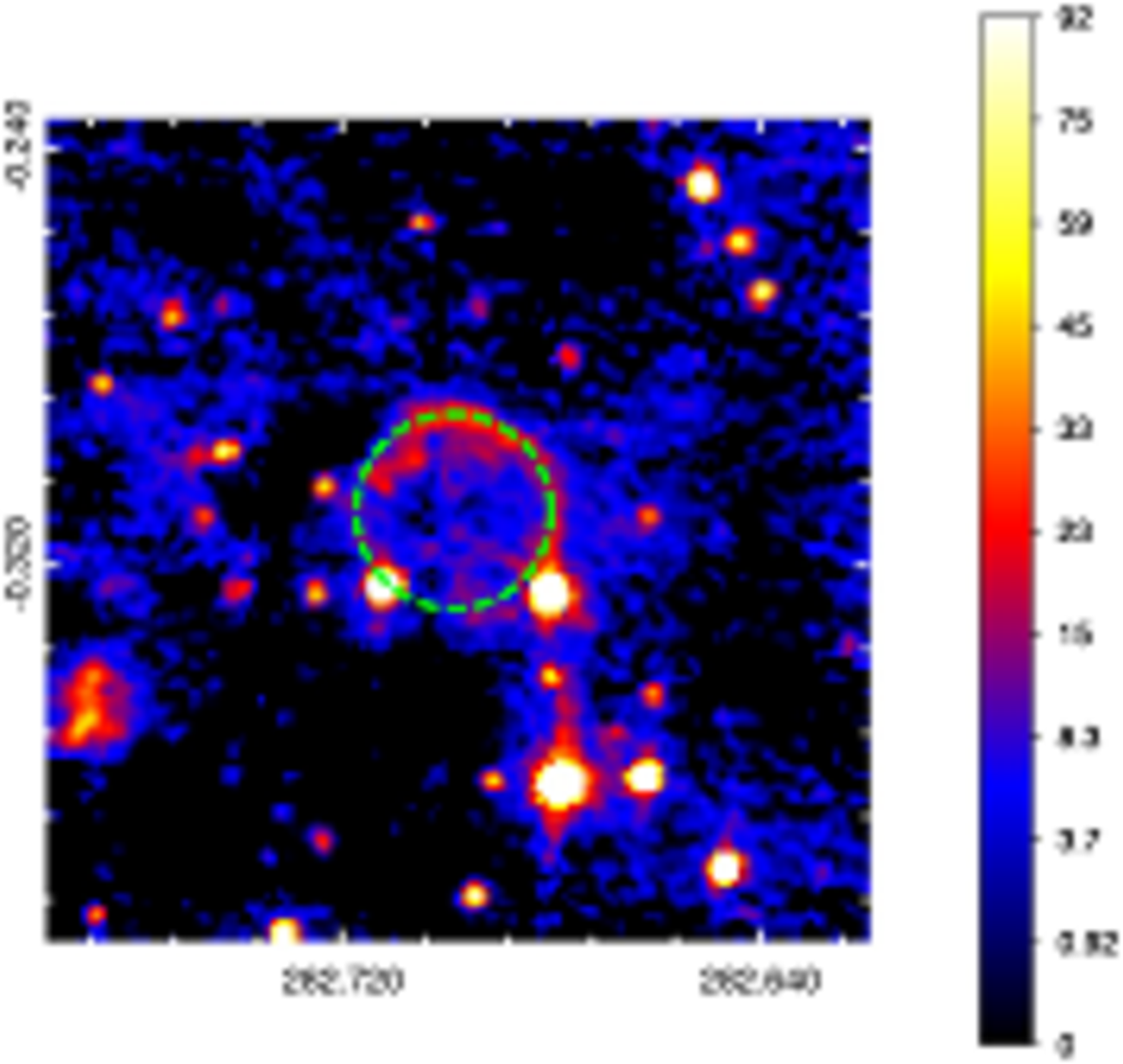}}}%
\mbox{\raisebox{0mm}{\includegraphics[width=40mm]{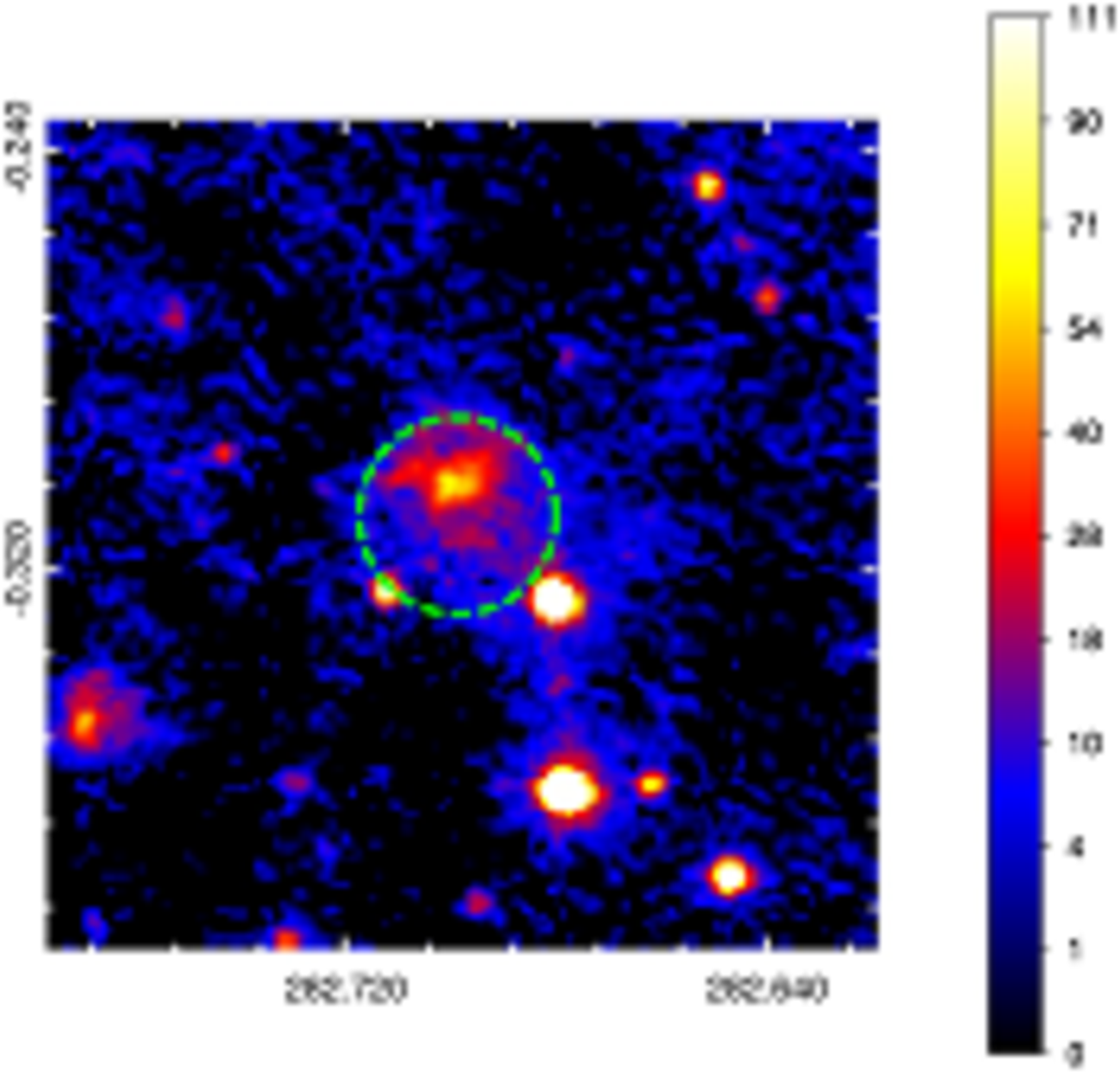}}}%
\mbox{\raisebox{0mm}{\includegraphics[width=40mm]{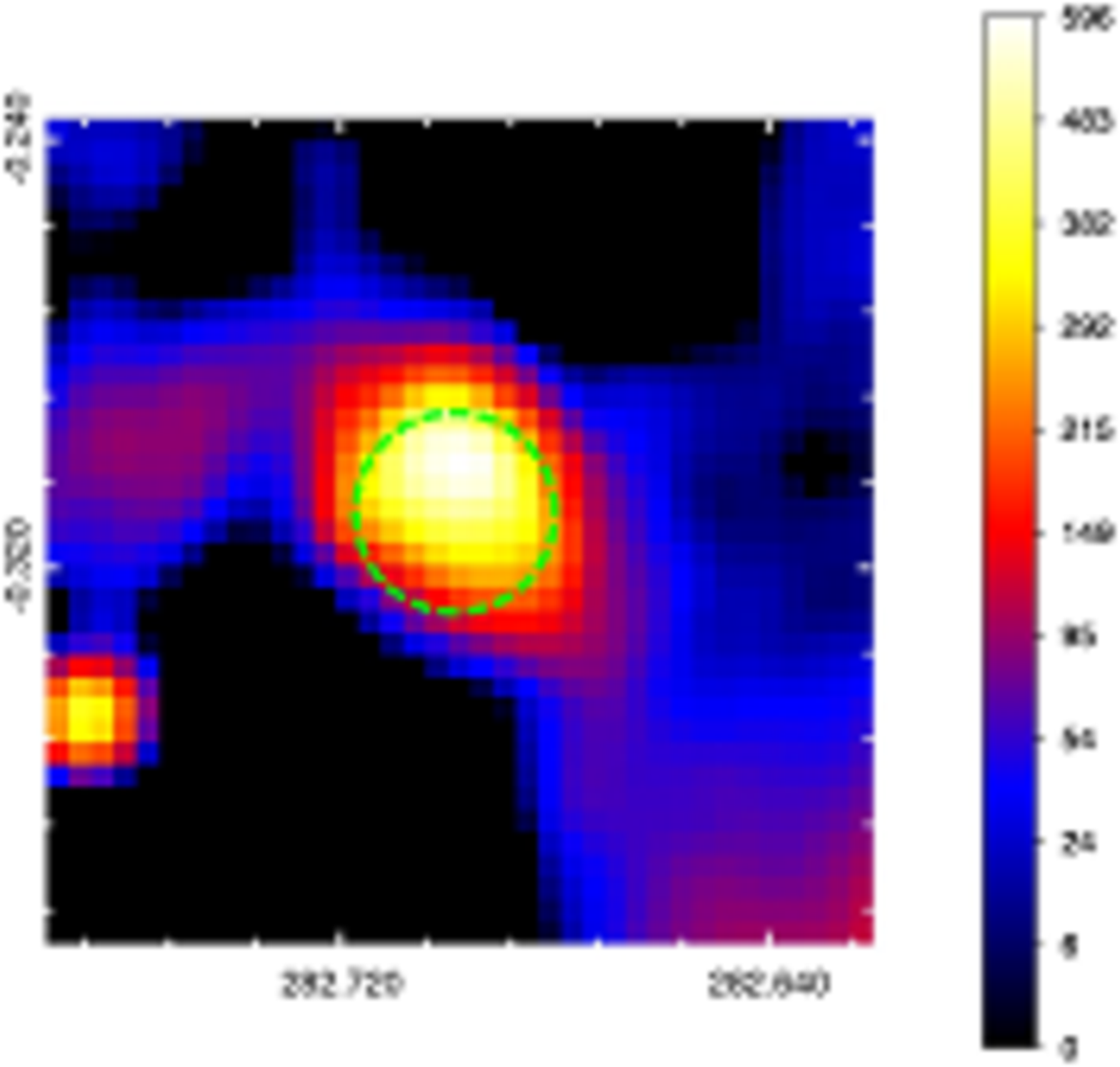}}}%
}
\subfigure{
\mbox{\raisebox{6mm}{\rotatebox{90}{\small{DEC (J2000)}}}}%
\mbox{\raisebox{0mm}{\includegraphics[width=40mm]{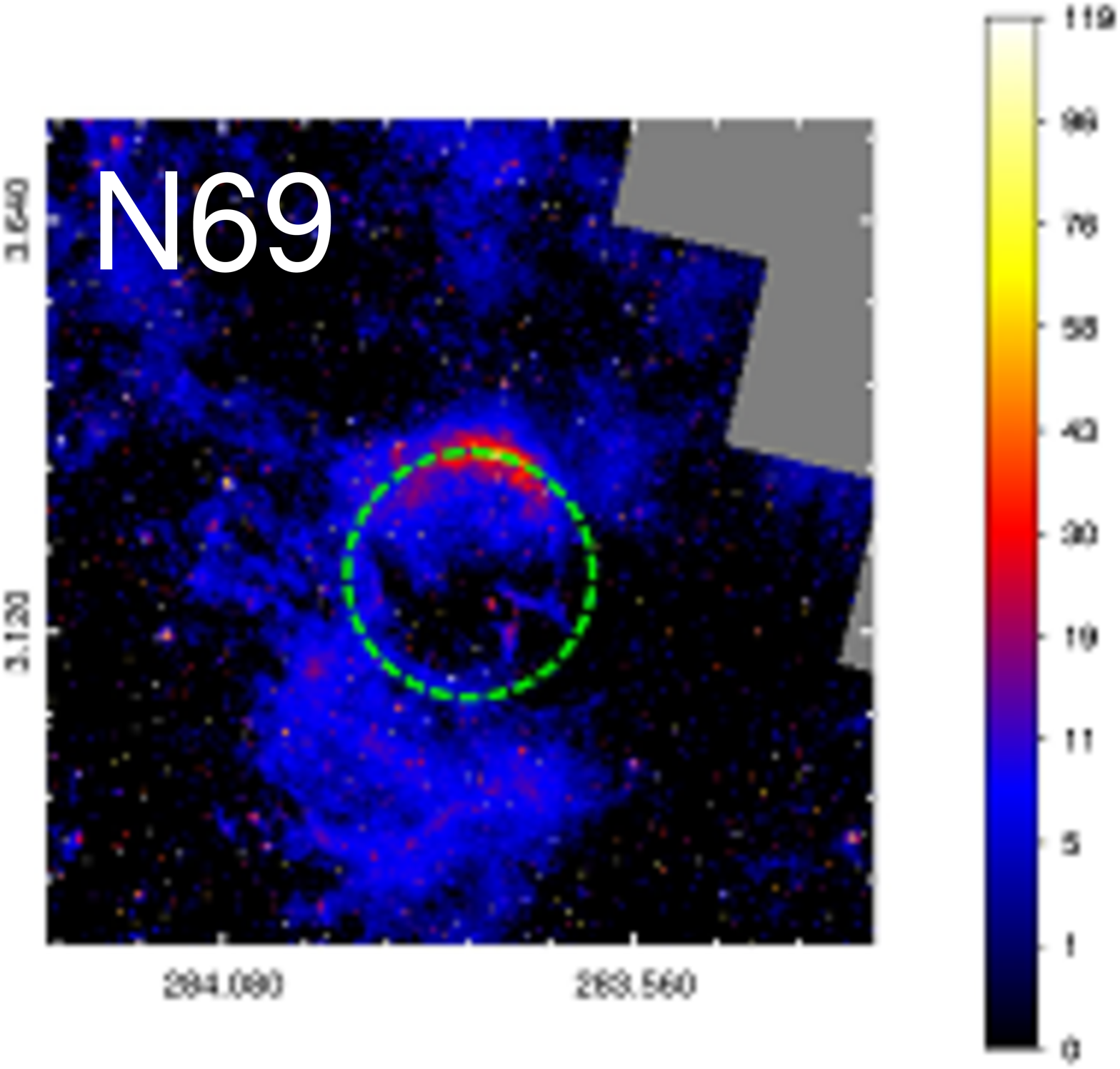}}}%
\mbox{\raisebox{0mm}{\includegraphics[width=40mm]{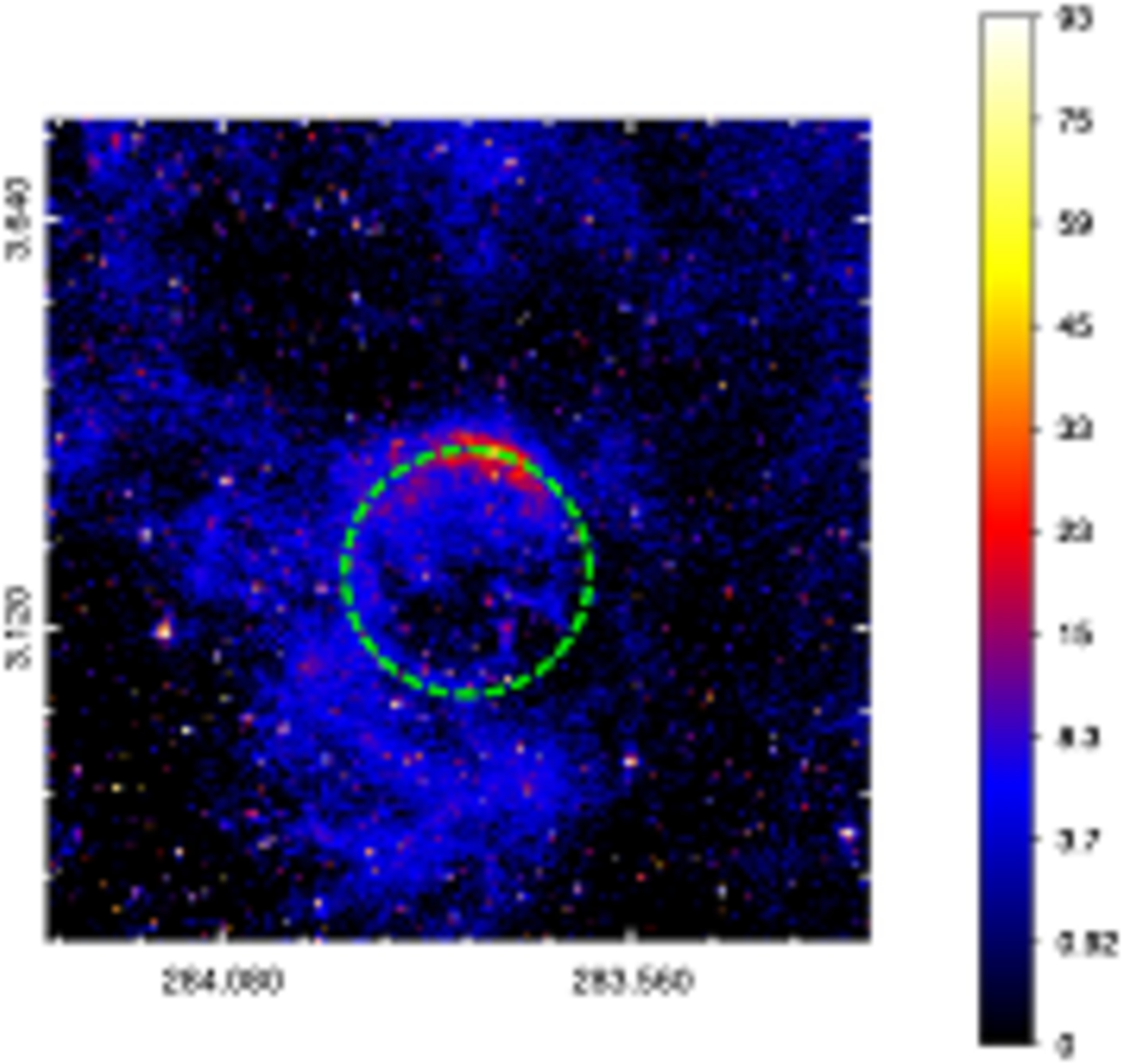}}}%
\mbox{\raisebox{0mm}{\includegraphics[width=40mm]{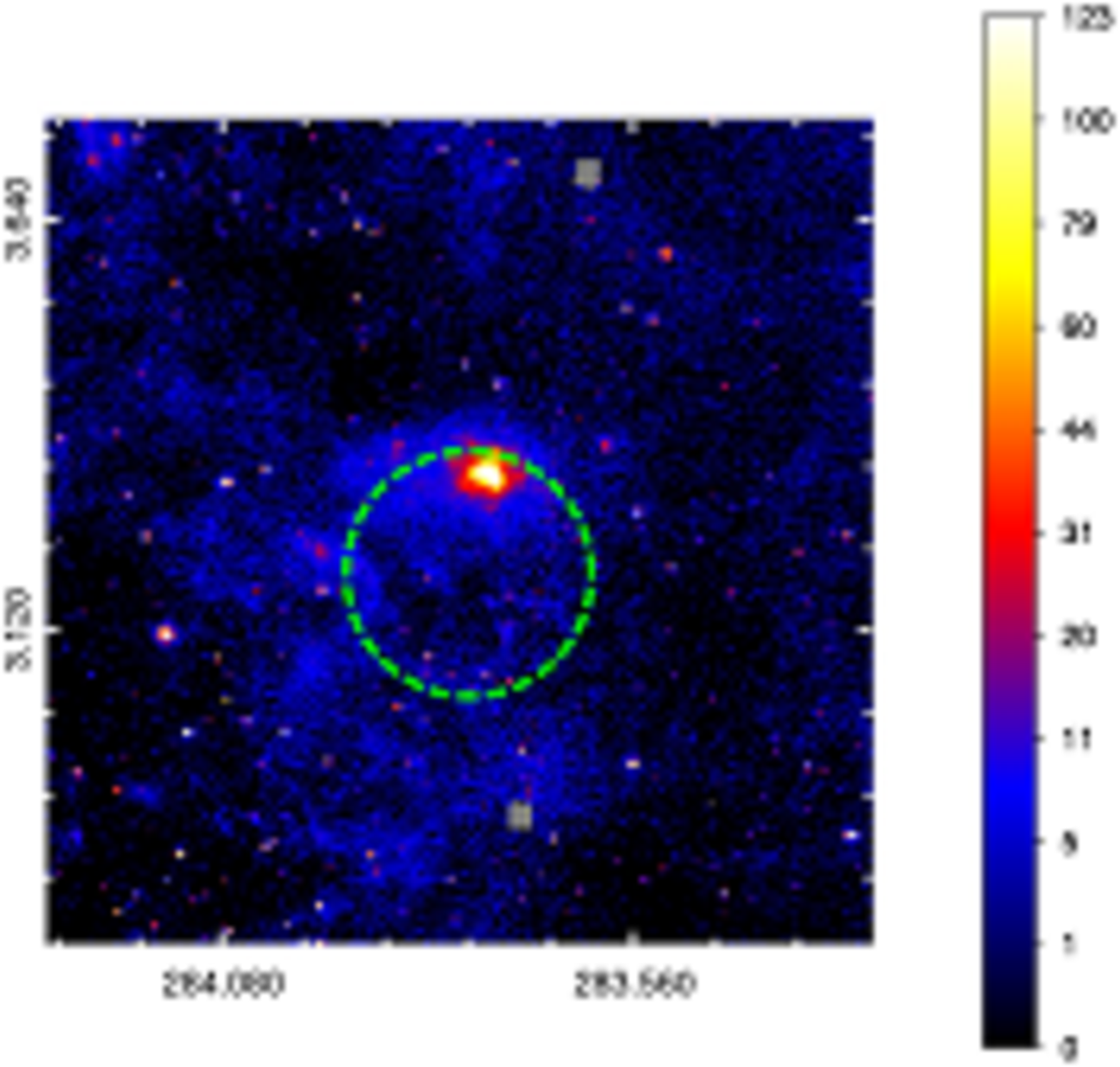}}}%
\mbox{\raisebox{0mm}{\includegraphics[width=40mm]{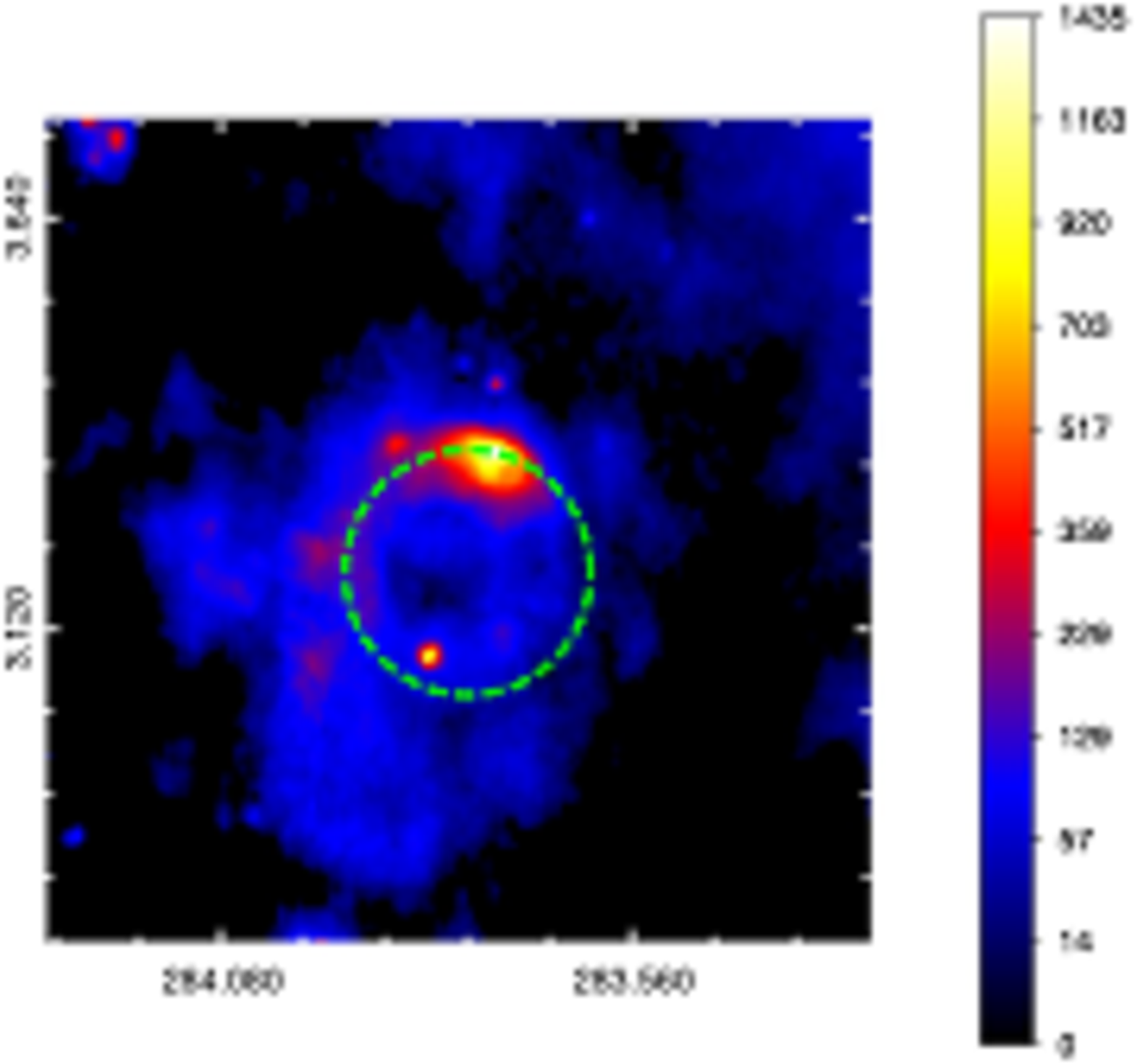}}}%
}
\subfigure{
\mbox{\raisebox{6mm}{\rotatebox{90}{\small{DEC (J2000)}}}}%
\mbox{\raisebox{0mm}{\includegraphics[width=40mm]{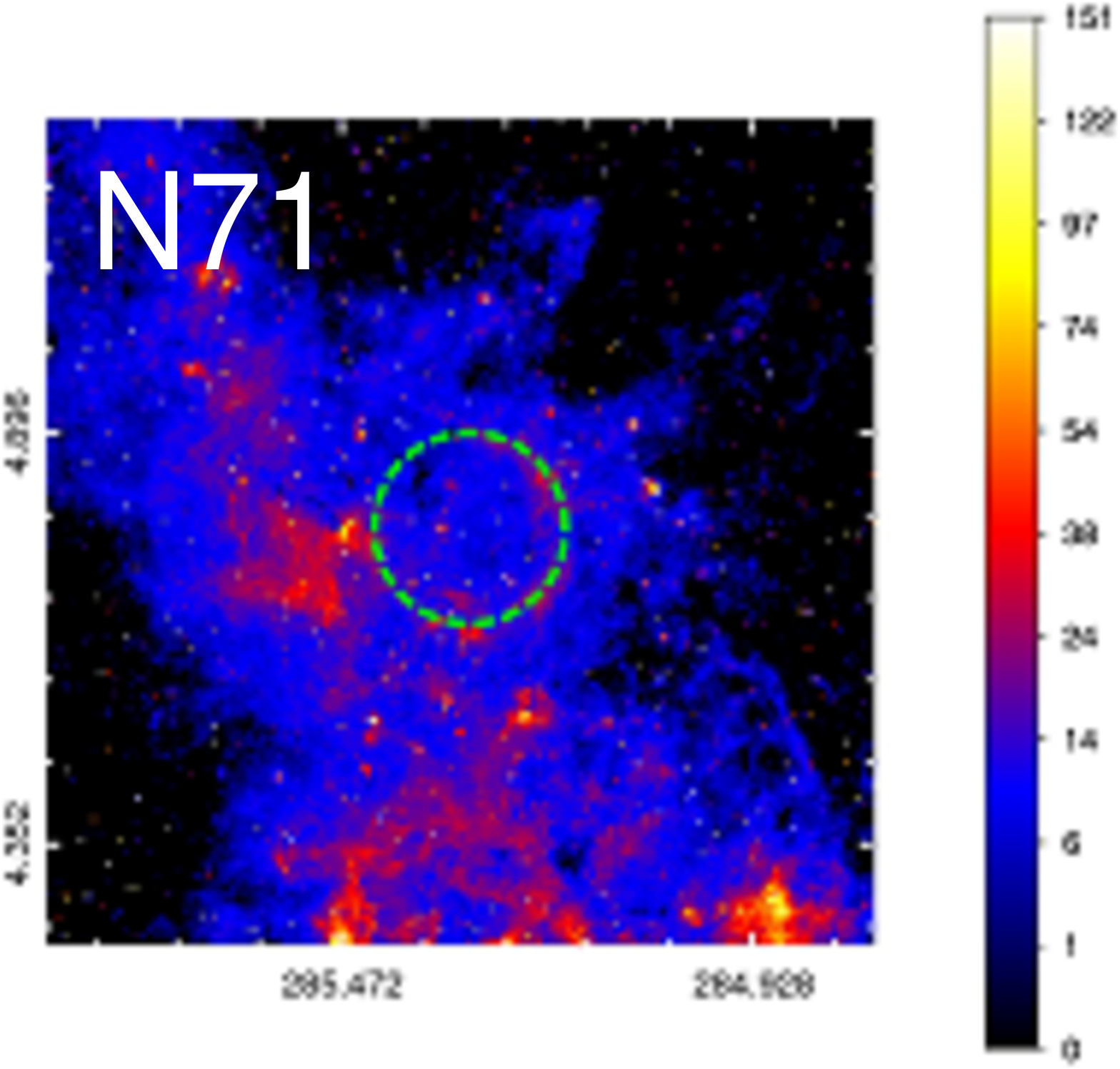}}}%
\mbox{\raisebox{0mm}{\includegraphics[width=40mm]{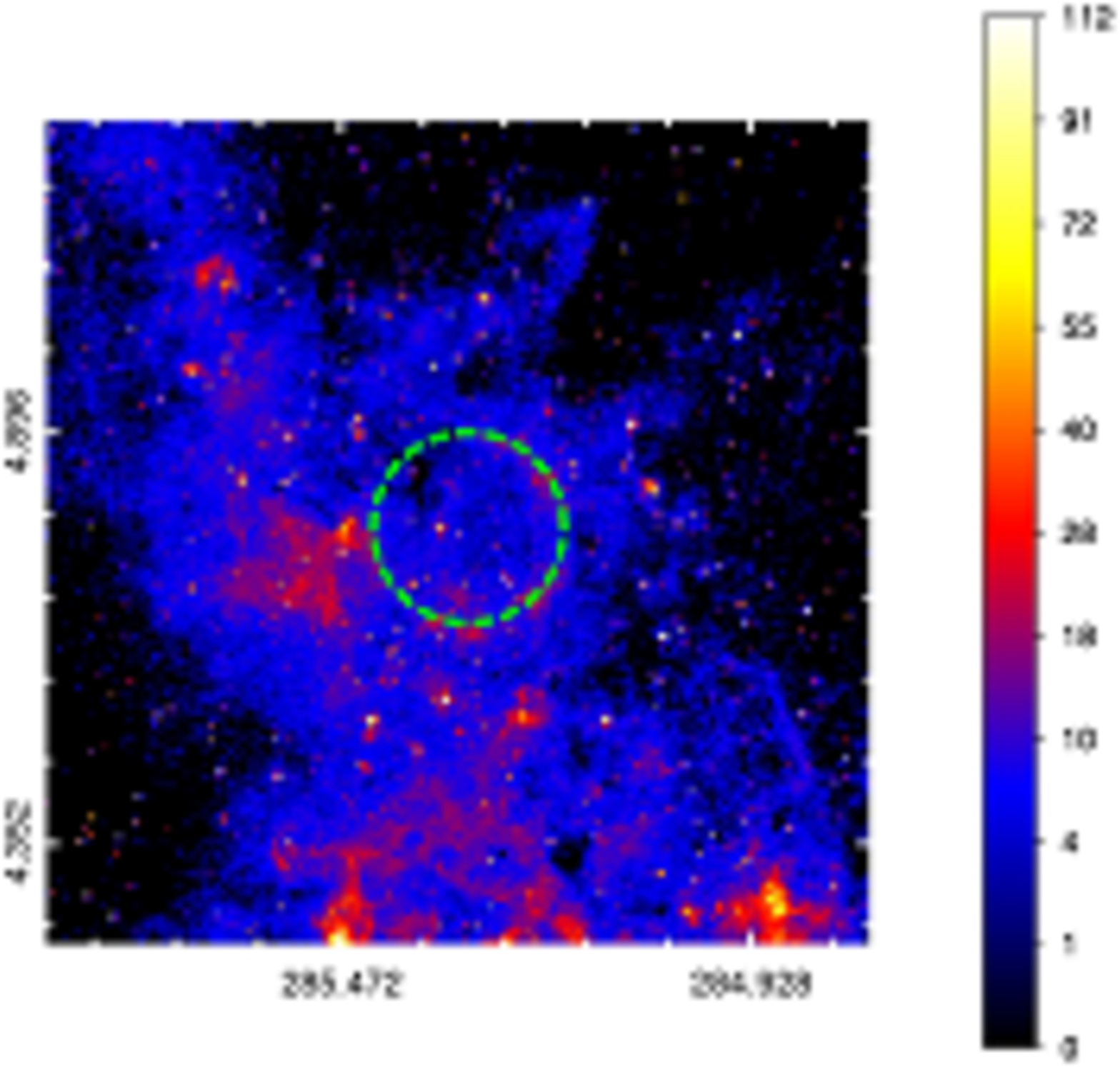}}}%
\mbox{\raisebox{0mm}{\includegraphics[width=40mm]{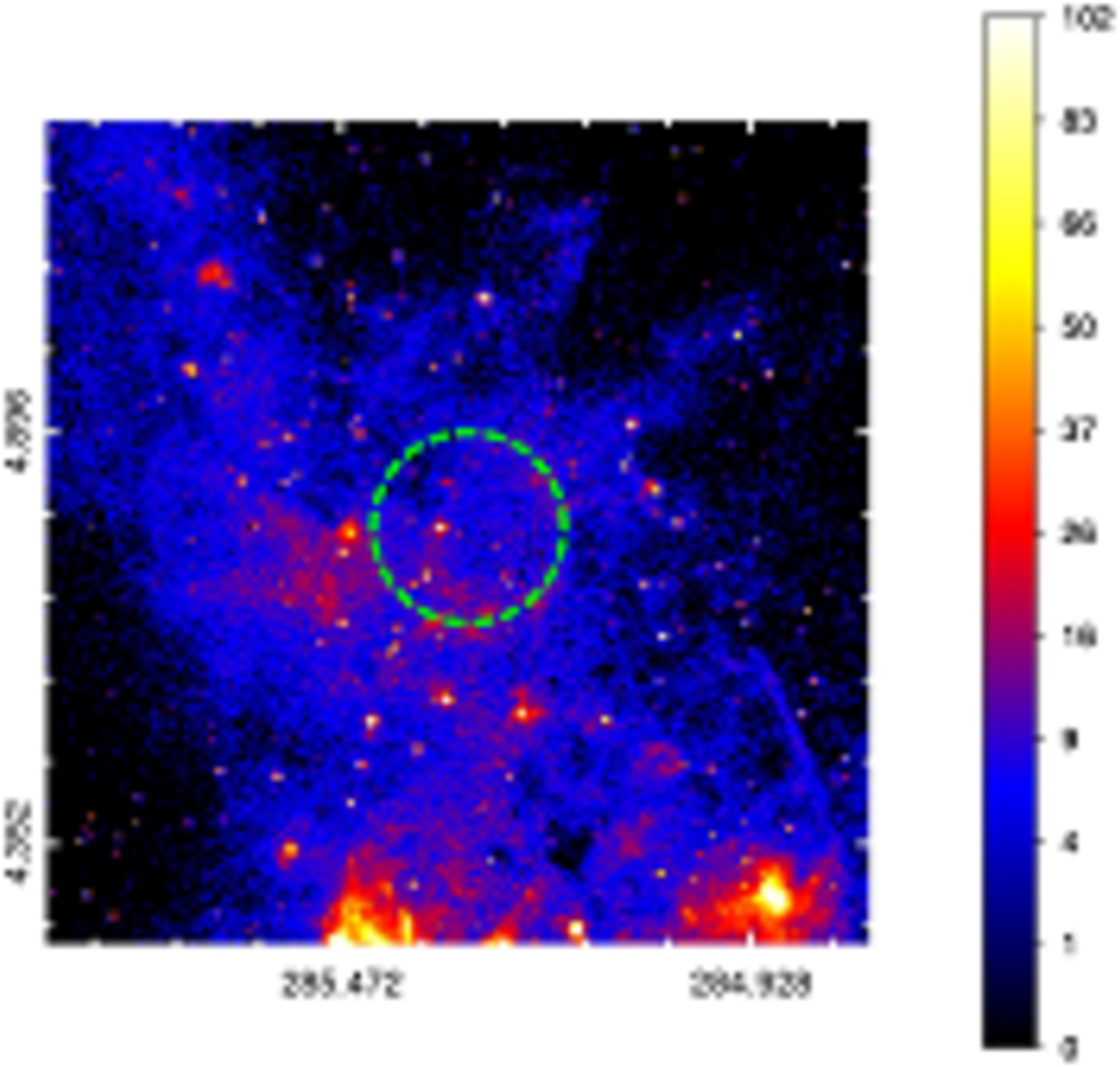}}}%
\mbox{\raisebox{0mm}{\includegraphics[width=40mm]{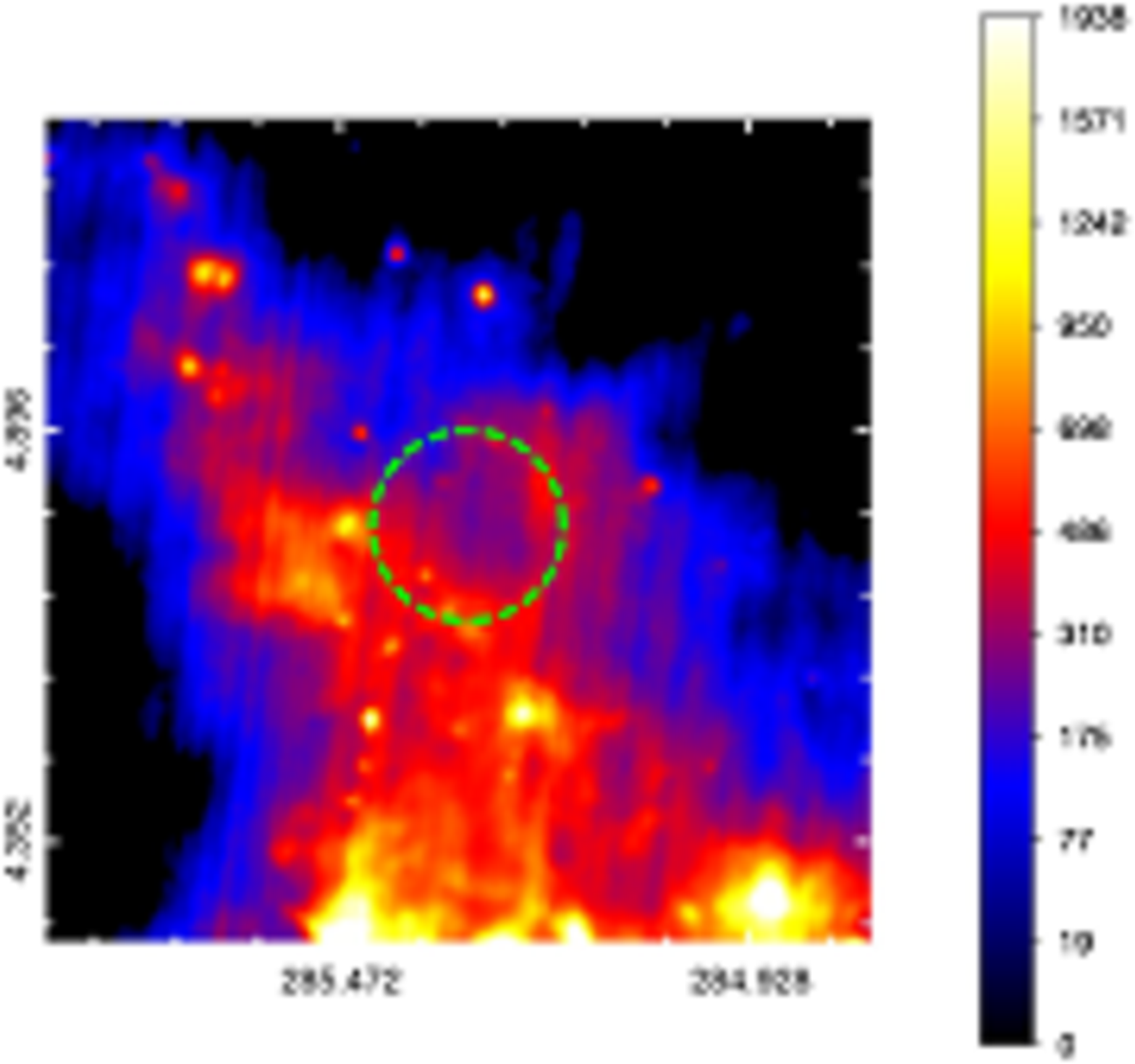}}}%
}
\caption{Continued.} \label{fig:Introfig1:c}
\end{figure*}

\addtocounter{figure}{-1}
\begin{figure*}[ht]
\addtocounter{subfigure}{1}
\centering
\subfigure{
\makebox[180mm][l]{\raisebox{0mm}[0mm][0mm]{ \hspace{15mm} \small{8 \mic}} \hspace{29.5mm} \small{9 \mic} \hspace{27mm} \small{18 \mic} \hspace{26.5mm} \small{90 \mic}}%
}
\subfigure{
\makebox[180mm][l]{\raisebox{0mm}[0mm][0mm]{ \hspace{11mm} \small{RA (J2000)}} \hspace{19.5mm} \small{RA (J2000)} \hspace{20mm} \small{RA (J2000)} \hspace{20mm} \small{RA (J2000)}}%
}
\subfigure{
\mbox{\raisebox{6mm}{\rotatebox{90}{\small{DEC (J2000)}}}}%
\mbox{\raisebox{0mm}{\includegraphics[width=40mm]{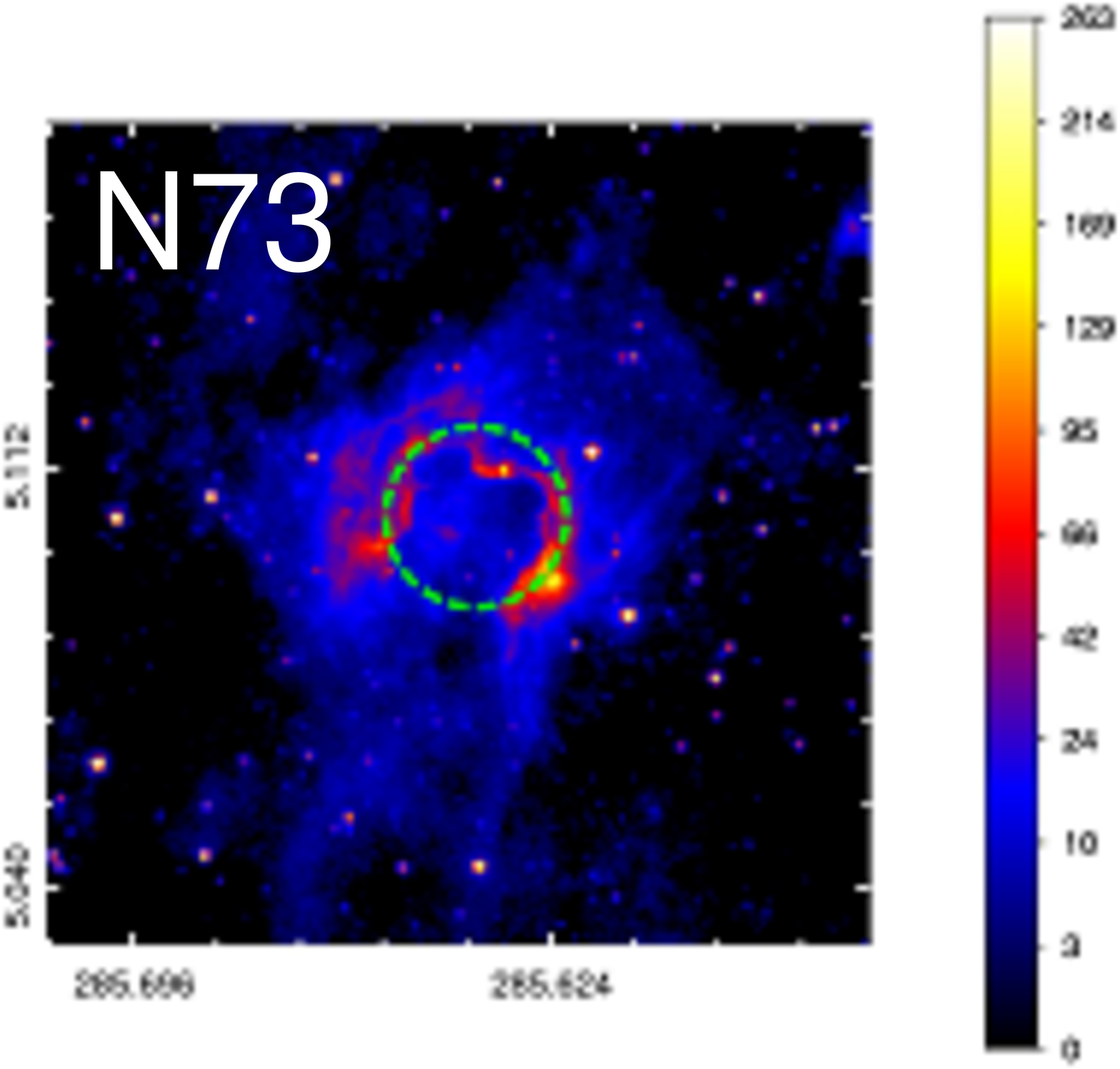}}}%
\mbox{\raisebox{0mm}{\includegraphics[width=40mm]{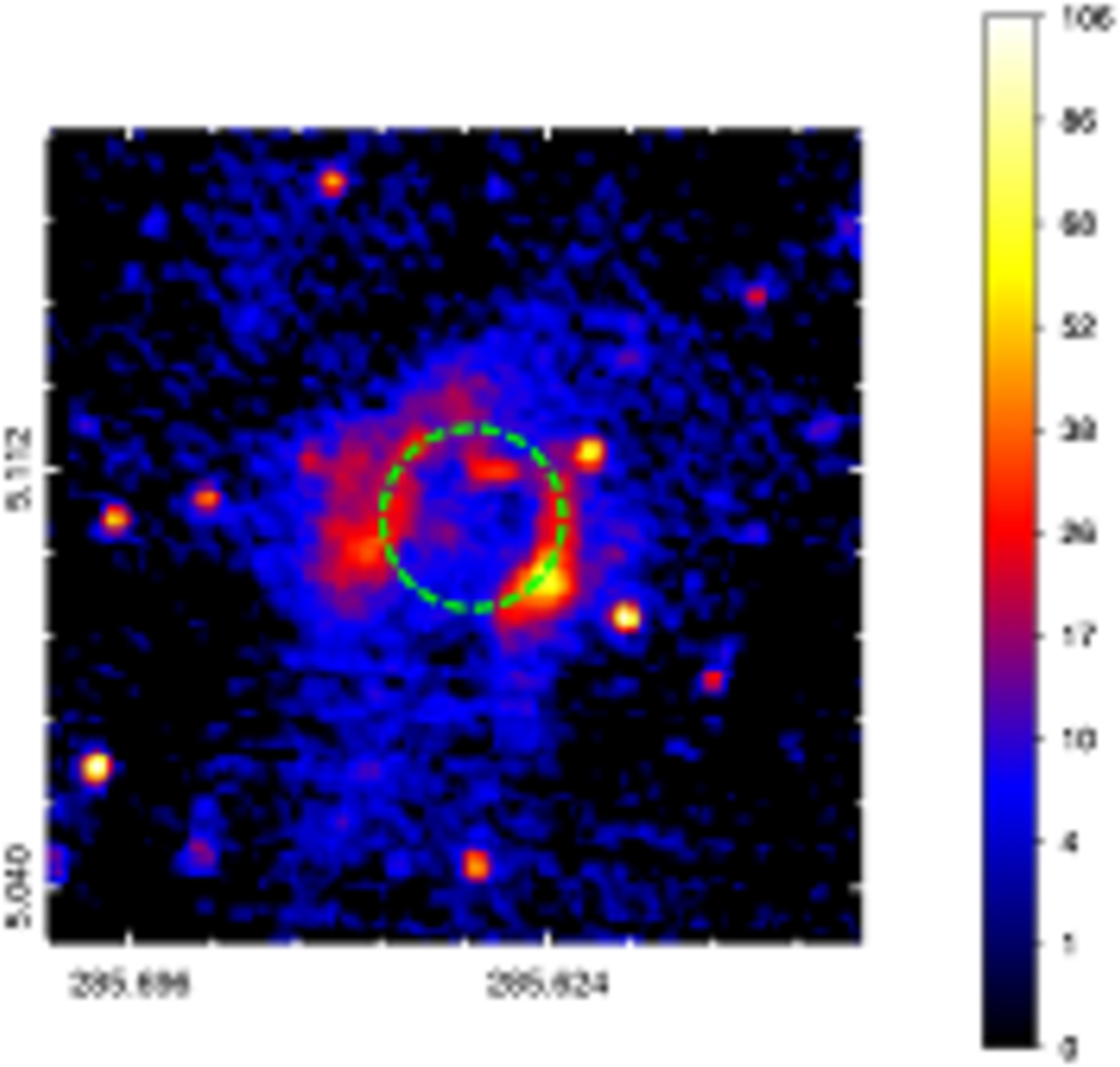}}}%
\mbox{\raisebox{0mm}{\includegraphics[width=40mm]{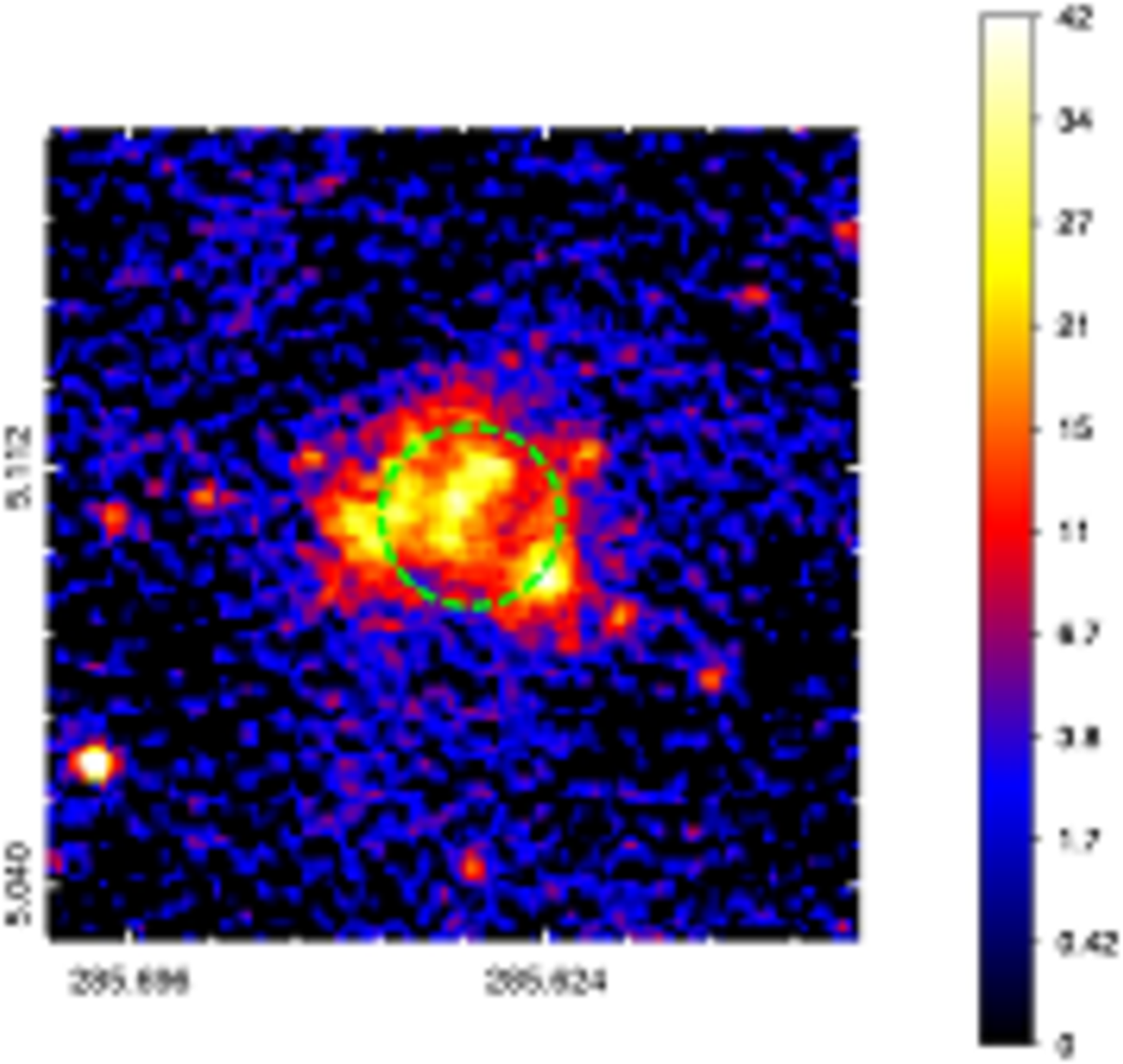}}}%
\mbox{\raisebox{0mm}{\includegraphics[width=40mm]{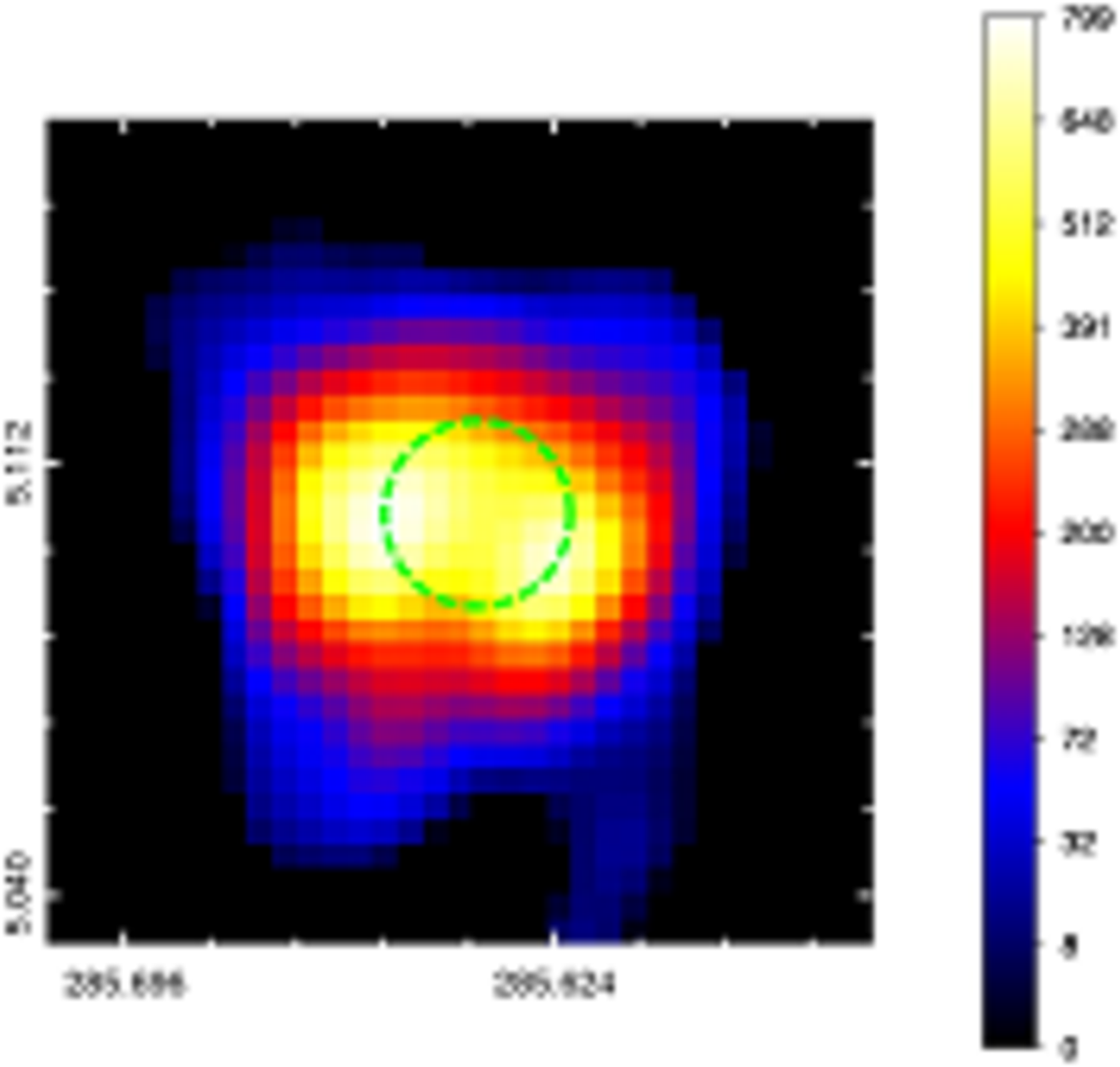}}}%
}
\subfigure{
\mbox{\raisebox{6mm}{\rotatebox{90}{\small{DEC (J2000)}}}}%
\mbox{\raisebox{0mm}{\includegraphics[width=40mm]{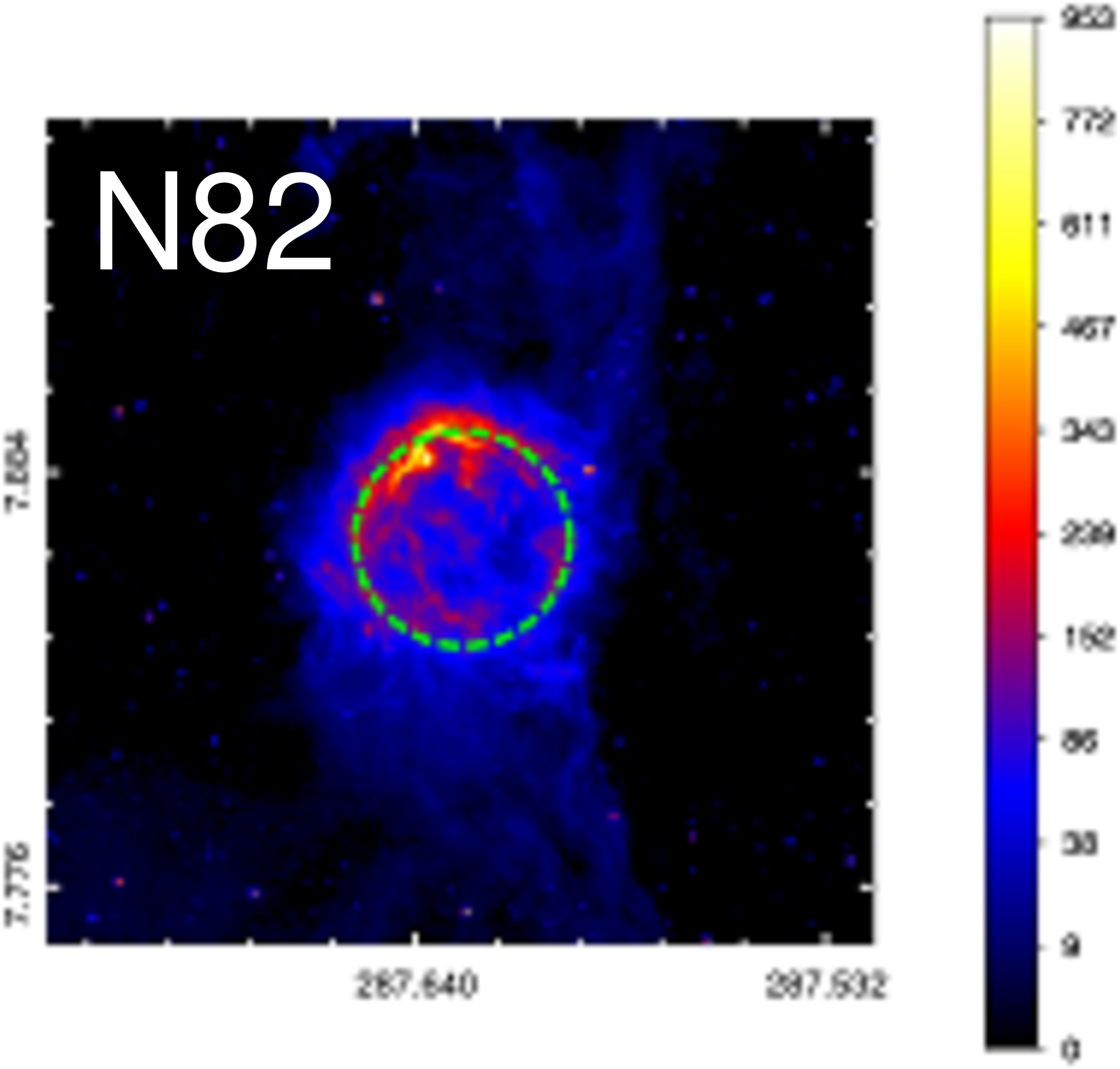}}}%
\mbox{\raisebox{0mm}{\includegraphics[width=40mm]{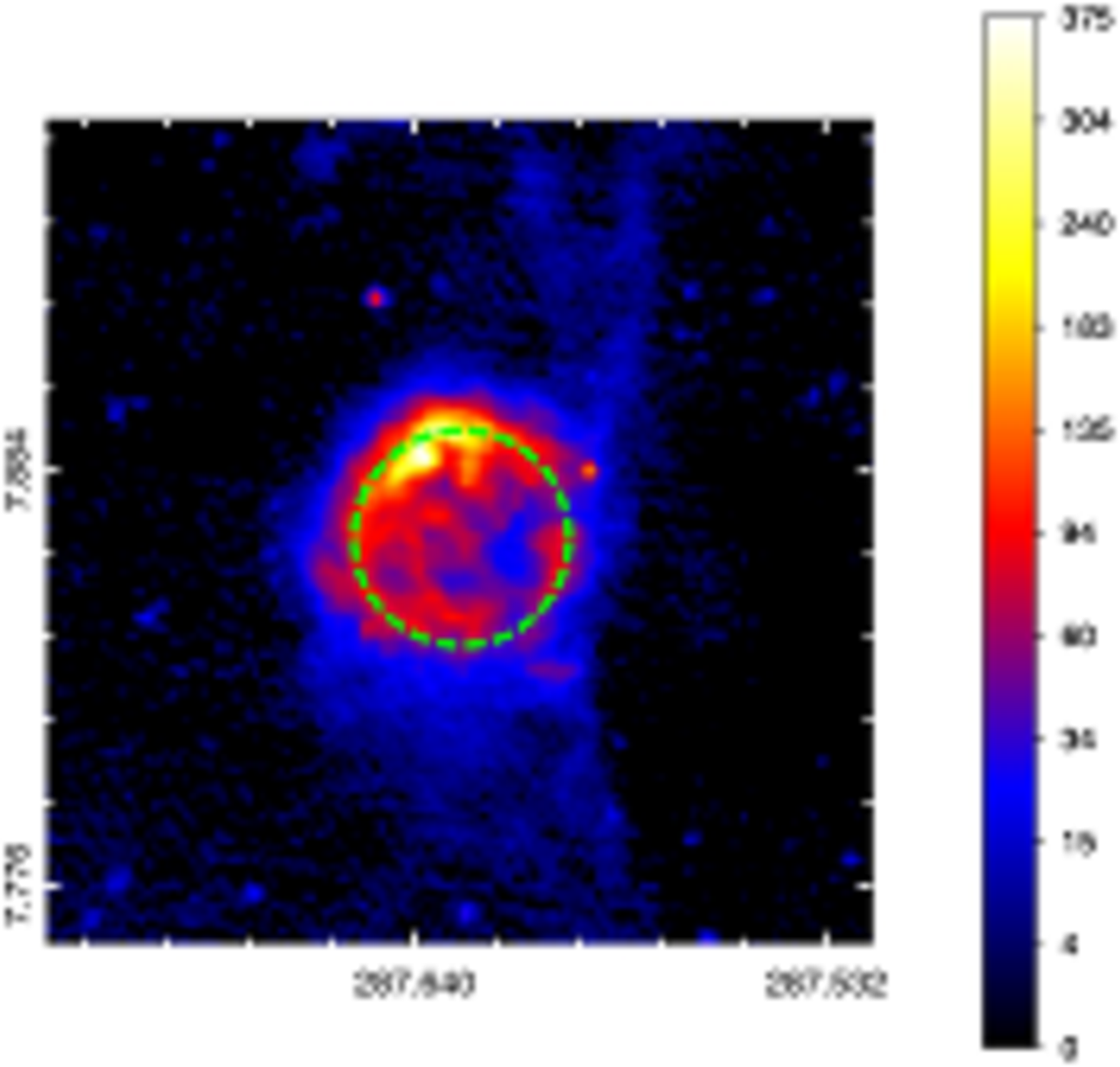}}}%
\mbox{\raisebox{0mm}{\includegraphics[width=40mm]{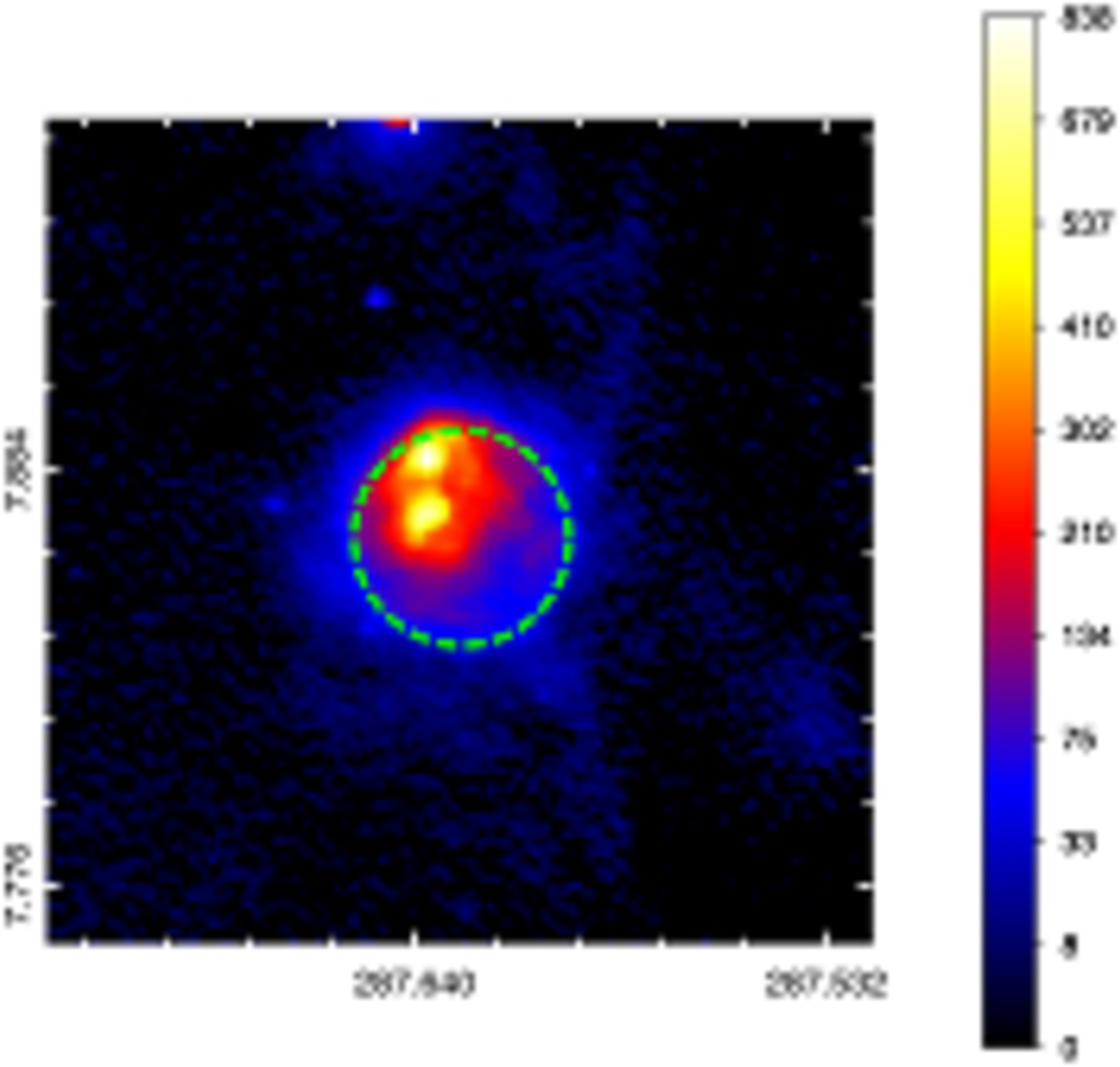}}}%
\mbox{\raisebox{0mm}{\includegraphics[width=40mm]{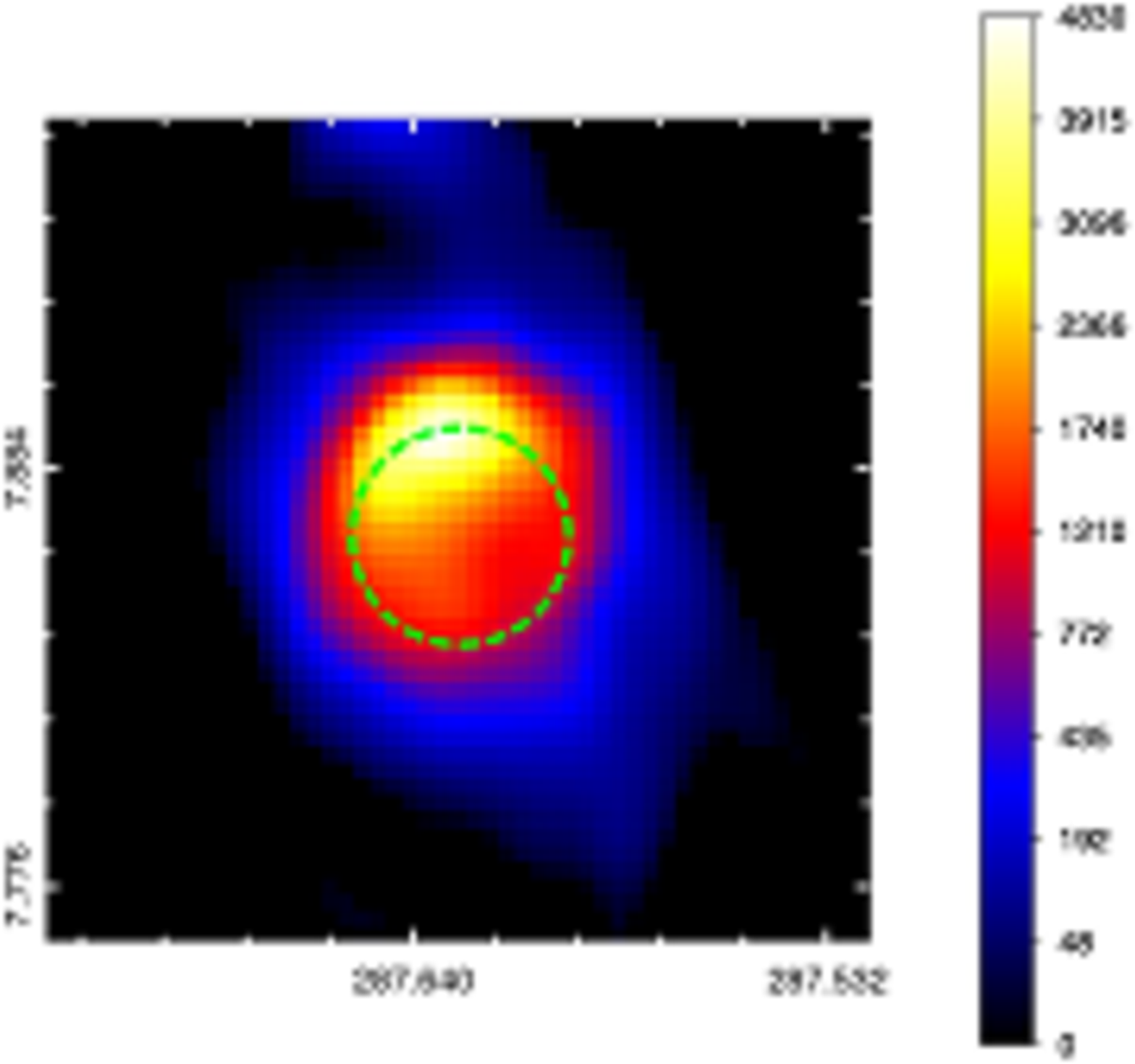}}}%
}
\subfigure{
\mbox{\raisebox{6mm}{\rotatebox{90}{\small{DEC (J2000)}}}}%
\mbox{\raisebox{0mm}{\includegraphics[width=40mm]{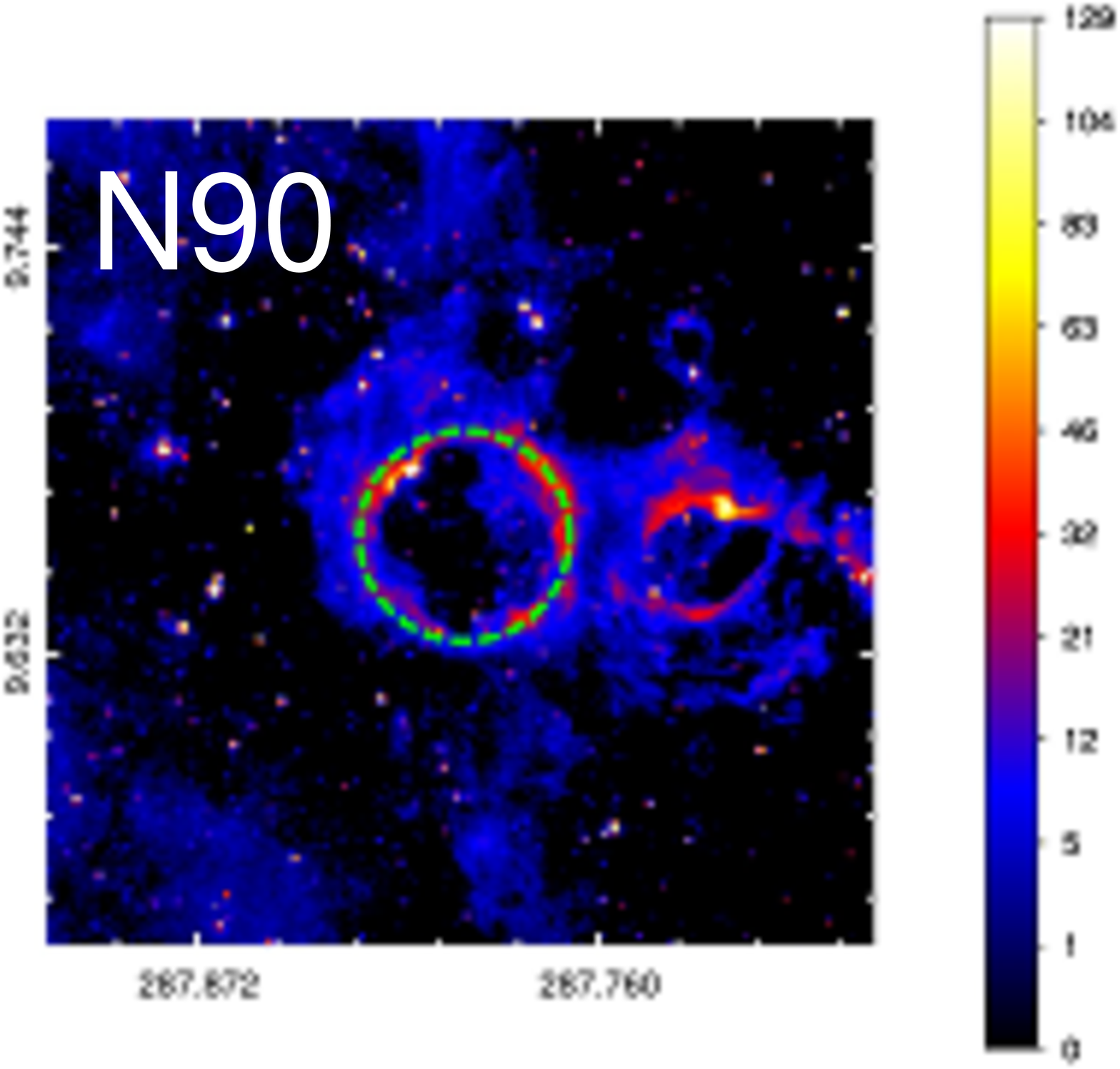}}}%
\mbox{\raisebox{0mm}{\includegraphics[width=40mm]{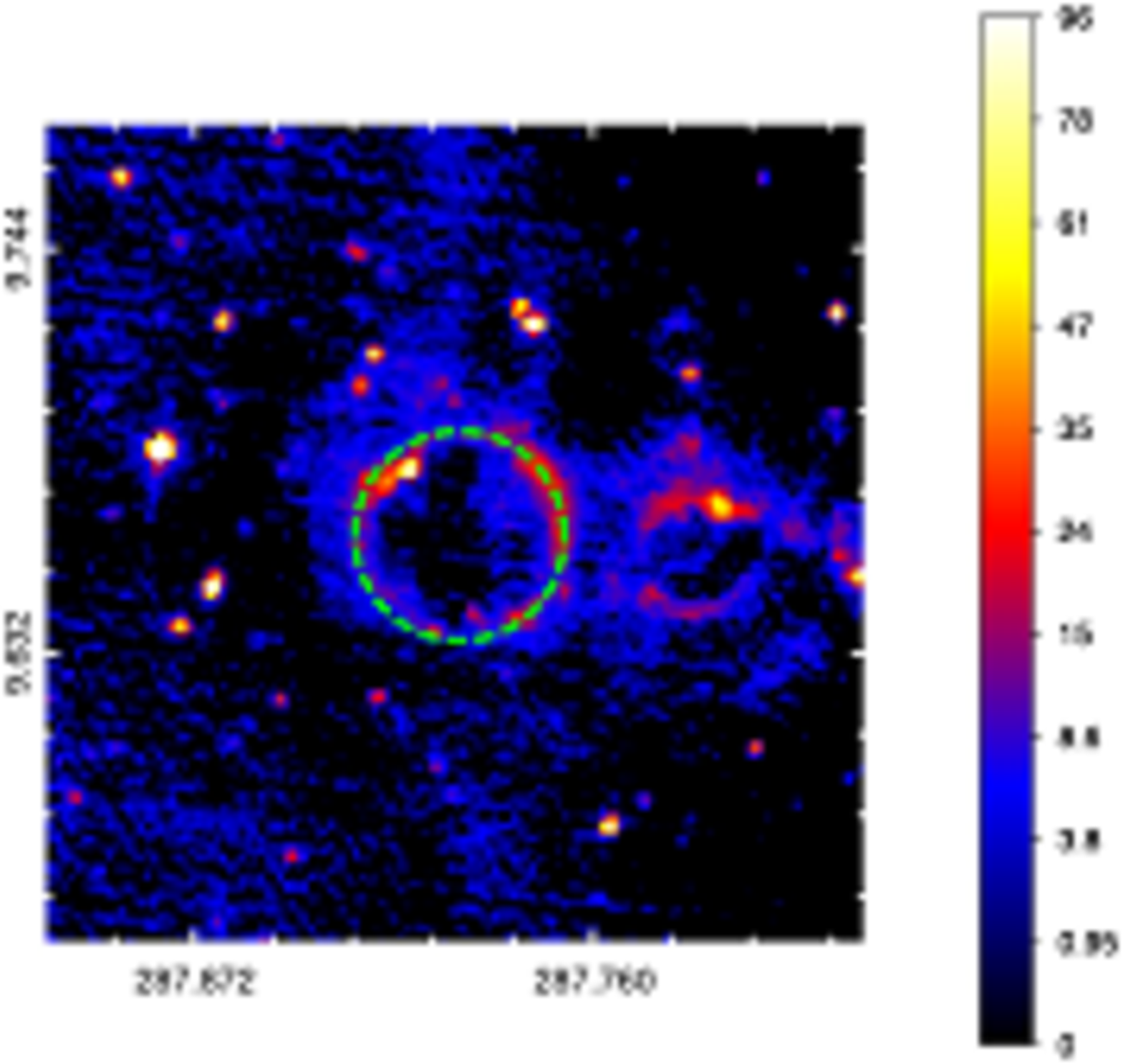}}}%
\mbox{\raisebox{0mm}{\includegraphics[width=40mm]{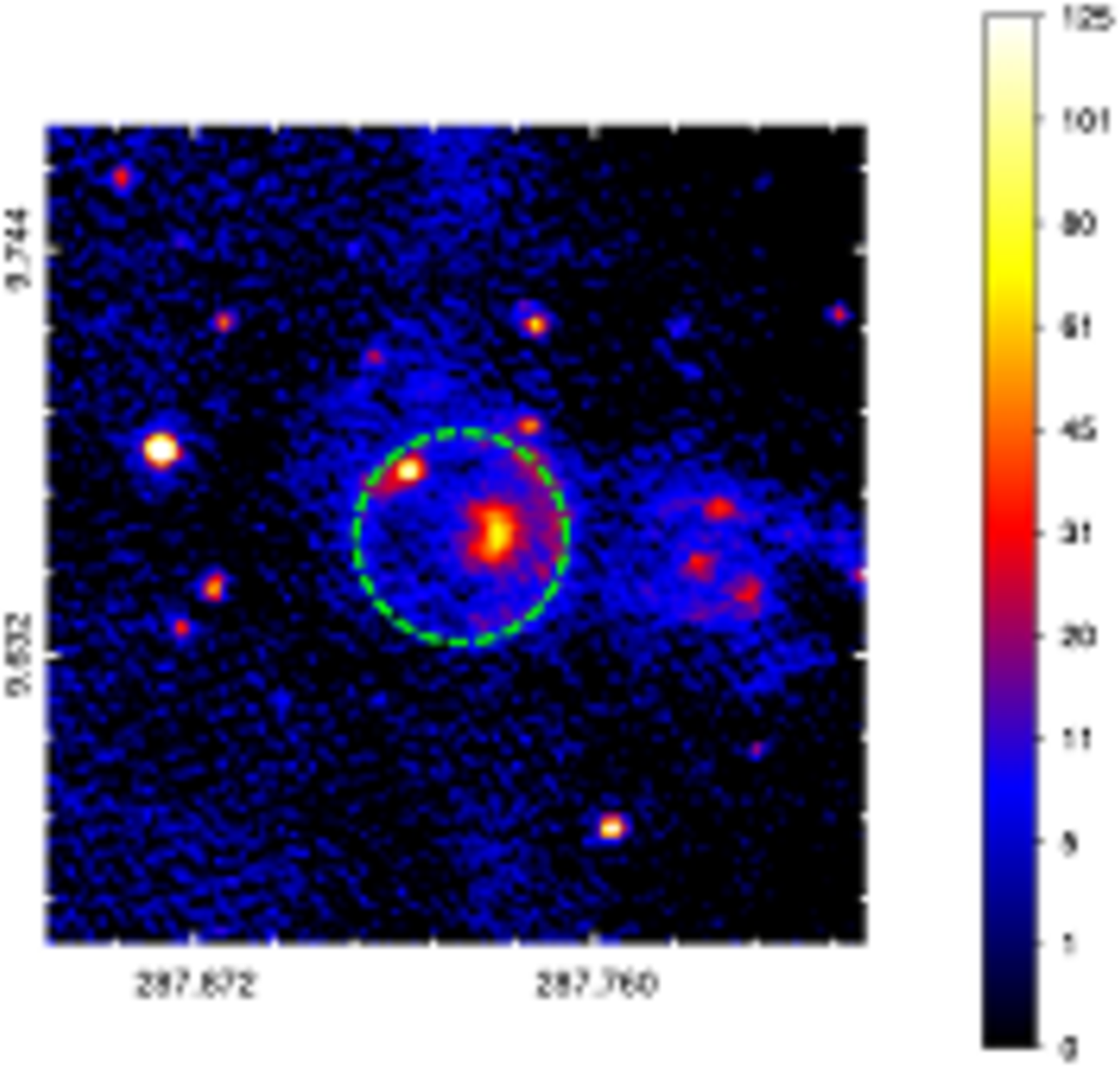}}}%
\mbox{\raisebox{0mm}{\includegraphics[width=40mm]{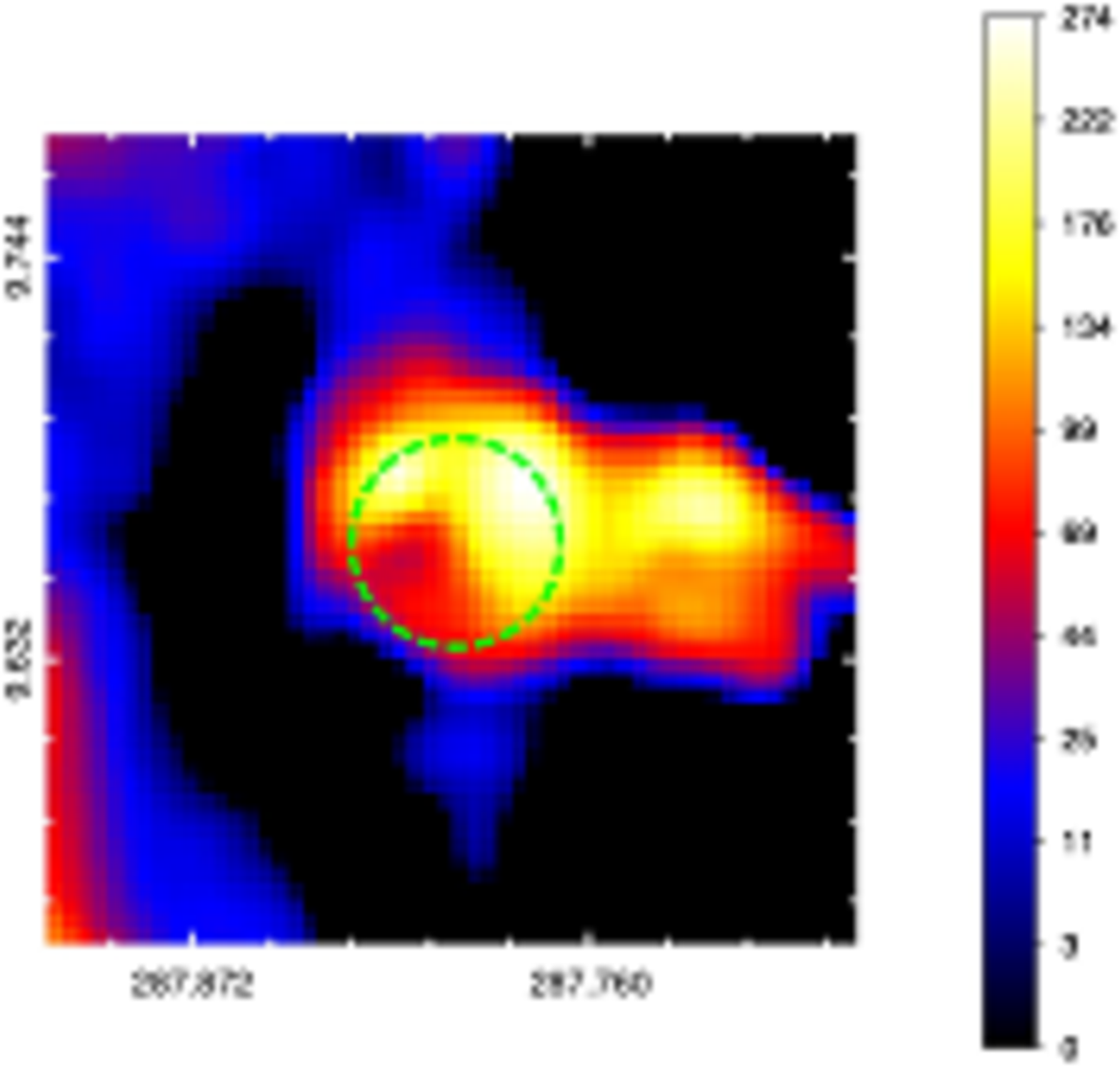}}}%
}
\subfigure{
\mbox{\raisebox{6mm}{\rotatebox{90}{\small{DEC (J2000)}}}}%
\mbox{\raisebox{0mm}{\includegraphics[width=40mm]{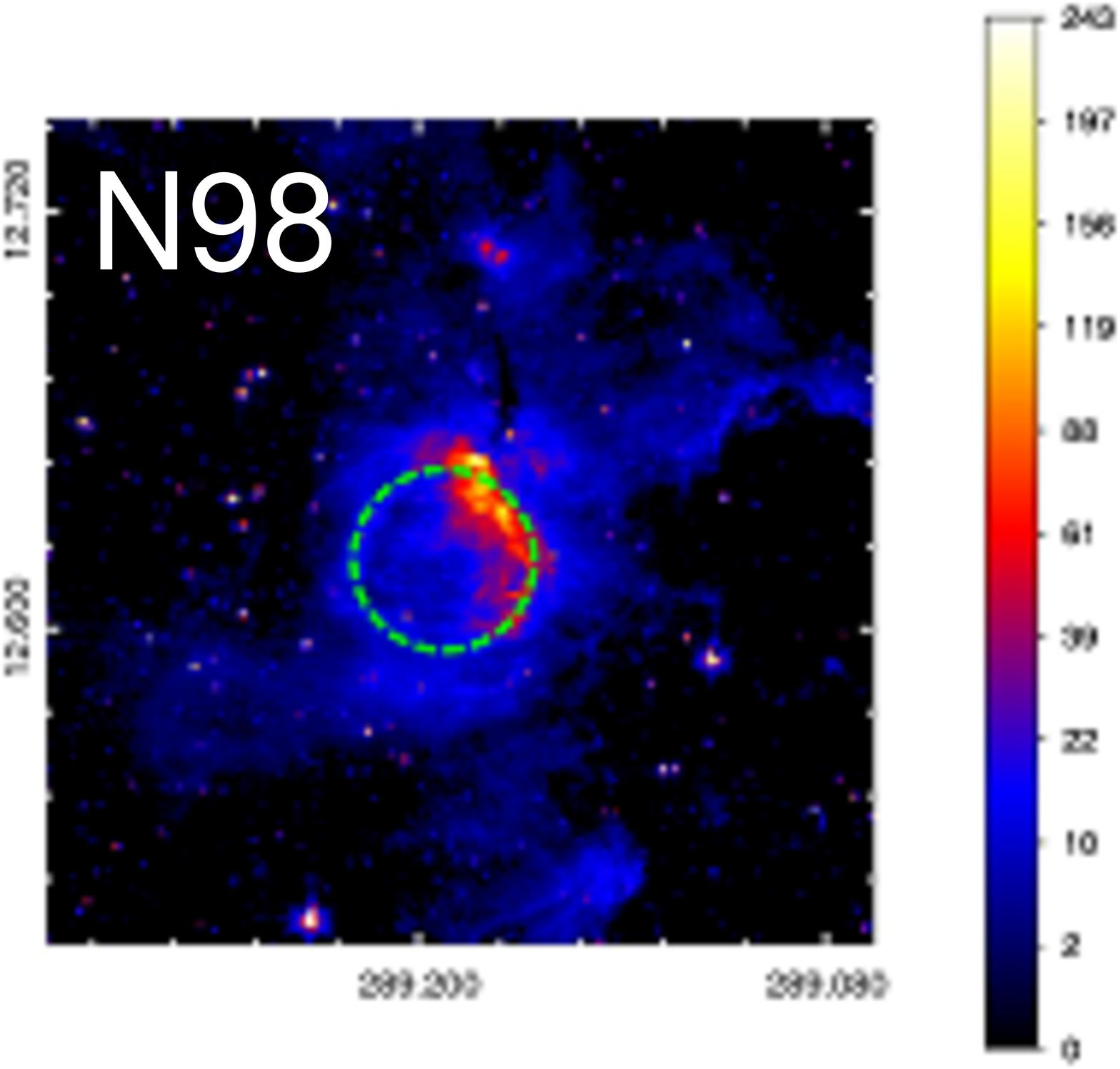}}}%
\mbox{\raisebox{0mm}{\includegraphics[width=40mm]{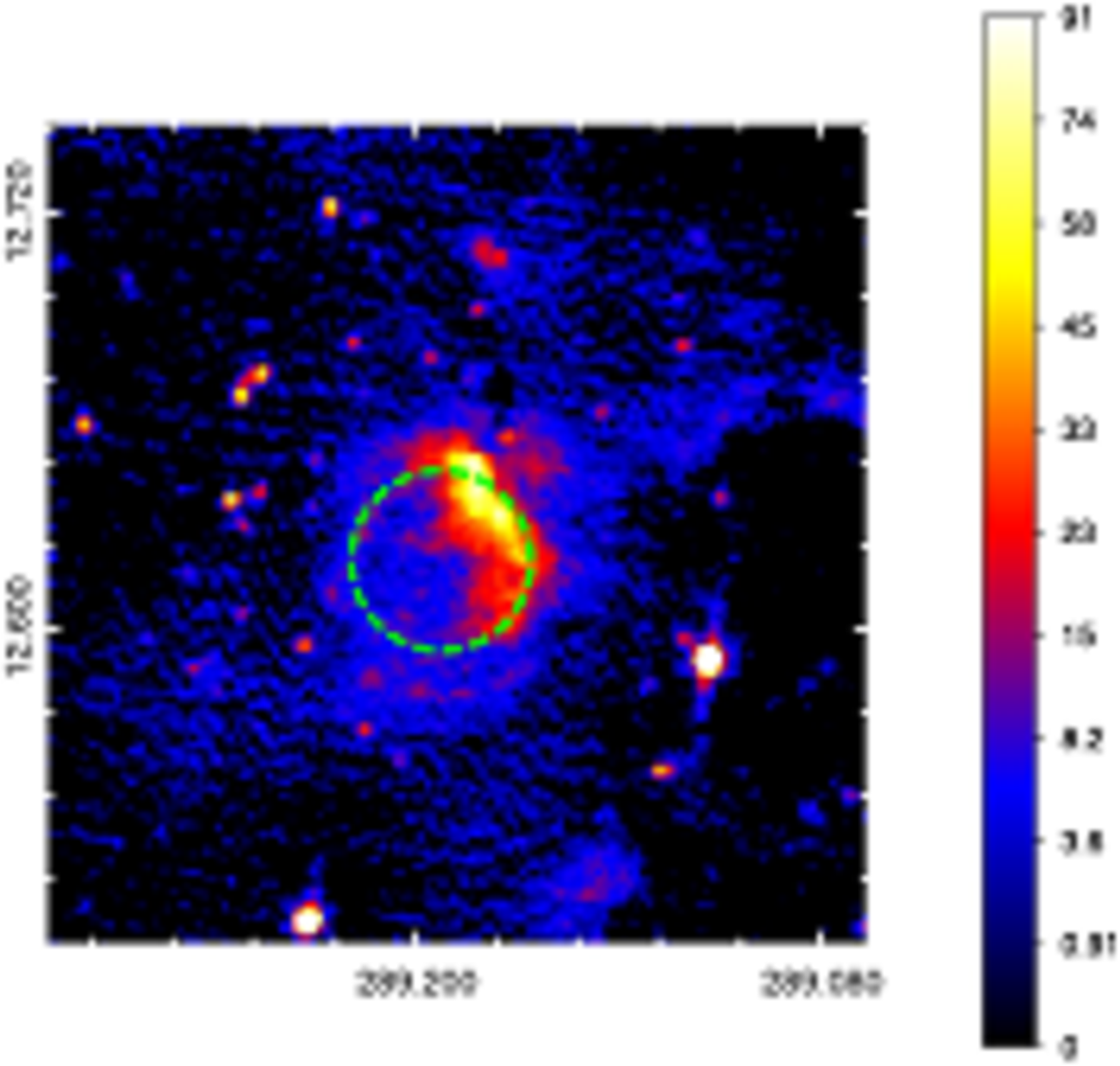}}}%
\mbox{\raisebox{0mm}{\includegraphics[width=40mm]{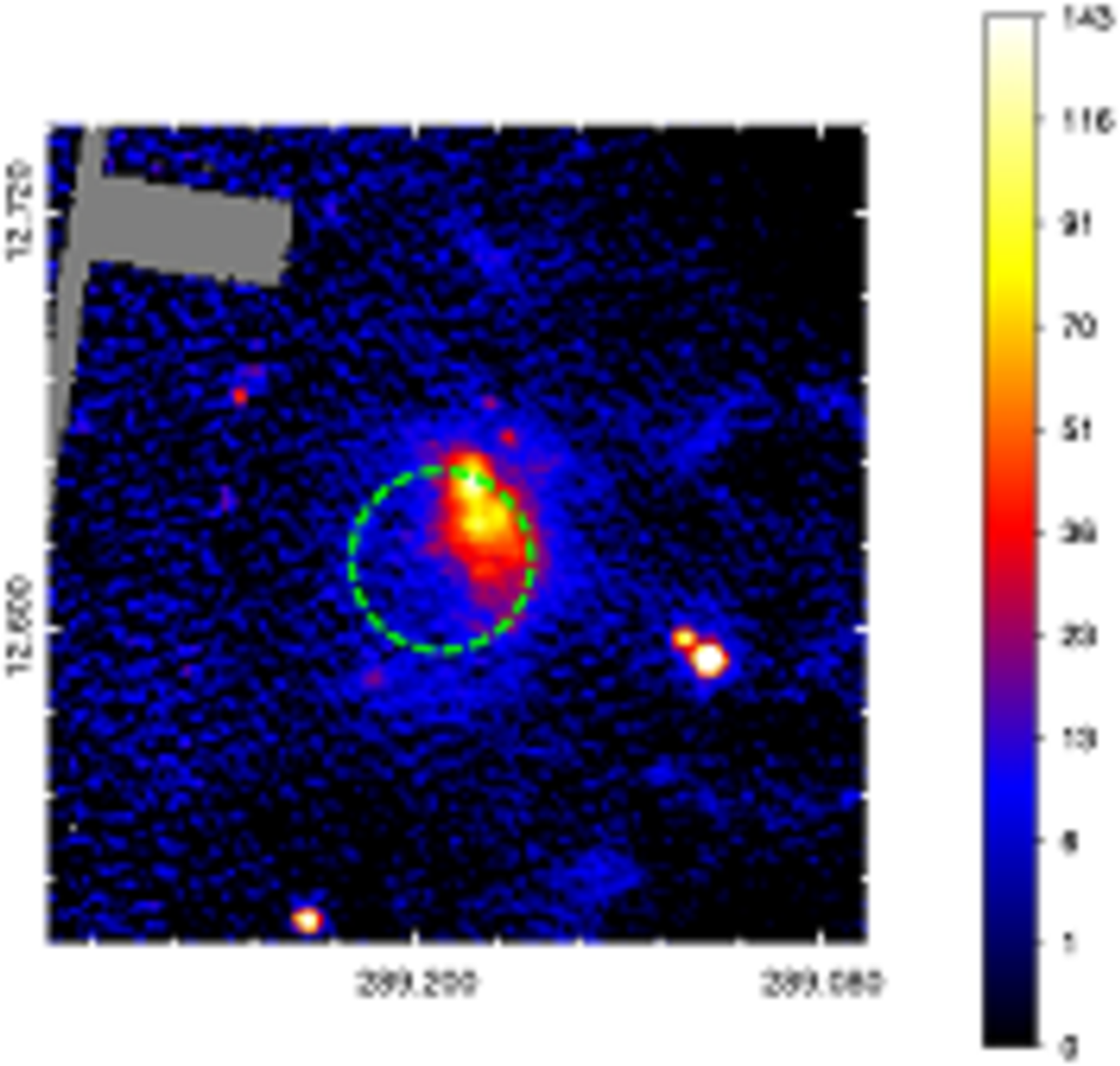}}}%
\mbox{\raisebox{0mm}{\includegraphics[width=40mm]{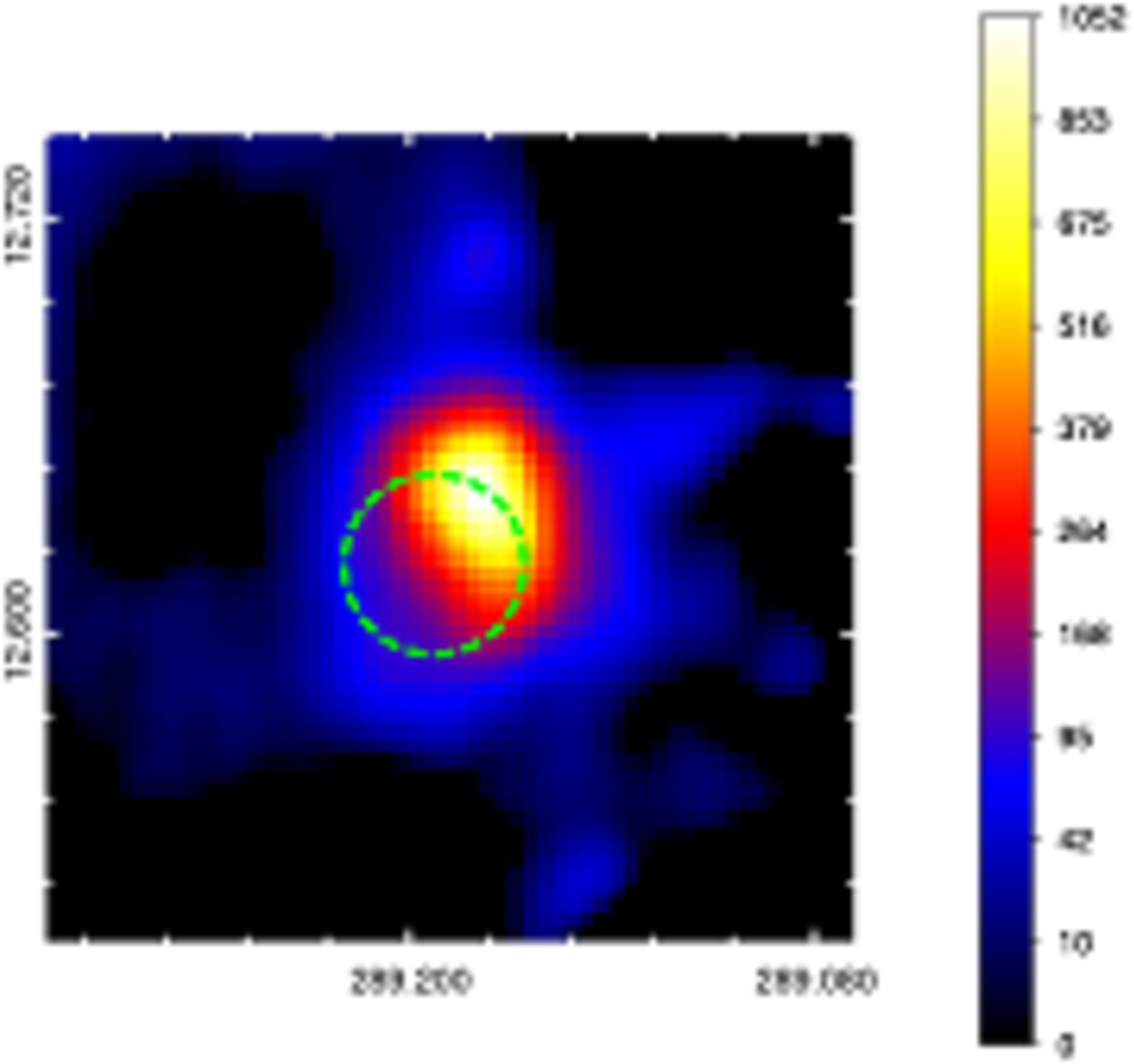}}}%
}
\subfigure{
\mbox{\raisebox{6mm}{\rotatebox{90}{\small{DEC (J2000)}}}}%
\mbox{\raisebox{0mm}{\includegraphics[width=40mm]{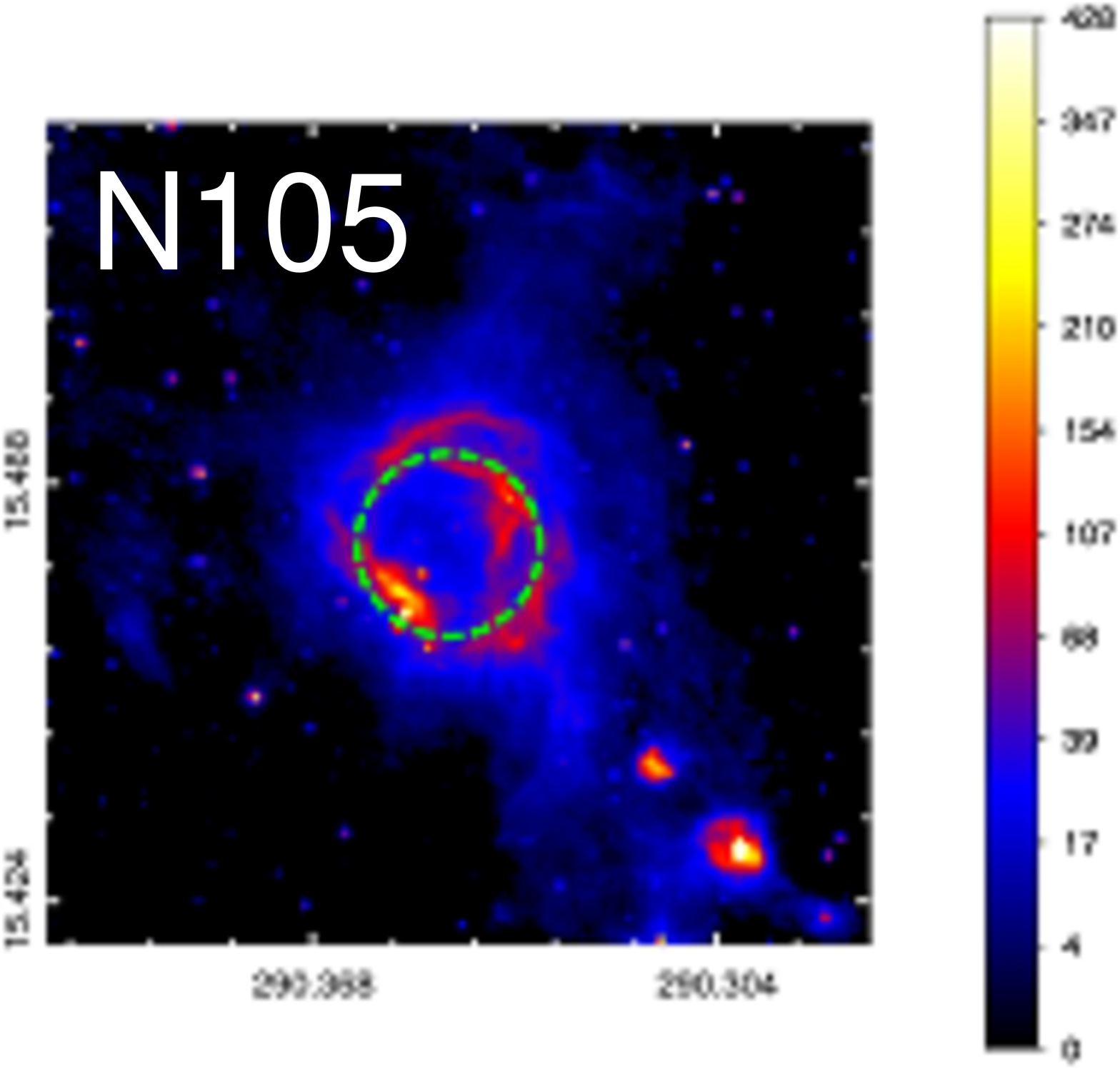}}}%
\mbox{\raisebox{0mm}{\includegraphics[width=40mm]{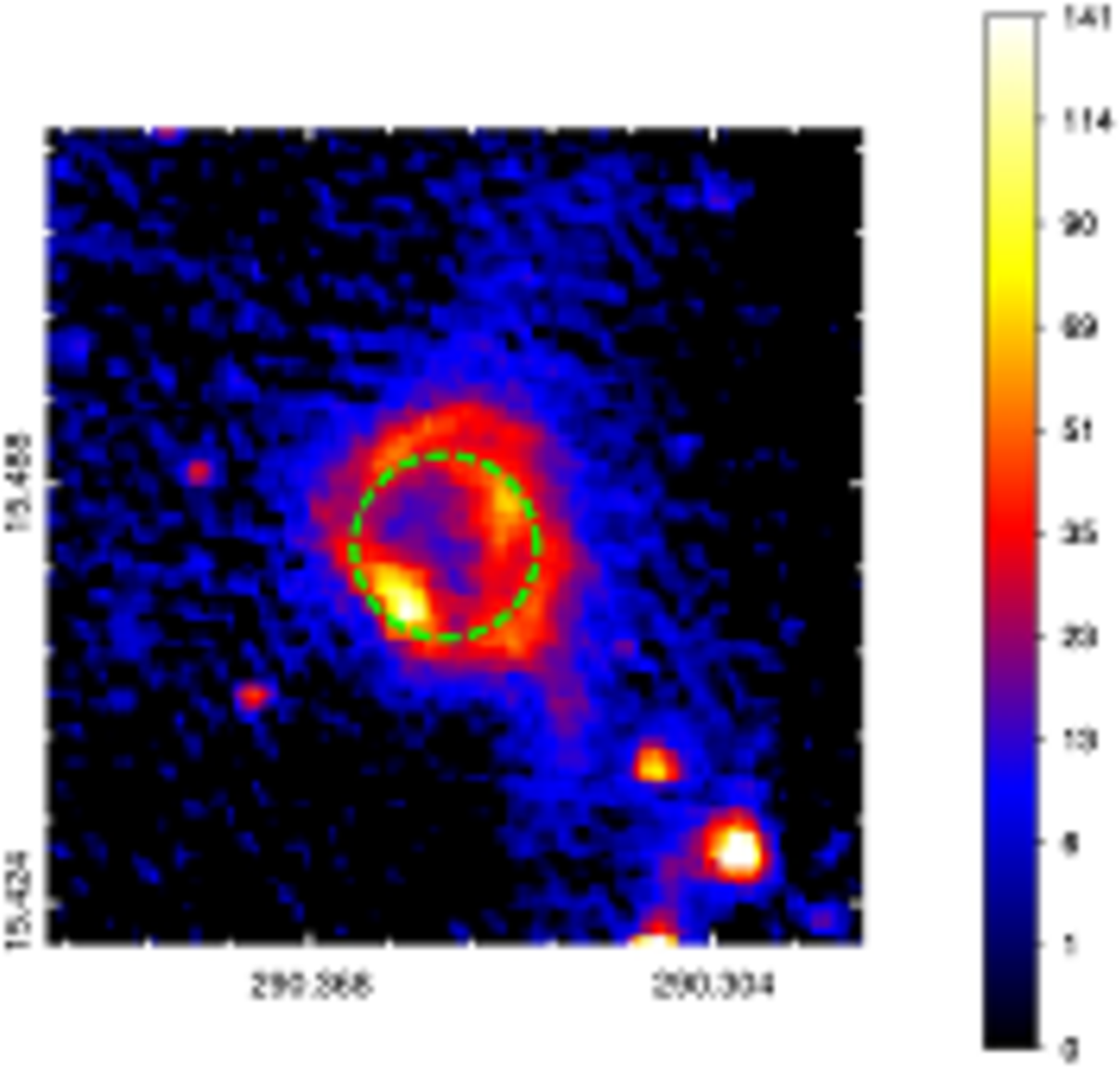}}}%
\mbox{\raisebox{0mm}{\includegraphics[width=40mm]{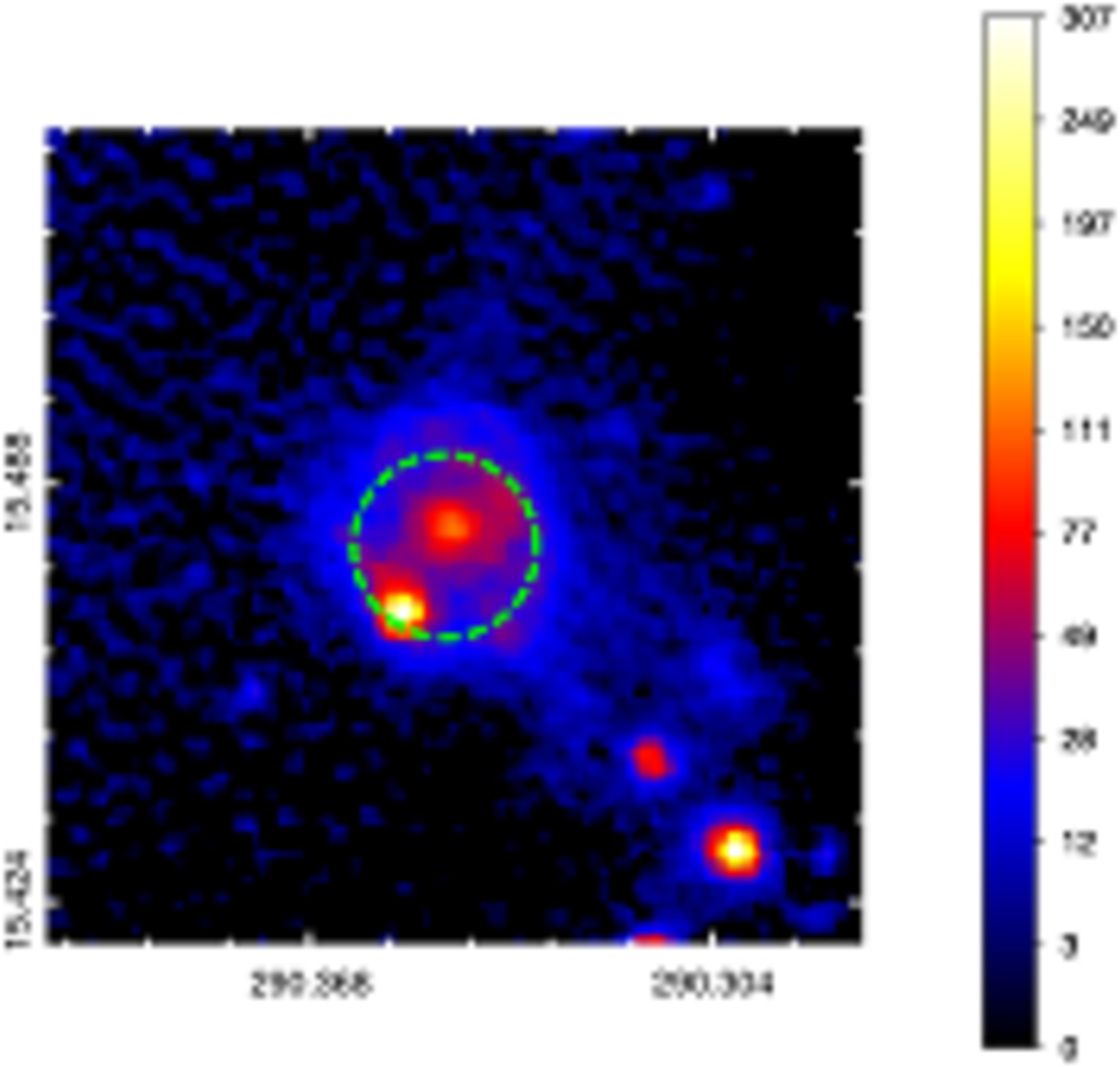}}}%
\mbox{\raisebox{0mm}{\includegraphics[width=40mm]{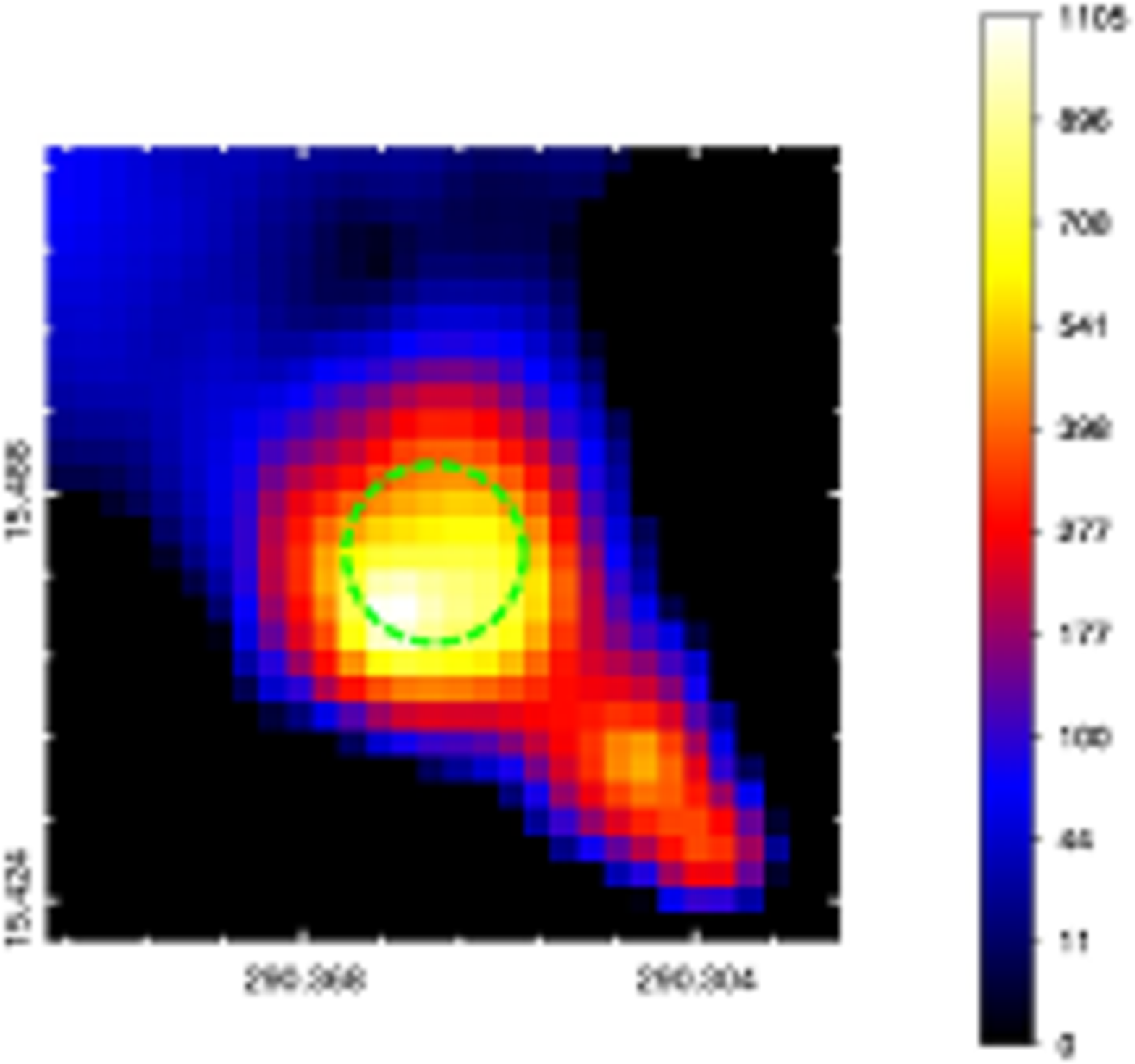}}}%
}
\caption{Continued.} \label{fig:Introfig1:d}
\end{figure*}

\addtocounter{figure}{-1}
\begin{figure*}[ht]
\addtocounter{subfigure}{1}
\centering
\subfigure{
\makebox[180mm][l]{\raisebox{0mm}[0mm][0mm]{ \hspace{15mm} \small{8 \mic}} \hspace{29.5mm} \small{9 \mic} \hspace{27mm} \small{18 \mic} \hspace{26.5mm} \small{90 \mic}}%
}
\subfigure{
\makebox[180mm][l]{\raisebox{0mm}[0mm][0mm]{ \hspace{11mm} \small{RA (J2000)}} \hspace{19.5mm} \small{RA (J2000)} \hspace{20mm} \small{RA (J2000)} \hspace{20mm} \small{RA (J2000)}}%
}
\subfigure{
\mbox{\raisebox{6mm}{\rotatebox{90}{\small{DEC (J2000)}}}}%
\mbox{\raisebox{0mm}{\includegraphics[width=40mm]{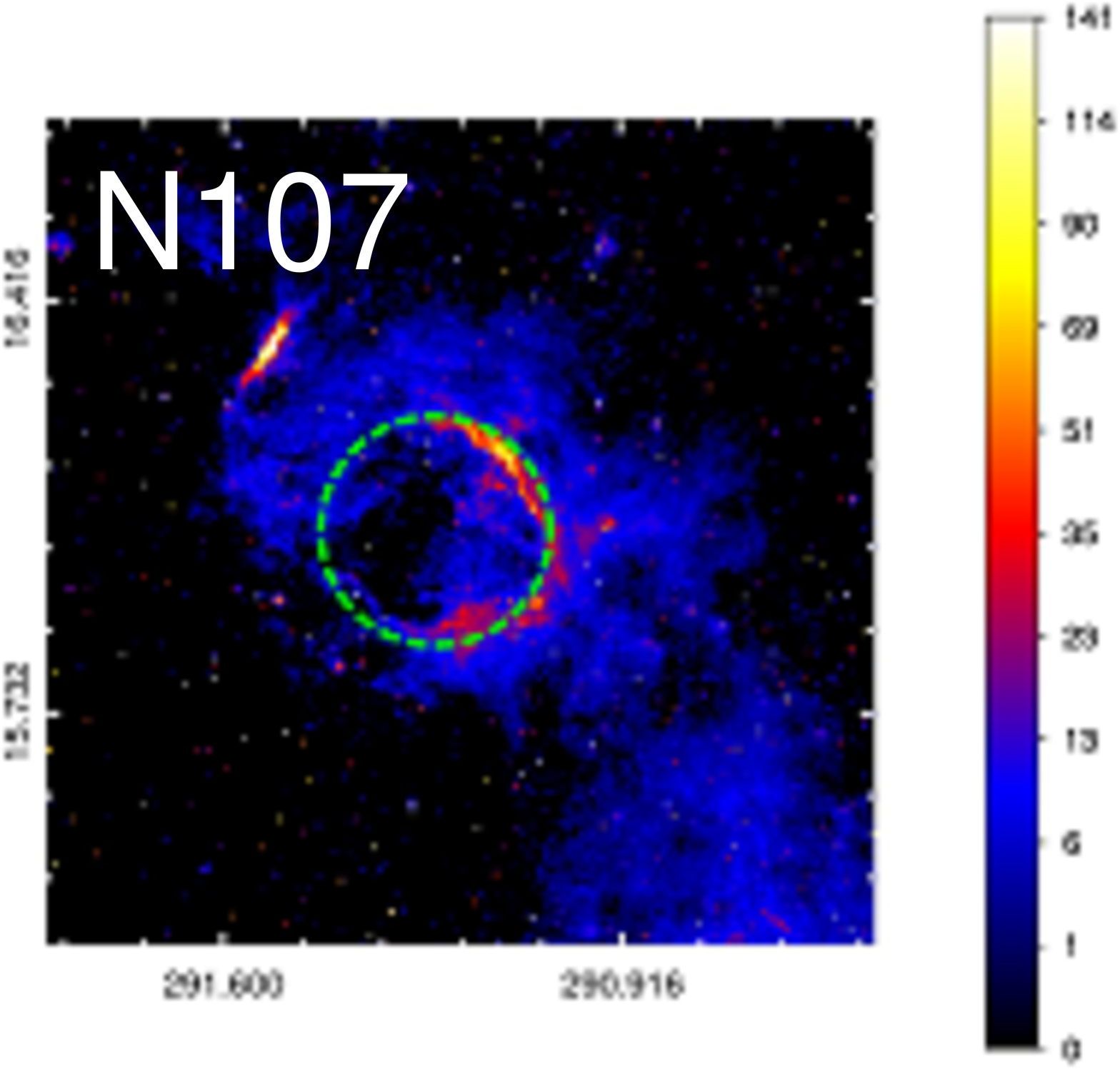}}}%
\mbox{\raisebox{0mm}{\includegraphics[width=40mm]{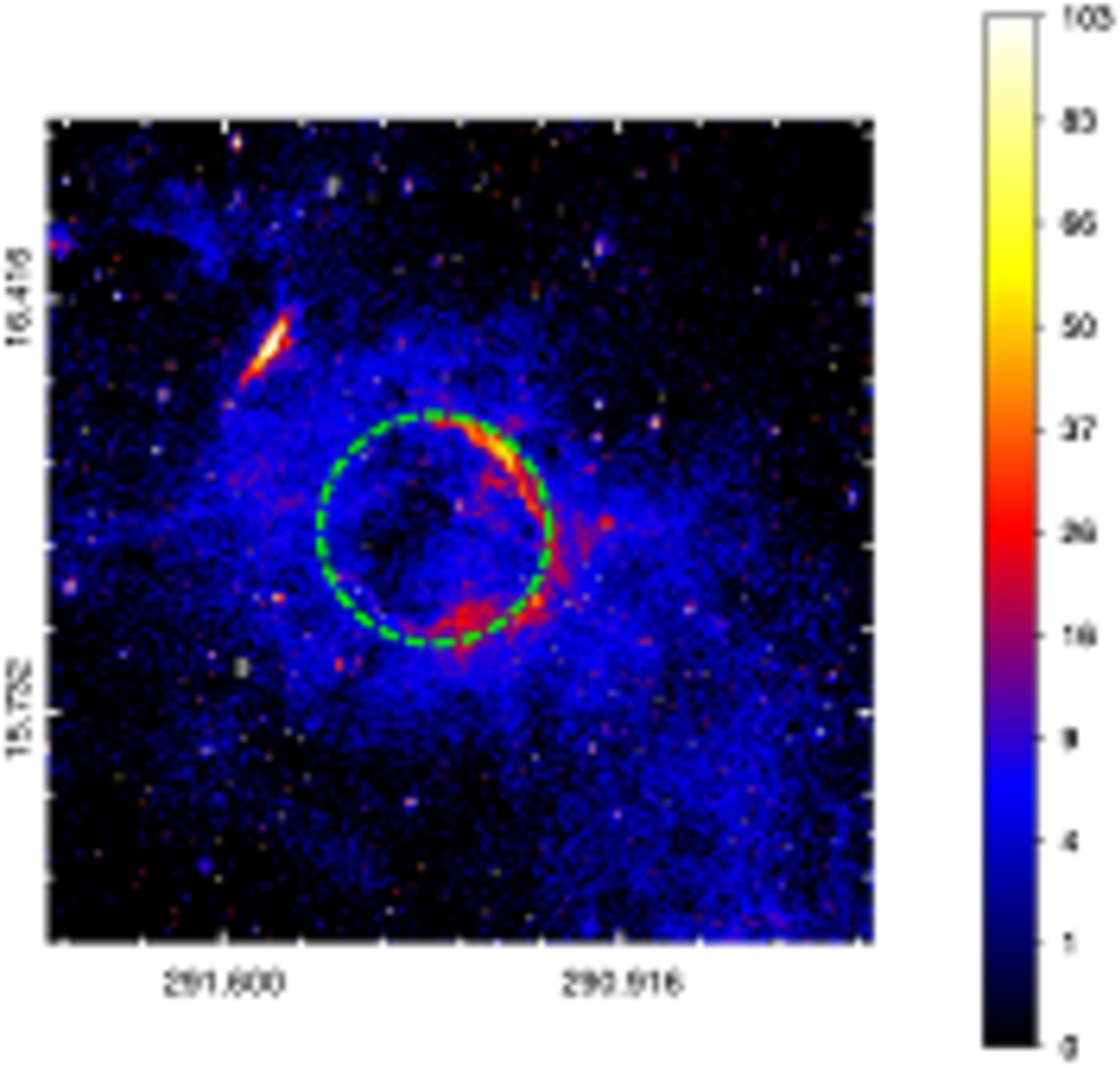}}}%
\mbox{\raisebox{0mm}{\includegraphics[width=40mm]{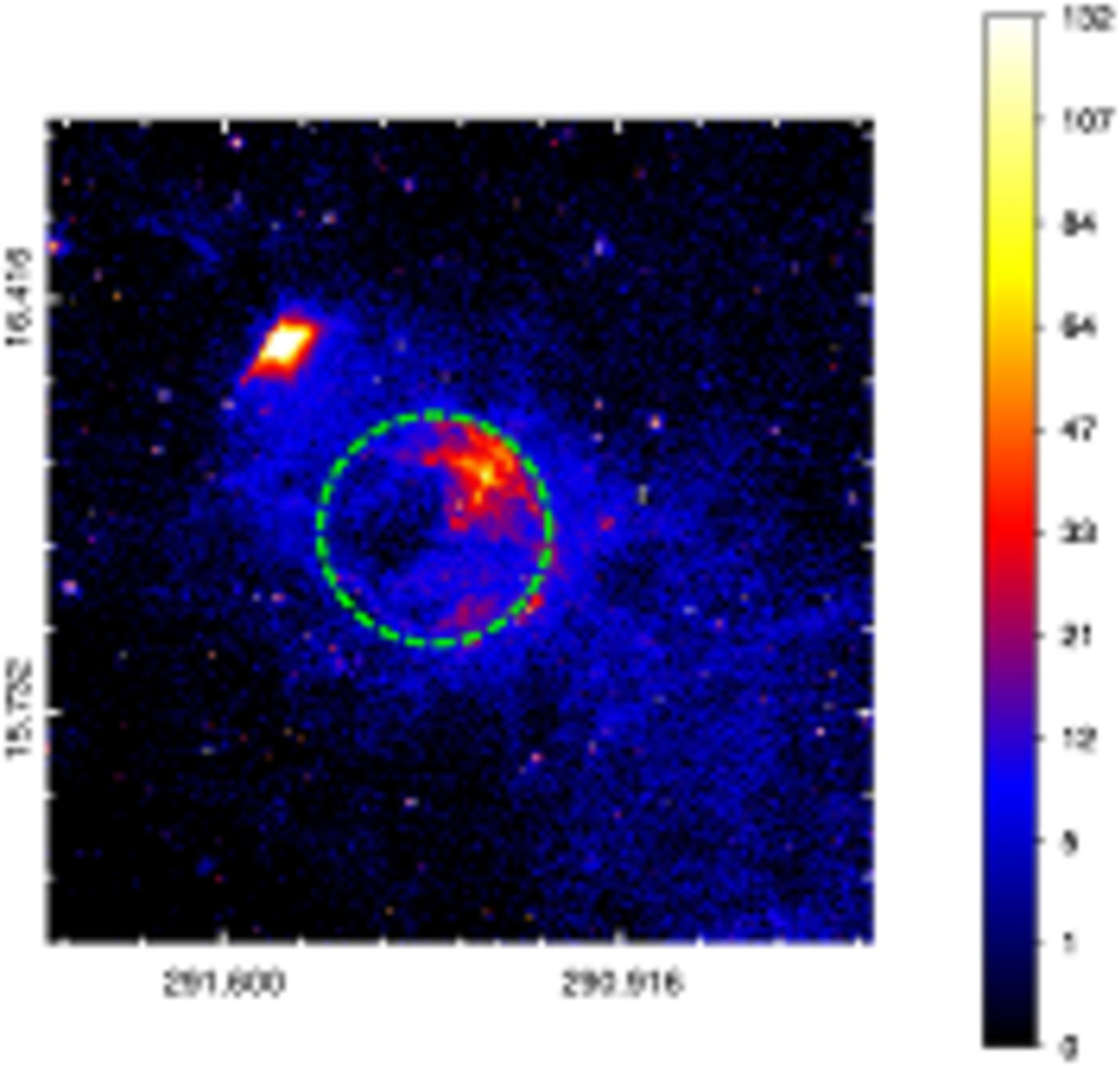}}}%
\mbox{\raisebox{0mm}{\includegraphics[width=40mm]{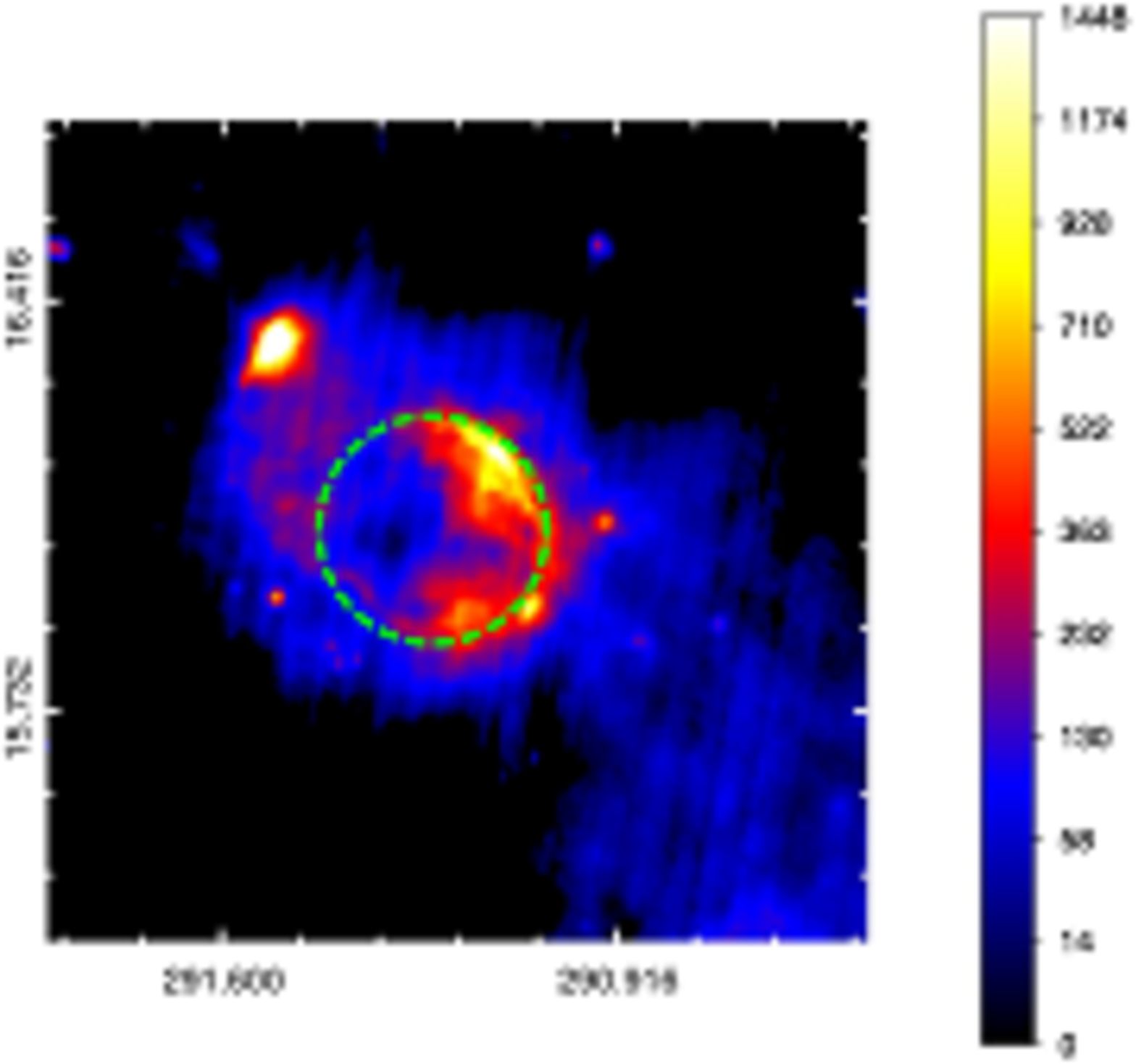}}}%
}
\subfigure{
\mbox{\raisebox{6mm}{\rotatebox{90}{\small{DEC (J2000)}}}}%
\mbox{\raisebox{0mm}{\includegraphics[width=40mm]{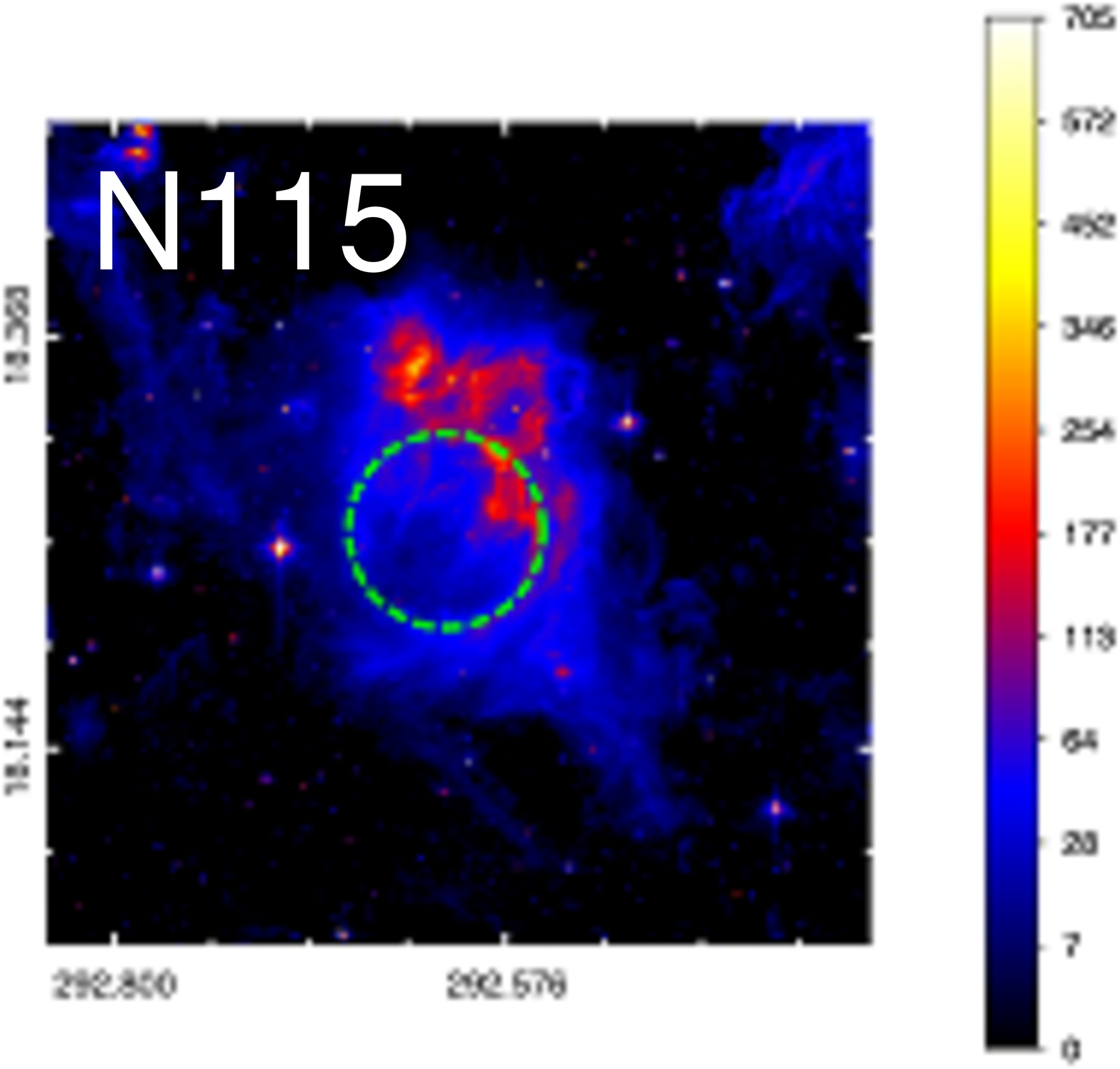}}}%
\mbox{\raisebox{0mm}{\includegraphics[width=40mm]{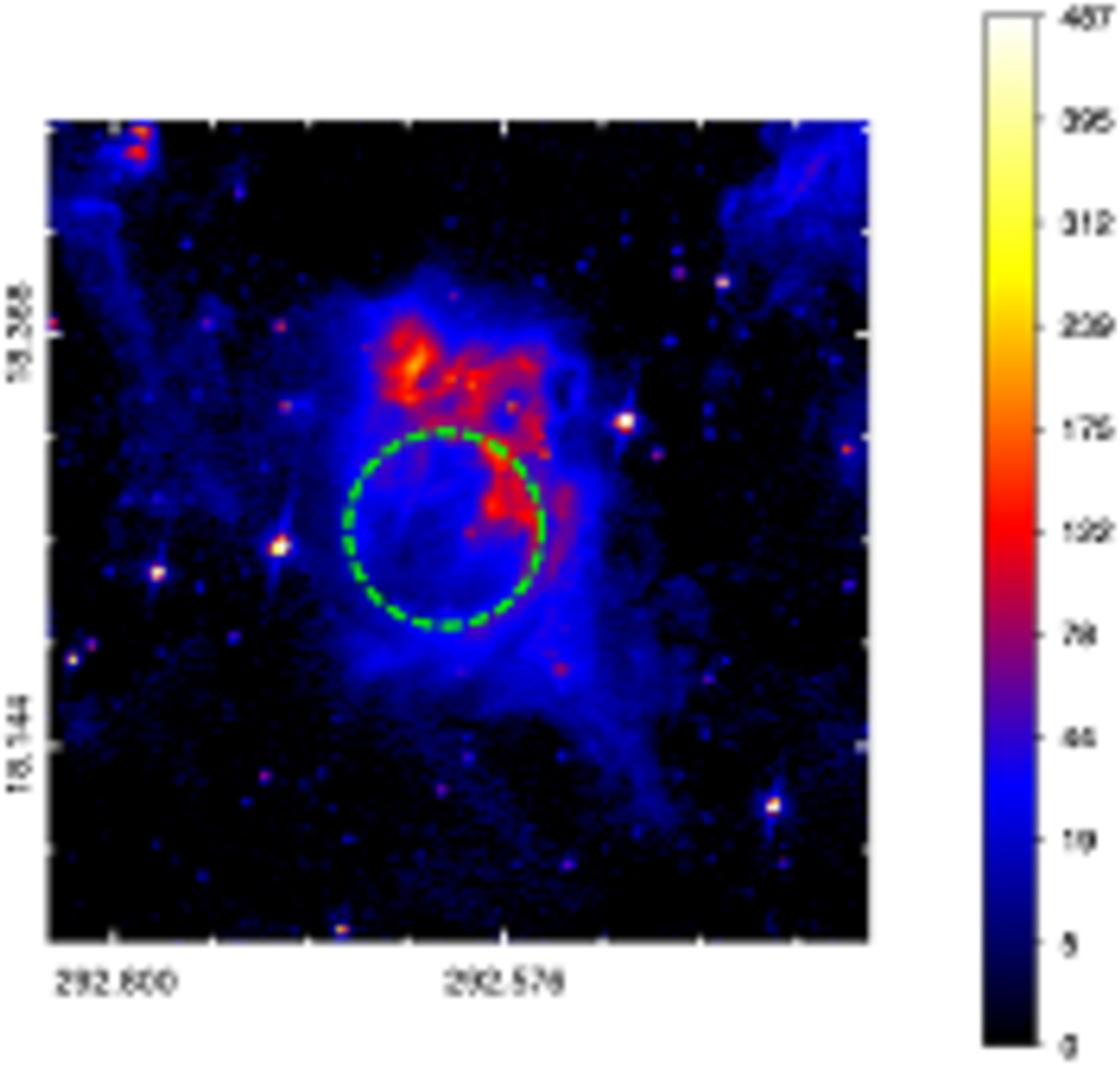}}}%
\mbox{\raisebox{0mm}{\includegraphics[width=40mm]{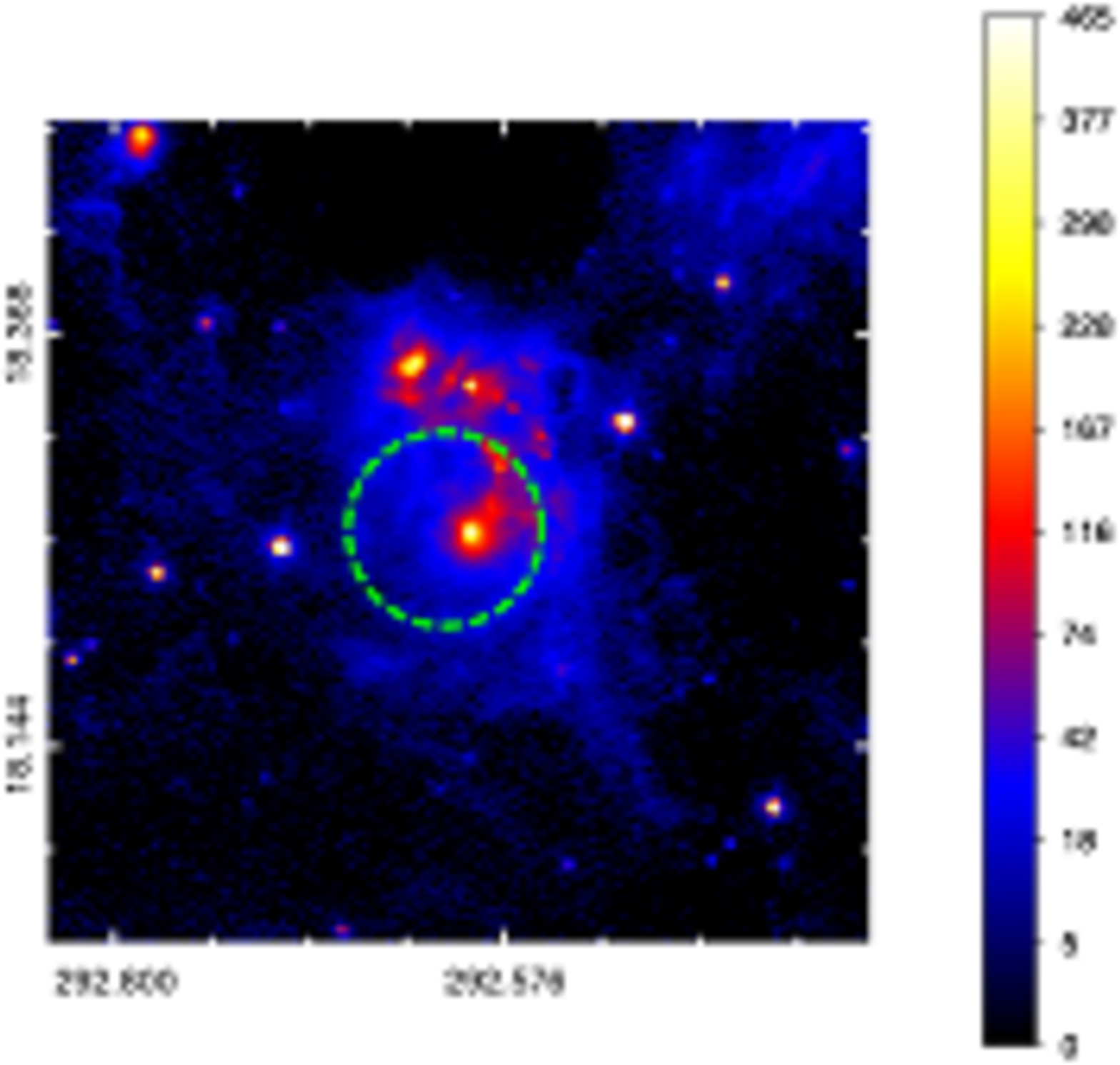}}}%
\mbox{\raisebox{0mm}{\includegraphics[width=40mm]{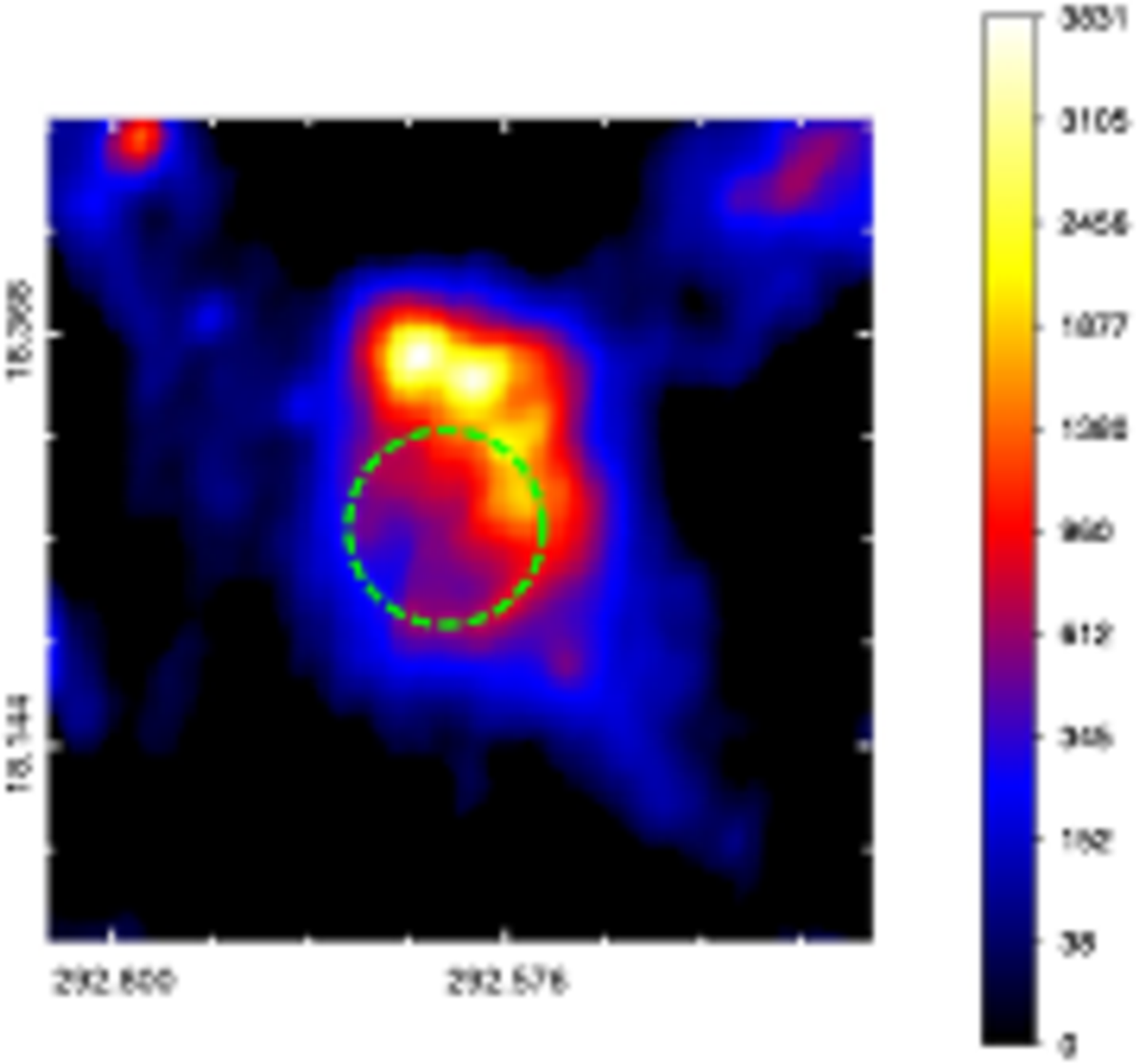}}}%
}
\subfigure{
\mbox{\raisebox{6mm}{\rotatebox{90}{\small{DEC (J2000)}}}}%
\mbox{\raisebox{0mm}{\includegraphics[width=40mm]{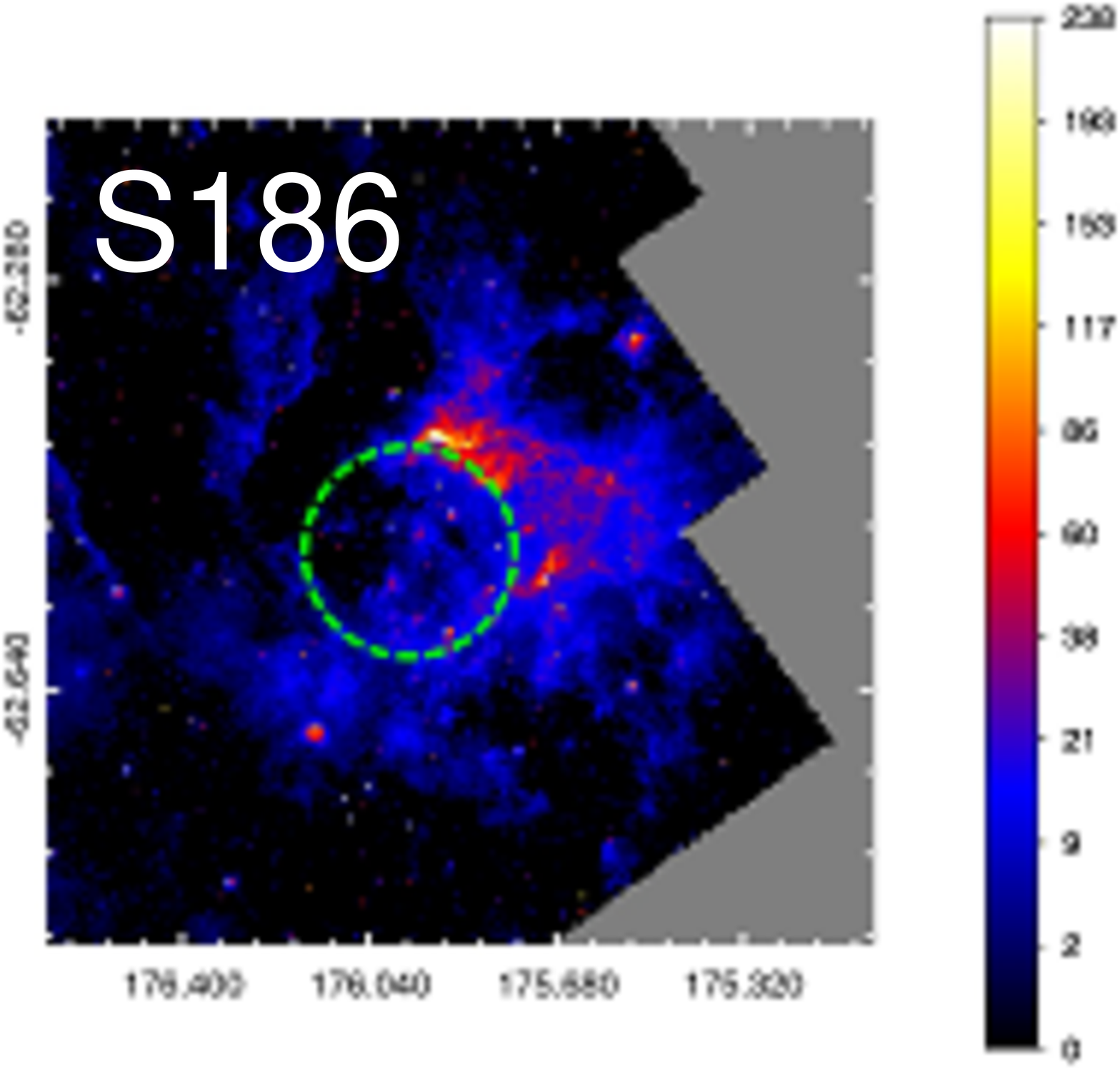}}}%
\mbox{\raisebox{0mm}{\includegraphics[width=40mm]{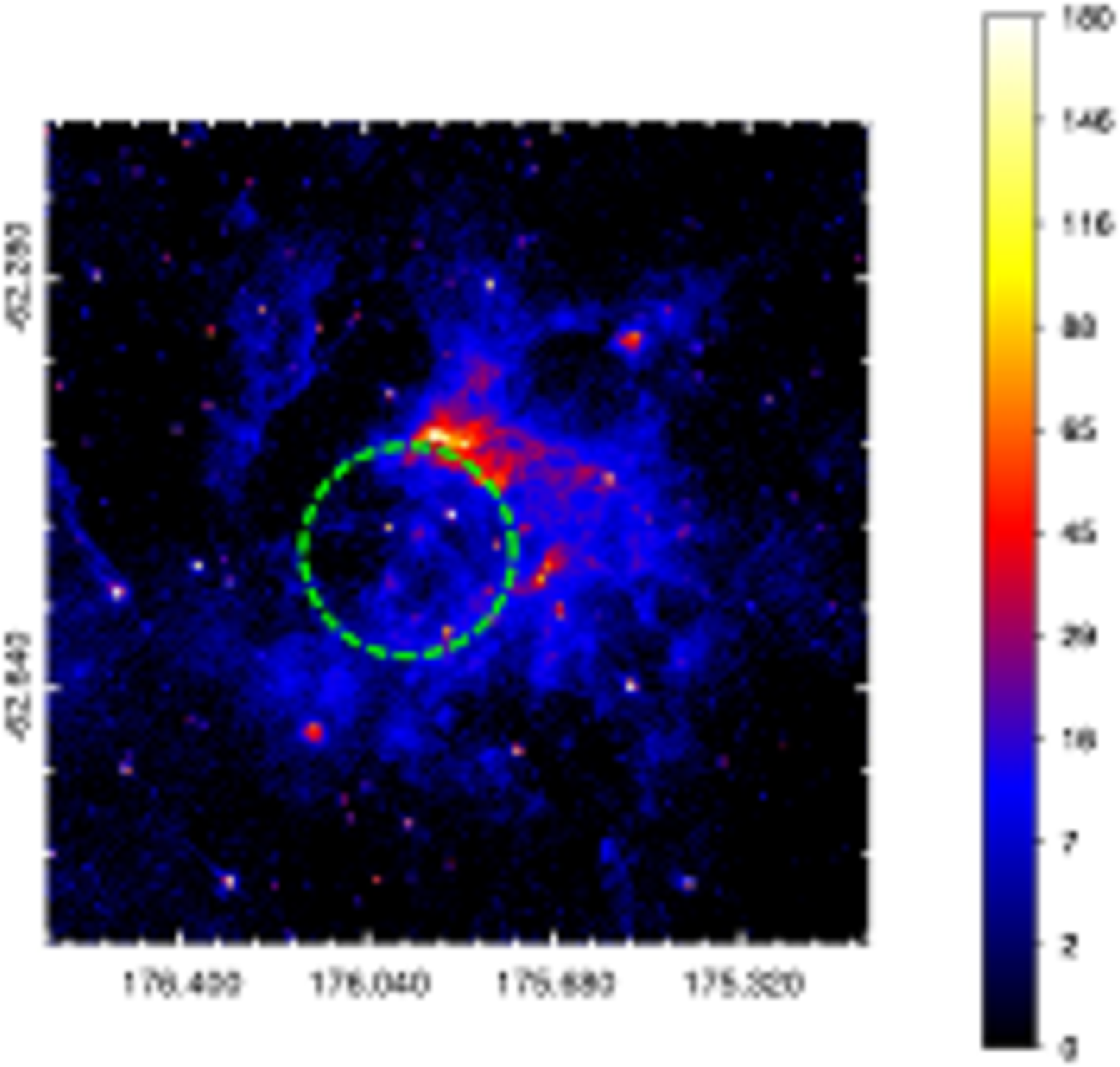}}}%
\mbox{\raisebox{0mm}{\includegraphics[width=40mm]{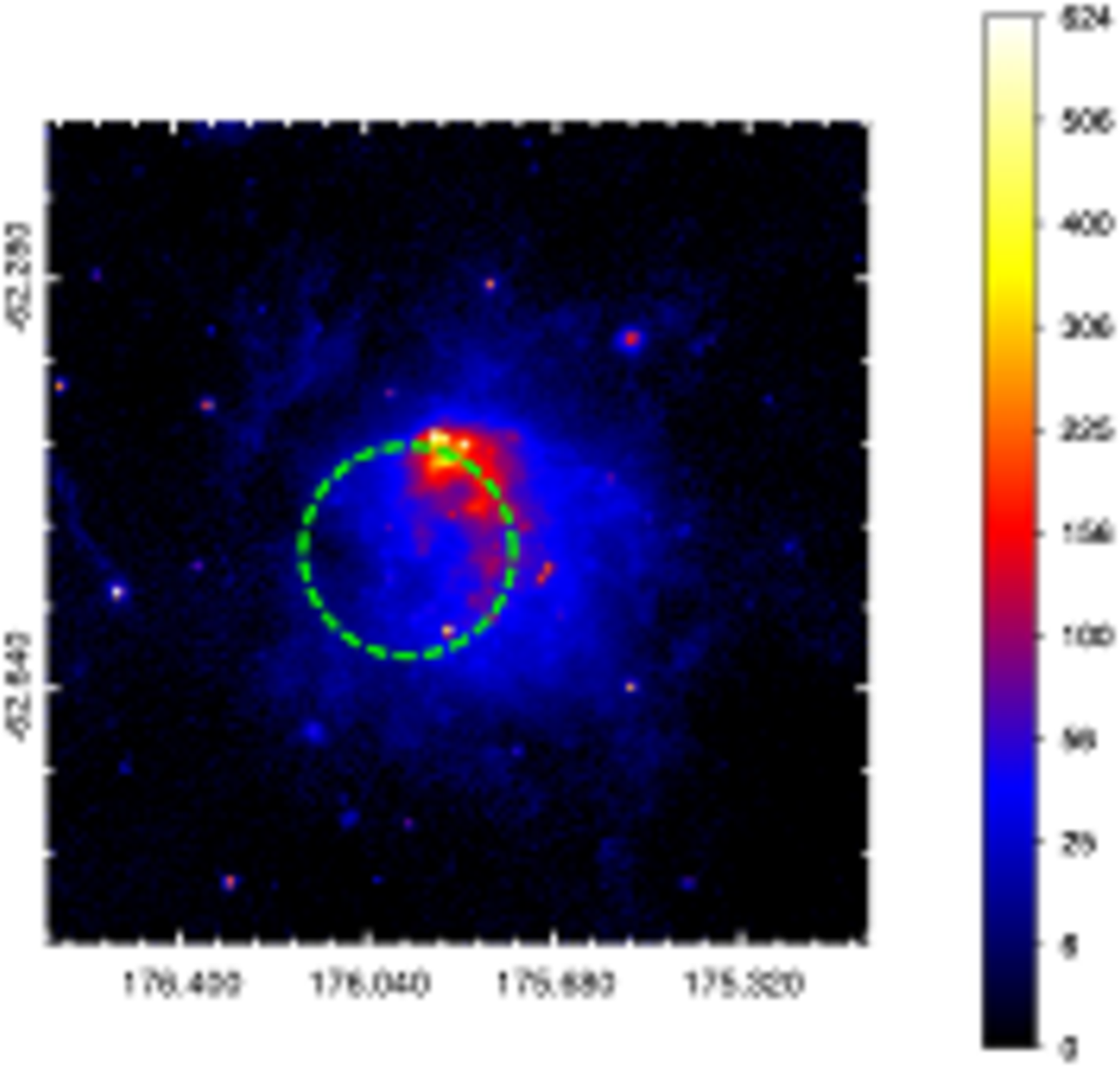}}}%
\mbox{\raisebox{0mm}{\includegraphics[width=40mm]{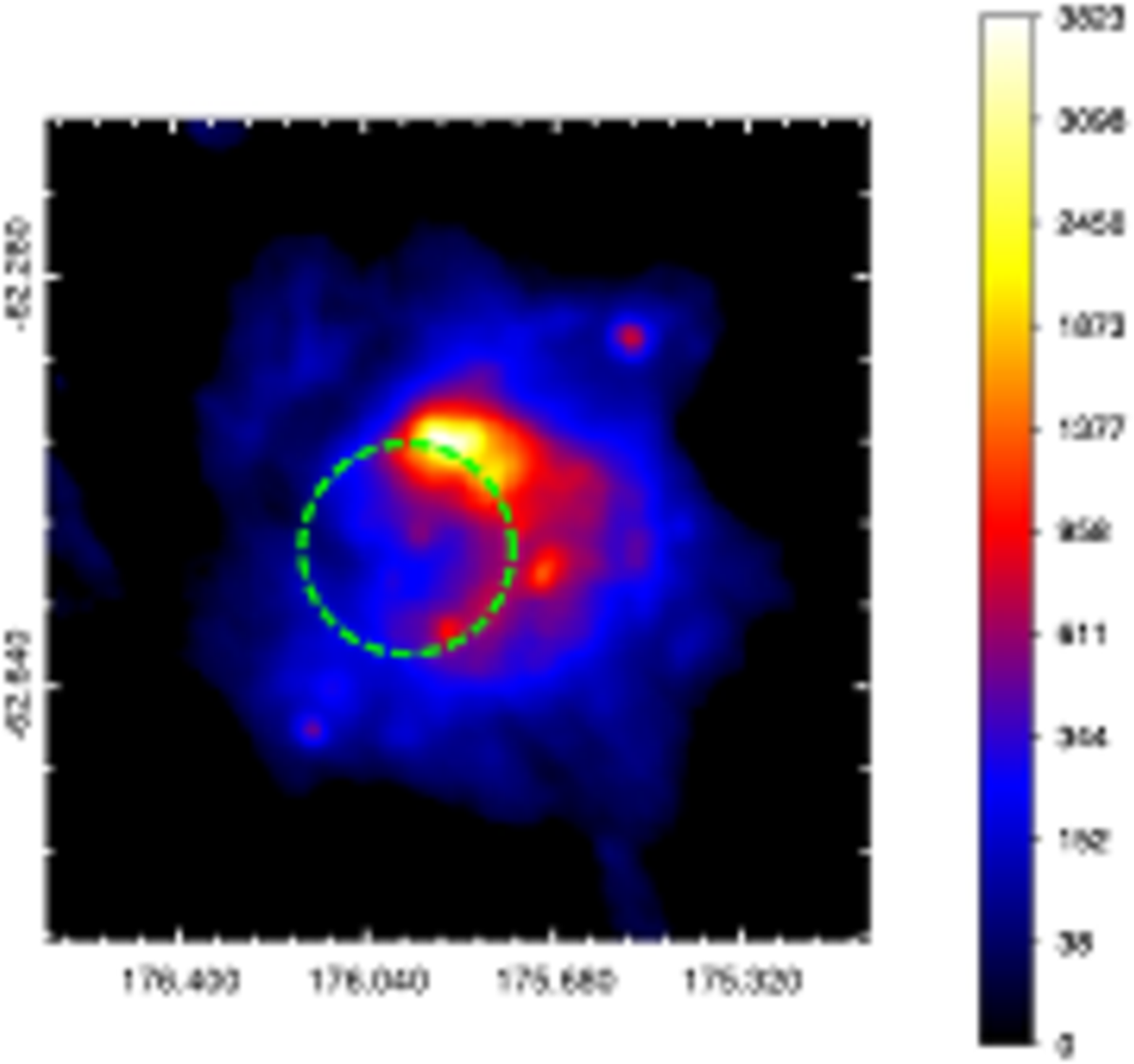}}}%
}
\subfigure{
\mbox{\raisebox{6mm}{\rotatebox{90}{\small{DEC (J2000)}}}}%
\mbox{\raisebox{0mm}{\includegraphics[width=40mm]{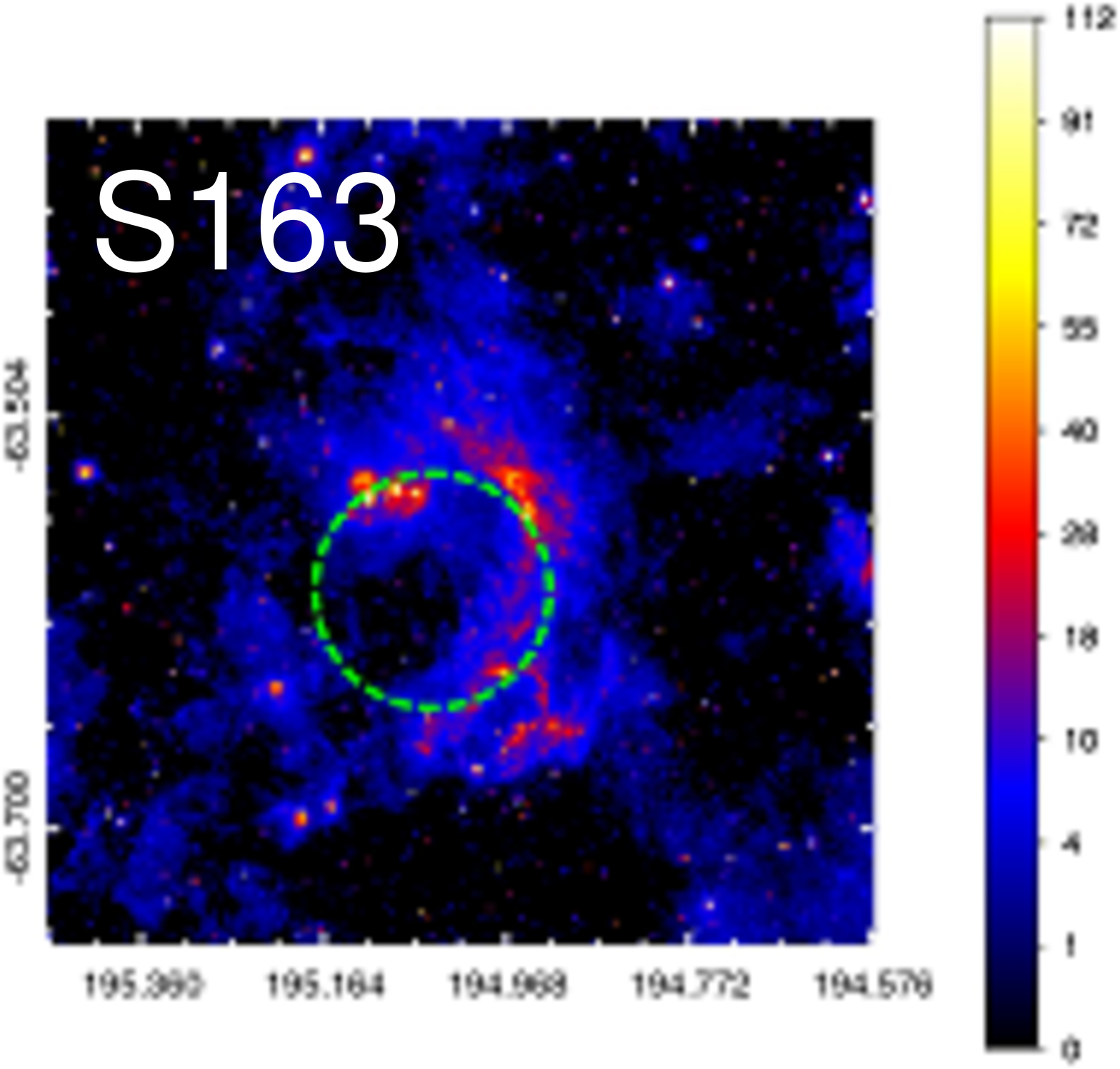}}}%
\mbox{\raisebox{0mm}{\includegraphics[width=40mm]{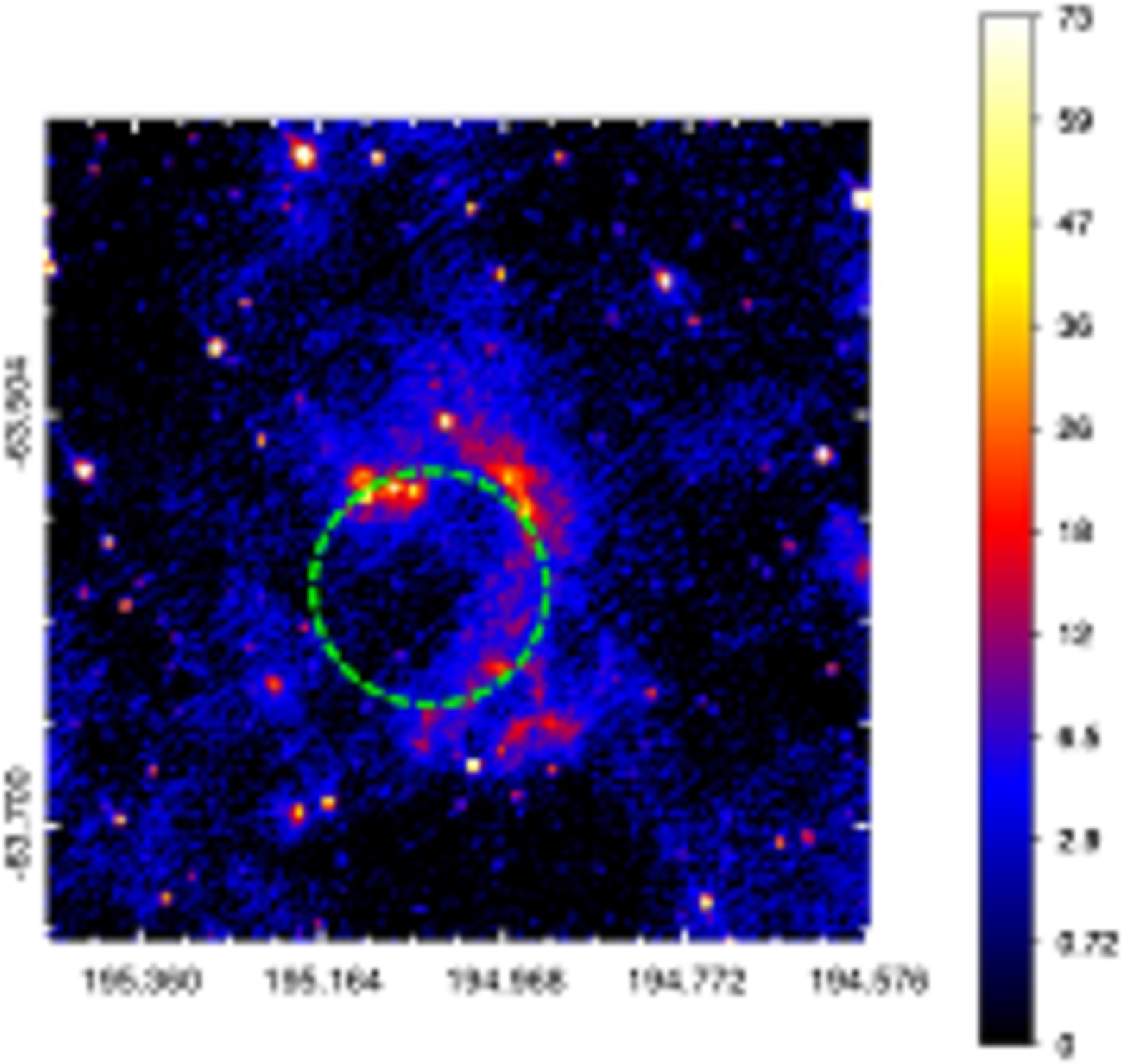}}}%
\mbox{\raisebox{0mm}{\includegraphics[width=40mm]{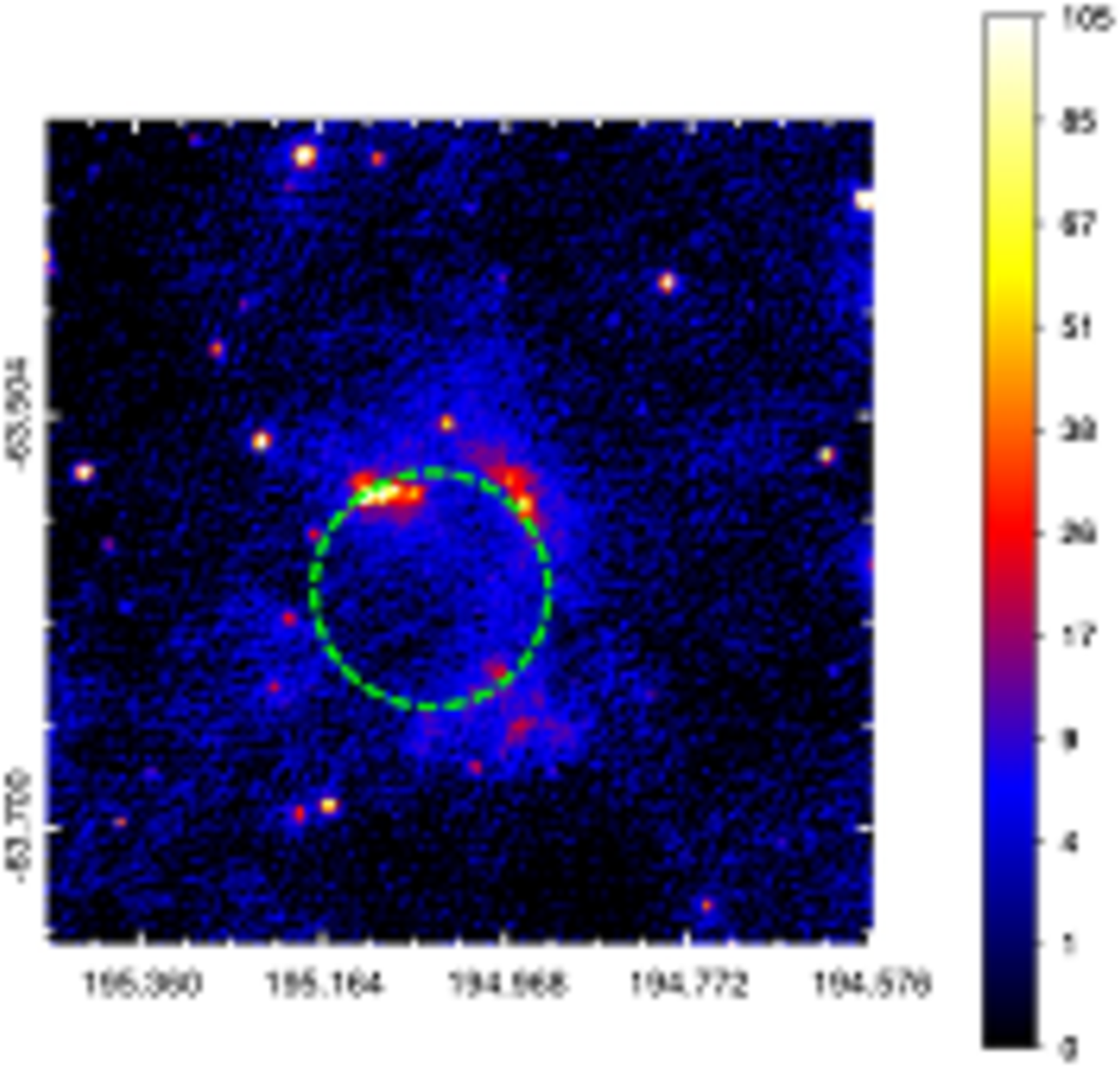}}}%
\mbox{\raisebox{0mm}{\includegraphics[width=40mm]{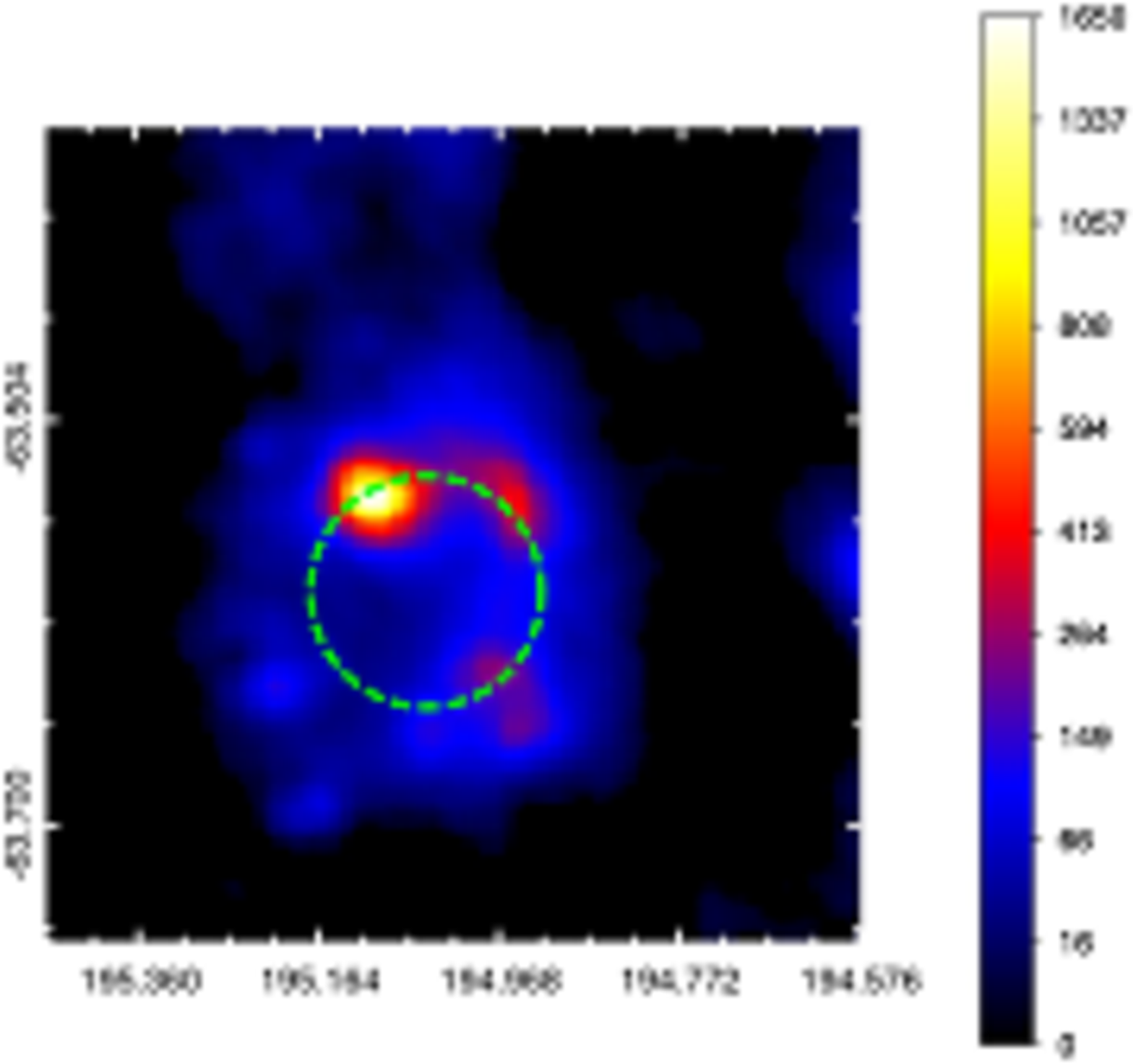}}}%
}
\subfigure{
\mbox{\raisebox{6mm}{\rotatebox{90}{\small{DEC (J2000)}}}}%
\mbox{\raisebox{0mm}{\includegraphics[width=40mm]{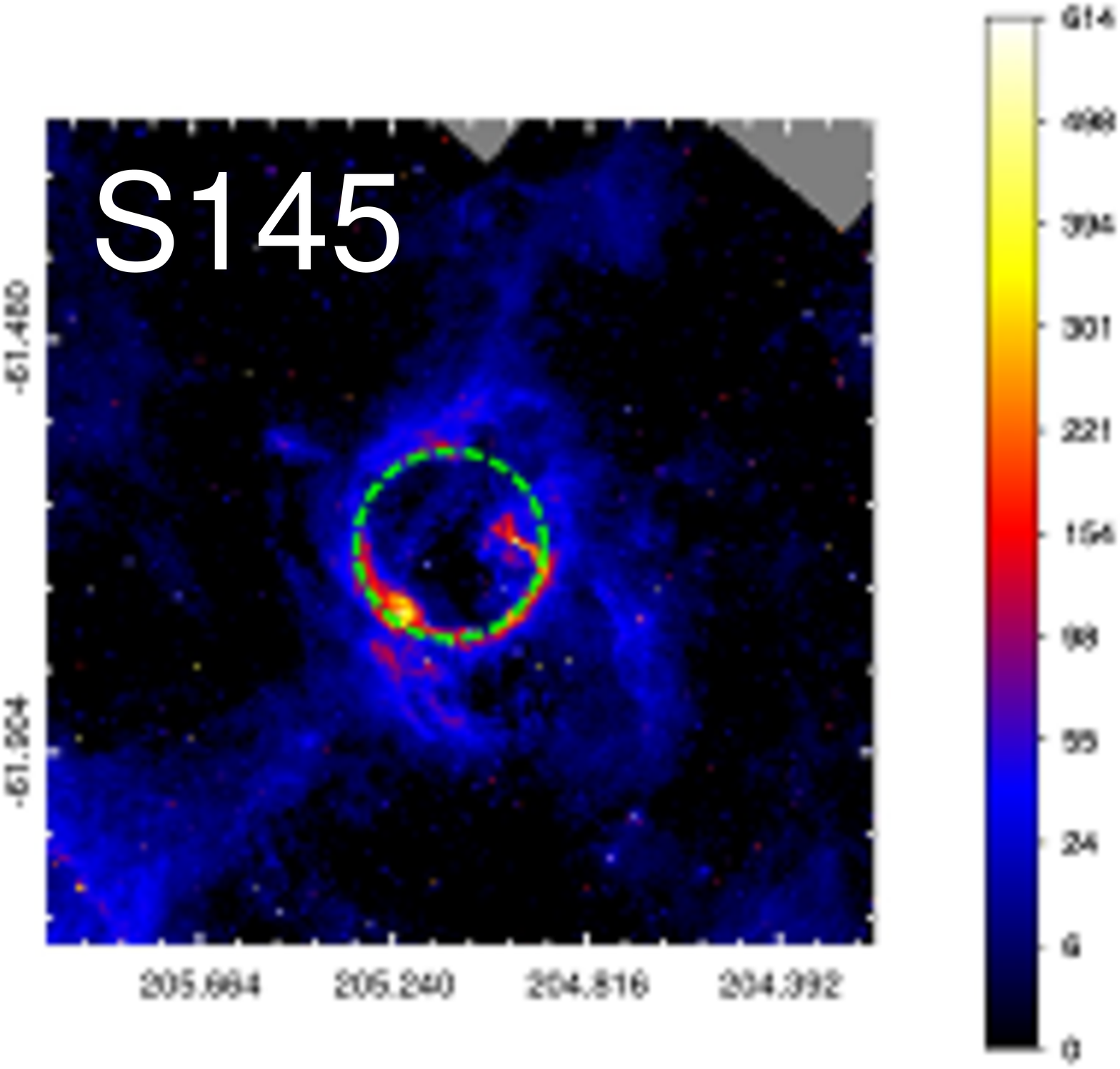}}}%
\mbox{\raisebox{0mm}{\includegraphics[width=40mm]{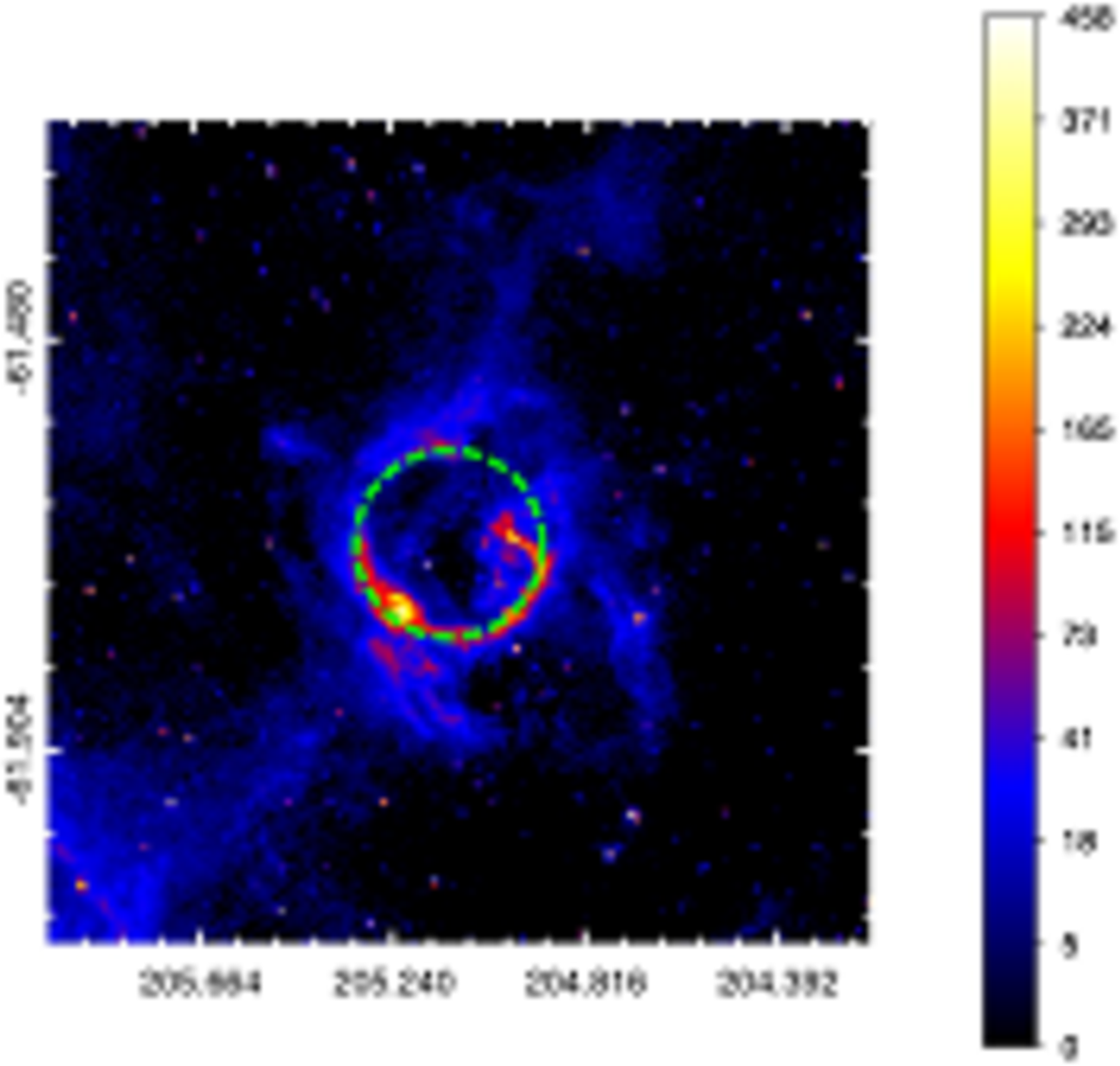}}}%
\mbox{\raisebox{0mm}{\includegraphics[width=40mm]{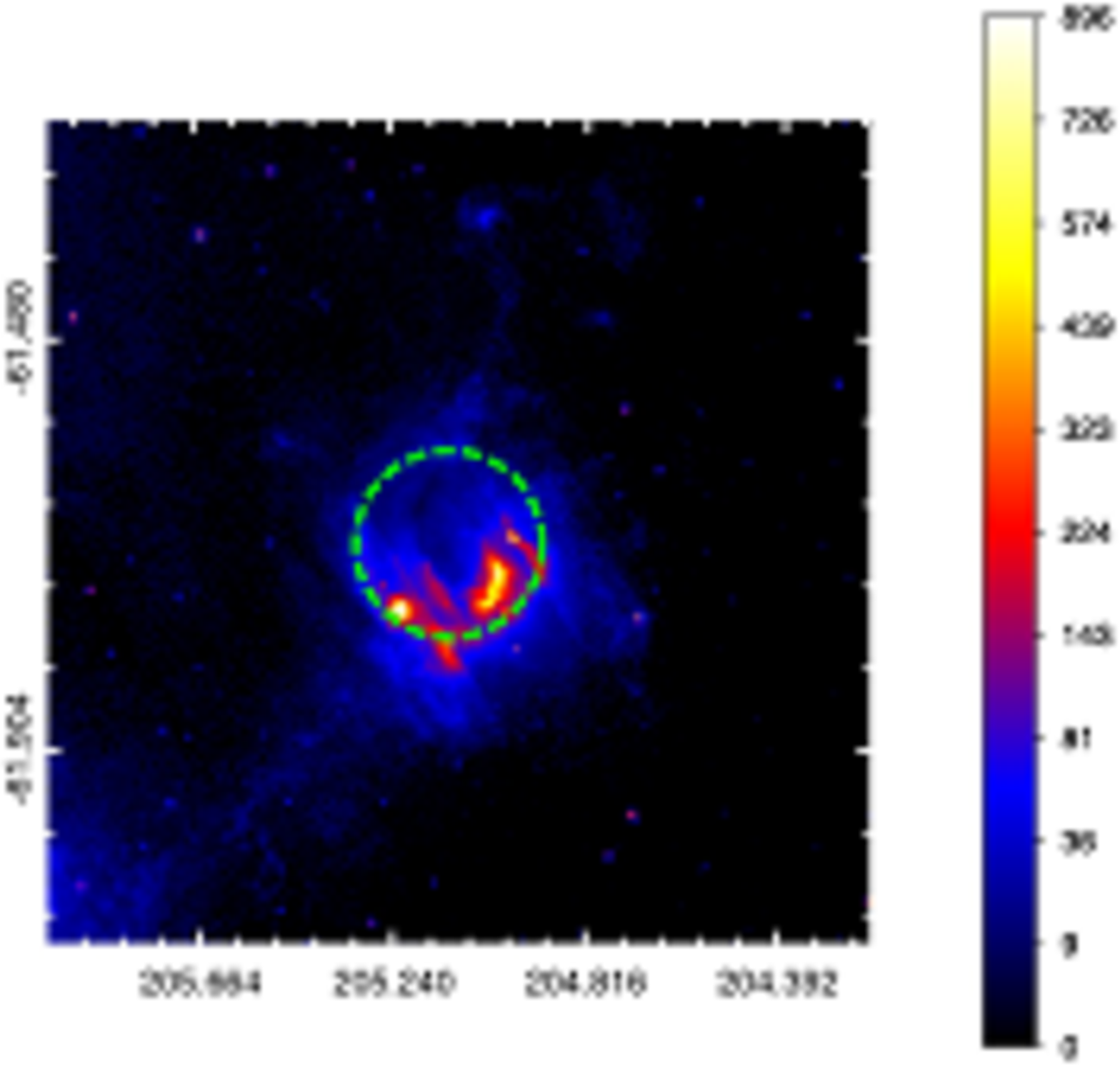}}}%
\mbox{\raisebox{0mm}{\includegraphics[width=40mm]{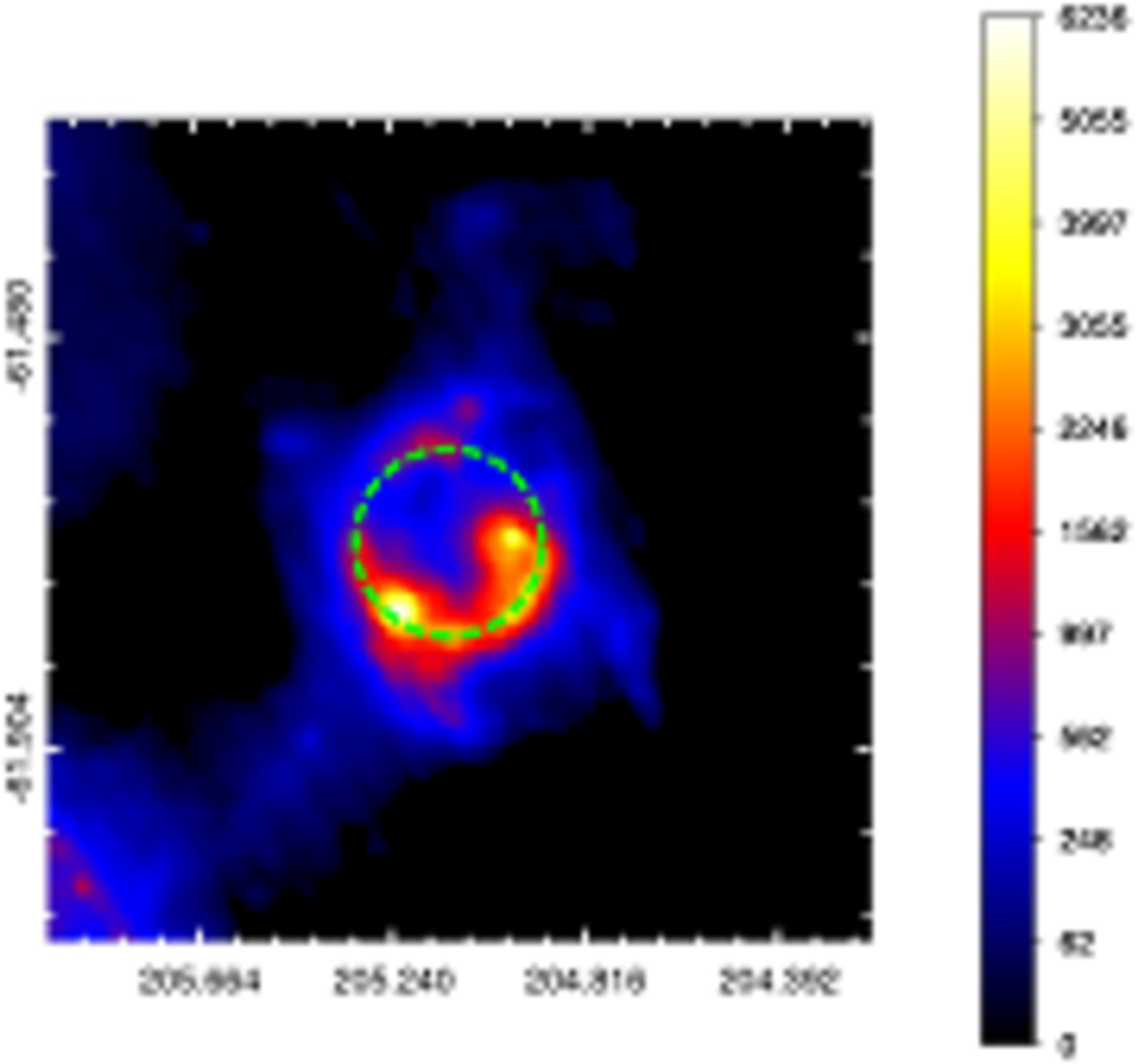}}}%
}
\caption{Continued.} \label{fig:Introfig1:e}
\end{figure*}

\addtocounter{figure}{-1}
\begin{figure*}[ht]
\addtocounter{subfigure}{1}
\centering
\subfigure{
\makebox[180mm][l]{\raisebox{0mm}[0mm][0mm]{ \hspace{15mm} \small{8 \mic}} \hspace{29.5mm} \small{9 \mic} \hspace{27mm} \small{18 \mic} \hspace{26.5mm} \small{90 \mic}}%
}
\subfigure{
\makebox[180mm][l]{\raisebox{0mm}[0mm][0mm]{ \hspace{11mm} \small{RA (J2000)}} \hspace{19.5mm} \small{RA (J2000)} \hspace{20mm} \small{RA (J2000)} \hspace{20mm} \small{RA (J2000)}}%
}
\subfigure{
\mbox{\raisebox{6mm}{\rotatebox{90}{\small{DEC (J2000)}}}}%
\mbox{\raisebox{0mm}{\includegraphics[width=40mm]{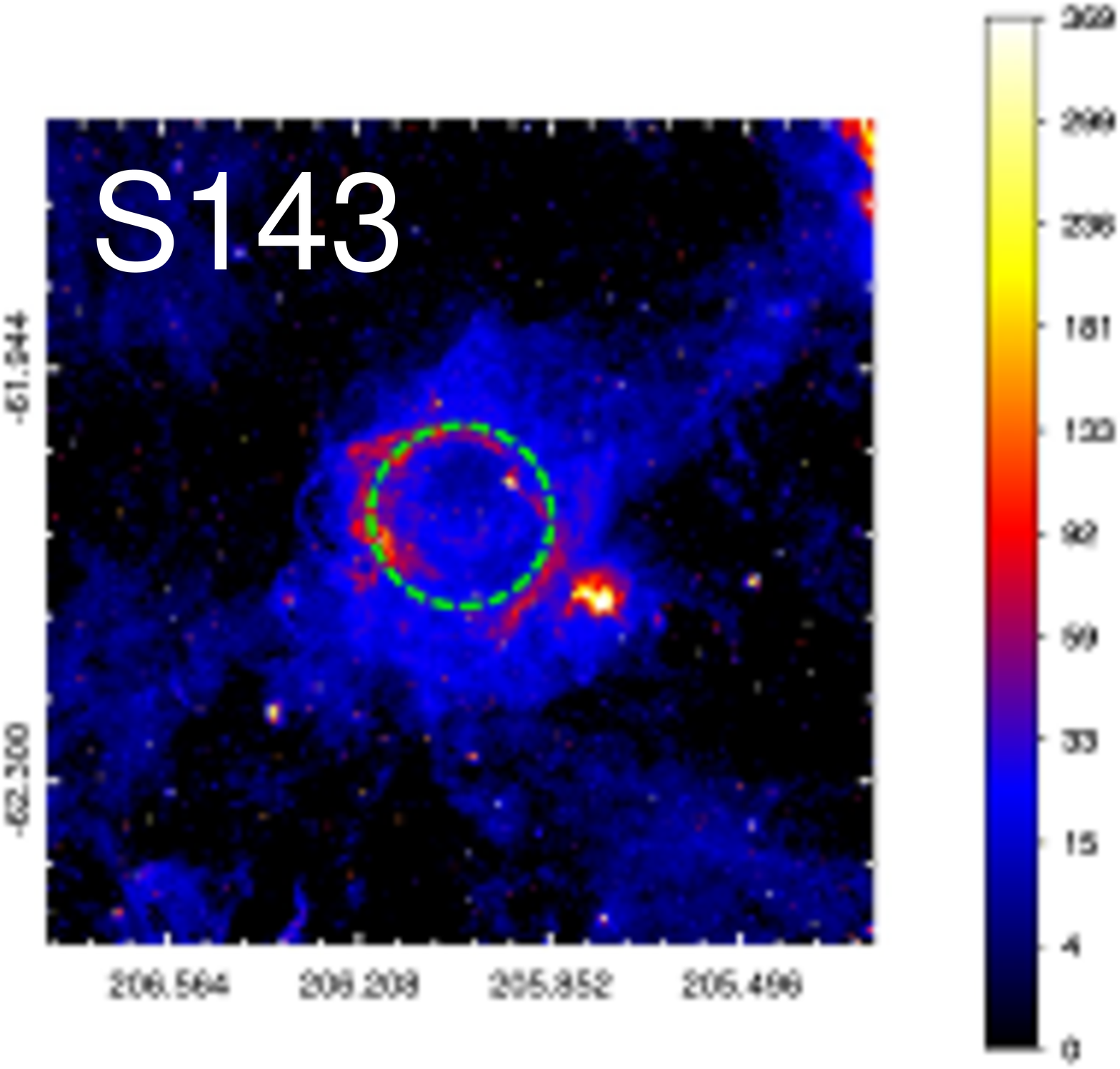}}}%
\mbox{\raisebox{0mm}{\includegraphics[width=40mm]{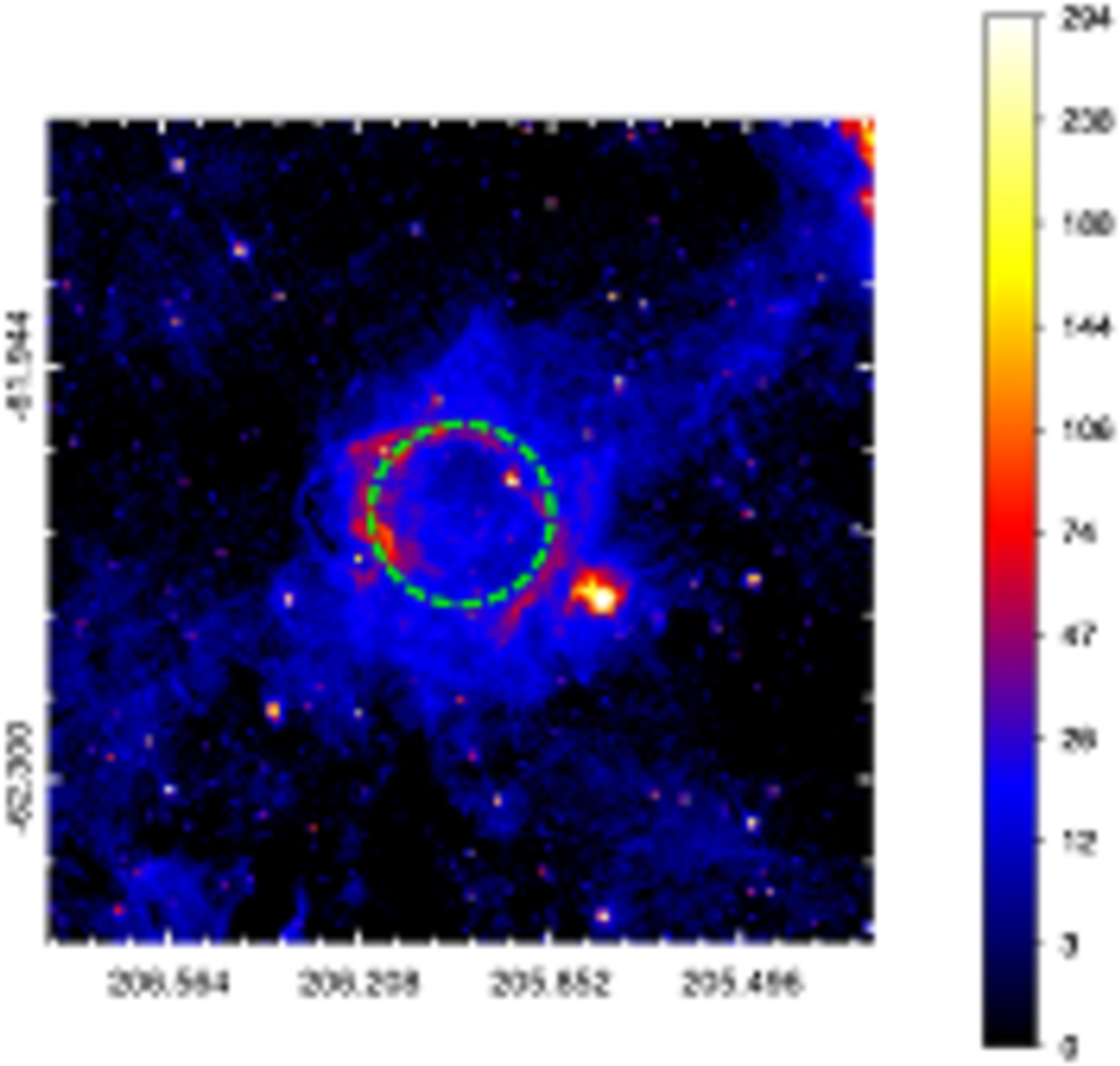}}}%
\mbox{\raisebox{0mm}{\includegraphics[width=40mm]{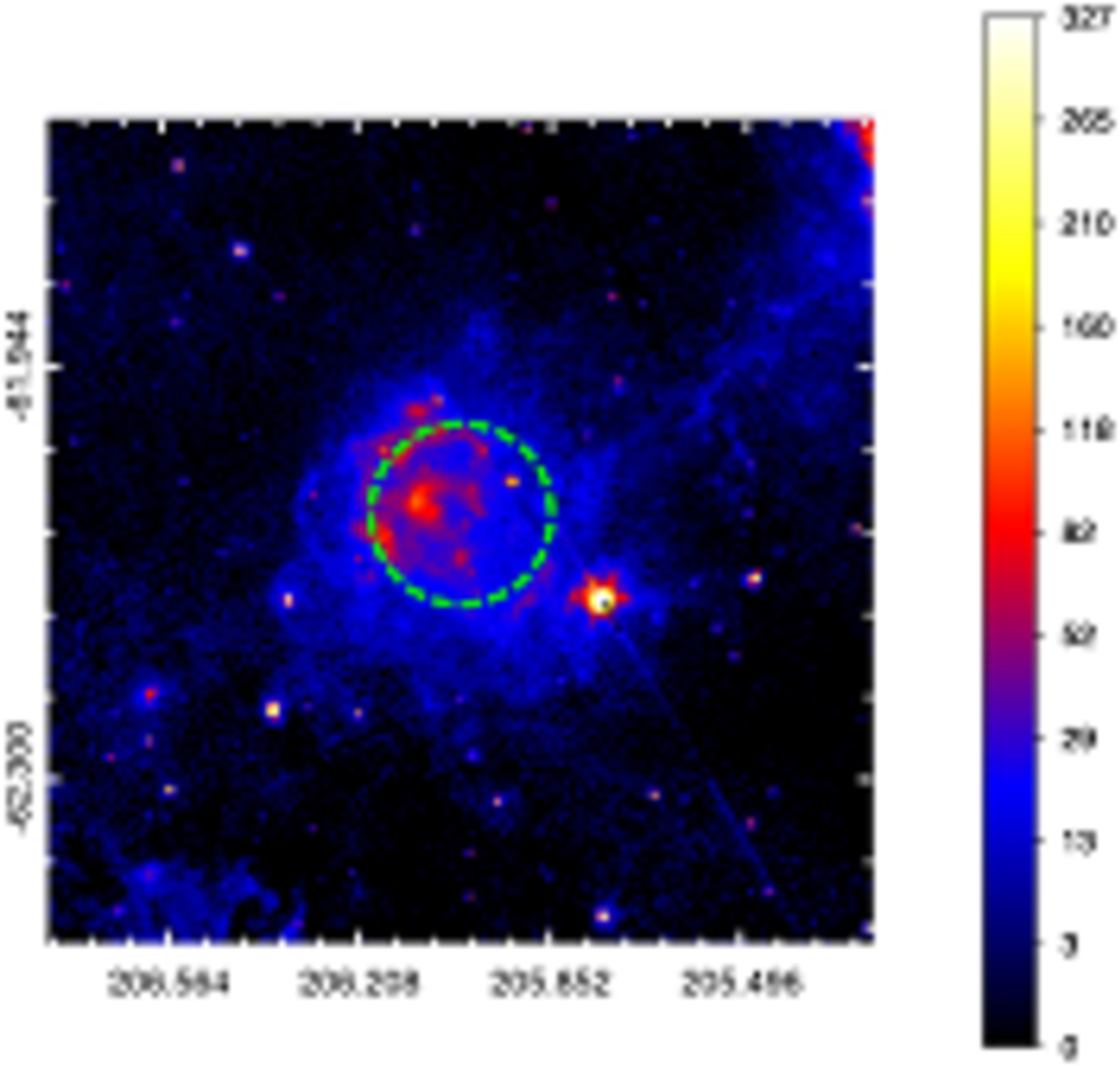}}}%
\mbox{\raisebox{0mm}{\includegraphics[width=40mm]{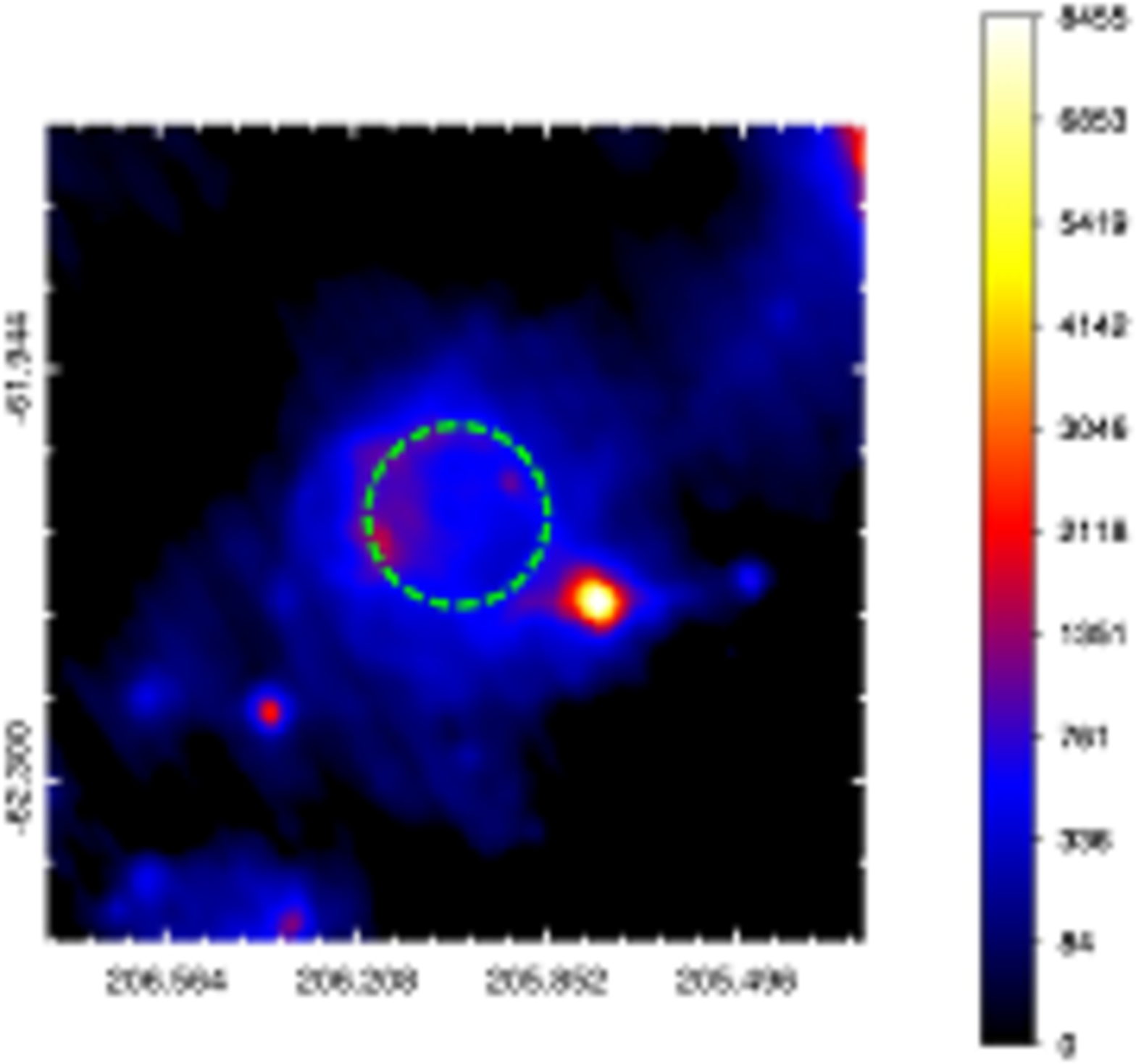}}}%
}
\subfigure{
\mbox{\raisebox{6mm}{\rotatebox{90}{\small{DEC (J2000)}}}}%
\mbox{\raisebox{0mm}{\includegraphics[width=40mm]{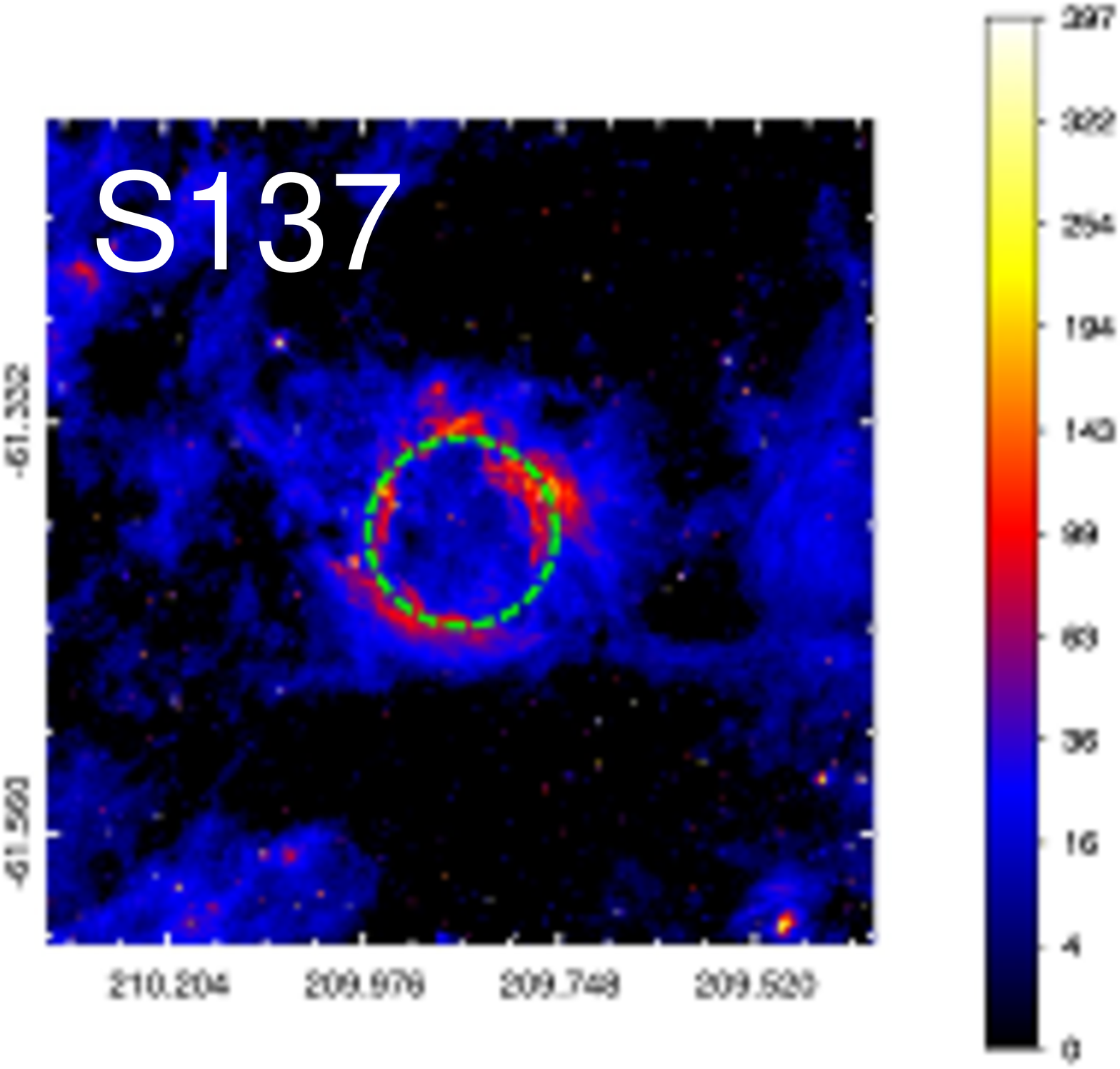}}}%
\mbox{\raisebox{0mm}{\includegraphics[width=40mm]{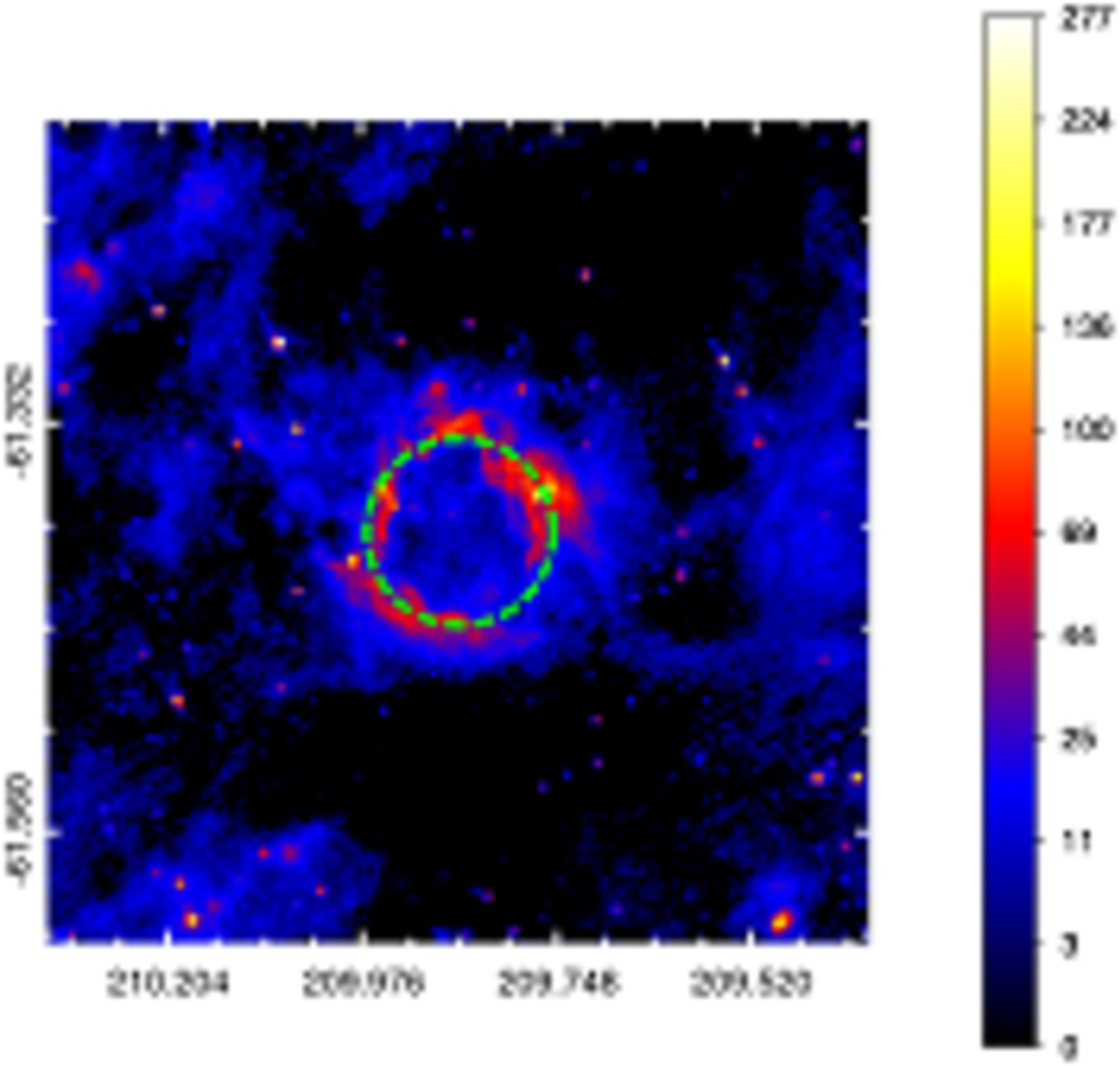}}}%
\mbox{\raisebox{0mm}{\includegraphics[width=40mm]{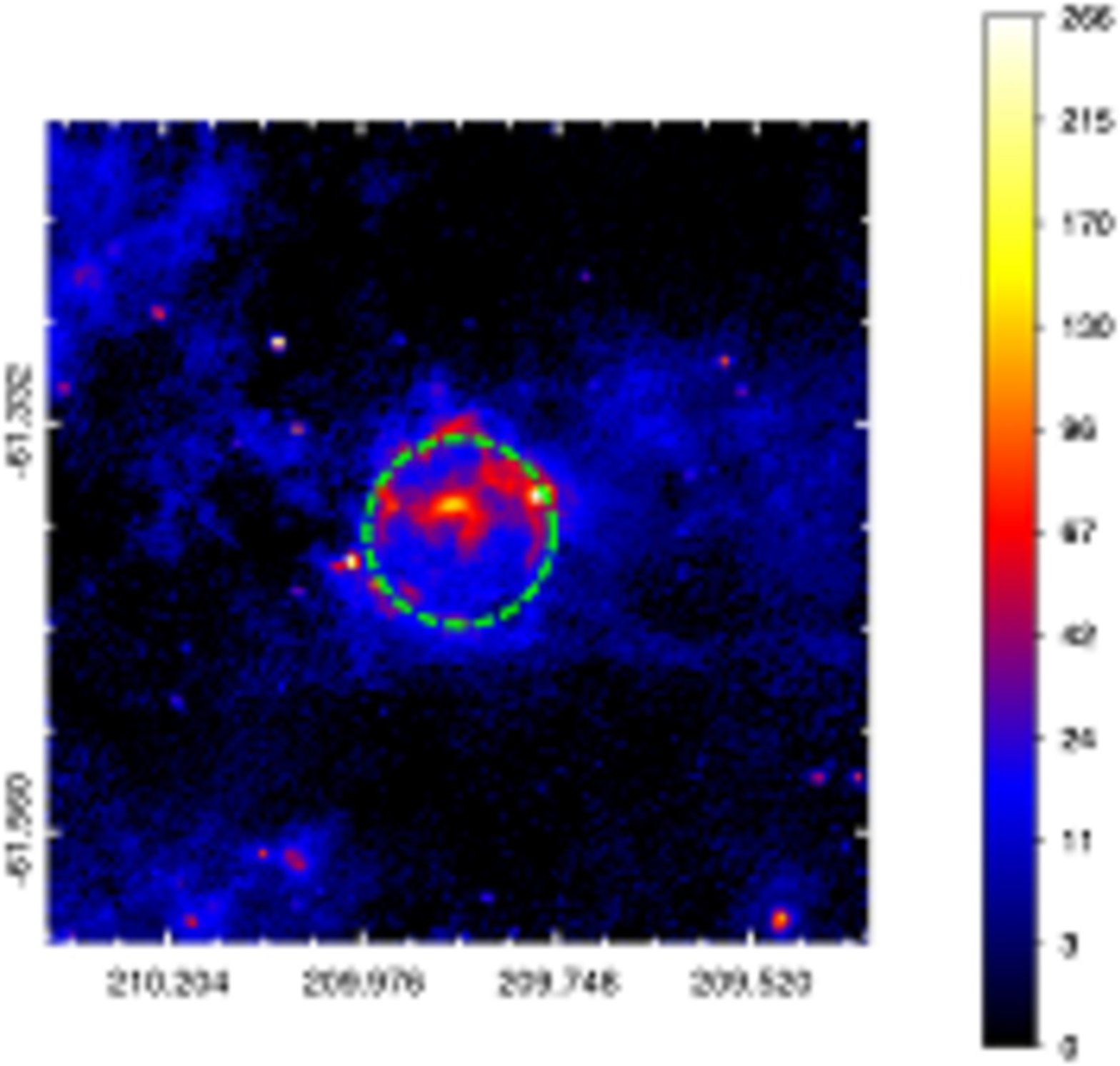}}}%
\mbox{\raisebox{0mm}{\includegraphics[width=40mm]{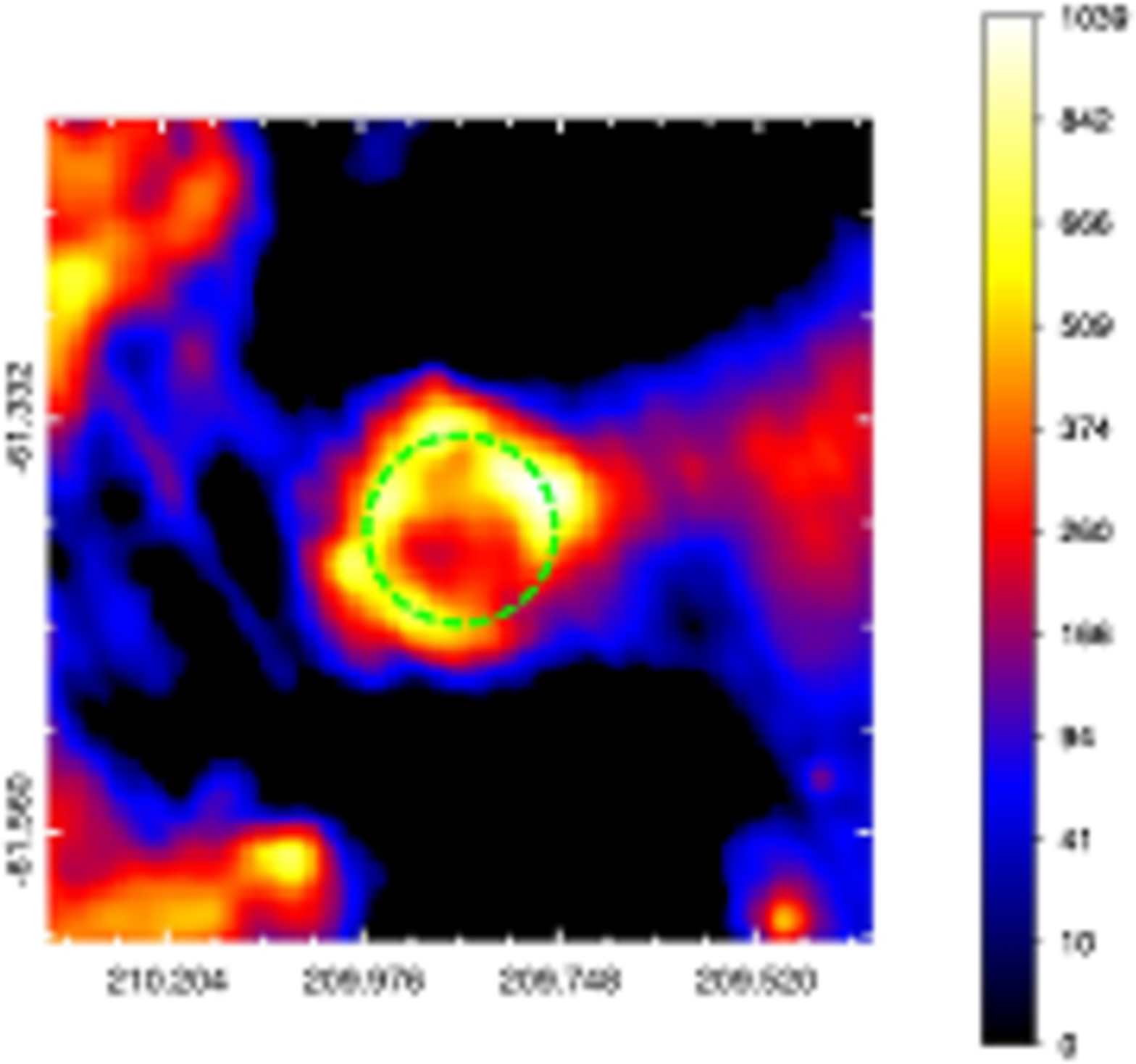}}}%
}
\subfigure{
\mbox{\raisebox{6mm}{\rotatebox{90}{\small{DEC (J2000)}}}}%
\mbox{\raisebox{0mm}{\includegraphics[width=40mm]{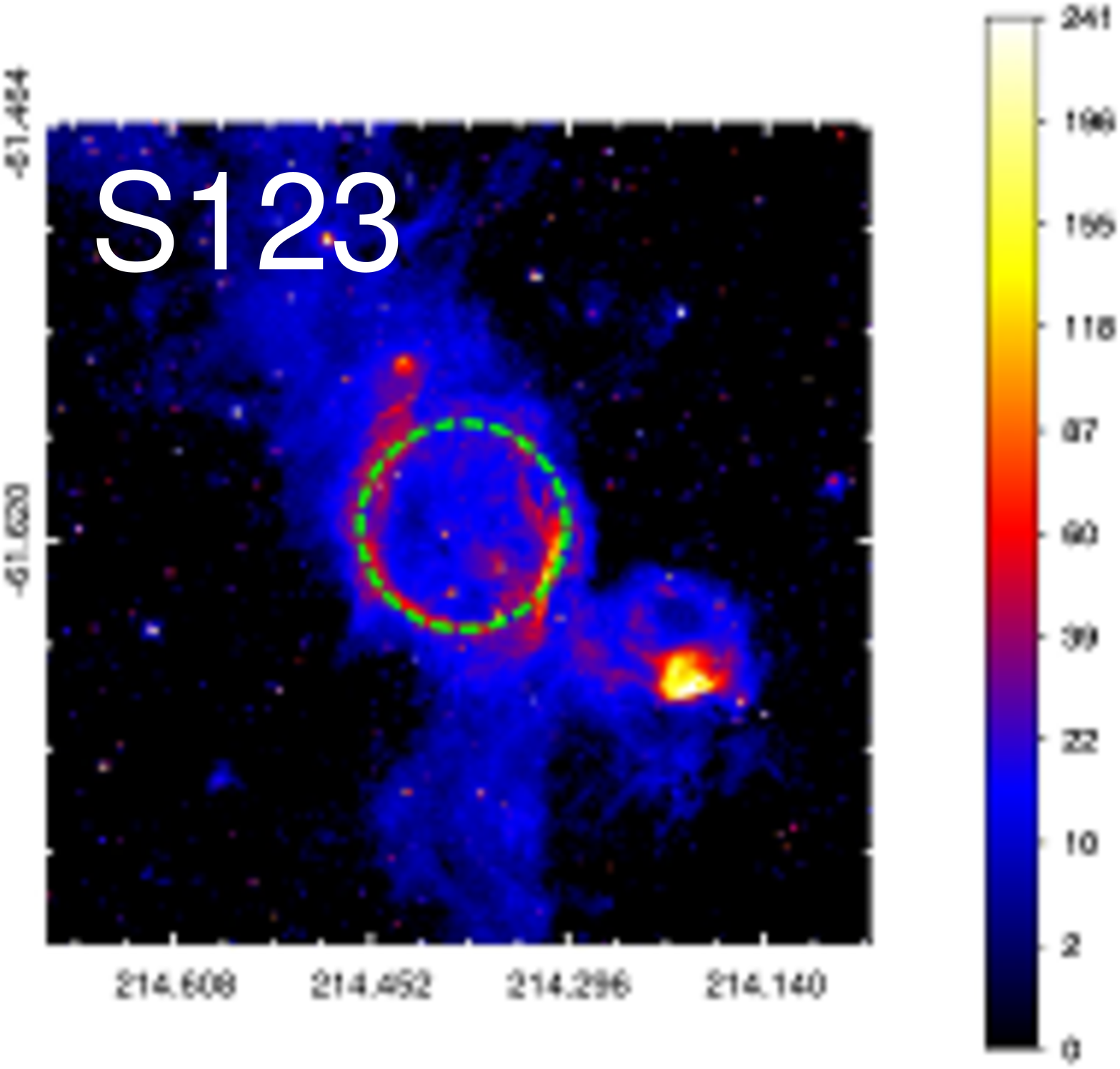}}}%
\mbox{\raisebox{0mm}{\includegraphics[width=40mm]{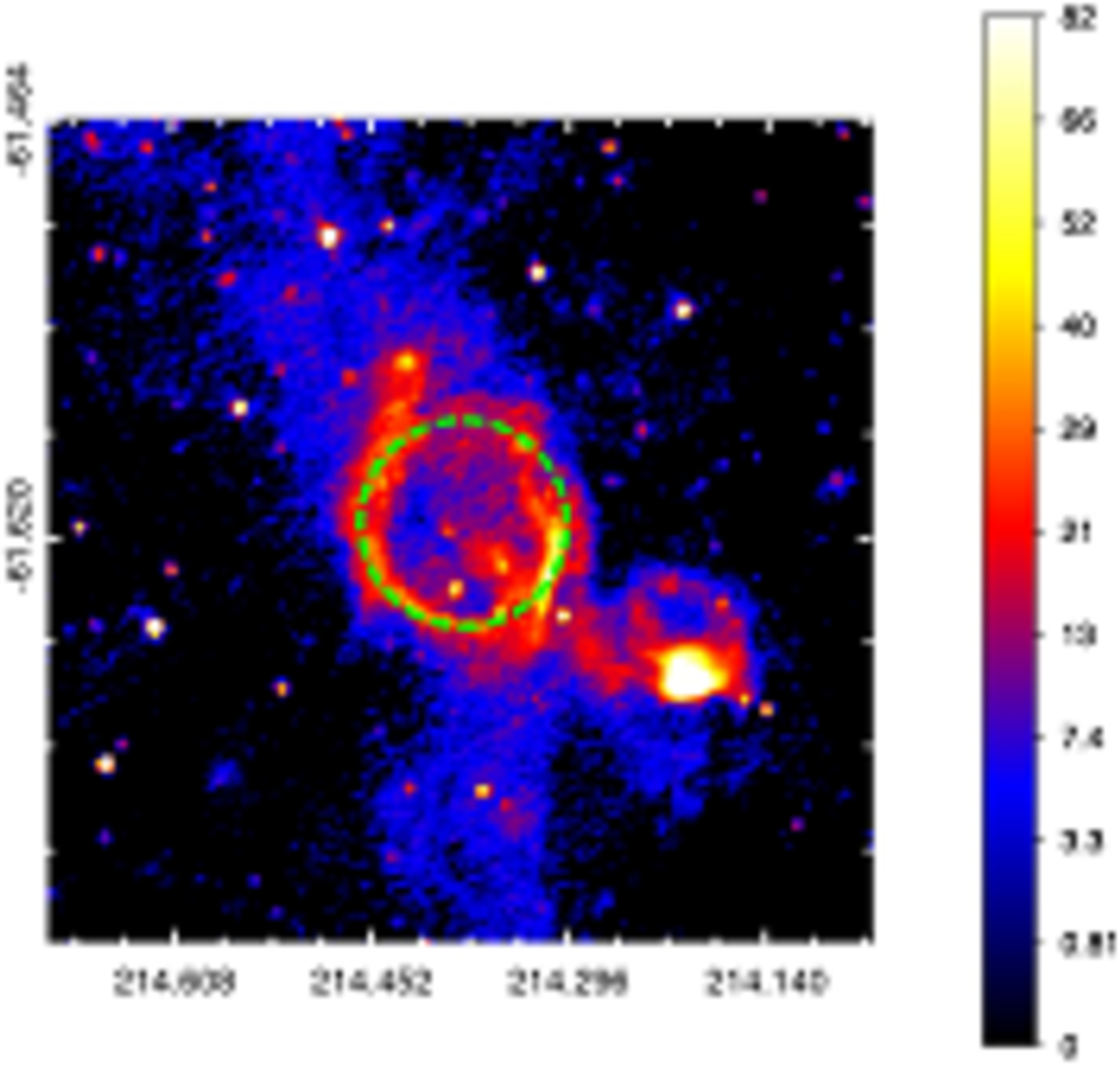}}}%
\mbox{\raisebox{0mm}{\includegraphics[width=40mm]{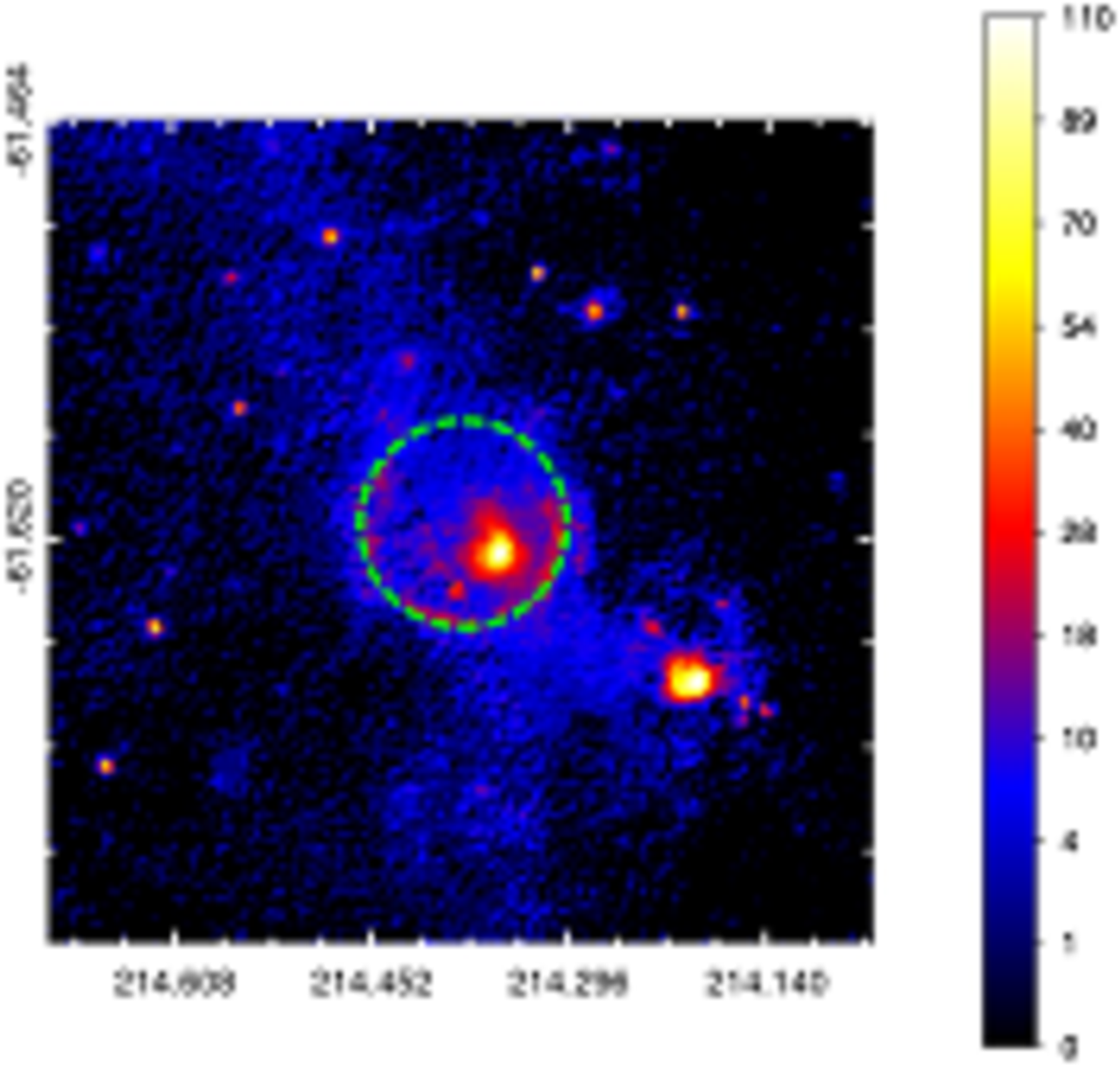}}}%
\mbox{\raisebox{0mm}{\includegraphics[width=40mm]{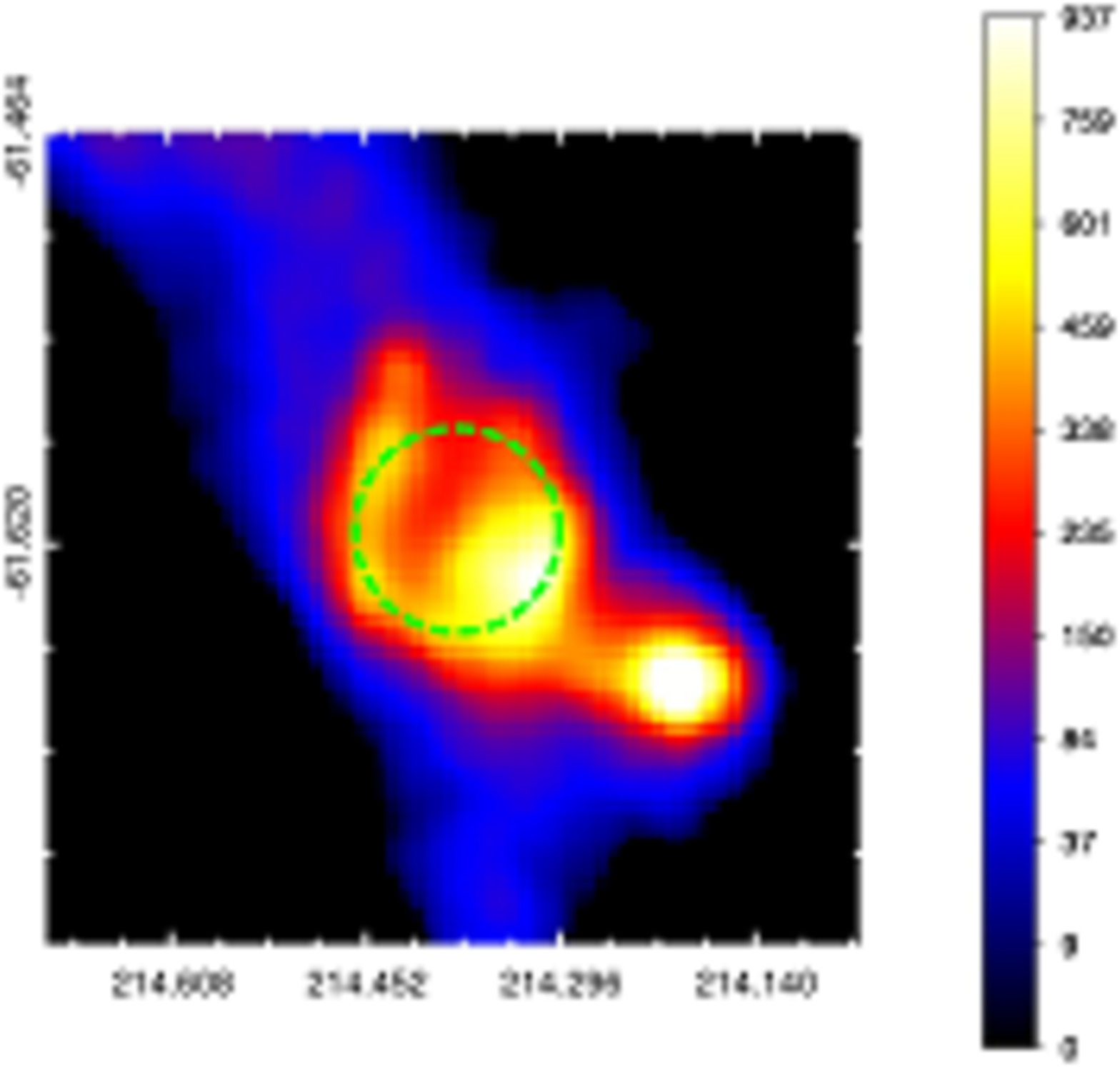}}}%
}
\subfigure{
\mbox{\raisebox{6mm}{\rotatebox{90}{\small{DEC (J2000)}}}}%
\mbox{\raisebox{0mm}{\includegraphics[width=40mm]{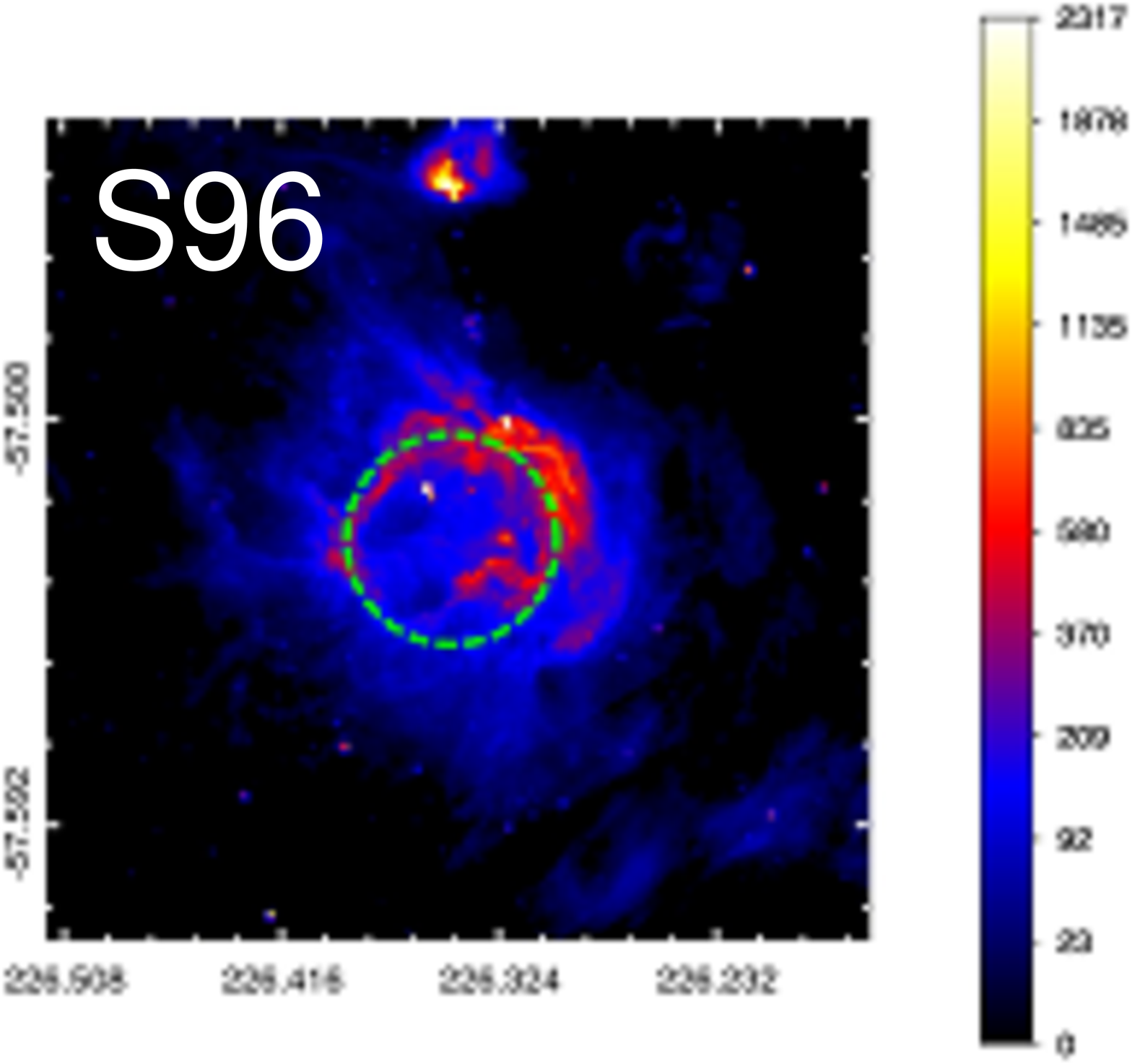}}}%
\mbox{\raisebox{0mm}{\includegraphics[width=40mm]{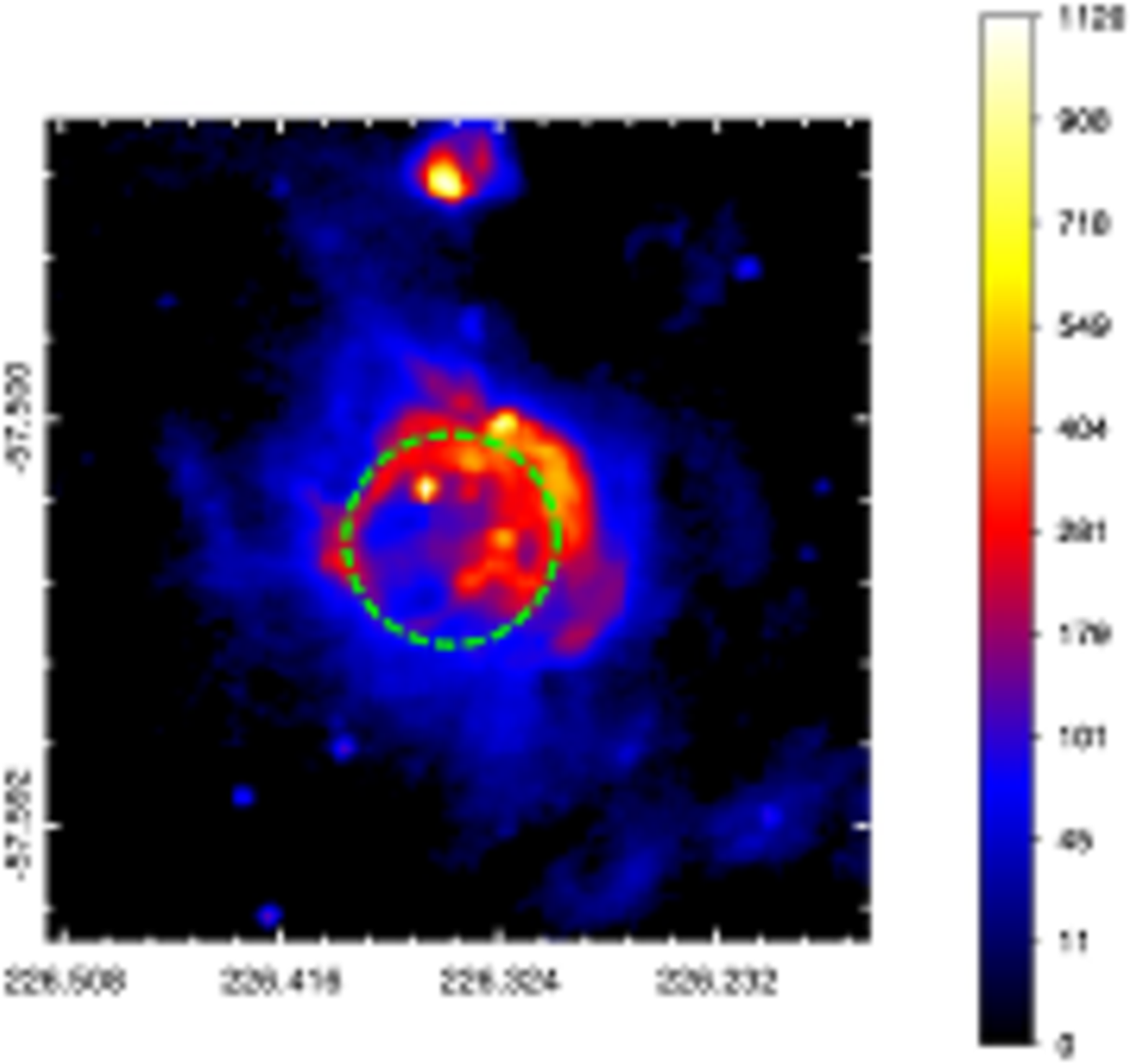}}}%
\mbox{\raisebox{0mm}{\includegraphics[width=40mm]{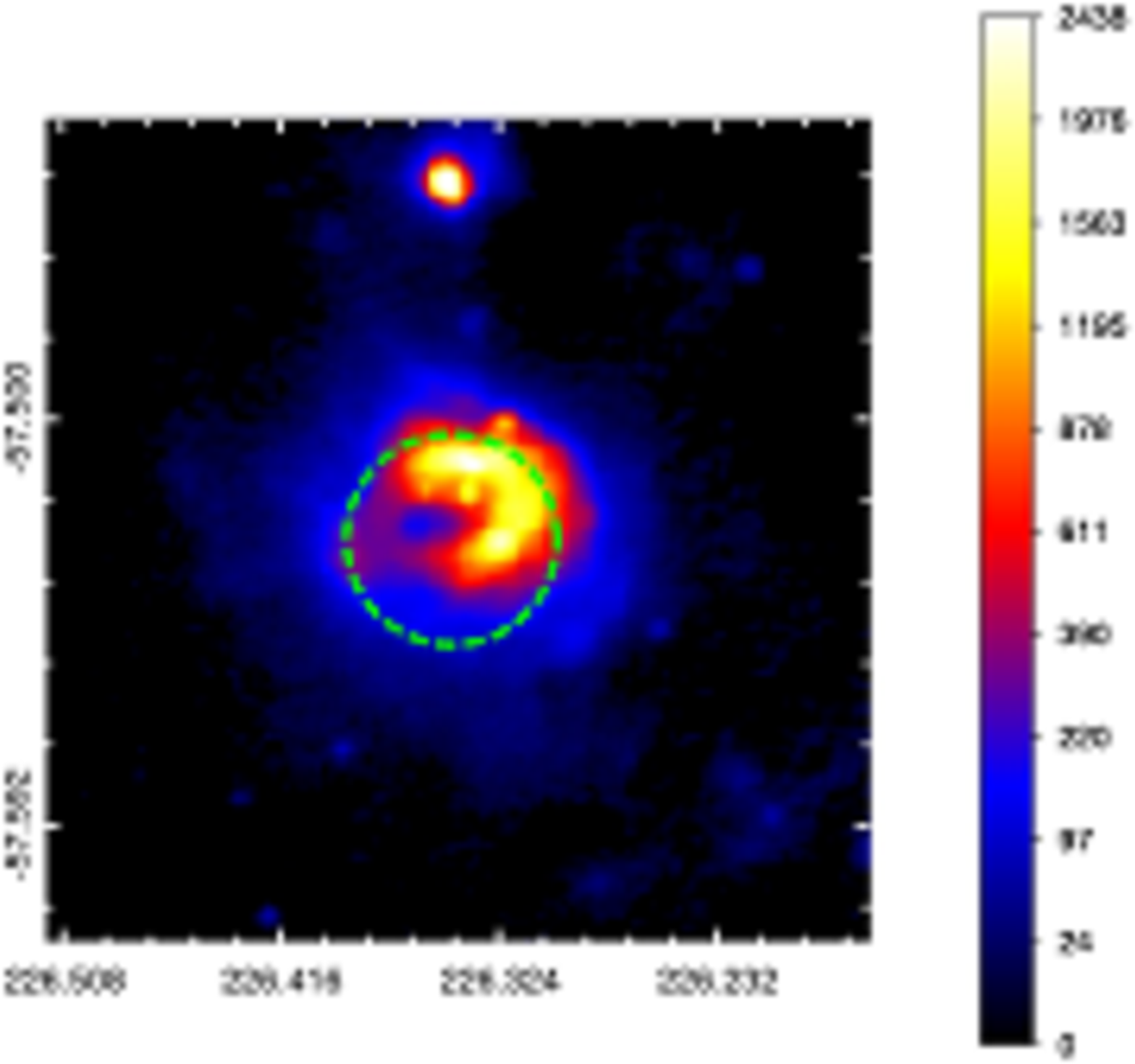}}}%
\mbox{\raisebox{0mm}{\includegraphics[width=40mm]{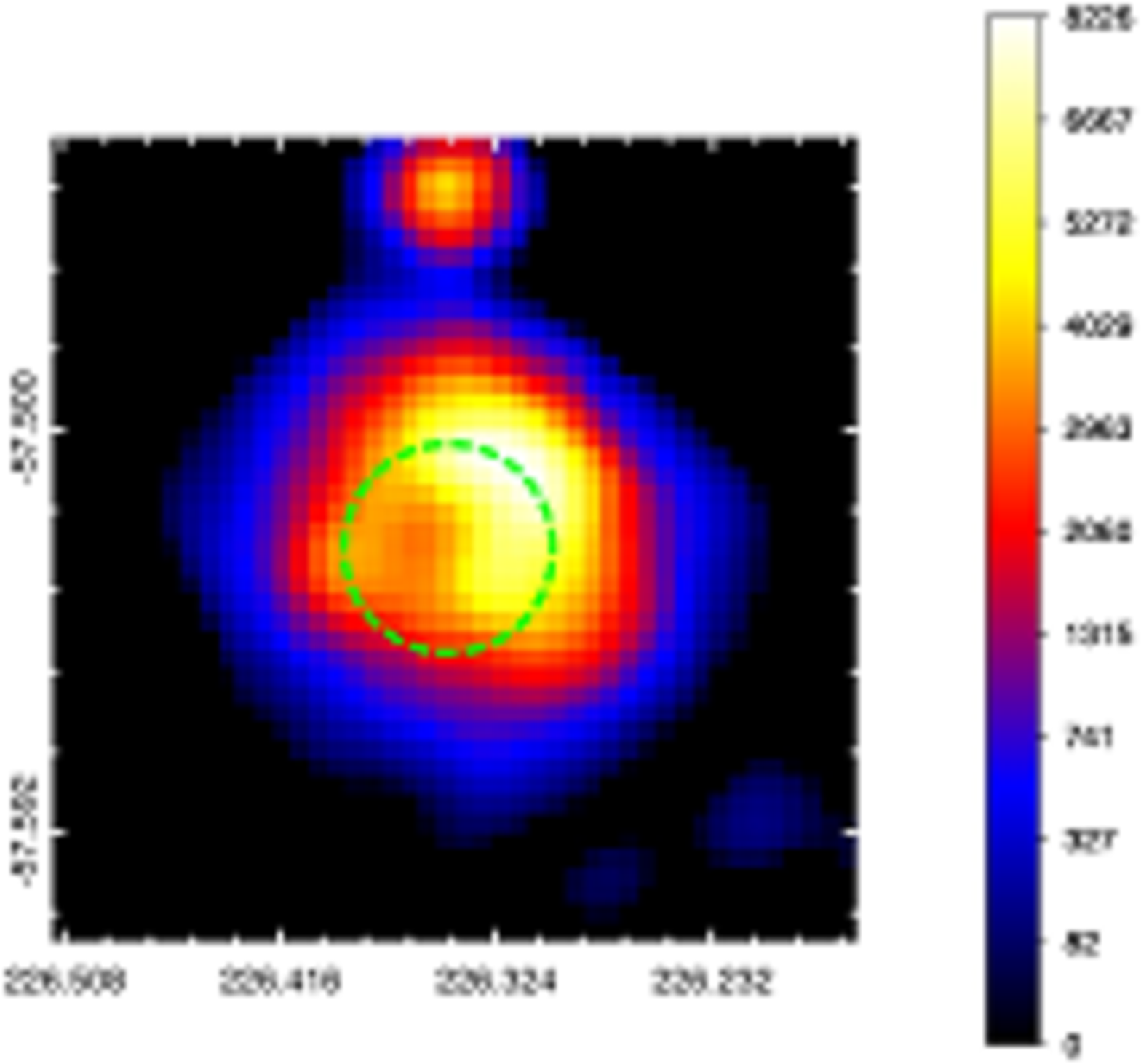}}}%
}
\subfigure{
\mbox{\raisebox{6mm}{\rotatebox{90}{\small{DEC (J2000)}}}}%
\mbox{\raisebox{0mm}{\includegraphics[width=40mm]{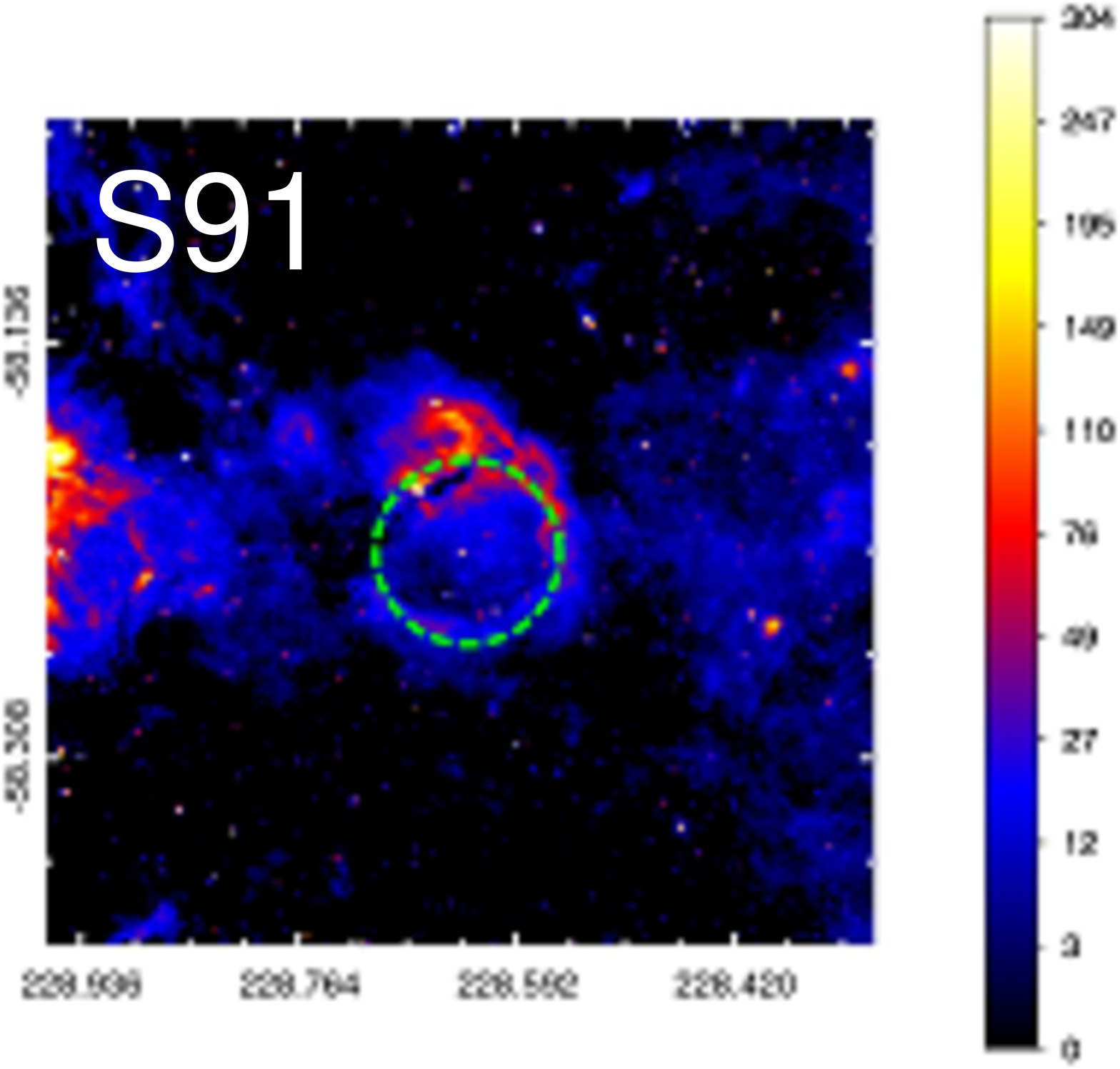}}}%
\mbox{\raisebox{0mm}{\includegraphics[width=40mm]{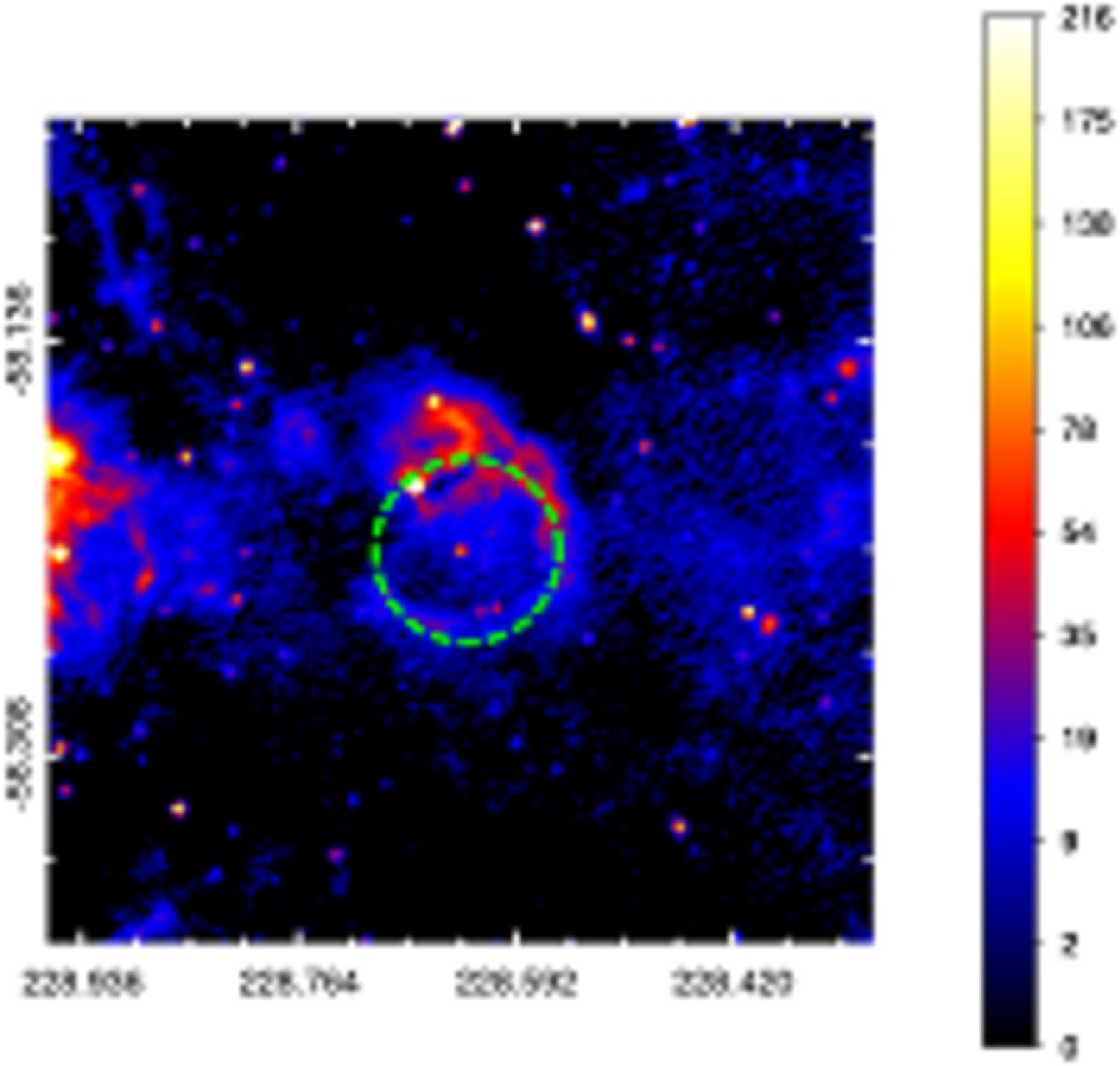}}}%
\mbox{\raisebox{0mm}{\includegraphics[width=40mm]{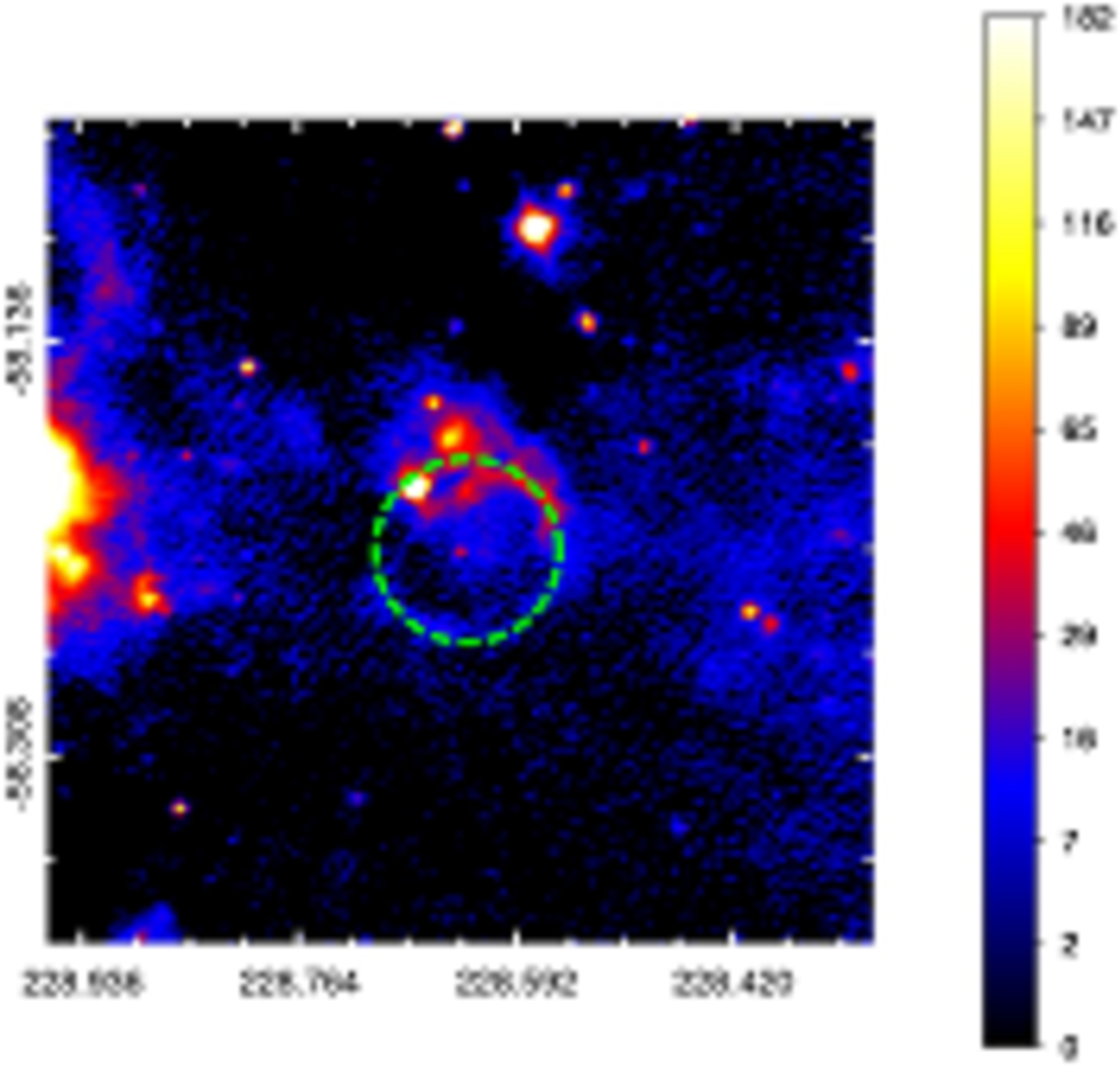}}}%
\mbox{\raisebox{0mm}{\includegraphics[width=40mm]{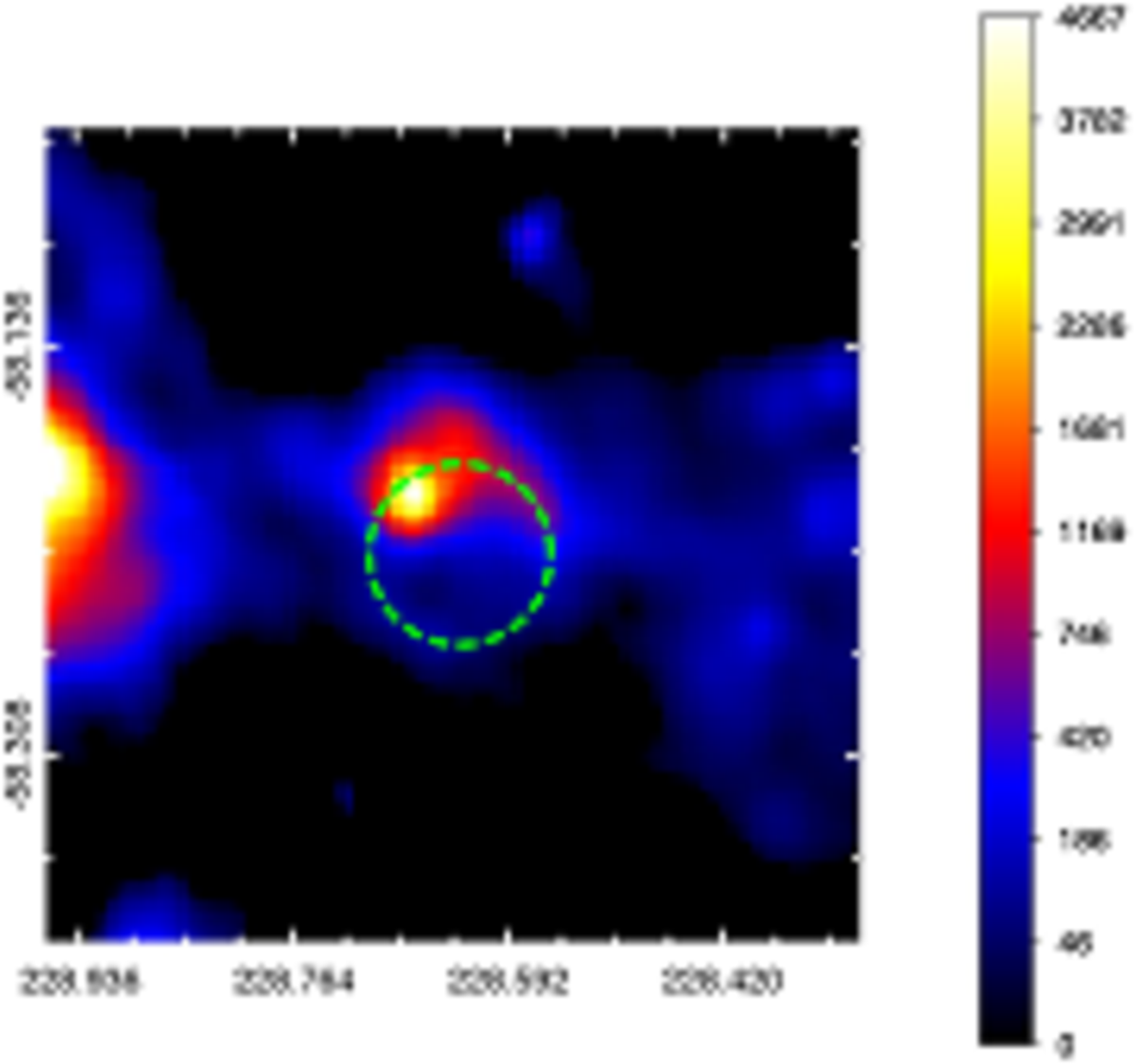}}}%
}
\caption{Continued.} \label{fig:Introfig1:f}
\end{figure*}

\addtocounter{figure}{-1}
\begin{figure*}[ht]
\addtocounter{subfigure}{1}
\centering
\subfigure{
\makebox[180mm][l]{\raisebox{0mm}[0mm][0mm]{ \hspace{15mm} \small{8 \mic}} \hspace{29.5mm} \small{9 \mic} \hspace{27mm} \small{18 \mic} \hspace{26.5mm} \small{90 \mic}}%
}
\subfigure{
\makebox[180mm][l]{\raisebox{0mm}[0mm][0mm]{ \hspace{11mm} \small{RA (J2000)}} \hspace{19.5mm} \small{RA (J2000)} \hspace{20mm} \small{RA (J2000)} \hspace{20mm} \small{RA (J2000)}}%
}
\subfigure{
\mbox{\raisebox{6mm}{\rotatebox{90}{\small{DEC (J2000)}}}}%
\mbox{\raisebox{0mm}{\includegraphics[width=40mm]{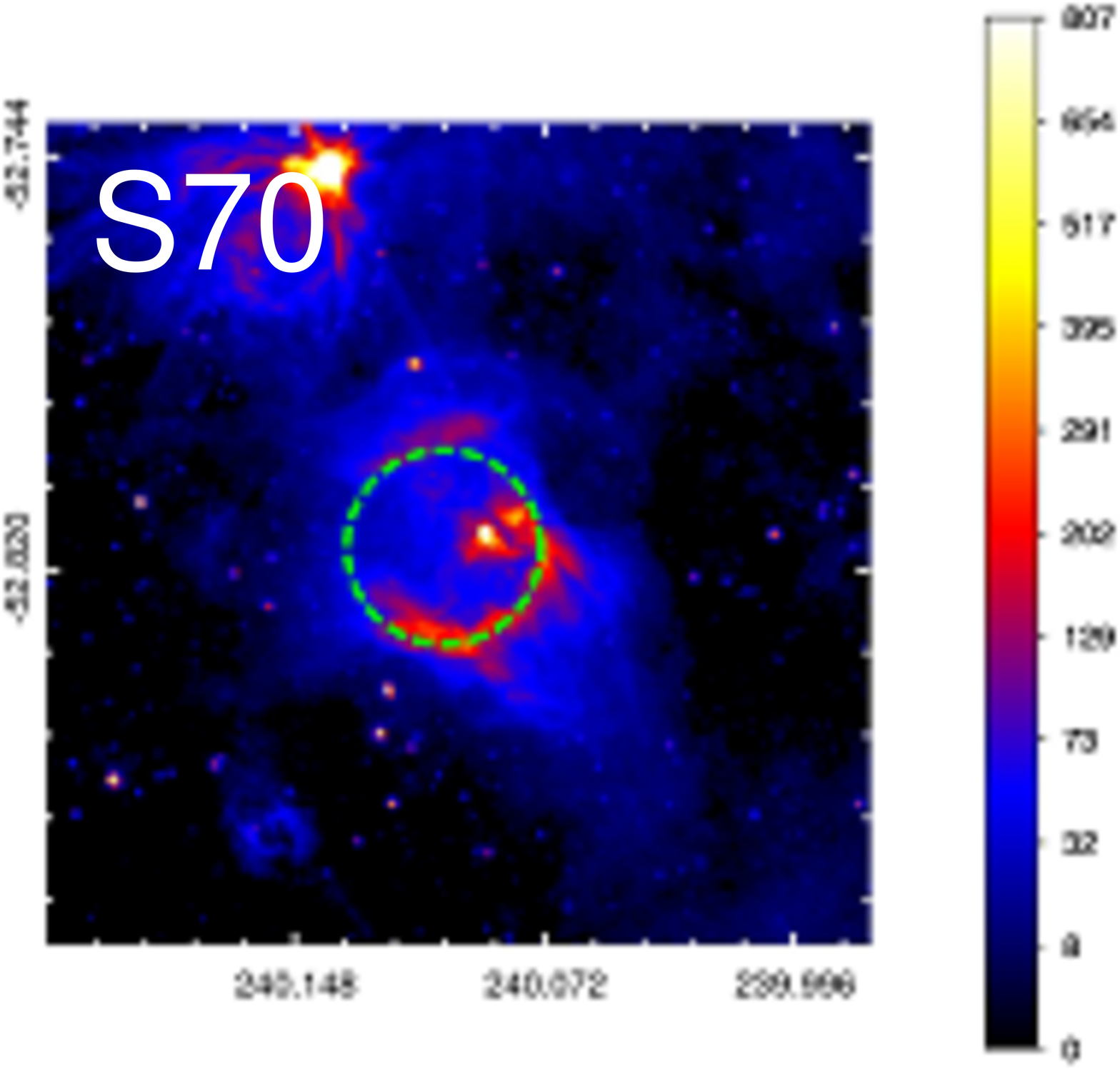}}}%
\mbox{\raisebox{0mm}{\includegraphics[width=40mm]{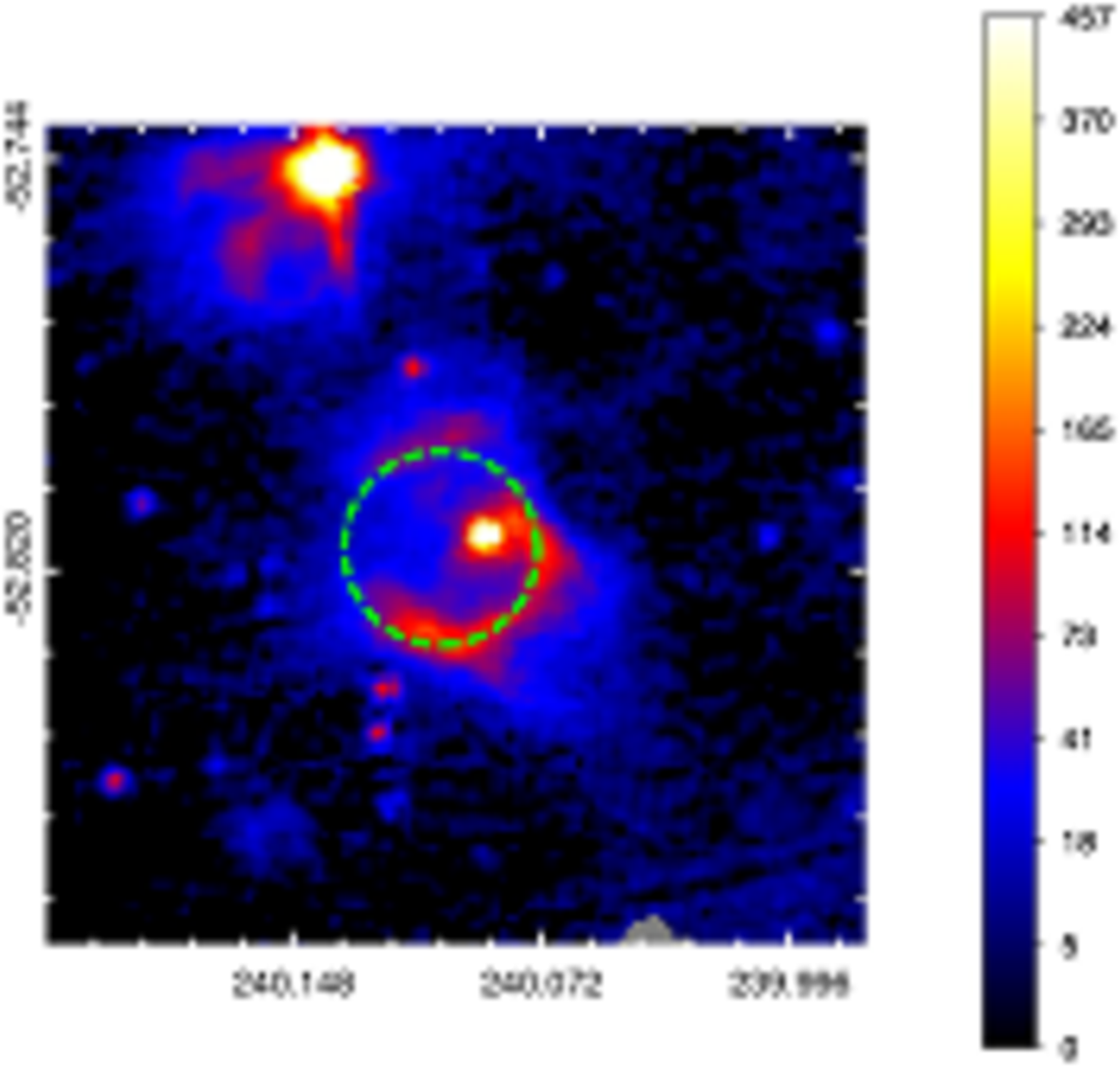}}}%
\mbox{\raisebox{0mm}{\includegraphics[width=40mm]{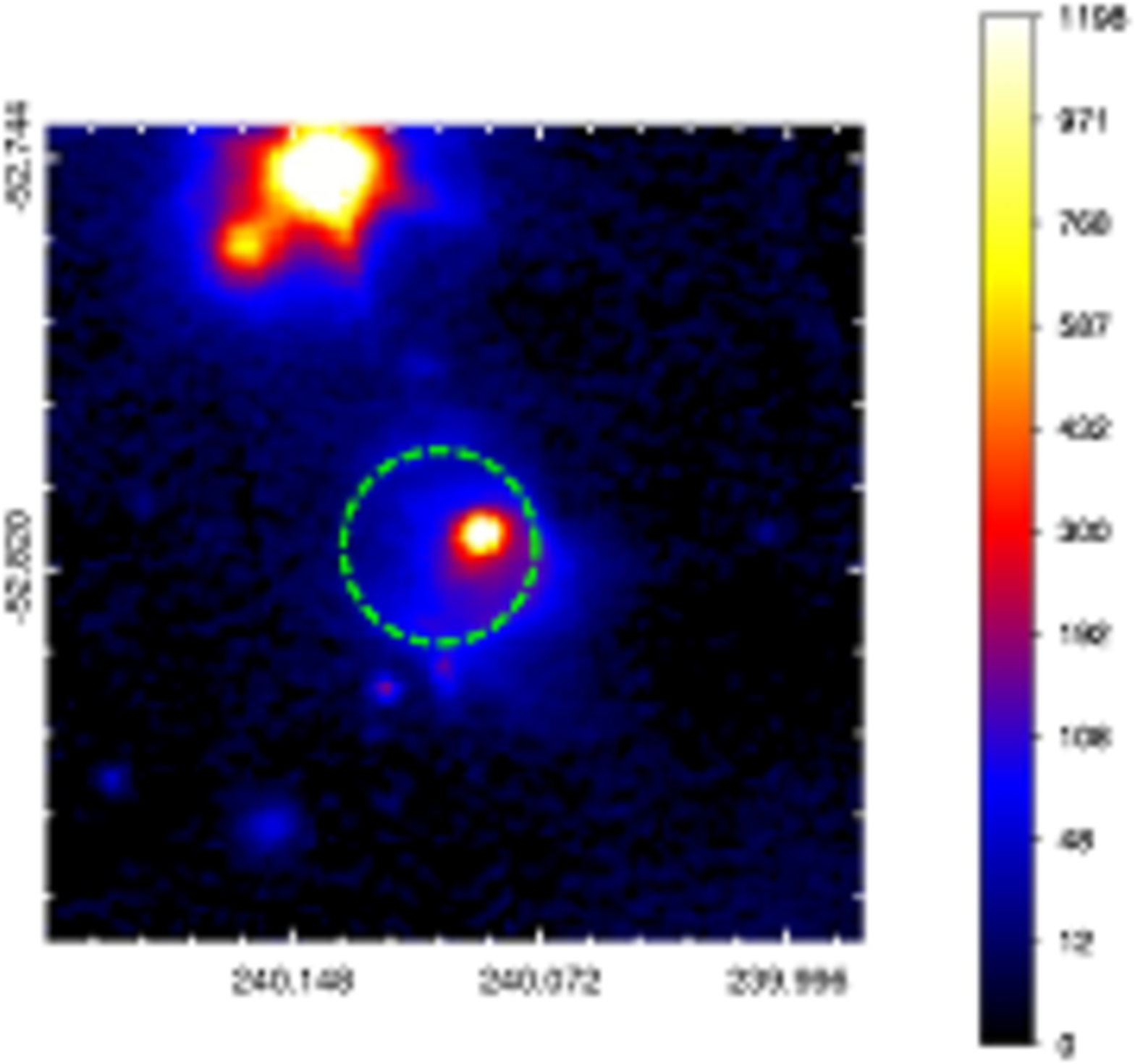}}}%
\mbox{\raisebox{0mm}{\includegraphics[width=40mm]{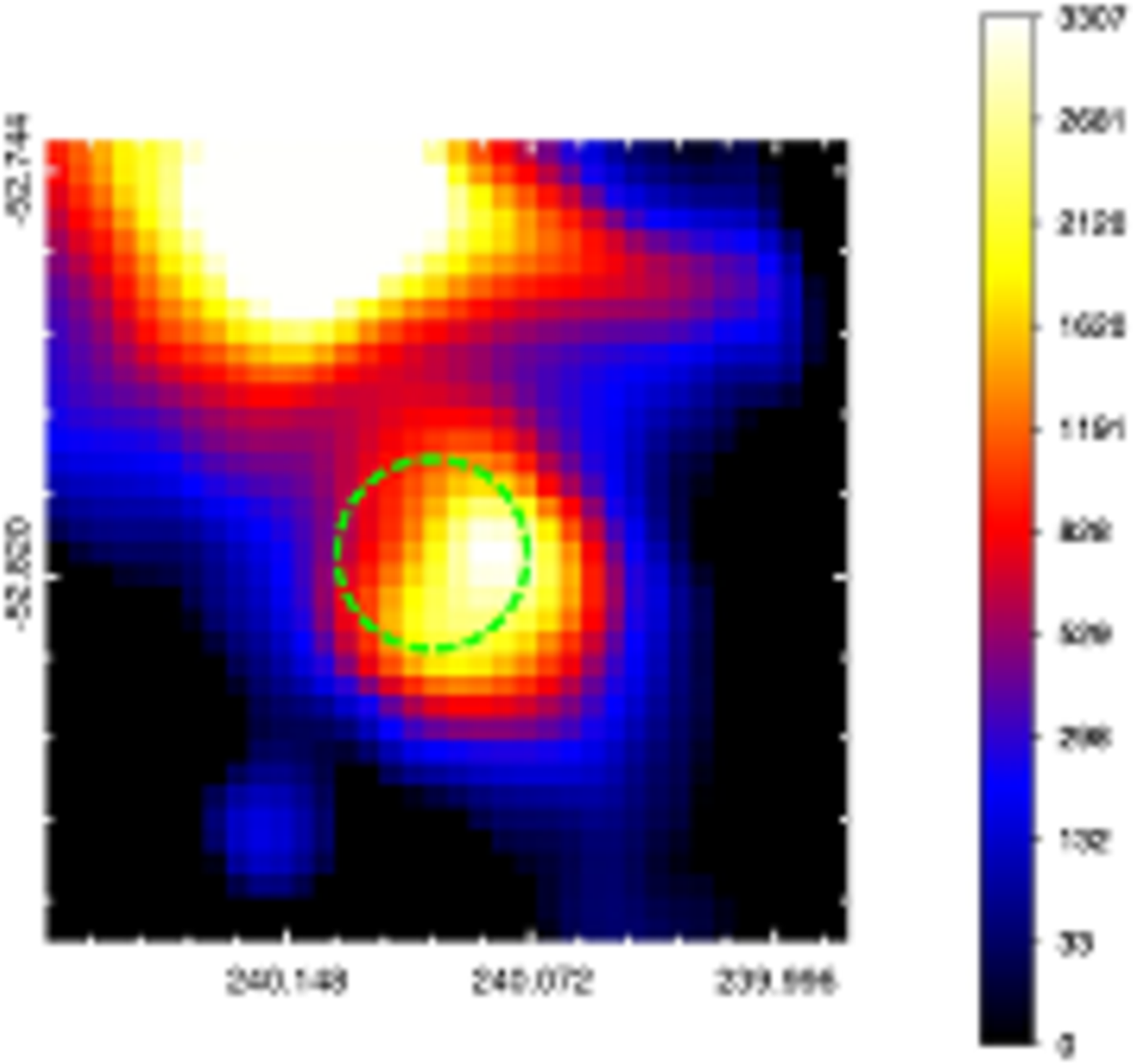}}}%
}
\subfigure{
\mbox{\raisebox{6mm}{\rotatebox{90}{\small{DEC (J2000)}}}}%
\mbox{\raisebox{0mm}{\includegraphics[width=40mm]{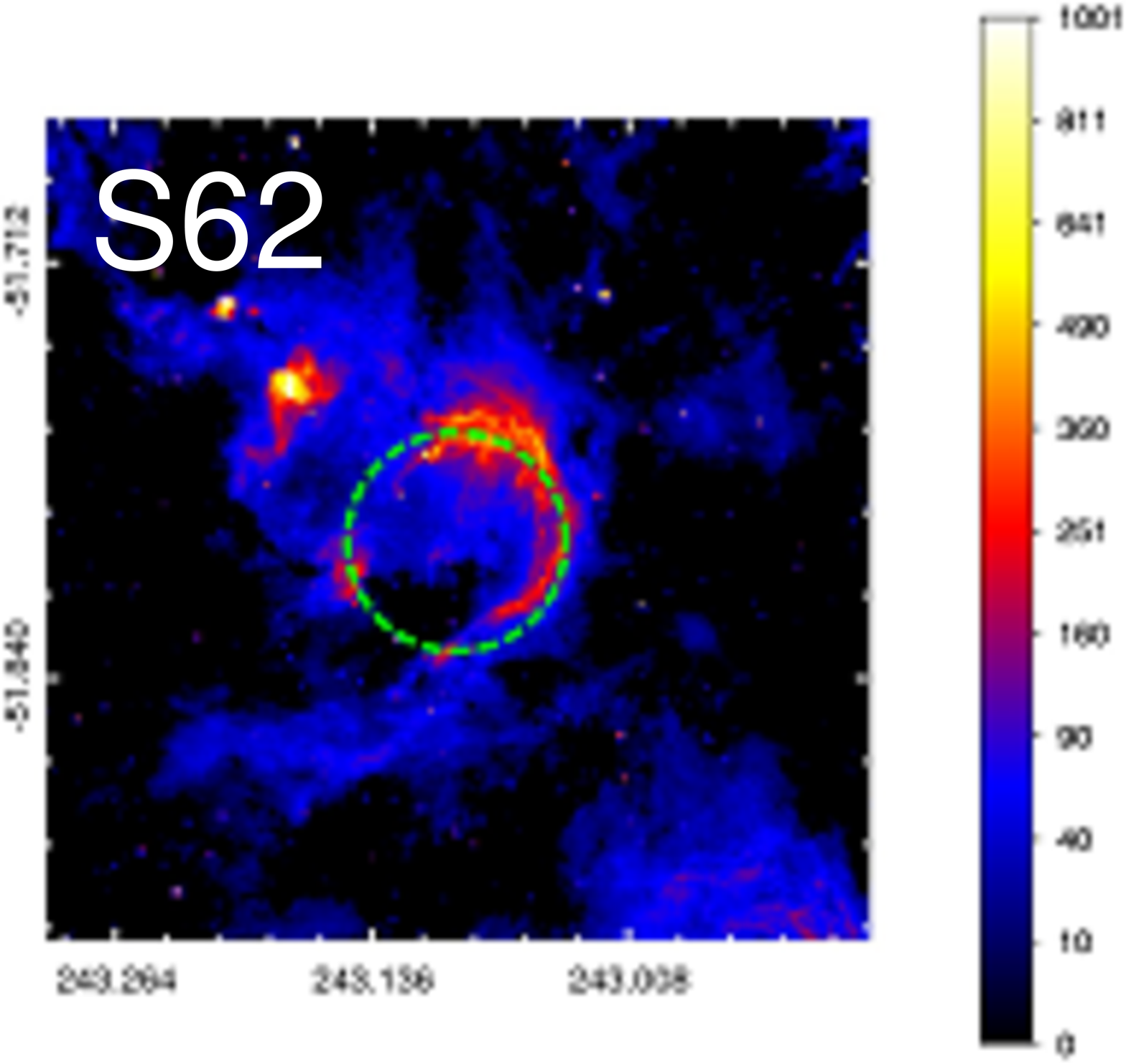}}}%
\mbox{\raisebox{0mm}{\includegraphics[width=40mm]{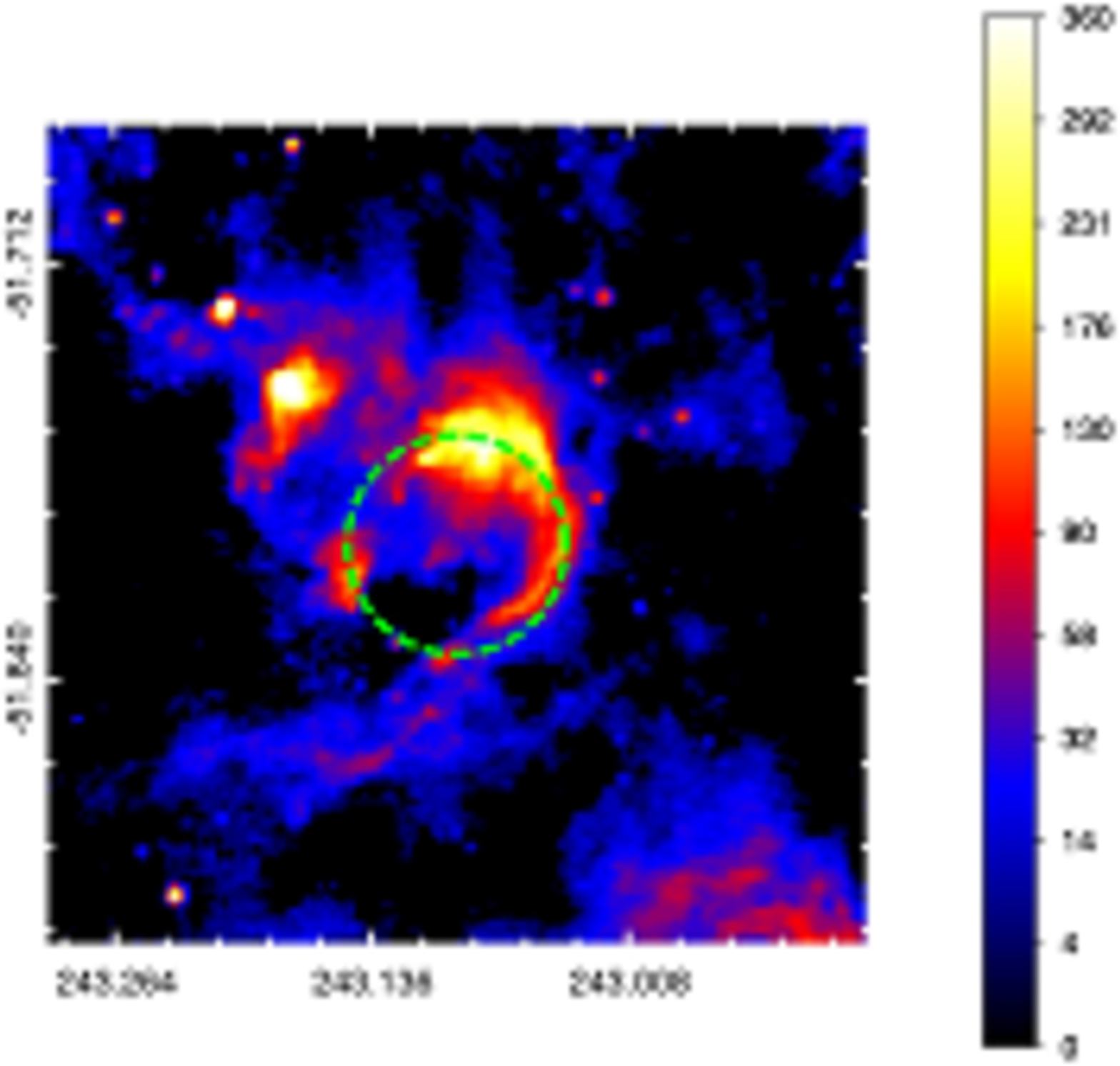}}}%
\mbox{\raisebox{0mm}{\includegraphics[width=40mm]{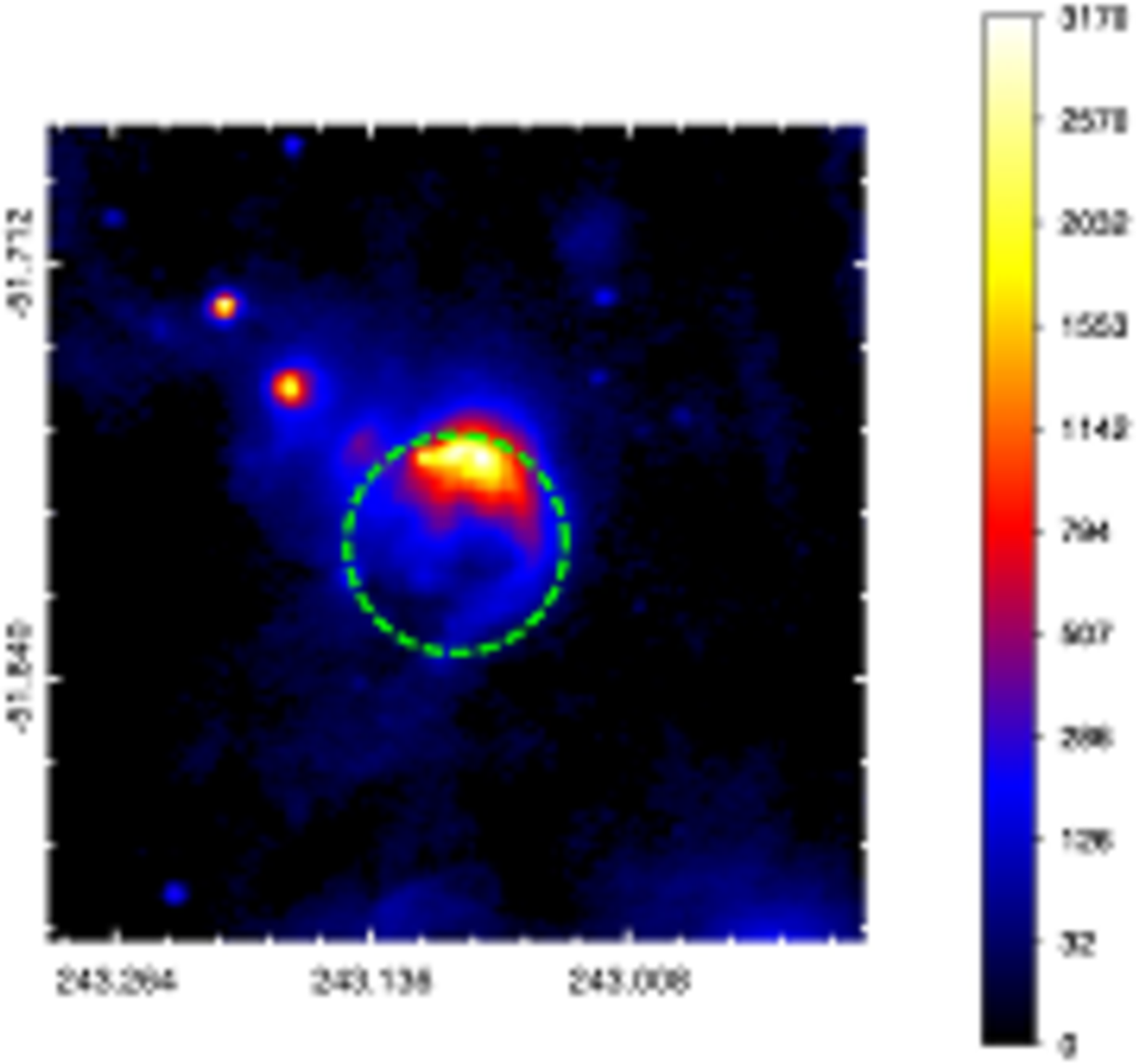}}}%
\mbox{\raisebox{0mm}{\includegraphics[width=40mm]{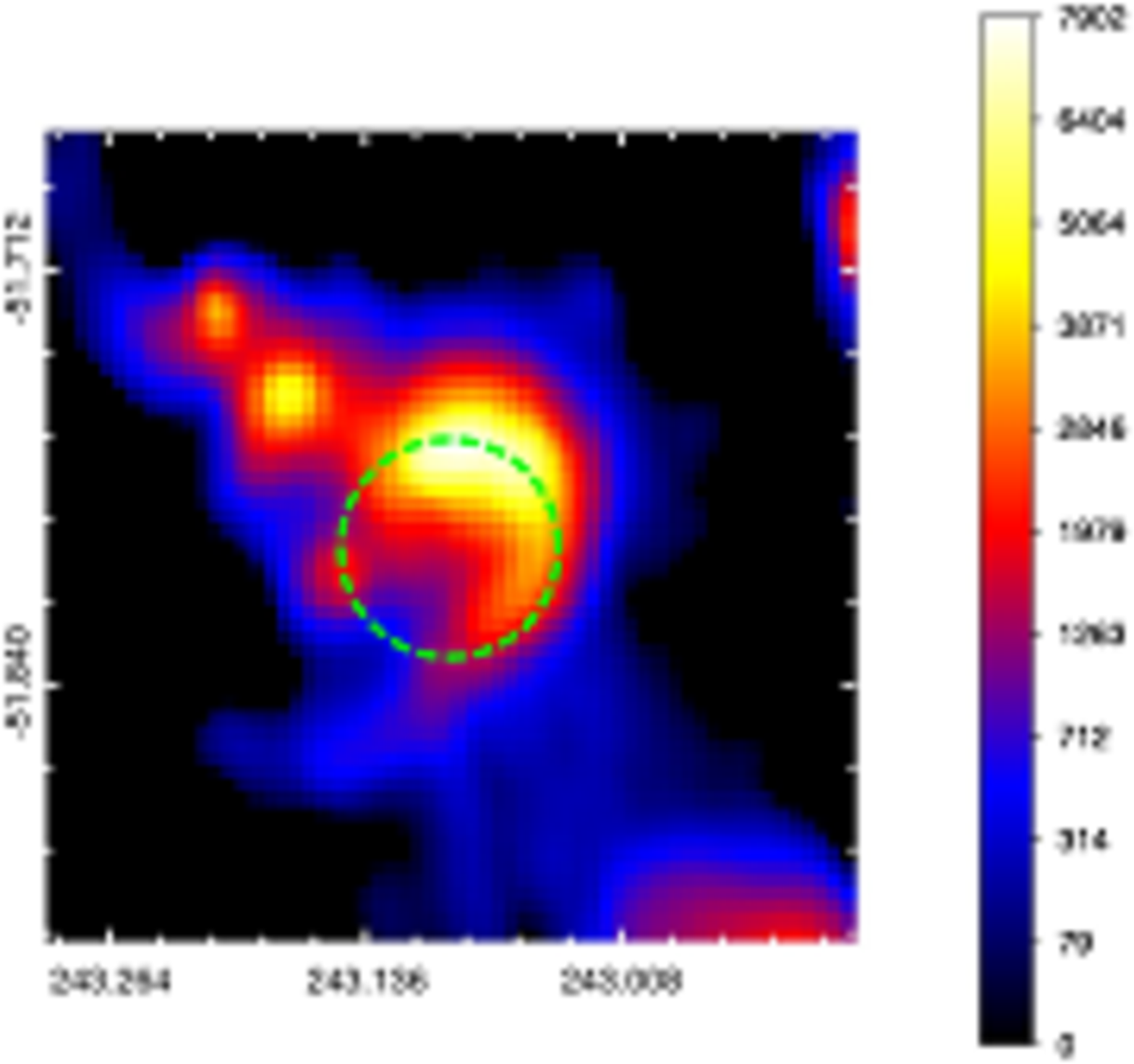}}}%
}
\subfigure{
\mbox{\raisebox{6mm}{\rotatebox{90}{\small{DEC (J2000)}}}}%
\mbox{\raisebox{0mm}{\includegraphics[width=40mm]{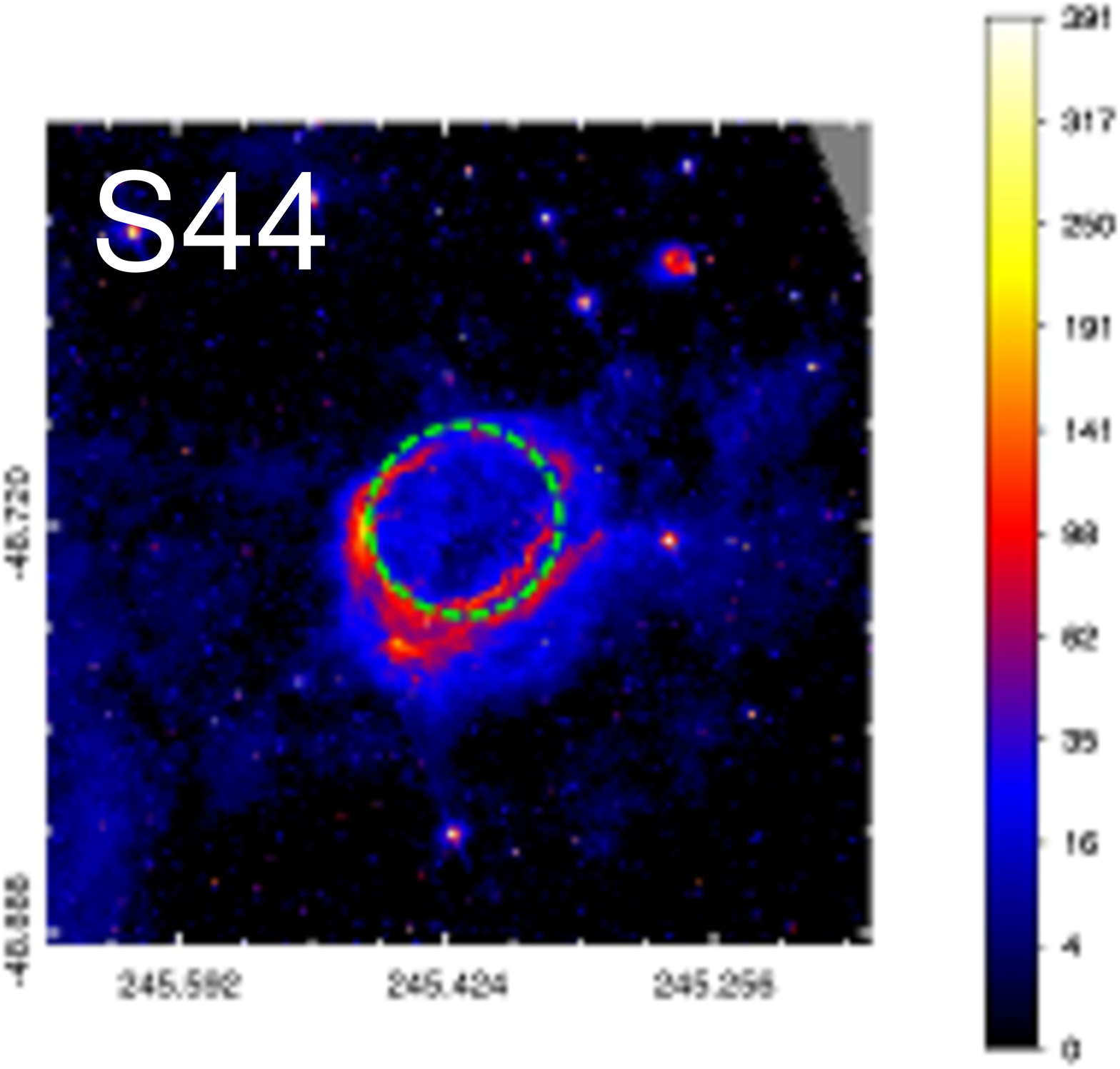}}}%
\mbox{\raisebox{0mm}{\includegraphics[width=40mm]{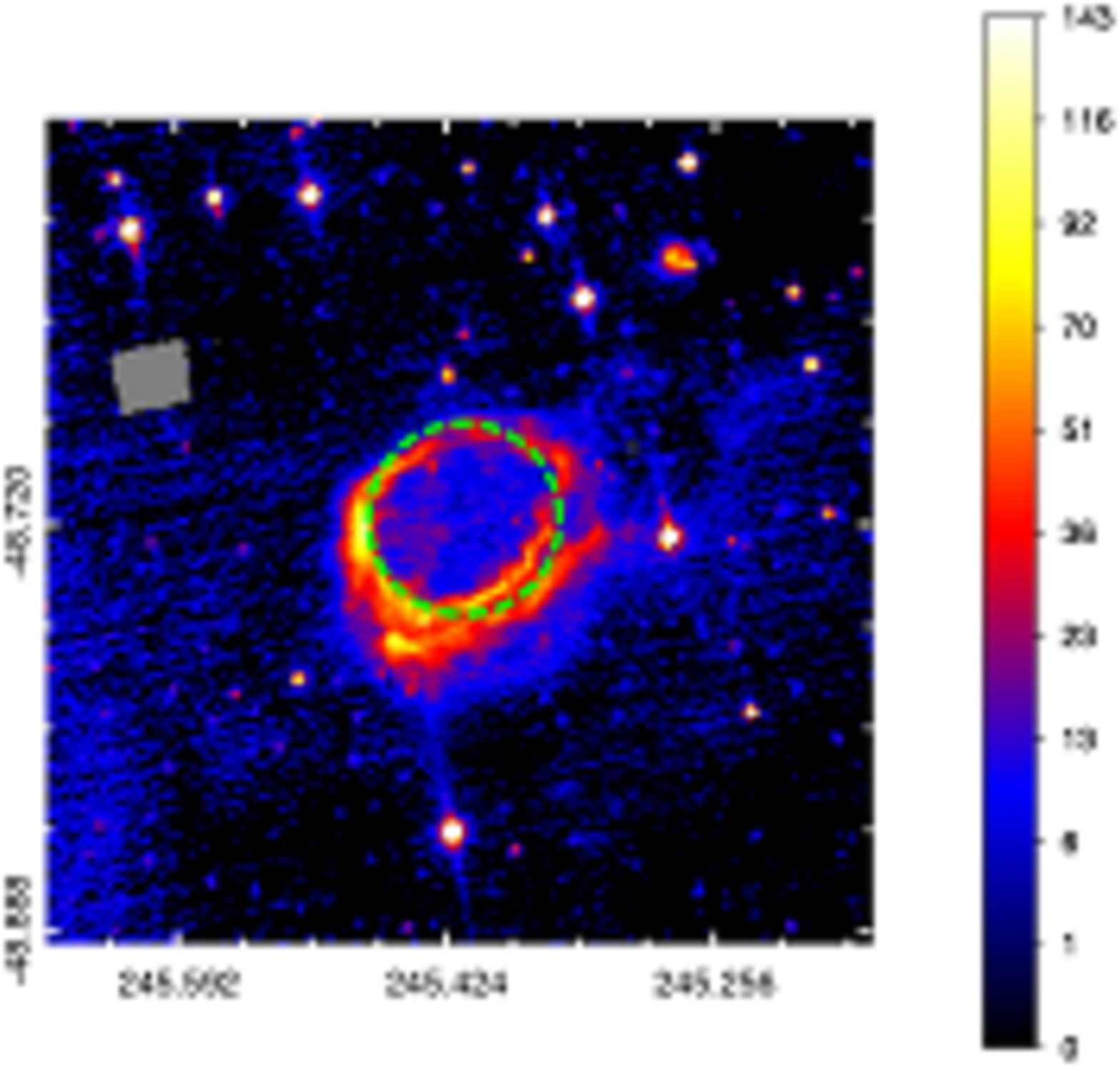}}}%
\mbox{\raisebox{0mm}{\includegraphics[width=40mm]{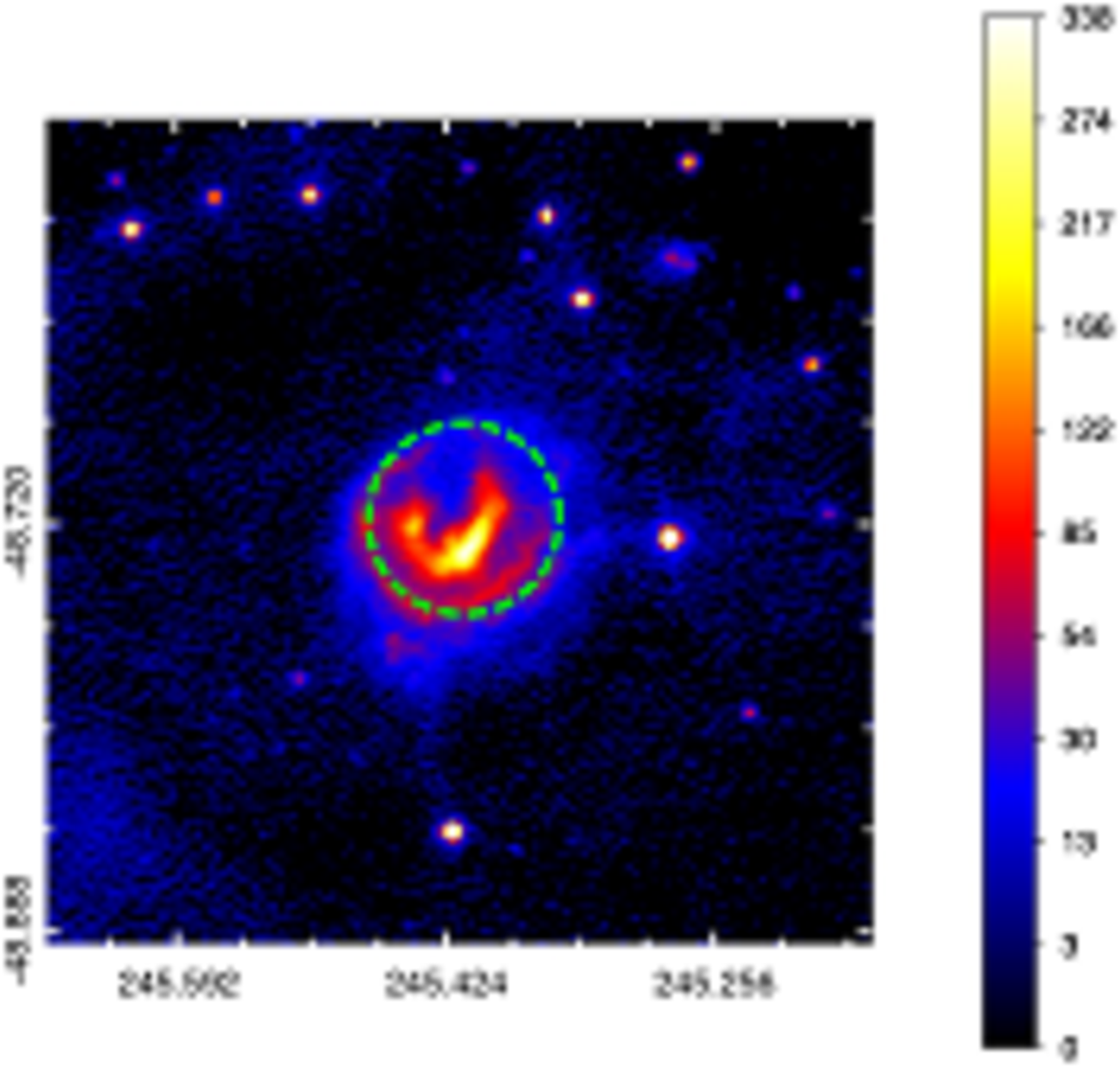}}}%
\mbox{\raisebox{0mm}{\includegraphics[width=40mm]{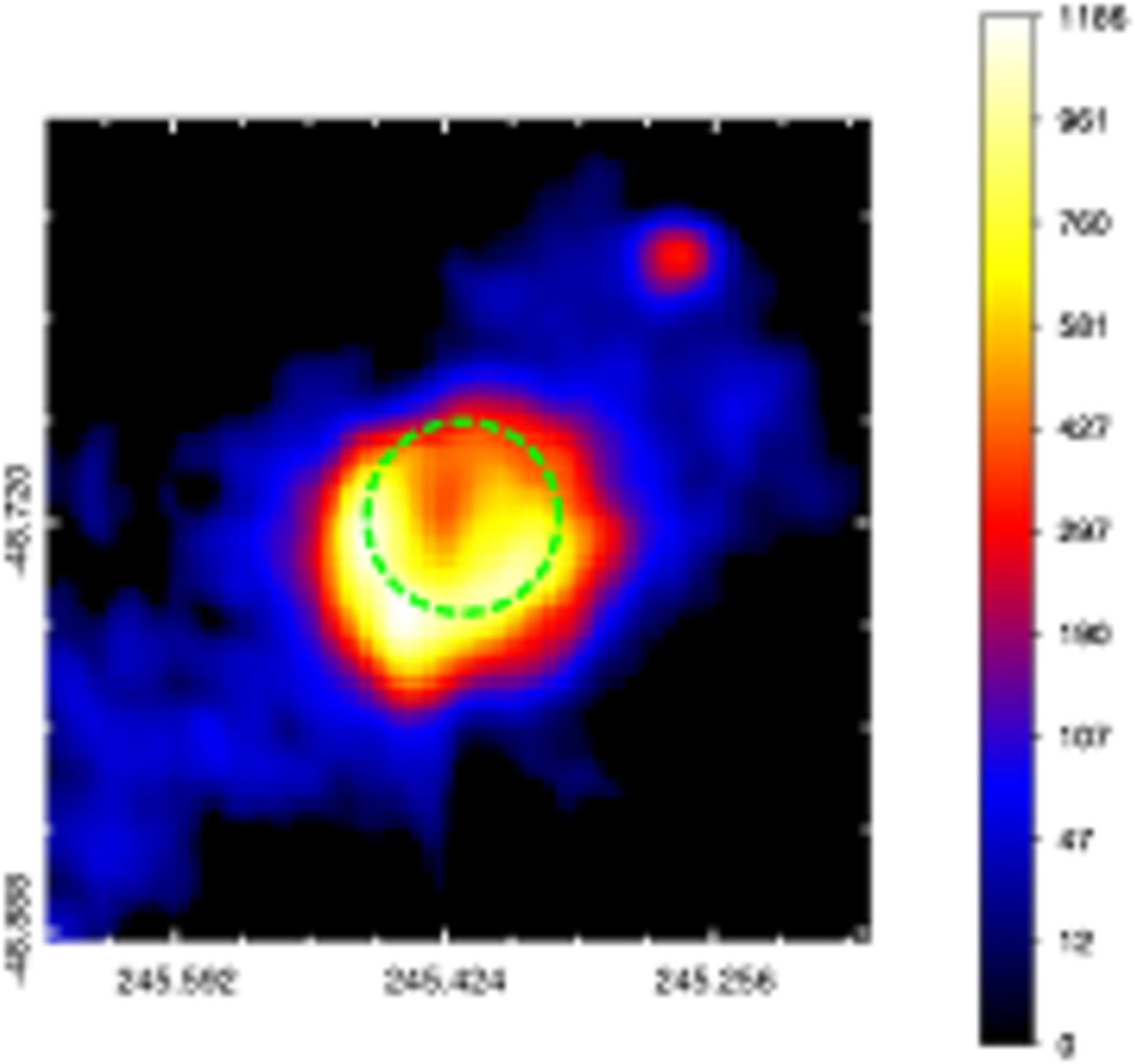}}}%
}
\subfigure{
\mbox{\raisebox{6mm}{\rotatebox{90}{\small{DEC (J2000)}}}}%
\mbox{\raisebox{0mm}{\includegraphics[width=40mm]{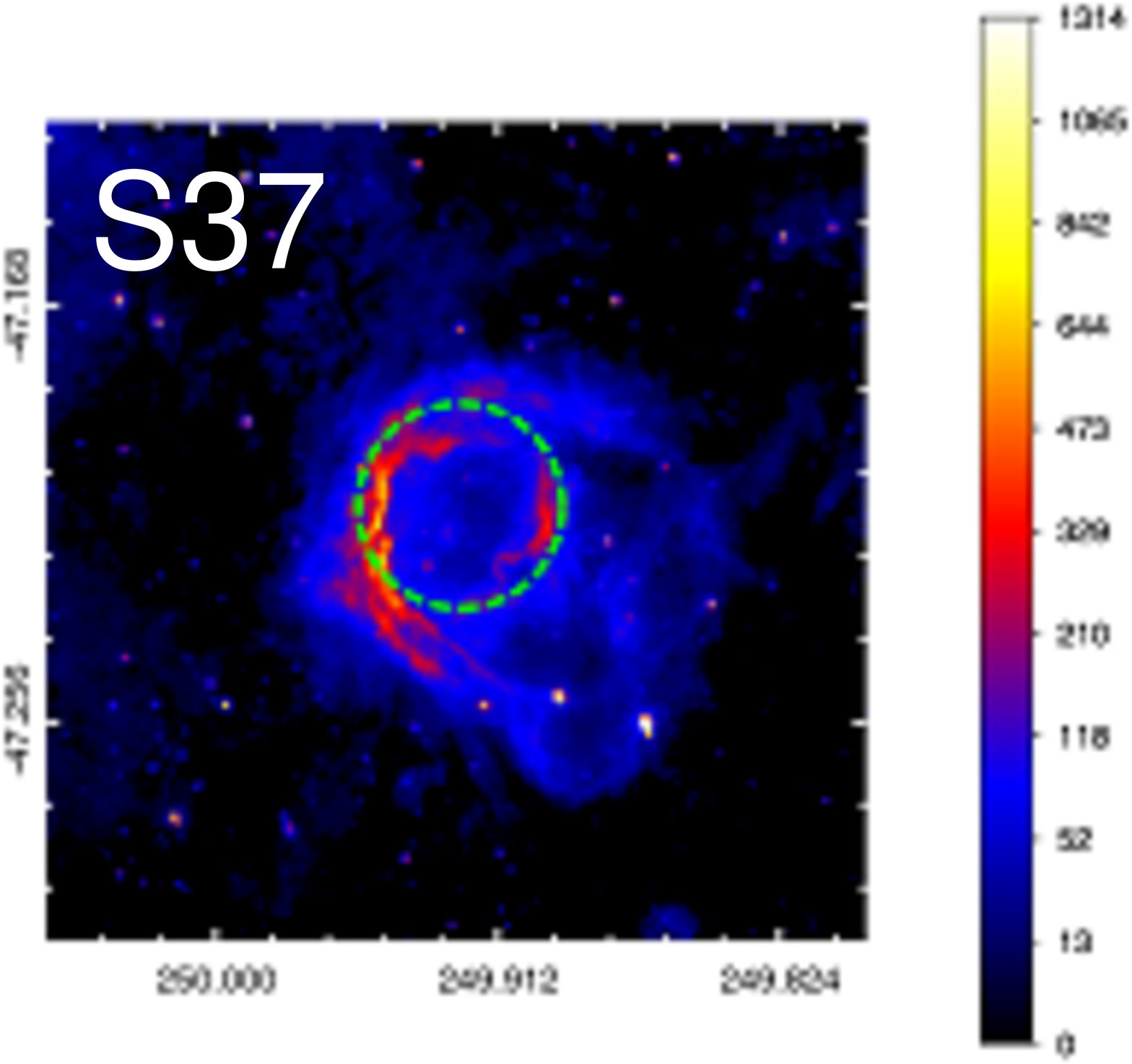}}}%
\mbox{\raisebox{0mm}{\includegraphics[width=40mm]{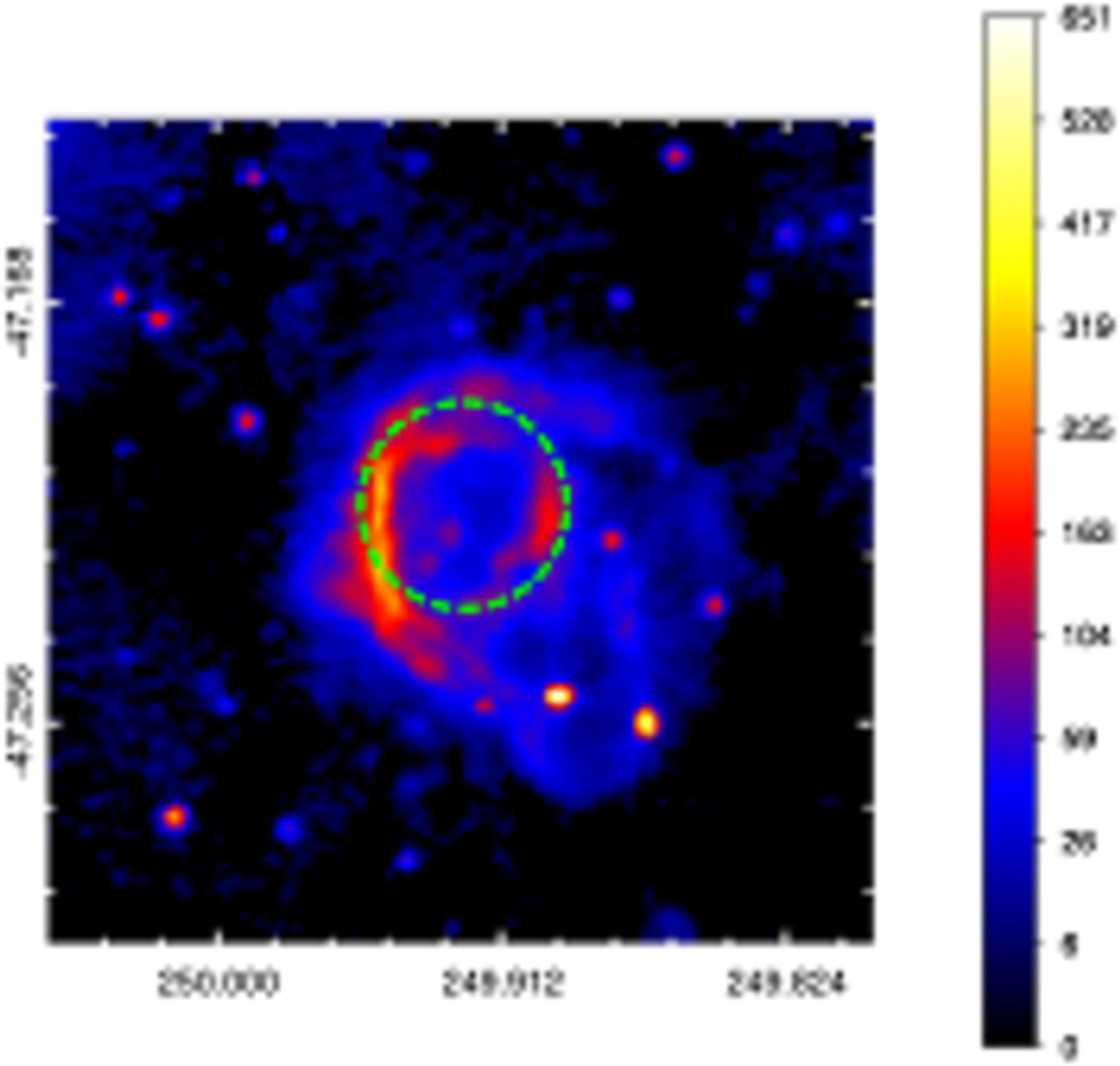}}}%
\mbox{\raisebox{0mm}{\includegraphics[width=40mm]{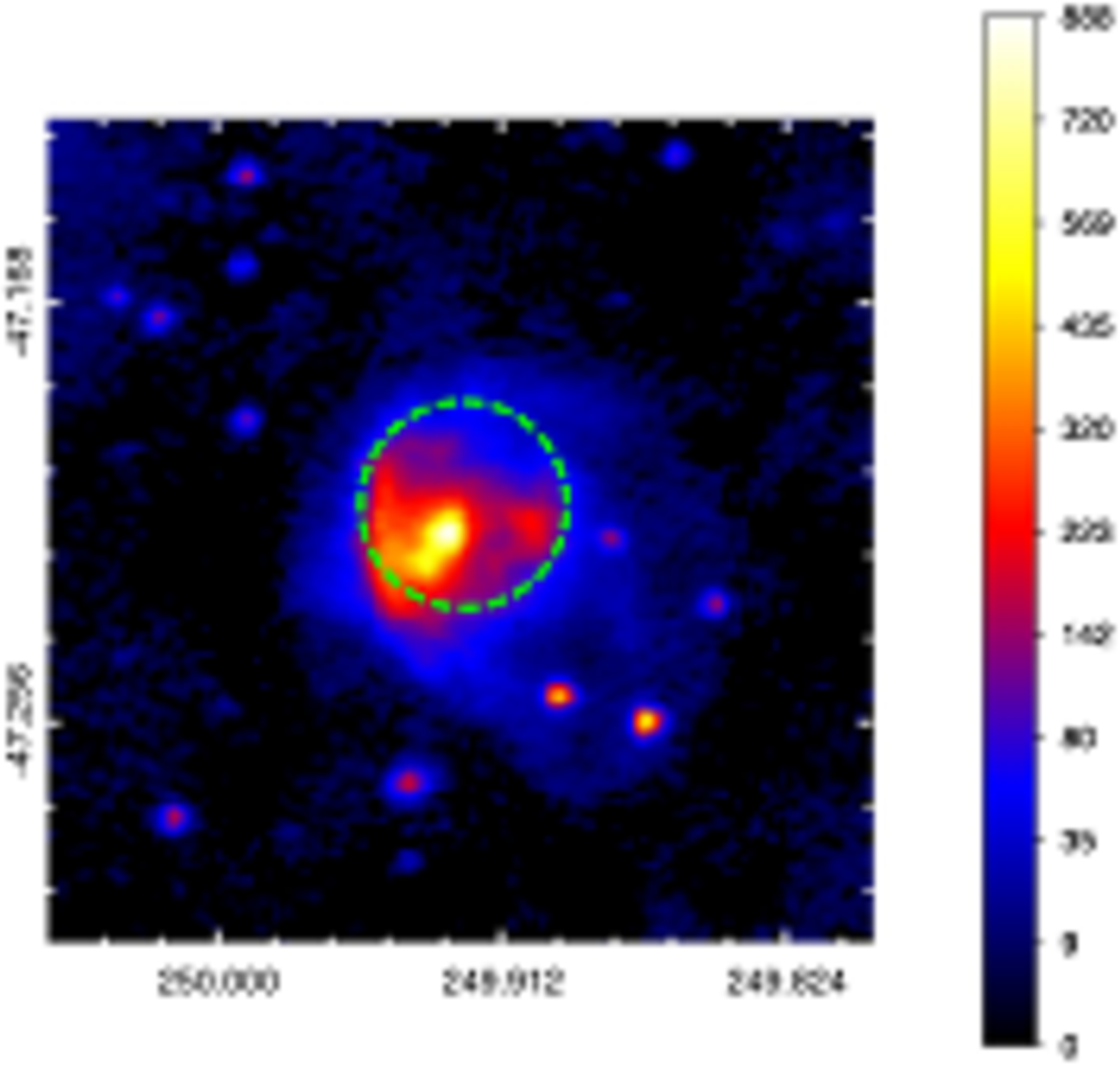}}}%
\mbox{\raisebox{0mm}{\includegraphics[width=40mm]{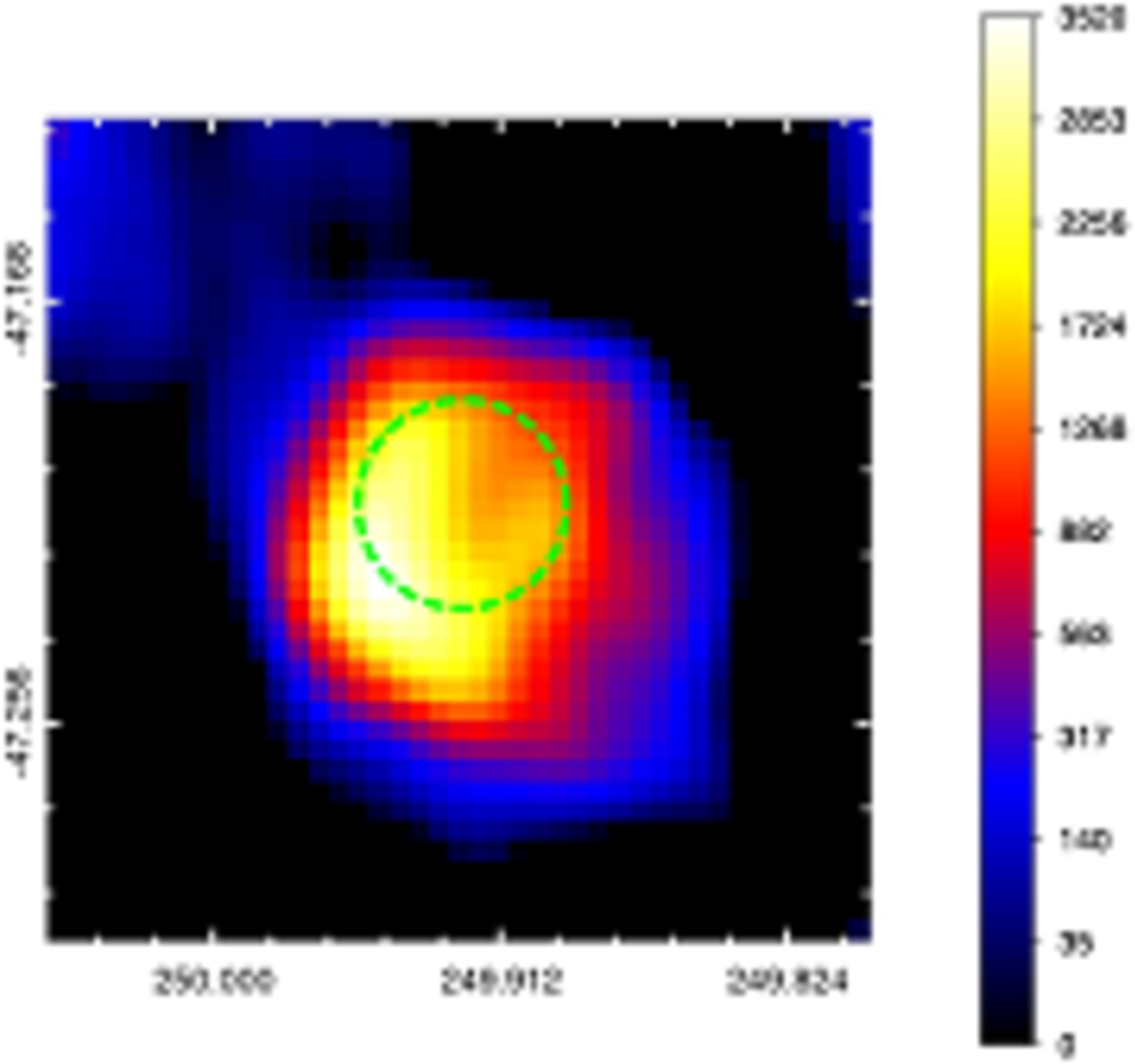}}}%
}
\subfigure{
\mbox{\raisebox{6mm}{\rotatebox{90}{\small{DEC (J2000)}}}}%
\mbox{\raisebox{0mm}{\includegraphics[width=40mm]{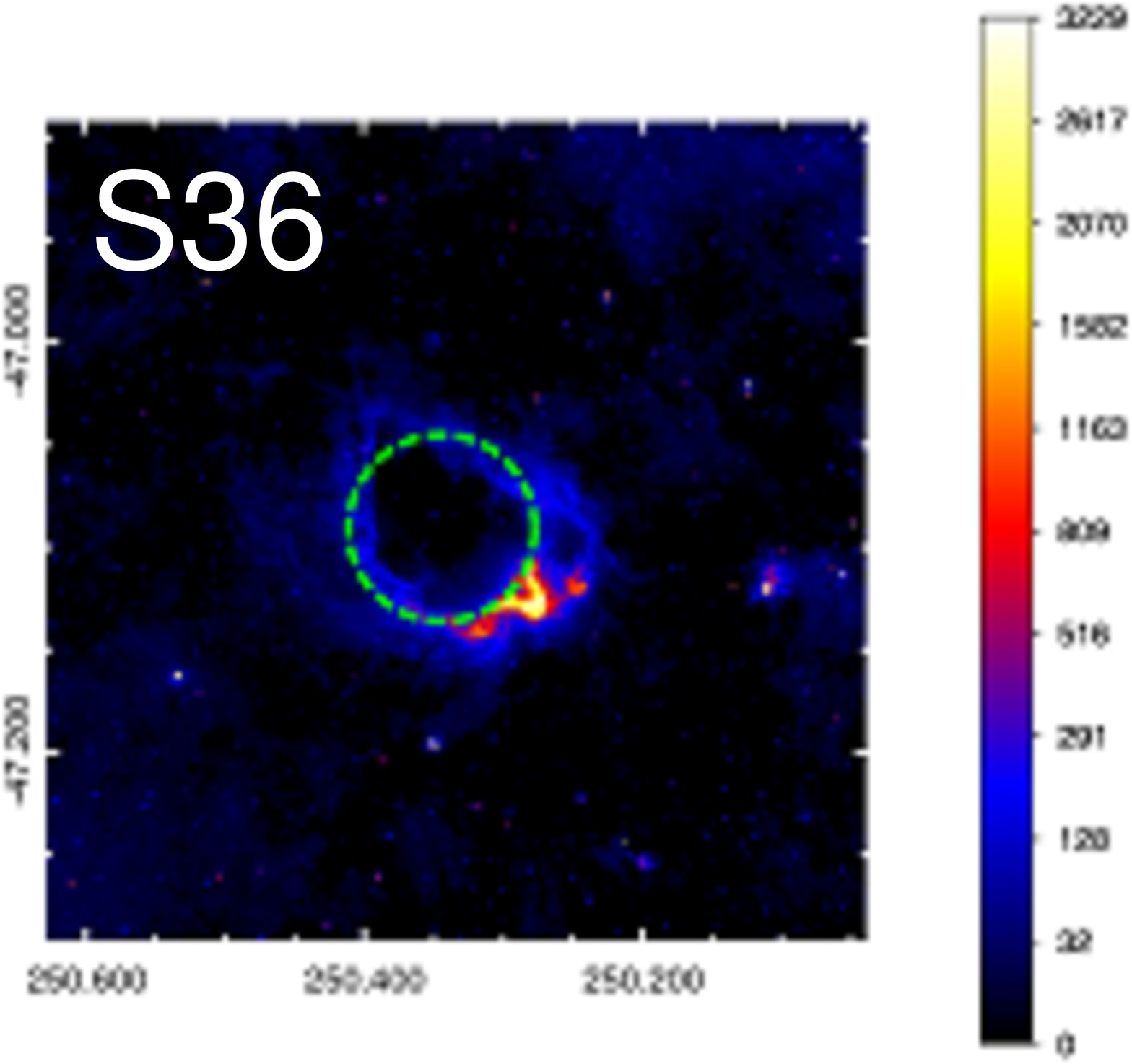}}}%
\mbox{\raisebox{0mm}{\includegraphics[width=40mm]{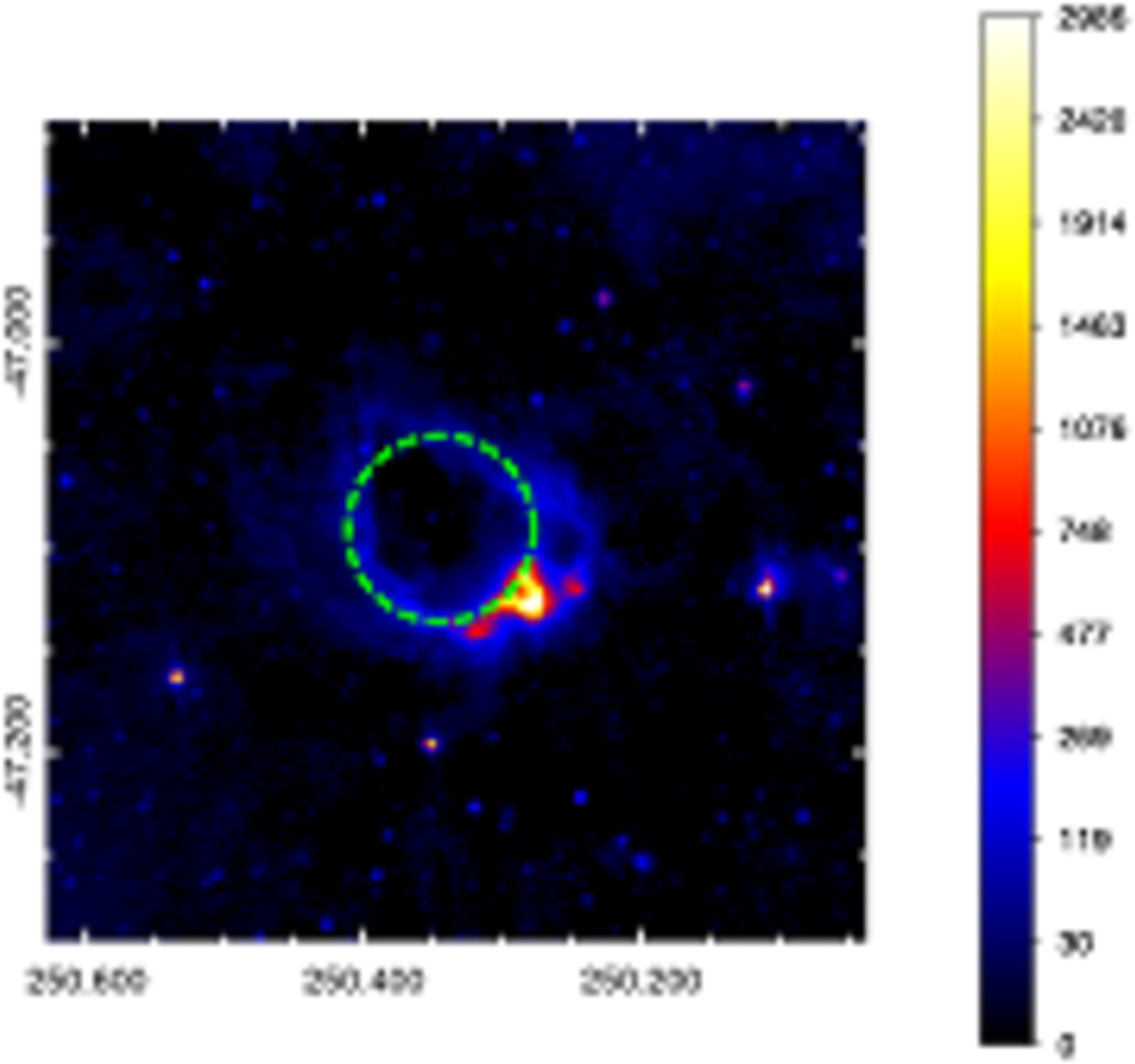}}}%
\mbox{\raisebox{0mm}{\includegraphics[width=40mm]{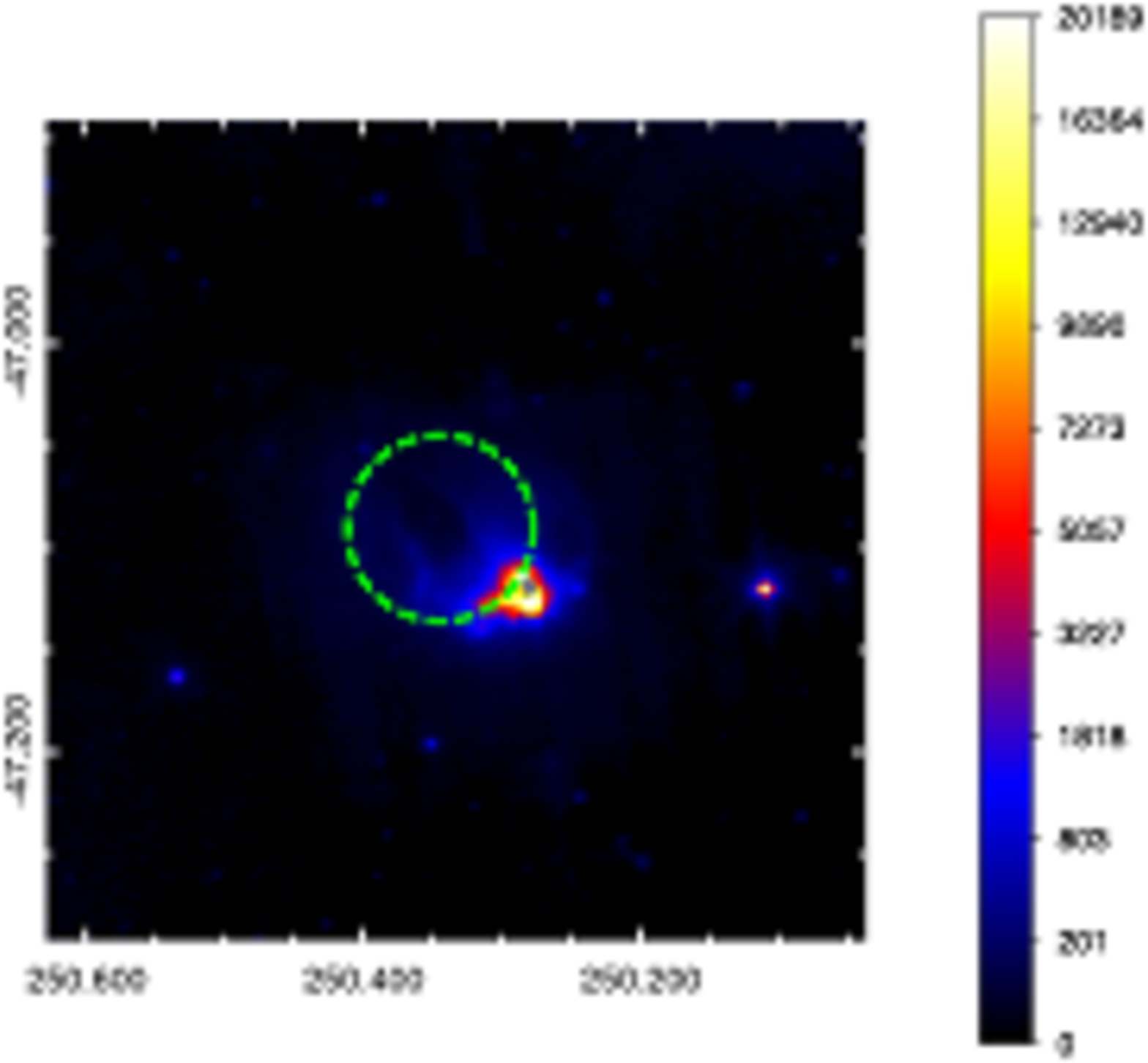}}}%
\mbox{\raisebox{0mm}{\includegraphics[width=40mm]{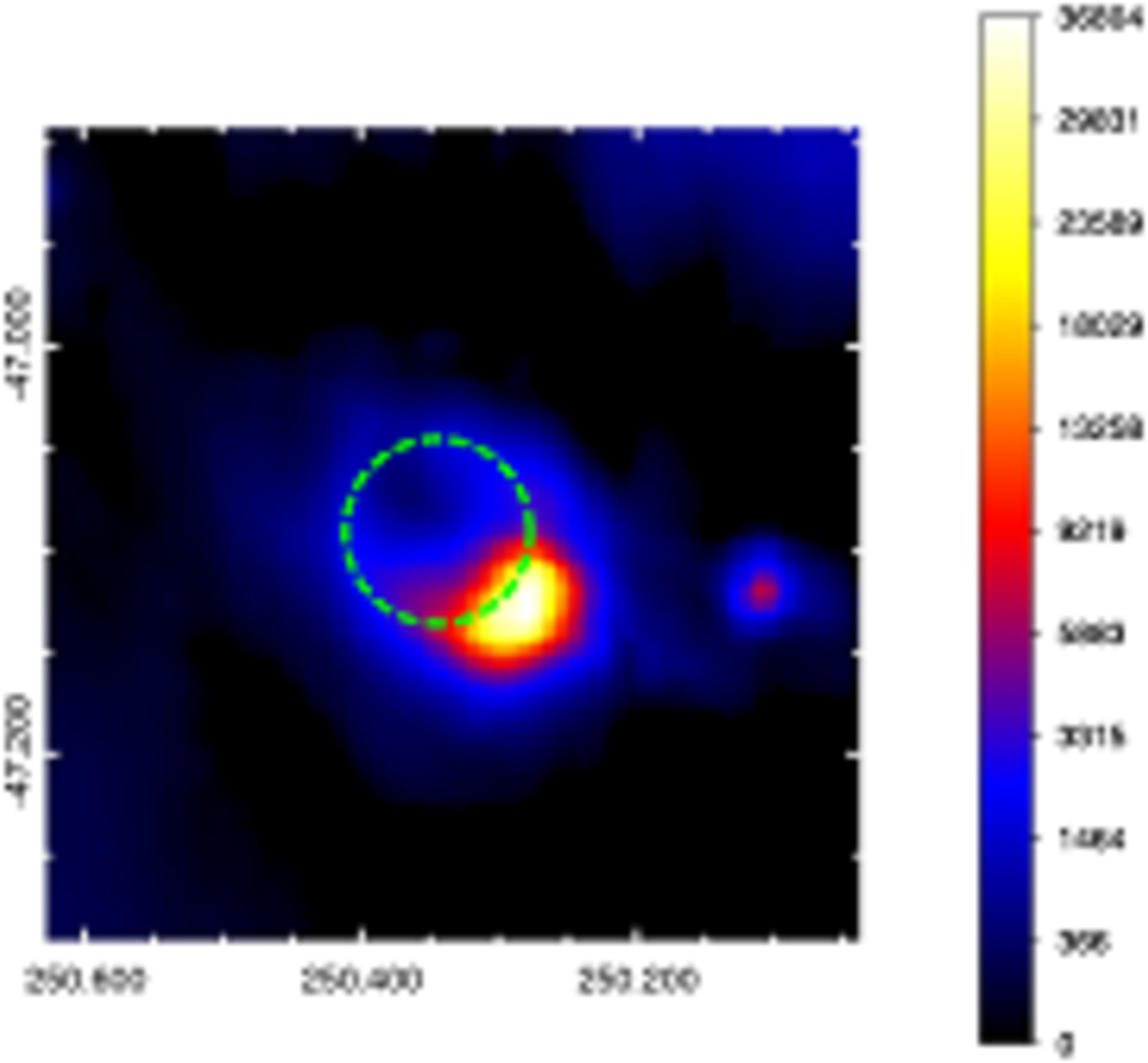}}}%
}
\caption{Continued.} \label{fig:Introfig1:g}
\end{figure*}

\addtocounter{figure}{-1}
\begin{figure*}[ht]
\addtocounter{subfigure}{1}
\centering
\subfigure{
\makebox[180mm][l]{\raisebox{0mm}[0mm][0mm]{ \hspace{15mm} \small{8 \mic}} \hspace{29.5mm} \small{9 \mic} \hspace{27mm} \small{18 \mic} \hspace{26.5mm} \small{90 \mic}}%
}
\subfigure{
\makebox[180mm][l]{\raisebox{0mm}[0mm][0mm]{ \hspace{11mm} \small{RA (J2000)}} \hspace{19.5mm} \small{RA (J2000)} \hspace{20mm} \small{RA (J2000)} \hspace{20mm} \small{RA (J2000)}}%
}
\subfigure{
\mbox{\raisebox{6mm}{\rotatebox{90}{\small{DEC (J2000)}}}}%
\mbox{\raisebox{0mm}{\includegraphics[width=40mm]{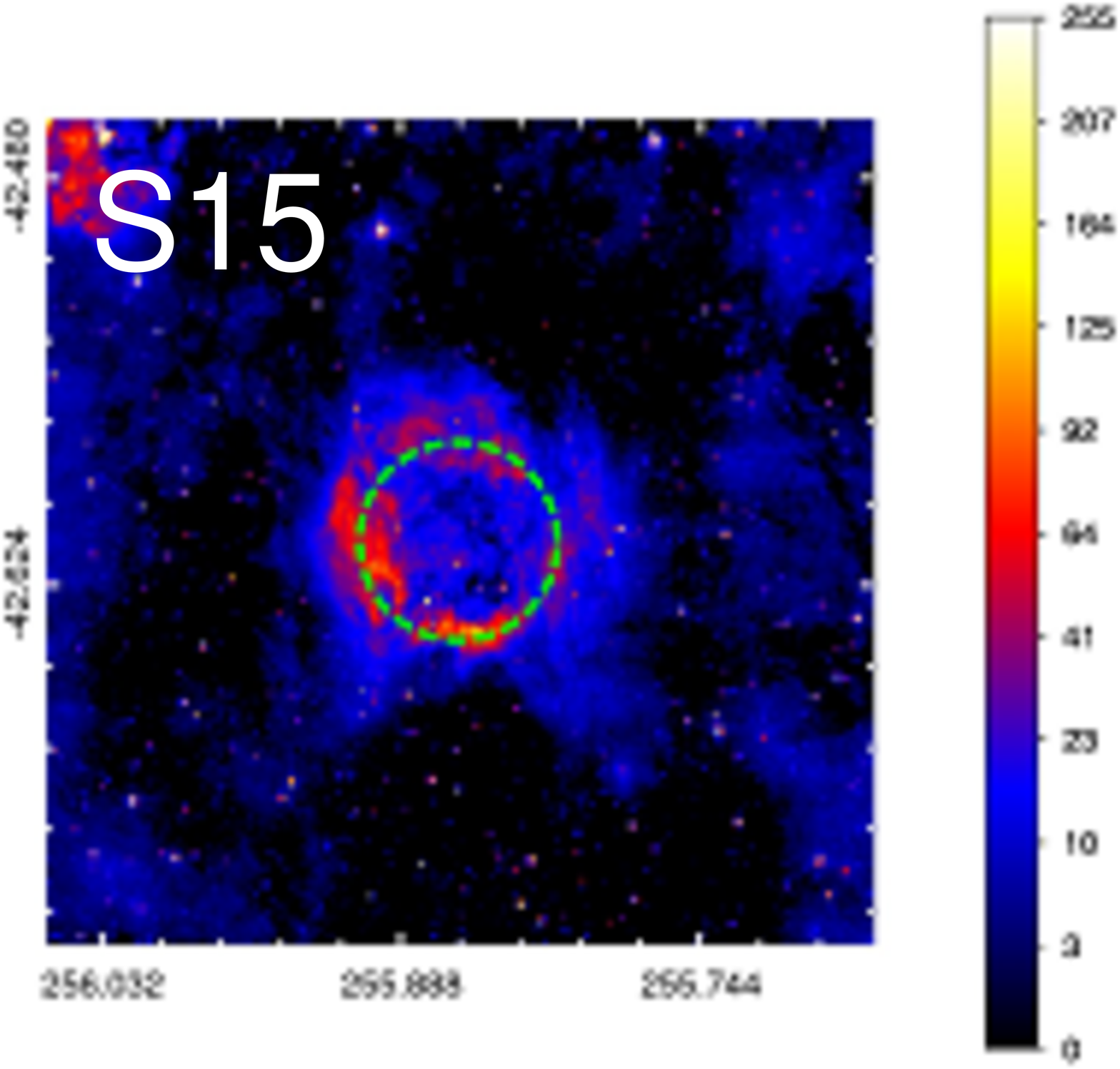}}}%
\mbox{\raisebox{0mm}{\includegraphics[width=40mm]{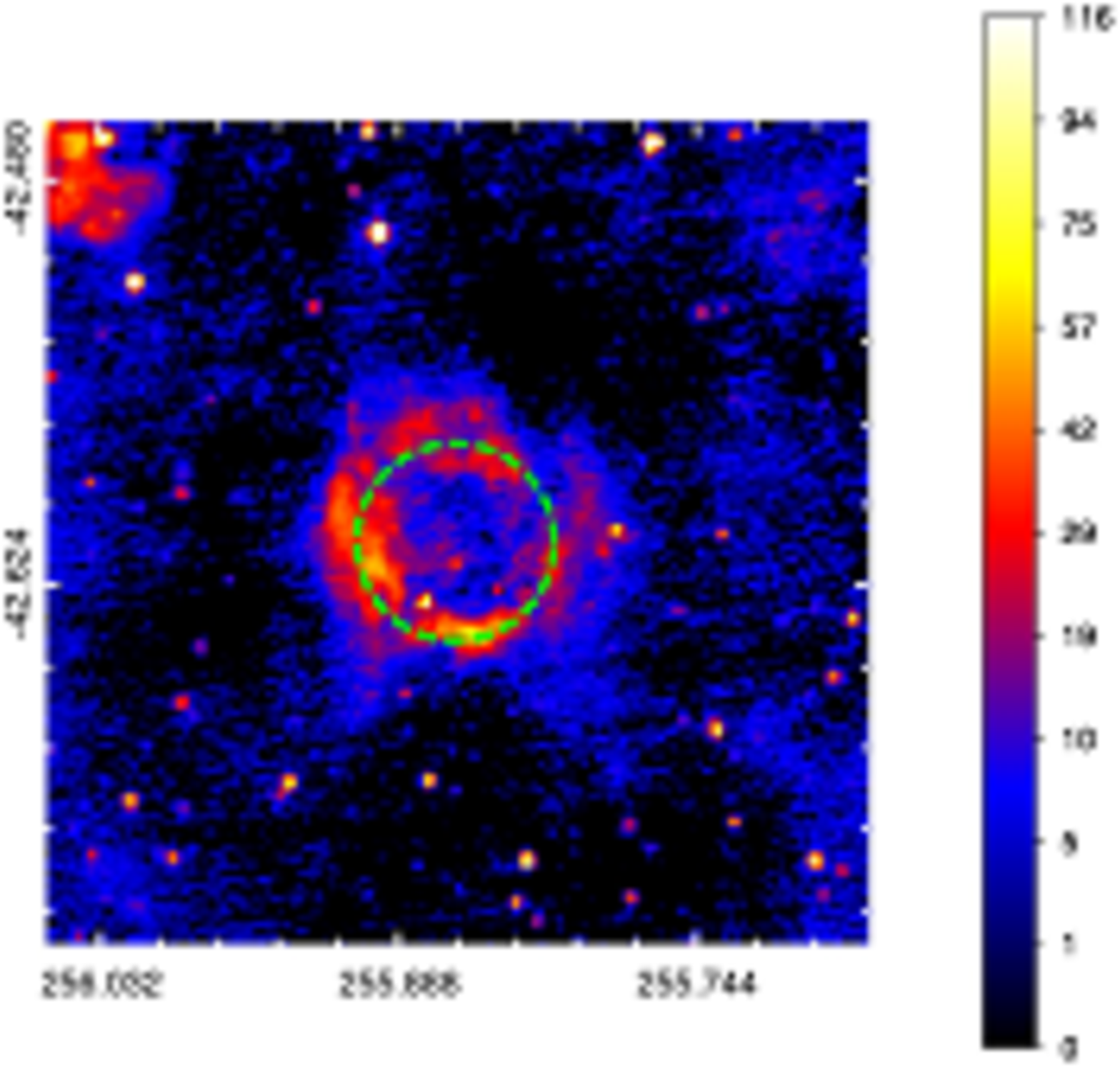}}}%
\mbox{\raisebox{0mm}{\includegraphics[width=40mm]{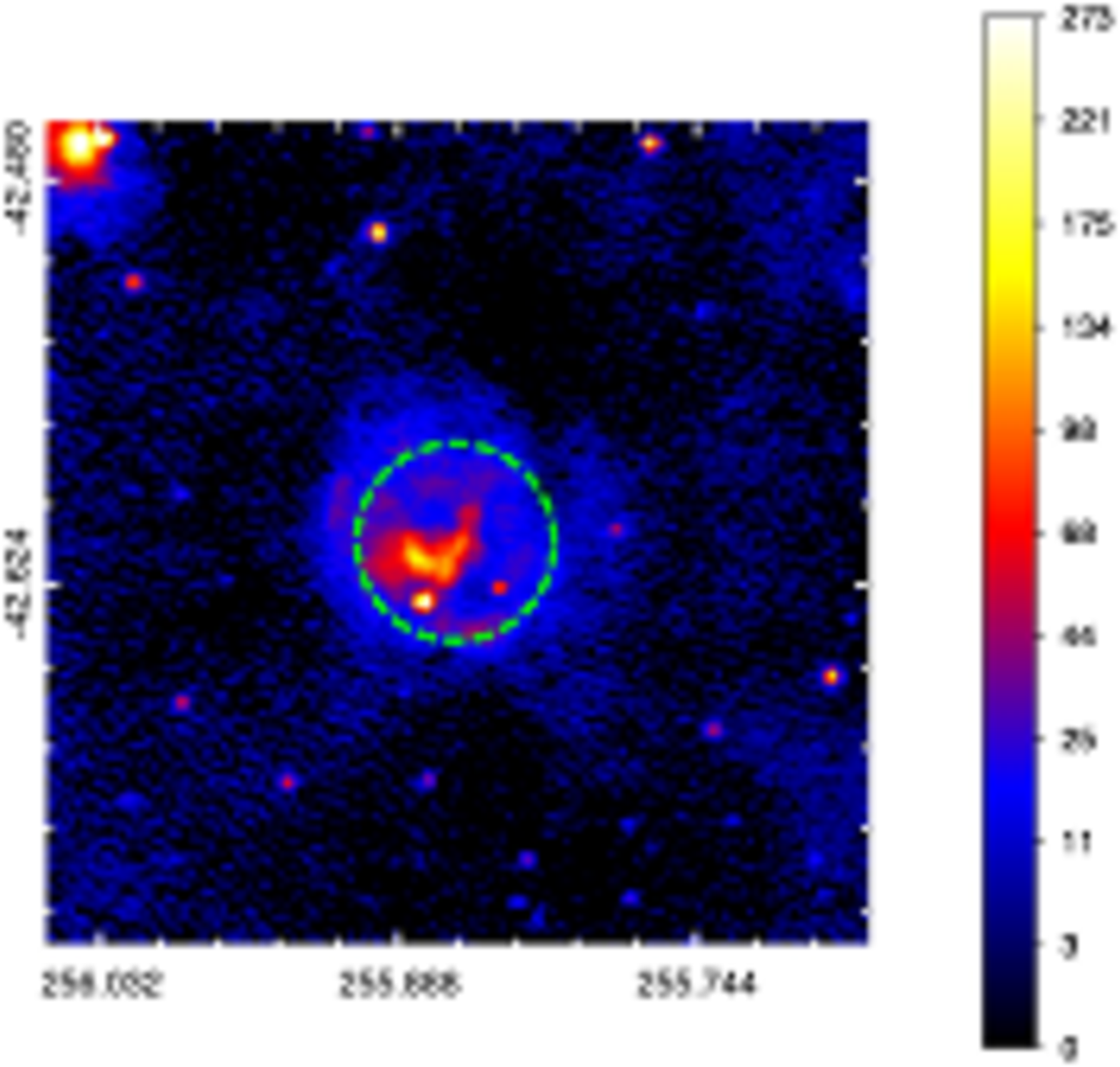}}}%
\mbox{\raisebox{0mm}{\includegraphics[width=40mm]{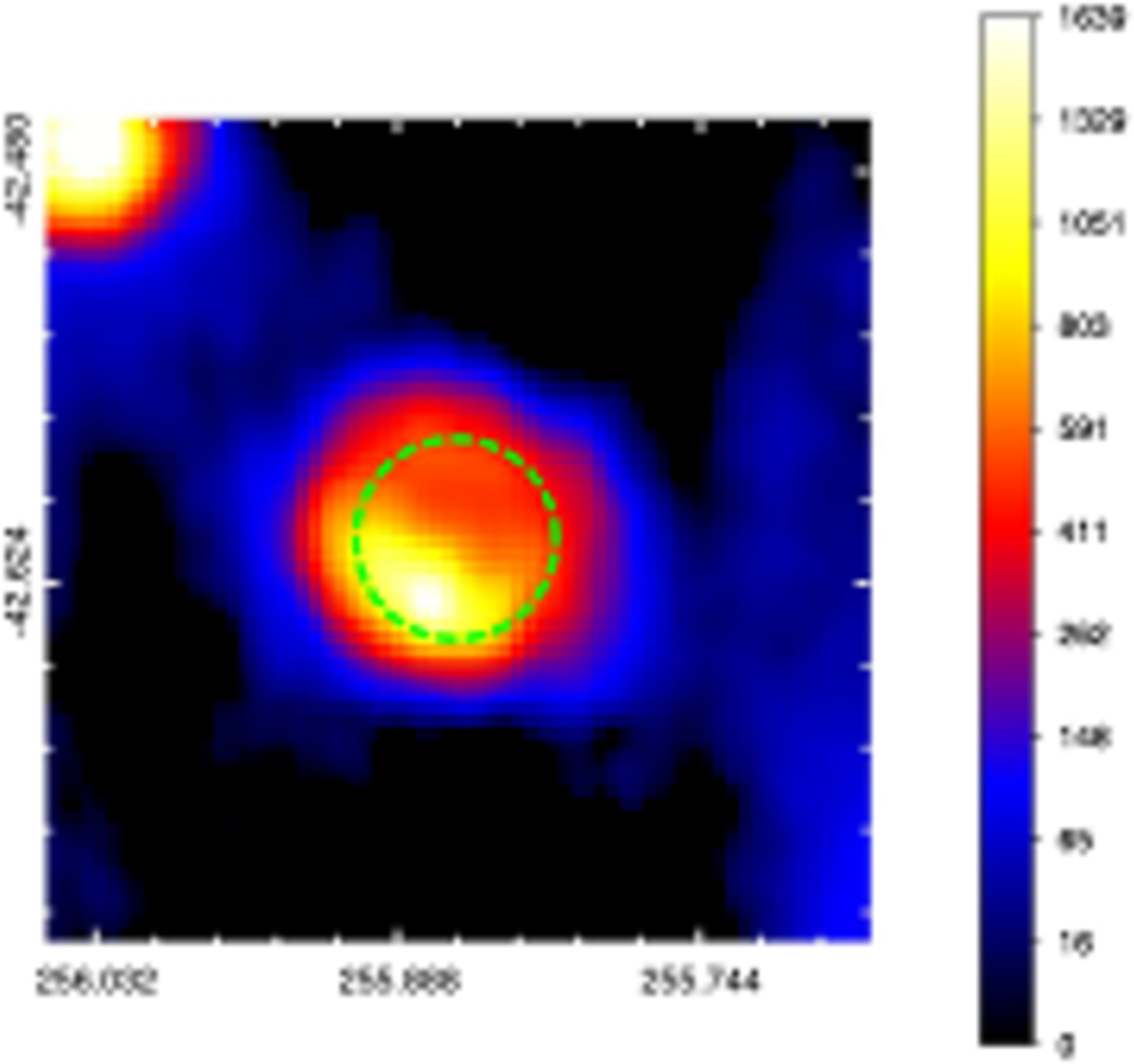}}}%
}
\subfigure{
\mbox{\raisebox{6mm}{\rotatebox{90}{\small{DEC (J2000)}}}}%
\mbox{\raisebox{0mm}{\includegraphics[width=40mm]{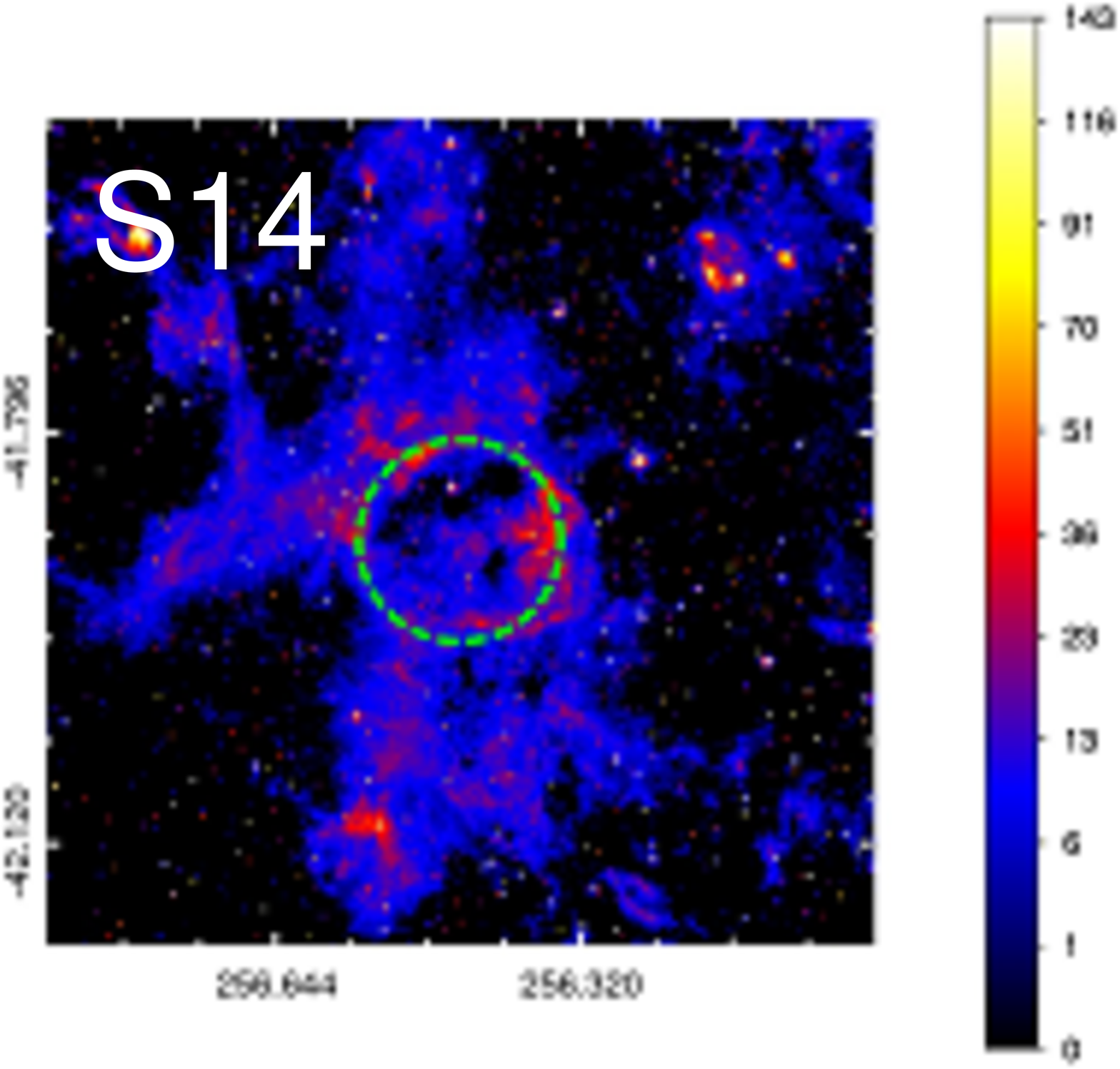}}}%
\mbox{\raisebox{0mm}{\includegraphics[width=40mm]{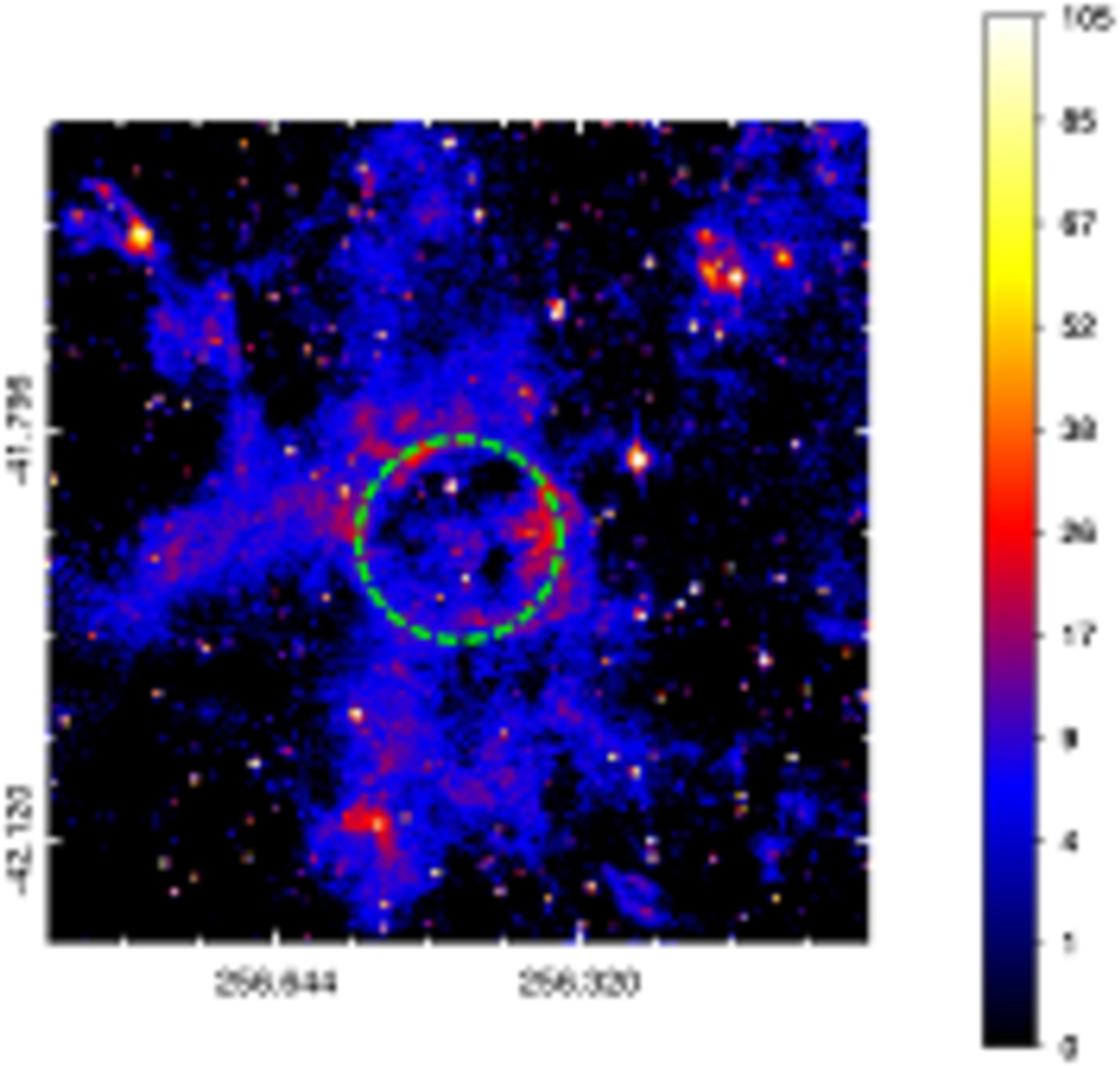}}}%
\mbox{\raisebox{0mm}{\includegraphics[width=40mm]{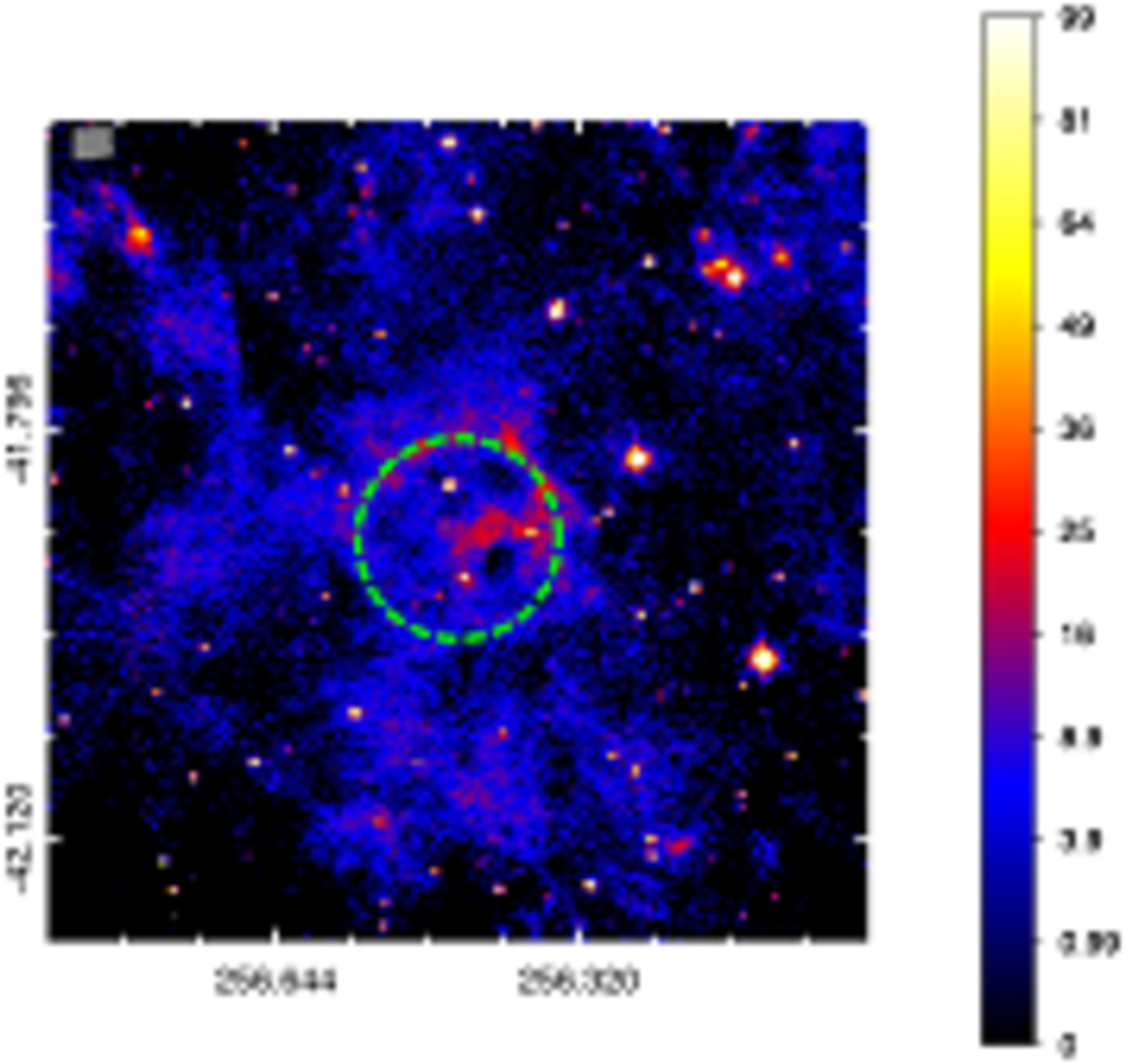}}}%
\mbox{\raisebox{0mm}{\includegraphics[width=40mm]{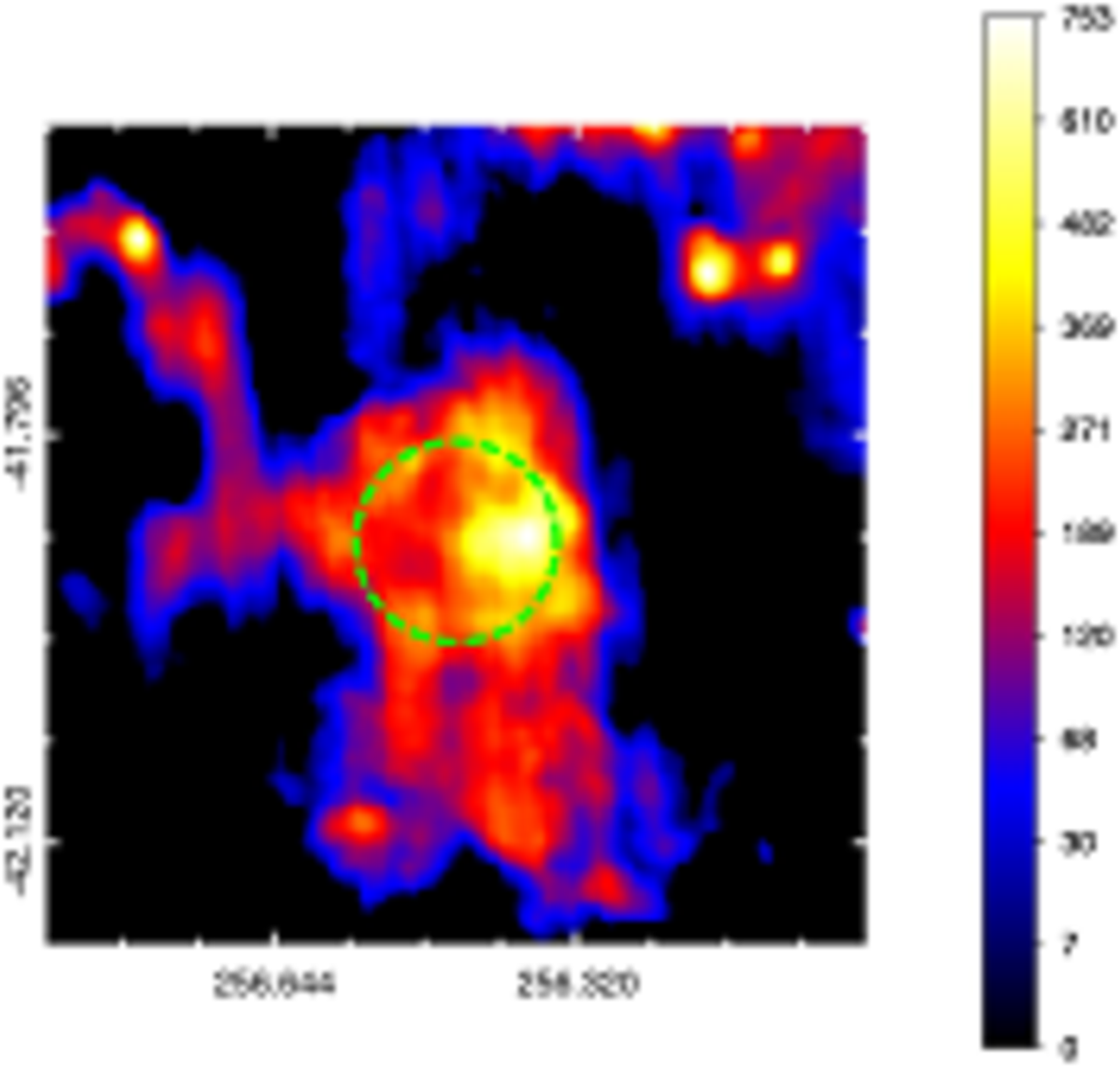}}}%
}
\subfigure{
\mbox{\raisebox{6mm}{\rotatebox{90}{\small{DEC (J2000)}}}}%
\mbox{\raisebox{0mm}{\includegraphics[width=40mm]{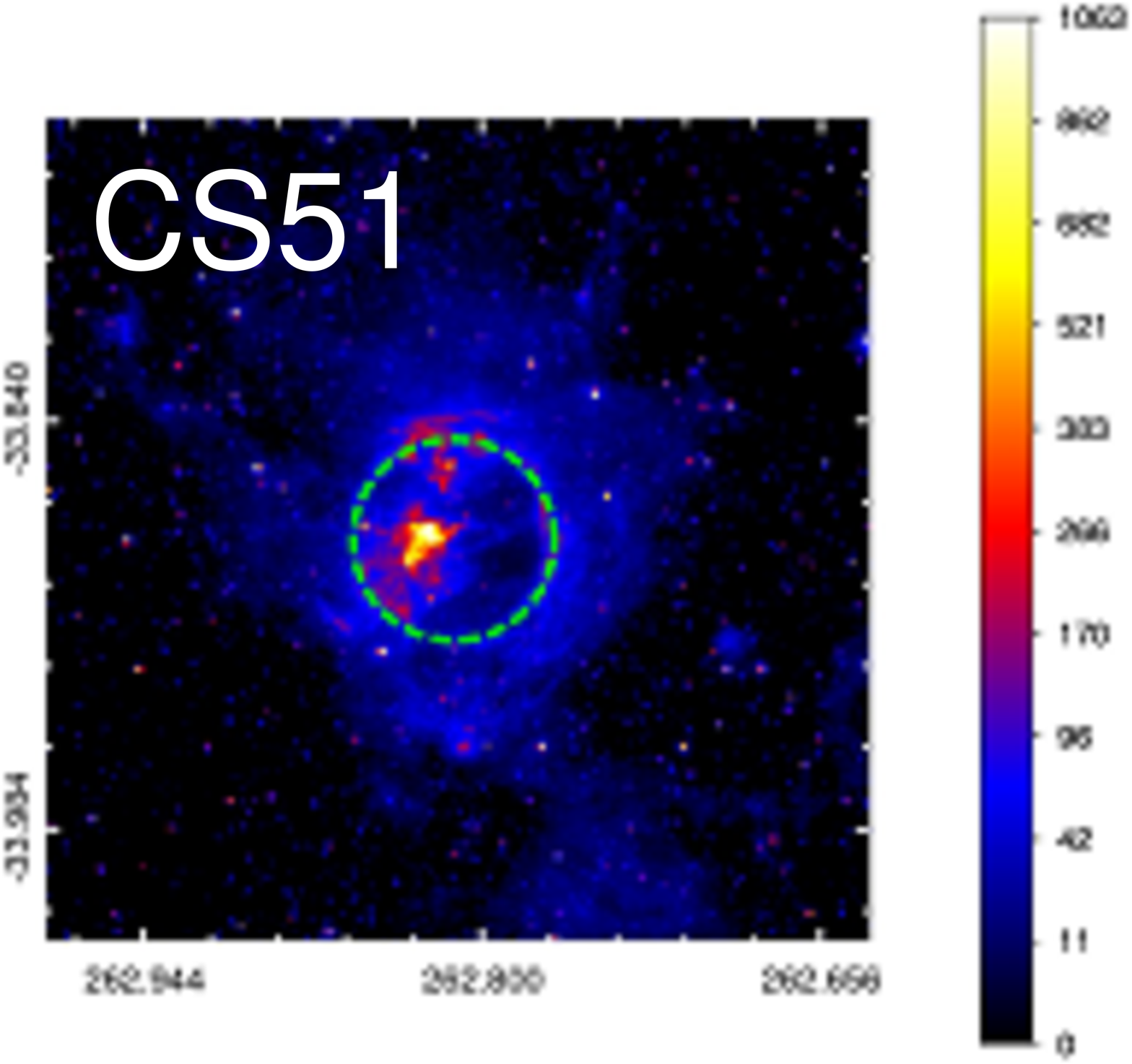}}}%
\mbox{\raisebox{0mm}{\includegraphics[width=40mm]{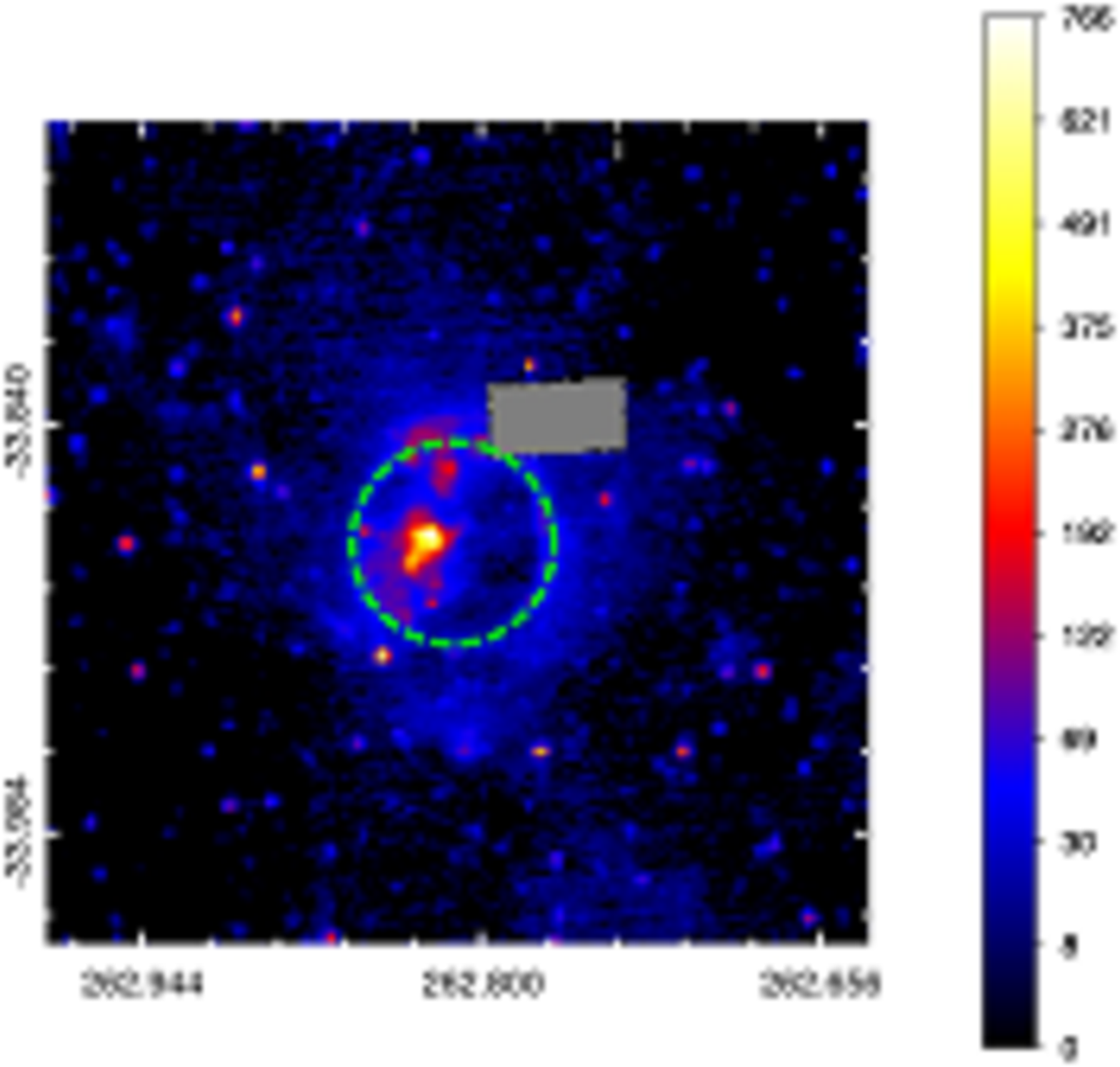}}}%
\mbox{\raisebox{0mm}{\includegraphics[width=40mm]{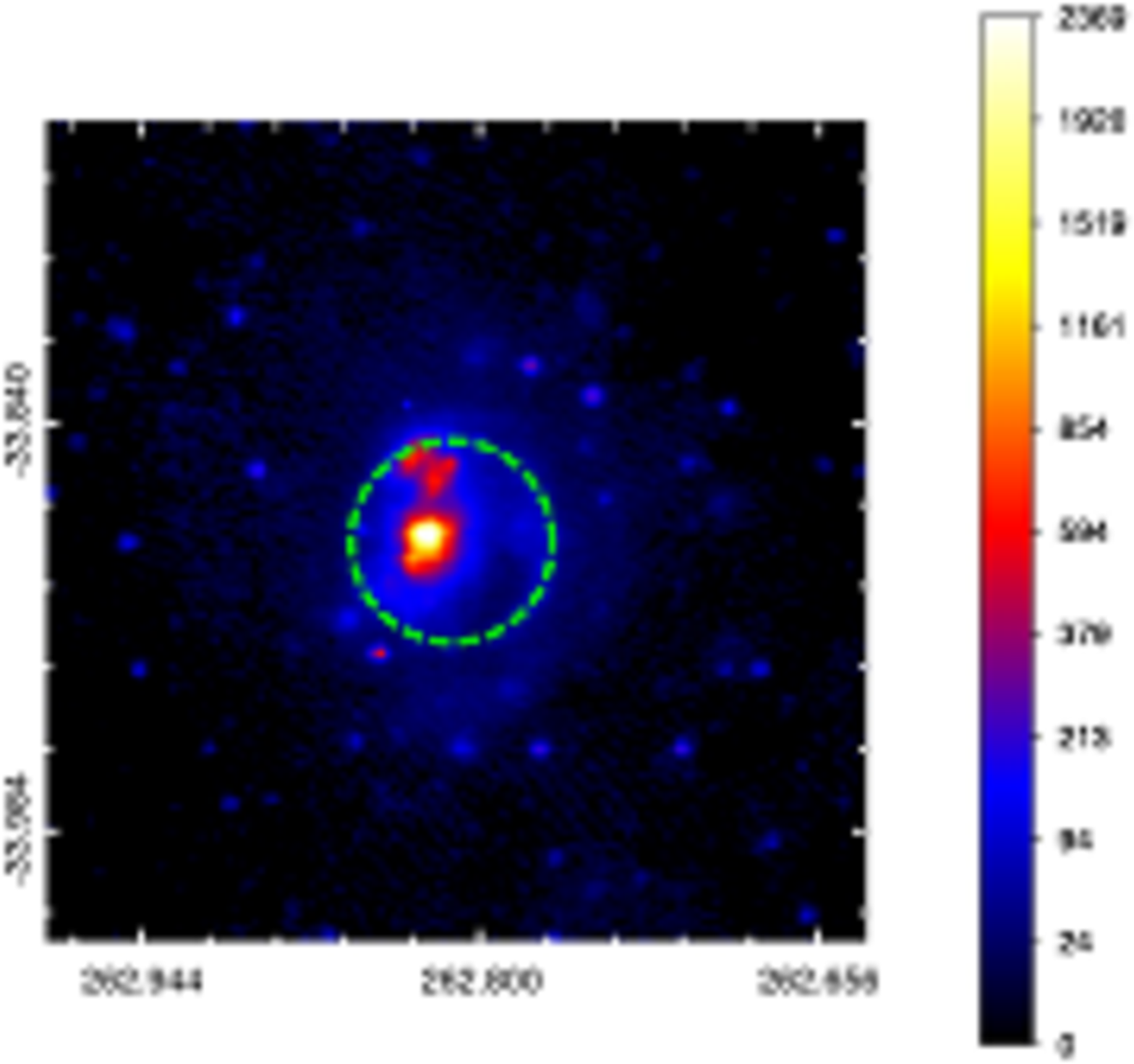}}}%
\mbox{\raisebox{0mm}{\includegraphics[width=40mm]{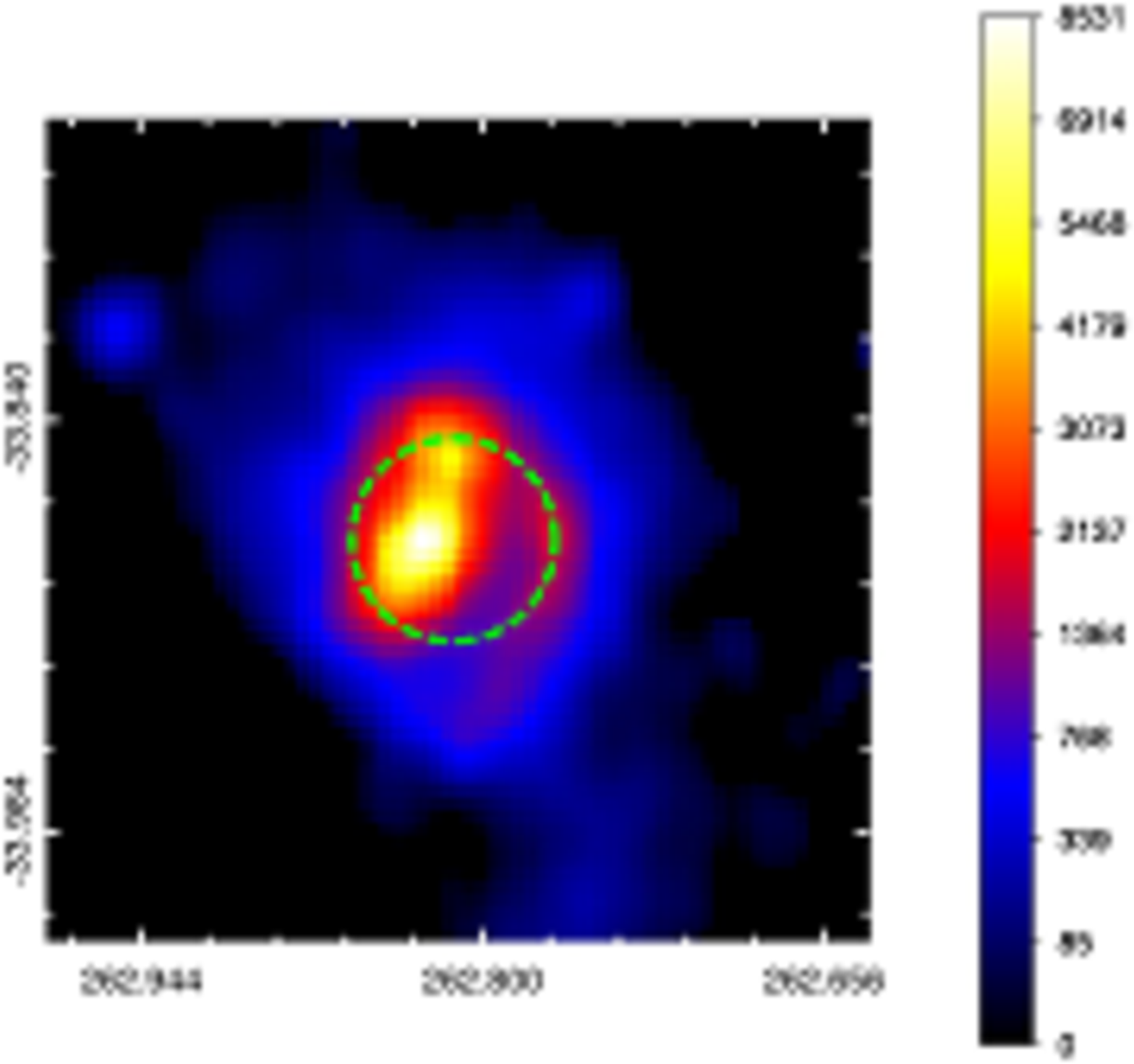}}}%
}
\subfigure{
\mbox{\raisebox{6mm}{\rotatebox{90}{\small{DEC (J2000)}}}}%
\mbox{\raisebox{0mm}{\includegraphics[width=40mm]{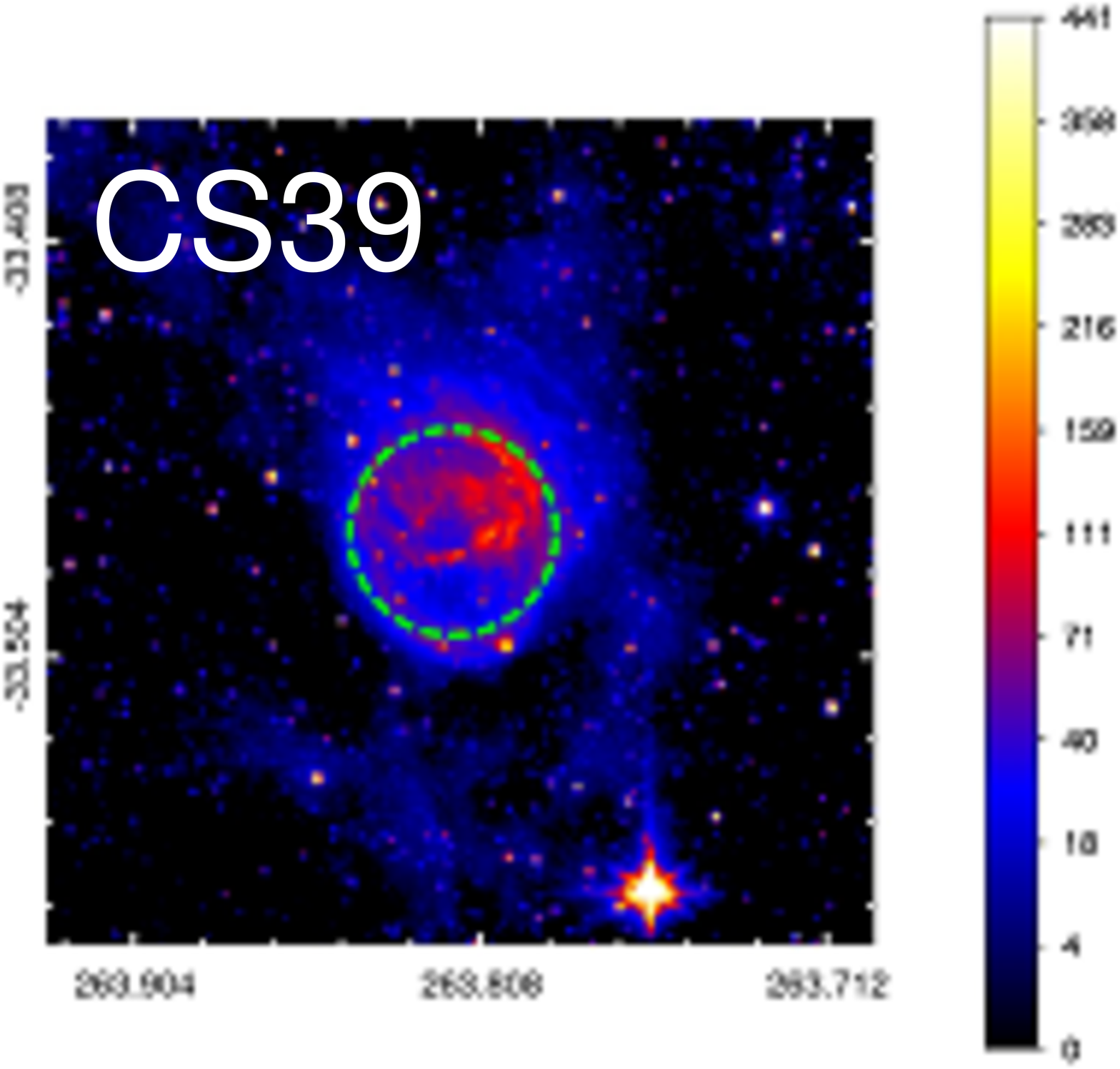}}}%
\mbox{\raisebox{0mm}{\includegraphics[width=40mm]{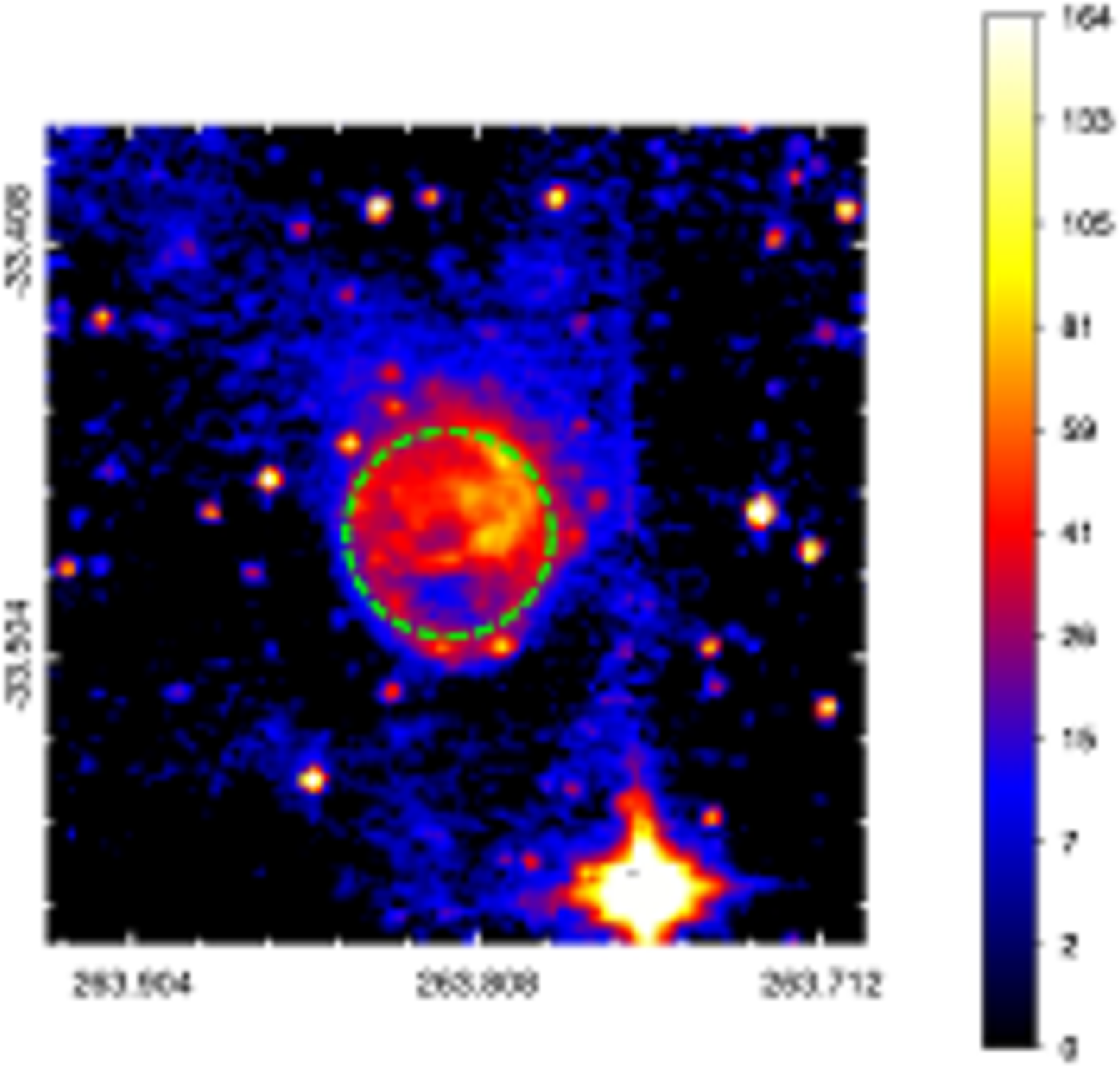}}}%
\mbox{\raisebox{0mm}{\includegraphics[width=40mm]{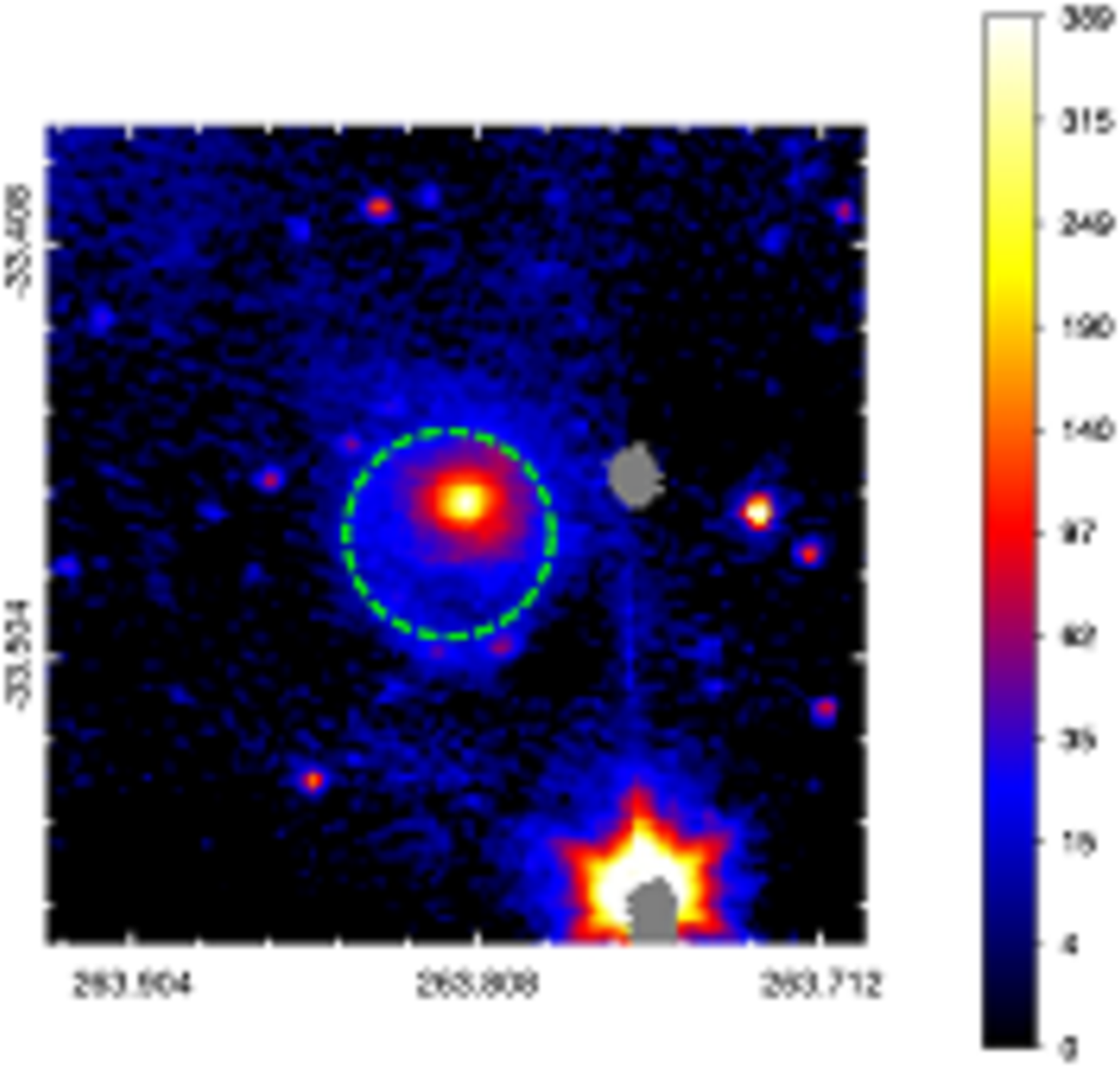}}}%
\mbox{\raisebox{0mm}{\includegraphics[width=40mm]{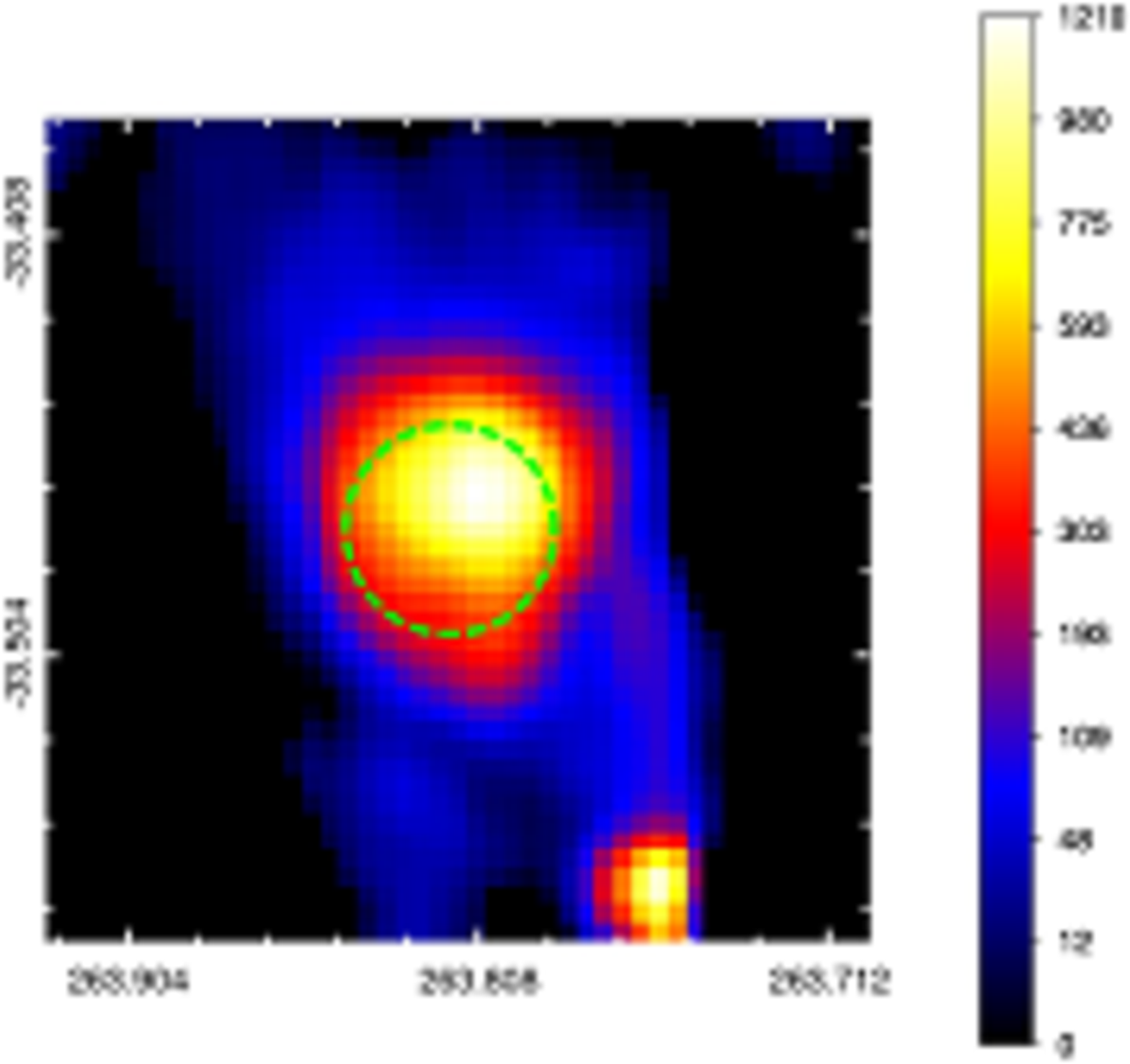}}}%
}
\subfigure{
\mbox{\raisebox{6mm}{\rotatebox{90}{\small{DEC (J2000)}}}}%
\mbox{\raisebox{0mm}{\includegraphics[width=40mm]{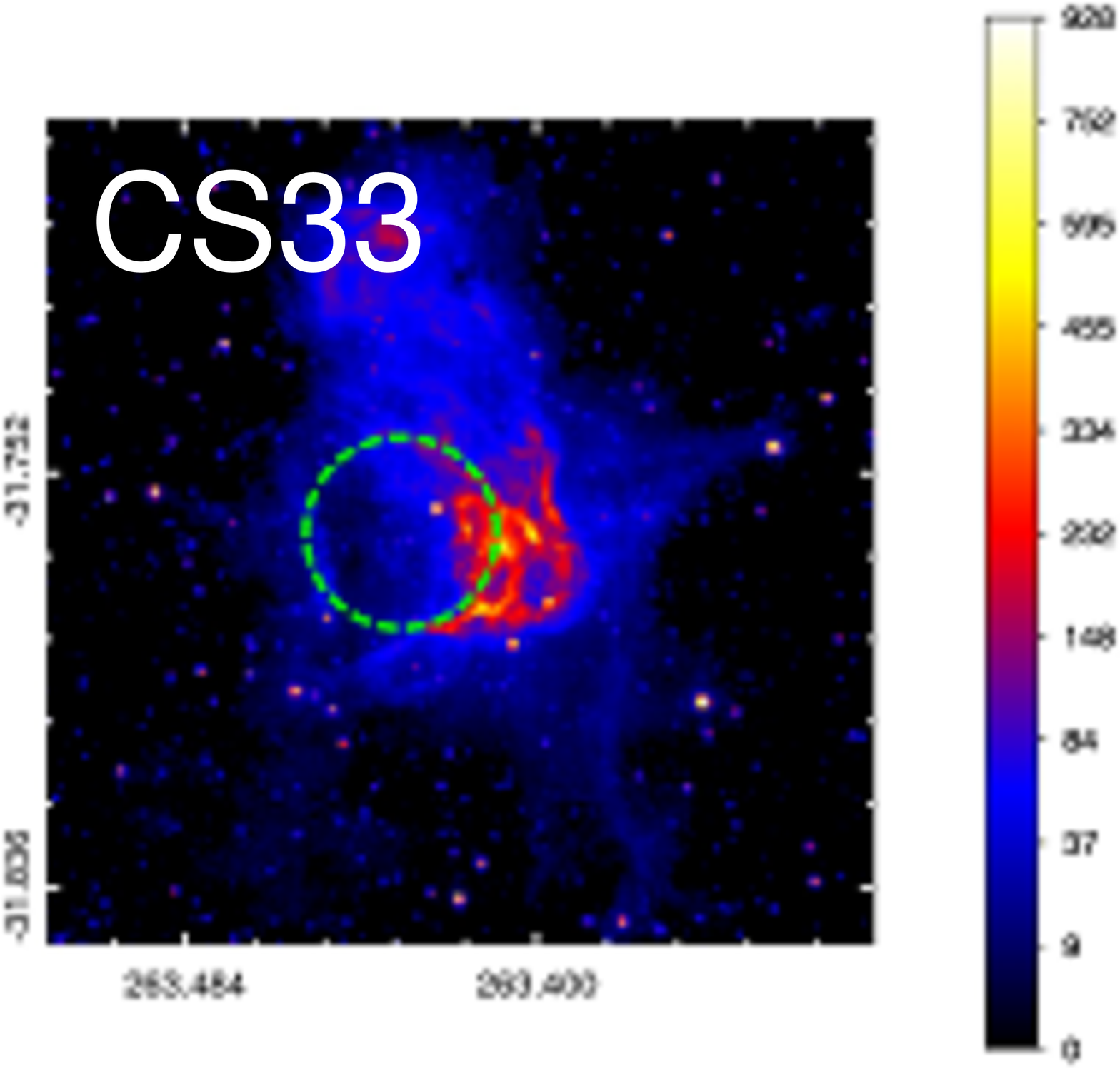}}}%
\mbox{\raisebox{0mm}{\includegraphics[width=40mm]{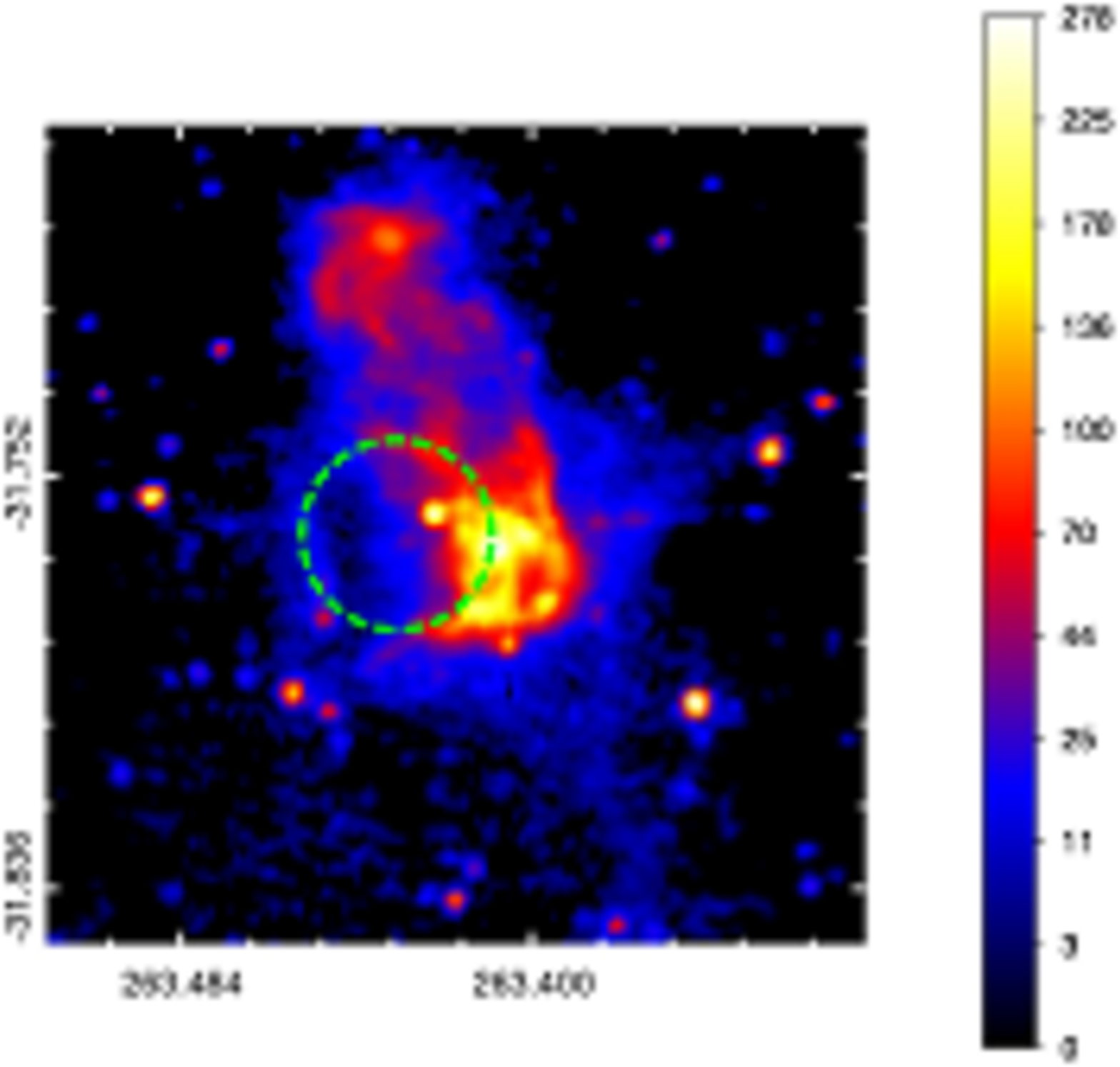}}}%
\mbox{\raisebox{0mm}{\includegraphics[width=40mm]{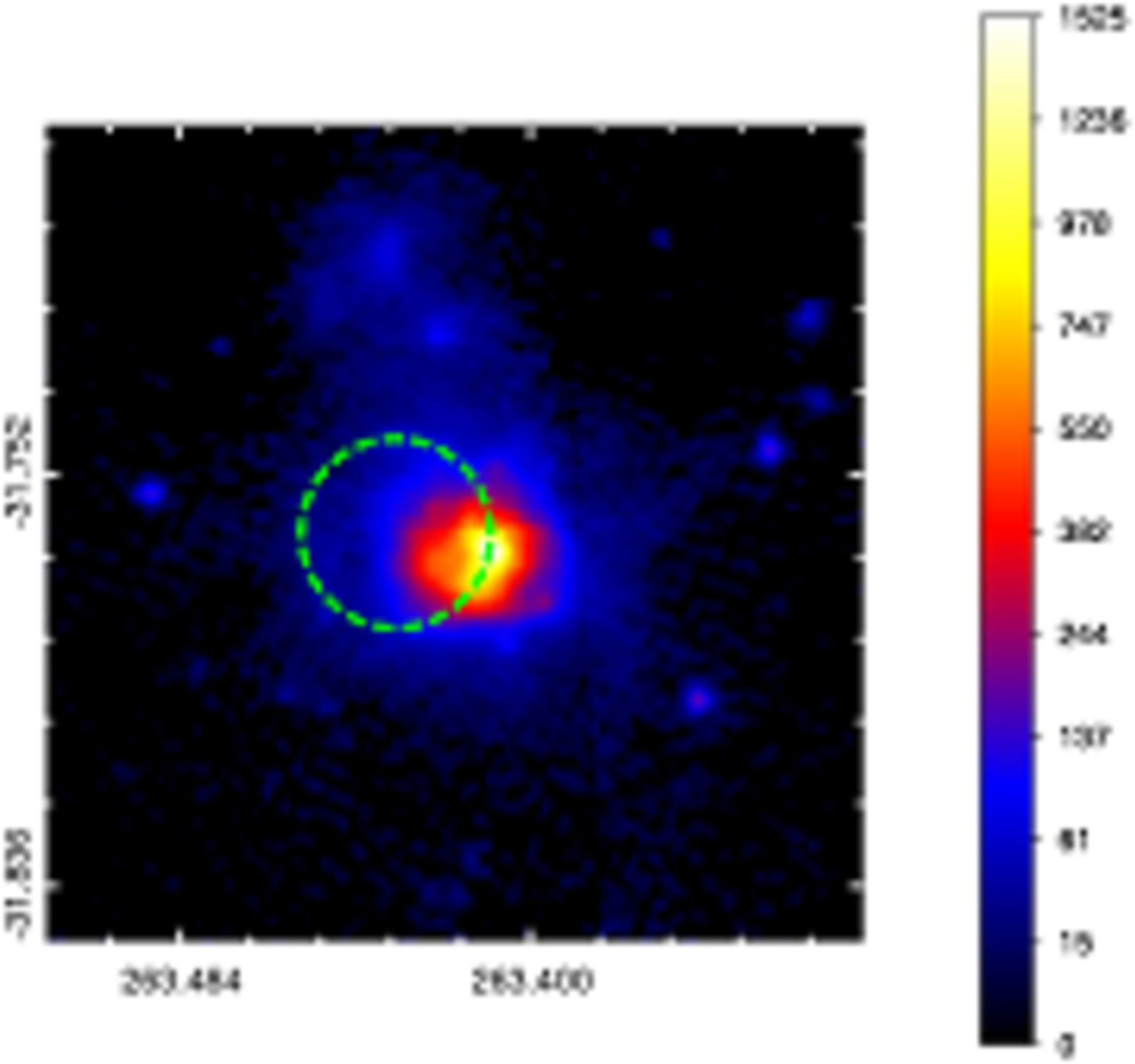}}}%
\mbox{\raisebox{0mm}{\includegraphics[width=40mm]{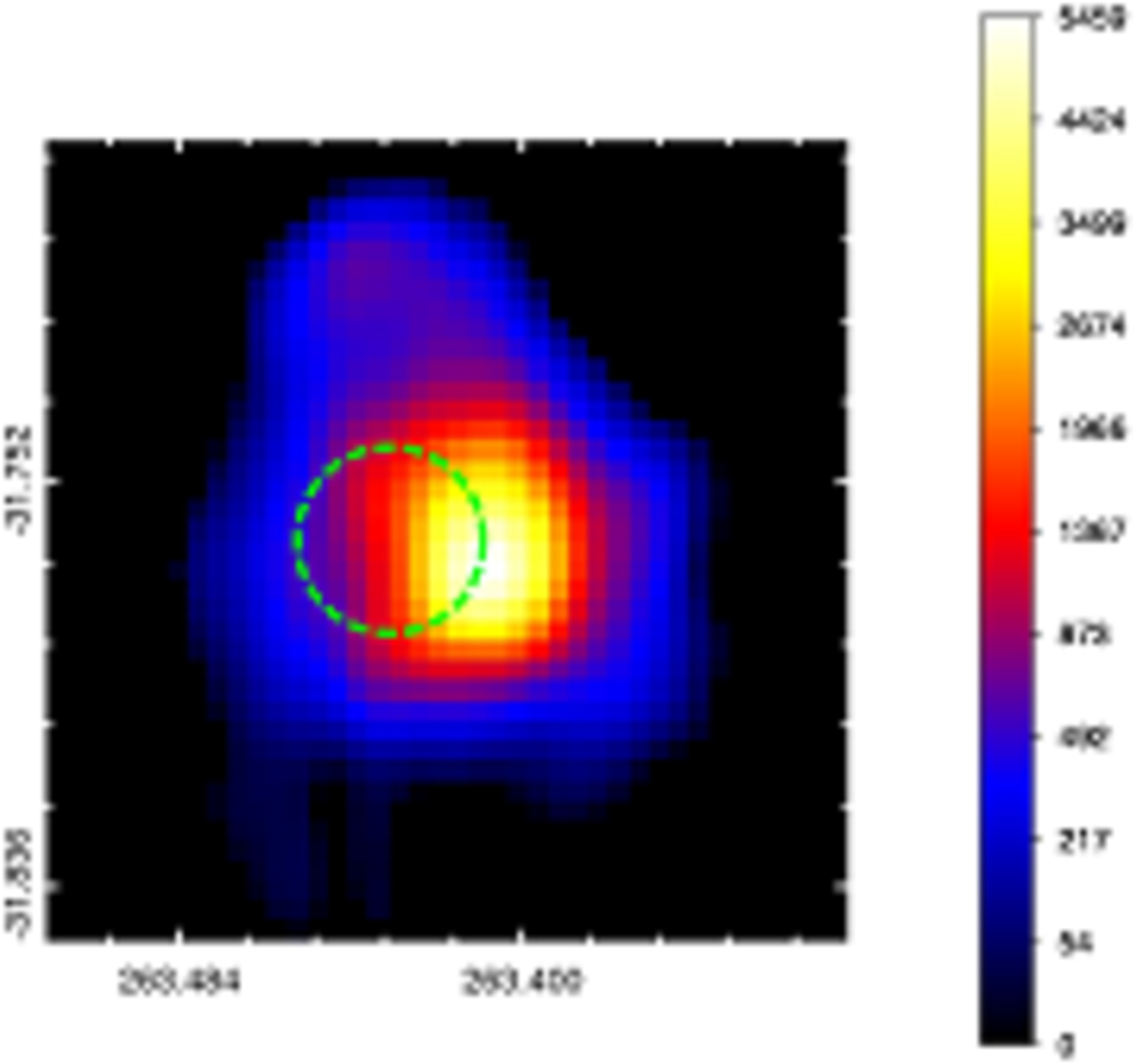}}}%
}
\caption{Continued.} \label{fig:Introfig1:h}
\end{figure*}

\clearpage

\section{Observations and data analyses}
The AKARI satellite (\citealt{Murakami2007}) performed all-sky surveys in the 6 photometric bands at the central wavelengths of 9, 18, 65, 90, 140 and 160 \mic and provided the all-sky maps with better spatial resolutions and a wider spectral range than IRAS. We used the mid-IR data (Ishihara et al. in preparation) and the far-IR archive data\footnote[1]{Taken from http://www.ir.isas.jaxa.jp/AKARI/Archive/Images/FISMAP/.} publicly released in 2015 (\citealt{Doi2015}). The full widths at half maxima (FWHMs) of the beam sizes in the 9, 18, 65, 90, 140 and 160 \mic bands are \timeform{5.5''}, \timeform{5.7''}, \timeform{63.4''}, \timeform{77.8''}, \timeform{88.3''} and \timeform{88.3''}, respectively, and the pixel scales of the mid-IR (9 and 18 \emic) and far-IR (65, 90, 140 and 160 \emic) all-sky images are \timeform{4.68''} and \timeform{15''}, respectively (\citealt{Onaka2007}; \citealt{Takita2015}; Ishihara et al. in preparation).

Among the Galactic IR bubbles cataloged in \citet{Churchwell2006} and (2007), we selected the 223 bubbles for which the distances were measured in the previous works (\citealt{Beaumont2010}; \citealt{Cappa2014}; \citealt{Churchwell2006}; \citealt{Churchwell2007}; \citealt{Deharveng2009}; \citealt{Deharveng2010}; \citealt{Dewangan2015}; \citealt{Gennaro2012}; \citealt{Hou2014}; \citealt{Pavel2012}; \citealt{Rahman2010}; \citealt{Rodriguez-Esnard2012}; \citealt{Watson2009}; \citealt{Watson2010}; \citealt{Zhang2013}). The distances and the references are summarized in columns 5 and 6 of table \ref{tab:Mettab1}, respectively. In some cases, the kinematic distances which are determined from the Galactic rotational model and the radial velocities of the objects have distance ambiguity problems, as discussed in \citet{Roman-Duval2009}. If there was ambiguity in the kinematic distance, we adopted shorter distances since nearer bubbles are more likely to be detected (\citealt{Churchwell2006}). For the bubbles with no distance errors given in table \ref{tab:Mettab1}, we adopted the typical distance error of 13{\%}, which is the median for the bubbles with known distance errors. Among the 223 bubbles, we found that 23 bubbles severely suffer source confusion and thus removed them from our sample. Hence we selected 200 bubbles in total as our initial sample. 

\begin{figure}[ht] %
\begin{tabular}{cc} %
\centering %
\begin{minipage}[t]{0.50\hsize} %
\begin{center} %
\includegraphics[clip, width=85mm]{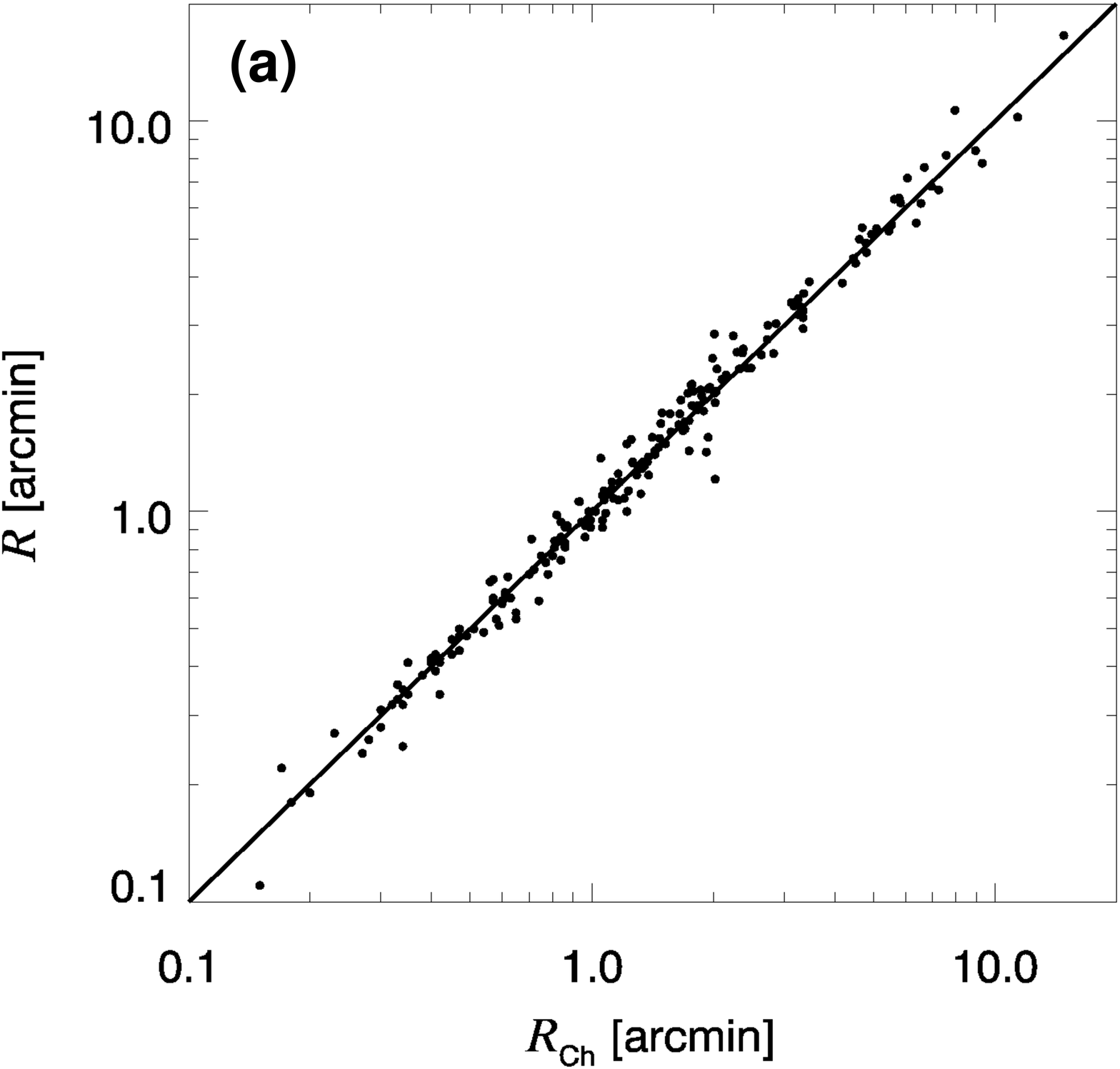} %
\end{center} %
\end{minipage} %
\begin{minipage}[t]{0.50\hsize} %
\begin{center} %
\includegraphics[clip, width=85mm]{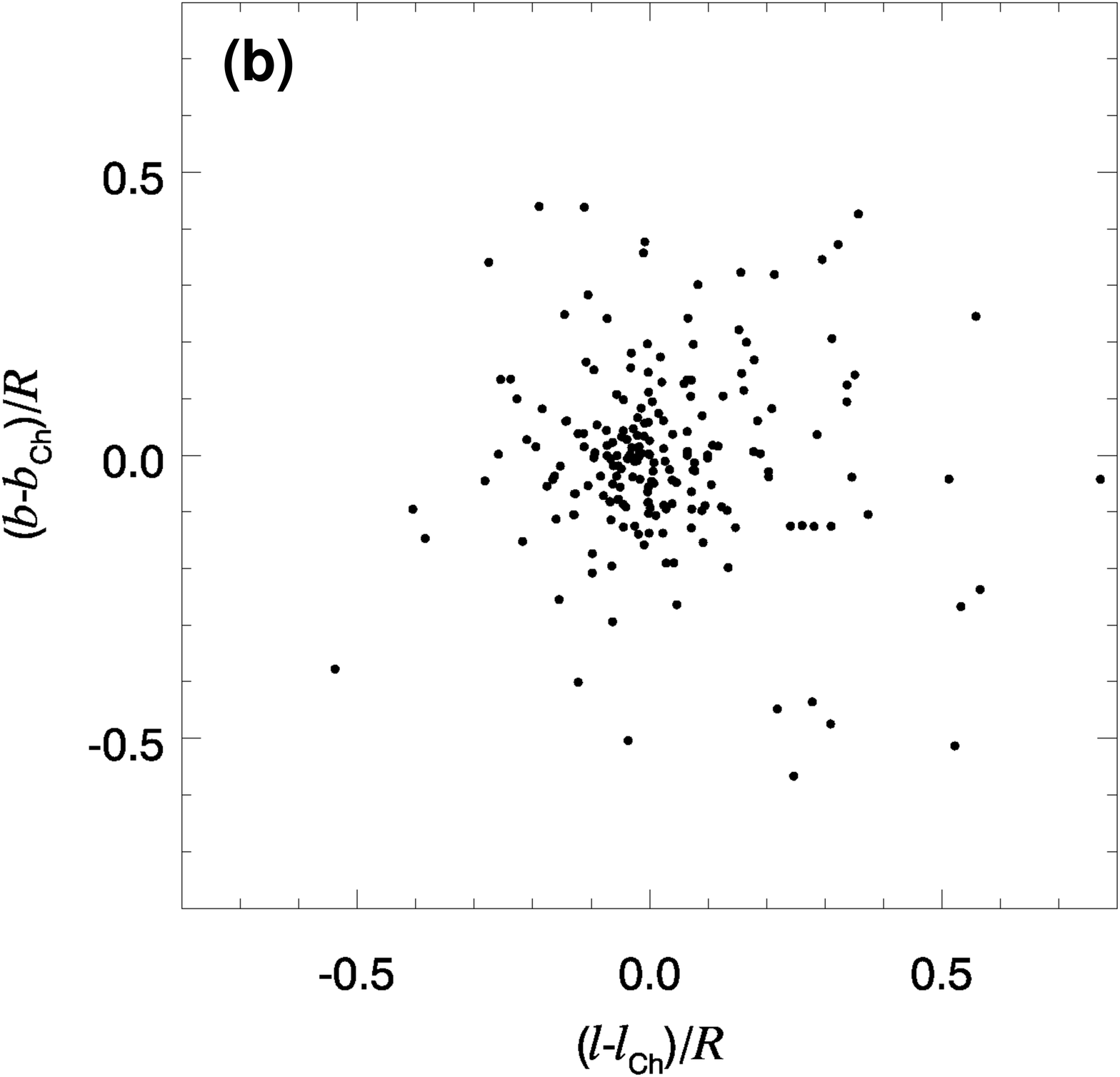} %
\end{center} %
\end{minipage} %
\end{tabular} %
\begin{flushleft} %
\caption{Scatter plots of (a) $R{_{\rm Ch}}$ versus $R$ and (b) $(l - l{_{\rm Ch}})/ R$ versus $(b - b{_{\rm Ch}})/ R$ for our sample of the Galactic IR bubbles. The line in panel (a) indicates $R$ = $R{_{\rm Ch}}$. \label{fig:Metfig1}} %
\end{flushleft} %
\end{figure}

Then we performed circular fitting to the selected sample to determine the central position and the radius of each bubble, using the Spitzer 8 \mic images\footnote[2]{Taken from http://irsa.ipac.caltech.edu/data/SPITZER/GLIMPSE/.}. For initial conditions of the fitting procedure, we considered the central positions ($l{_{\rm Ch}}$, $b{_{\rm Ch}}$) and radii ($R{_{\rm Ch}}$) provided by \citet{Churchwell2006} and (2007) based on their visual inspection. We estimated the background brightness level from the annular region at (2$-$4)$\times R{_{\rm Ch}}$ and subtracted it from the image of each bubble. Among the pixels contained in the area of $<$ 2$R{_{\rm Ch}}$, we selected the higher-brightness pixels which occupy 20$\%$ of the total. We carried out circular fitting to those pixels, where free parameters are the central position and the radius of a bubble. We used a least-square method to determine the best-fit parameters, calculating the following value:
\begin{equation}
\label{eq:crv}
\Delta^2 = \frac{\sum_{i}^{n} (\sqrt{(x_i-l)^2+(y_i-b)^2}-R)^{2}}{n},
\end{equation}
where ($x_i$, $y_i$) is the coordinates of the $i$-th pixel, ($l, b$) and $R$ are the central position and the radius of a fitting circle, respectively, and $n$ is the number of the pixels. \citet{Churchwell2006} investigated the relation between the shell thicknesses and the outer bubble radii, $R_{\rm out}$, for all the cataloged bubbles and found that almost every bubble has the thickness of $<$ 0.4$R_{\rm out}$. Therefore in the case that $|{\Delta}|$ is larger than 0.2$R$, we considered that the bubble is likely to contain unrelated objects and removed pixels with larger deviations from the fitted circle. We iterated the fitting procedure until $|{\Delta}|$ becomes $<$ 0.2$R$. The result is summarized in columns 7 to 9 of table \ref{tab:Mettab1}. We compared the best-fit $R$, $l$ and $b$ with  $R{_{\rm Ch}}$,  $l{_{\rm Ch}}$ and $b{_{\rm Ch}}$ in figure \ref{fig:Metfig1}. The figure shows that the standard deviations of the differences are $\delta(|R - R{_{\rm Ch}}|) = 0.11R$, $\delta(|l - l{_{\rm Ch}}|) = 0.17R$ and $\delta(|b - b{_{\rm Ch}}|) = 0.17R$. Thus we below consider potential systematic uncertainties of 11$\%$ and 17$\%$ of the radii and the central positions, respectively.

Now that the radii and the central positions are determined, we classify our sample into two types of morphologies (i.e., closed or broken). We estimated the covering fraction which represents the covering rate of the Spitzer 8 \mic shell for the whole direction viewed from the bubble center. We again considered the annular region centered at ($l, b$) with $\pm$ 0.2$R$ in radial thickness and divided it into 40 sectors equally (i.e., each subtended by 9\degreee). For each sector, we counted the number of pixels which have brightness higher than 20$\%$ of the brightness averaged over the annular region. If the filling factor, the ratio of those pixels to the total in a sector, was $<$ 20$\%$, we regarded the sector as broken. Hence we classified the bubbles where at least one sector was missing into broken bubbles and the others into closed bubbles. To secure reliable classification, we used two thresholds for the brightness level, 20$\%$ and 30$\%$, to identify a broken sector. Closed bubbles were identified on the basis of the higher threshold (30$\%$) for a broken sector, while broken bubbles were identified on the basis of the lower threshold (20$\%$) for a broken sector. As a result, the 200 sample bubbles are categorized into 81 closed, 58 broken and 61 unclassified bubbles, as summarized in column 10 of table \ref{tab:Mettab1}. In comparison with \citet{Churchwell2006} and (2007), their 8 closed bubbles are categorized into broken bubbles and their 1 broken bubble is into a closed bubble by our classification.

Next, we estimated the flux densities of each bubble in the AKARI 6 bands through aperture photometry of a circular region of $< 2R$ centered at ($l, b$). The result is summarized in table \ref{tab:Mettab2}. We created a global SED from those flux densities. In order to create a local SED, we smoothed the 6-band images with a common Gaussian kernel of \timeform{90''}, which approximately corresponds to the FWHMs of the beam sizes of the 140 and 160 \mic all-sky images, and set a common pixel scale of \timeform{15''} among the images. We subtracted the background levels estimated from the annular regions at (2$-$4)$\times R$ for each bubble. Flux uncertainties contain both random and systematic errors; the random errors were calculated from the background fluctuation, while the systematic errors were 10$\%$ (originating in the data processing such as corrections for the non-linearity of detector photoresponse; Ishihara et al. in preparation) for the 9 and 18 \mic bands and 15$\%$ for the 65, 90, 140 and 160 \mic bands. For the far-IR bands, we adopted 15$\%$ rather than 10\% given by \citet{Takita2015}, considering residual striping patterns along the scan direction which were recognized for about 40 bubbles in our sample.

The local SEDs created for every \timeform{15''}$\times$\timeform{15''} pixel of the AKARI 6-band images were decomposed into PAH, warm and cold dust components by model fitting as was performed in \citet{Suzuki2013} and \citet{Kaneda2013}. In the fitting, we adopted the PAH model spectrum in \citet{Draine2007} assuming the PAH size distribution and ionizing fraction typical of the diffuse ISM, and modified blackbody spectra with the emissivity power-law index $\beta$ of 2 for warm and cold dust components. Free parameters were the amplitudes of the dust components and the temperature of warm dust. We fixed the temperature of cold dust at 18 K, because the temperature could not be confined well for the local SEDs. The temperature of 18 K was estimated from the fitting to the SED averaged over all the bubbles. It should be noted that the three components are just representative ones to reproduce the observed SEDs as good as possible and the selection of the model components does not essentially affect the following results. To determine the initial parameters of the local SED fitting, we first carried out fitting to the global SED described above. Since 20 bubbles were not acceptable with a 90$\%$ confidence level, we excluded these bubbles from our sample. Thus our sample finally contains 74 closed, 49 broken and 57 unclassified bubbles. Among them, relatively large 20 closed and 20 broken bubbles are shown in figure \ref{fig:Introfig1} in increasing order of the Galactic longitude.

We obtained the brightness distributions of the PAH, warm and cold dust components, $I_{\rm{PAH}}$, $I_{\rm{warm}}$ and $I_{\rm{cold}}$, respectively, integrating each component over the wavelength range of 5$-$1000 \micron. Figure \ref{fig:Metfig2} shows examples of the $I_{\rm{PAH}}$, $I_{\rm{warm}}$ and $I_{\rm{cold}}$ maps thus obtained, which are the same bubbles as shown in figure \ref{fig:Introfig1}. We estimated the fluxes of each dust component, using the same aperture as used in creating the global SEDs (i.e., $< 2R$). Using the fluxes and the distance, we obtained the luminosity of each component ($L_{\rm{PAH}}$, $L_{\rm{warm}}$ and $L_{\rm{cold}}$) and added them to obtain the total IR luminosity, $L_{\rm TIR}$ ($=L_{\rm{PAH}}+L_{\rm{warm}}+L_{\rm{cold}}$). The luminosities are summarized in table \ref{tab:Luminosity}.

To evaluate the effects of a rather small number of the AKARI SED data points in obtaining the luminosities, we overplotted the Spitzer 5.8, 8.0, Herschel 70, 160, 250, 350 and 500 \mic photometric data points\footnote[3]{Taken from http://www.cosmos.esa.int/web/herschel/science-archive/.} in the global SEDs of two typical examples in figure \ref{fig:SEDwO}. We confirm that our estimates are consistent with the photometric data points except those in the submillimeter region. The result of the SED fitting to all the data points, with warm and cold dust temperatures allowed to vary, shows that $L_{\rm{PAH}}$, $L_{\rm{warm}}$ and $L_{\rm{cold}}$ change by about $-$3\%, $-$30\% and $+$20\%, respectively. Hence the effects on $L_{\rm{PAH}}$ are likely to be negligible, whereas the effects on $L_{\rm{warm}}$ and $L_{\rm{cold}}$ are not. In table \ref{tab:Luminosity}, we included 30\% and 20\% for the uncertainties of $L_{\rm{warm}}$ and $L_{\rm{cold}}$ in addition to the fitting and distance errors.

\begin{figure*}[ht]
\centering
\subfigure{
\makebox[180mm][l]{\raisebox{0mm}[0mm][0mm]{ \hspace{16mm} \small{SED}} \hspace{27.5mm} \small{$I_{\rm{PAH}}$} \hspace{29.5mm} \small{$I_{\rm{warm}}$} \hspace{29.5mm} \small{$I_{\rm{cold}}$}}%
}
\subfigure{
\makebox[180mm][l]{\raisebox{0mm}{\hspace{49.5mm} \small{RA (J2000)} \hspace{19.5mm} \small{RA (J2000)} \hspace{19.5mm} \small{RA (J2000)}}}
}
\subfigure{
\mbox{\raisebox{0mm}{\includegraphics[width=40mm]{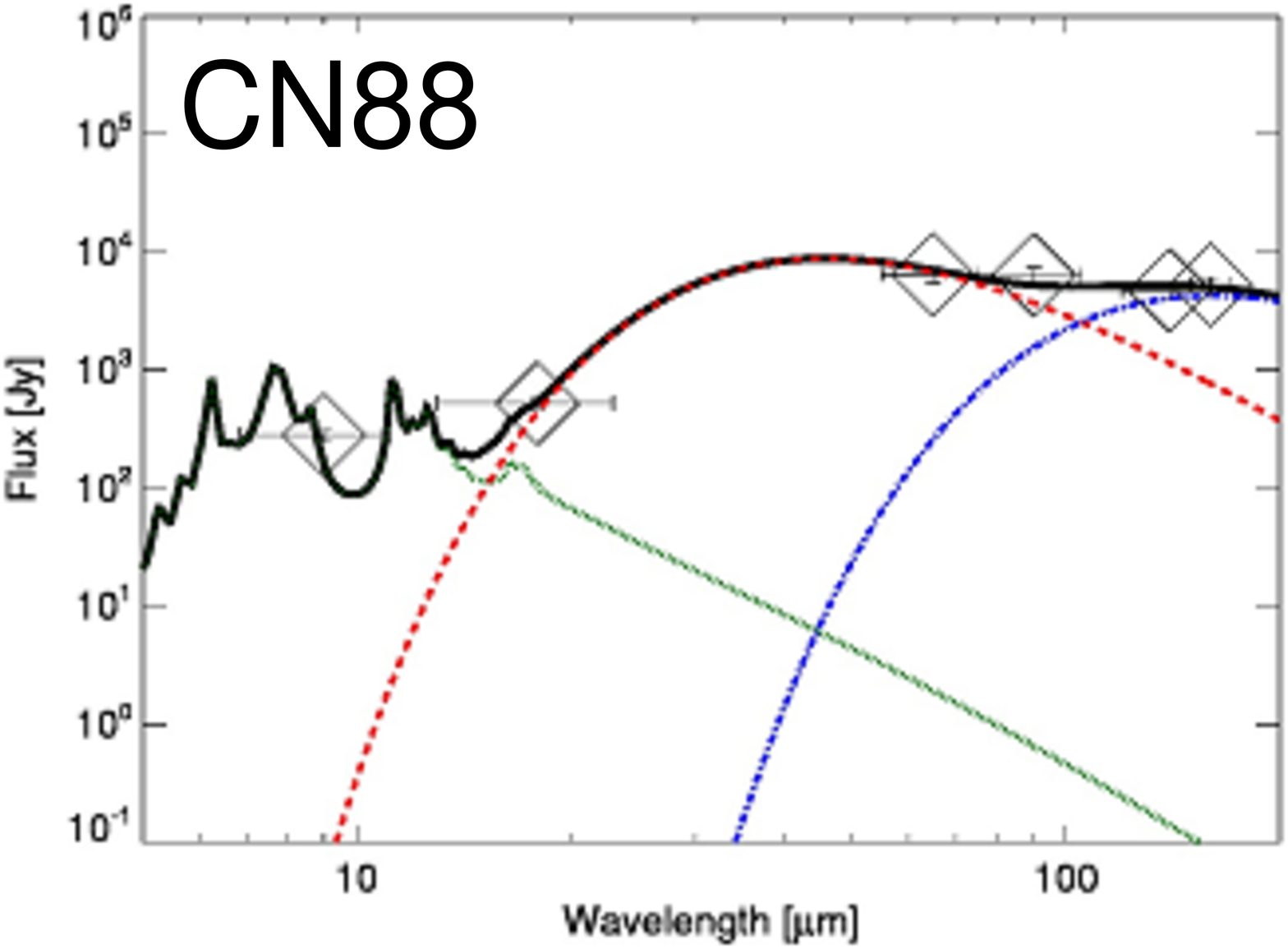}}}%
\mbox{\raisebox{6mm}{\rotatebox{90}{\small{DEC (J2000)}}}}%
\mbox{\raisebox{0mm}{\includegraphics[width=40mm]{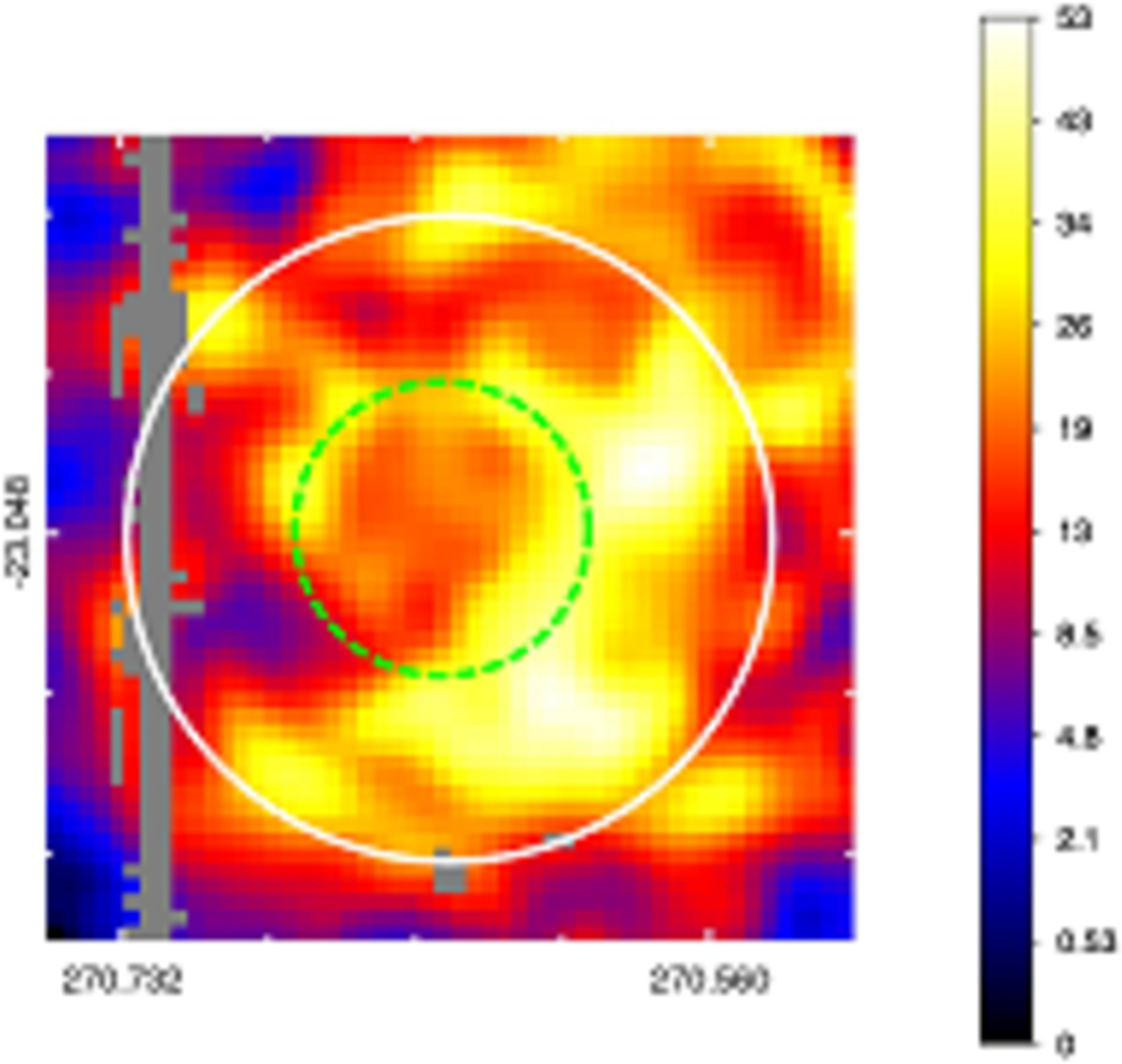}}}%
\mbox{\raisebox{0mm}{\includegraphics[width=40mm]{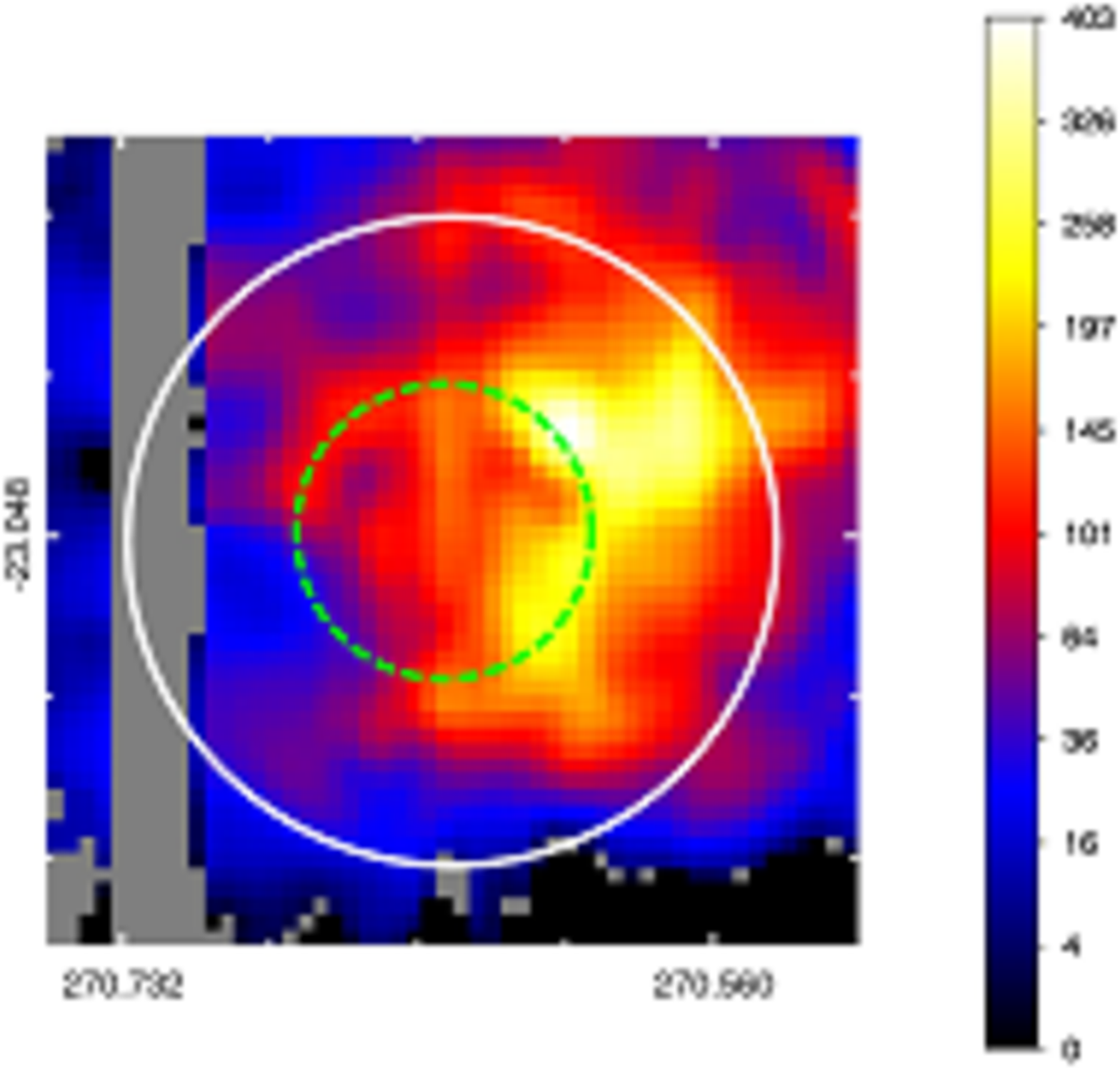}}}%
\mbox{\raisebox{0mm}{\includegraphics[width=40mm]{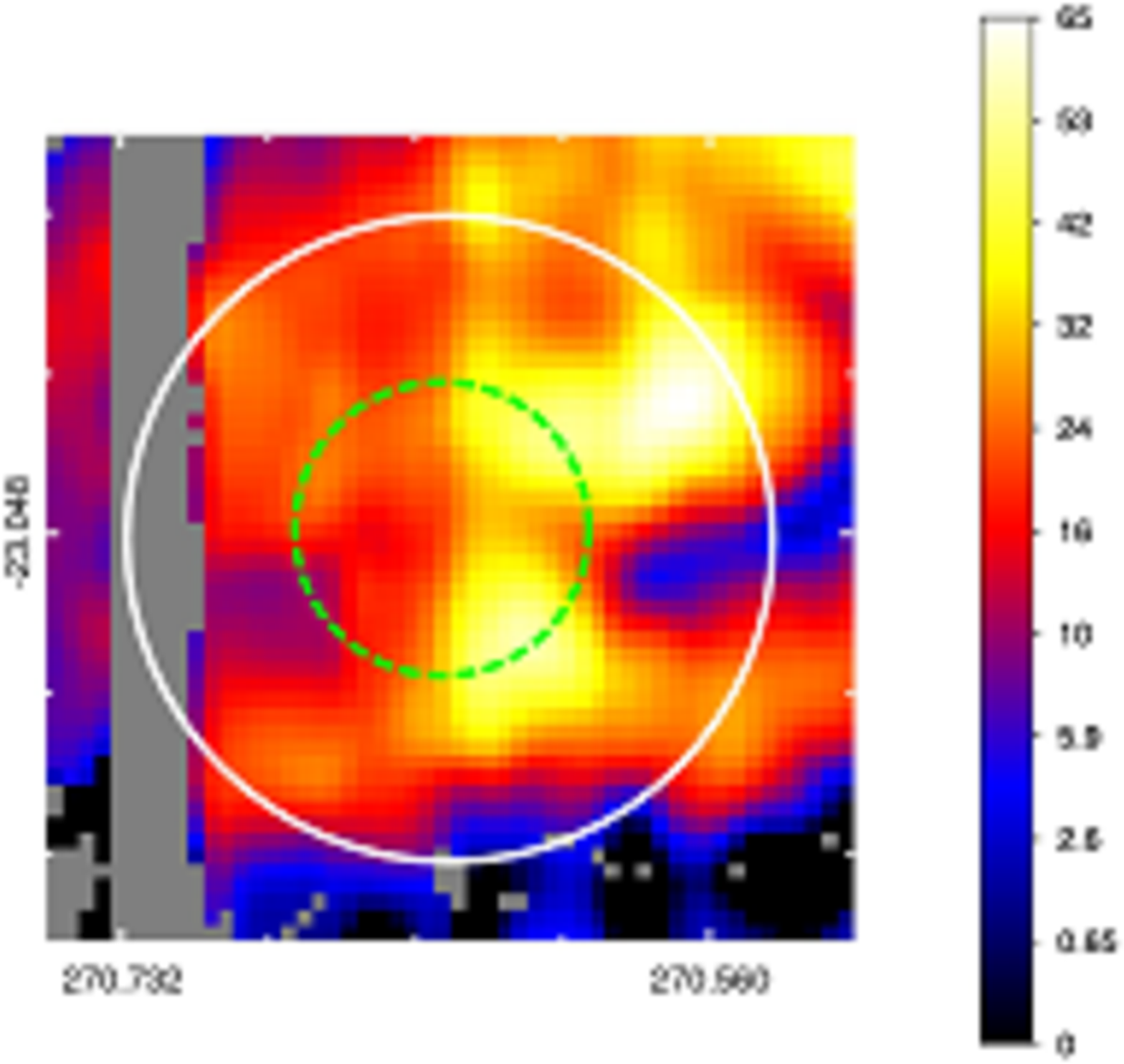}}}%
}
\subfigure{
\mbox{\raisebox{0mm}{\includegraphics[width=40mm]{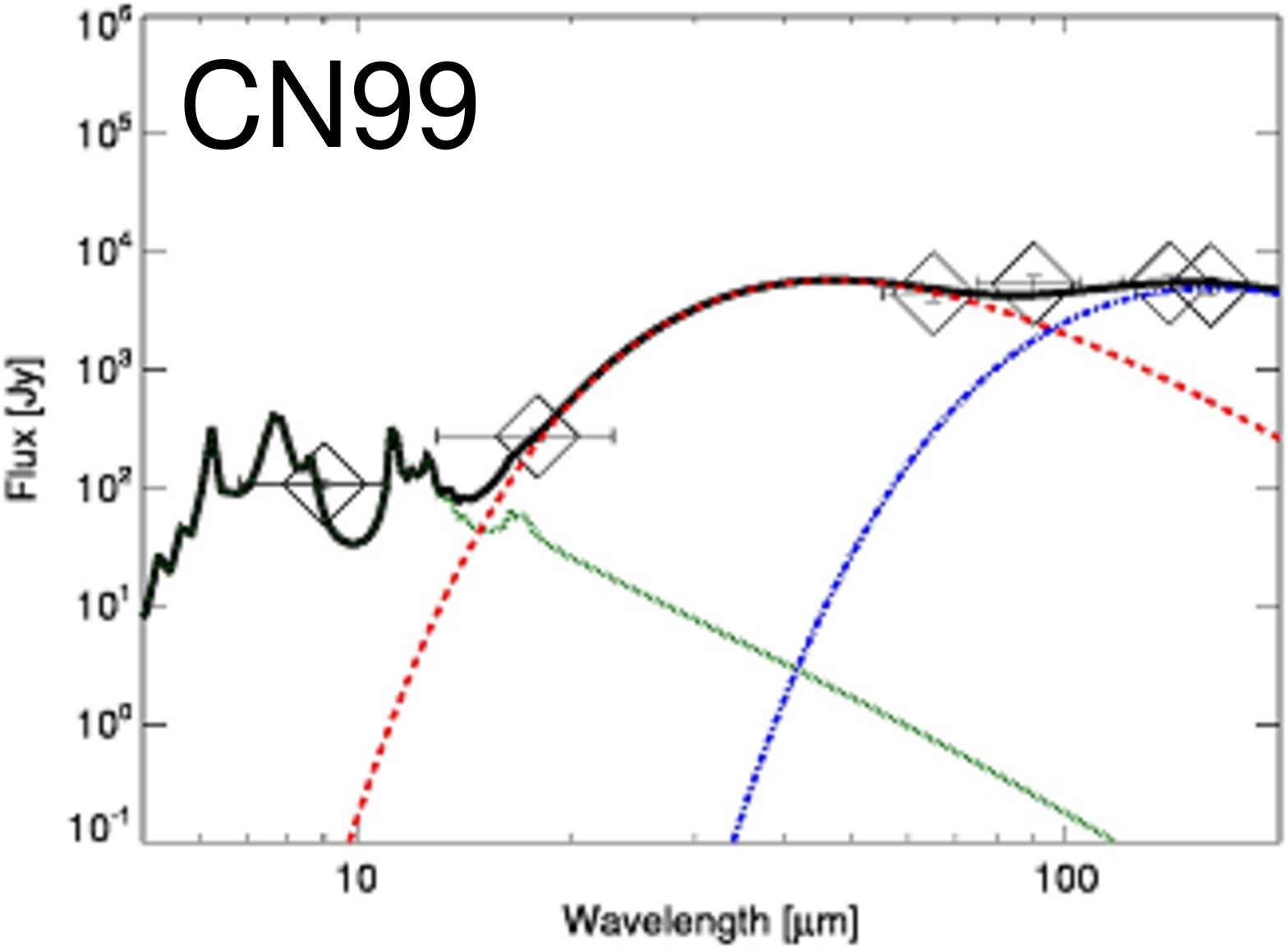}}}%
\mbox{\raisebox{6mm}{\rotatebox{90}{\small{DEC (J2000)}}}}%
\mbox{\raisebox{0mm}{\includegraphics[width=40mm]{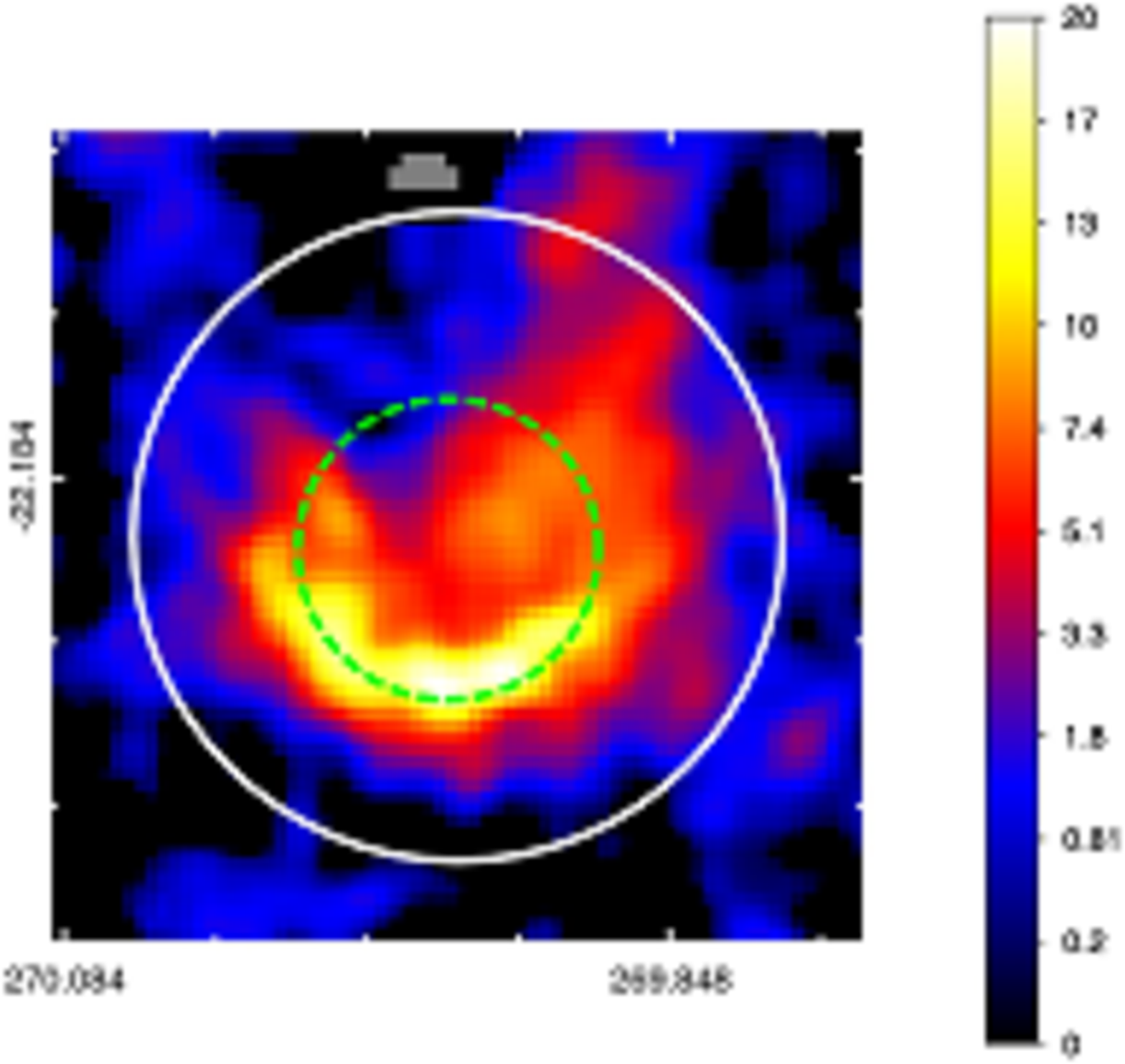}}}%
\mbox{\raisebox{0mm}{\includegraphics[width=40mm]{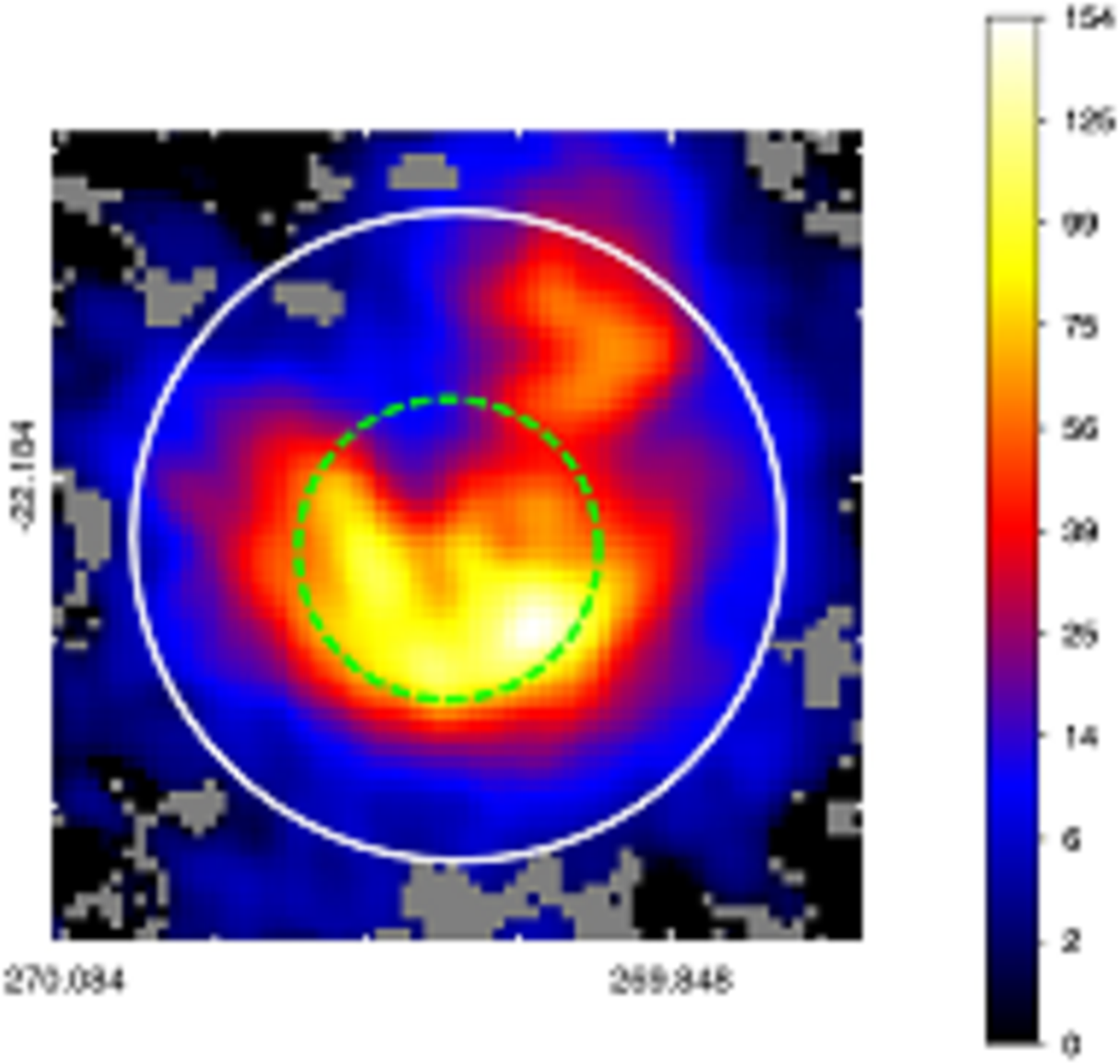}}}%
\mbox{\raisebox{0mm}{\includegraphics[width=40mm]{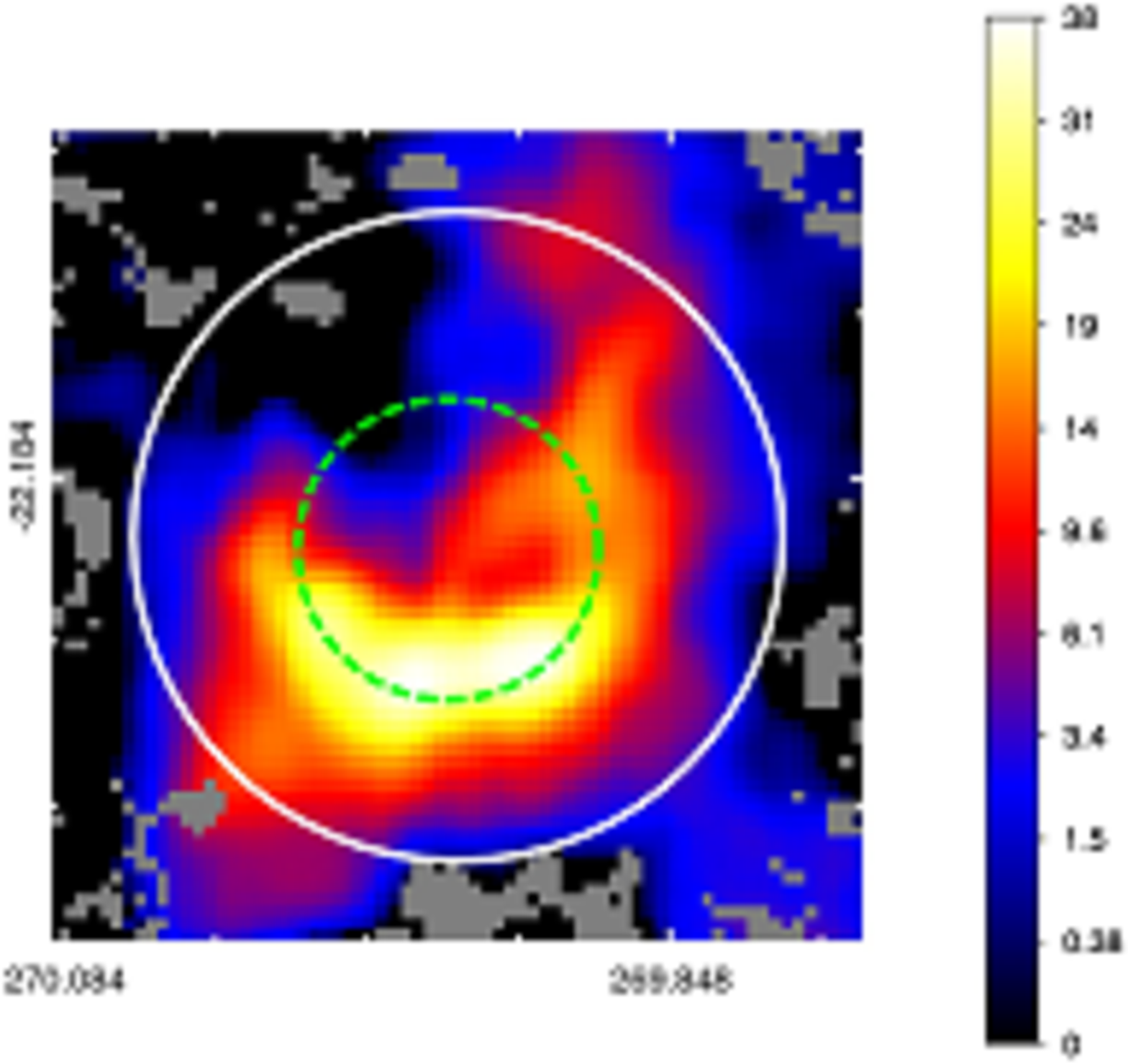}}}%
}
\subfigure{
\mbox{\raisebox{0mm}{\includegraphics[width=40mm]{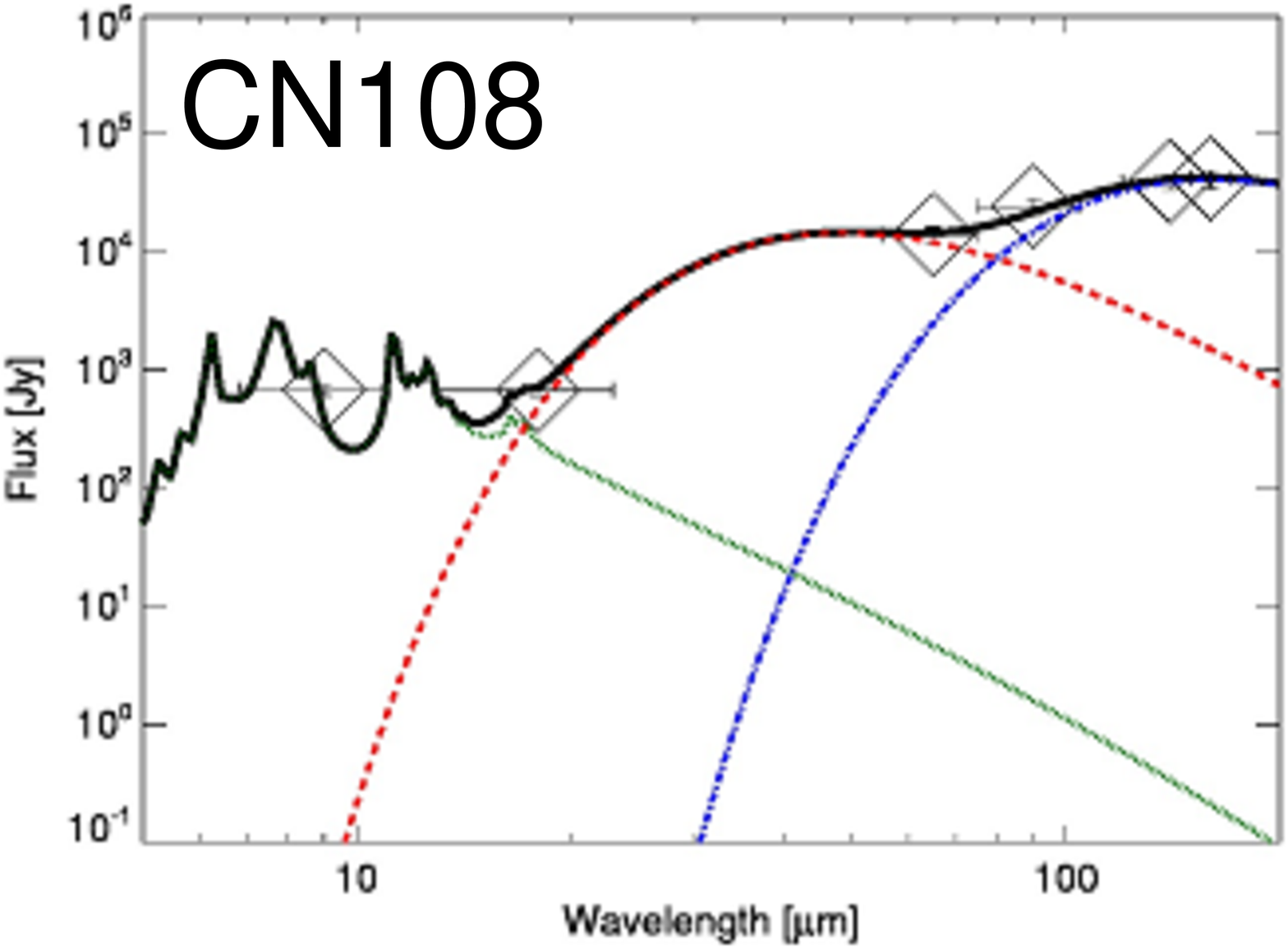}}}%
\mbox{\raisebox{6mm}{\rotatebox{90}{\small{DEC (J2000)}}}}%
\mbox{\raisebox{0mm}{\includegraphics[width=40mm]{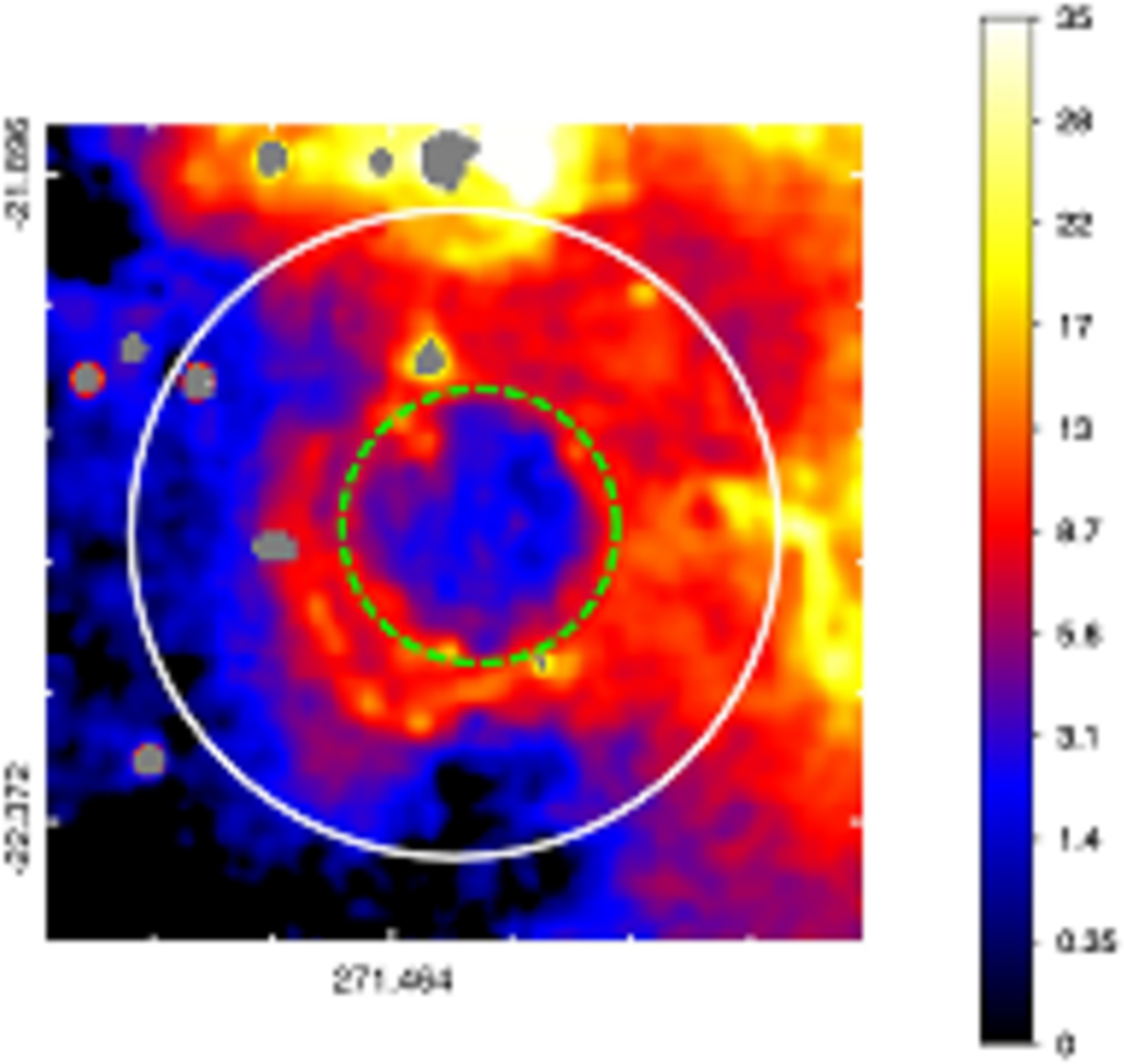}}}%
\mbox{\raisebox{0mm}{\includegraphics[width=40mm]{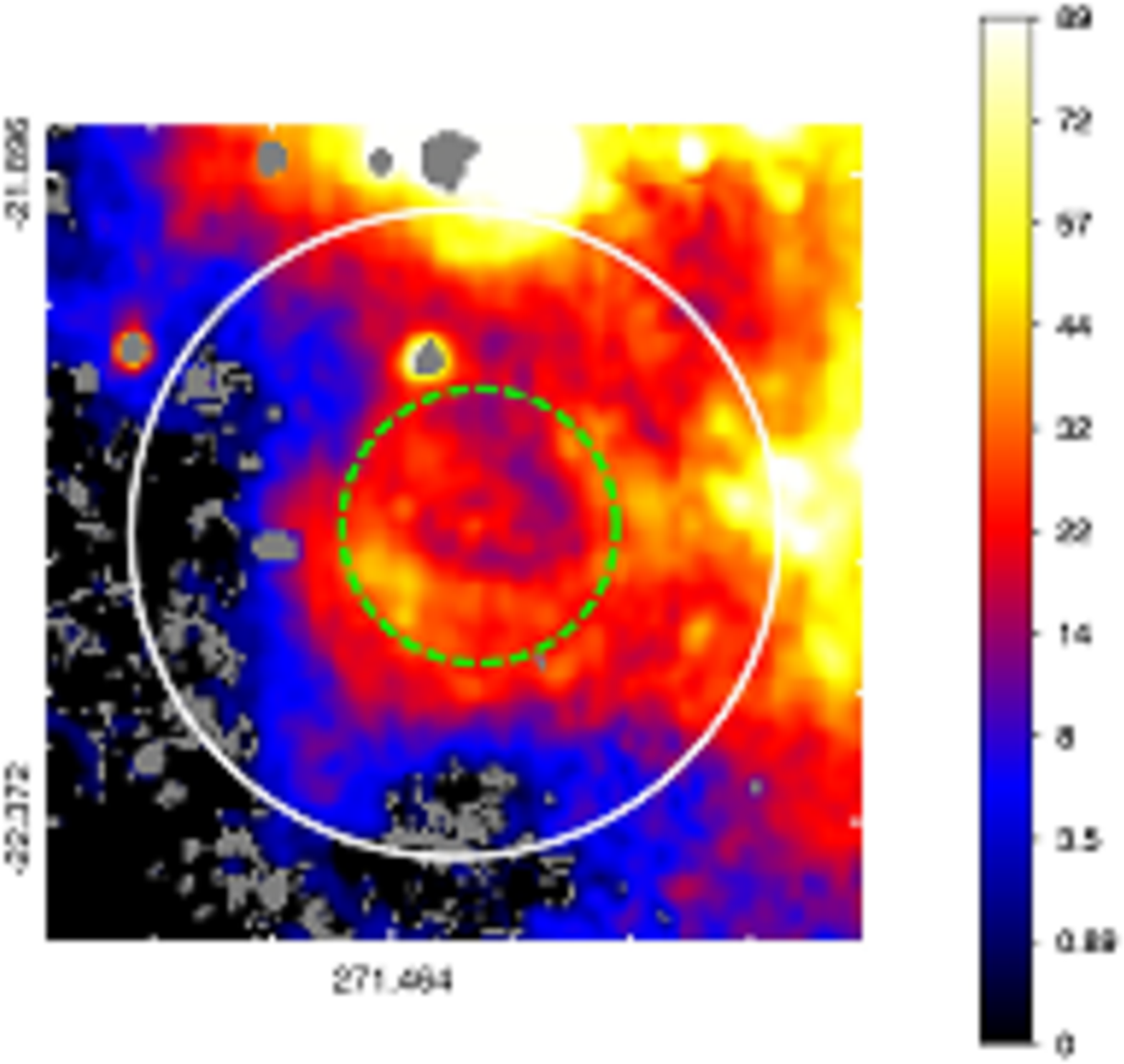}}}%
\mbox{\raisebox{0mm}{\includegraphics[width=40mm]{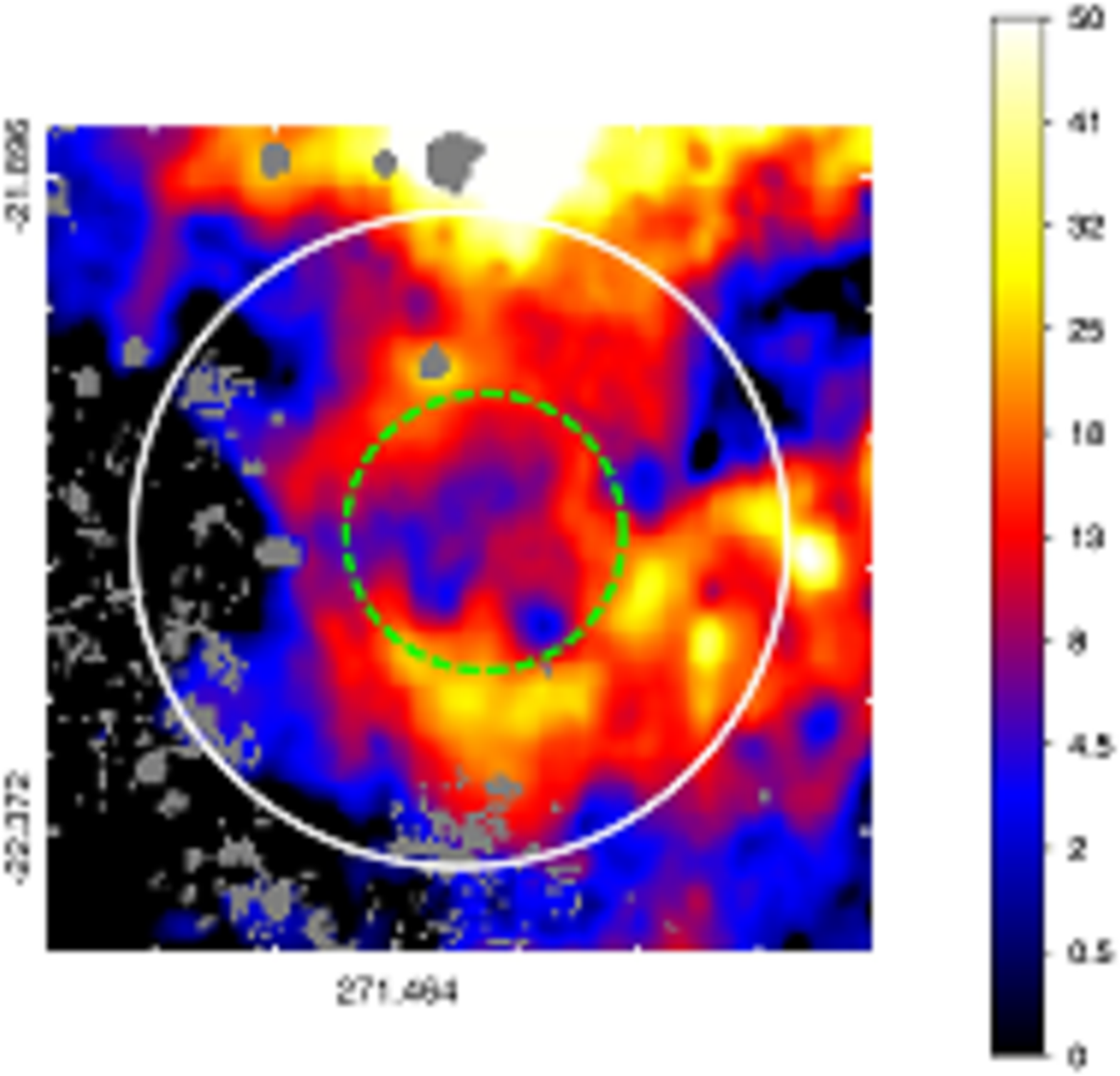}}}%
}
\subfigure{
\mbox{\raisebox{0mm}{\includegraphics[width=40mm]{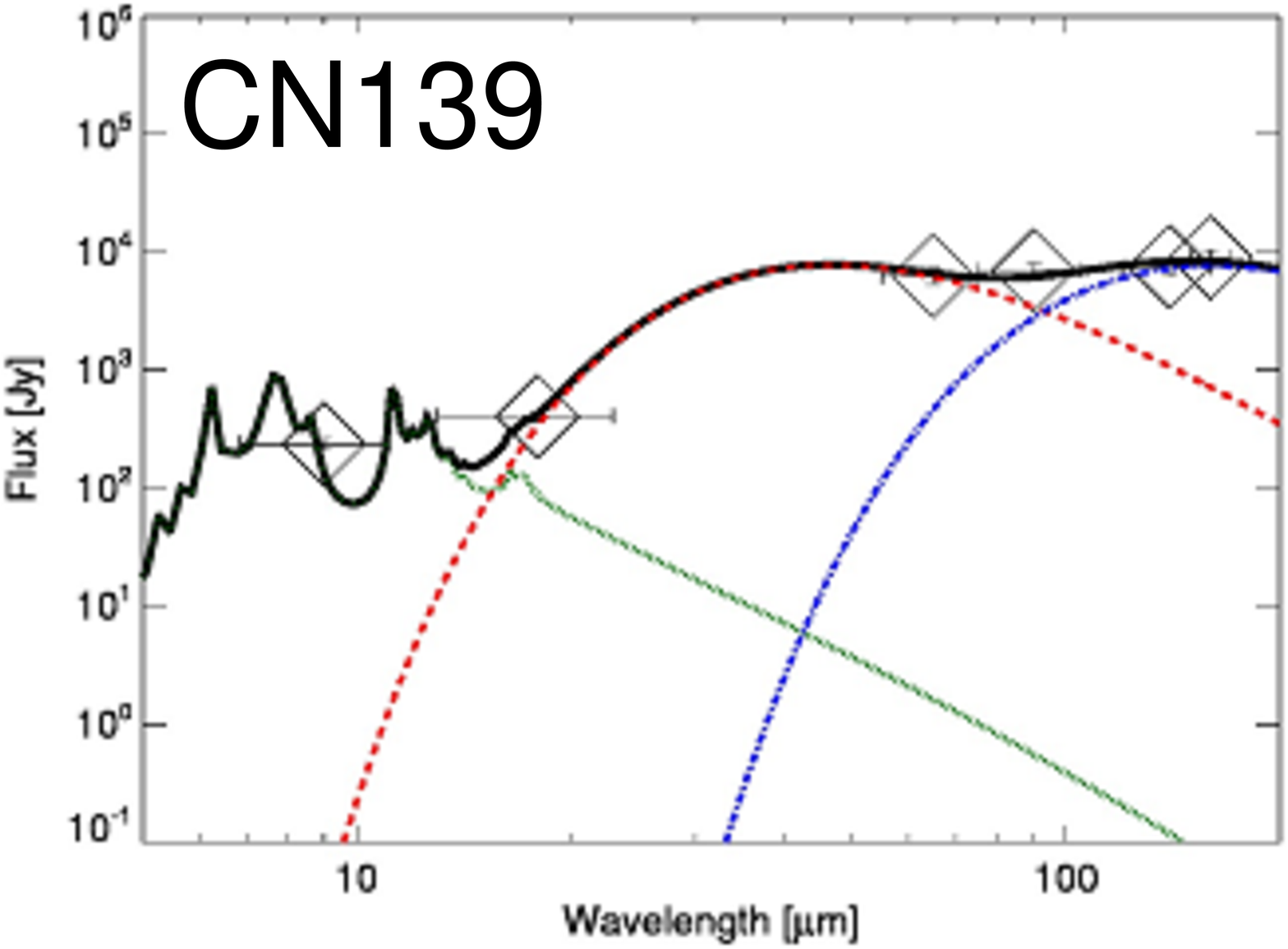}}}%
\mbox{\raisebox{6mm}{\rotatebox{90}{\small{DEC (J2000)}}}}%
\mbox{\raisebox{0mm}{\includegraphics[width=40mm]{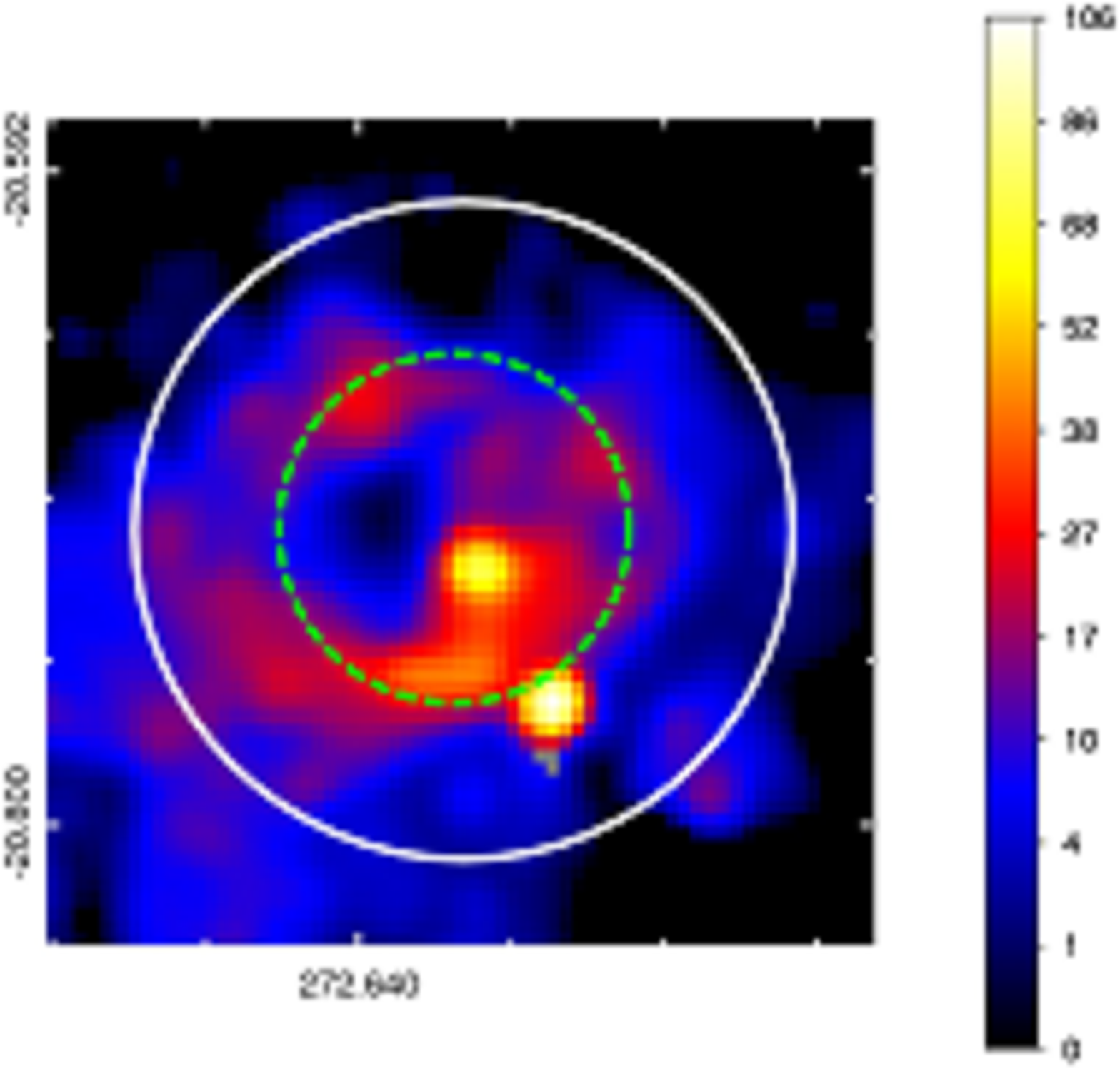}}}%
\mbox{\raisebox{0mm}{\includegraphics[width=40mm]{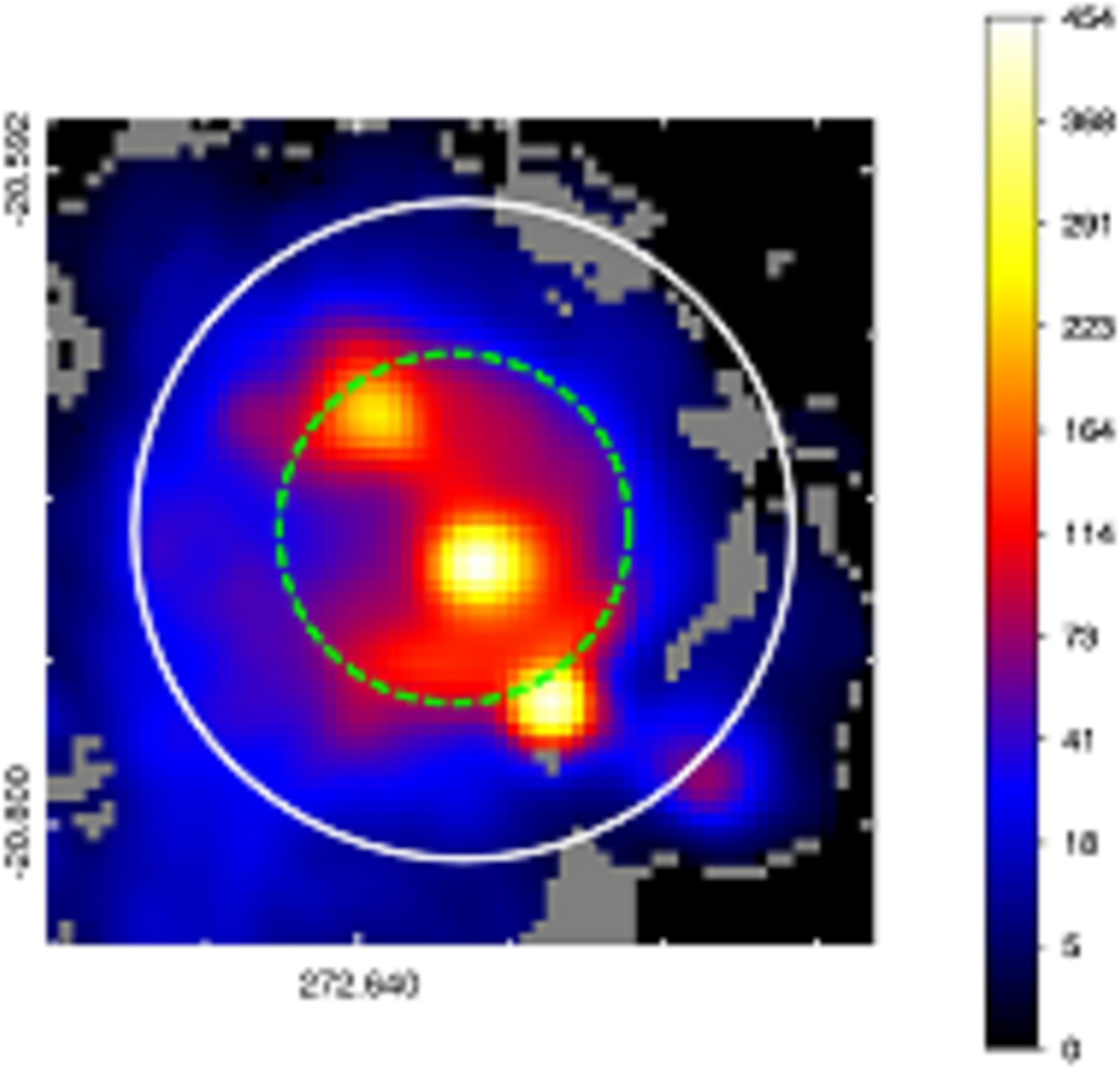}}}%
\mbox{\raisebox{0mm}{\includegraphics[width=40mm]{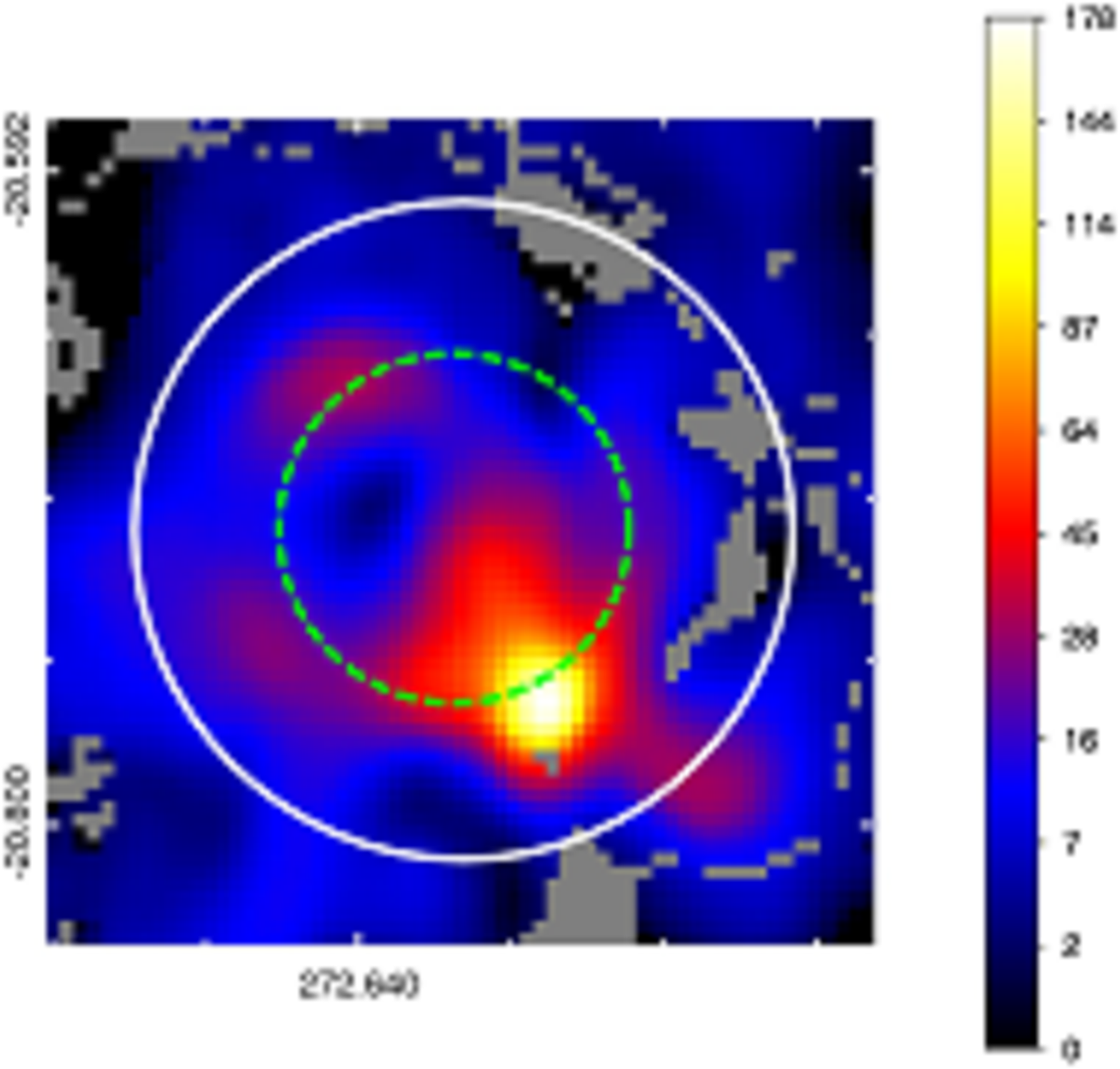}}}%
}
\subfigure{
\mbox{\raisebox{0mm}{\includegraphics[width=40mm]{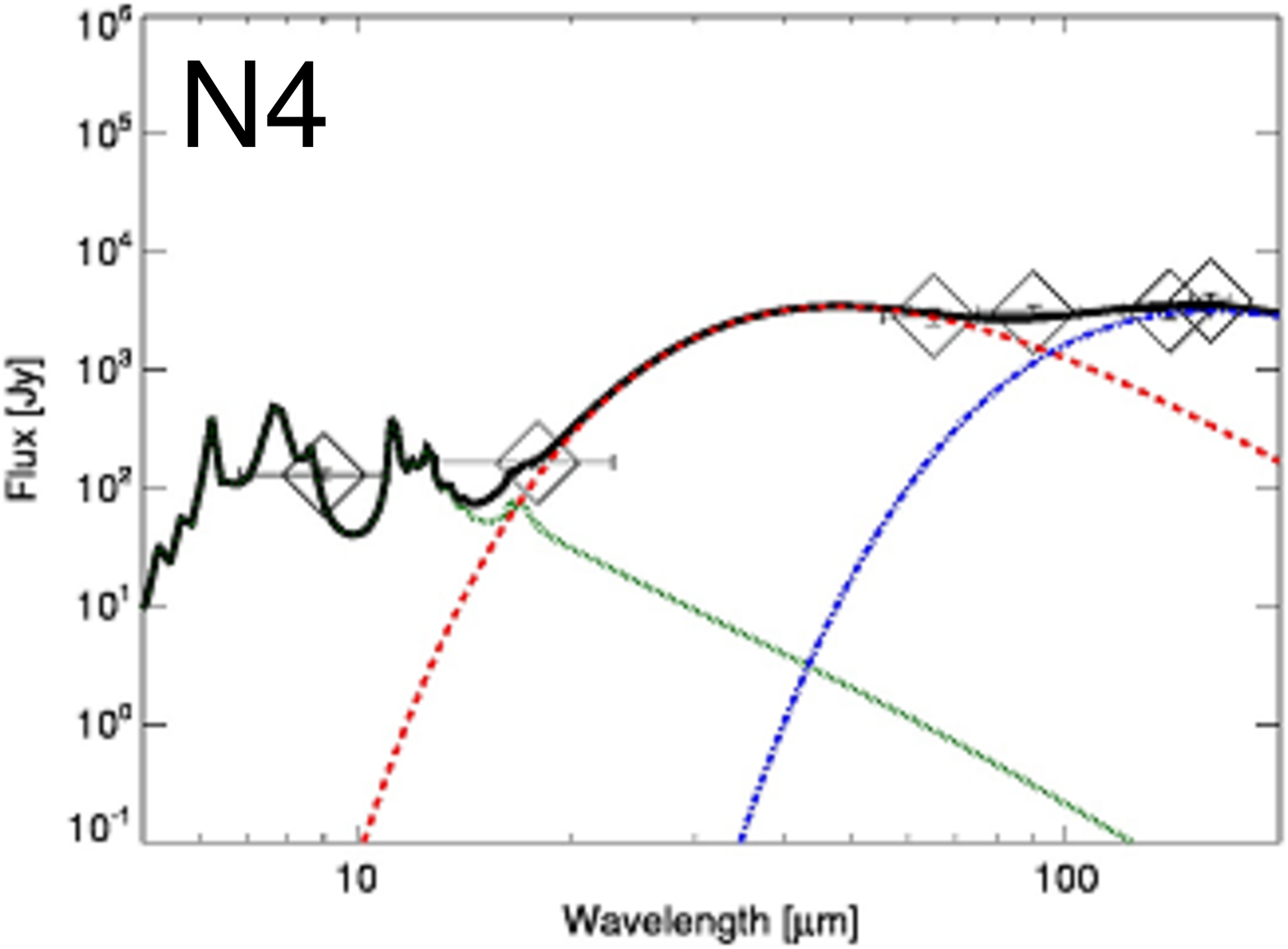}}}%
\mbox{\raisebox{6mm}{\rotatebox{90}{\small{DEC (J2000)}}}}%
\mbox{\raisebox{0mm}{\includegraphics[width=40mm]{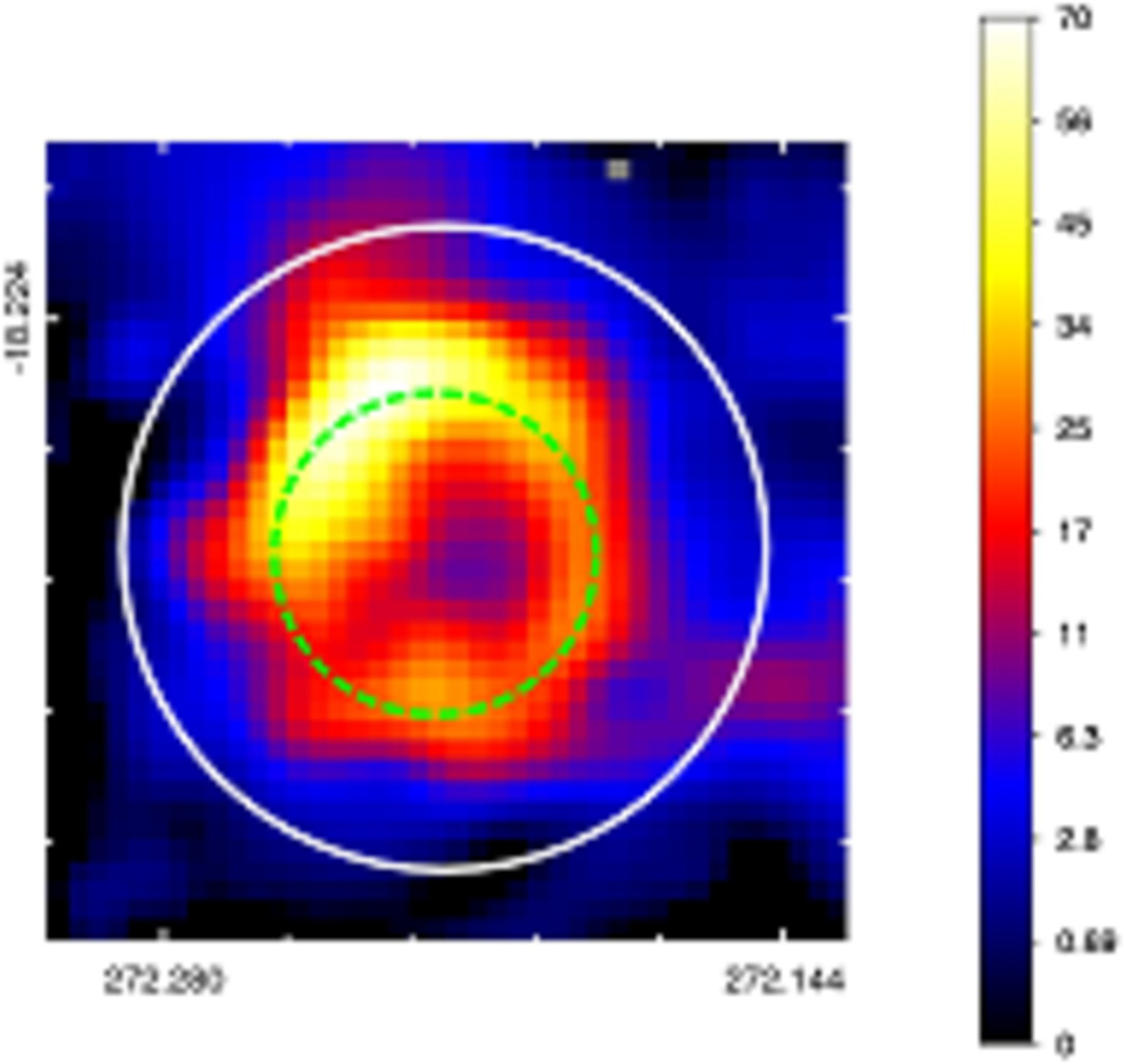}}}%
\mbox{\raisebox{0mm}{\includegraphics[width=40mm]{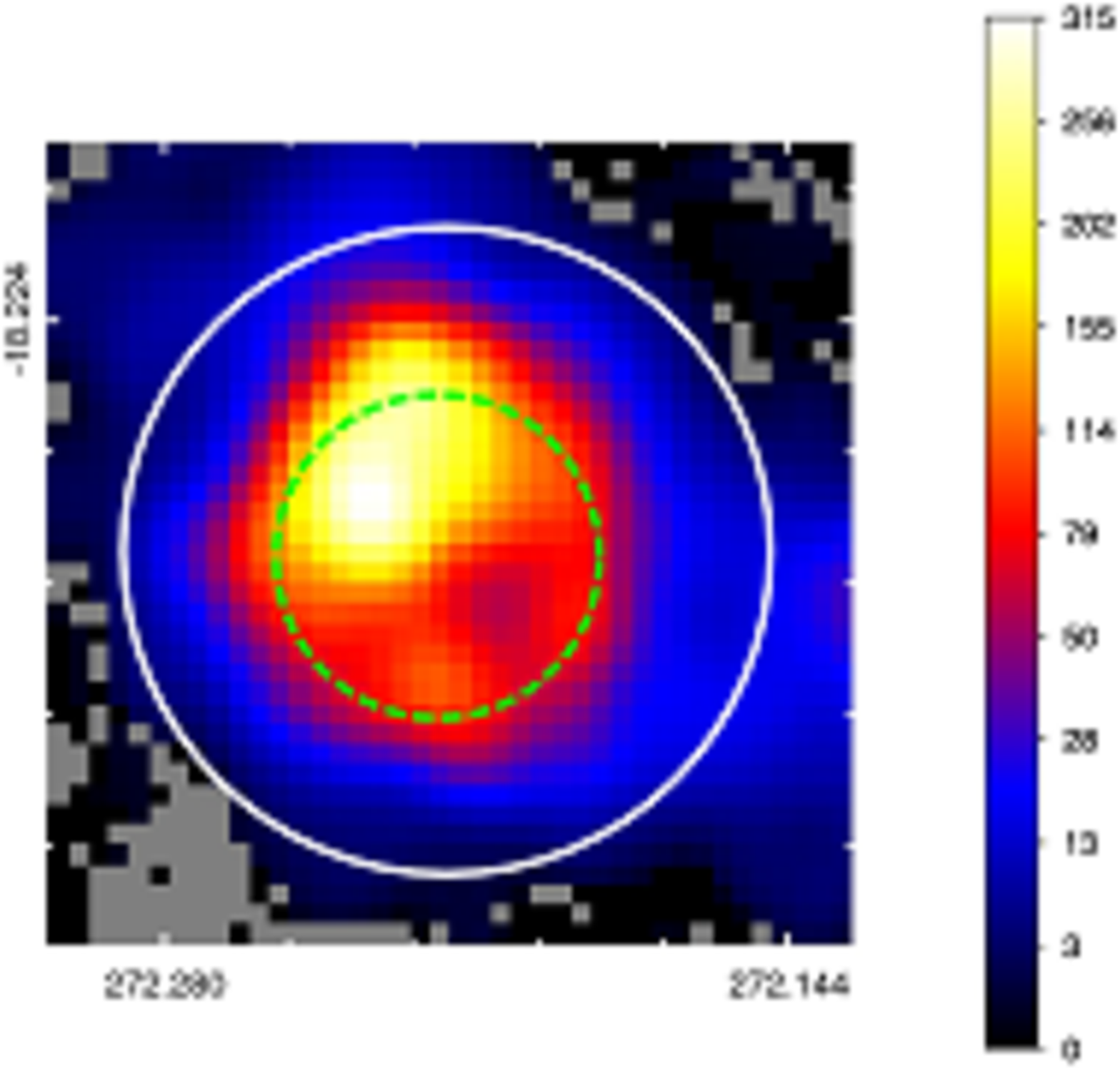}}}%
\mbox{\raisebox{0mm}{\includegraphics[width=40mm]{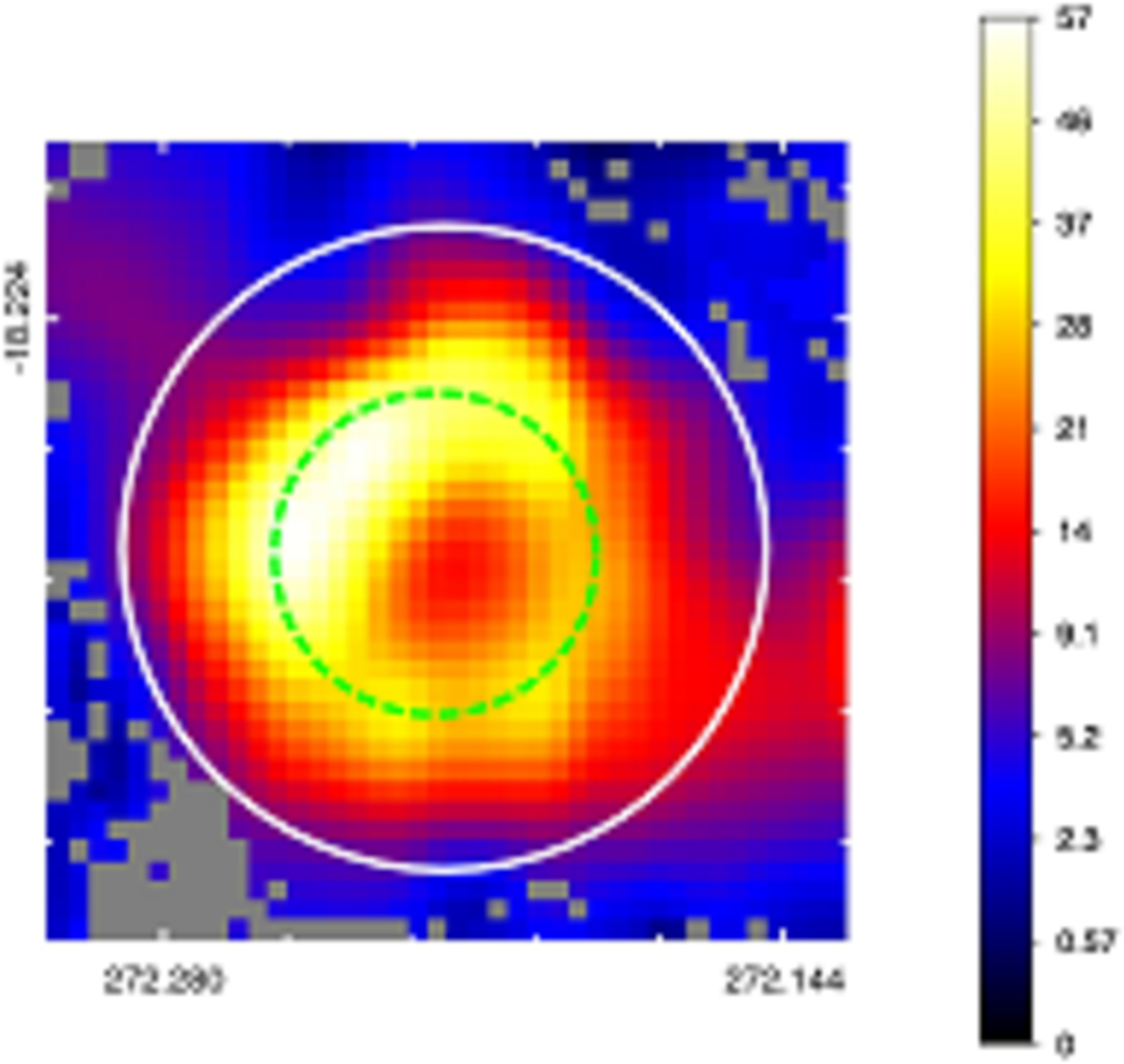}}}%
}
\raggedright
\caption{From left to right, the global SED of the AKARI 6 photometric data points fitted by the dust model, the brightness maps of the PAH, warm and cold dust components for the sample bubbles shown in figures \ref{fig:Introfig1}. In the SED plot, diamonds represent the photometric data points and the black solid curve indicates the best-fit result. The green dotted, red dashed and blue dashed-dotted curves indicate the best-fit PAH, warm and cold dust components, respectively. In the brightness maps, the green dashed circles are the same as those in figure \ref{fig:Introfig1}, while the white solid circles are drawn with ($l, b$) and $2R$ from our fitting result. The color levels in the maps are given in units of $\mu$W $\rm{m^{-2}}$ $\rm{sr^{-1}}$. \label{fig:Metfig2}} %
\end{figure*}

\addtocounter{figure}{-1}
\begin{figure*}[ht]
\addtocounter{subfigure}{1}
\centering
\subfigure{
\makebox[180mm][l]{\raisebox{0mm}[0mm][0mm]{ \hspace{20mm} \small{SED}} \hspace{27.5mm} \small{$I_{\rm{PAH}}$} \hspace{29.5mm} \small{$I_{\rm{warm}}$} \hspace{29.5mm} \small{$I_{\rm{cold}}$}}%
}
\subfigure{
\makebox[180mm][l]{\raisebox{0mm}{\hspace{52mm} \small{RA (J2000)} \hspace{20mm} \small{RA (J2000)} \hspace{20mm} \small{RA (J2000)}}}
}
\subfigure{
\mbox{\raisebox{0mm}{\includegraphics[width=40mm]{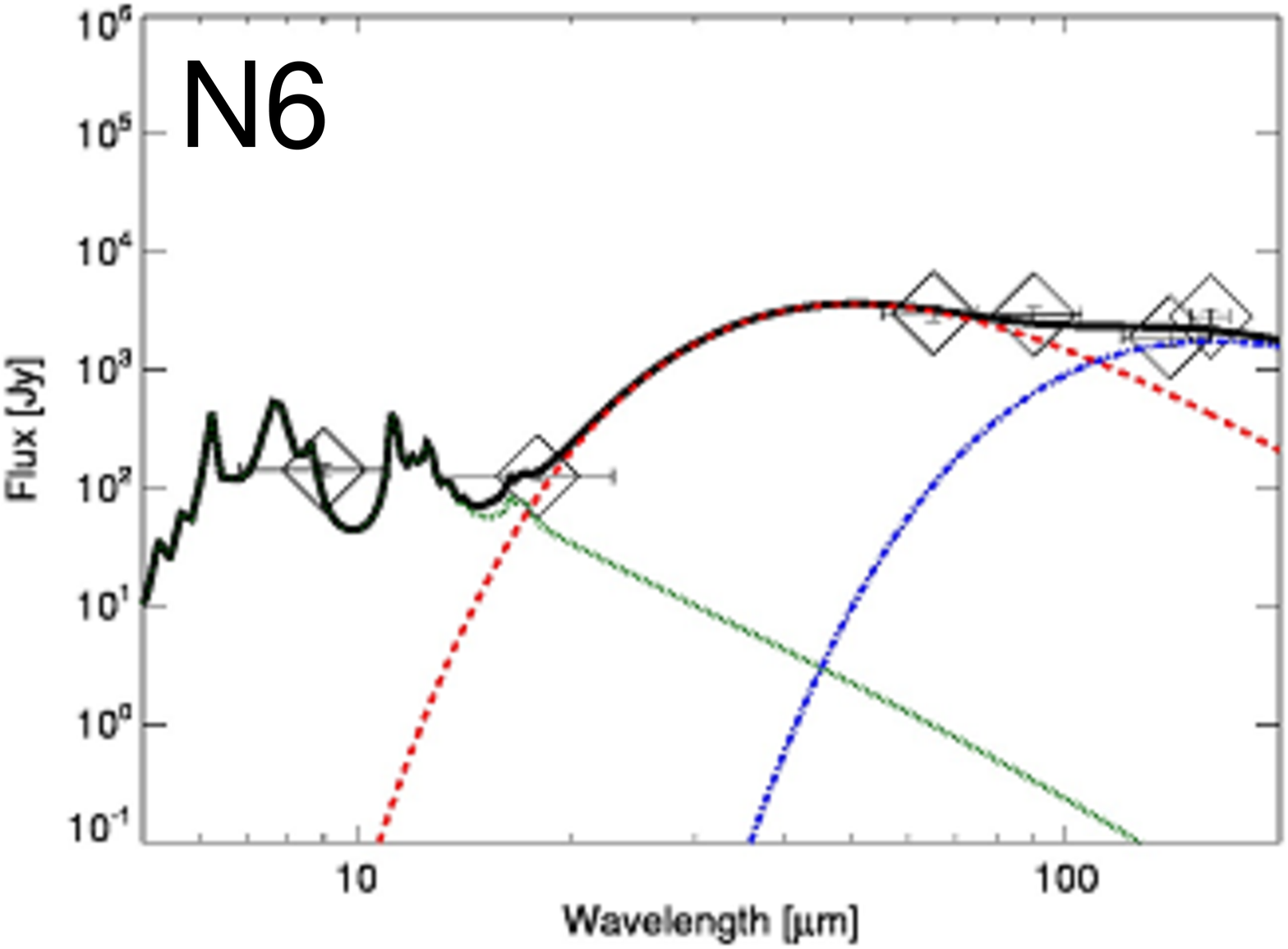}}}%
\mbox{\raisebox{6mm}{\rotatebox{90}{\small{DEC (J2000)}}}}%
\mbox{\raisebox{0mm}{\includegraphics[width=40mm]{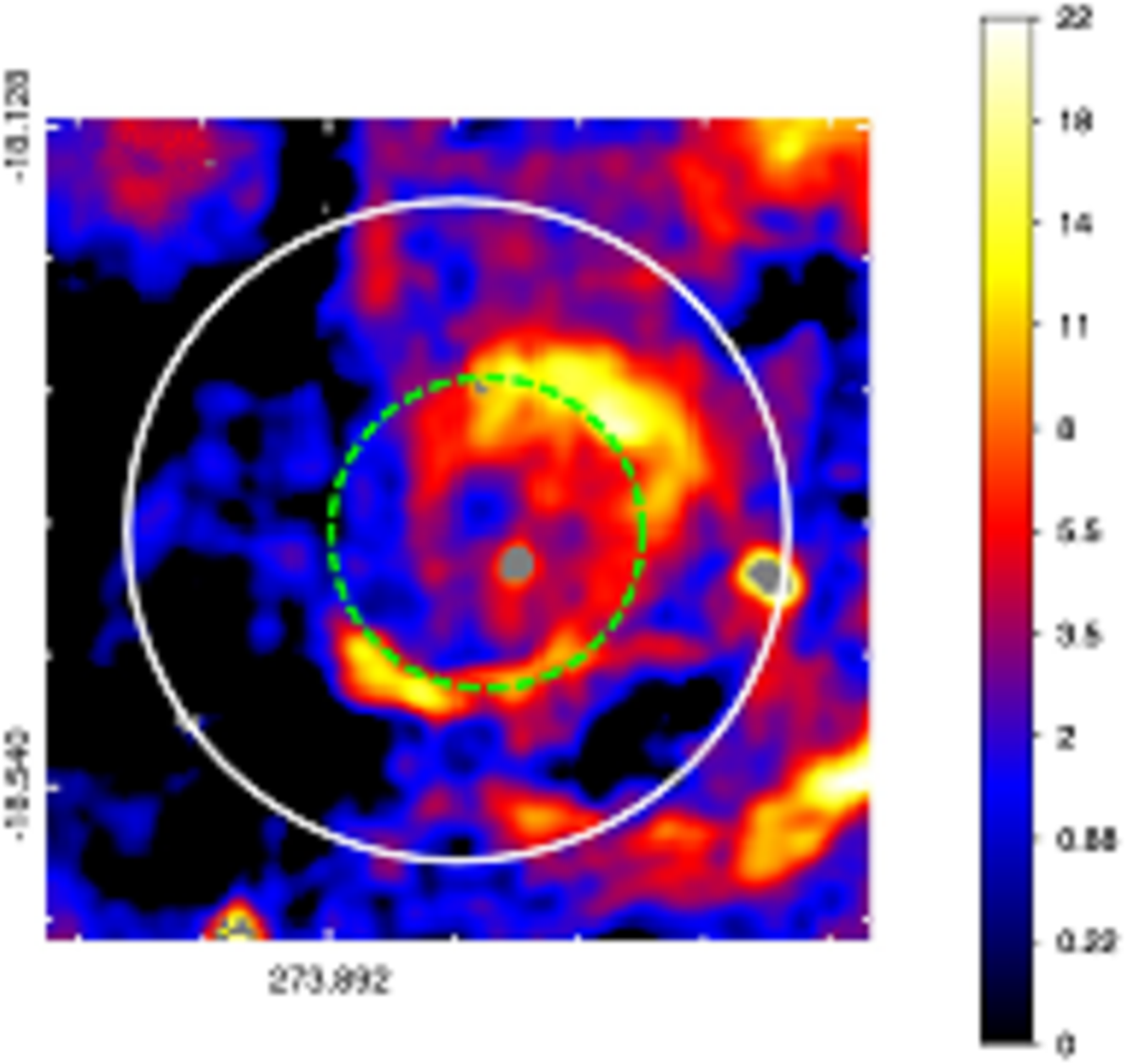}}}%
\mbox{\raisebox{0mm}{\includegraphics[width=40mm]{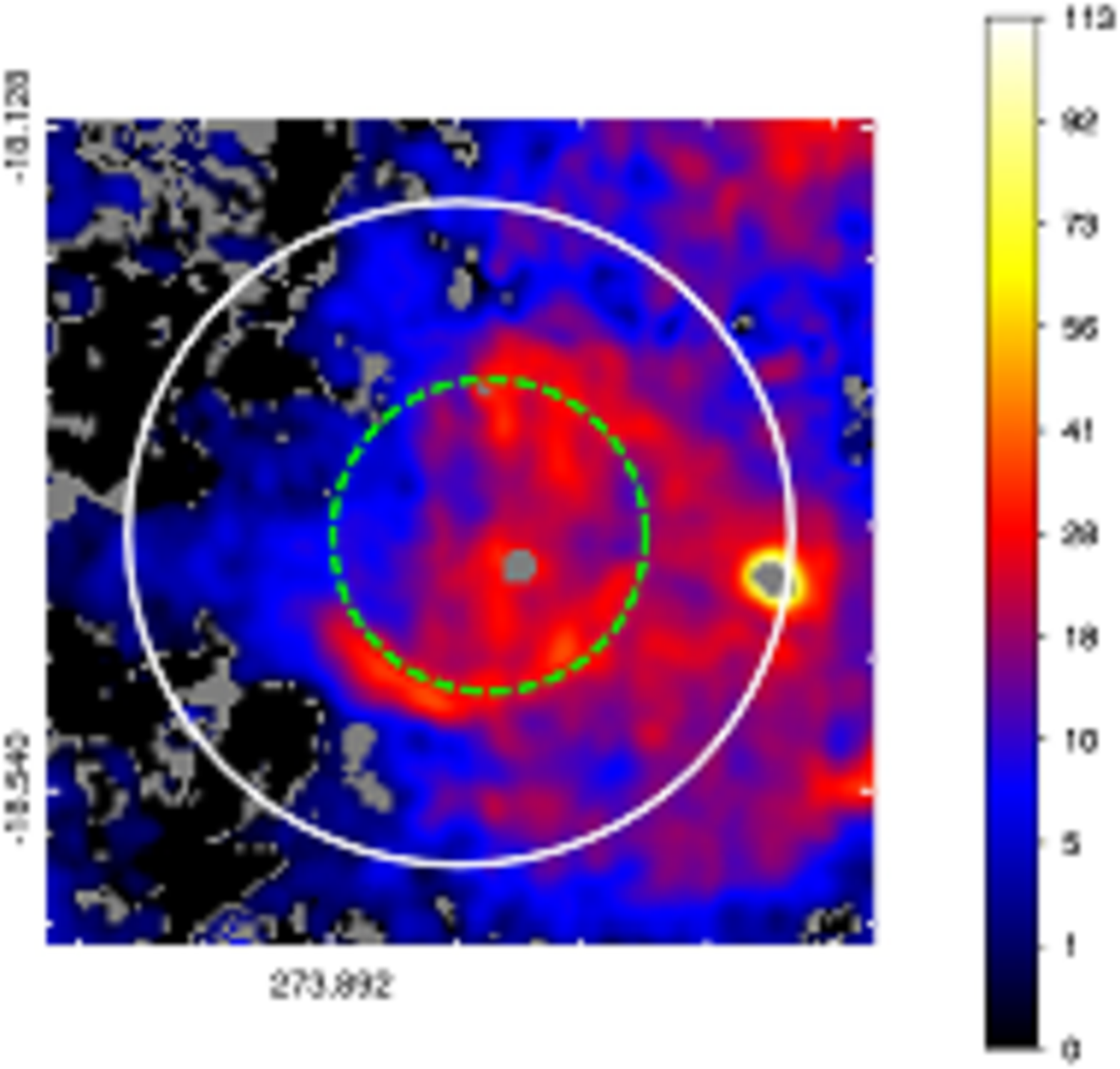}}}%
\mbox{\raisebox{0mm}{\includegraphics[width=40mm]{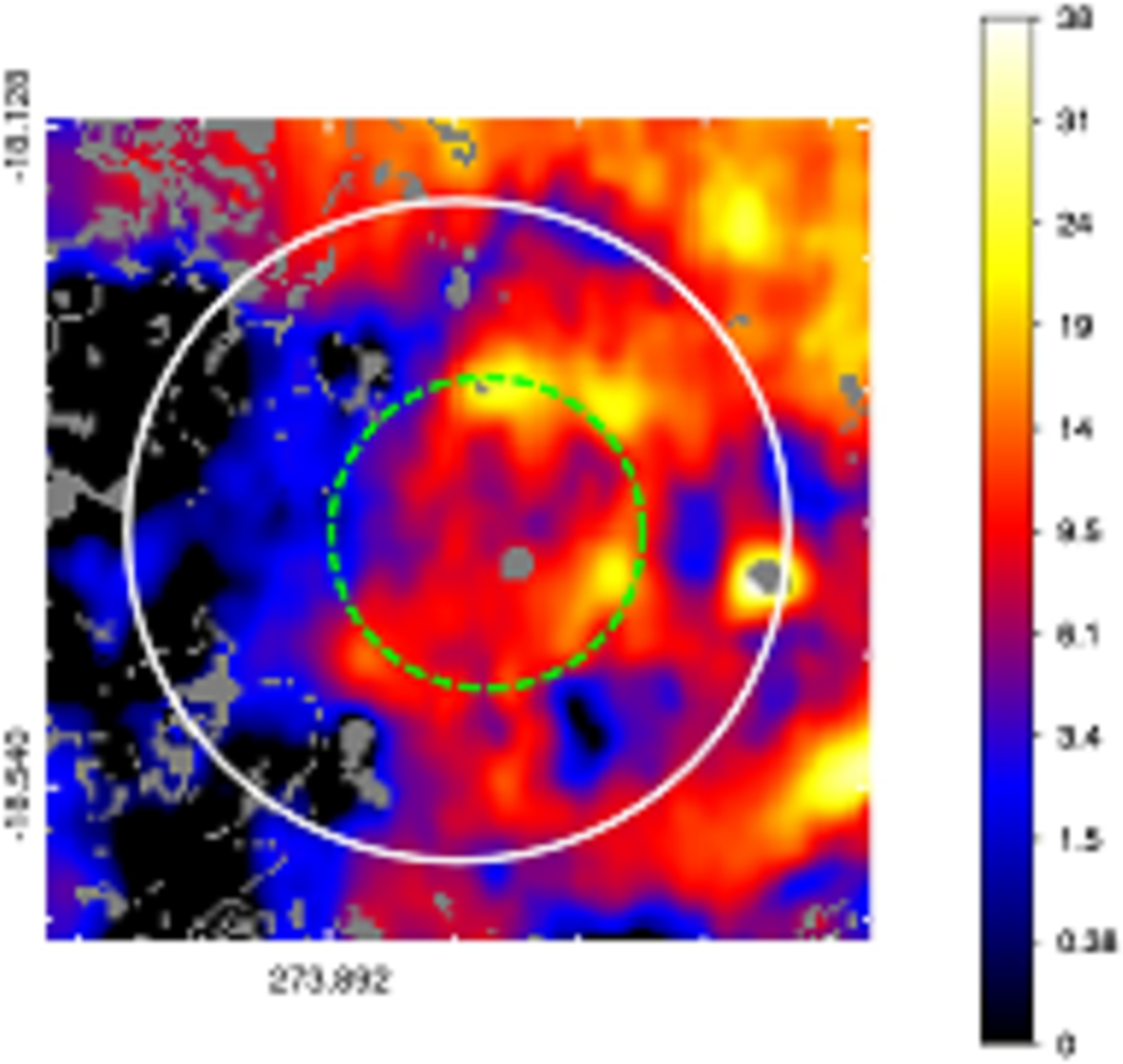}}}%
}
\subfigure{
\mbox{\raisebox{0mm}{\includegraphics[width=40mm]{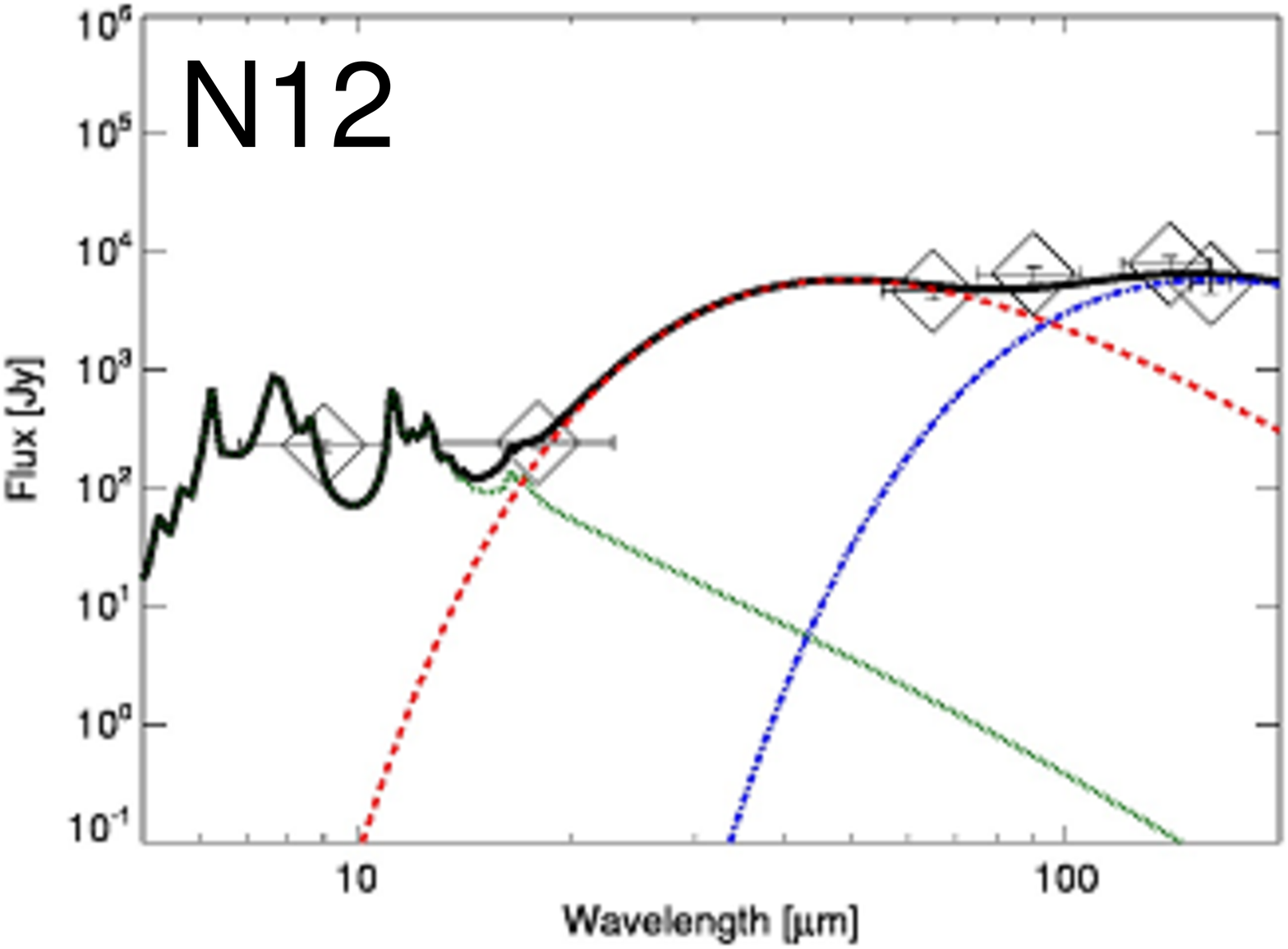}}}%
\mbox{\raisebox{6mm}{\rotatebox{90}{\small{DEC (J2000)}}}}%
\mbox{\raisebox{0mm}{\includegraphics[width=40mm]{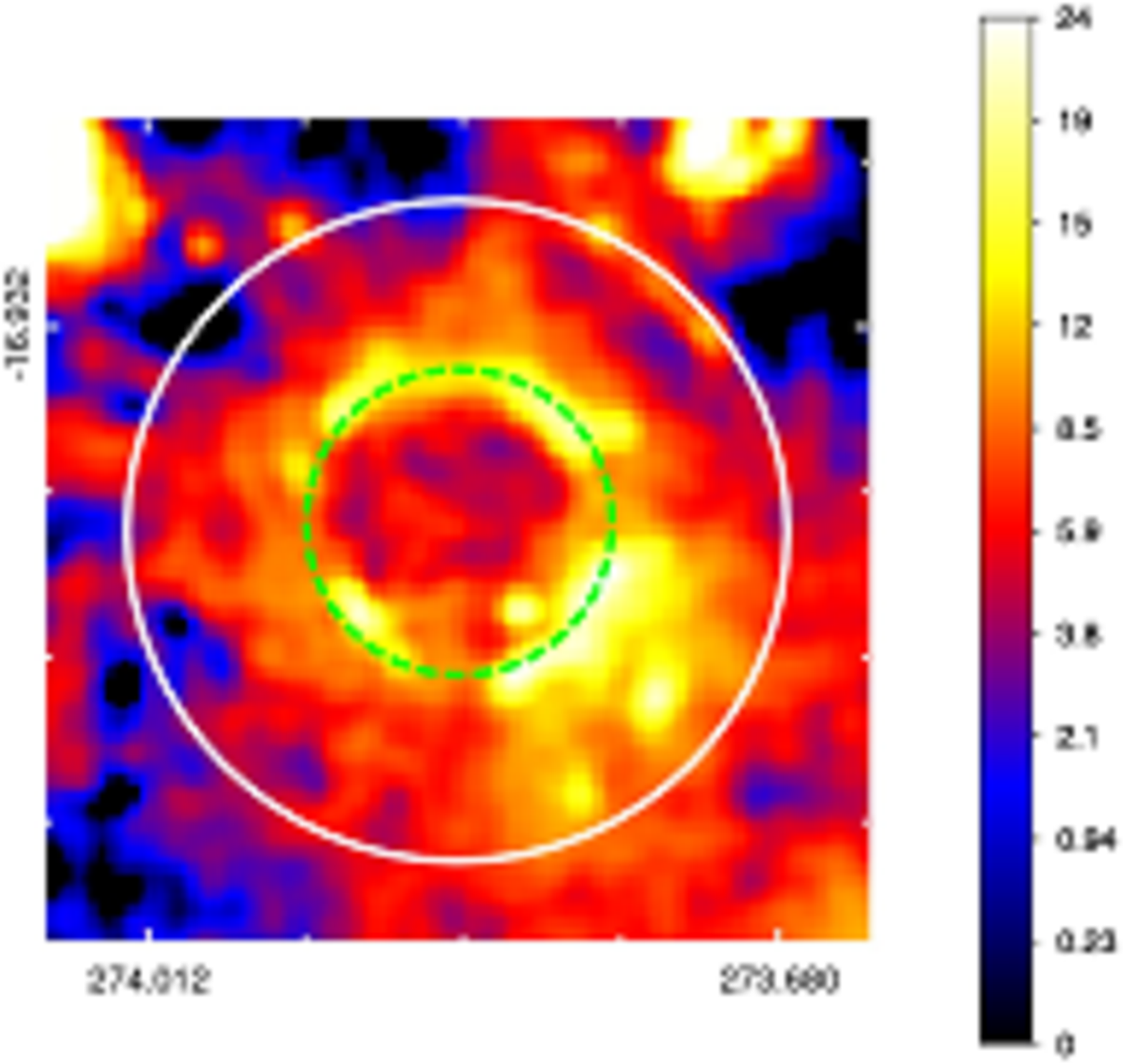}}}%
\mbox{\raisebox{0mm}{\includegraphics[width=40mm]{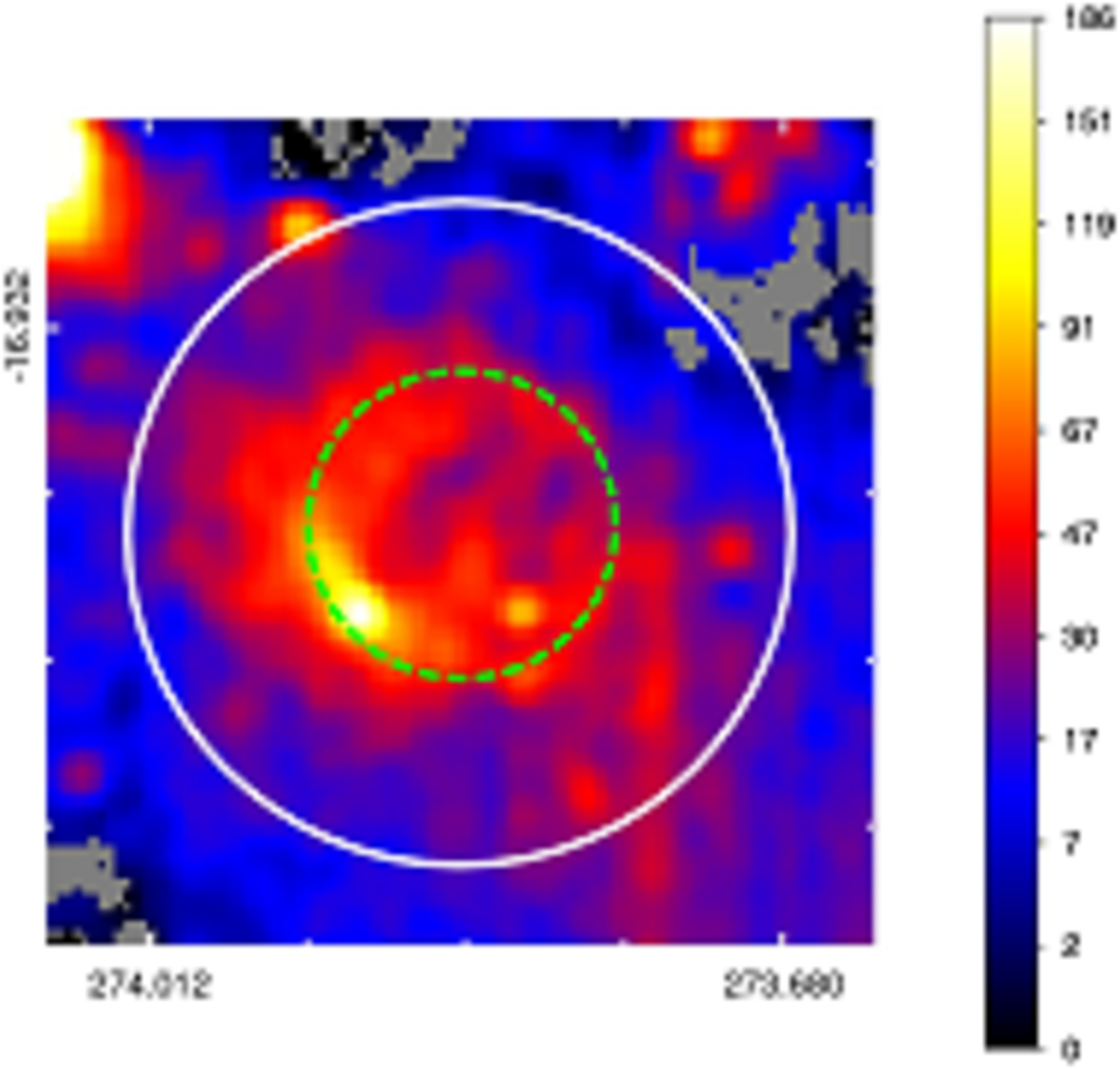}}}%
\mbox{\raisebox{0mm}{\includegraphics[width=40mm]{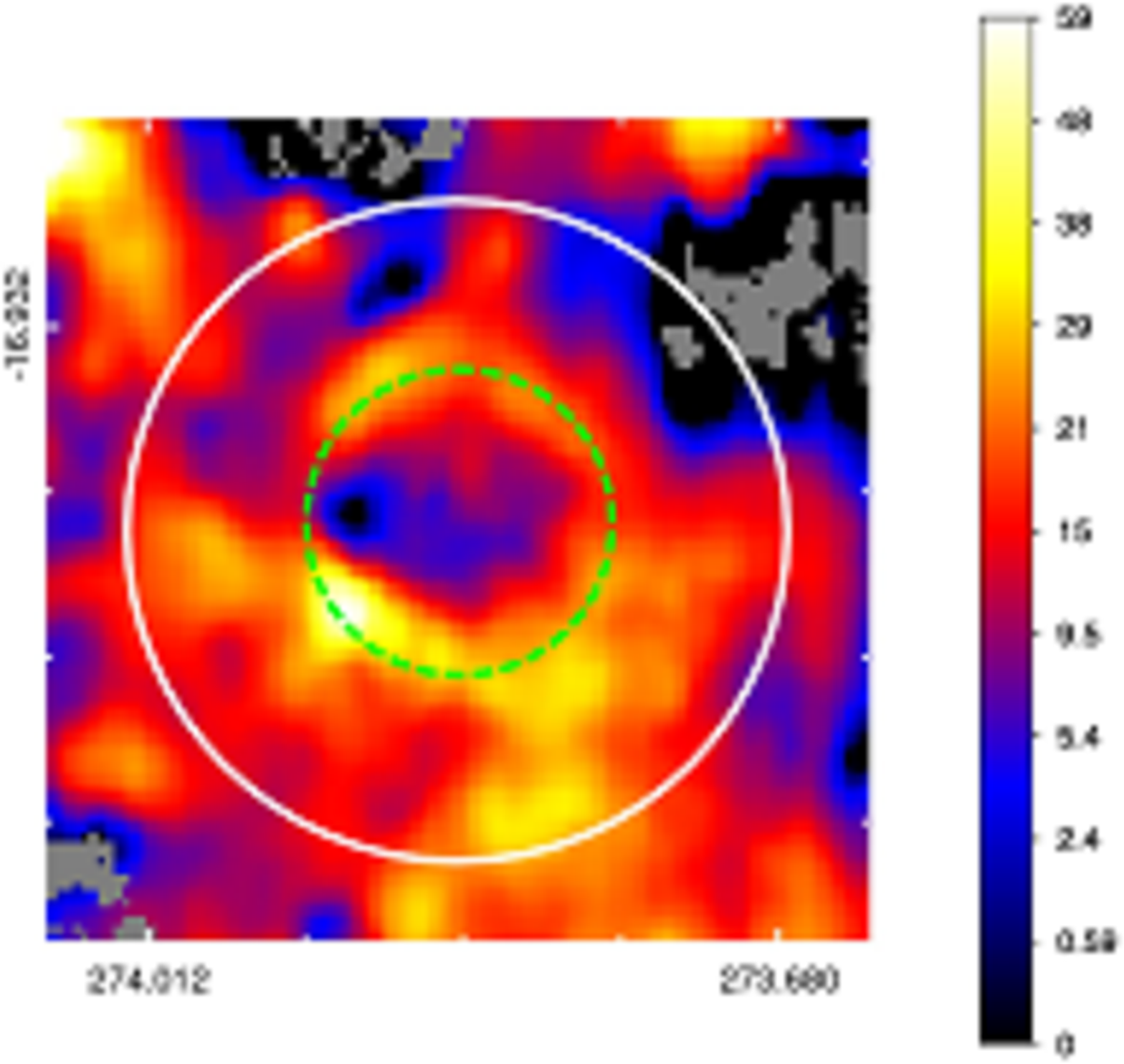}}}%
}
\subfigure{
\mbox{\raisebox{0mm}{\includegraphics[width=40mm]{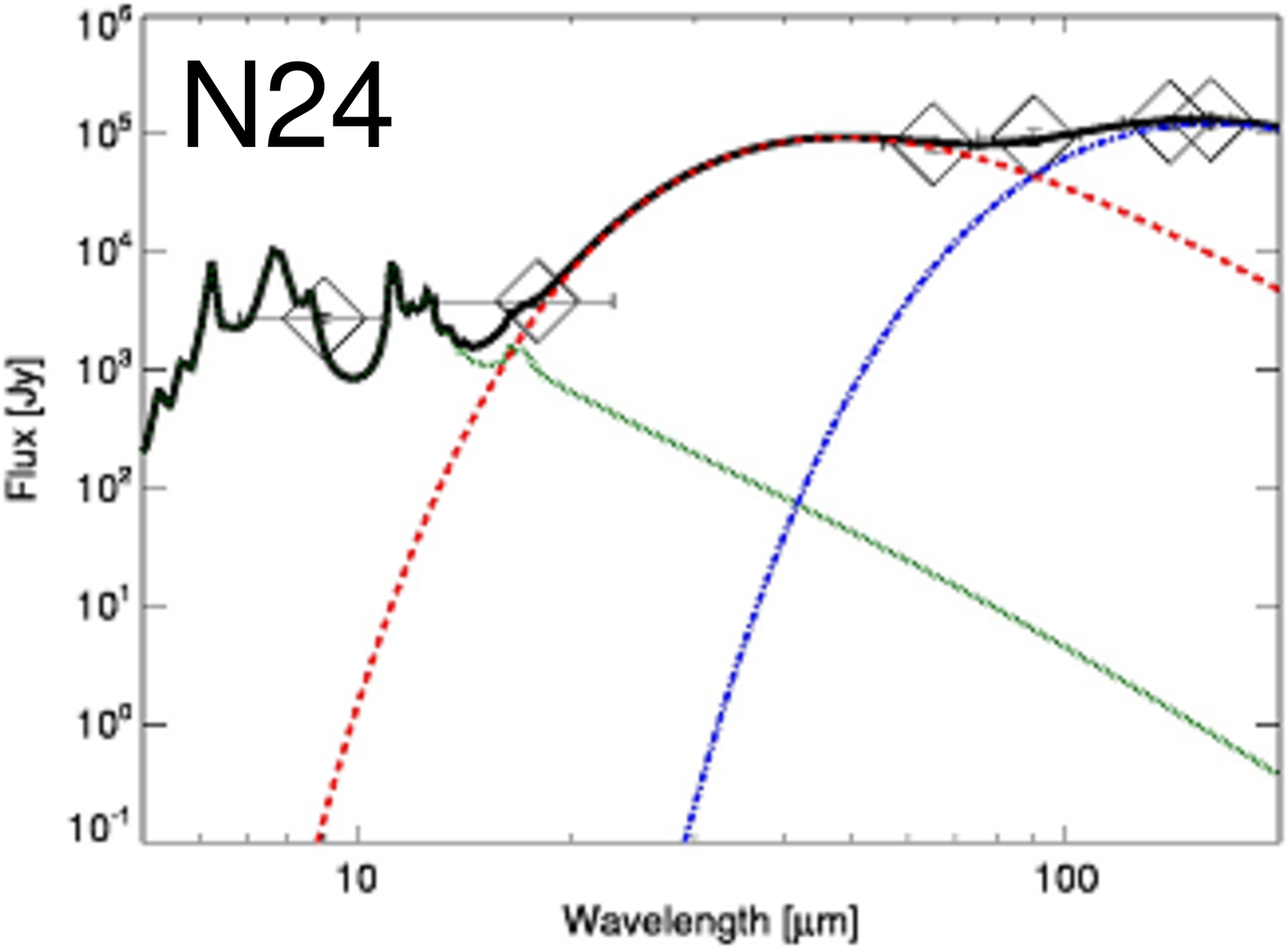}}}%
\mbox{\raisebox{6mm}{\rotatebox{90}{\small{DEC (J2000)}}}}%
\mbox{\raisebox{0mm}{\includegraphics[width=40mm]{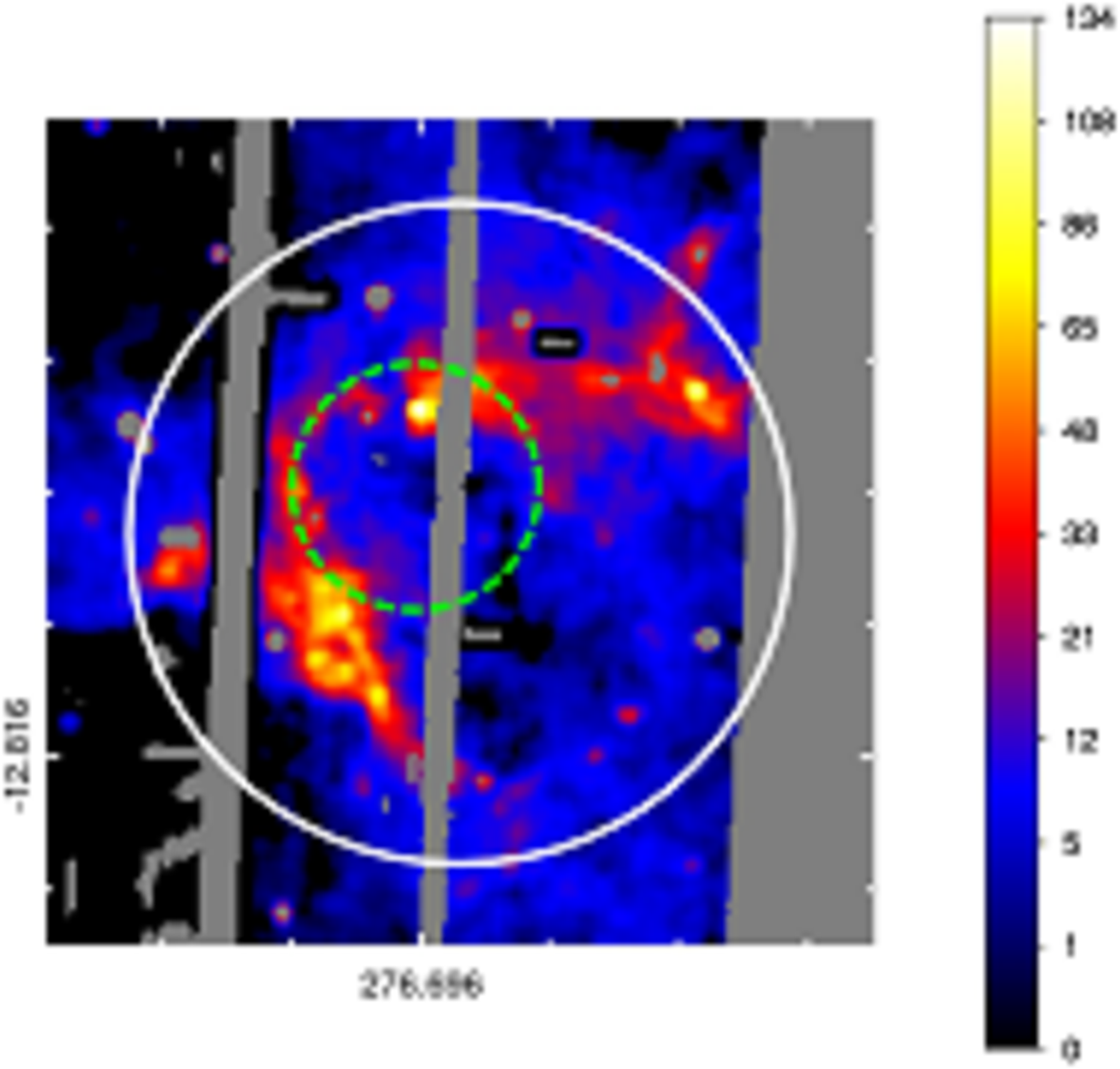}}}%
\mbox{\raisebox{0mm}{\includegraphics[width=40mm]{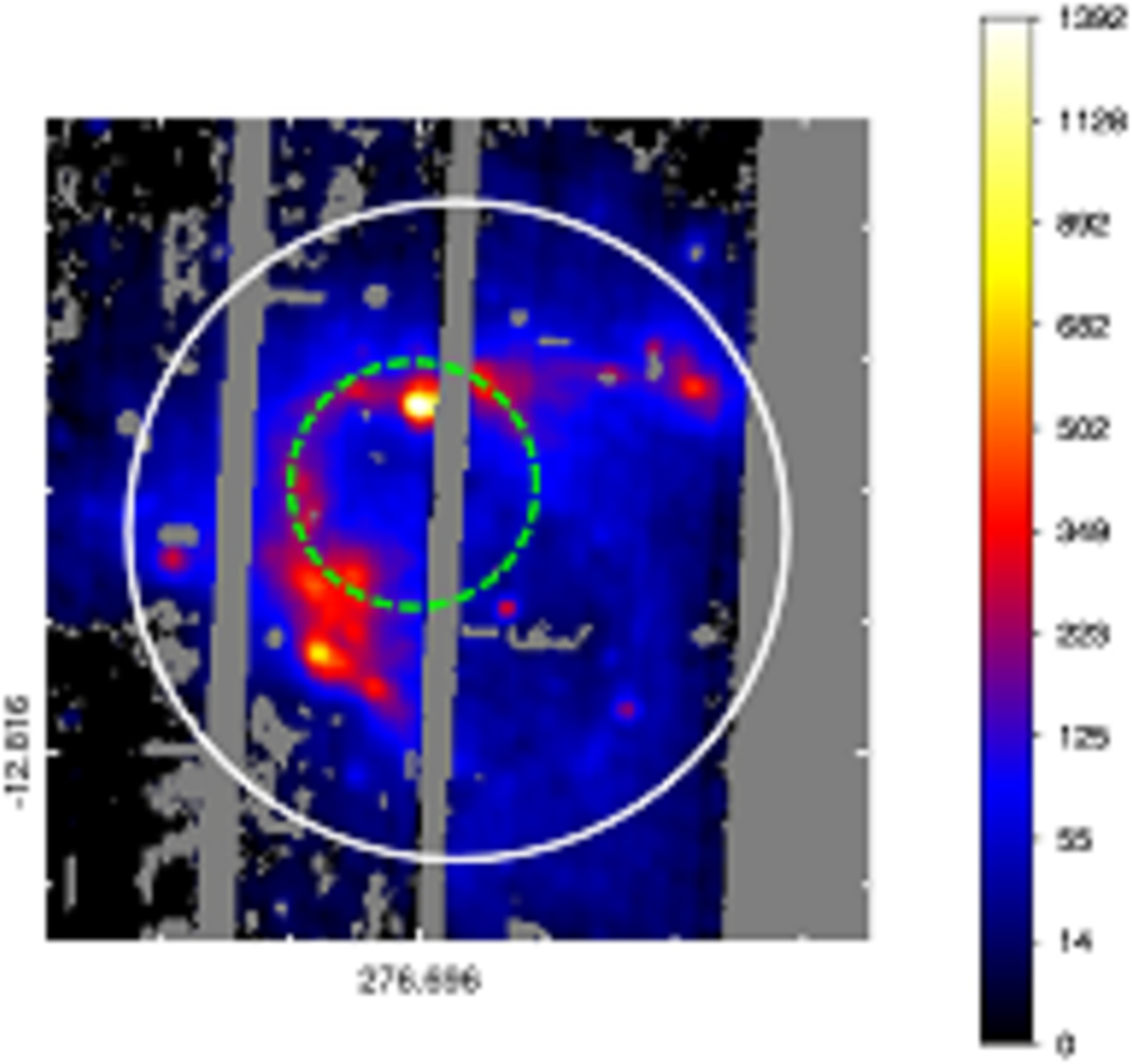}}}%
\mbox{\raisebox{0mm}{\includegraphics[width=40mm]{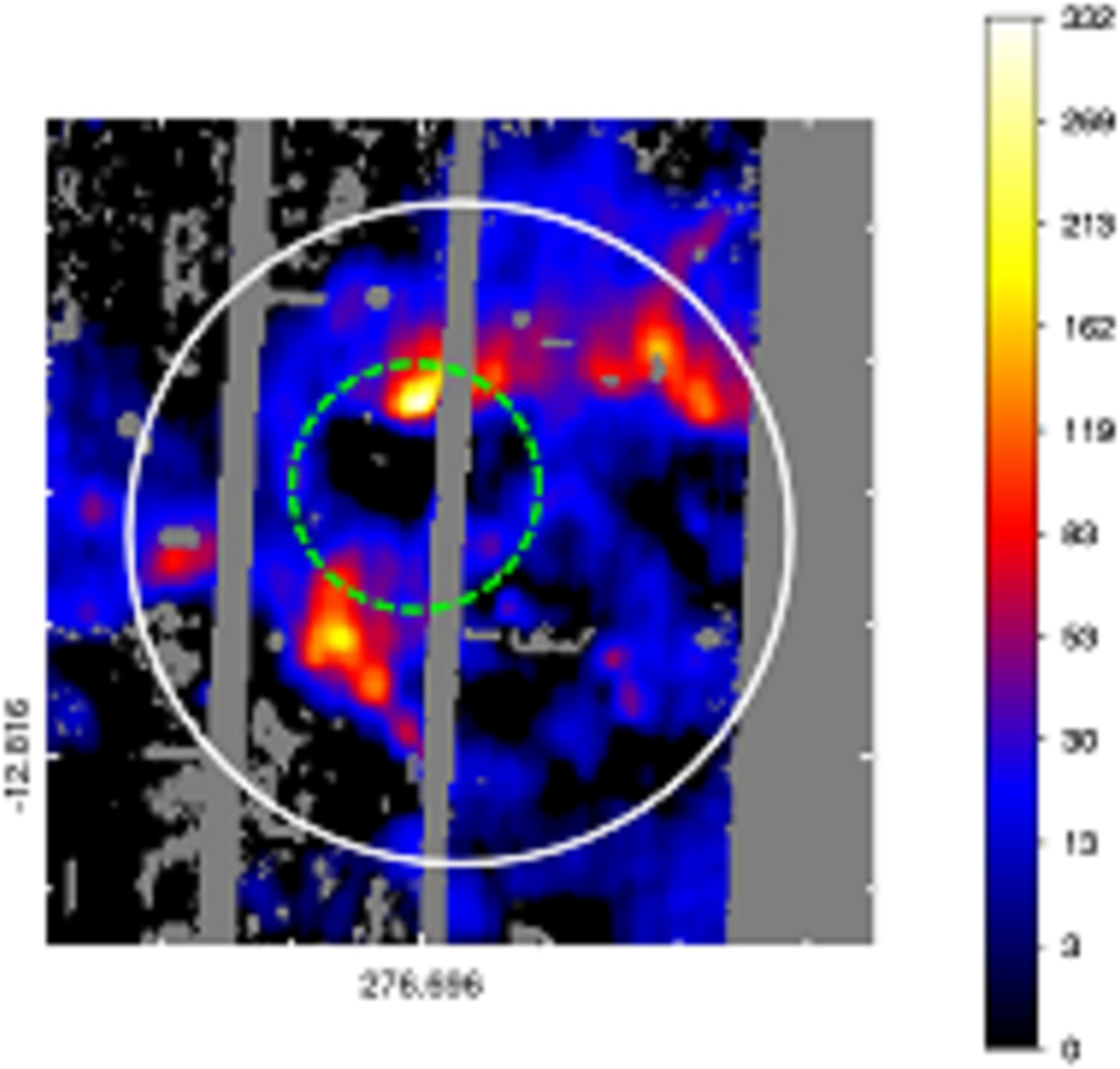}}}%
}
\subfigure{
\mbox{\raisebox{0mm}{\includegraphics[width=40mm]{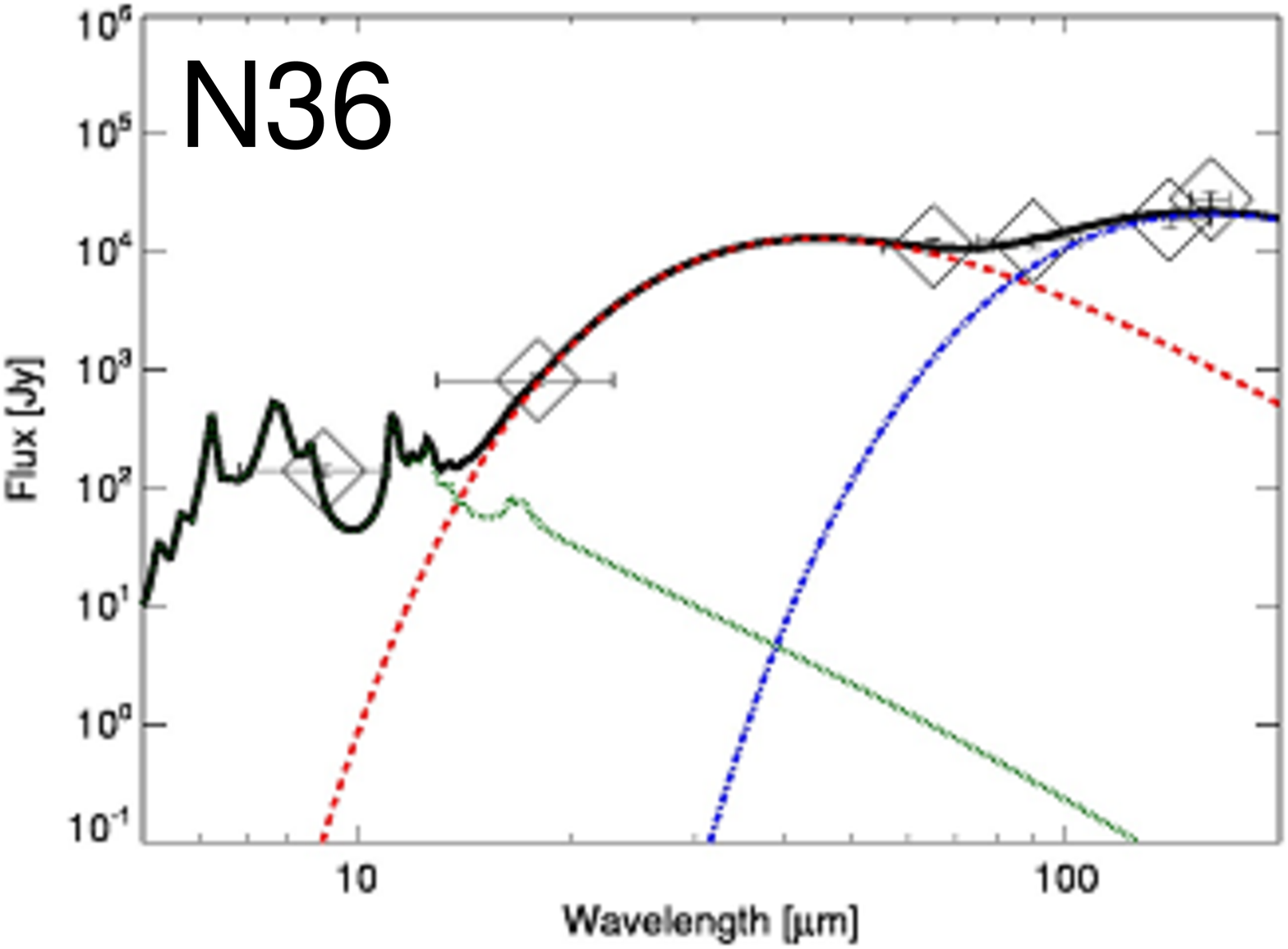}}}%
\mbox{\raisebox{6mm}{\rotatebox{90}{\small{DEC (J2000)}}}}%
\mbox{\raisebox{0mm}{\includegraphics[width=40mm]{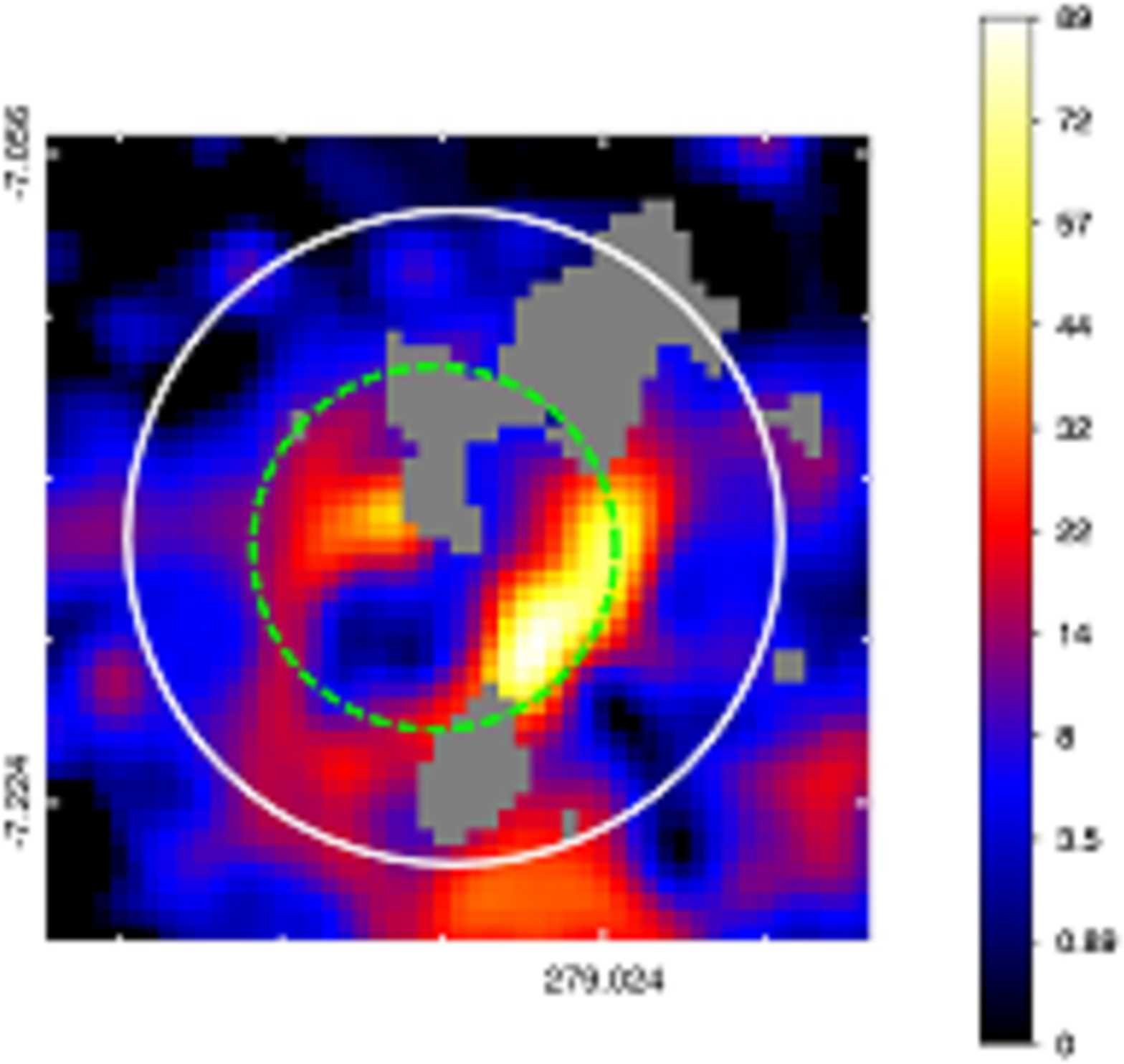}}}%
\mbox{\raisebox{0mm}{\includegraphics[width=40mm]{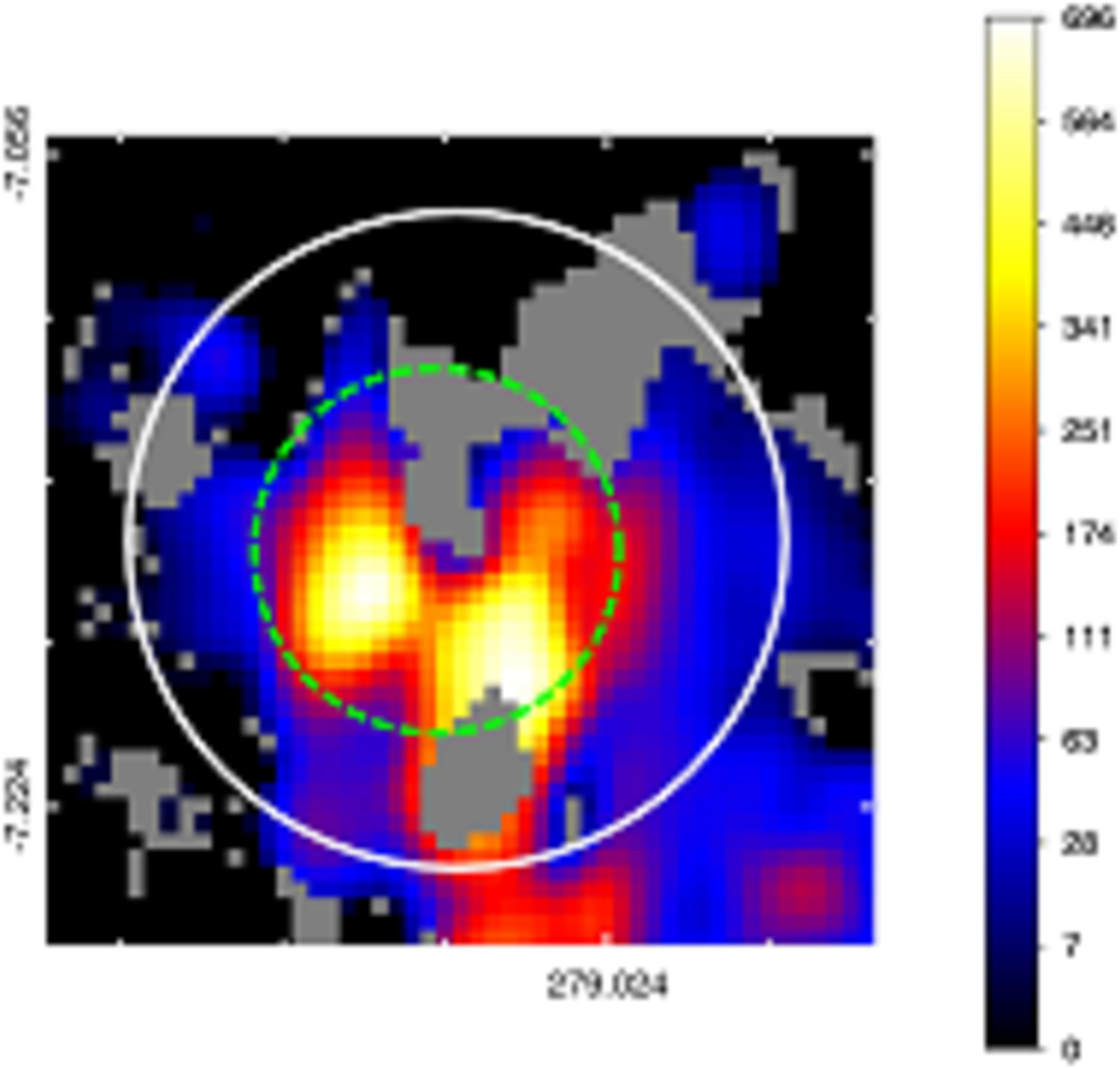}}}%
\mbox{\raisebox{0mm}{\includegraphics[width=40mm]{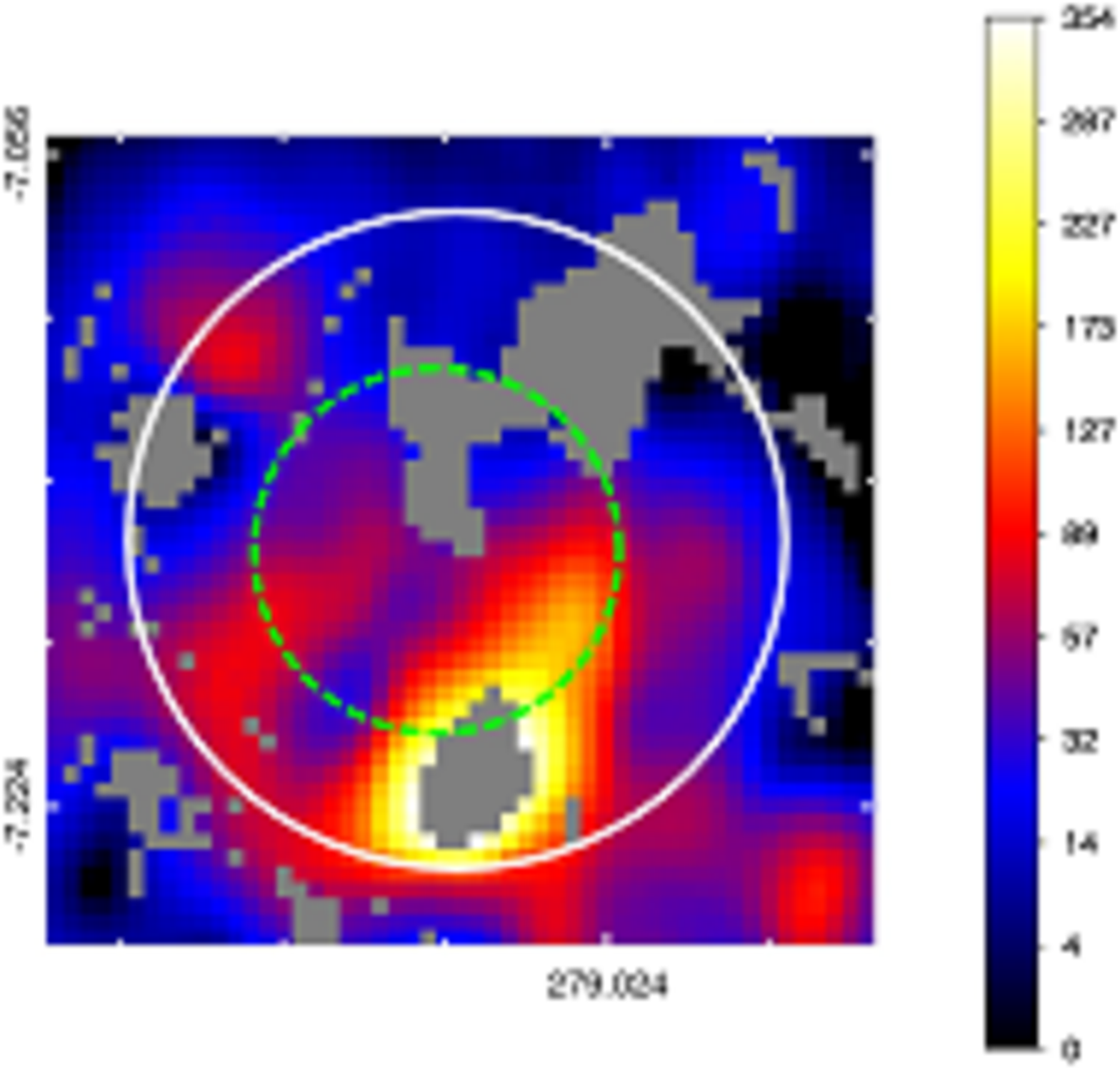}}}%
}
\subfigure{
\mbox{\raisebox{0mm}{\includegraphics[width=40mm]{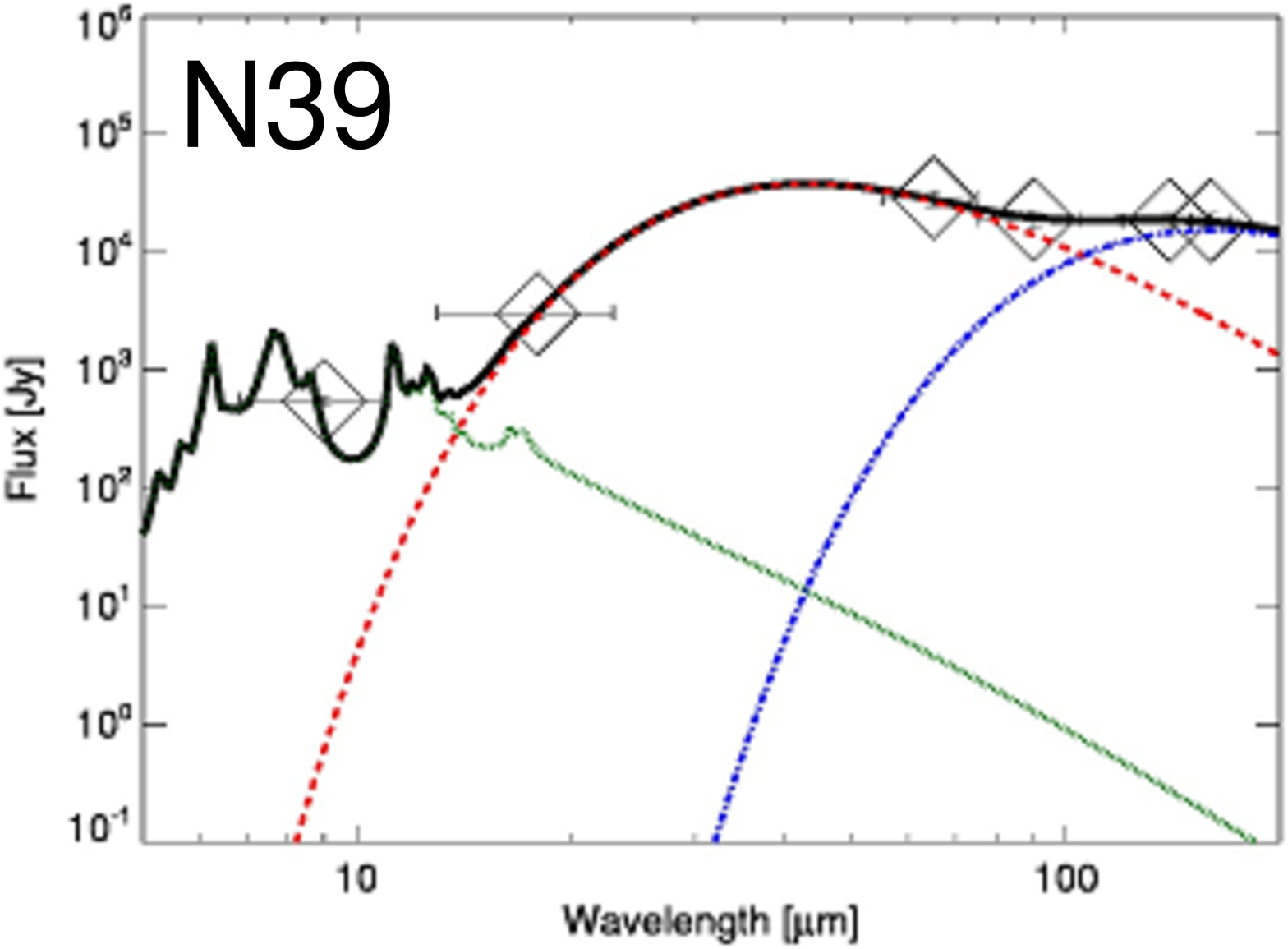}}}%
\mbox{\raisebox{6mm}{\rotatebox{90}{\small{DEC (J2000)}}}}%
\mbox{\raisebox{0mm}{\includegraphics[width=40mm]{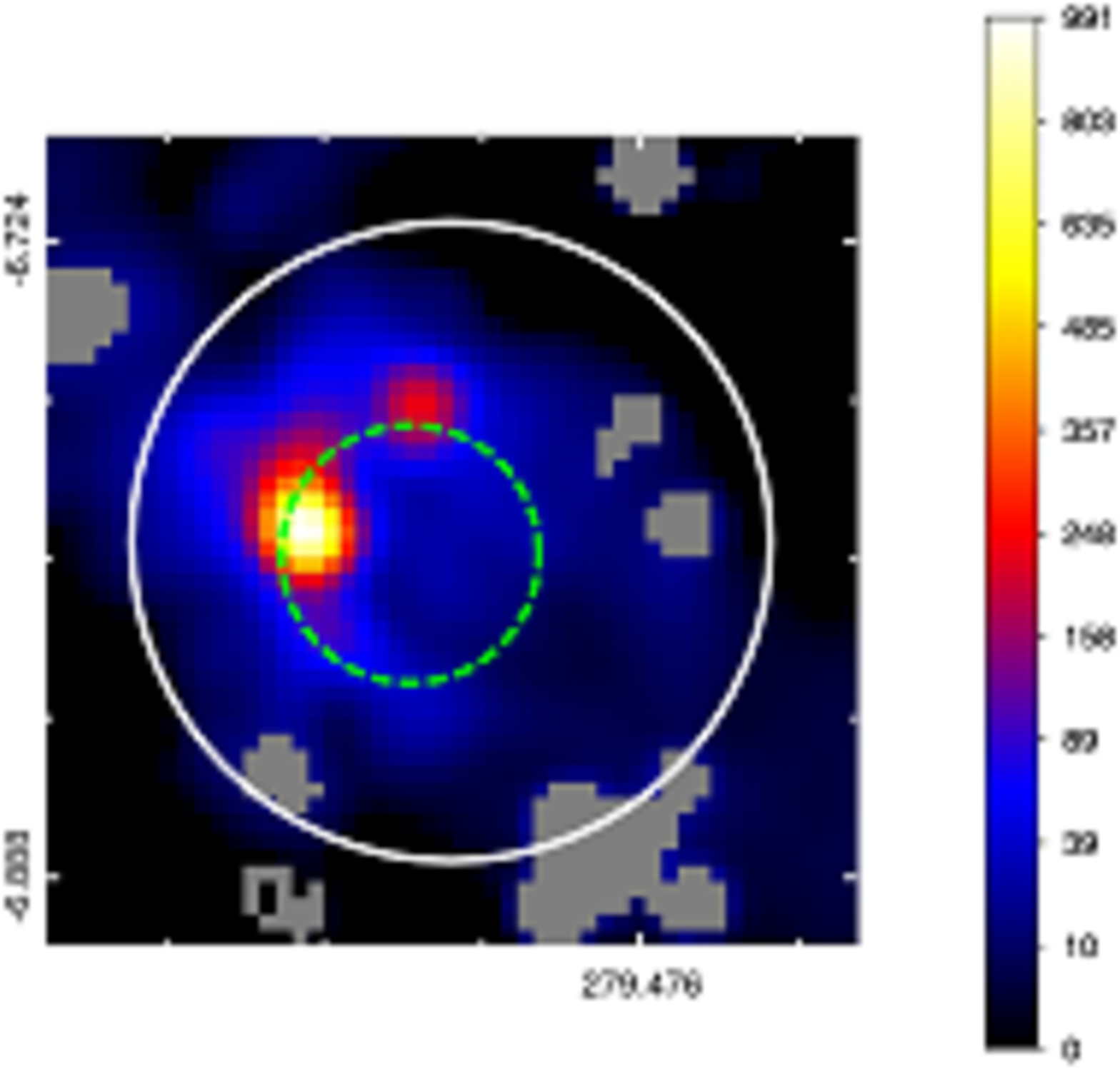}}}%
\mbox{\raisebox{0mm}{\includegraphics[width=40mm]{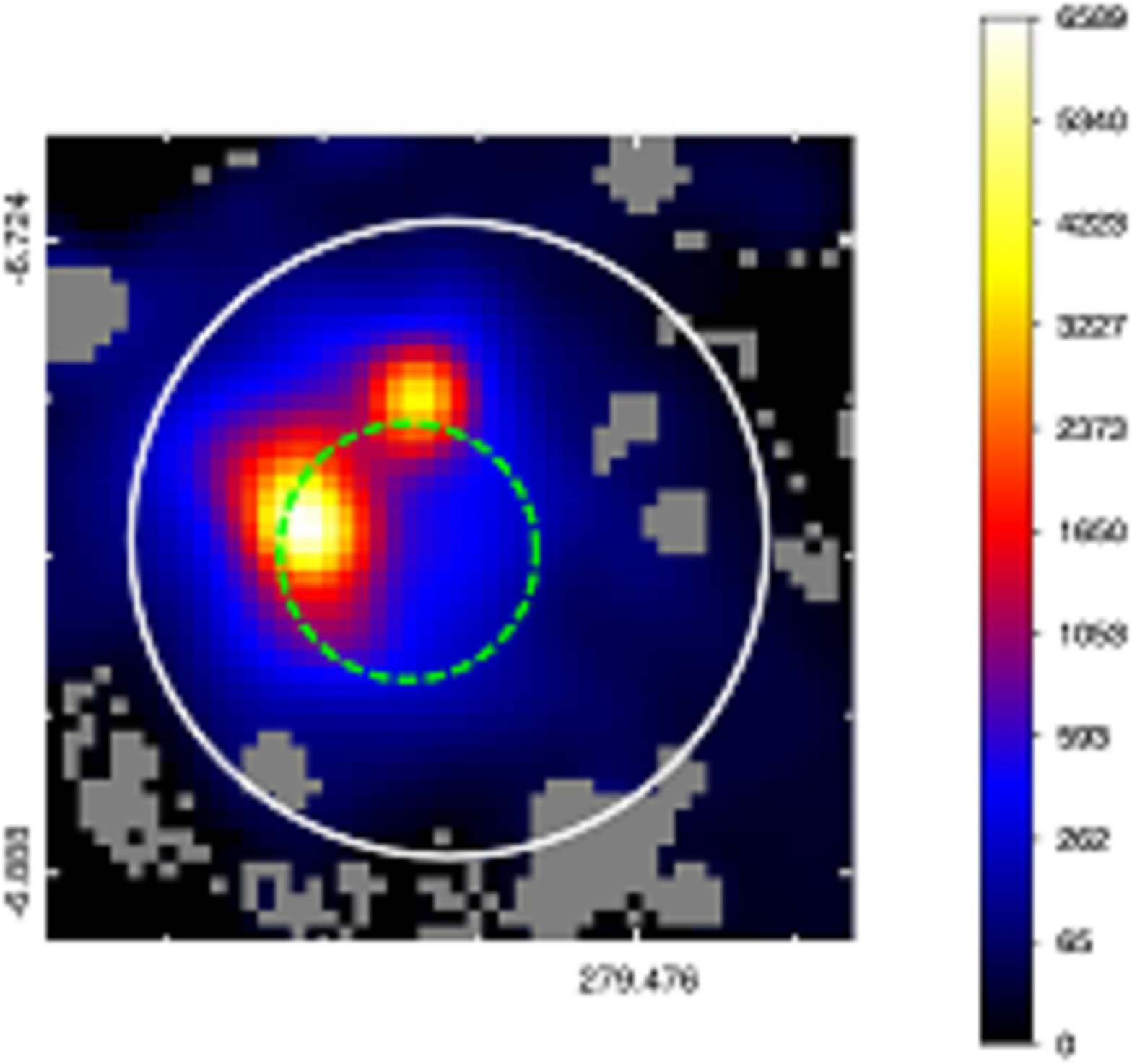}}}%
\mbox{\raisebox{0mm}{\includegraphics[width=40mm]{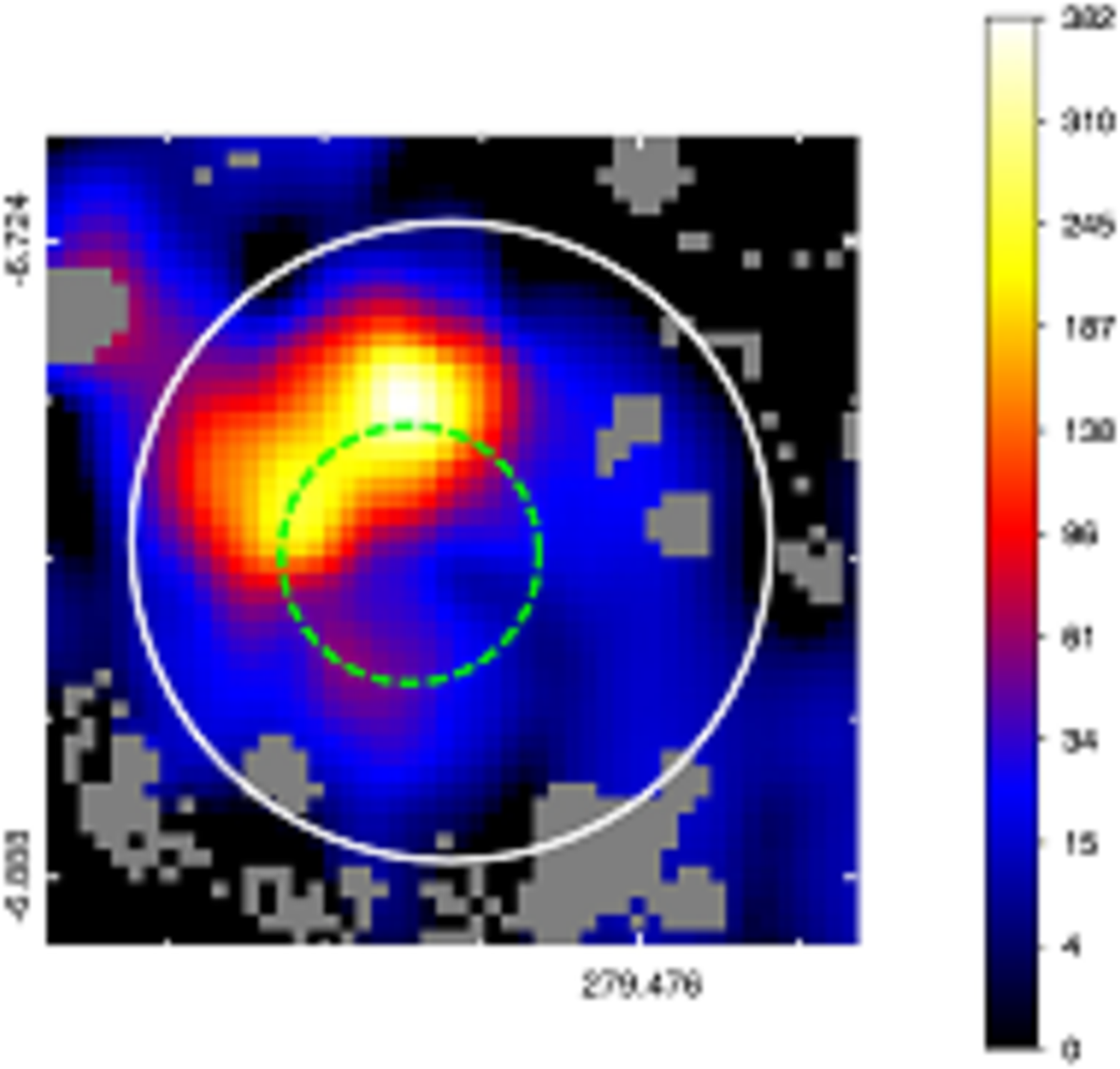}}}%
}
\caption{Continued.} \label{fig:Metfig2:b}
\end{figure*}

\addtocounter{figure}{-1}
\begin{figure*}[ht]
\addtocounter{subfigure}{1}
\centering
\subfigure{
\makebox[180mm][l]{\raisebox{0mm}[0mm][0mm]{ \hspace{20mm} \small{SED}} \hspace{27.5mm} \small{$I_{\rm{PAH}}$} \hspace{29.5mm} \small{$I_{\rm{warm}}$} \hspace{29.5mm} \small{$I_{\rm{cold}}$}}%
}
\subfigure{
\makebox[180mm][l]{\raisebox{0mm}{\hspace{52mm} \small{RA (J2000)} \hspace{20mm} \small{RA (J2000)} \hspace{20mm} \small{RA (J2000)}}}
}
\subfigure{
\mbox{\raisebox{0mm}{\includegraphics[width=40mm]{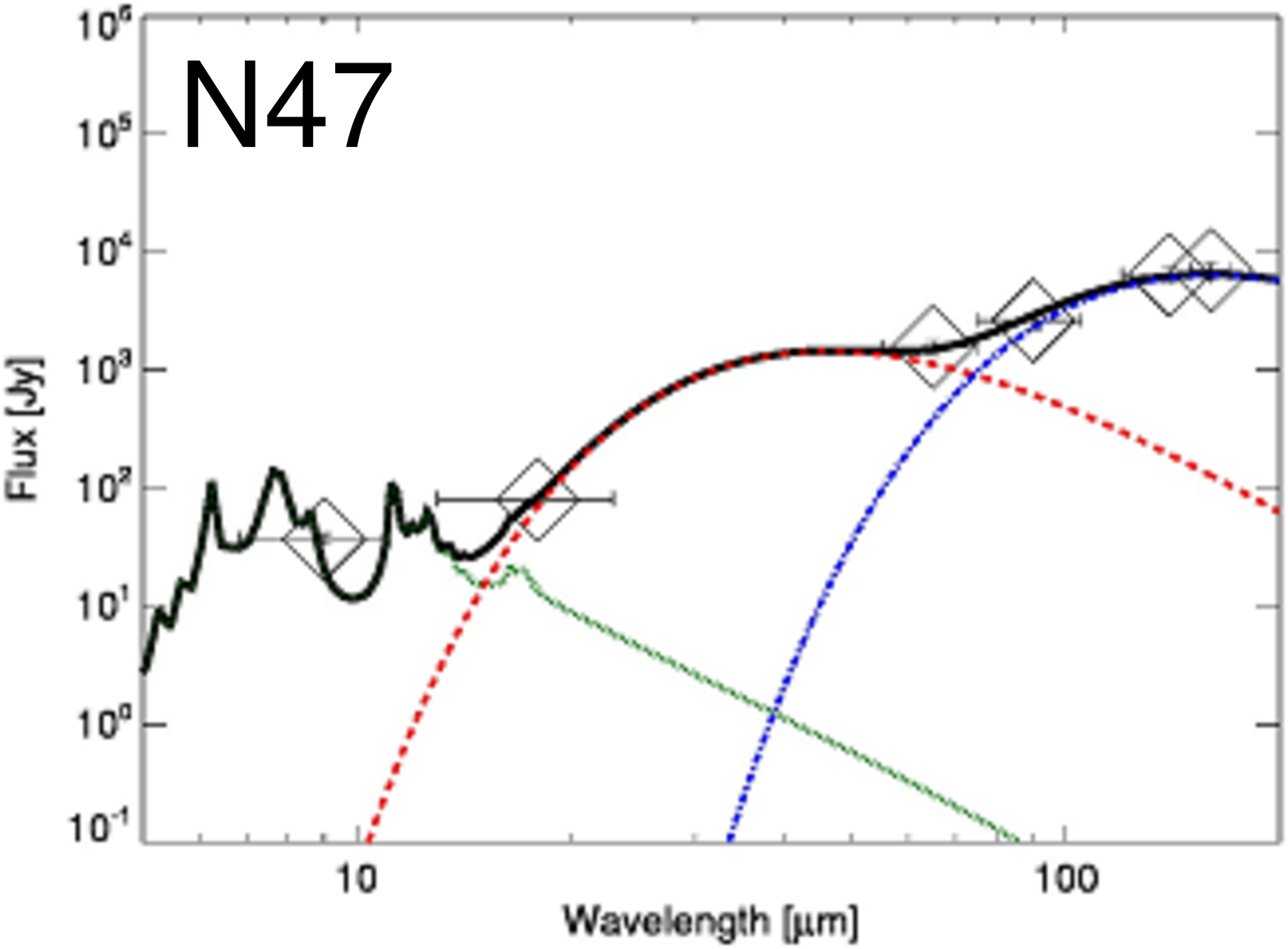}}}%
\mbox{\raisebox{6mm}{\rotatebox{90}{\small{DEC (J2000)}}}}%
\mbox{\raisebox{0mm}{\includegraphics[width=40mm]{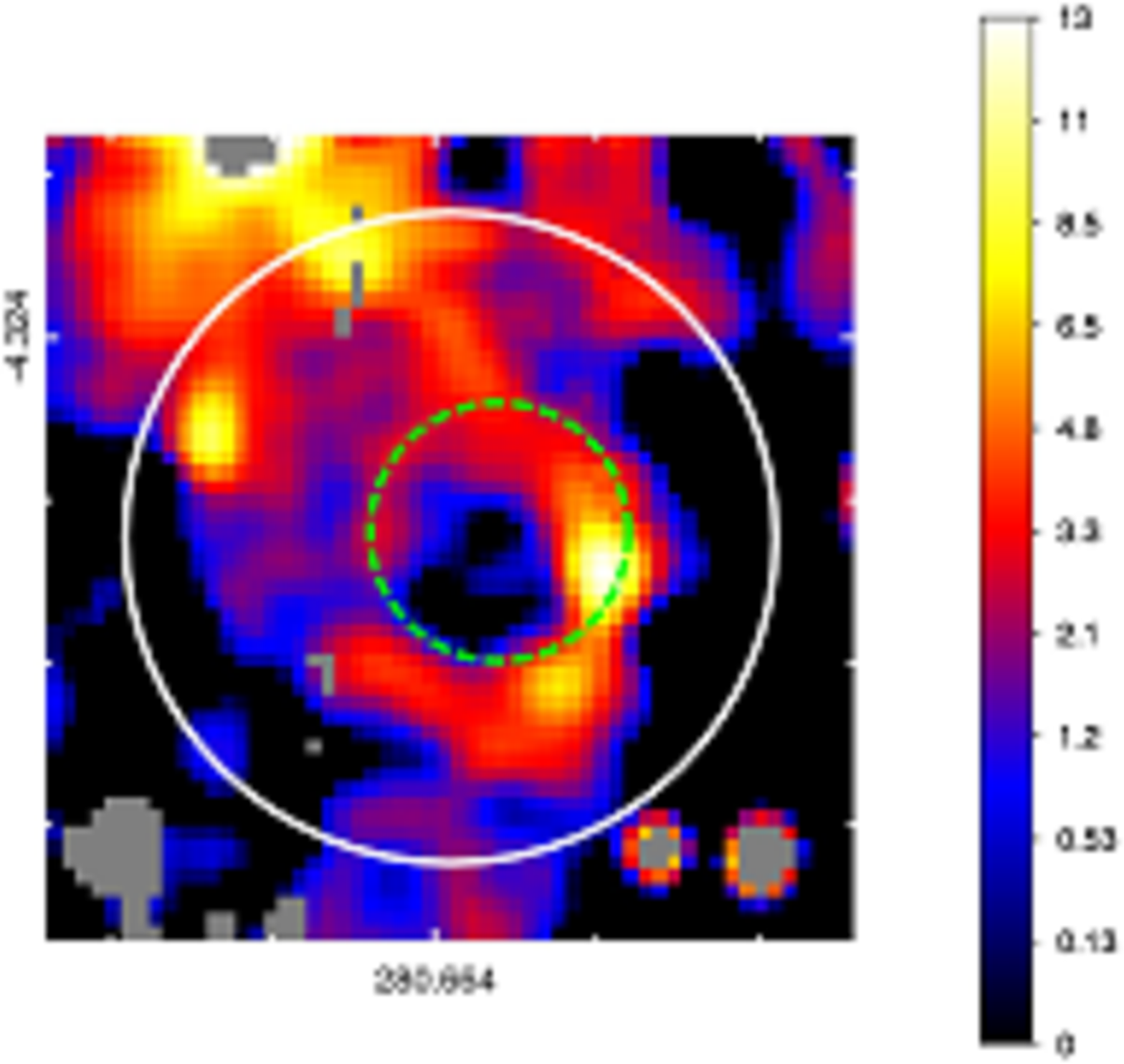}}}%
\mbox{\raisebox{0mm}{\includegraphics[width=40mm]{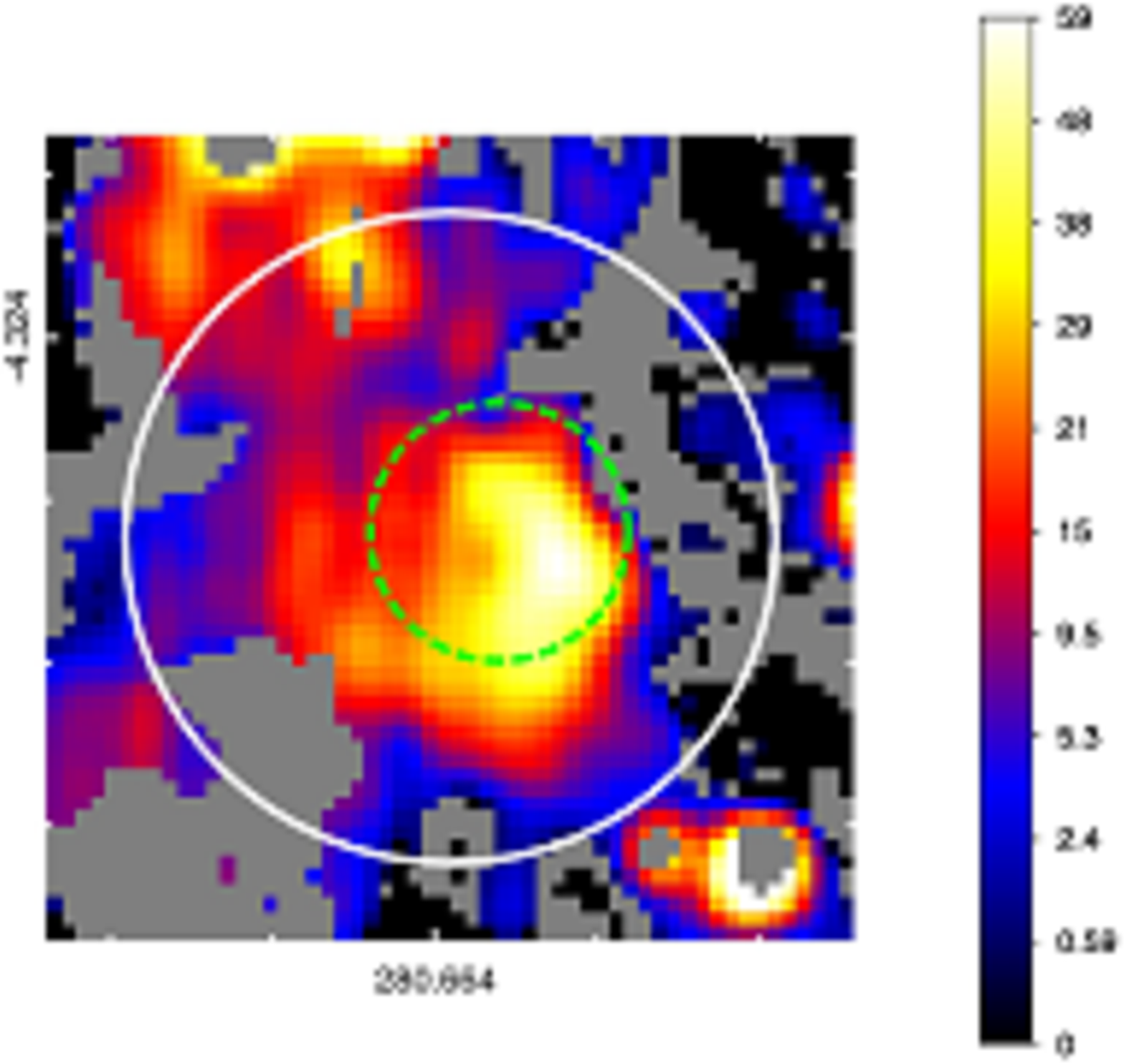}}}%
\mbox{\raisebox{0mm}{\includegraphics[width=40mm]{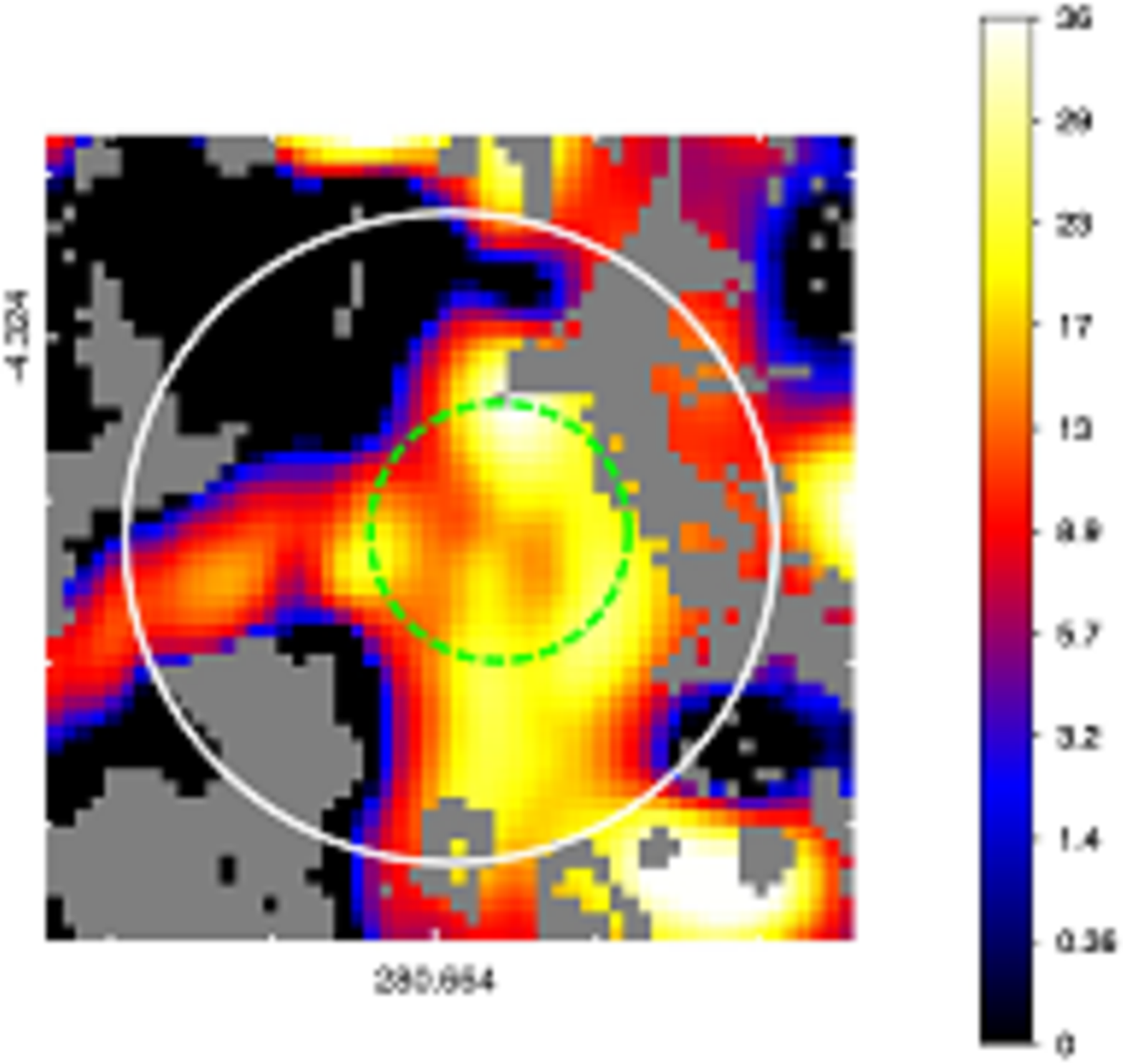}}}%
}
\subfigure{
\mbox{\raisebox{0mm}{\includegraphics[width=40mm]{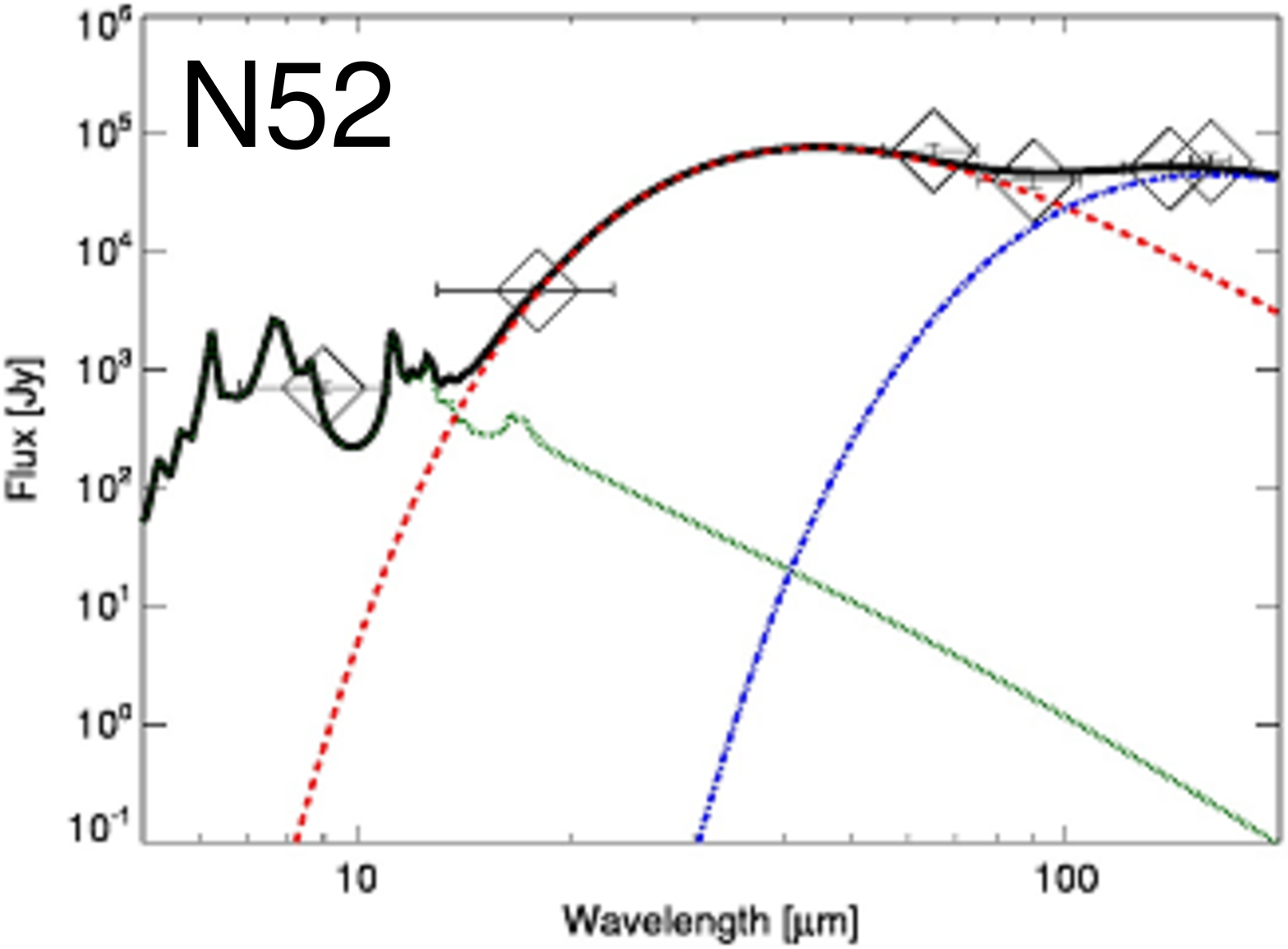}}}%
\mbox{\raisebox{6mm}{\rotatebox{90}{\small{DEC (J2000)}}}}%
\mbox{\raisebox{0mm}{\includegraphics[width=40mm]{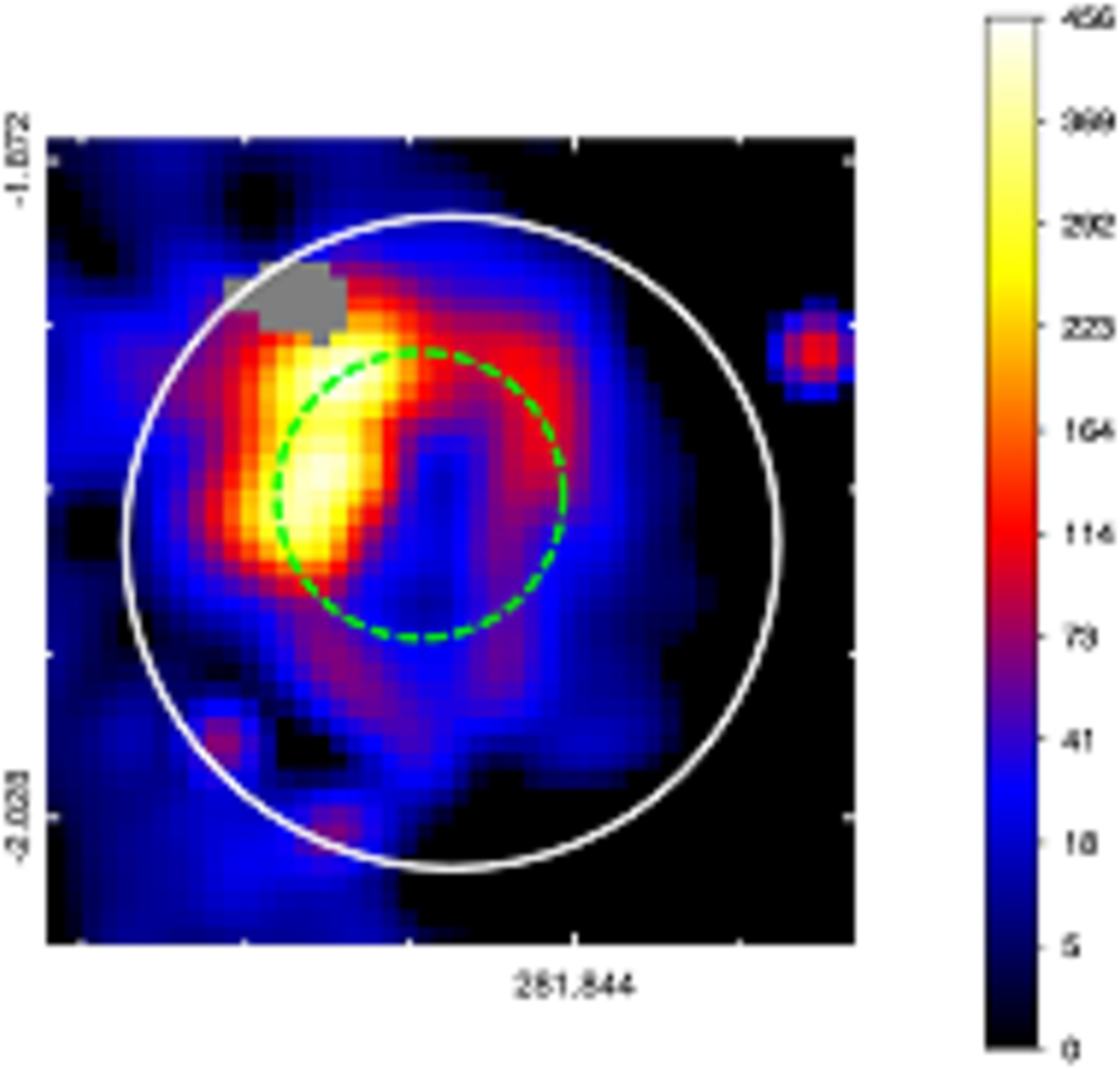}}}%
\mbox{\raisebox{0mm}{\includegraphics[width=40mm]{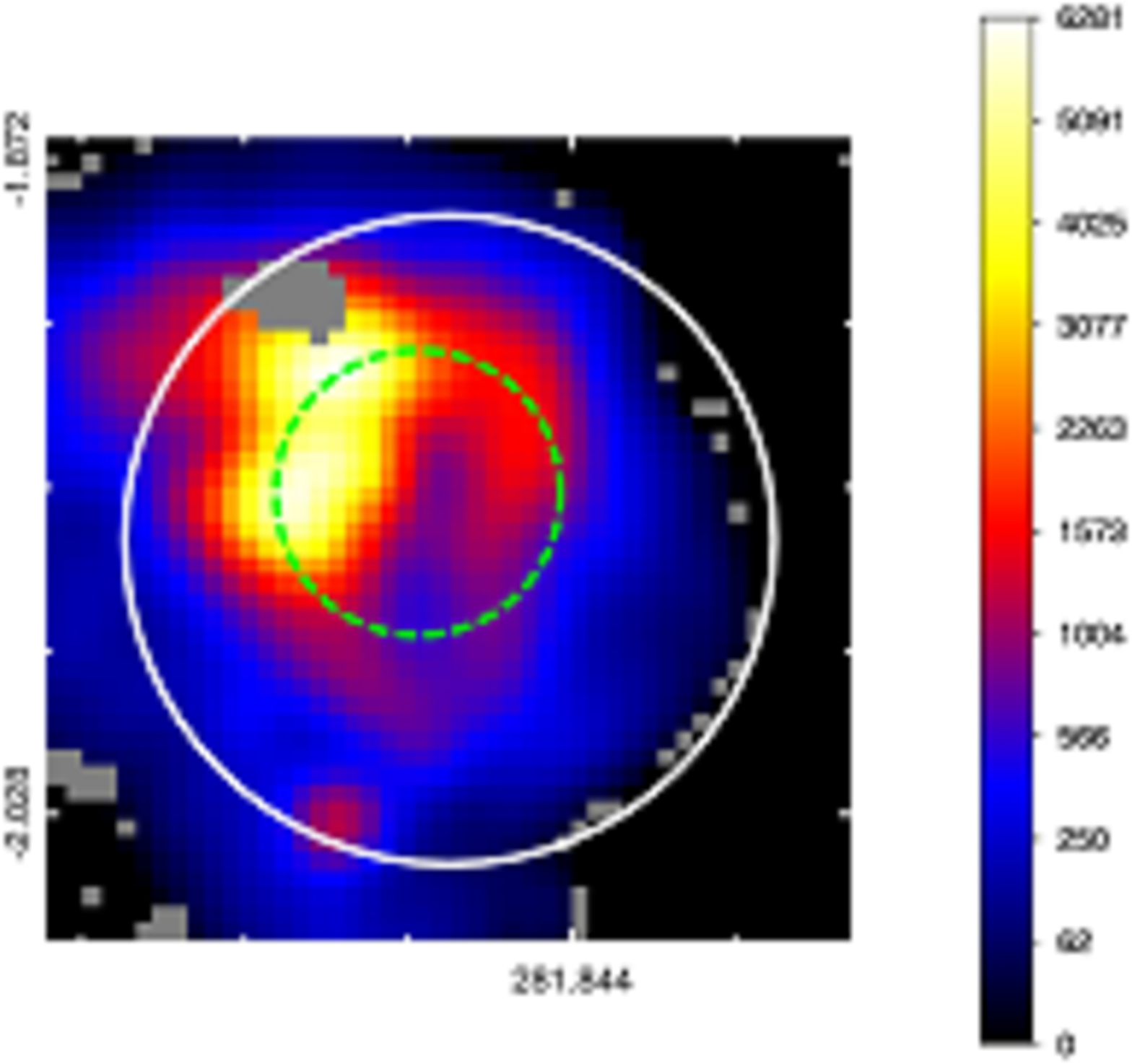}}}%
\mbox{\raisebox{0mm}{\includegraphics[width=40mm]{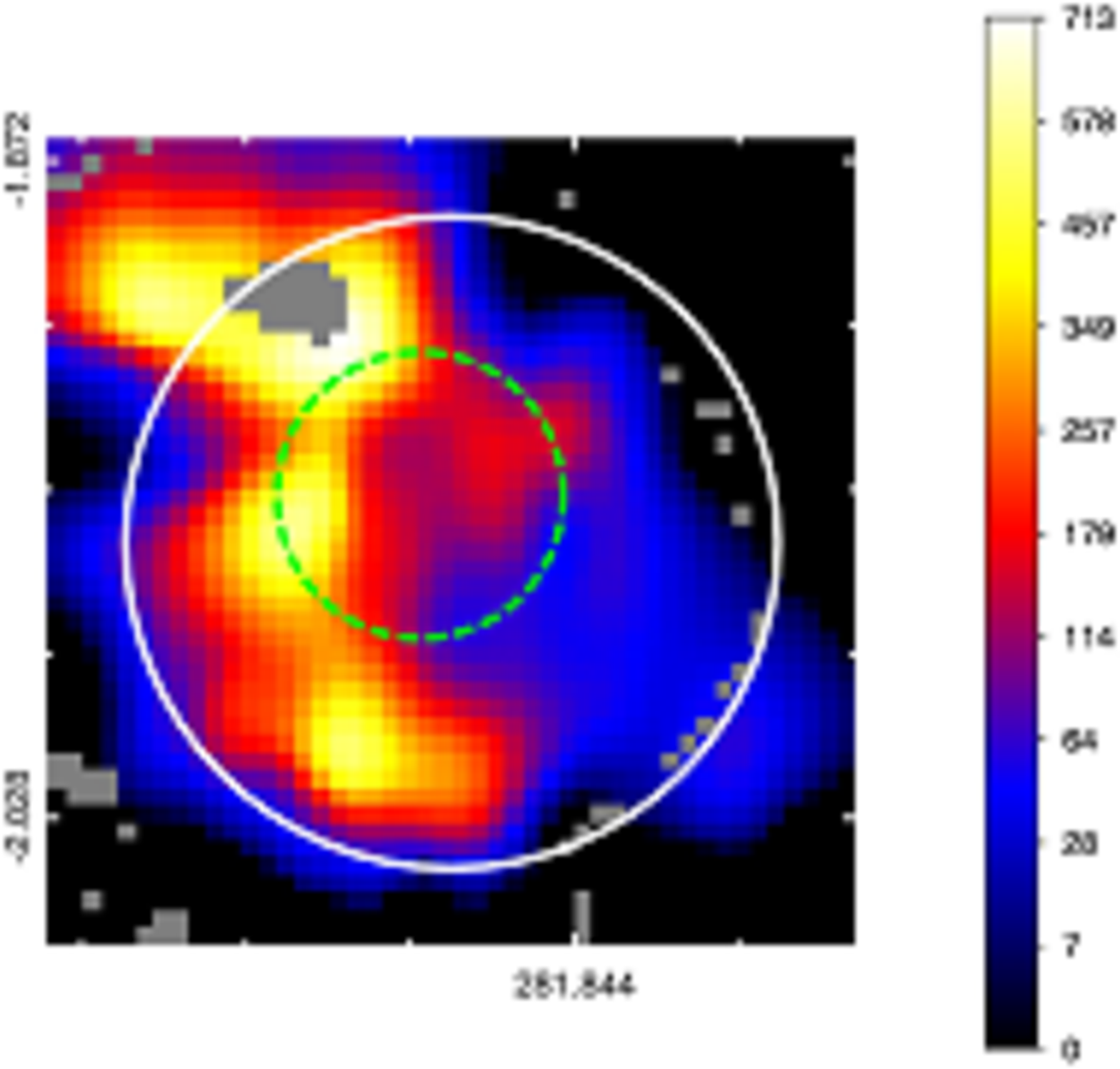}}}%
}
\subfigure{
\mbox{\raisebox{0mm}{\includegraphics[width=40mm]{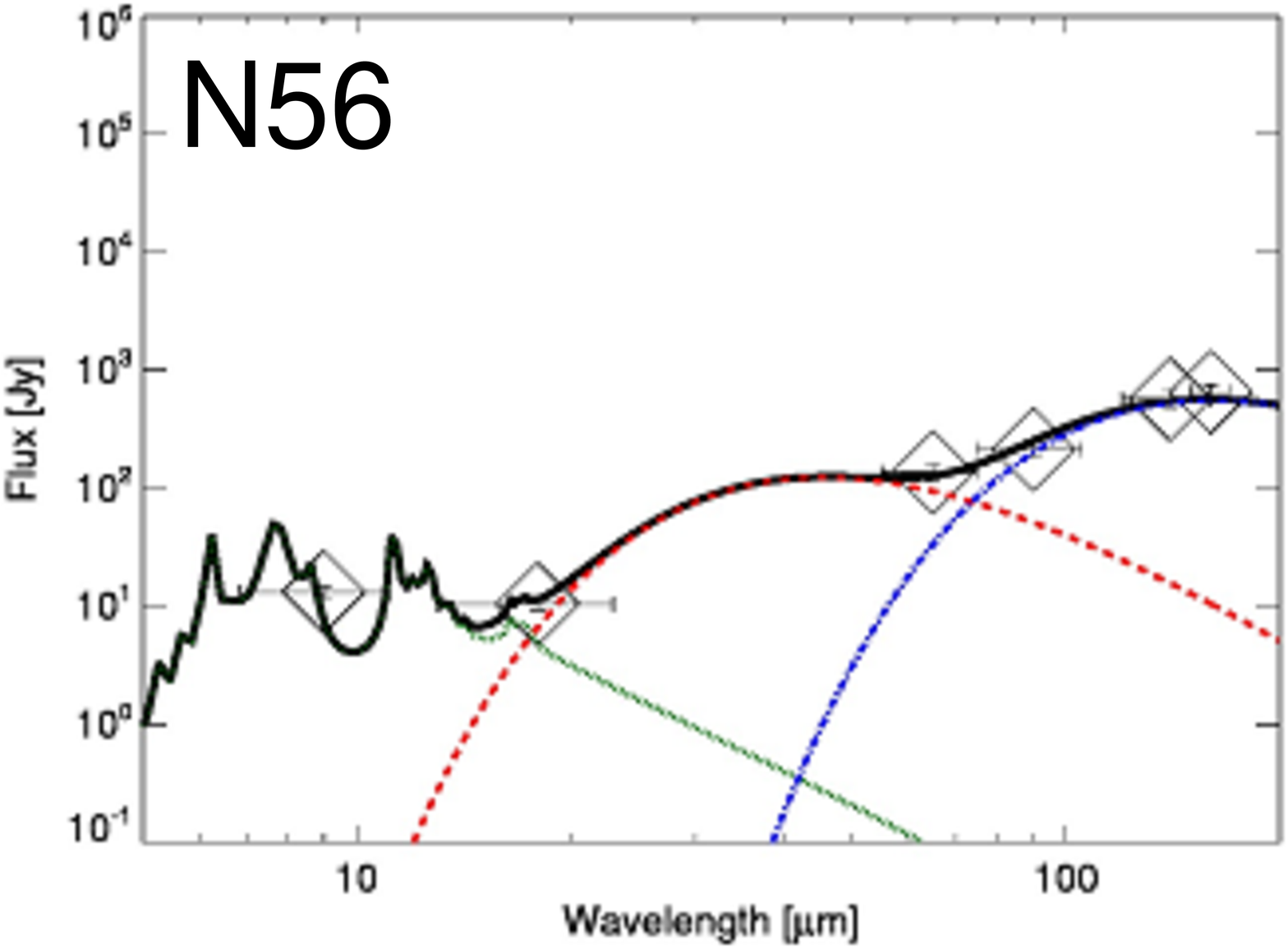}}}%
\mbox{\raisebox{6mm}{\rotatebox{90}{\small{DEC (J2000)}}}}%
\mbox{\raisebox{0mm}{\includegraphics[width=40mm]{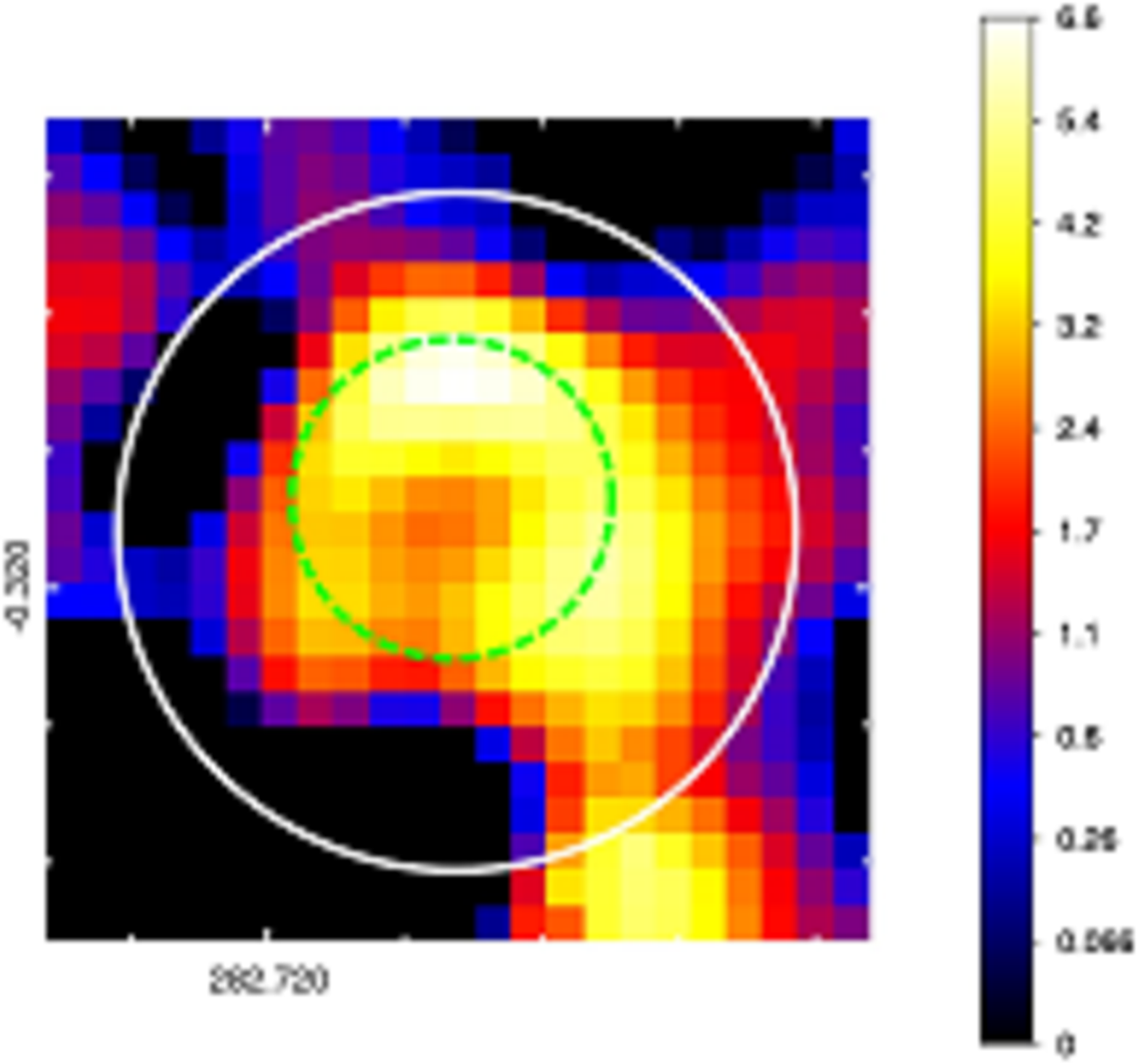}}}%
\mbox{\raisebox{0mm}{\includegraphics[width=40mm]{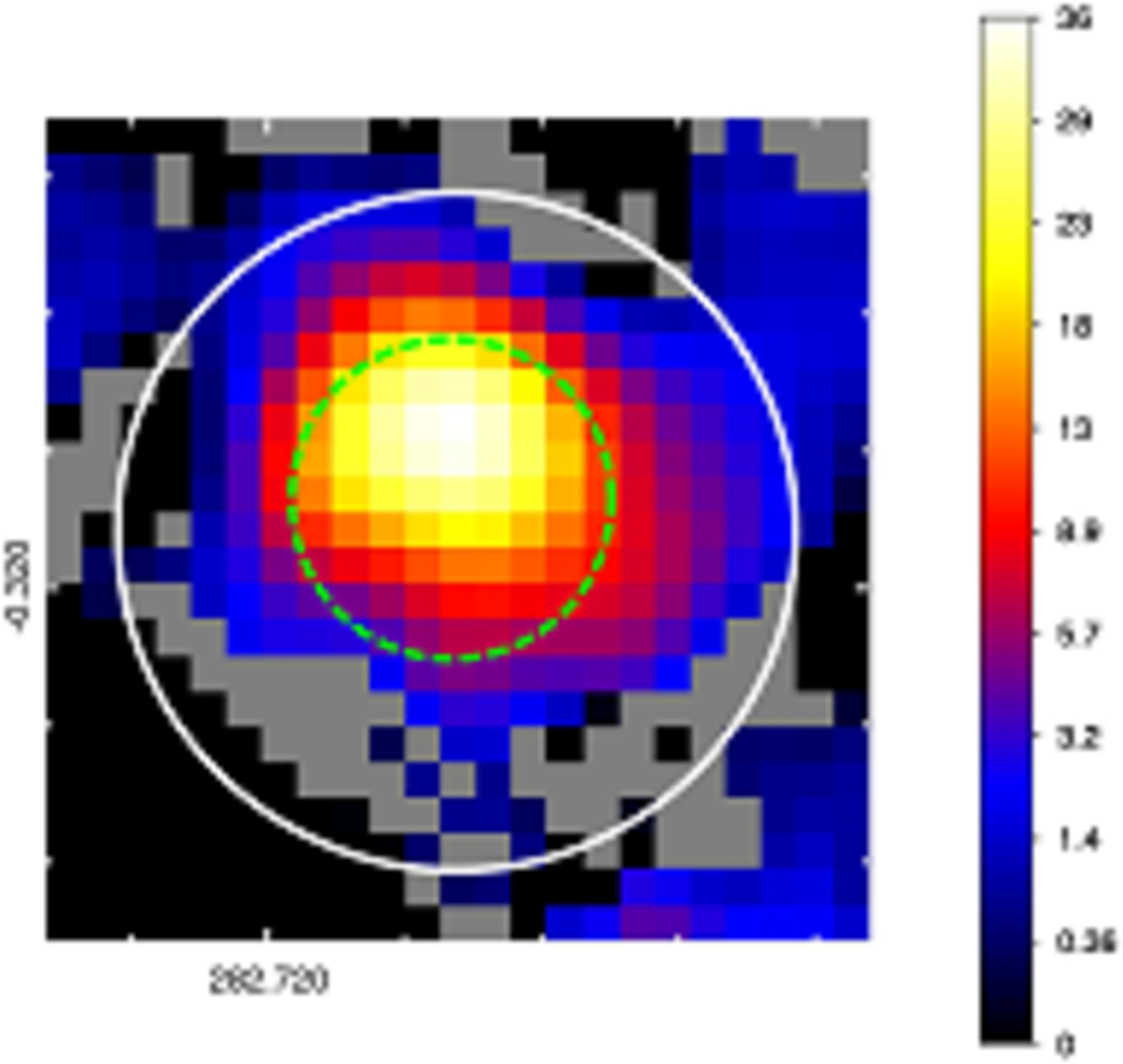}}}%
\mbox{\raisebox{0mm}{\includegraphics[width=40mm]{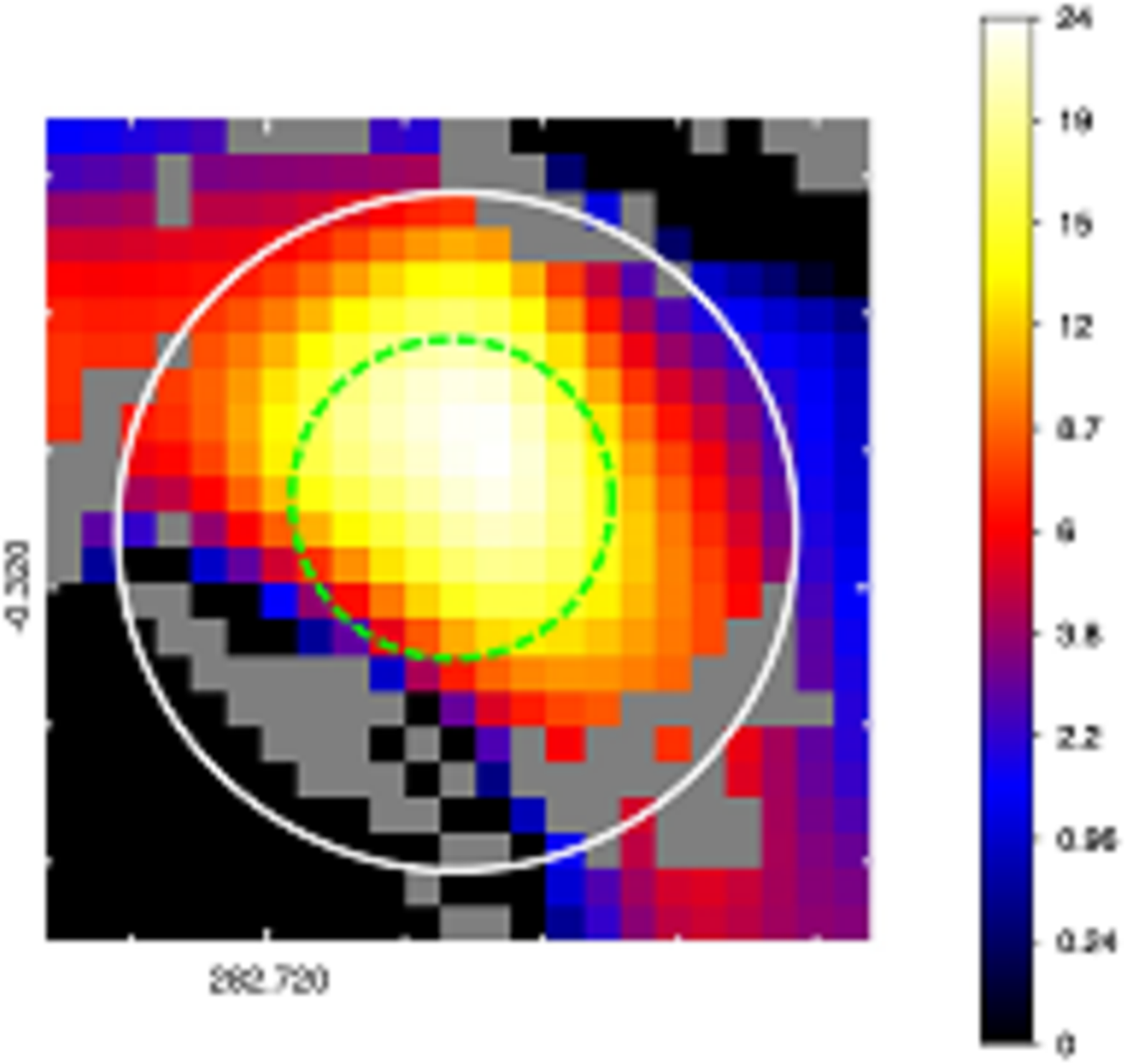}}}%
}
\subfigure{
\mbox{\raisebox{0mm}{\includegraphics[width=40mm]{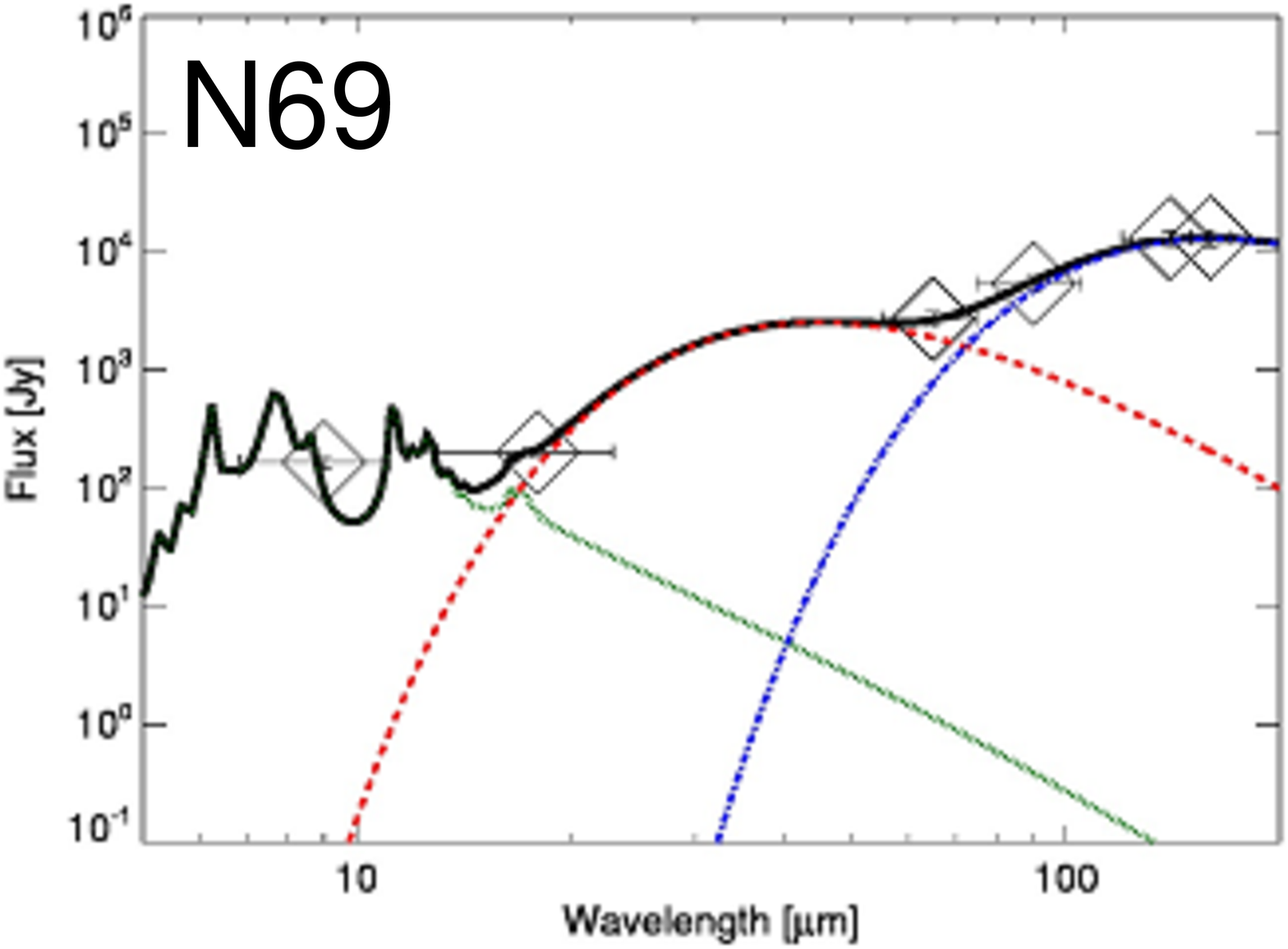}}}%
\mbox{\raisebox{6mm}{\rotatebox{90}{\small{DEC (J2000)}}}}%
\mbox{\raisebox{0mm}{\includegraphics[width=40mm]{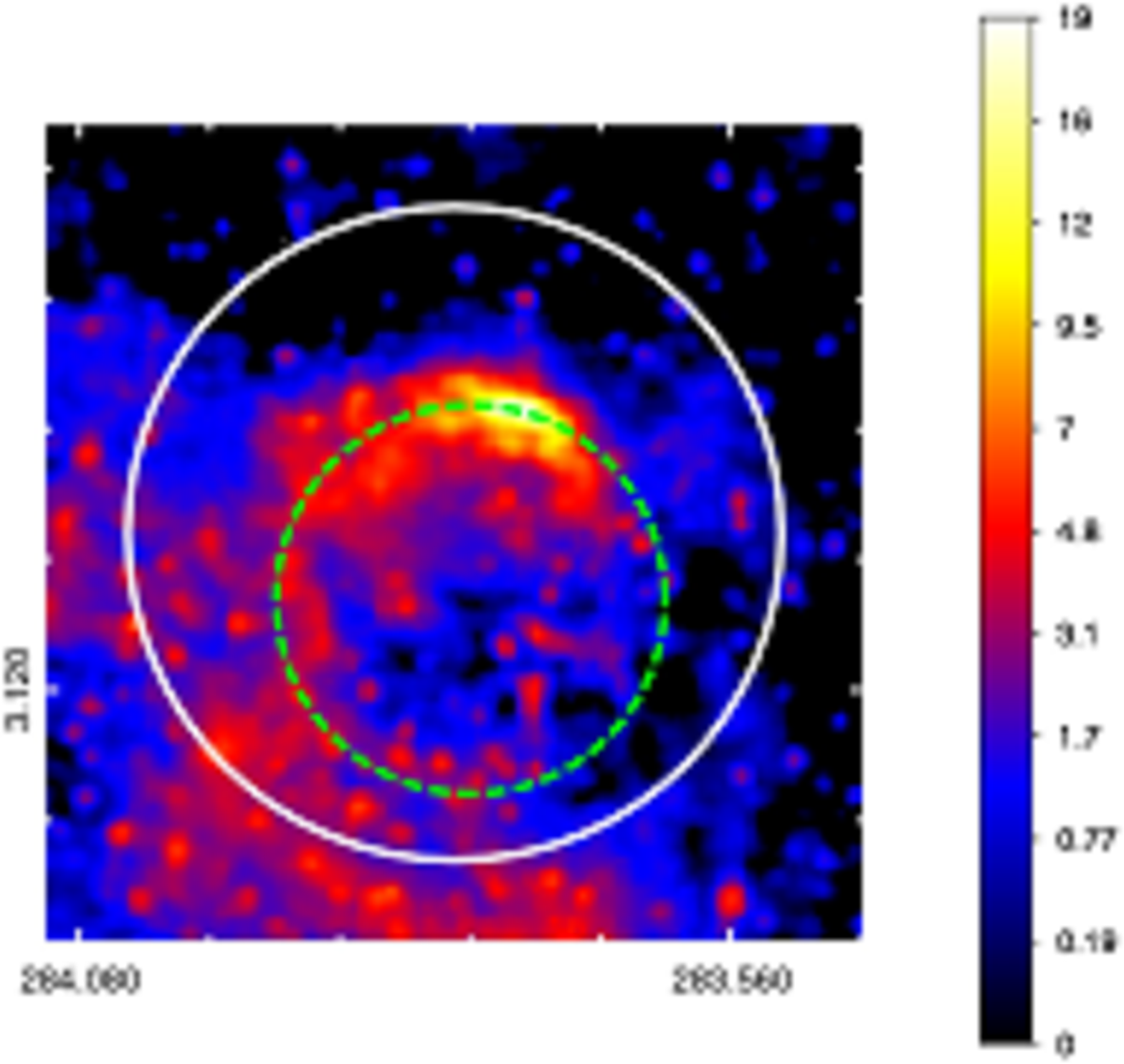}}}%
\mbox{\raisebox{0mm}{\includegraphics[width=40mm]{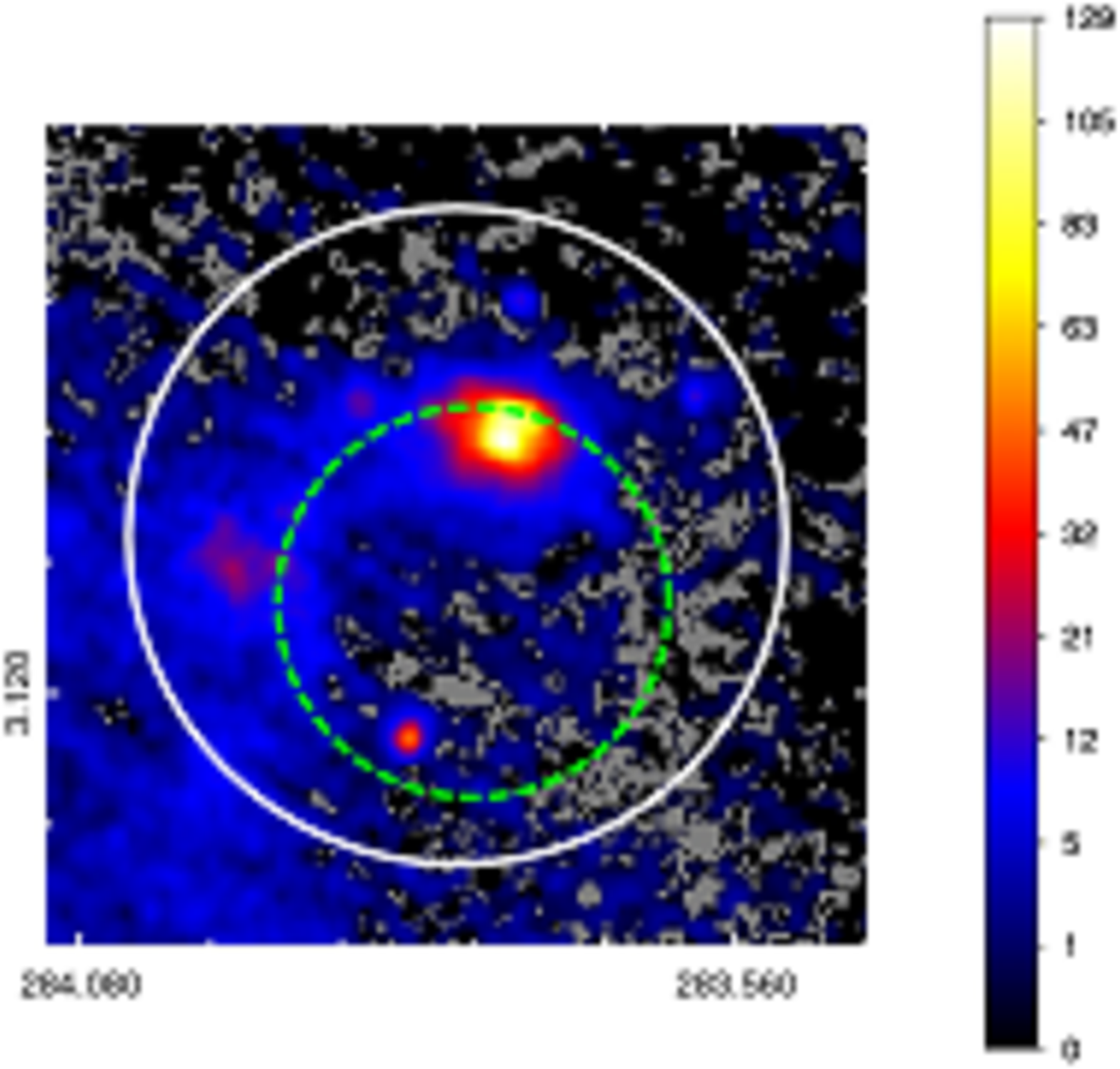}}}%
\mbox{\raisebox{0mm}{\includegraphics[width=40mm]{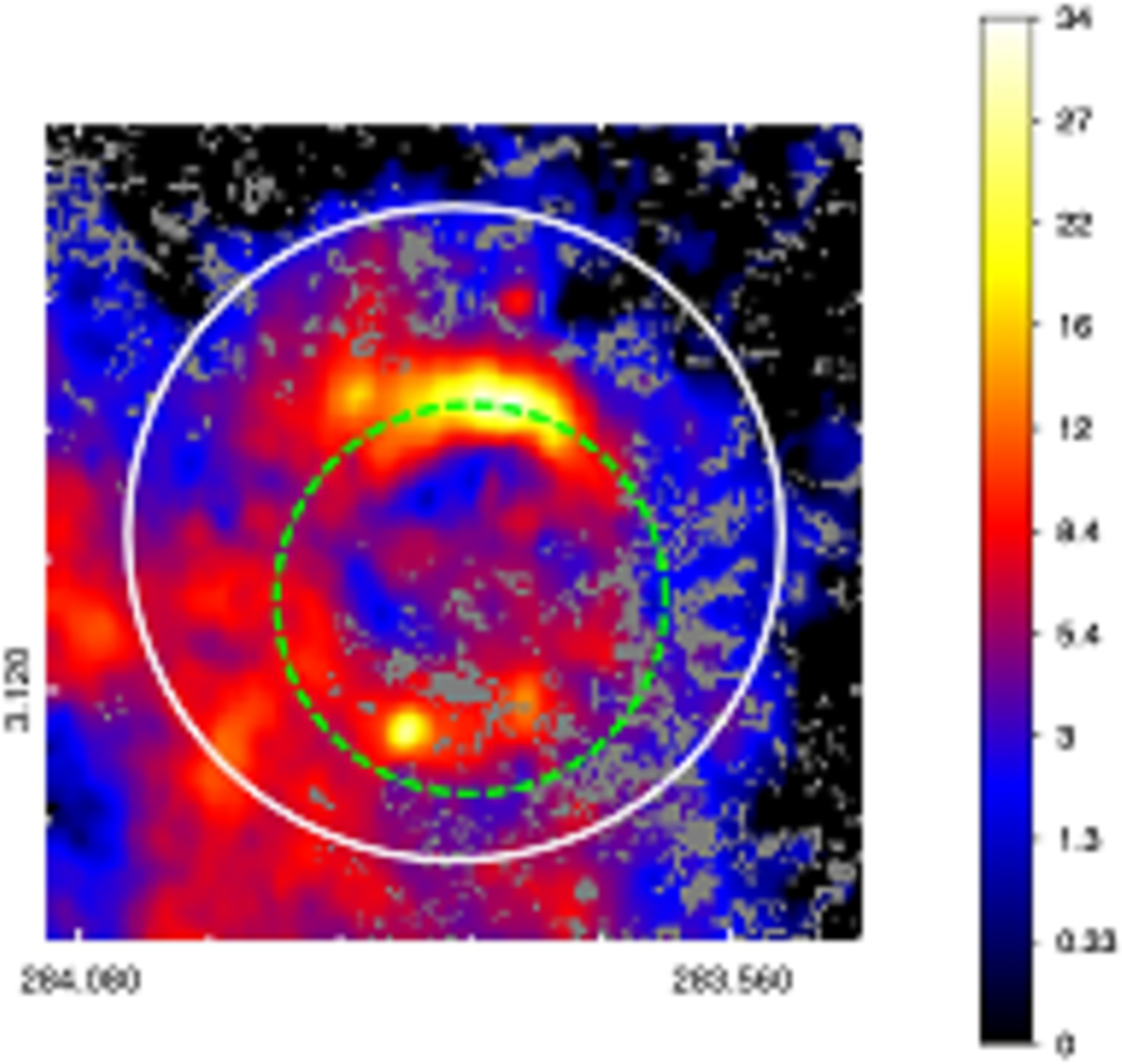}}}%
}
\subfigure{
\mbox{\raisebox{0mm}{\includegraphics[width=40mm]{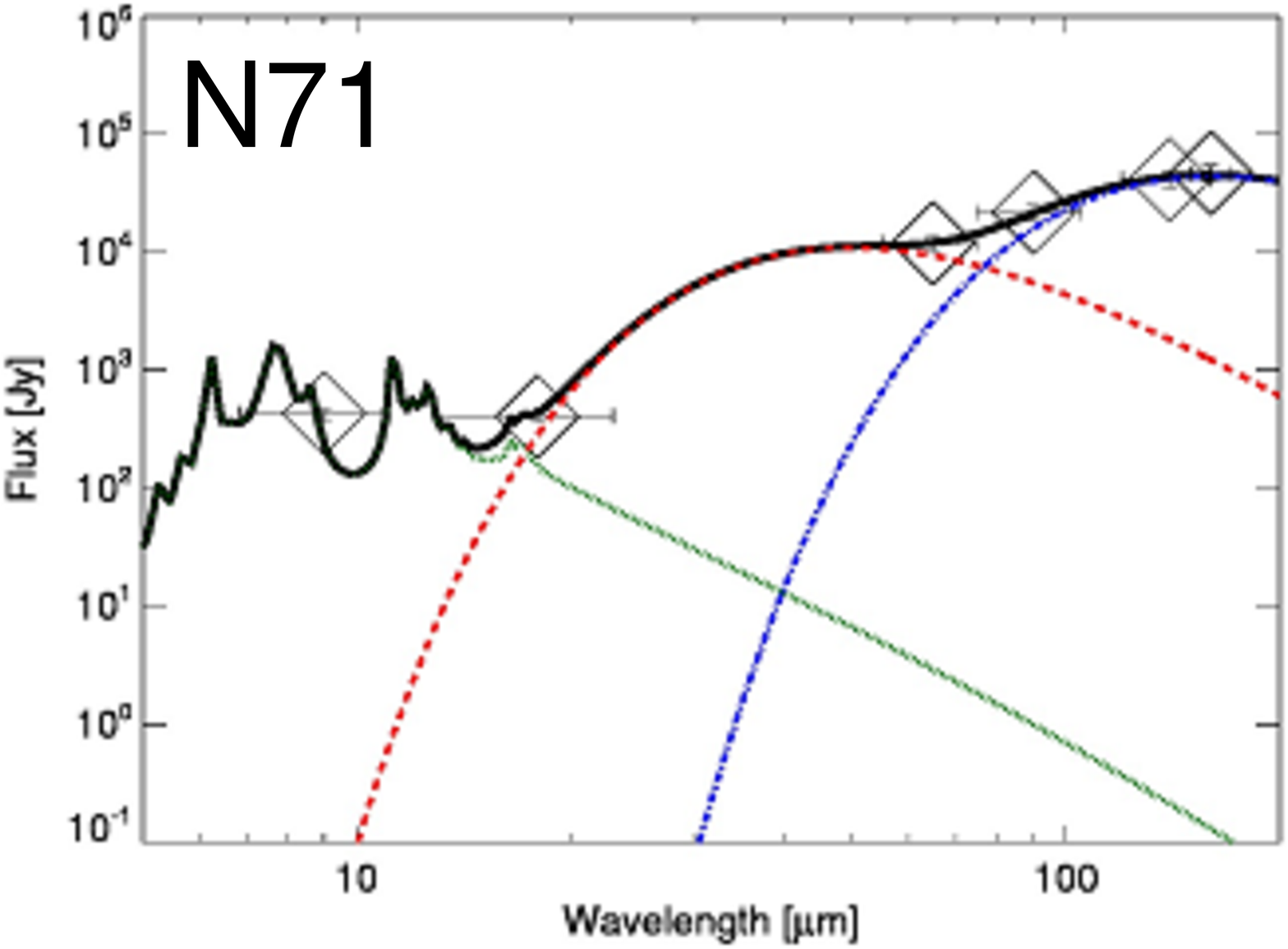}}}%
\mbox{\raisebox{6mm}{\rotatebox{90}{\small{DEC (J2000)}}}}%
\mbox{\raisebox{0mm}{\includegraphics[width=40mm]{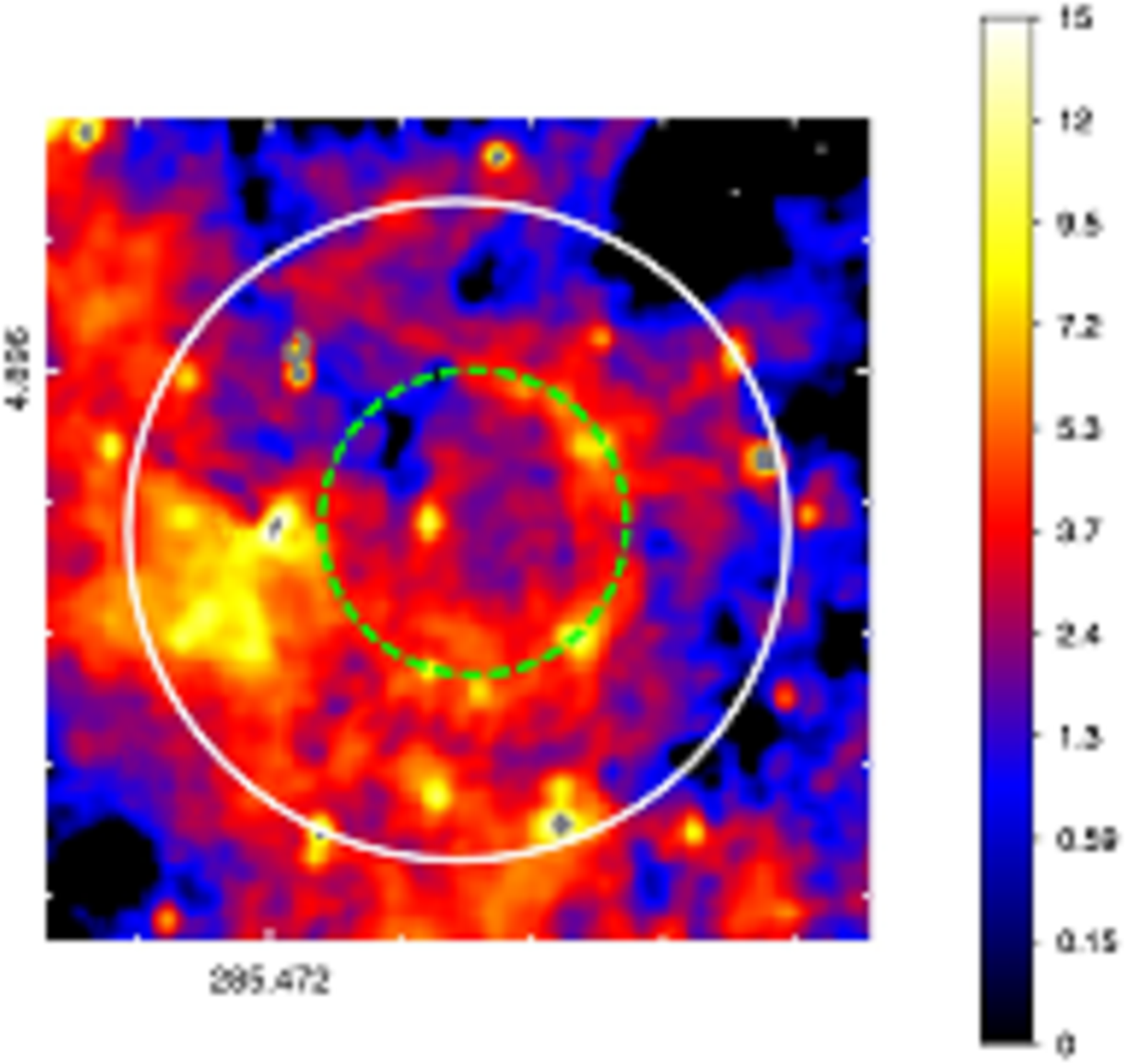}}}%
\mbox{\raisebox{0mm}{\includegraphics[width=40mm]{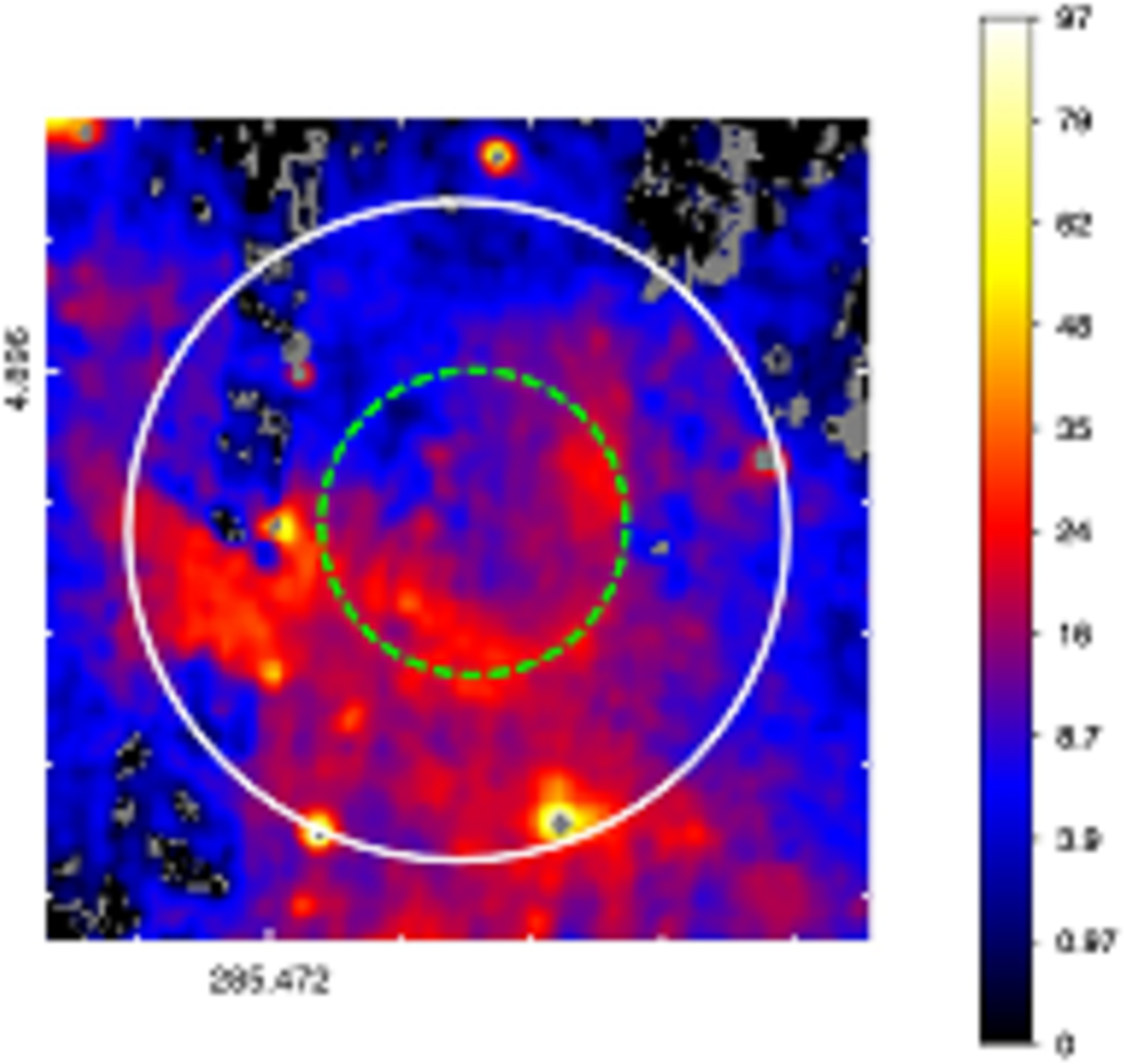}}}%
\mbox{\raisebox{0mm}{\includegraphics[width=40mm]{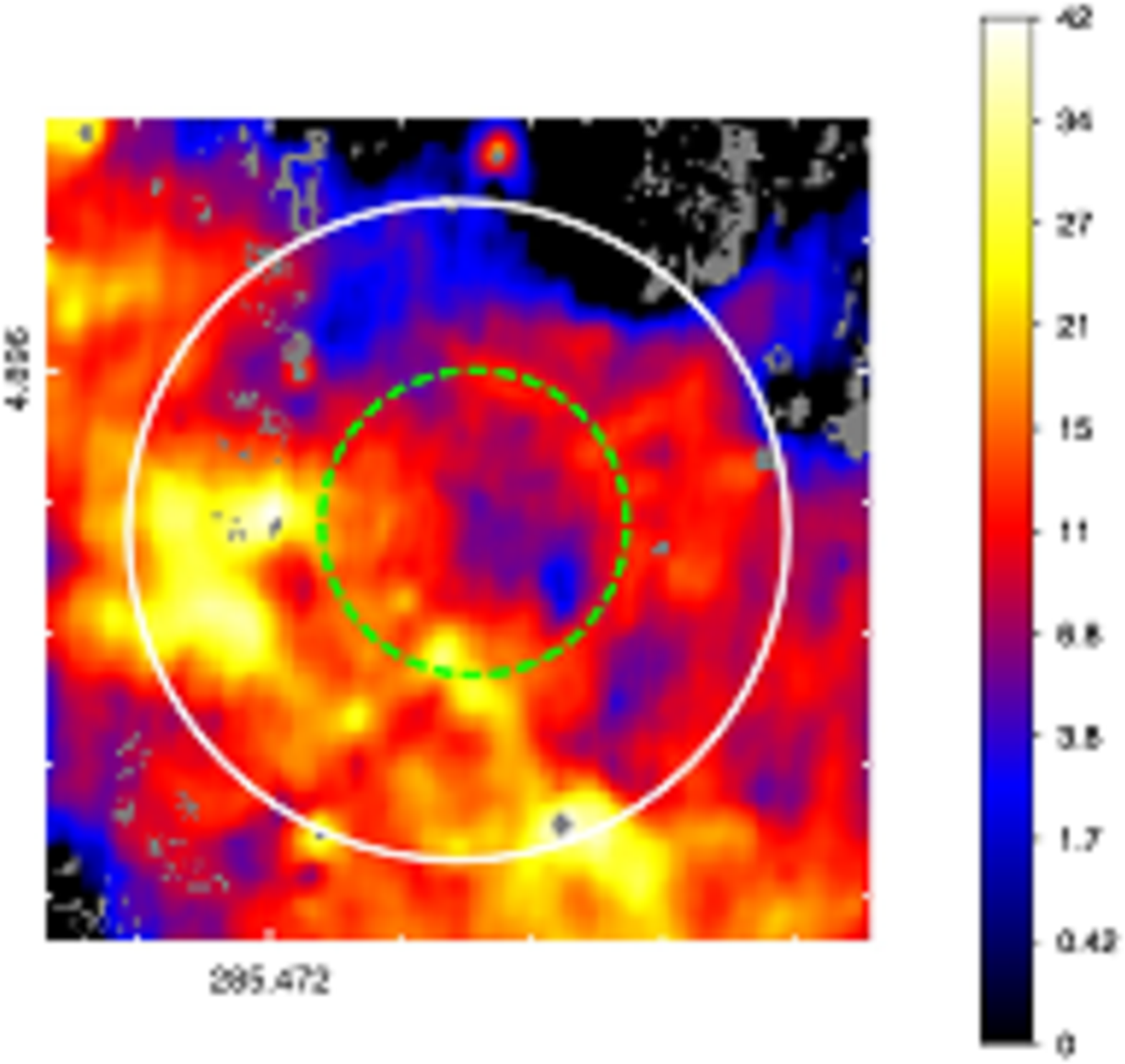}}}%
}
\caption{Continued.} \label{fig:Metfig2:c}
\end{figure*}

\addtocounter{figure}{-1}
\begin{figure*}[ht]
\addtocounter{subfigure}{1}
\centering
\subfigure{
\makebox[180mm][l]{\raisebox{0mm}[0mm][0mm]{ \hspace{20mm} \small{SED}} \hspace{27.5mm} \small{$I_{\rm{PAH}}$} \hspace{29.5mm} \small{$I_{\rm{warm}}$} \hspace{29.5mm} \small{$I_{\rm{cold}}$}}%
}
\subfigure{
\makebox[180mm][l]{\raisebox{0mm}{\hspace{52mm} \small{RA (J2000)} \hspace{20mm} \small{RA (J2000)} \hspace{20mm} \small{RA (J2000)}}}
}
\subfigure{
\mbox{\raisebox{0mm}{\includegraphics[width=40mm]{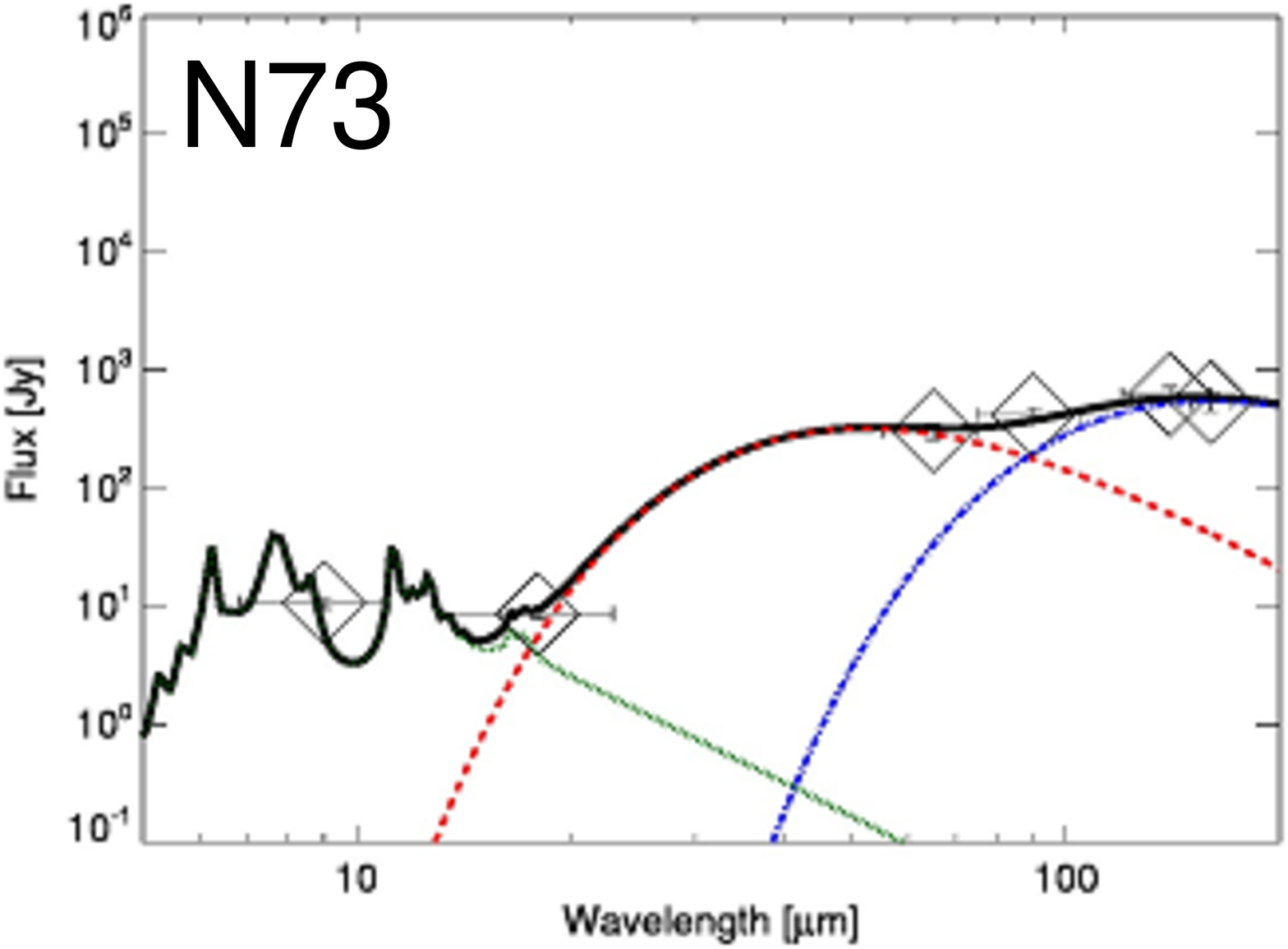}}}%
\mbox{\raisebox{6mm}{\rotatebox{90}{\small{DEC (J2000)}}}}%
\mbox{\raisebox{0mm}{\includegraphics[width=40mm]{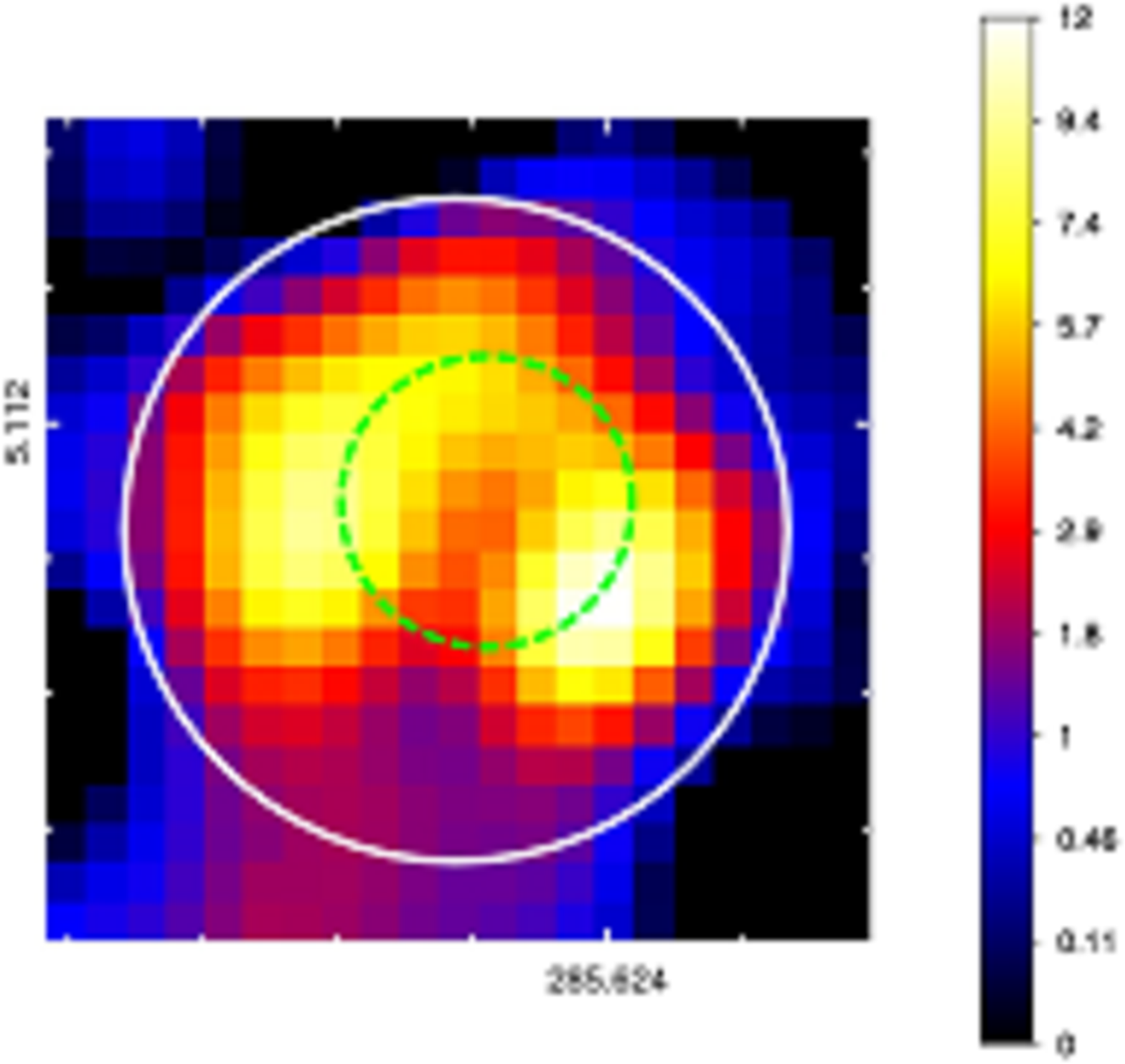}}}%
\mbox{\raisebox{0mm}{\includegraphics[width=40mm]{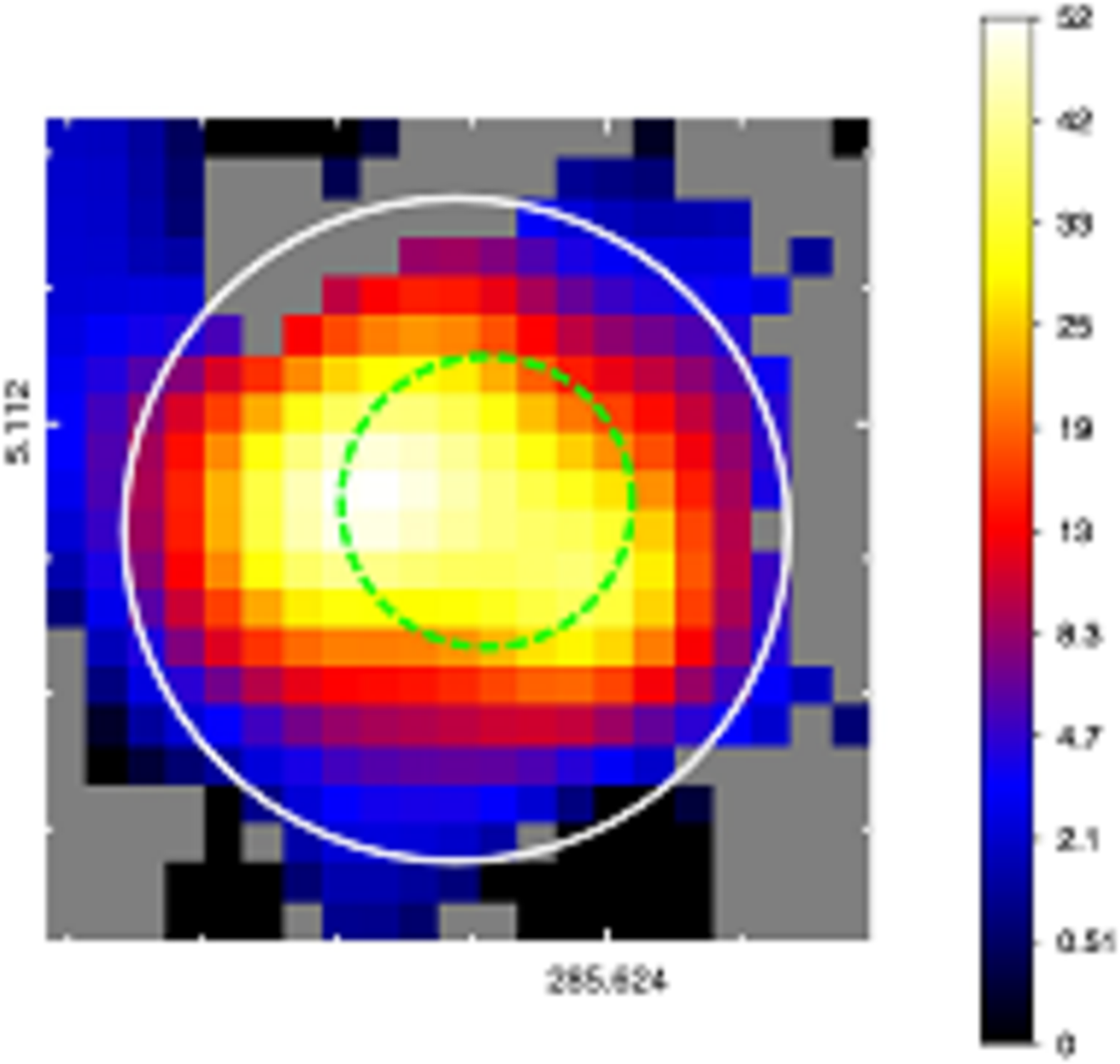}}}%
\mbox{\raisebox{0mm}{\includegraphics[width=40mm]{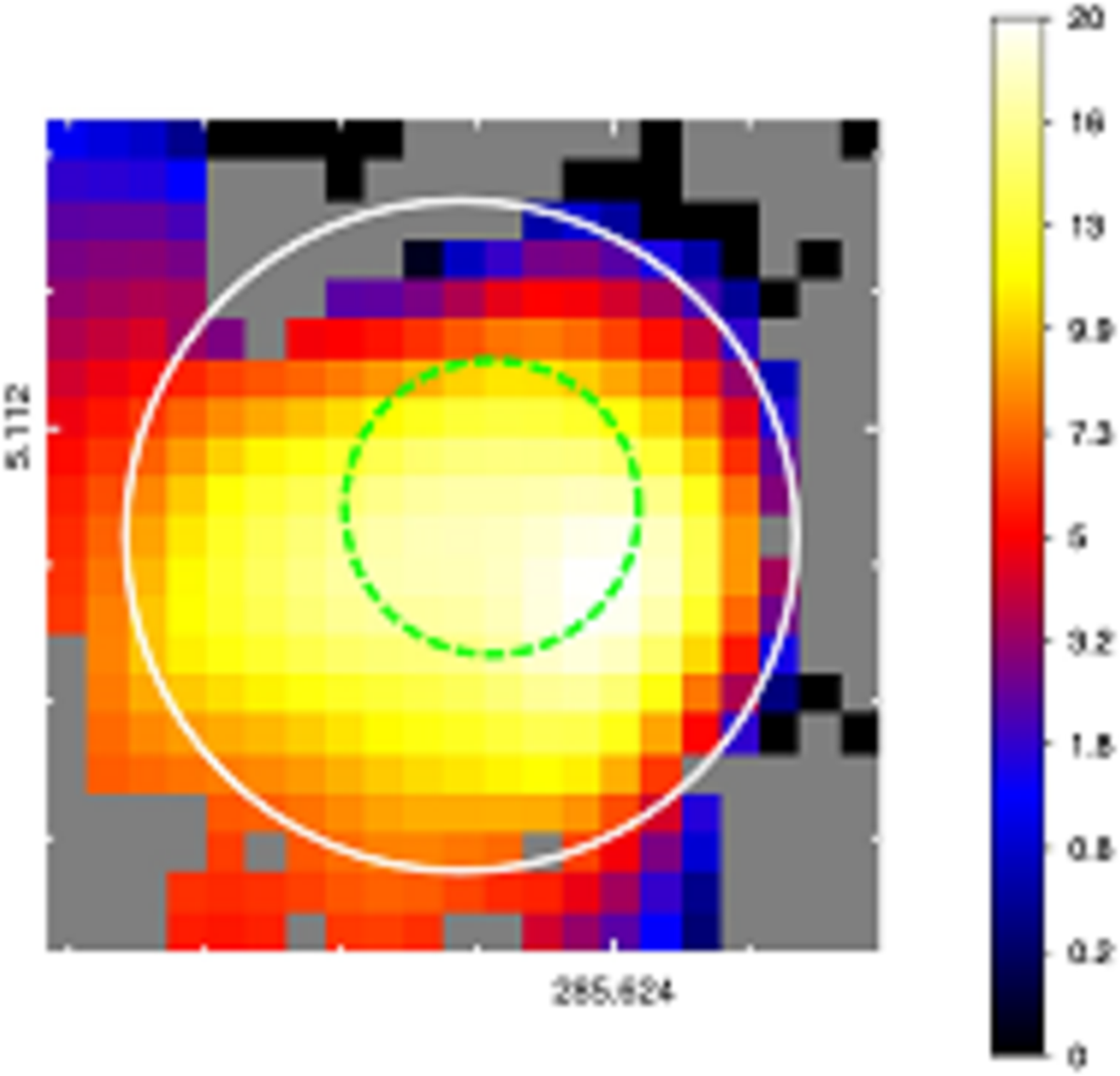}}}%
}
\subfigure{
\mbox{\raisebox{0mm}{\includegraphics[width=40mm]{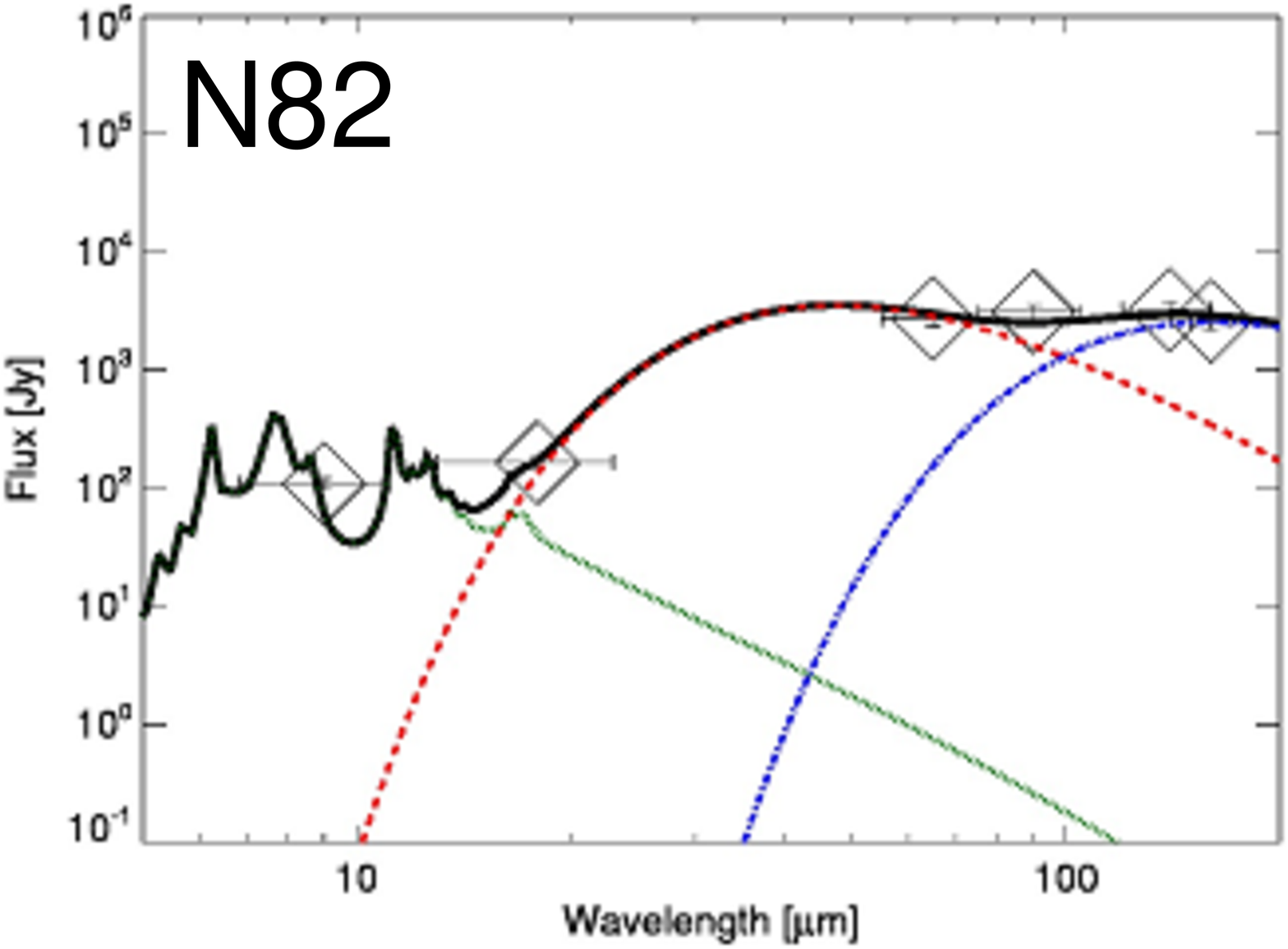}}}%
\mbox{\raisebox{6mm}{\rotatebox{90}{\small{DEC (J2000)}}}}%
\mbox{\raisebox{0mm}{\includegraphics[width=40mm]{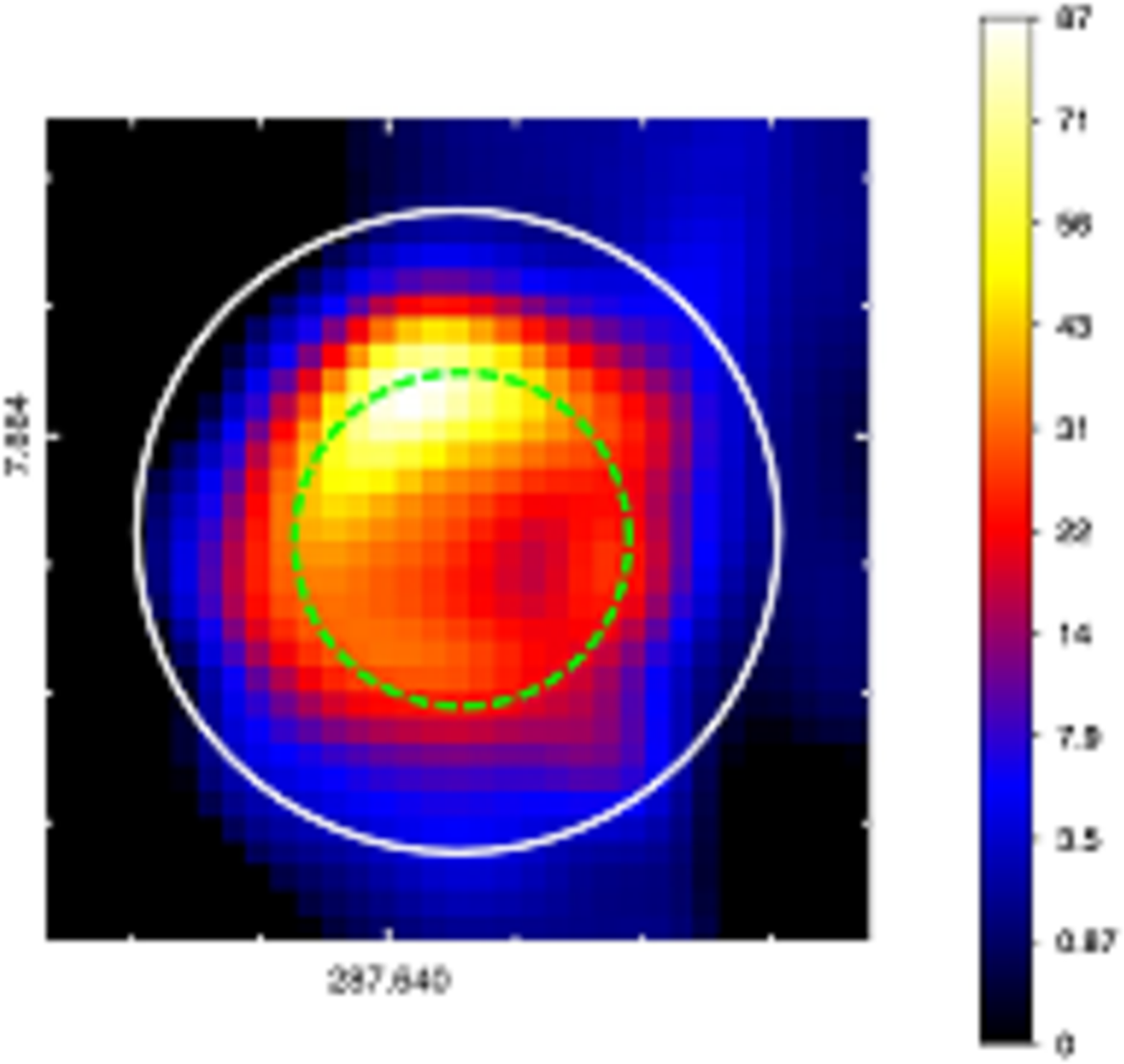}}}%
\mbox{\raisebox{0mm}{\includegraphics[width=40mm]{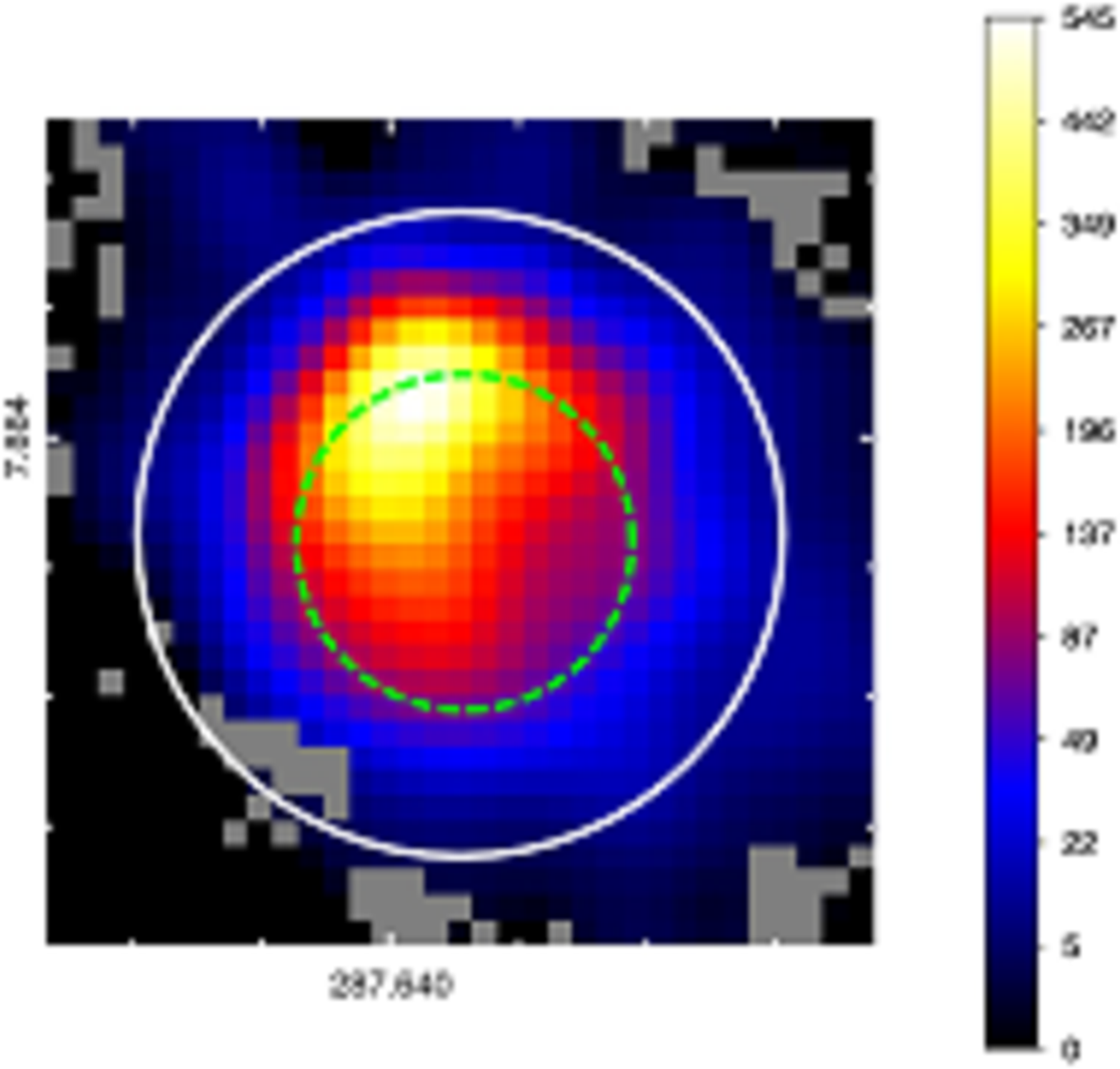}}}%
\mbox{\raisebox{0mm}{\includegraphics[width=40mm]{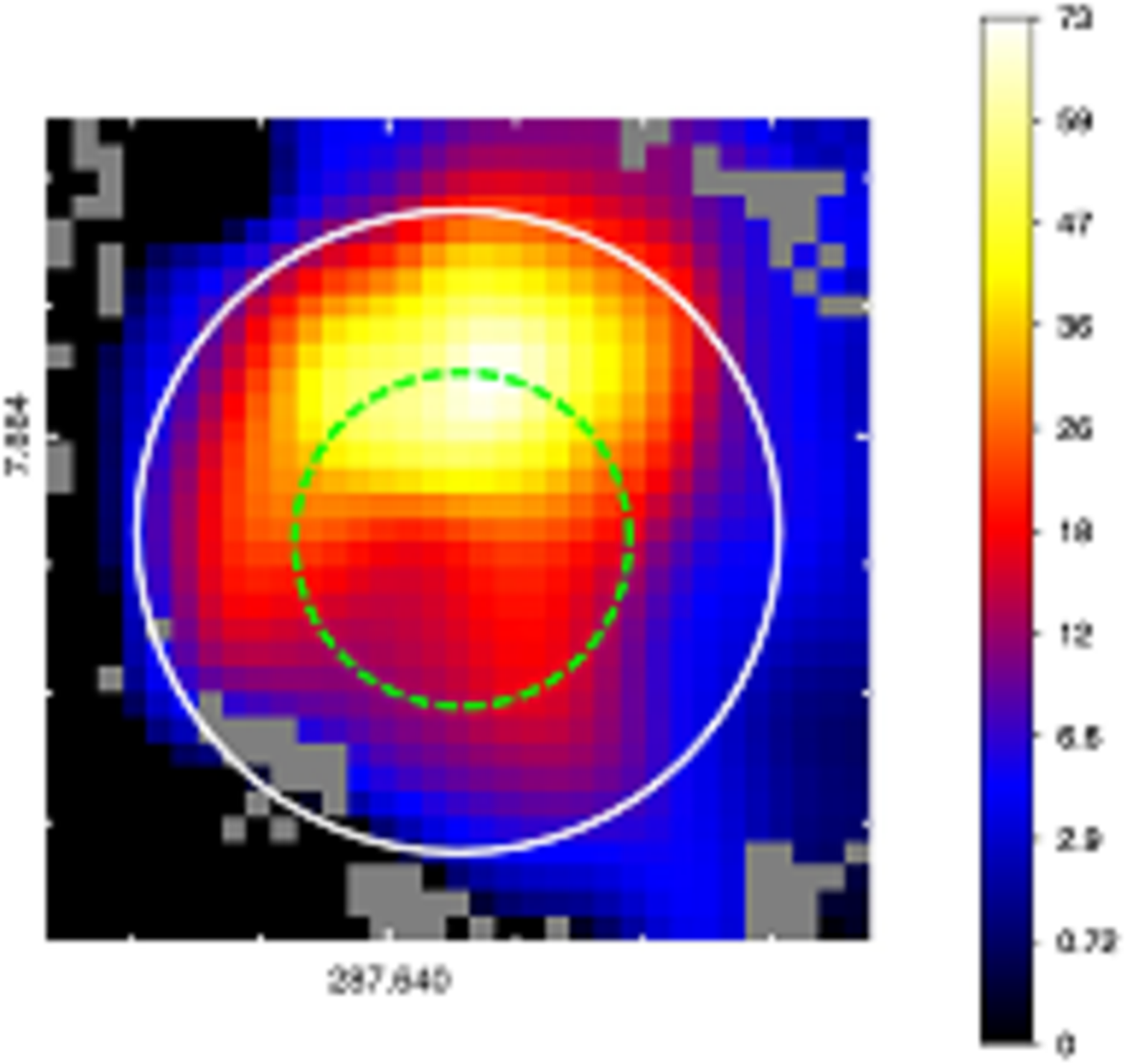}}}%
}
\subfigure{
\mbox{\raisebox{0mm}{\includegraphics[width=40mm]{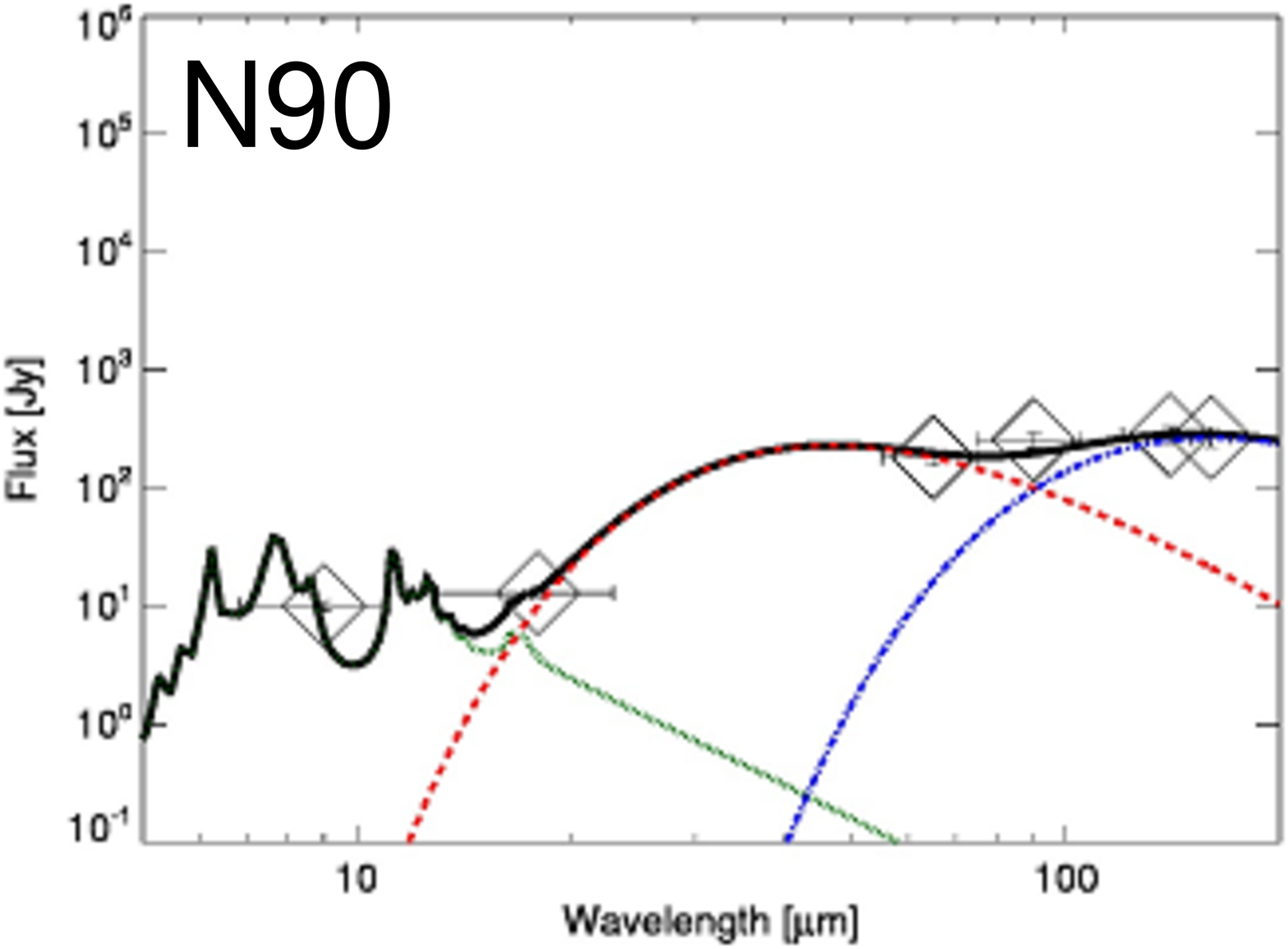}}}%
\mbox{\raisebox{6mm}{\rotatebox{90}{\small{DEC (J2000)}}}}%
\mbox{\raisebox{0mm}{\includegraphics[width=40mm]{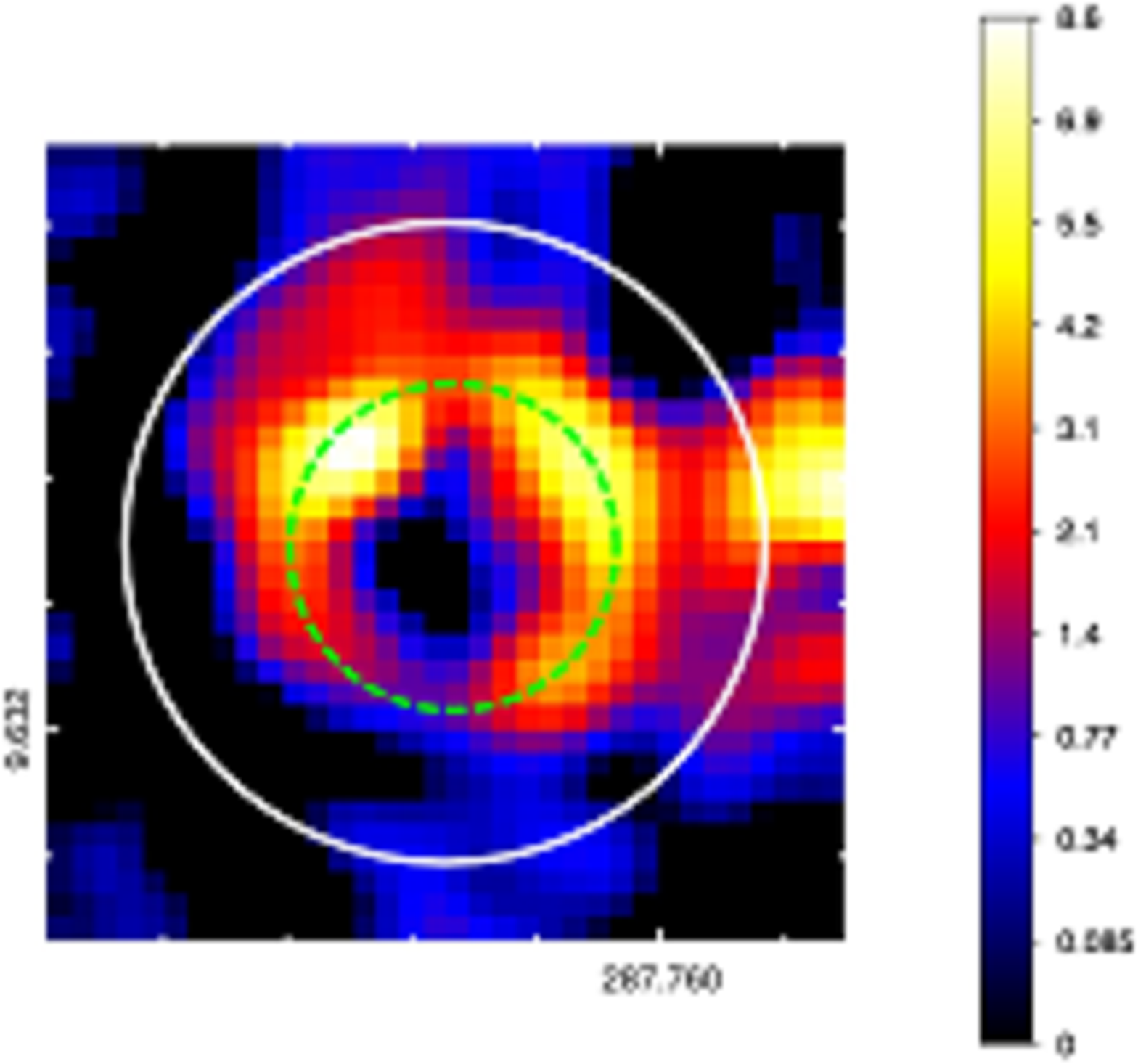}}}%
\mbox{\raisebox{0mm}{\includegraphics[width=40mm]{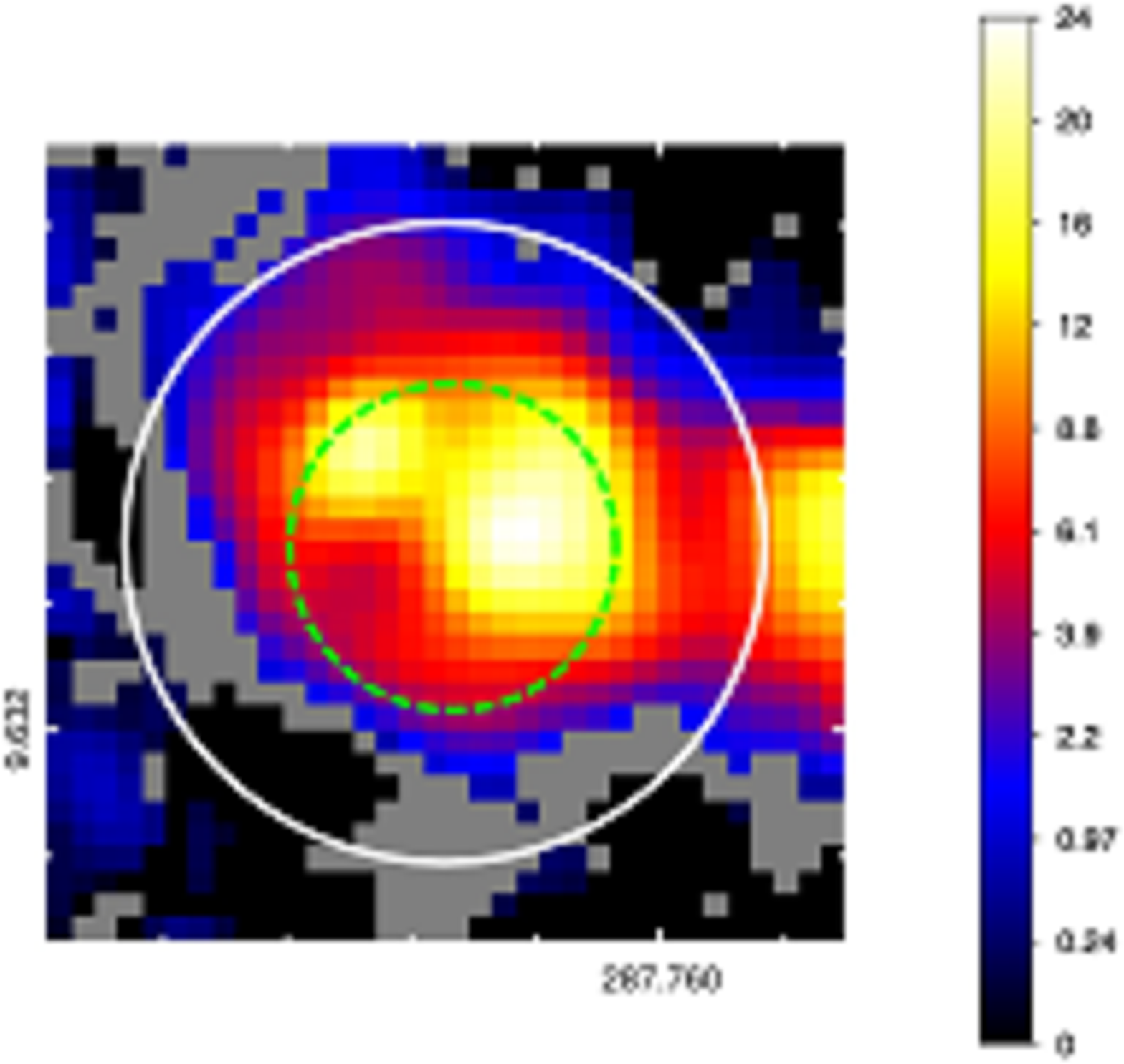}}}%
\mbox{\raisebox{0mm}{\includegraphics[width=40mm]{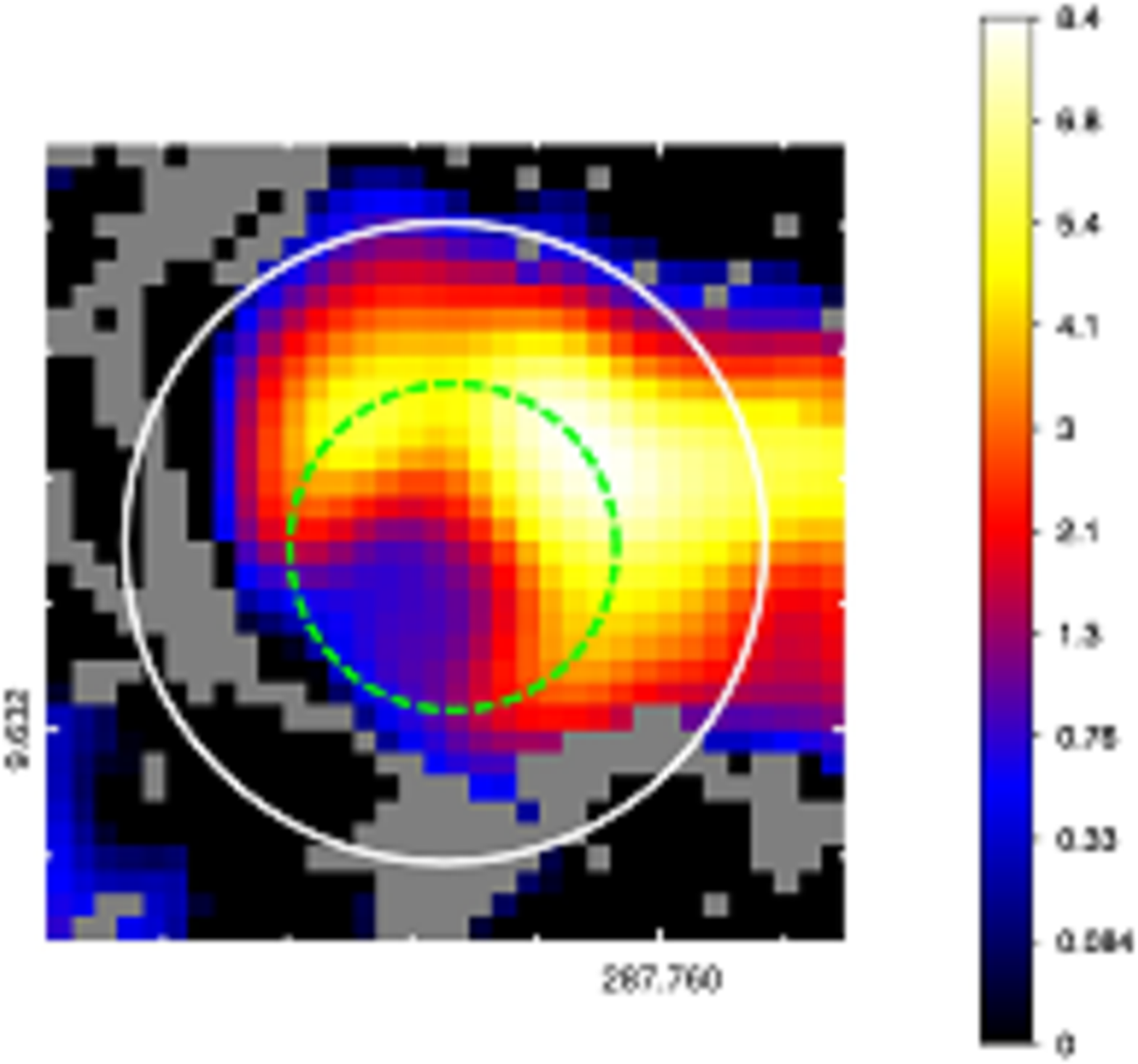}}}%
}
\subfigure{
\mbox{\raisebox{0mm}{\includegraphics[width=40mm]{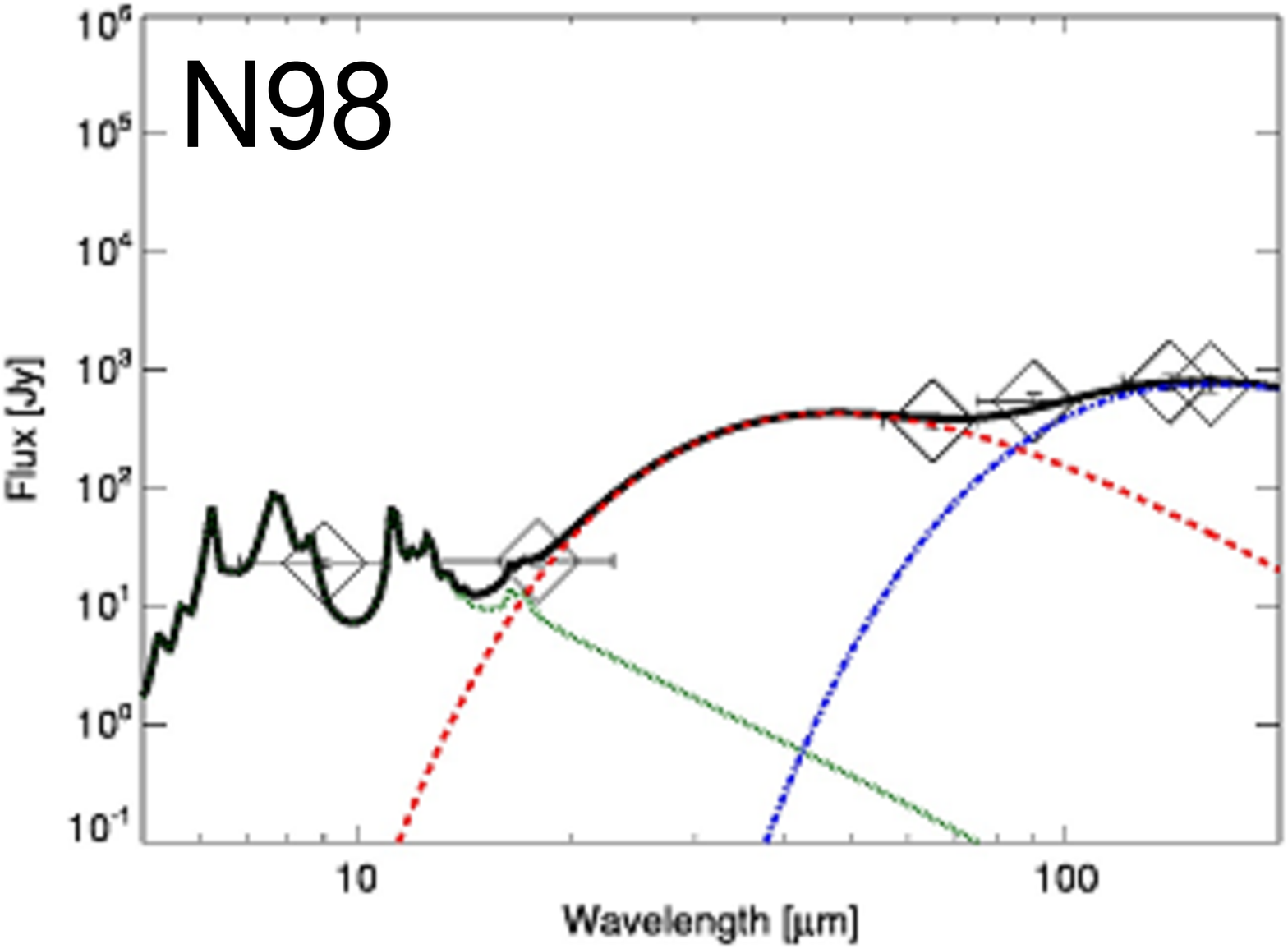}}}%
\mbox{\raisebox{6mm}{\rotatebox{90}{\small{DEC (J2000)}}}}%
\mbox{\raisebox{0mm}{\includegraphics[width=40mm]{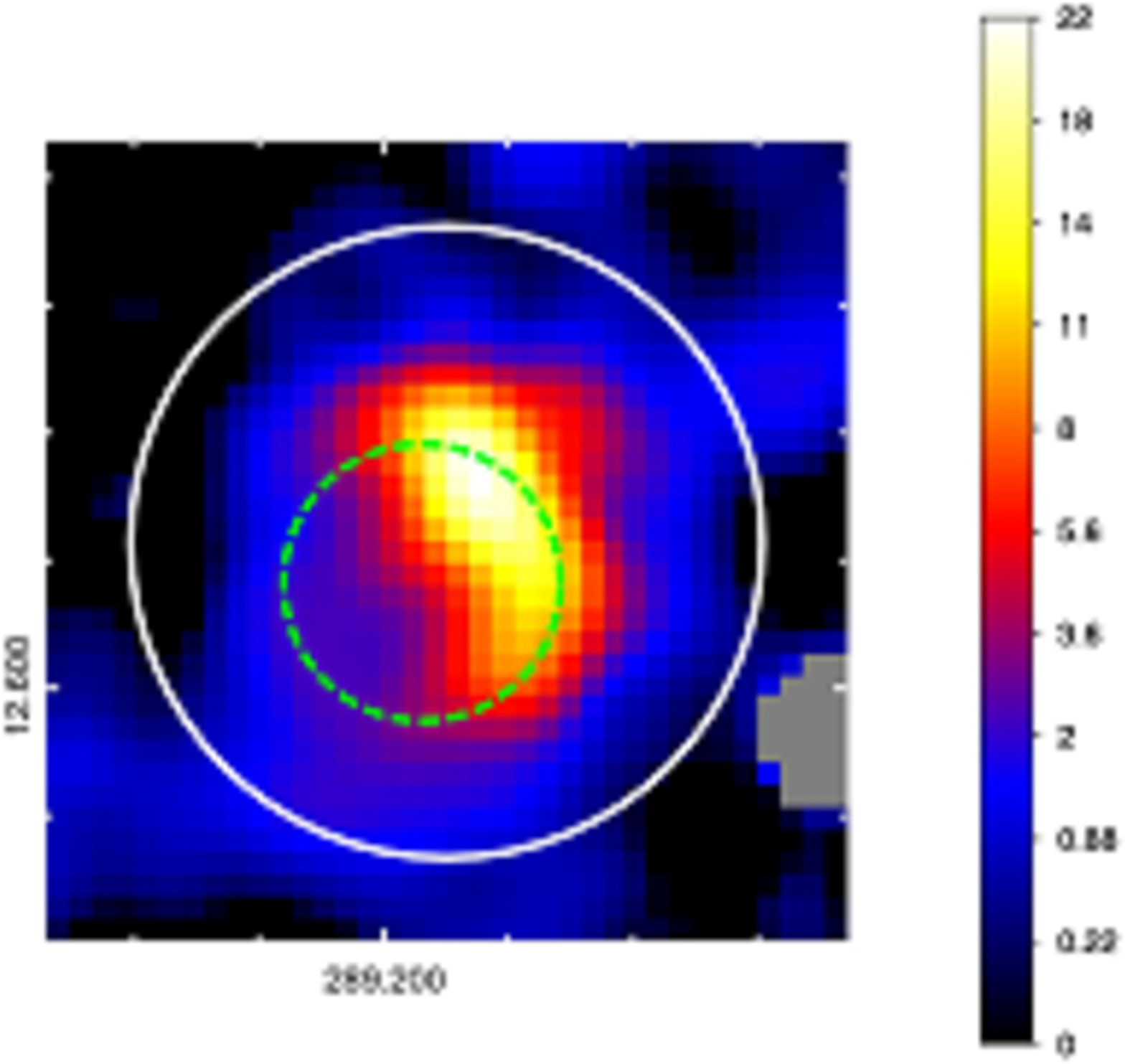}}}%
\mbox{\raisebox{0mm}{\includegraphics[width=40mm]{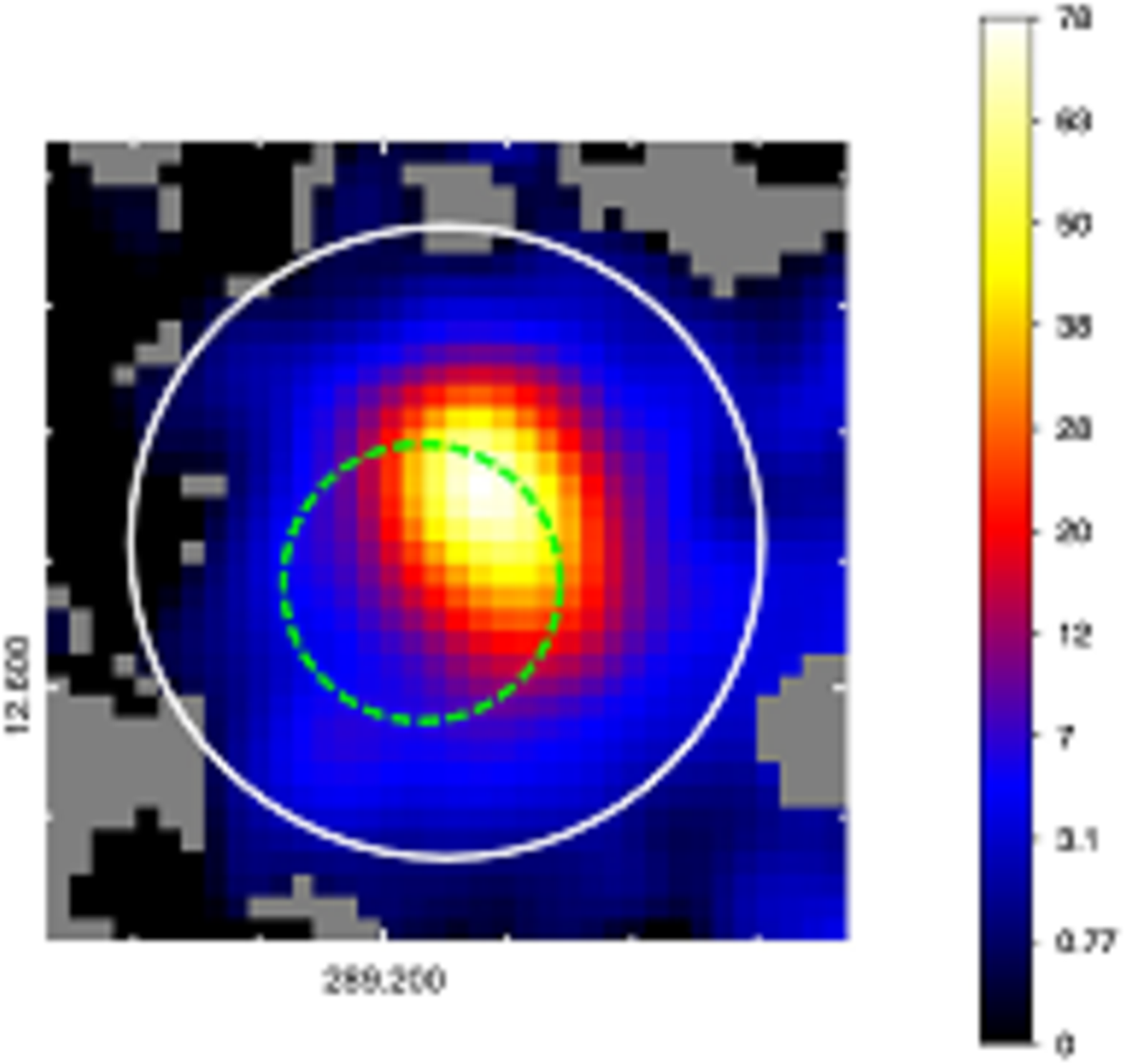}}}%
\mbox{\raisebox{0mm}{\includegraphics[width=40mm]{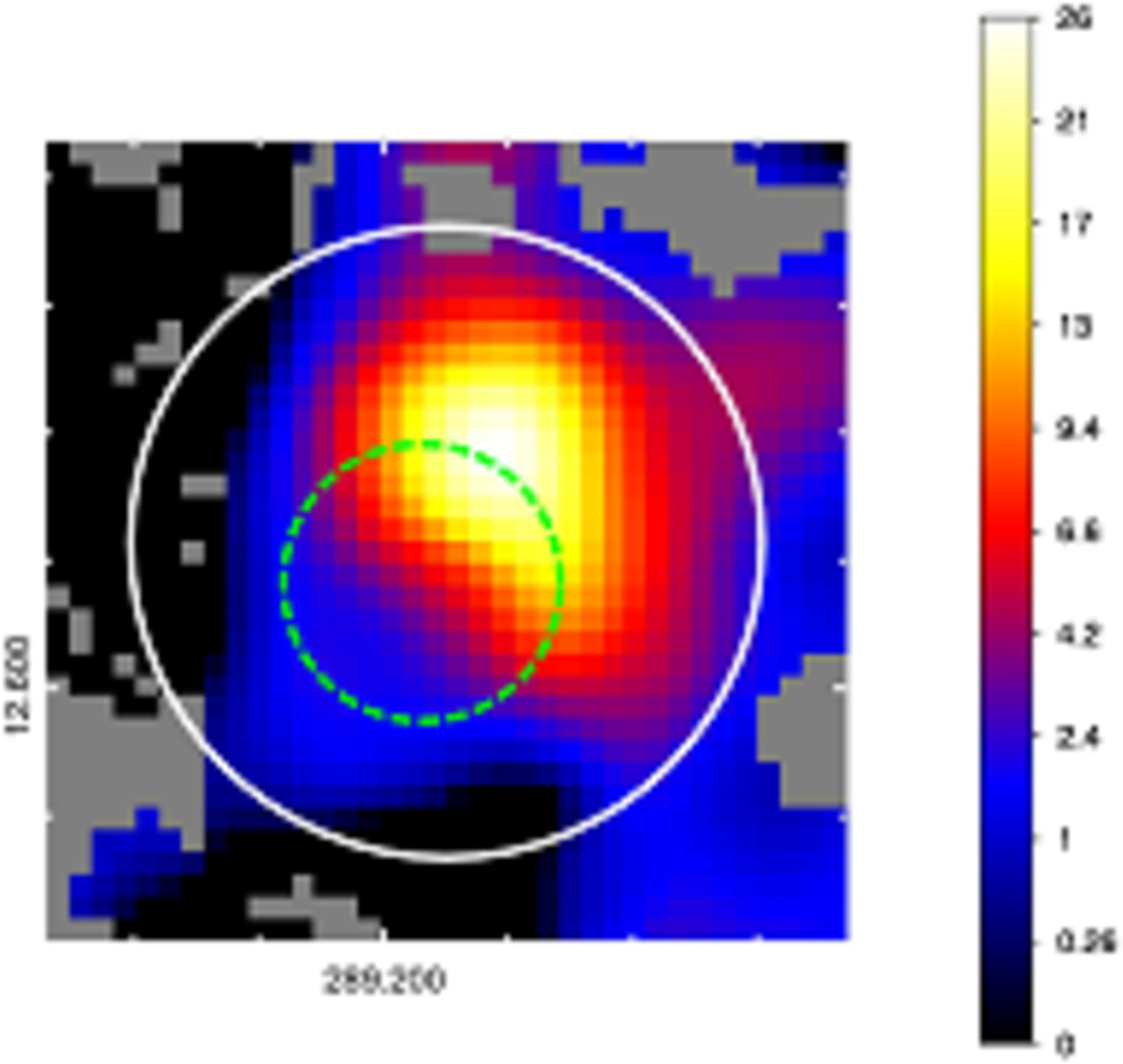}}}%
}
\subfigure{
\mbox{\raisebox{0mm}{\includegraphics[width=40mm]{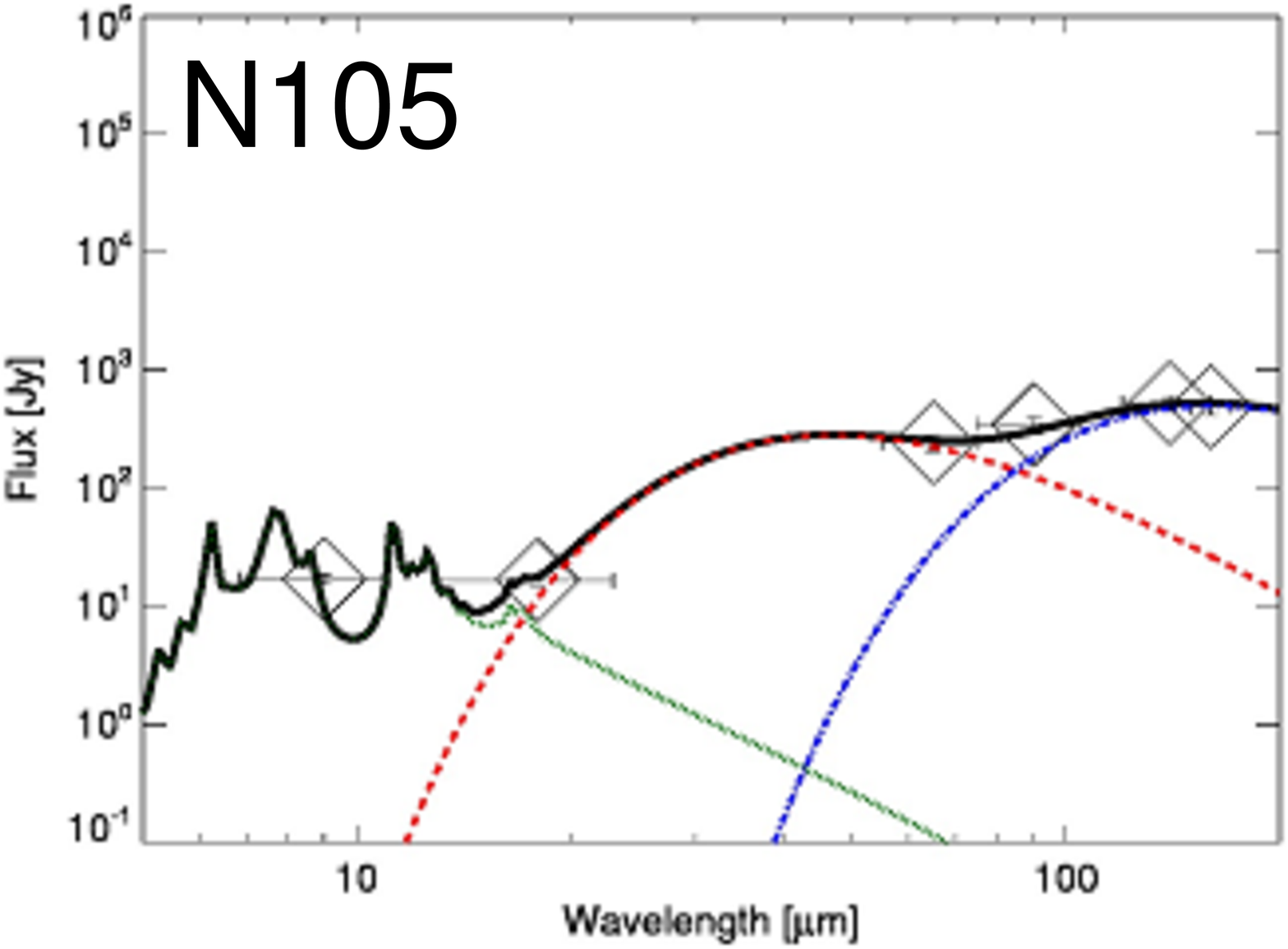}}}%
\mbox{\raisebox{6mm}{\rotatebox{90}{\small{DEC (J2000)}}}}%
\mbox{\raisebox{0mm}{\includegraphics[width=40mm]{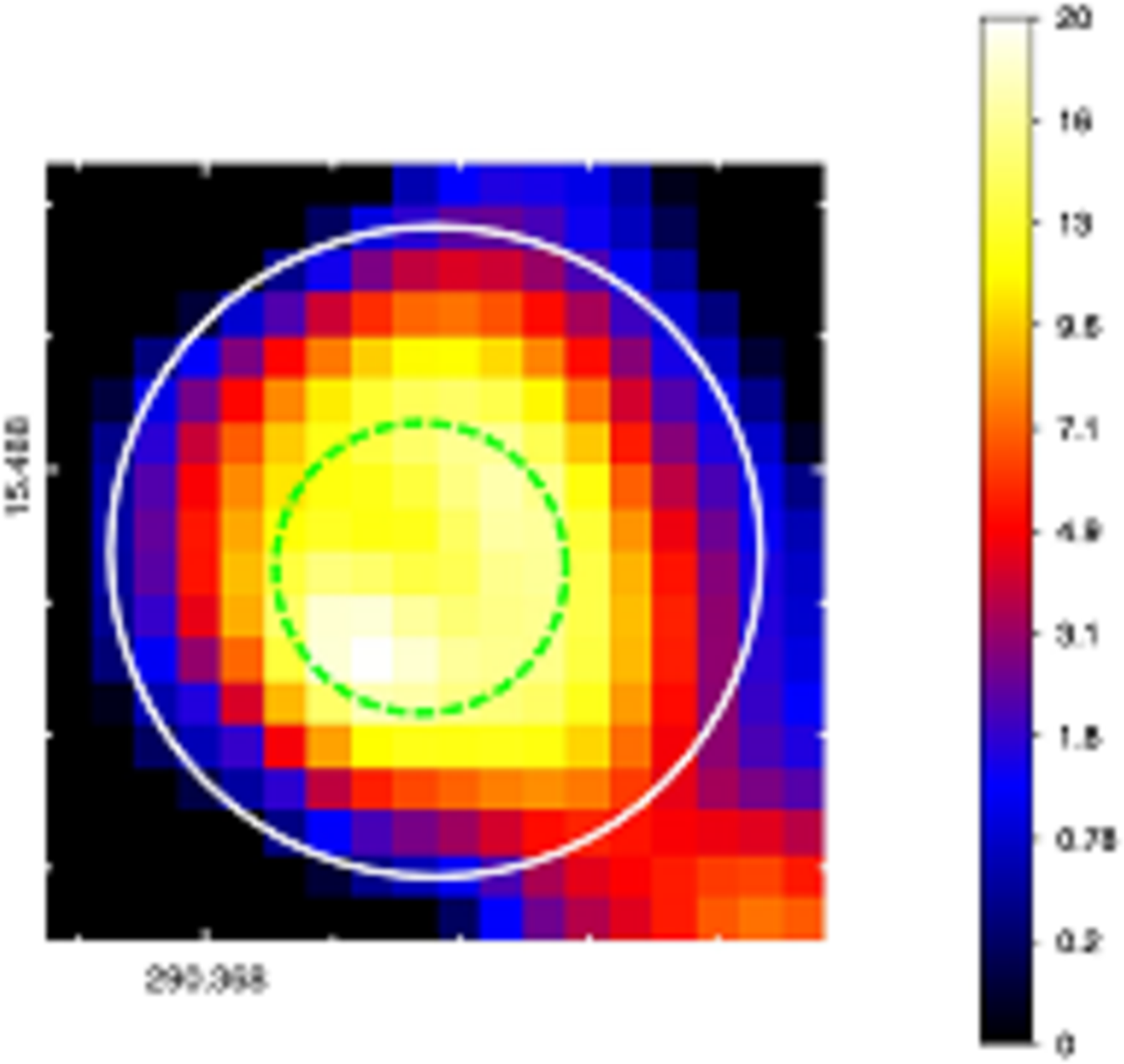}}}%
\mbox{\raisebox{0mm}{\includegraphics[width=40mm]{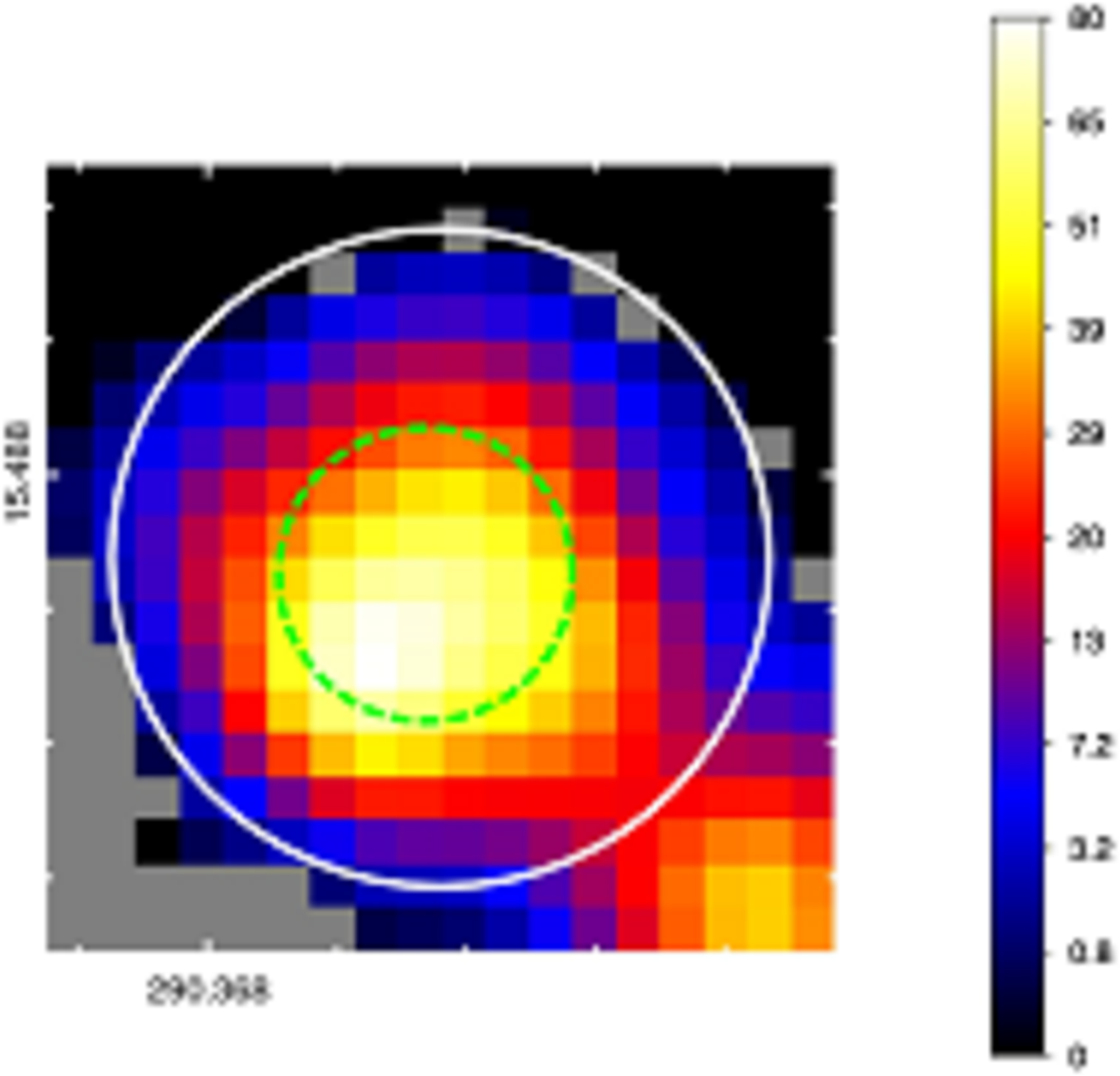}}}%
\mbox{\raisebox{0mm}{\includegraphics[width=40mm]{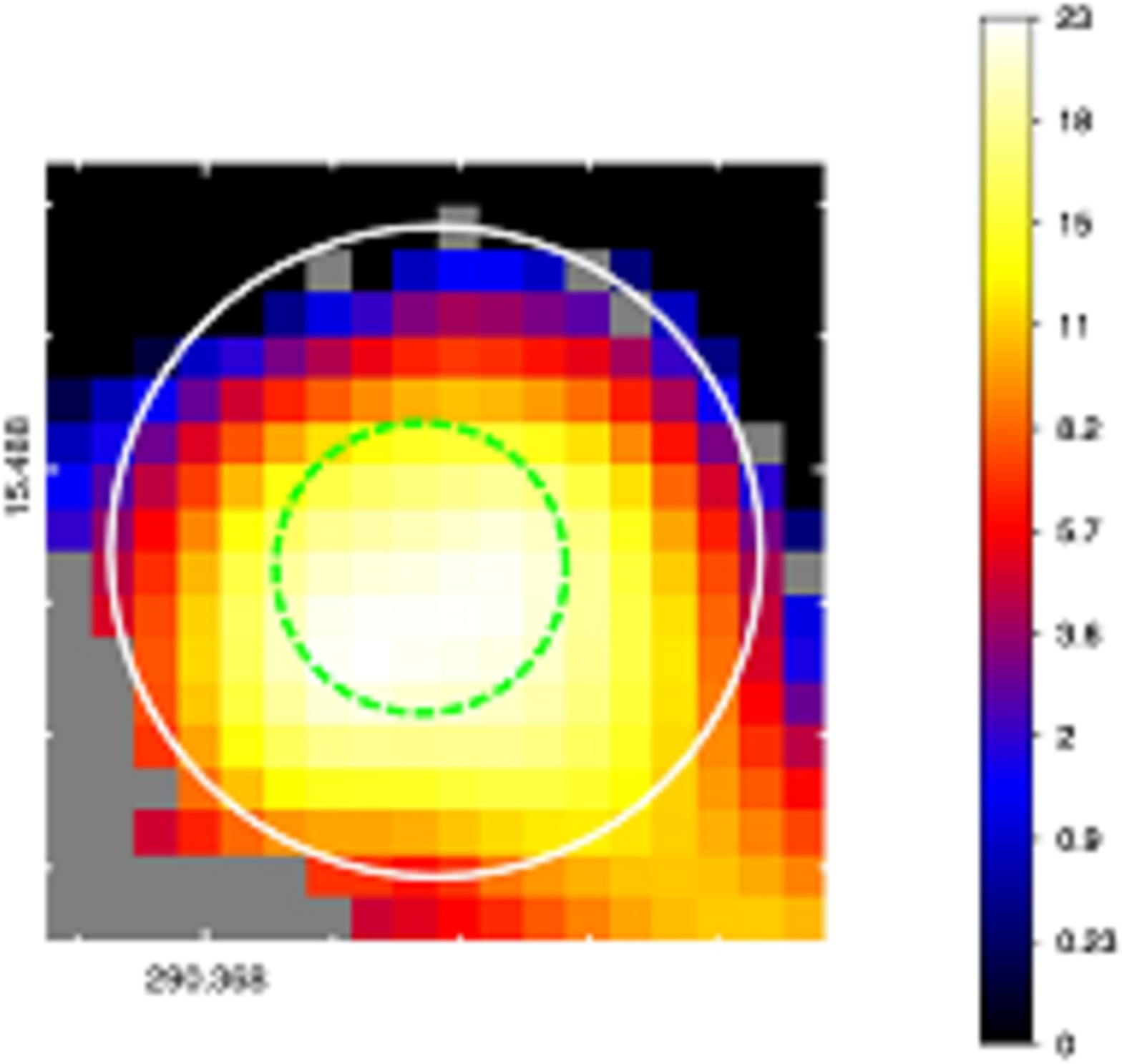}}}%
}
\caption{Continued.} \label{fig:Metfig2:d}
\end{figure*}

\addtocounter{figure}{-1}
\begin{figure*}[ht]
\addtocounter{subfigure}{1}
\centering
\subfigure{
\makebox[180mm][l]{\raisebox{0mm}[0mm][0mm]{ \hspace{20mm} \small{SED}} \hspace{27.5mm} \small{$I_{\rm{PAH}}$} \hspace{29.5mm} \small{$I_{\rm{warm}}$} \hspace{29.5mm} \small{$I_{\rm{cold}}$}}%
}
\subfigure{
\makebox[180mm][l]{\raisebox{0mm}{\hspace{52mm} \small{RA (J2000)} \hspace{20mm} \small{RA (J2000)} \hspace{20mm} \small{RA (J2000)}}}
}
\subfigure{
\mbox{\raisebox{0mm}{\includegraphics[width=40mm]{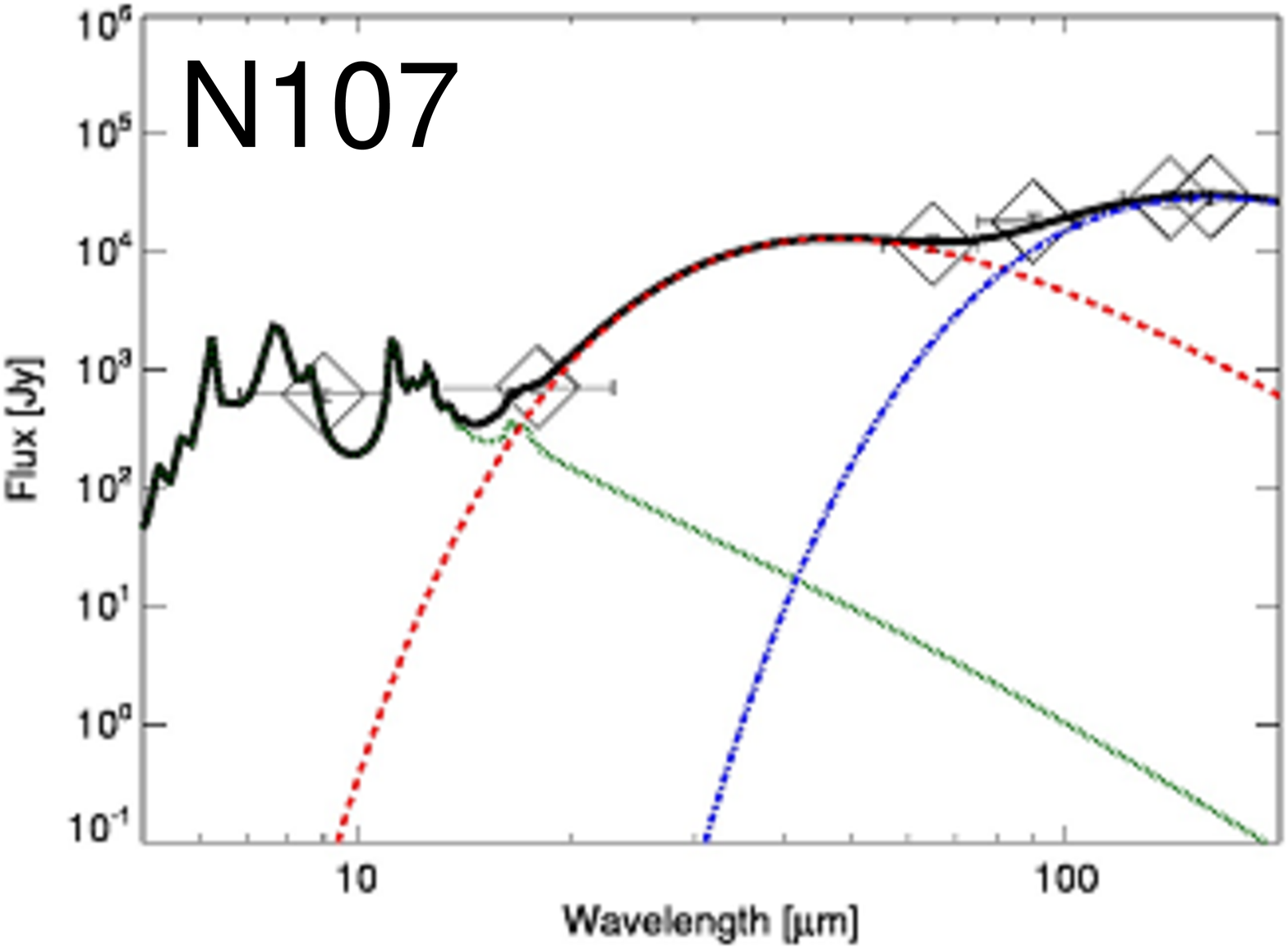}}}%
\mbox{\raisebox{6mm}{\rotatebox{90}{\small{DEC (J2000)}}}}%
\mbox{\raisebox{0mm}{\includegraphics[width=40mm]{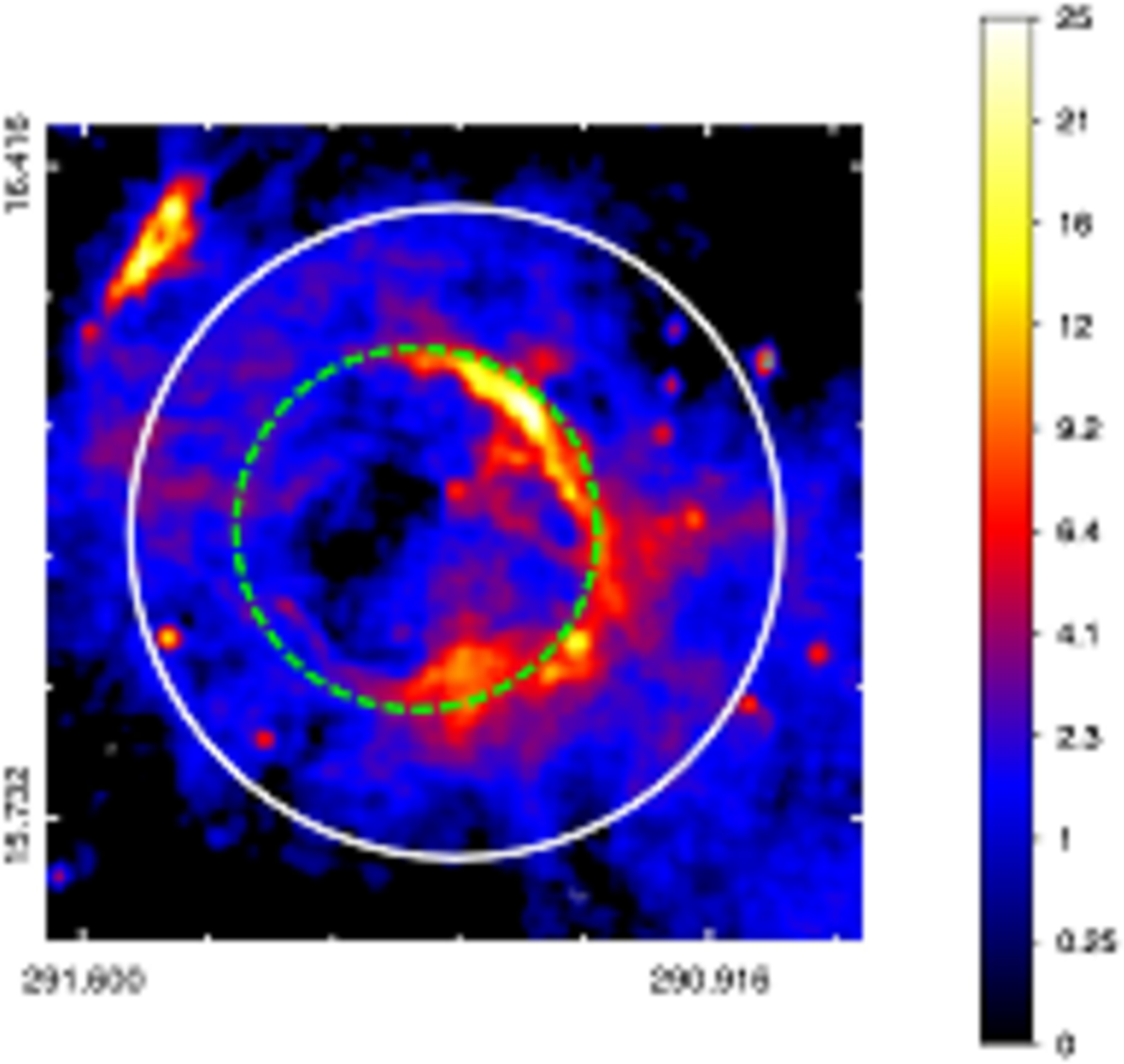}}}%
\mbox{\raisebox{0mm}{\includegraphics[width=40mm]{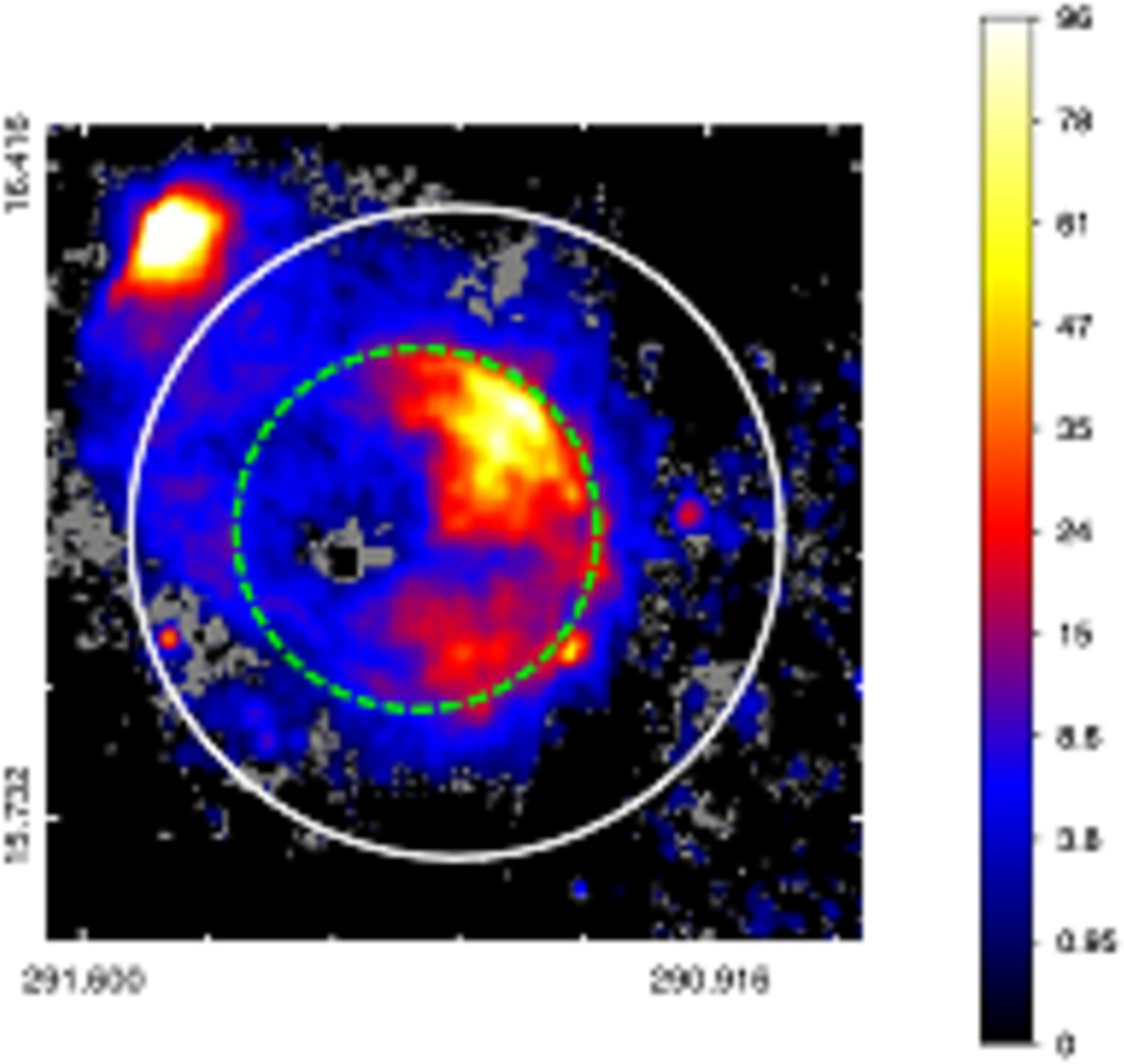}}}%
\mbox{\raisebox{0mm}{\includegraphics[width=40mm]{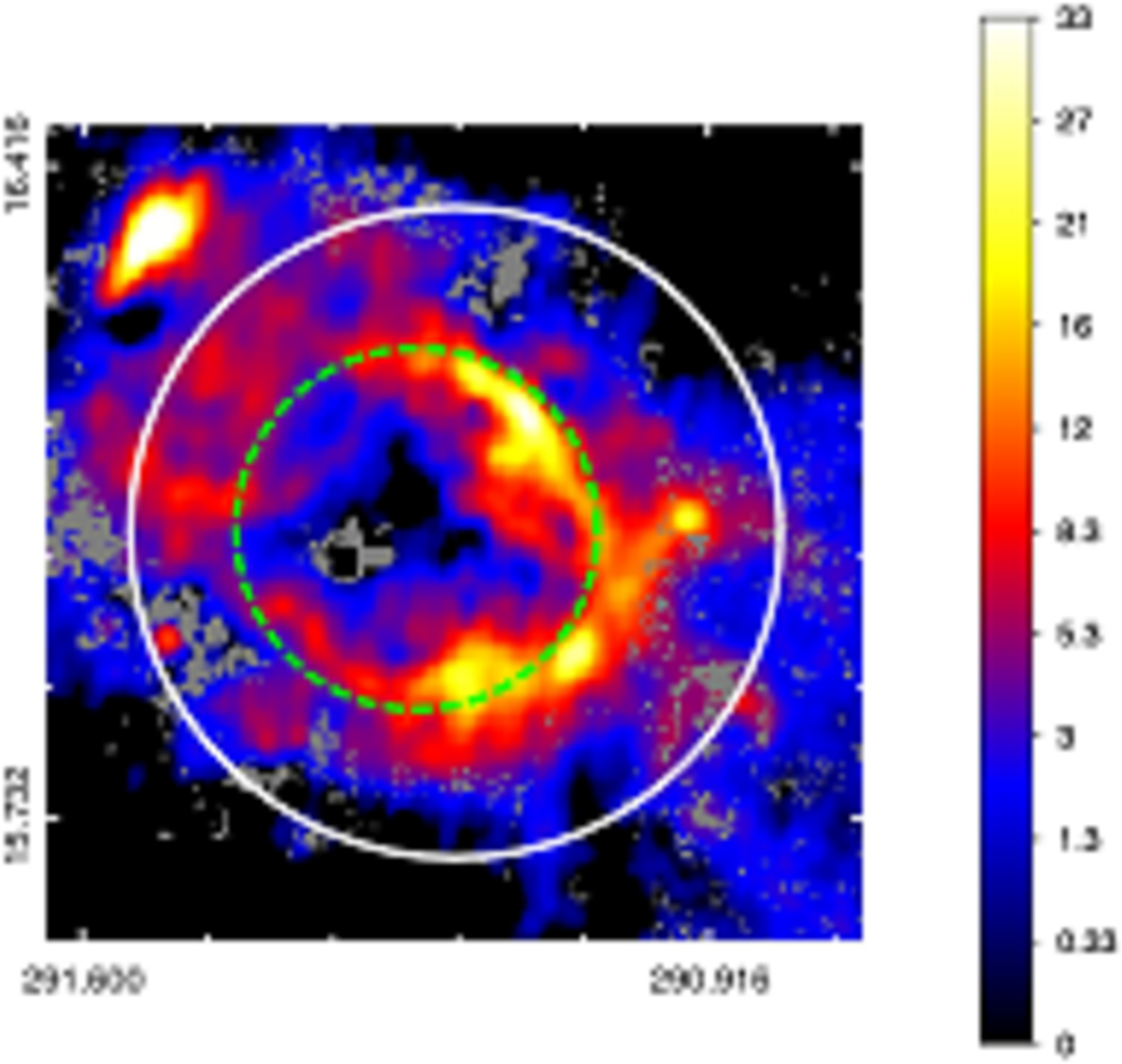}}}%
}
\subfigure{
\mbox{\raisebox{0mm}{\includegraphics[width=40mm]{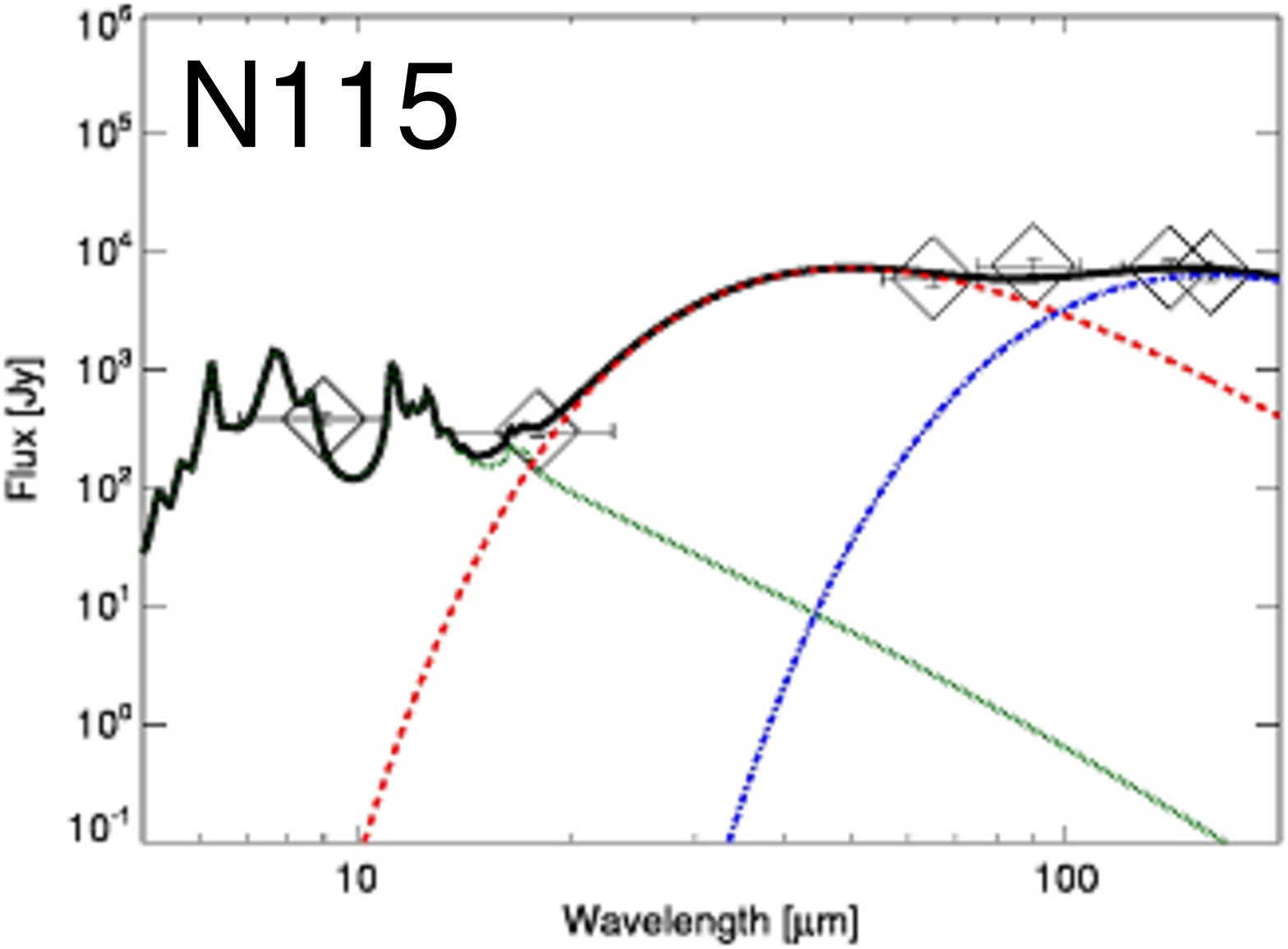}}}%
\mbox{\raisebox{6mm}{\rotatebox{90}{\small{DEC (J2000)}}}}%
\mbox{\raisebox{0mm}{\includegraphics[width=40mm]{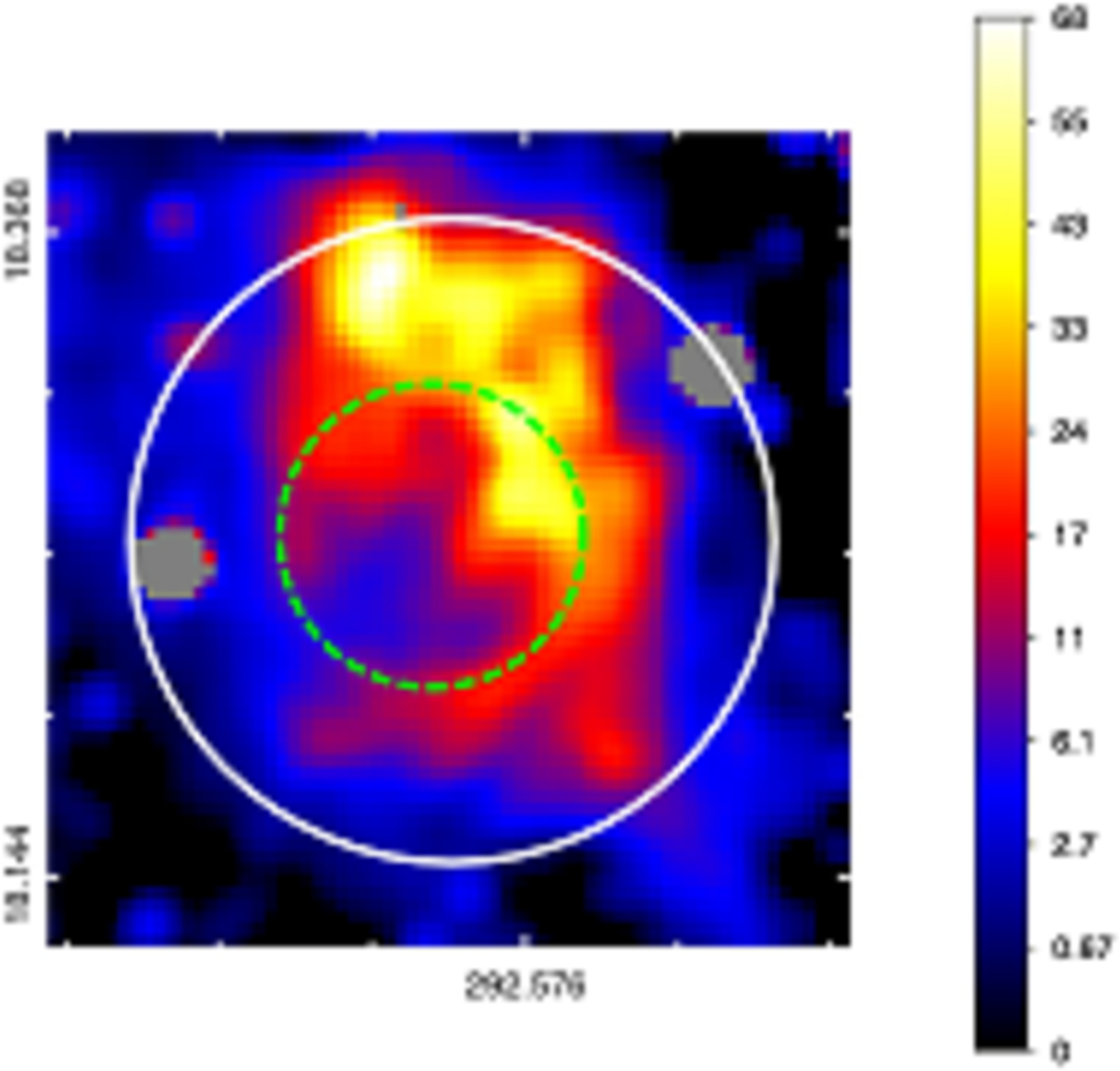}}}%
\mbox{\raisebox{0mm}{\includegraphics[width=40mm]{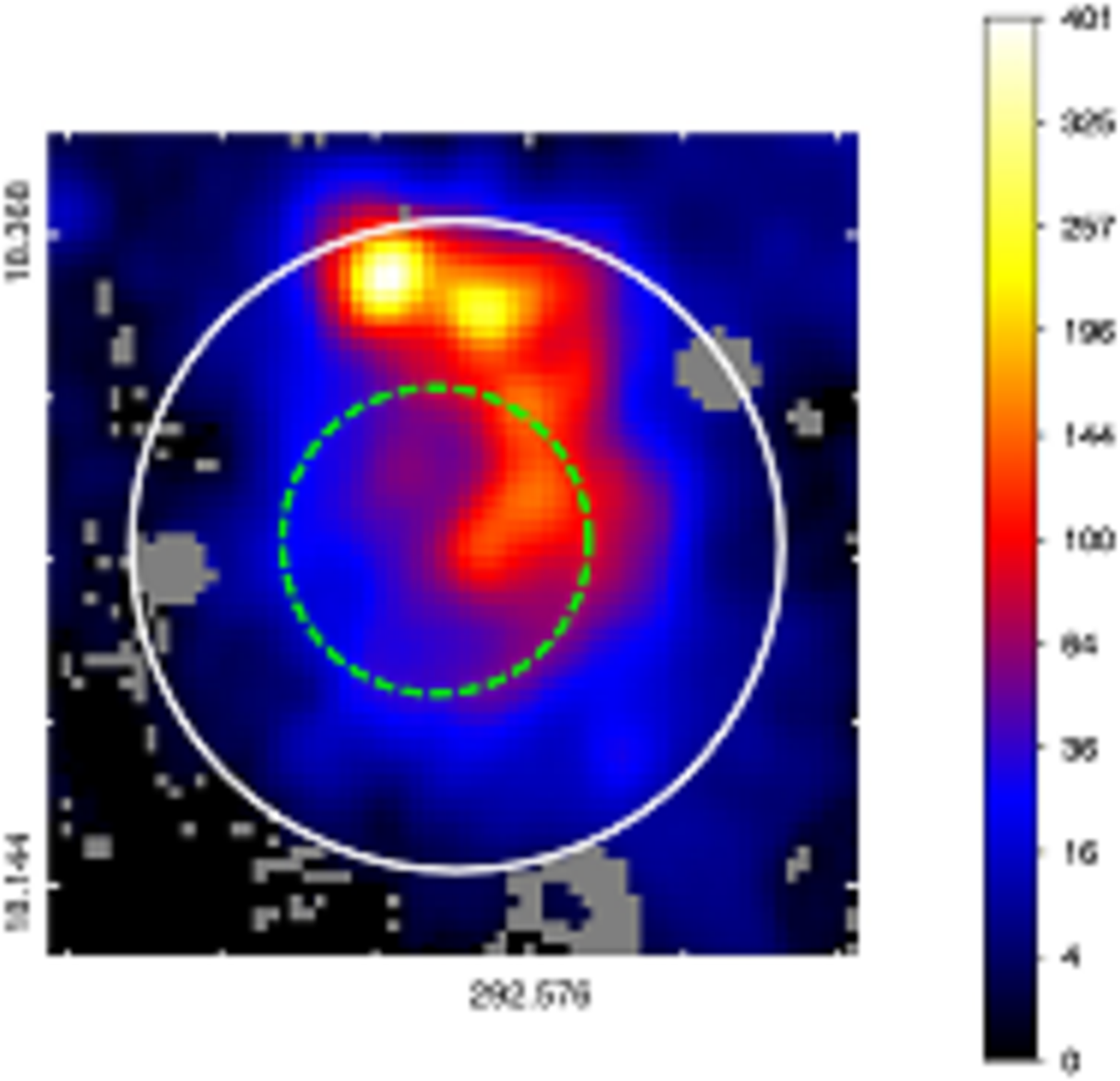}}}%
\mbox{\raisebox{0mm}{\includegraphics[width=40mm]{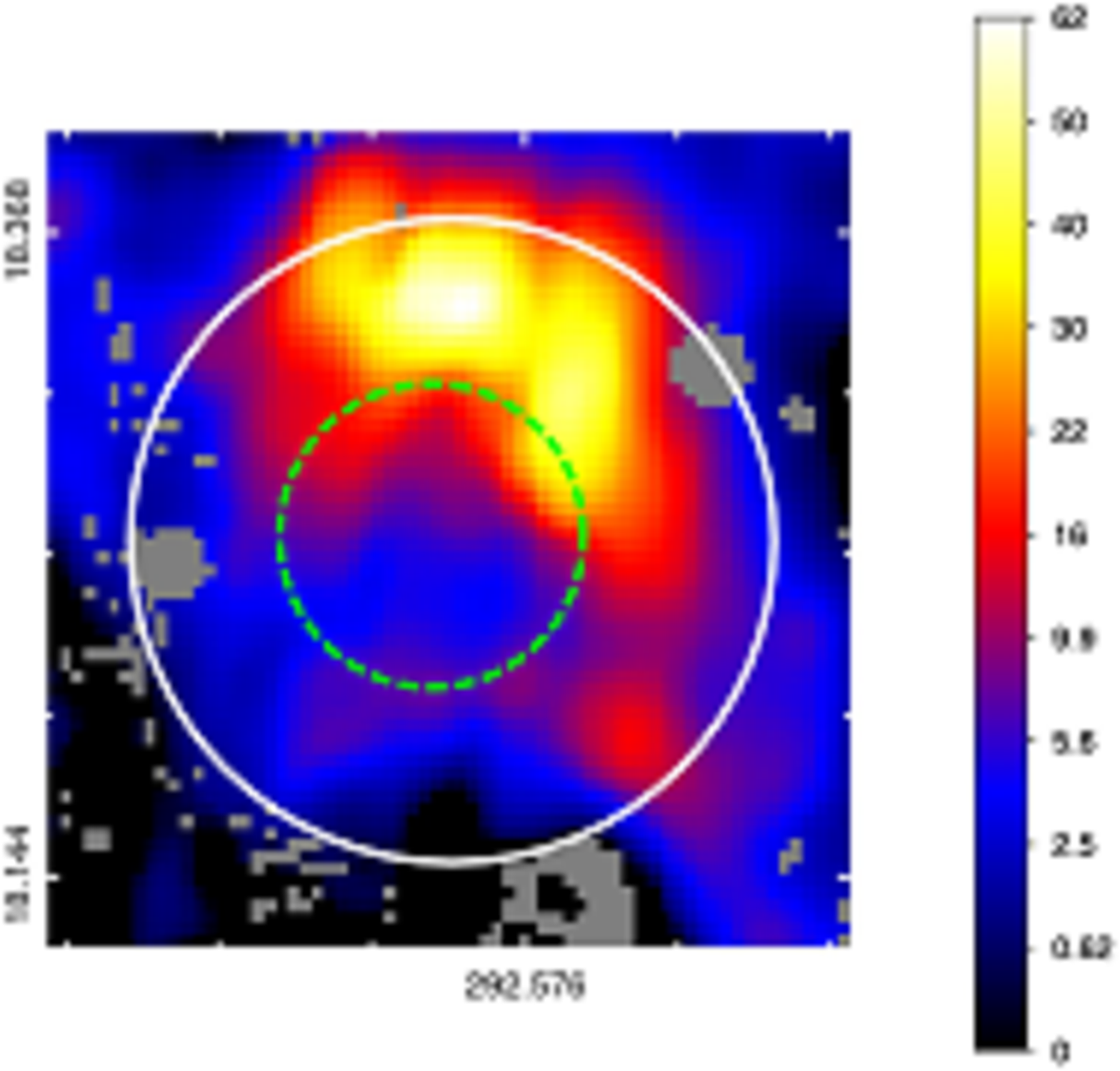}}}%
}
\subfigure{
\mbox{\raisebox{0mm}{\includegraphics[width=40mm]{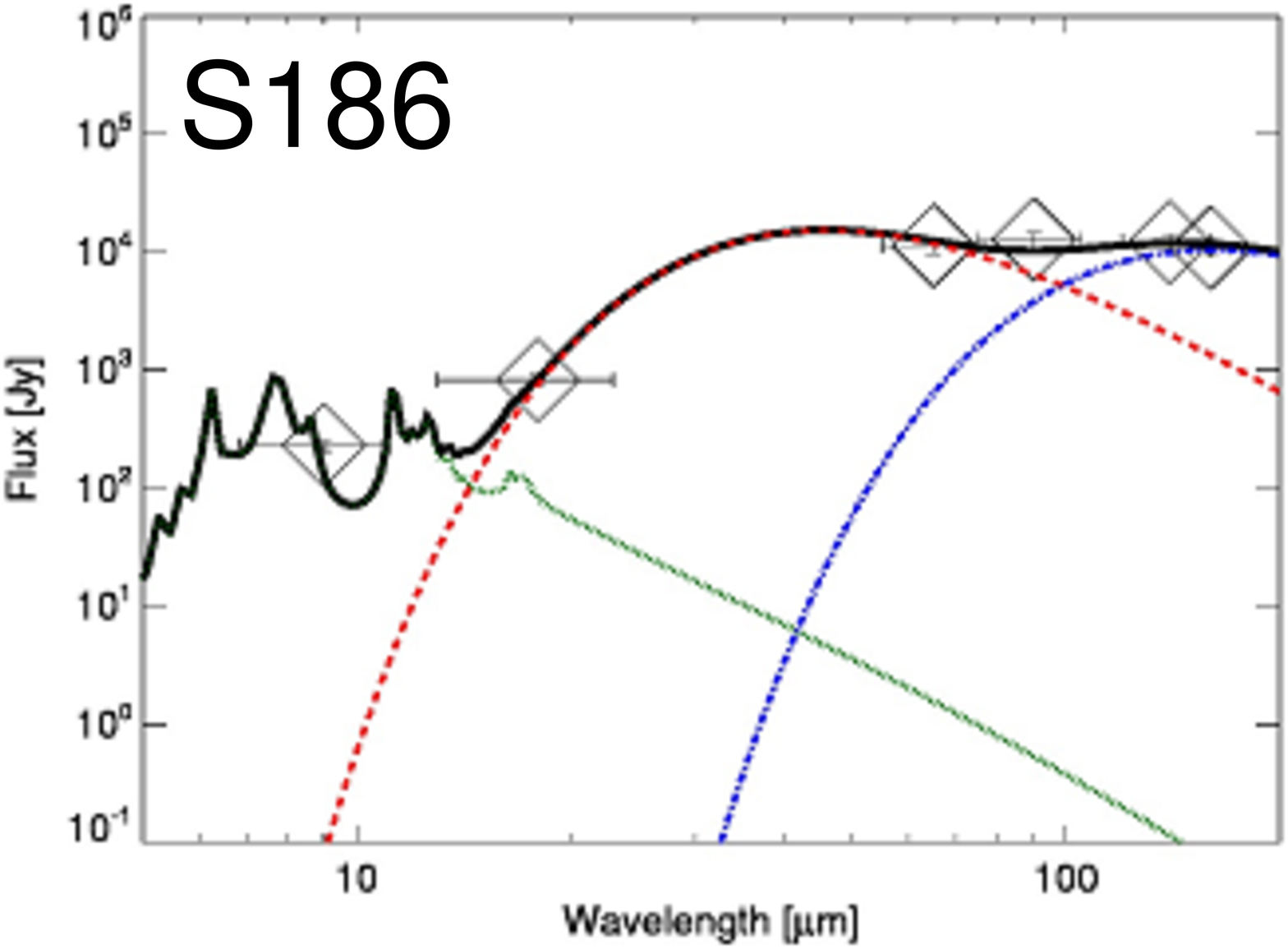}}}%
\mbox{\raisebox{6mm}{\rotatebox{90}{\small{DEC (J2000)}}}}%
\mbox{\raisebox{0mm}{\includegraphics[width=40mm]{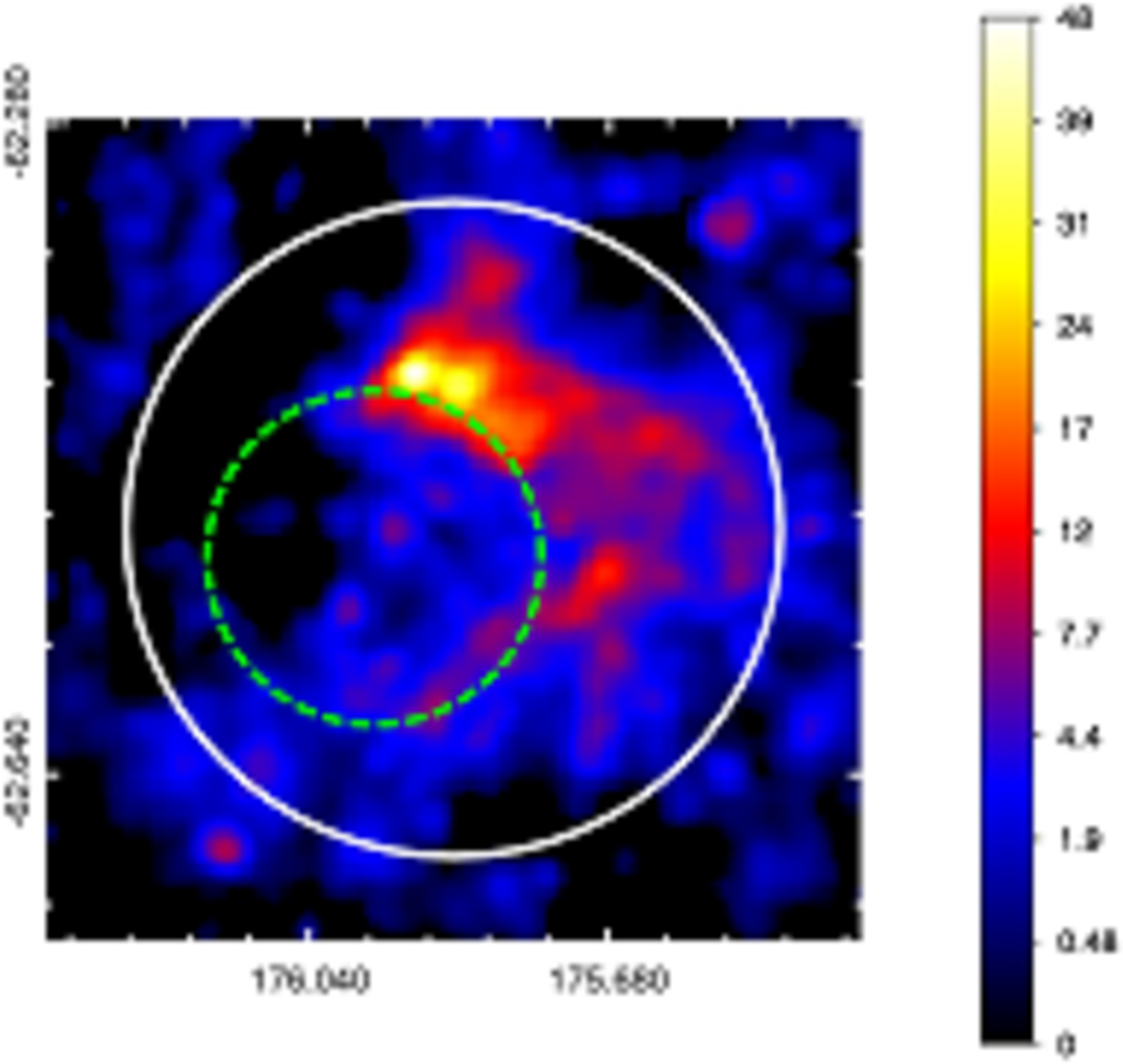}}}%
\mbox{\raisebox{0mm}{\includegraphics[width=40mm]{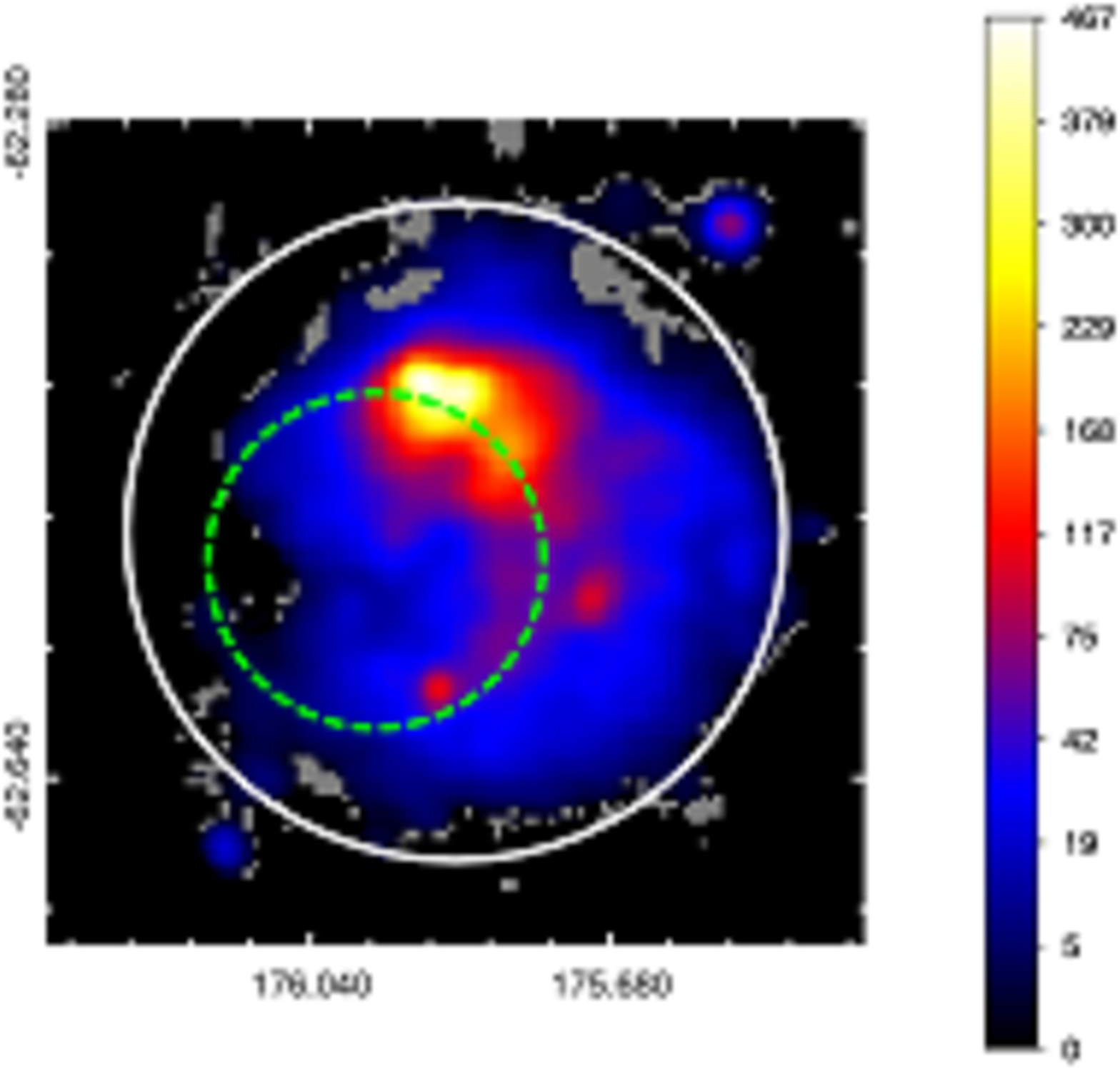}}}%
\mbox{\raisebox{0mm}{\includegraphics[width=40mm]{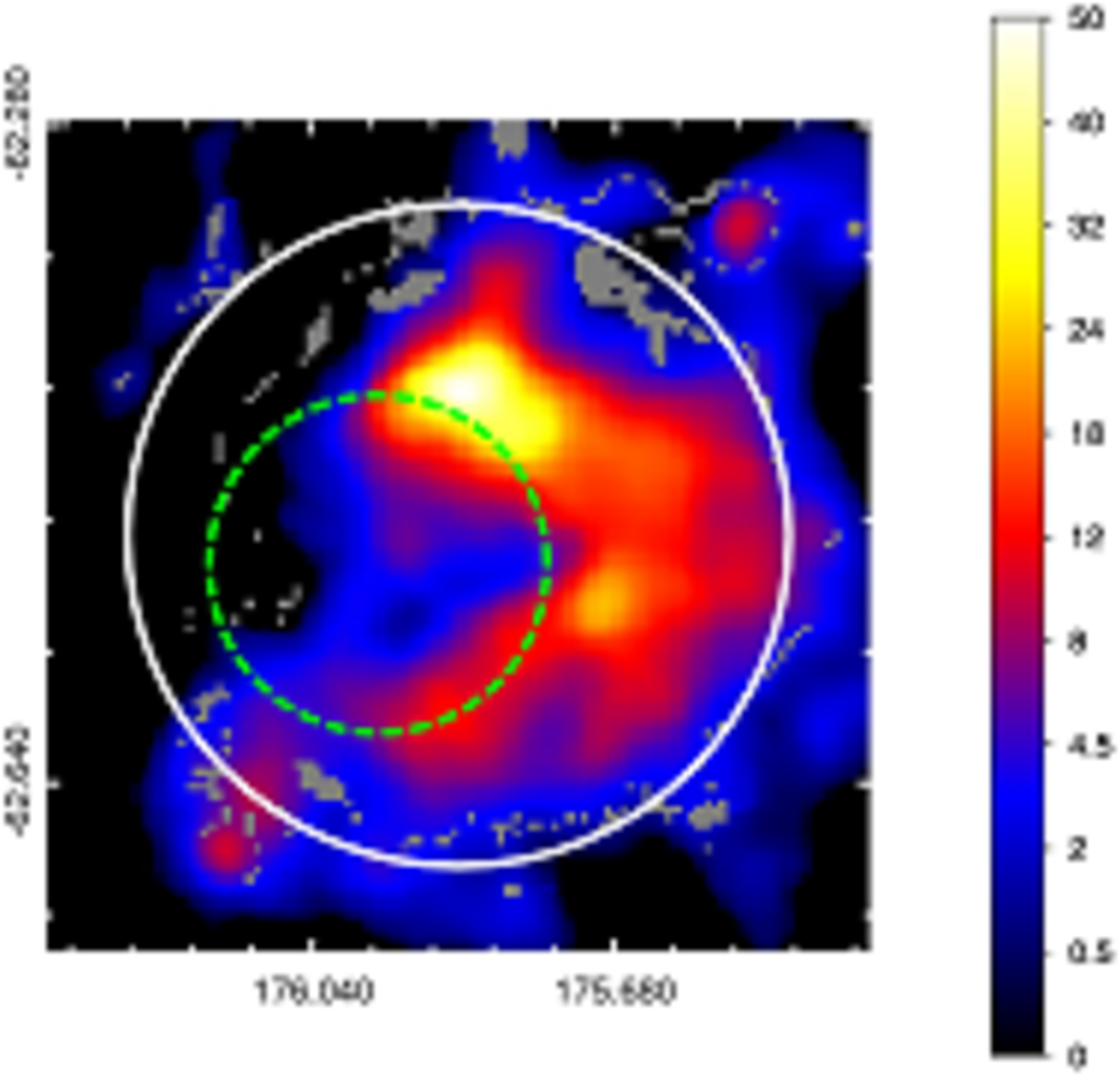}}}%
}
\subfigure{
\mbox{\raisebox{0mm}{\includegraphics[width=40mm]{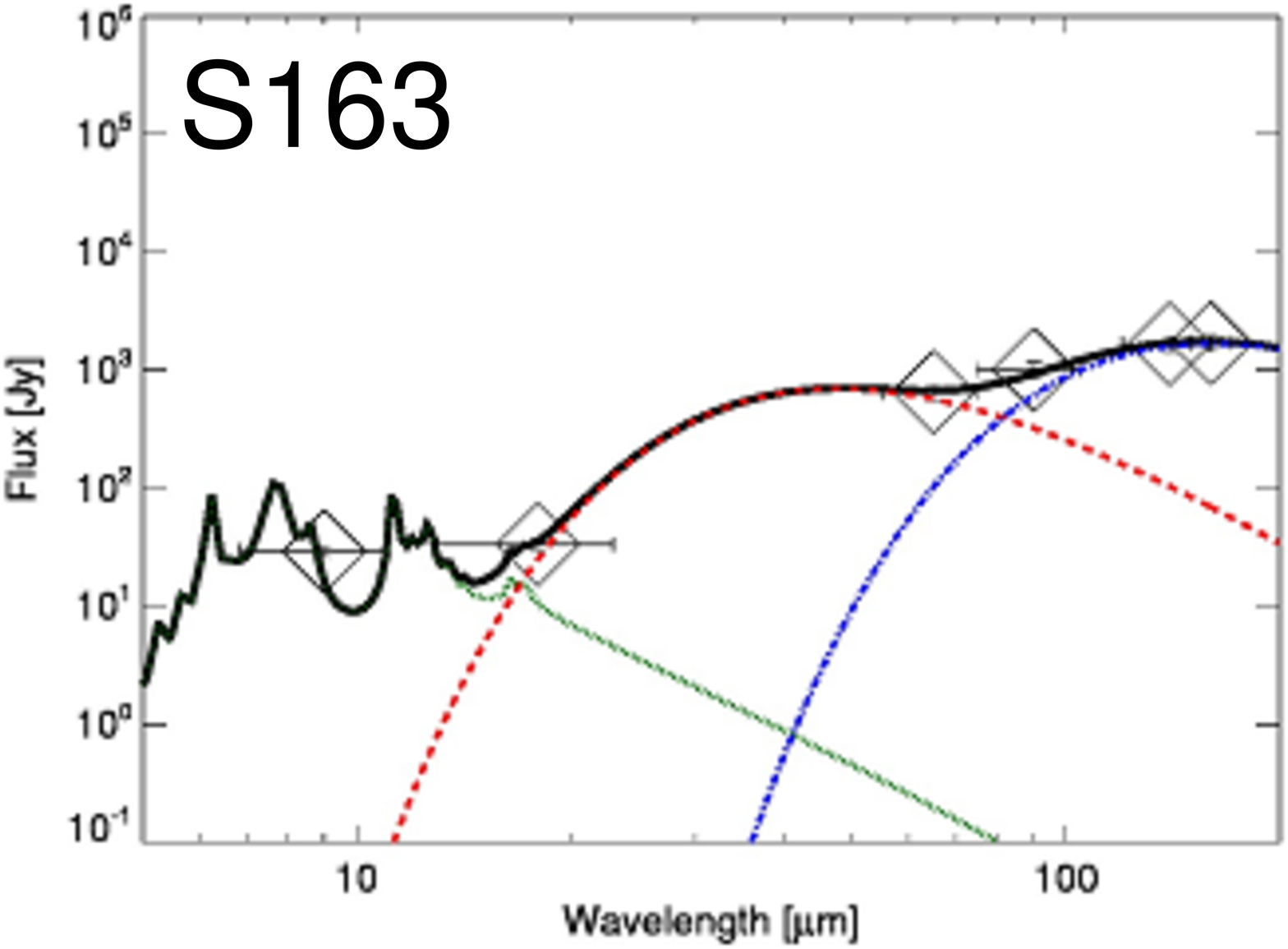}}}%
\mbox{\raisebox{6mm}{\rotatebox{90}{\small{DEC (J2000)}}}}%
\mbox{\raisebox{0mm}{\includegraphics[width=40mm]{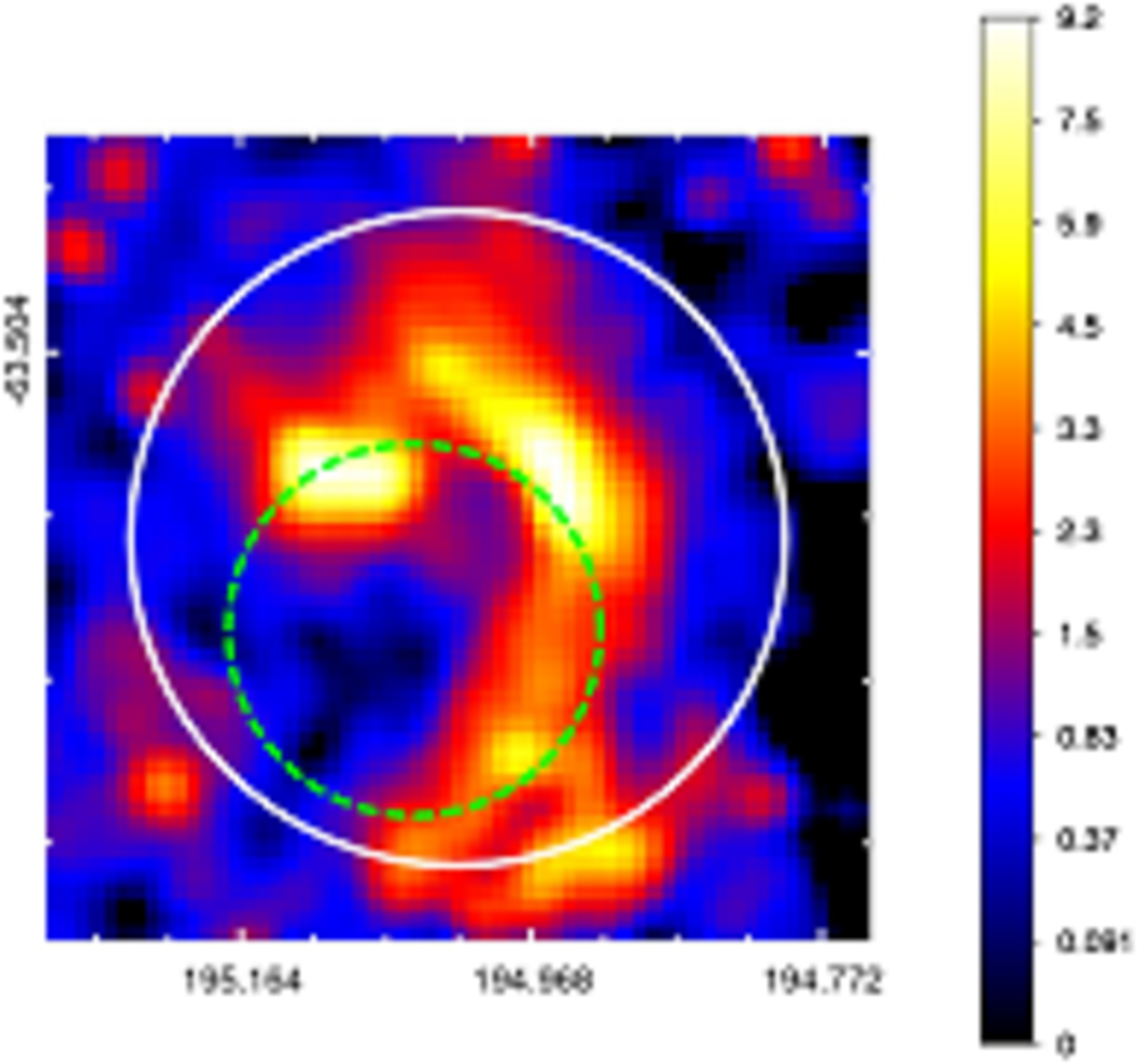}}}%
\mbox{\raisebox{0mm}{\includegraphics[width=40mm]{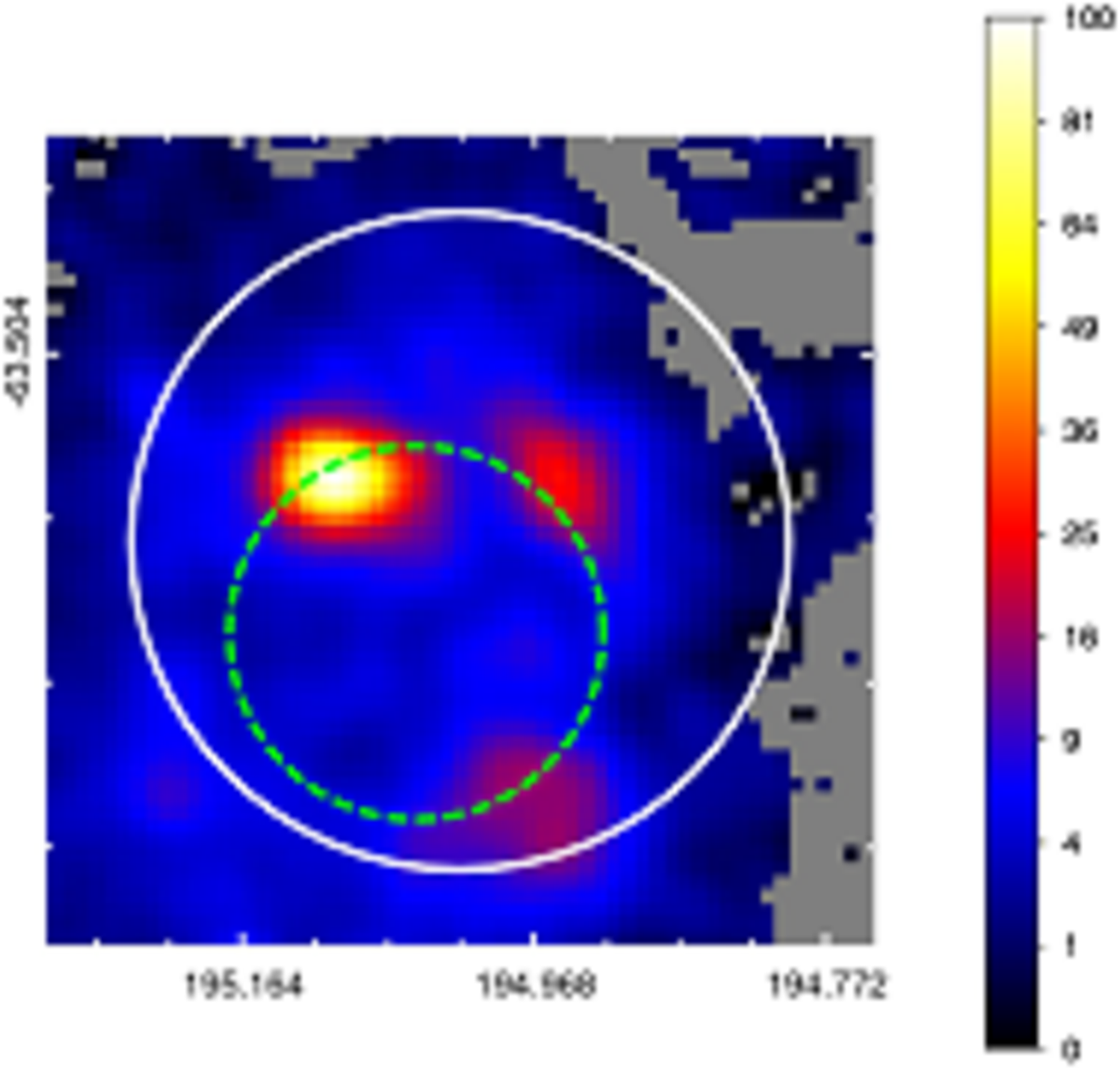}}}%
\mbox{\raisebox{0mm}{\includegraphics[width=40mm]{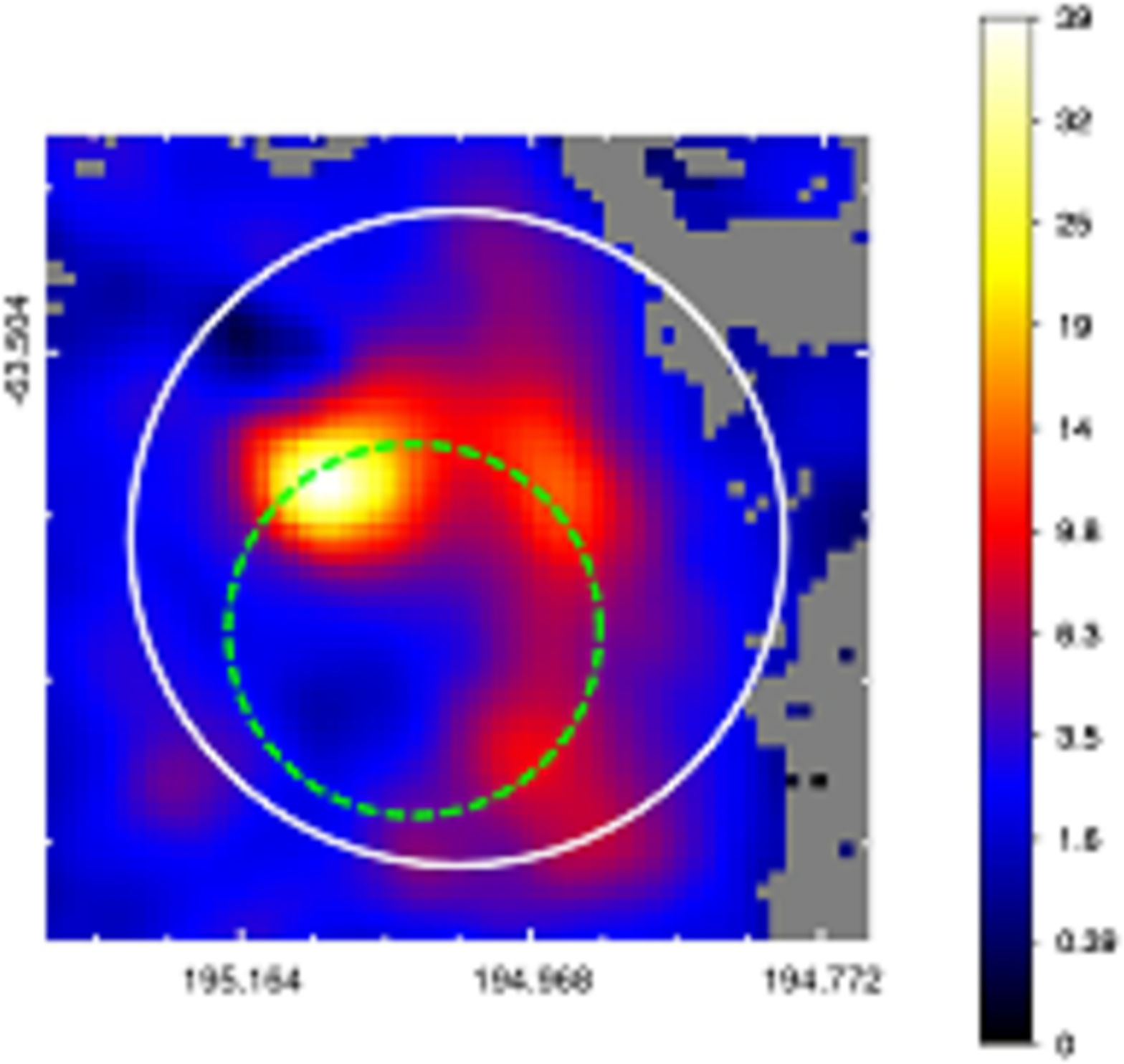}}}%
}
\subfigure{
\mbox{\raisebox{0mm}{\includegraphics[width=40mm]{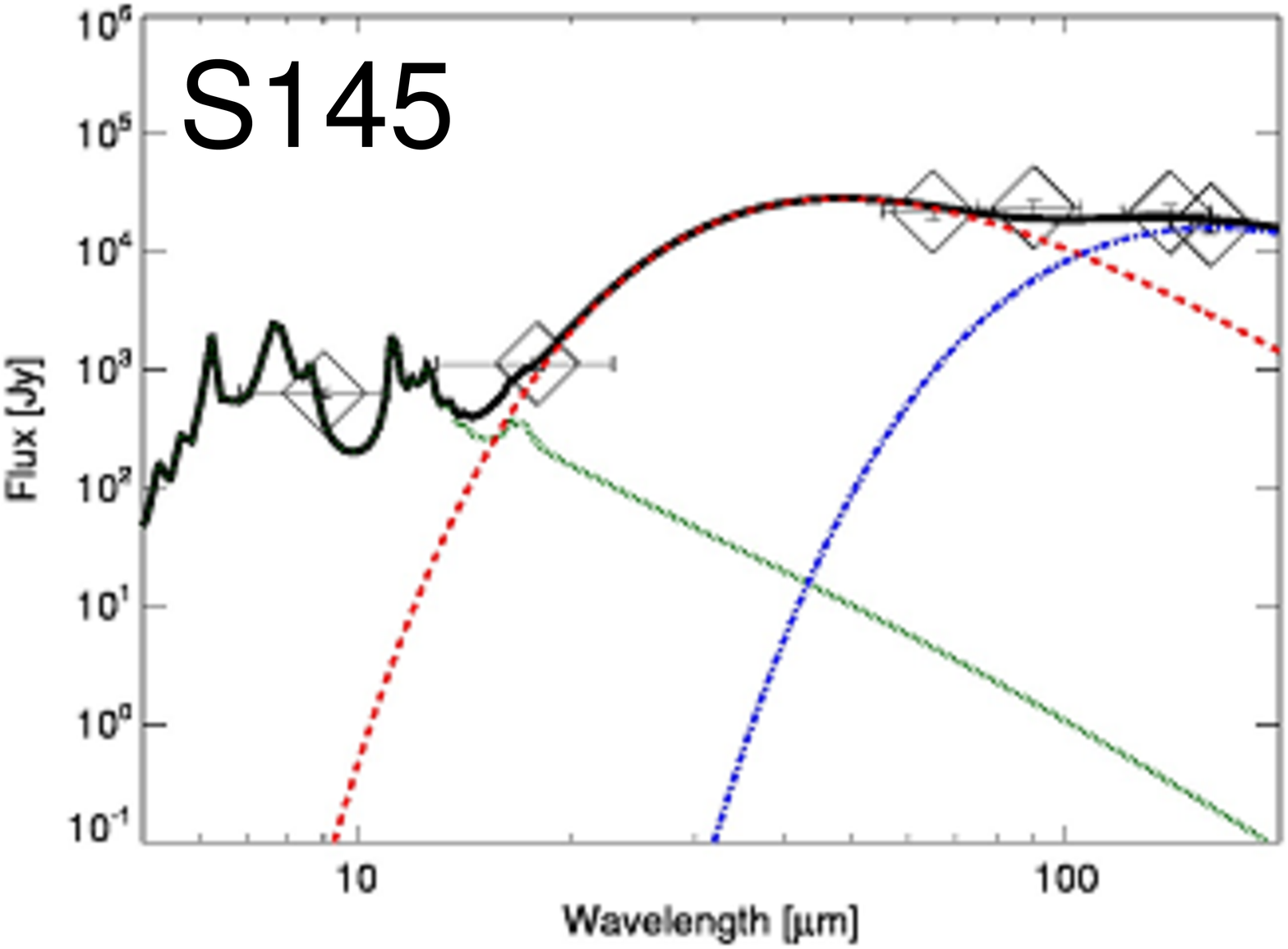}}}%
\mbox{\raisebox{6mm}{\rotatebox{90}{\small{DEC (J2000)}}}}%
\mbox{\raisebox{0mm}{\includegraphics[width=40mm]{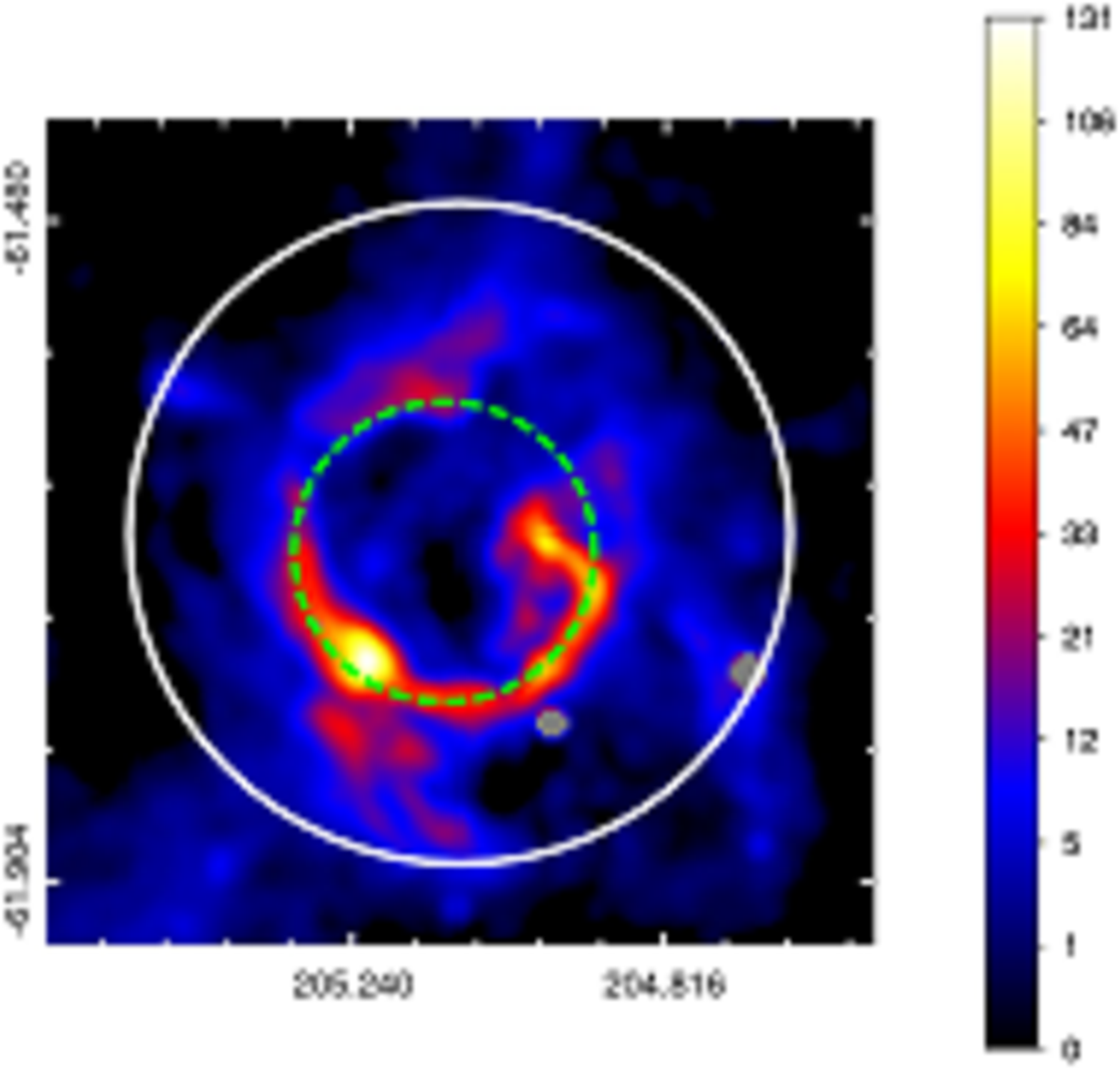}}}%
\mbox{\raisebox{0mm}{\includegraphics[width=40mm]{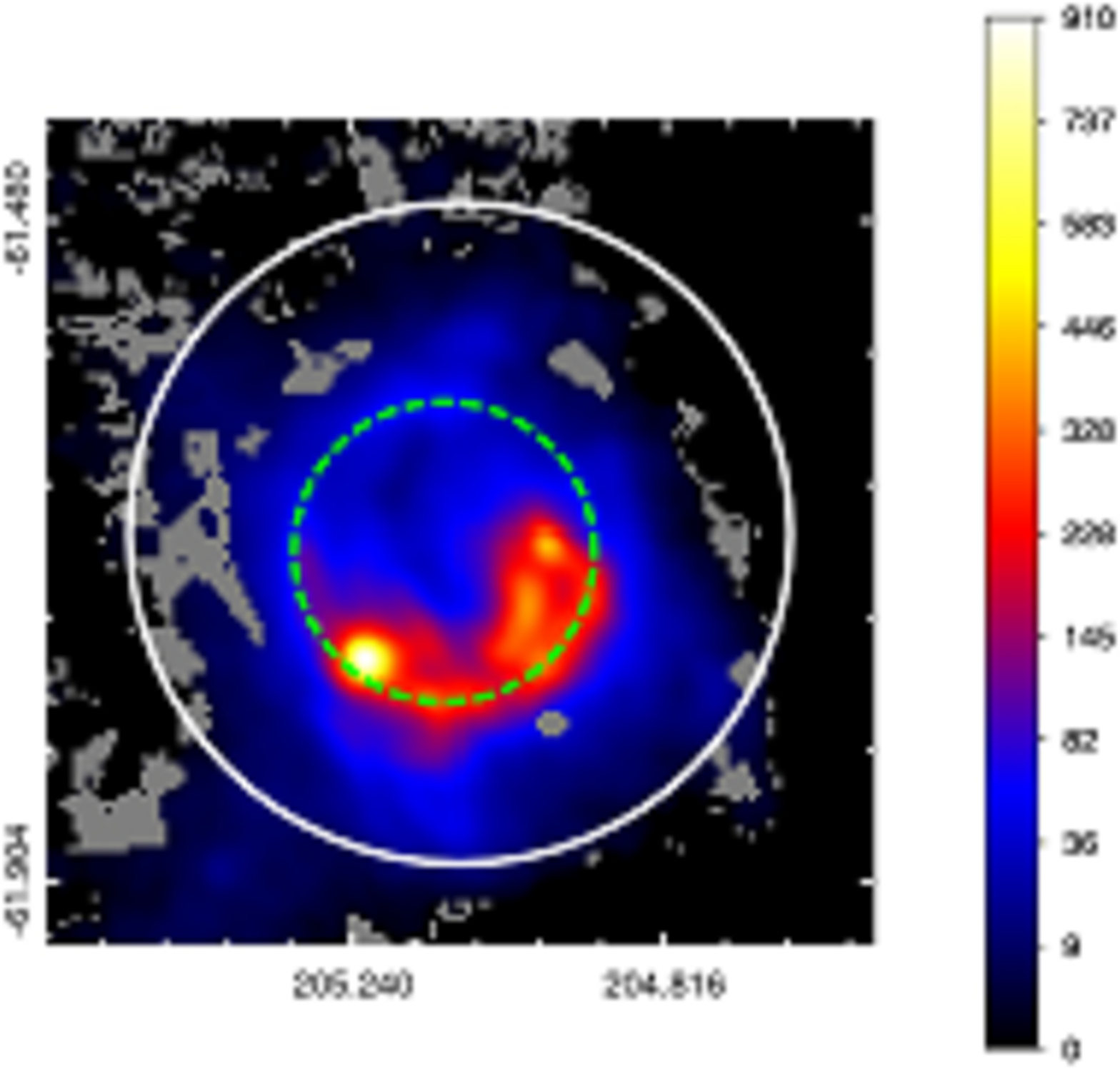}}}%
\mbox{\raisebox{0mm}{\includegraphics[width=40mm]{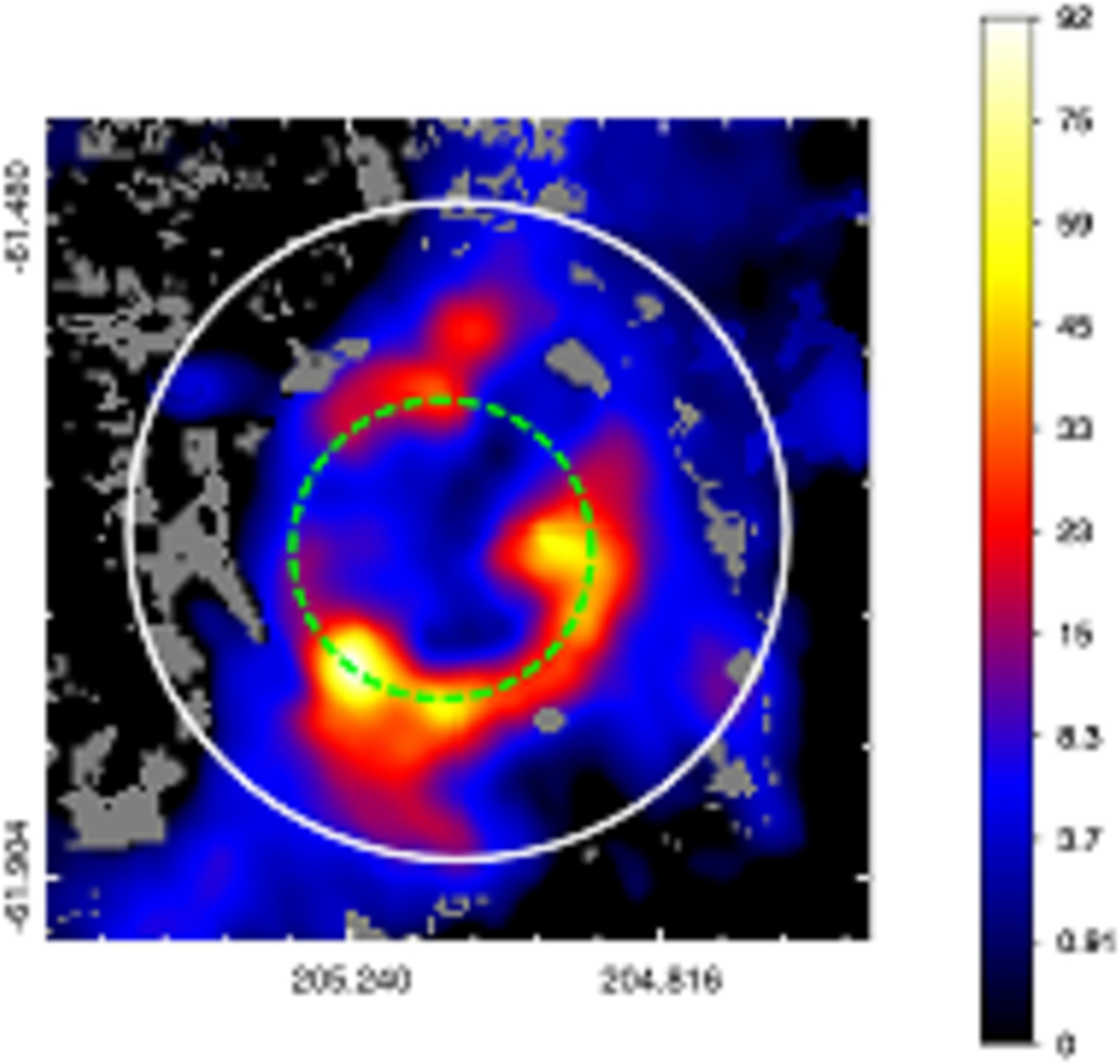}}}%
}
\caption{Continued.} \label{fig:Metfig2:e}
\end{figure*}

\addtocounter{figure}{-1}
\begin{figure*}[ht]
\addtocounter{subfigure}{1}
\centering
\subfigure{
\makebox[180mm][l]{\raisebox{0mm}[0mm][0mm]{ \hspace{20mm} \small{SED}} \hspace{27.5mm} \small{$I_{\rm{PAH}}$} \hspace{29.5mm} \small{$I_{\rm{warm}}$} \hspace{29.5mm} \small{$I_{\rm{cold}}$}}%
}
\subfigure{
\makebox[180mm][l]{\raisebox{0mm}{\hspace{52mm} \small{RA (J2000)} \hspace{20mm} \small{RA (J2000)} \hspace{20mm} \small{RA (J2000)}}}
}
\subfigure{
\mbox{\raisebox{0mm}{\includegraphics[width=40mm]{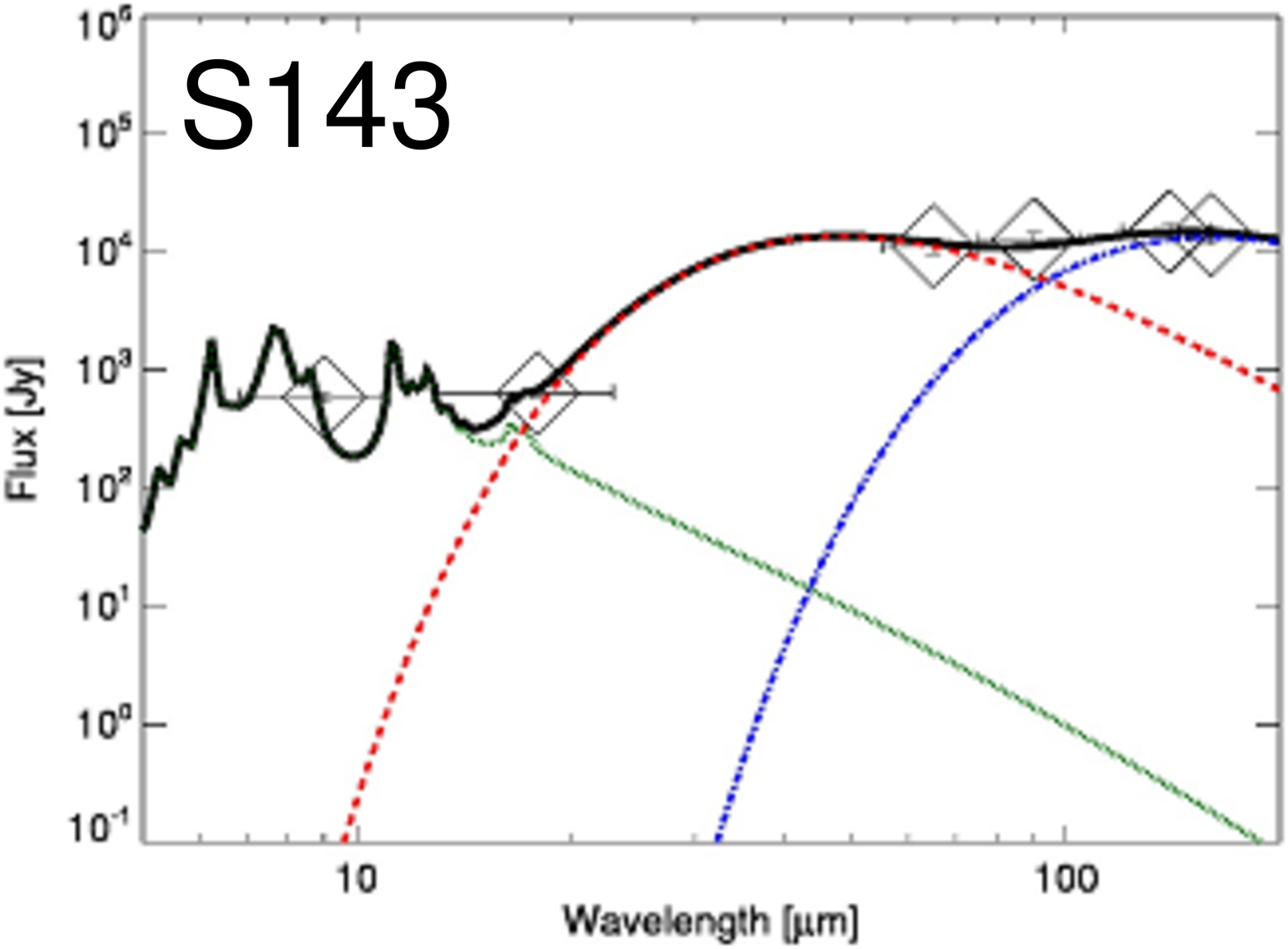}}}%
\mbox{\raisebox{6mm}{\rotatebox{90}{\small{DEC (J2000)}}}}%
\mbox{\raisebox{0mm}{\includegraphics[width=40mm]{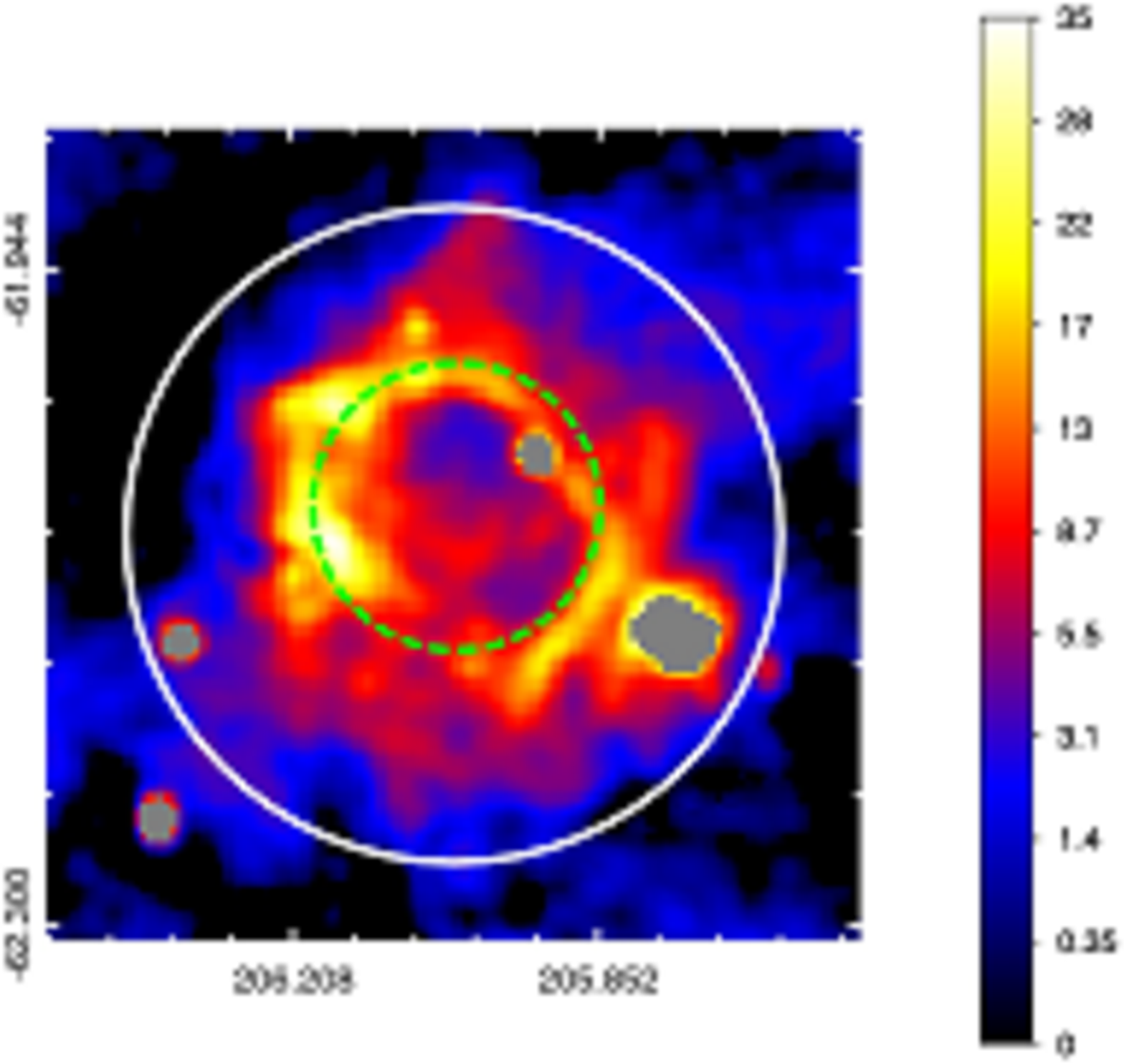}}}%
\mbox{\raisebox{0mm}{\includegraphics[width=40mm]{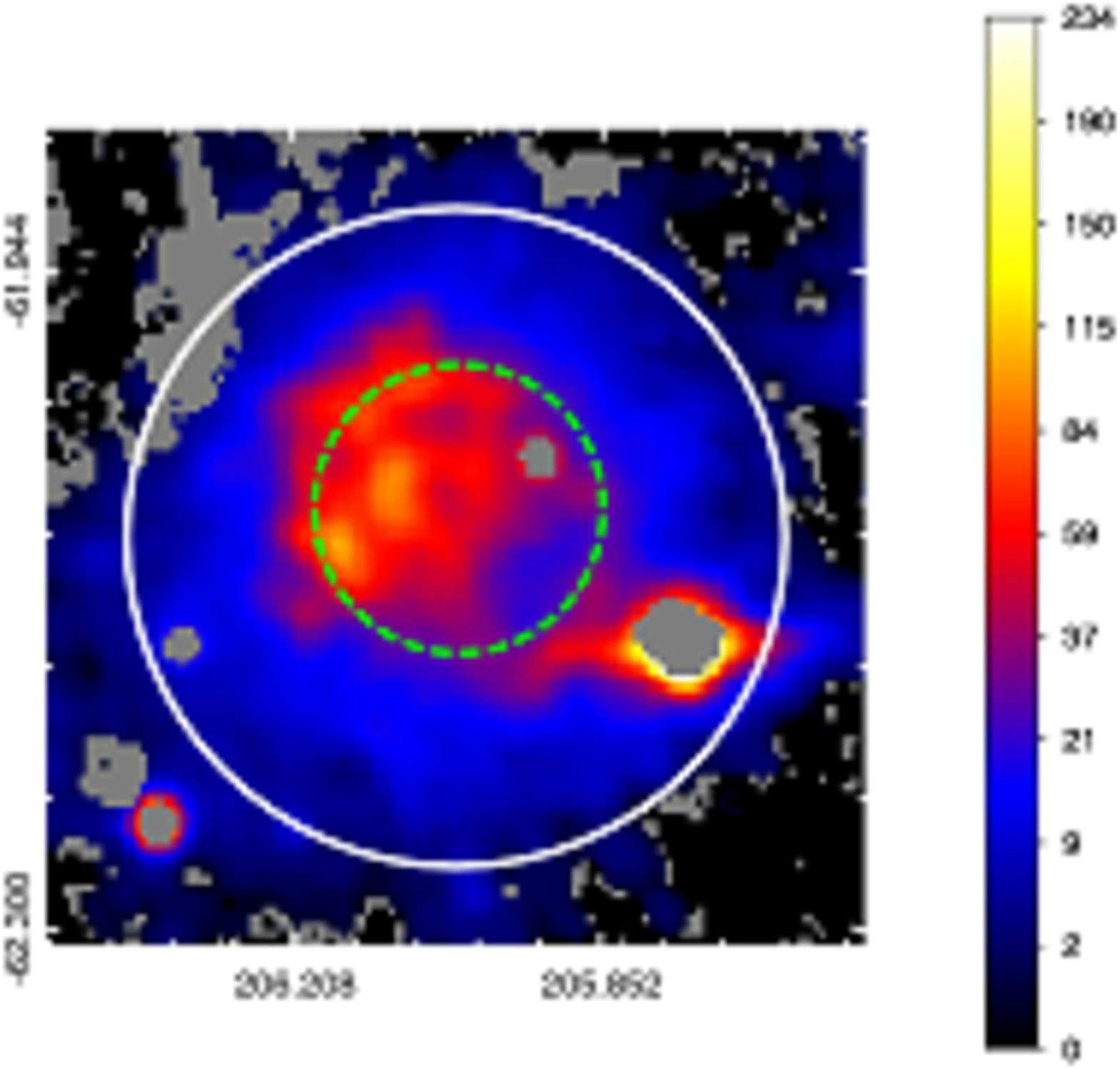}}}%
\mbox{\raisebox{0mm}{\includegraphics[width=40mm]{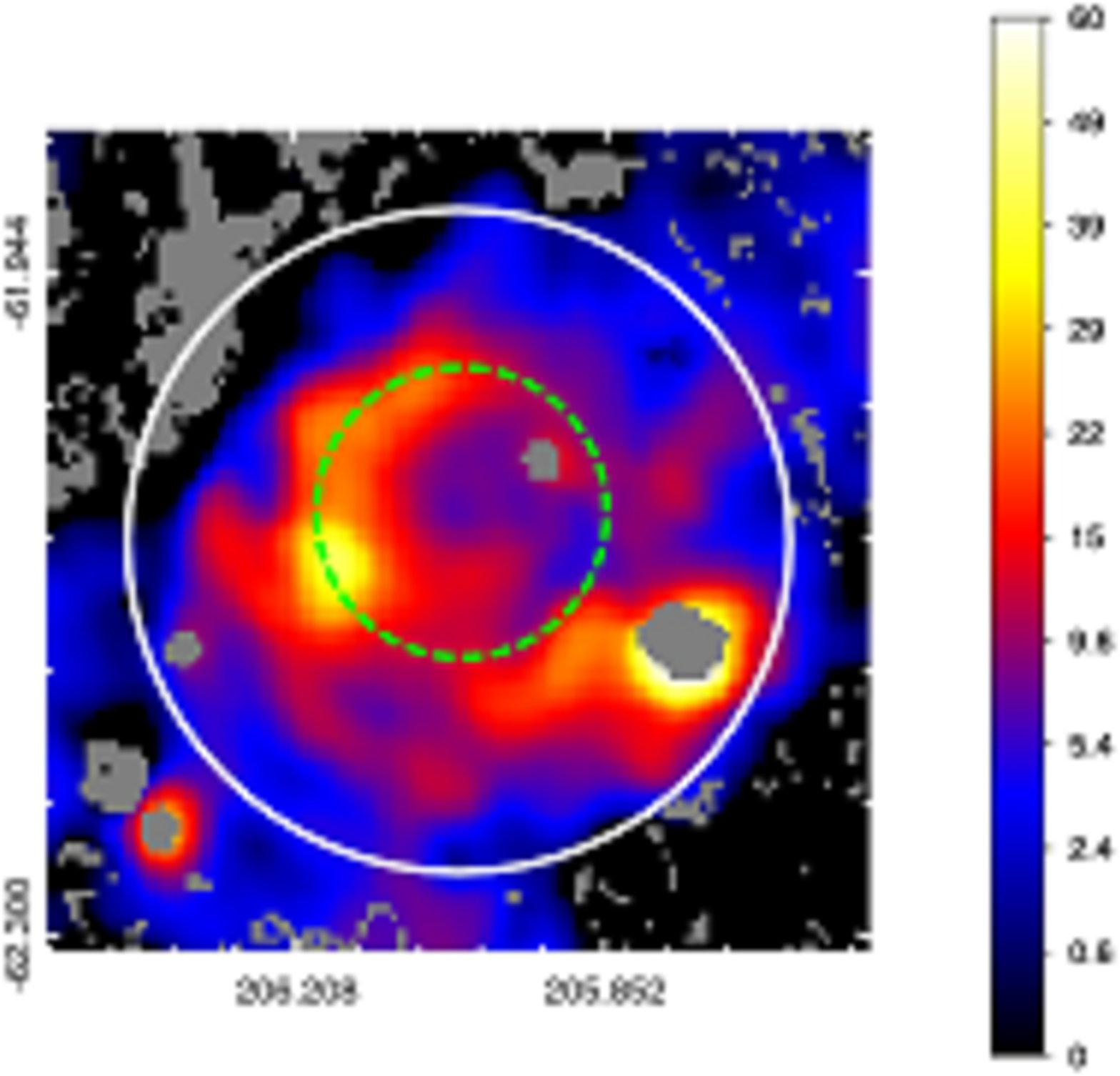}}}%
}
\subfigure{
\mbox{\raisebox{0mm}{\includegraphics[width=40mm]{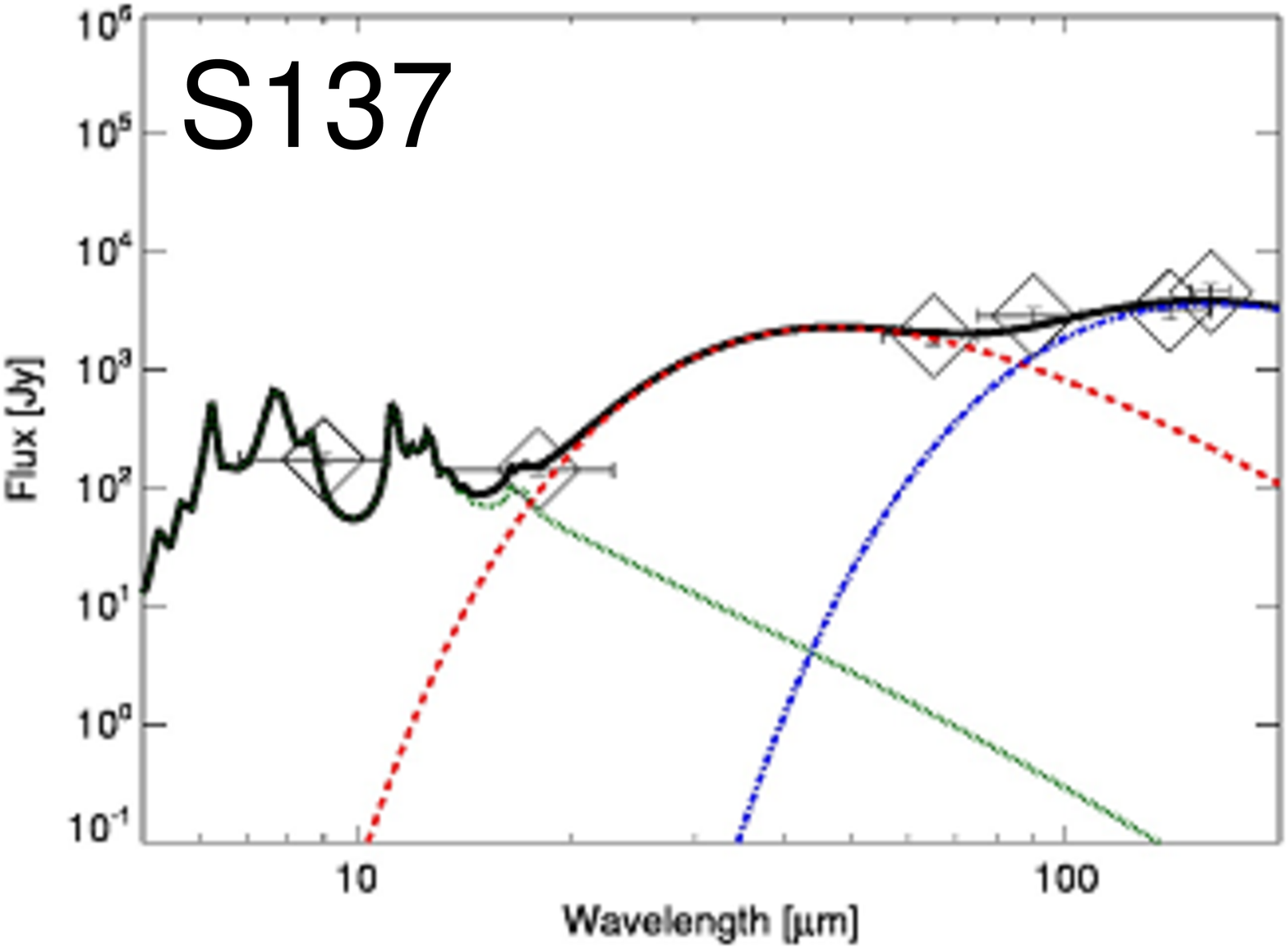}}}%
\mbox{\raisebox{6mm}{\rotatebox{90}{\small{DEC (J2000)}}}}%
\mbox{\raisebox{0mm}{\includegraphics[width=40mm]{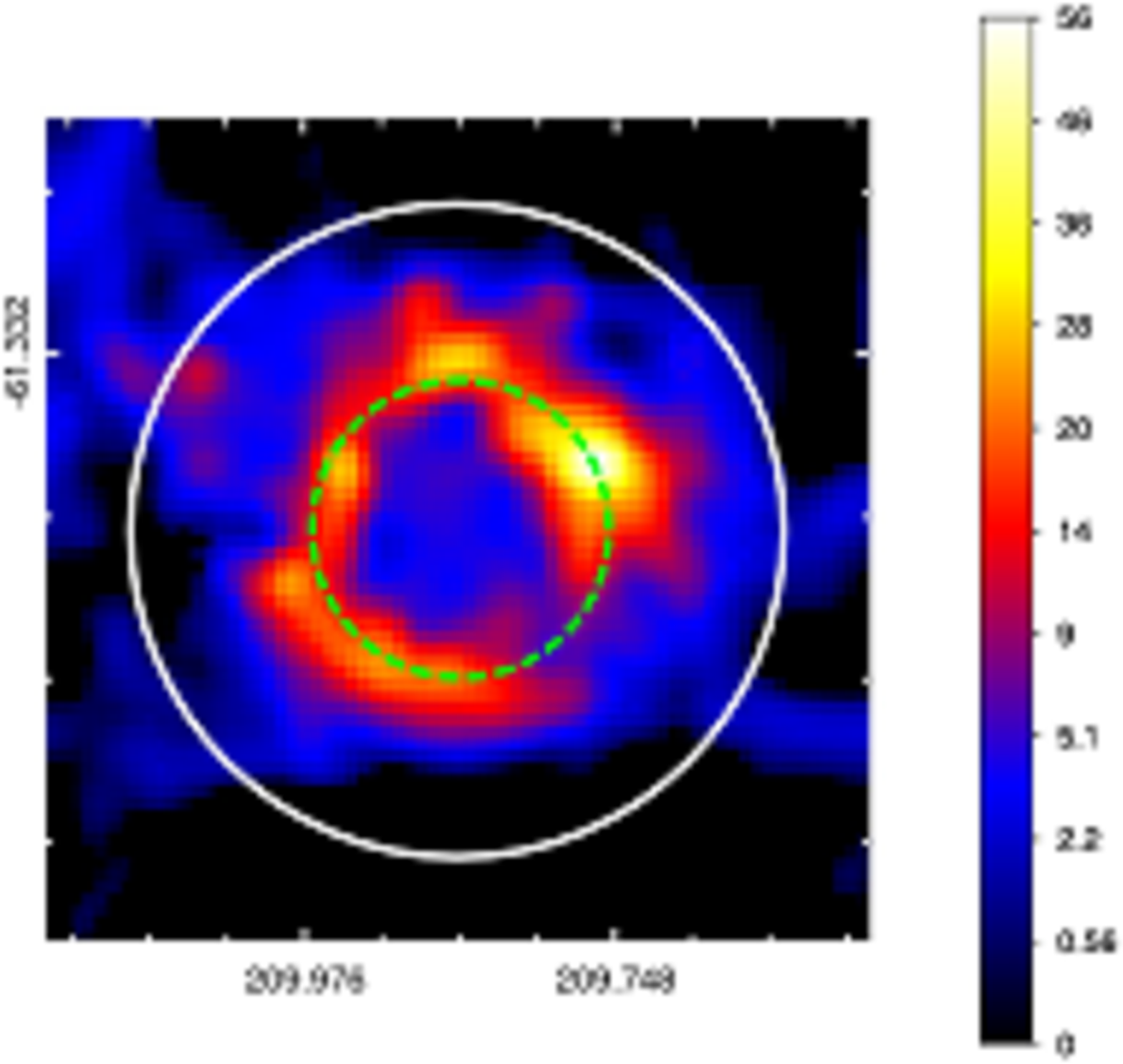}}}%
\mbox{\raisebox{0mm}{\includegraphics[width=40mm]{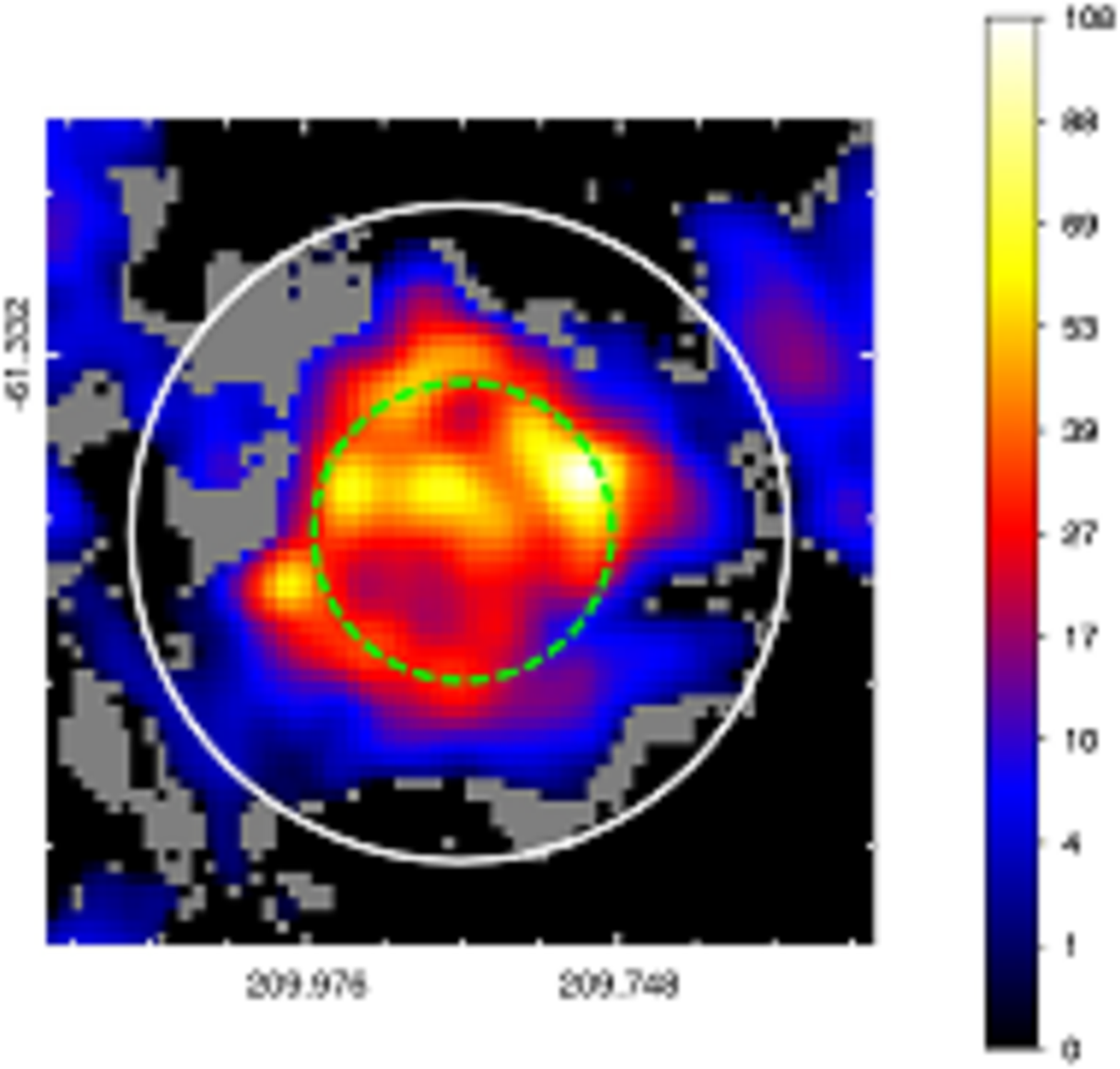}}}%
\mbox{\raisebox{0mm}{\includegraphics[width=40mm]{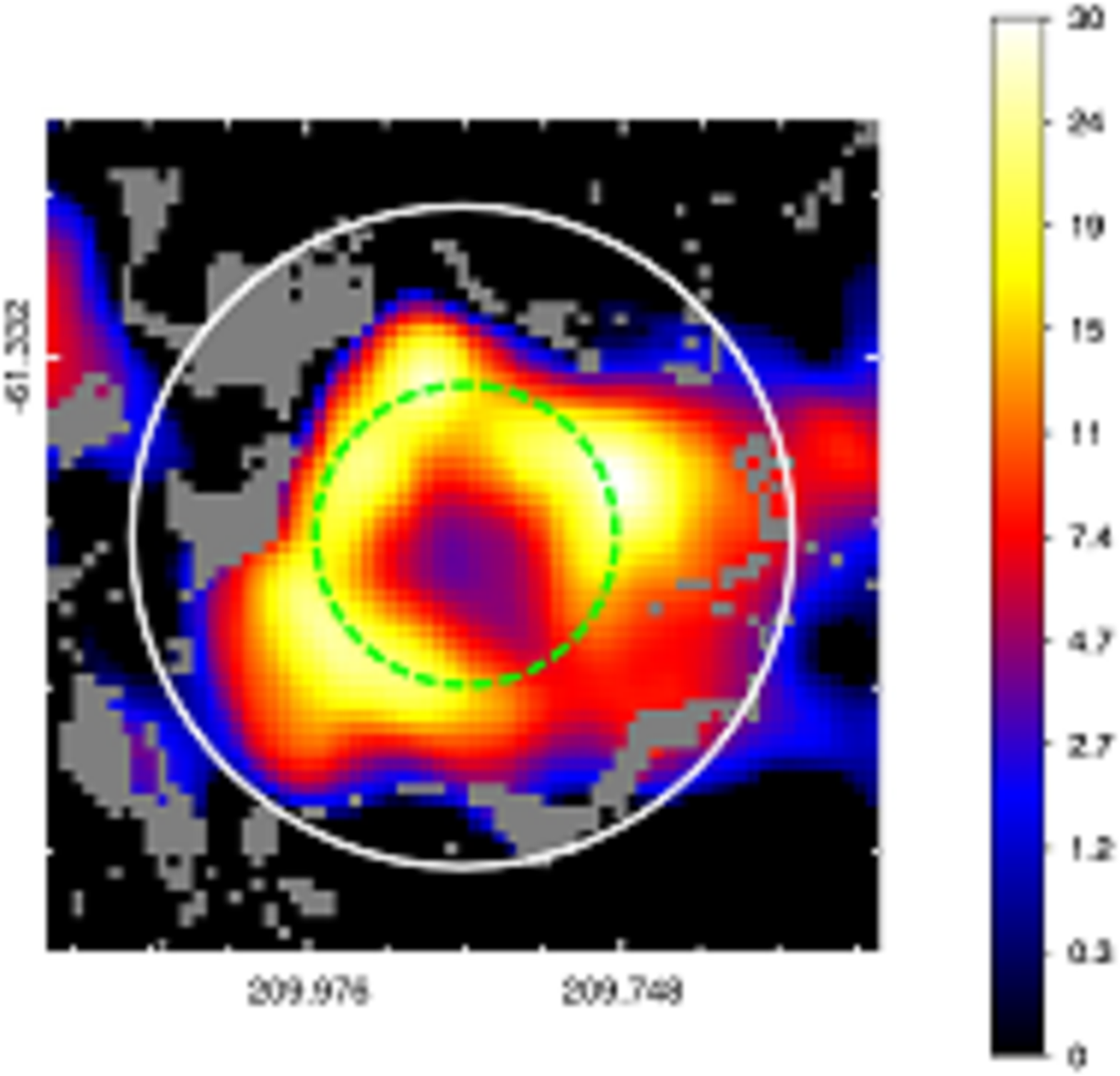}}}%
}
\subfigure{
\mbox{\raisebox{0mm}{\includegraphics[width=40mm]{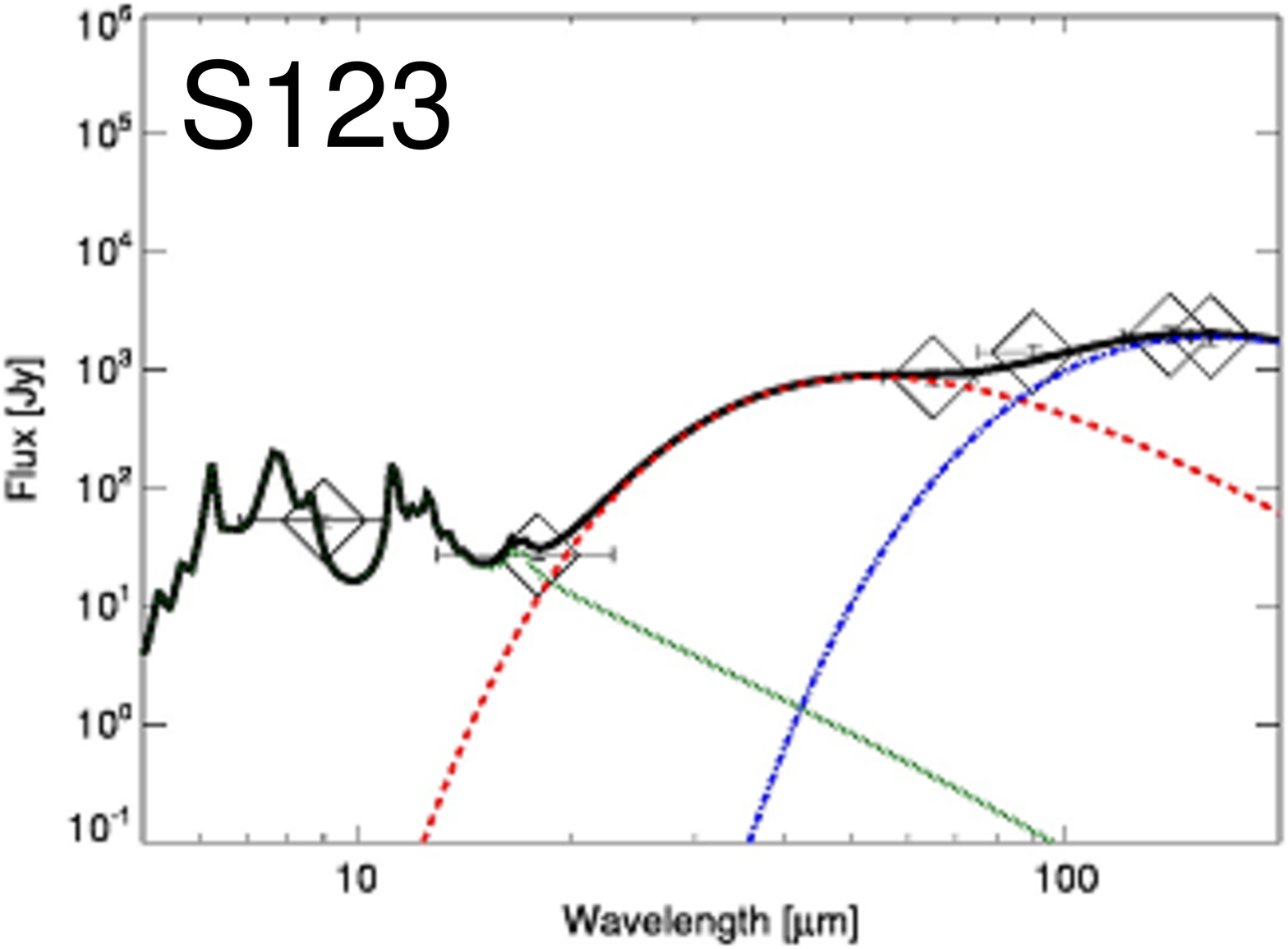}}}%
\mbox{\raisebox{6mm}{\rotatebox{90}{\small{DEC (J2000)}}}}%
\mbox{\raisebox{0mm}{\includegraphics[width=40mm]{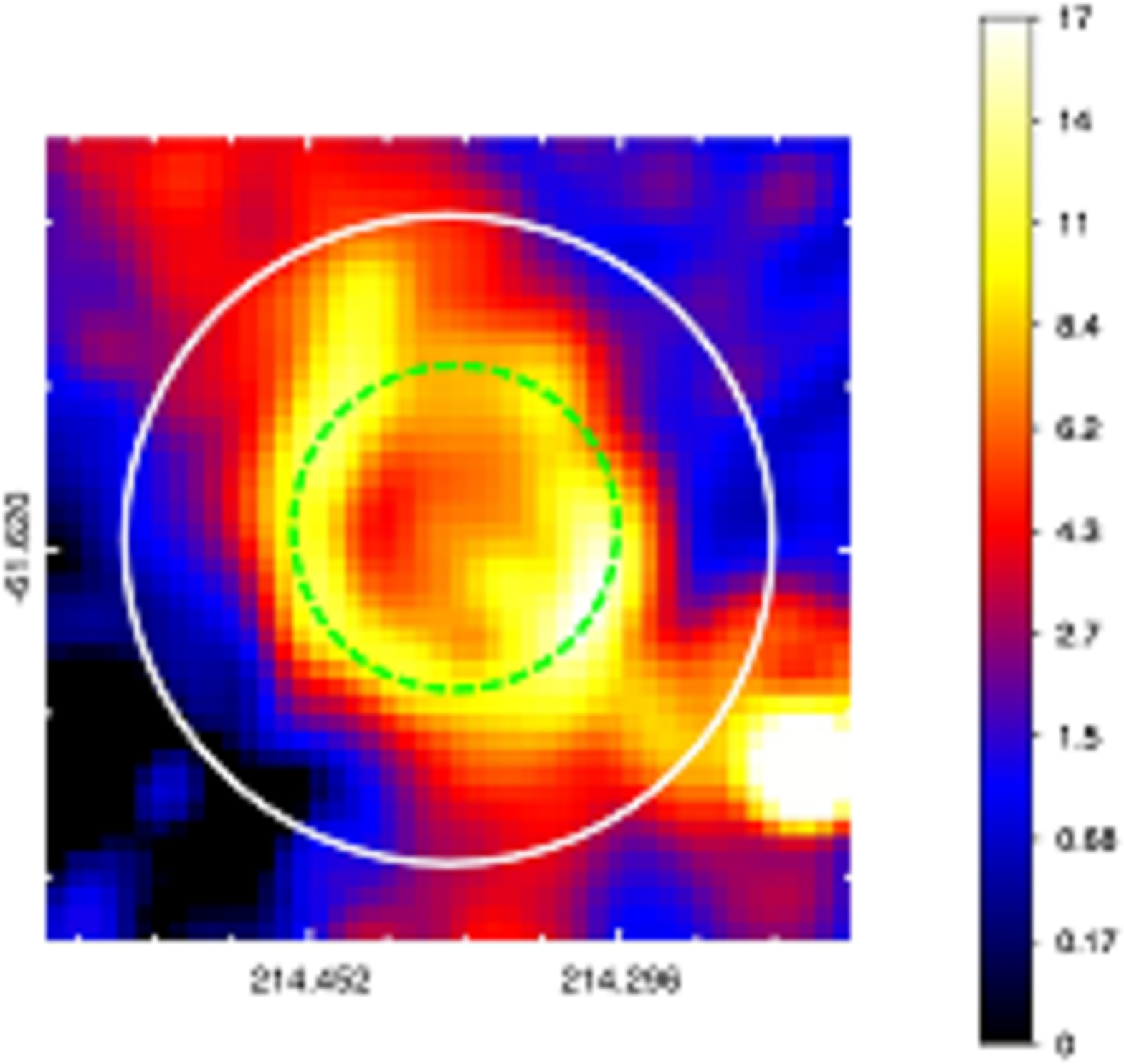}}}%
\mbox{\raisebox{0mm}{\includegraphics[width=40mm]{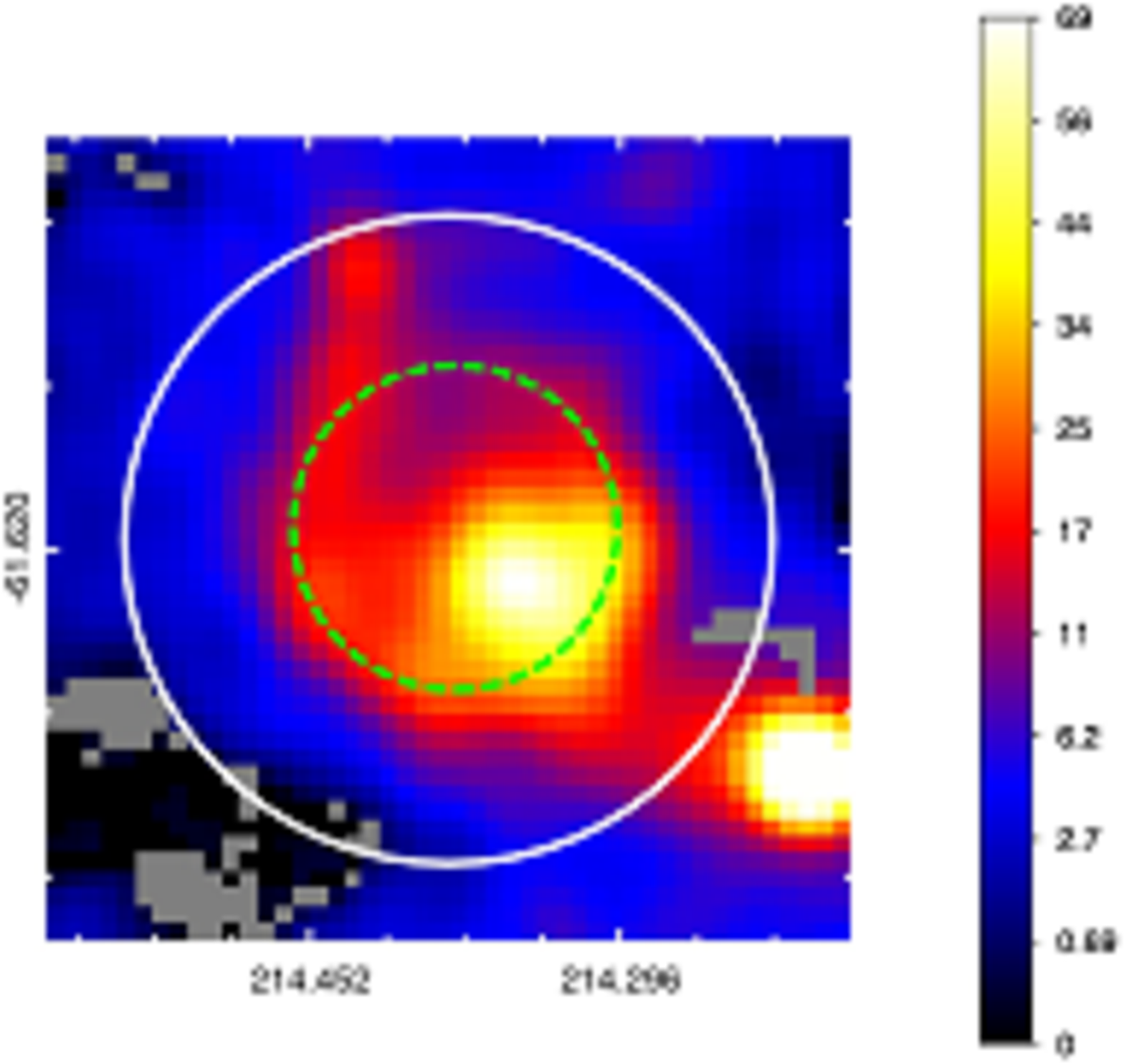}}}%
\mbox{\raisebox{0mm}{\includegraphics[width=40mm]{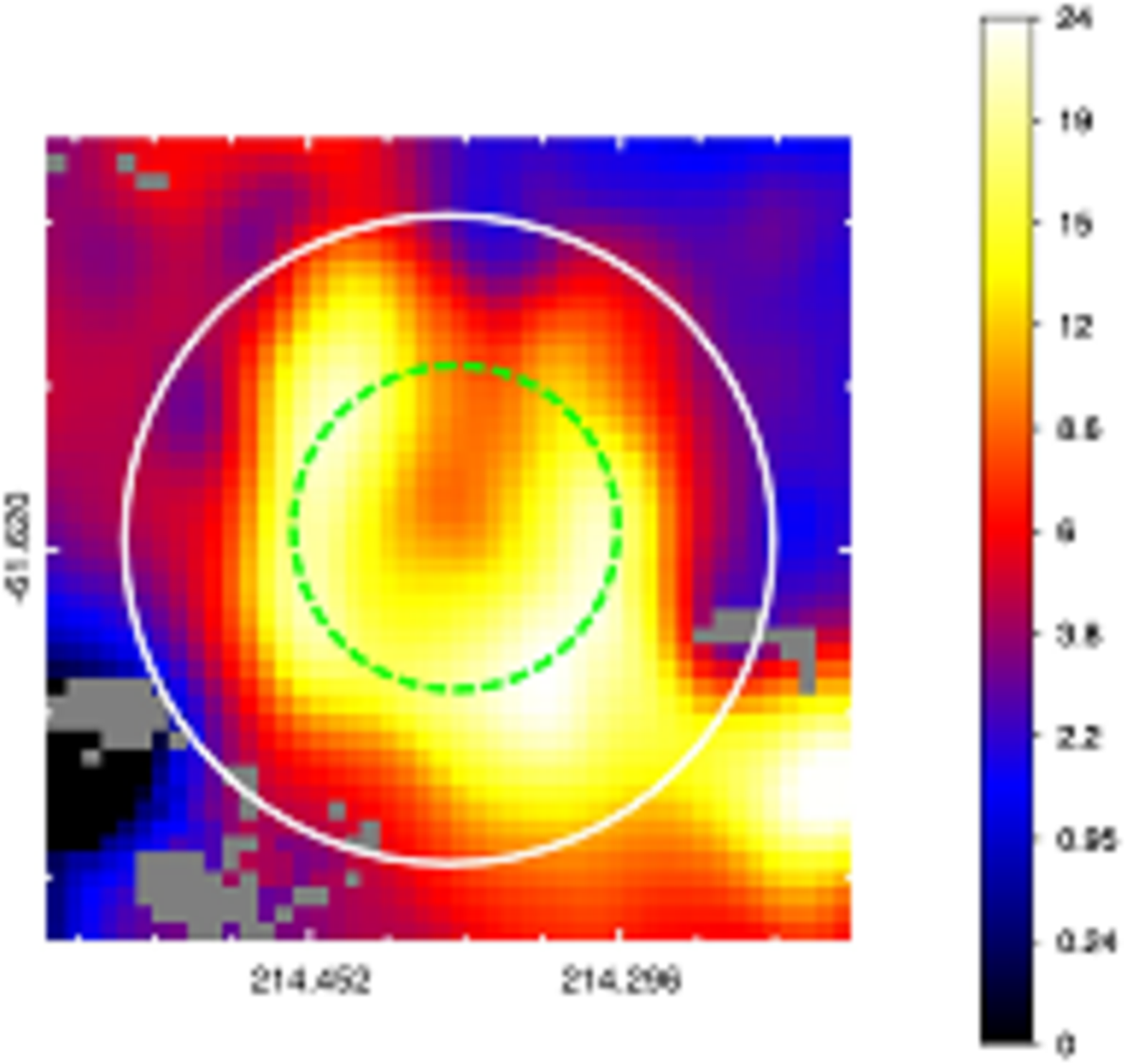}}}%
}
\subfigure{
\mbox{\raisebox{0mm}{\includegraphics[width=40mm]{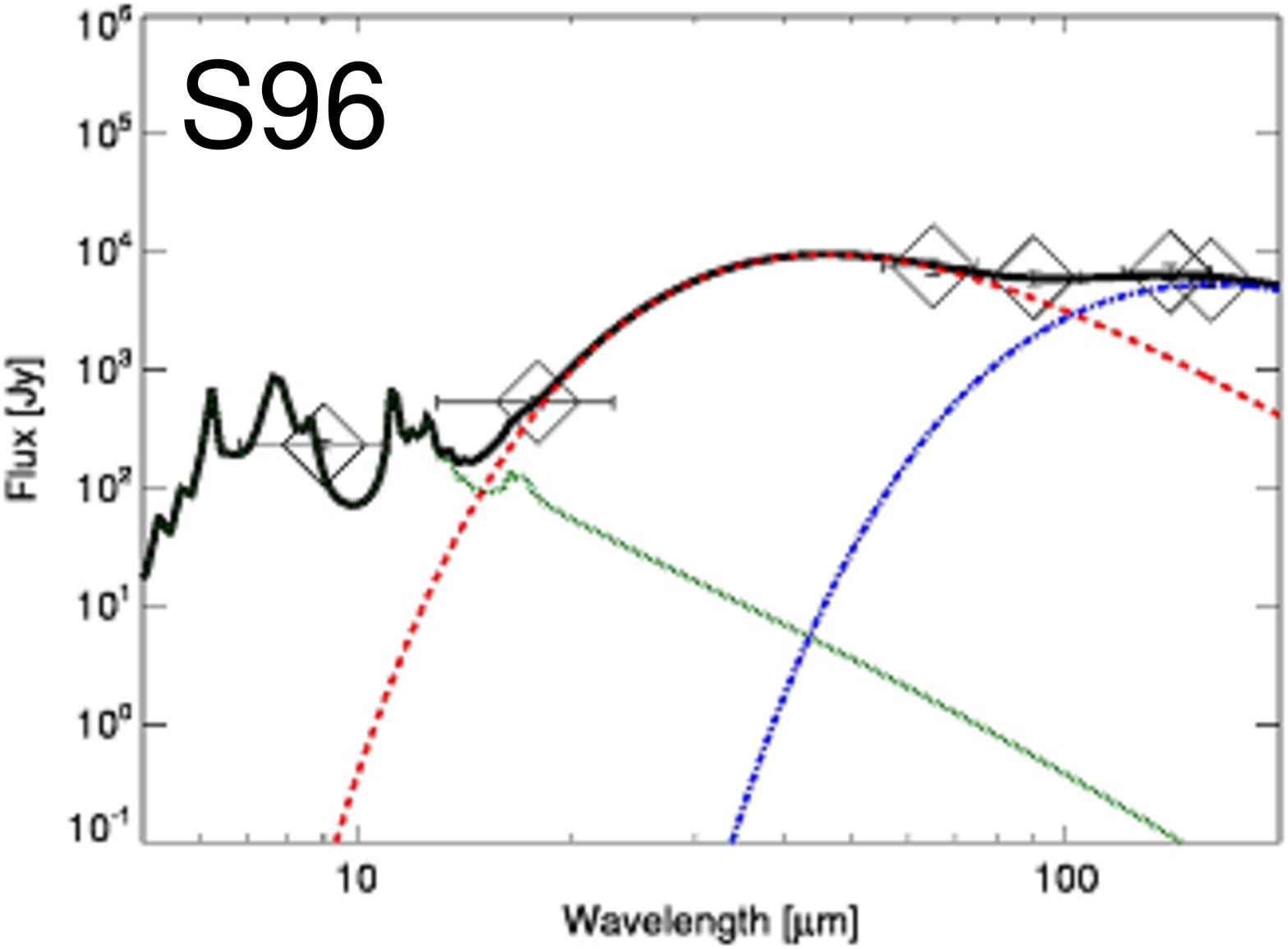}}}%
\mbox{\raisebox{6mm}{\rotatebox{90}{\small{DEC (J2000)}}}}%
\mbox{\raisebox{0mm}{\includegraphics[width=40mm]{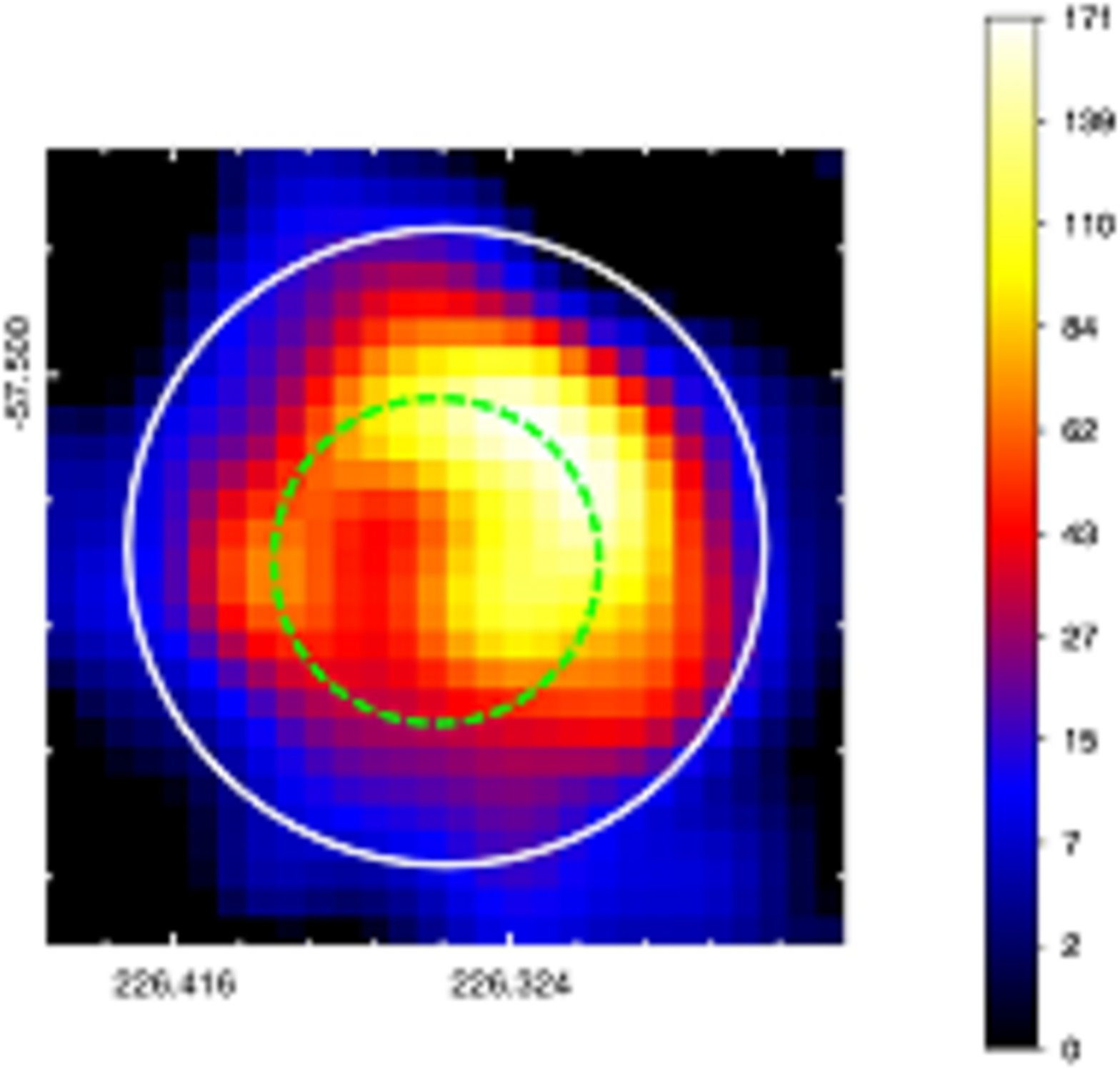}}}%
\mbox{\raisebox{0mm}{\includegraphics[width=40mm]{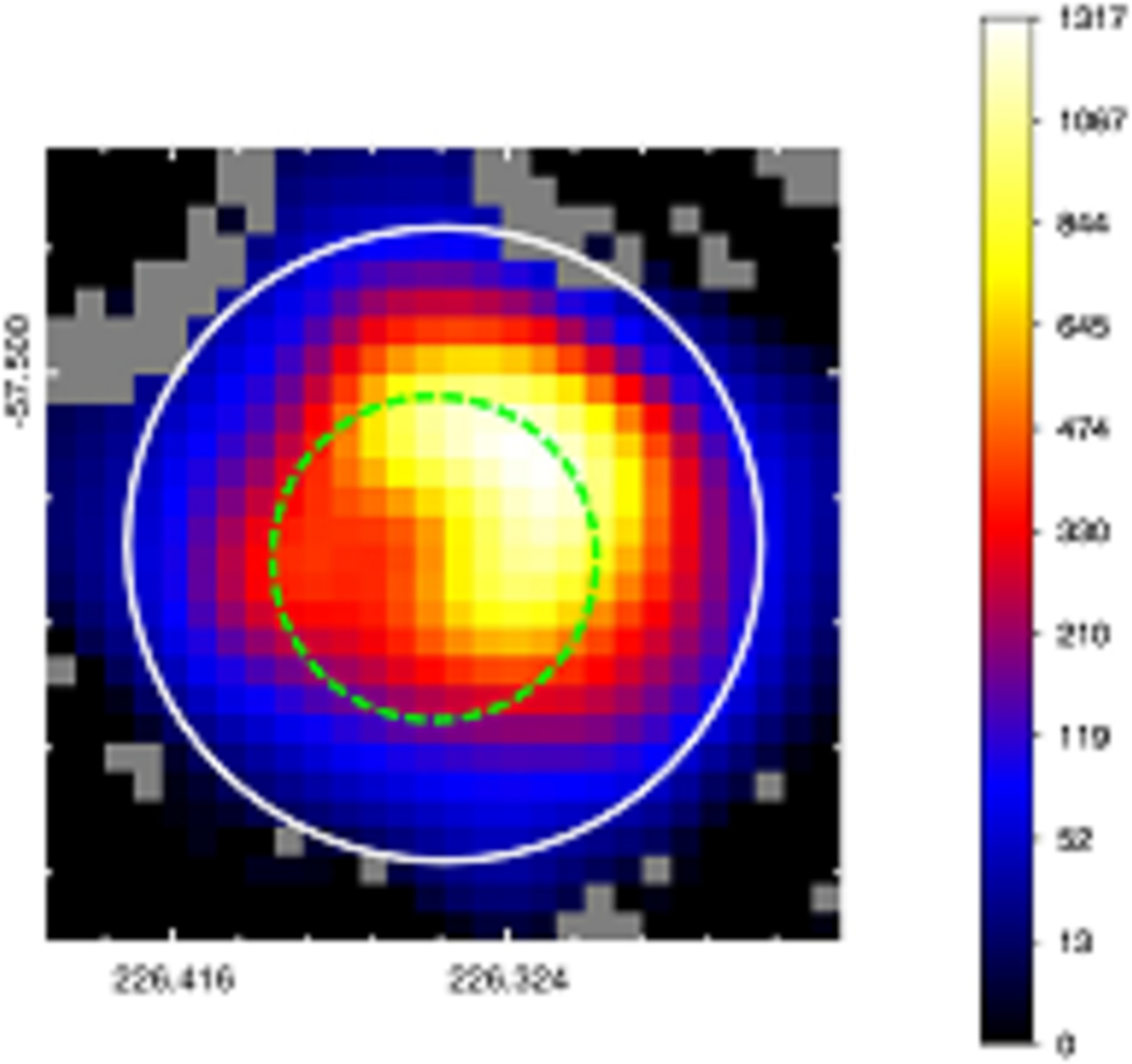}}}%
\mbox{\raisebox{0mm}{\includegraphics[width=40mm]{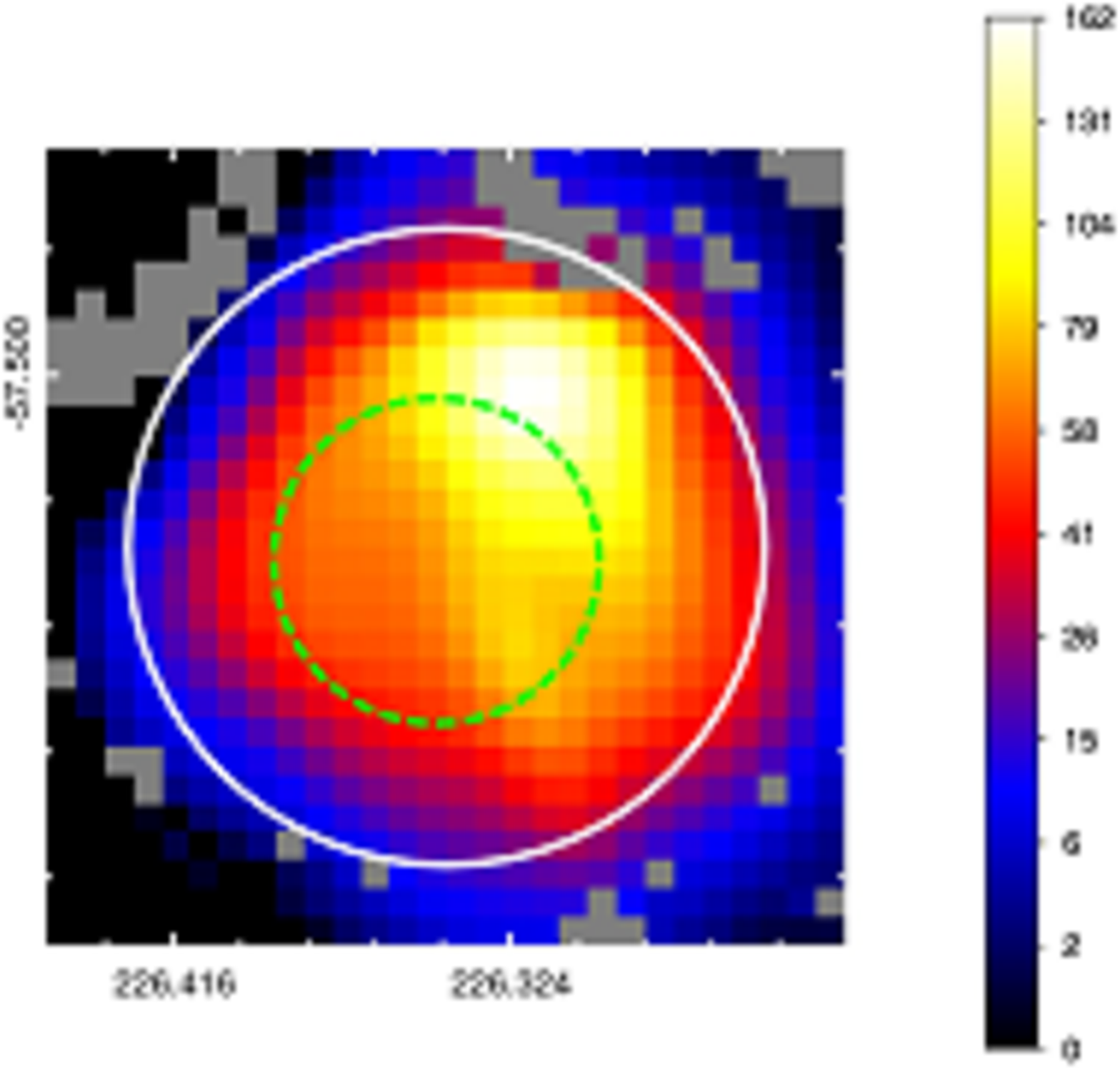}}}%
}
\subfigure{
\mbox{\raisebox{0mm}{\includegraphics[width=40mm]{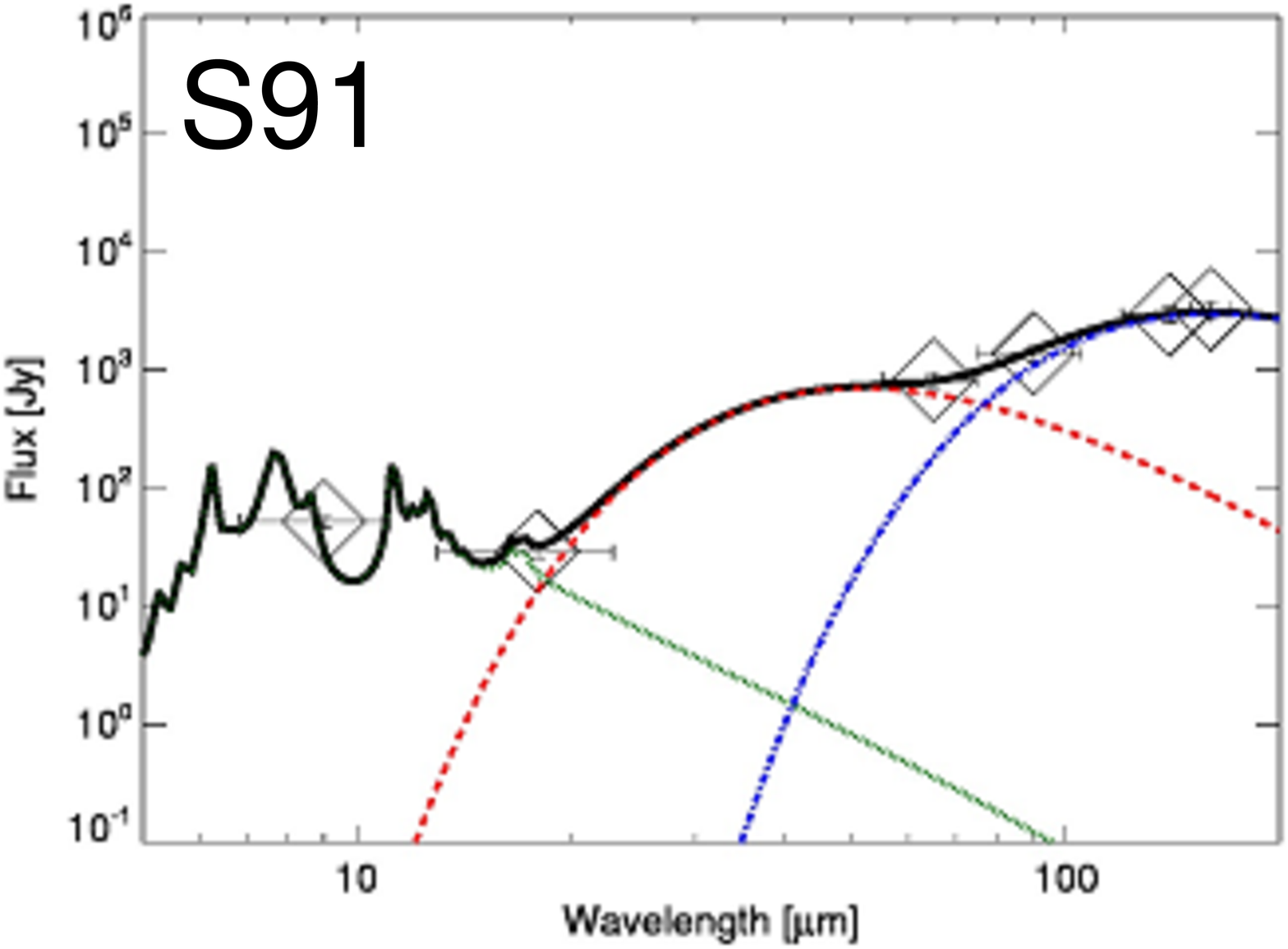}}}%
\mbox{\raisebox{6mm}{\rotatebox{90}{\small{DEC (J2000)}}}}%
\mbox{\raisebox{0mm}{\includegraphics[width=40mm]{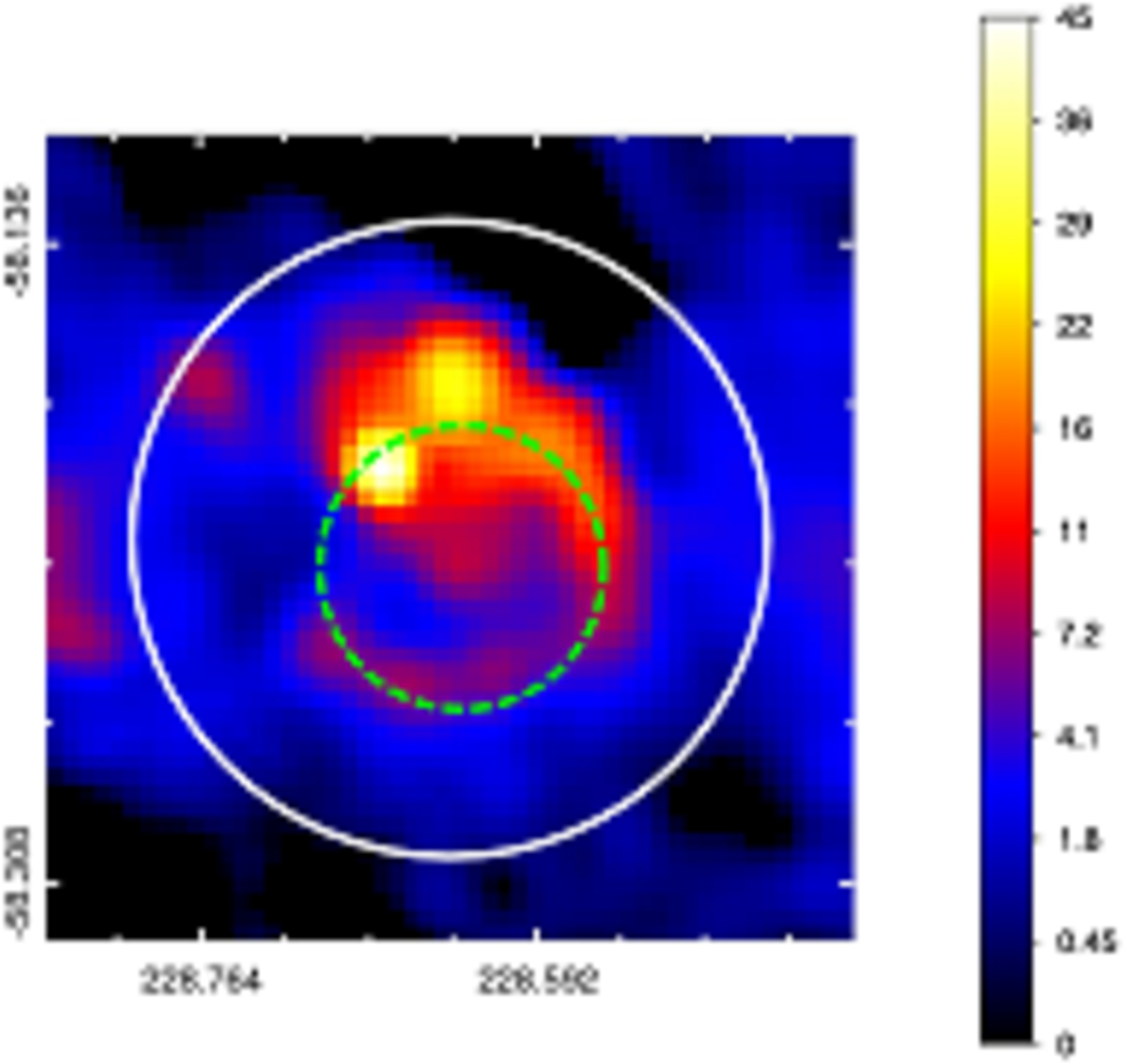}}}%
\mbox{\raisebox{0mm}{\includegraphics[width=40mm]{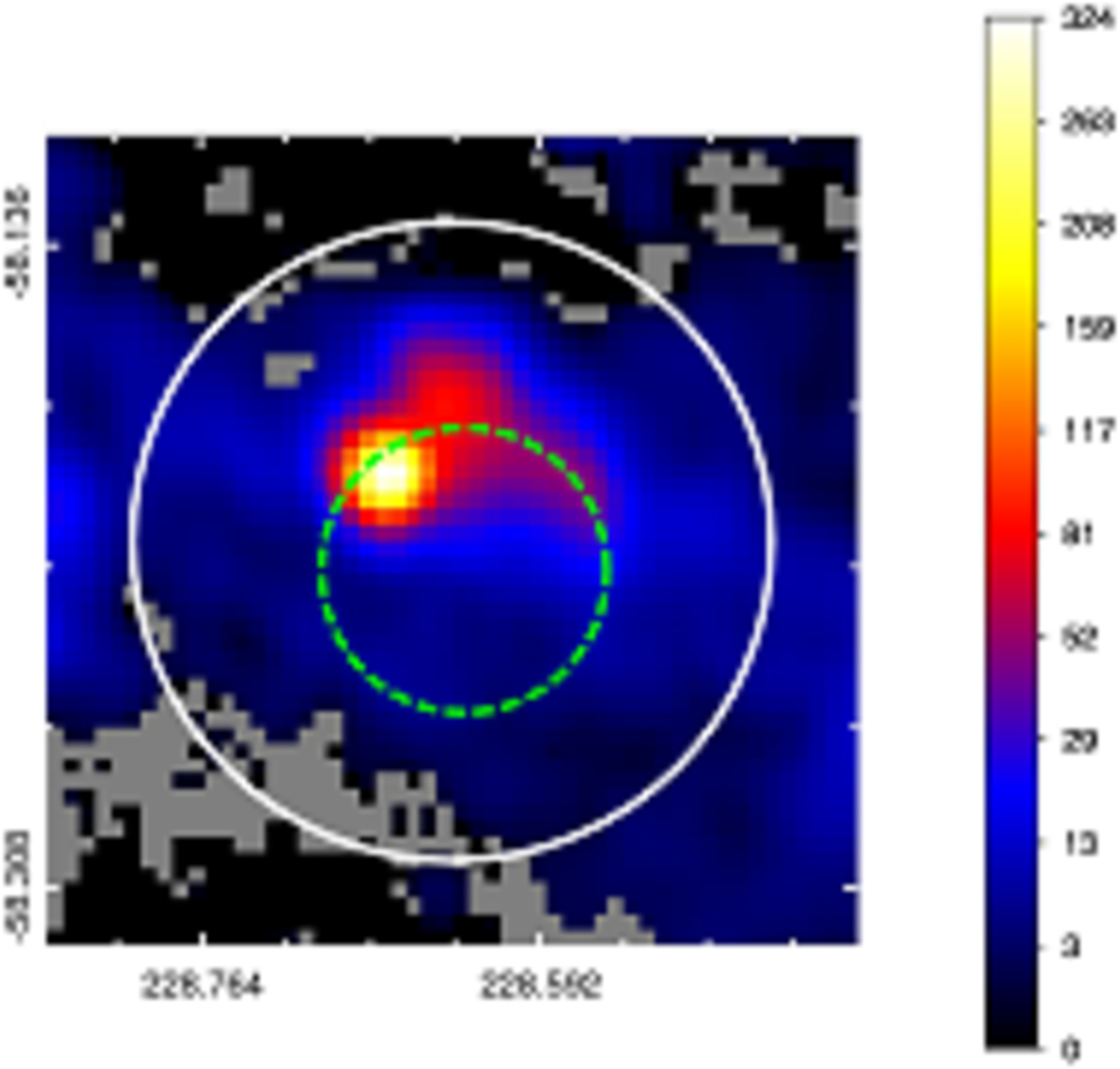}}}%
\mbox{\raisebox{0mm}{\includegraphics[width=40mm]{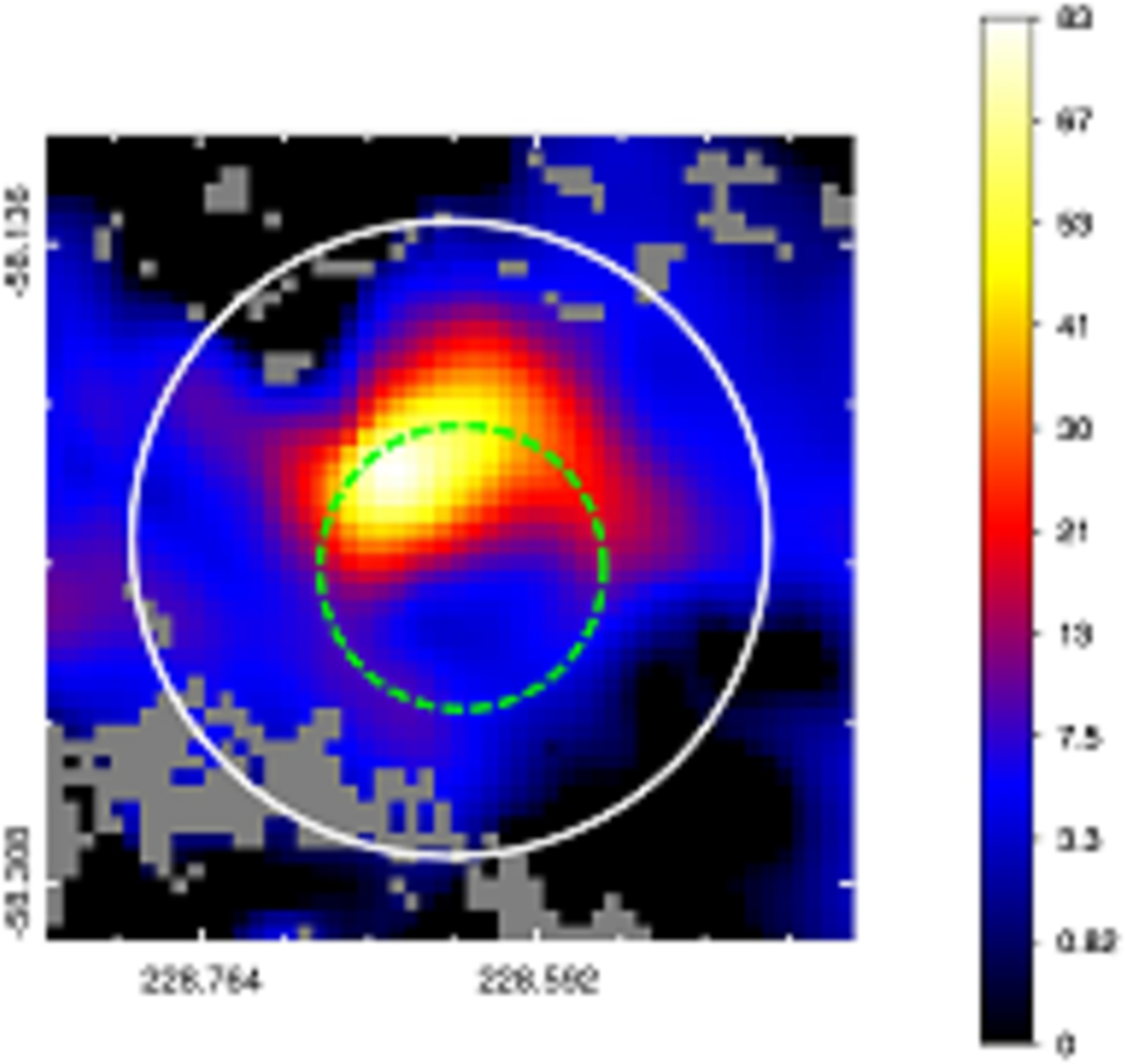}}}%
}
\caption{Continued.} \label{fig:Metfig2:f}
\end{figure*}

\addtocounter{figure}{-1}
\begin{figure*}[ht]
\addtocounter{subfigure}{1}
\centering
\subfigure{
\makebox[180mm][l]{\raisebox{0mm}[0mm][0mm]{ \hspace{20mm} \small{SED}} \hspace{27.5mm} \small{$I_{\rm{PAH}}$} \hspace{29.5mm} \small{$I_{\rm{warm}}$} \hspace{29.5mm} \small{$I_{\rm{cold}}$}}%
}
\subfigure{
\makebox[180mm][l]{\raisebox{0mm}{\hspace{52mm} \small{RA (J2000)} \hspace{20mm} \small{RA (J2000)} \hspace{20mm} \small{RA (J2000)}}}
}
\subfigure{
\mbox{\raisebox{0mm}{\includegraphics[width=40mm]{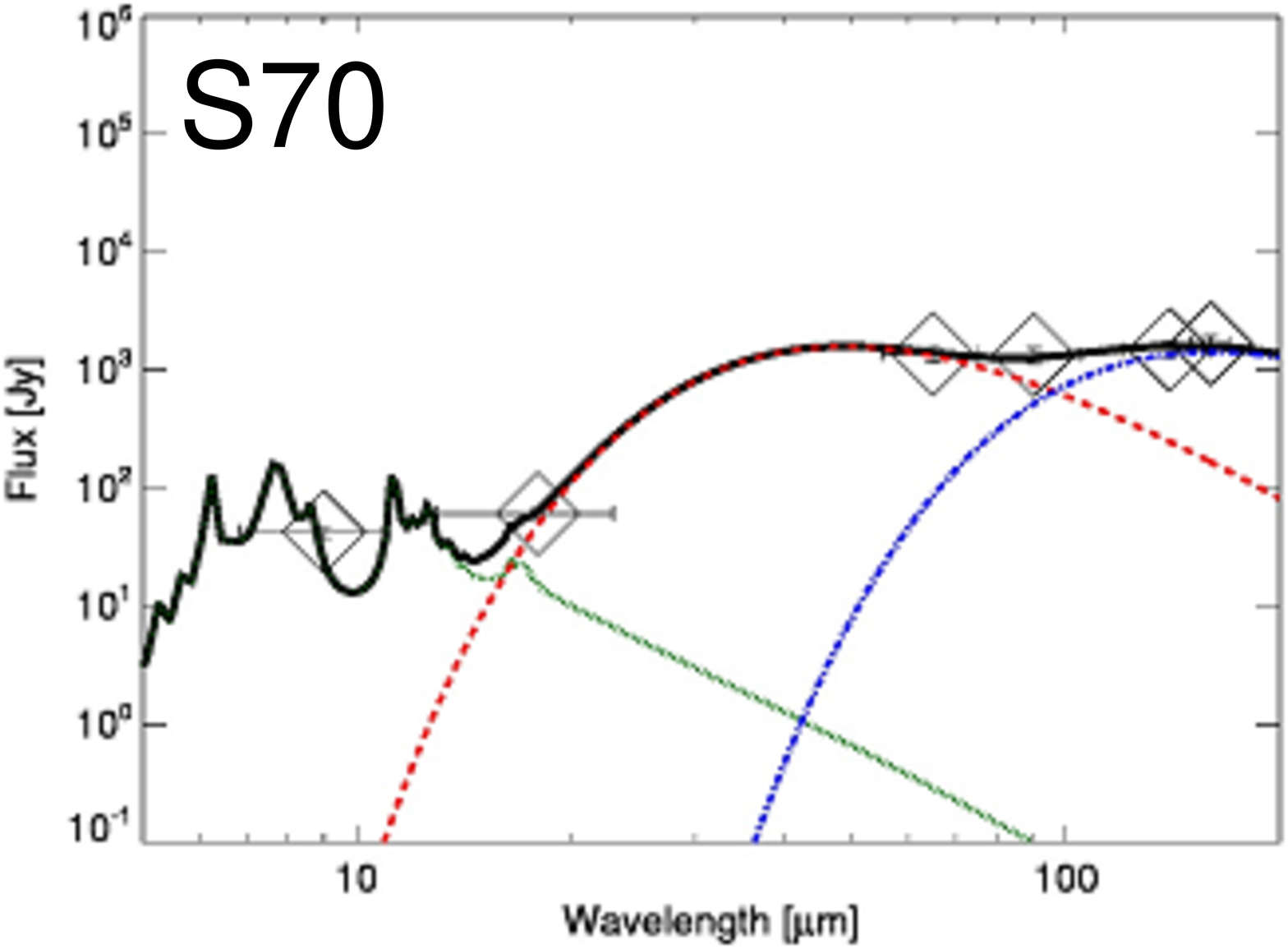}}}%
\mbox{\raisebox{6mm}{\rotatebox{90}{\small{DEC (J2000)}}}}%
\mbox{\raisebox{0mm}{\includegraphics[width=40mm]{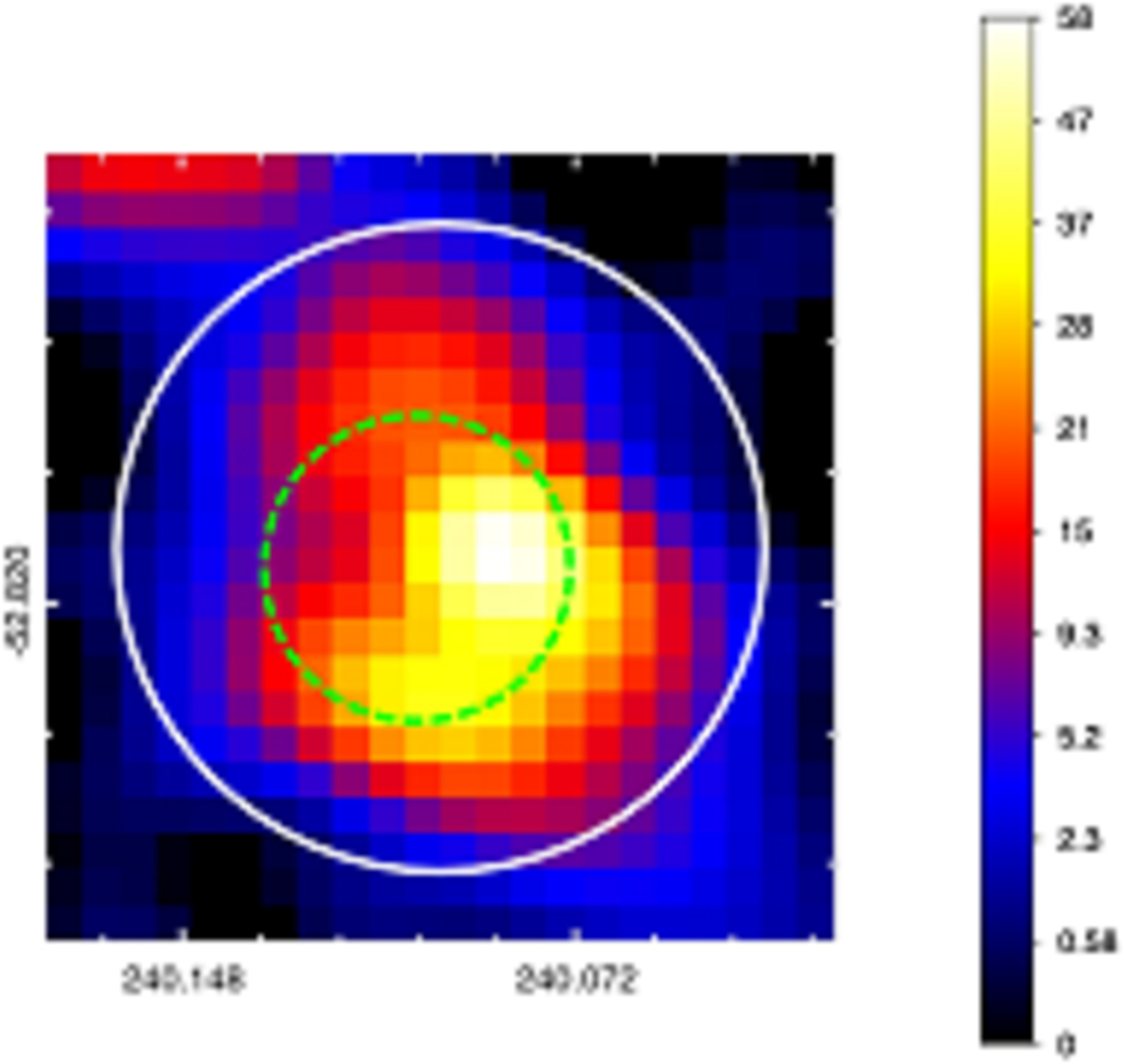}}}%
\mbox{\raisebox{0mm}{\includegraphics[width=40mm]{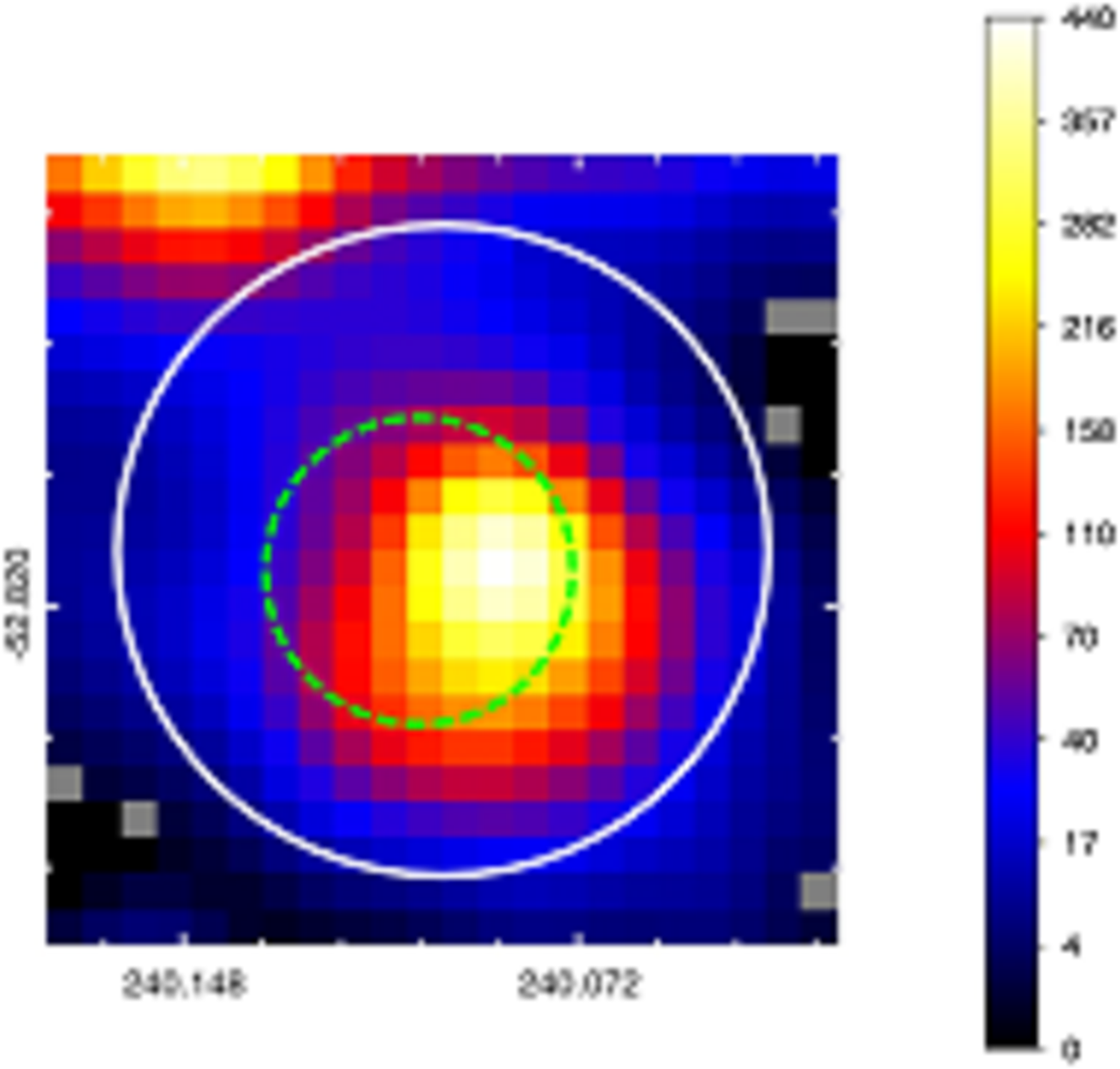}}}%
\mbox{\raisebox{0mm}{\includegraphics[width=40mm]{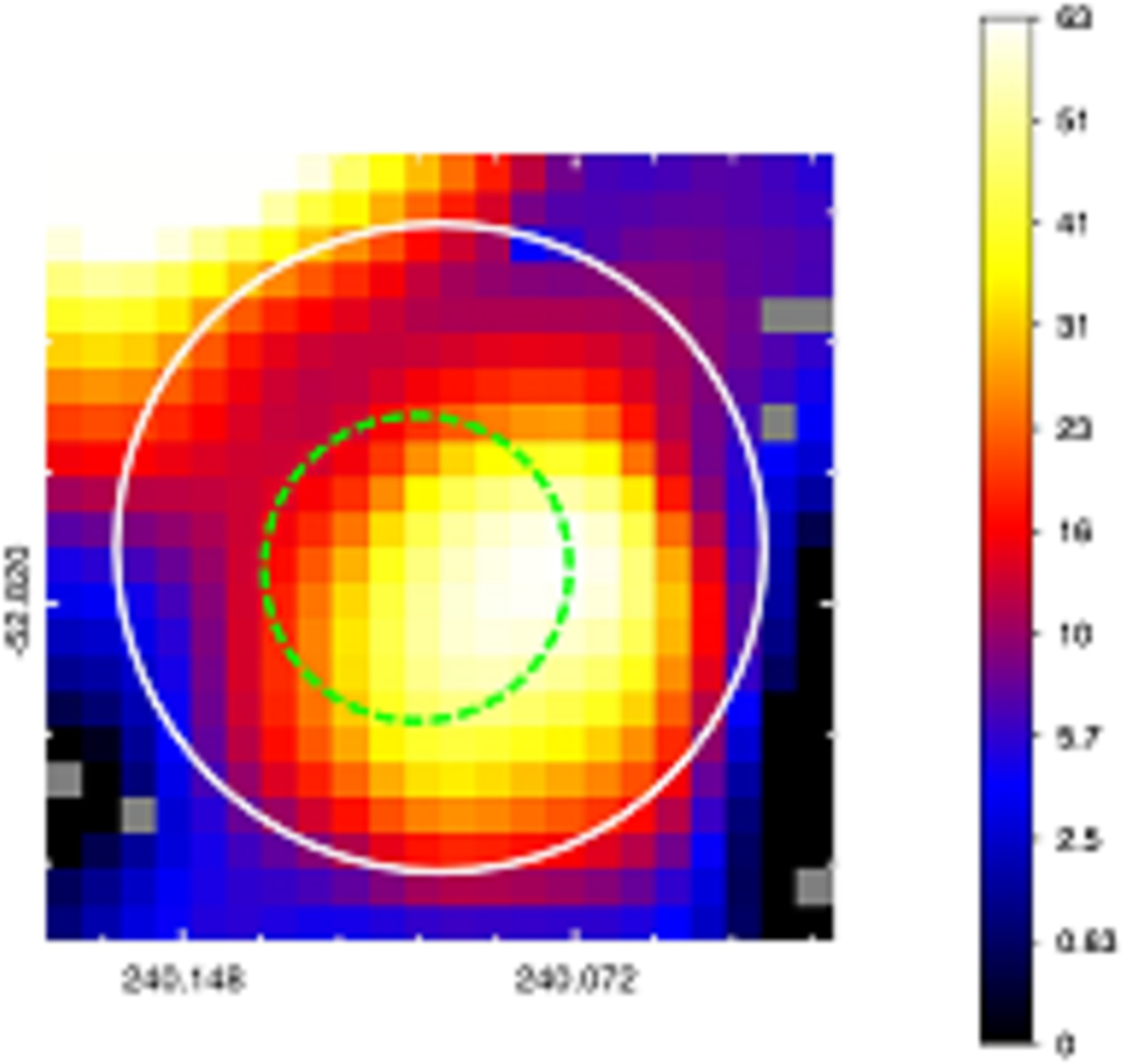}}}%
}
\subfigure{
\mbox{\raisebox{0mm}{\includegraphics[width=40mm]{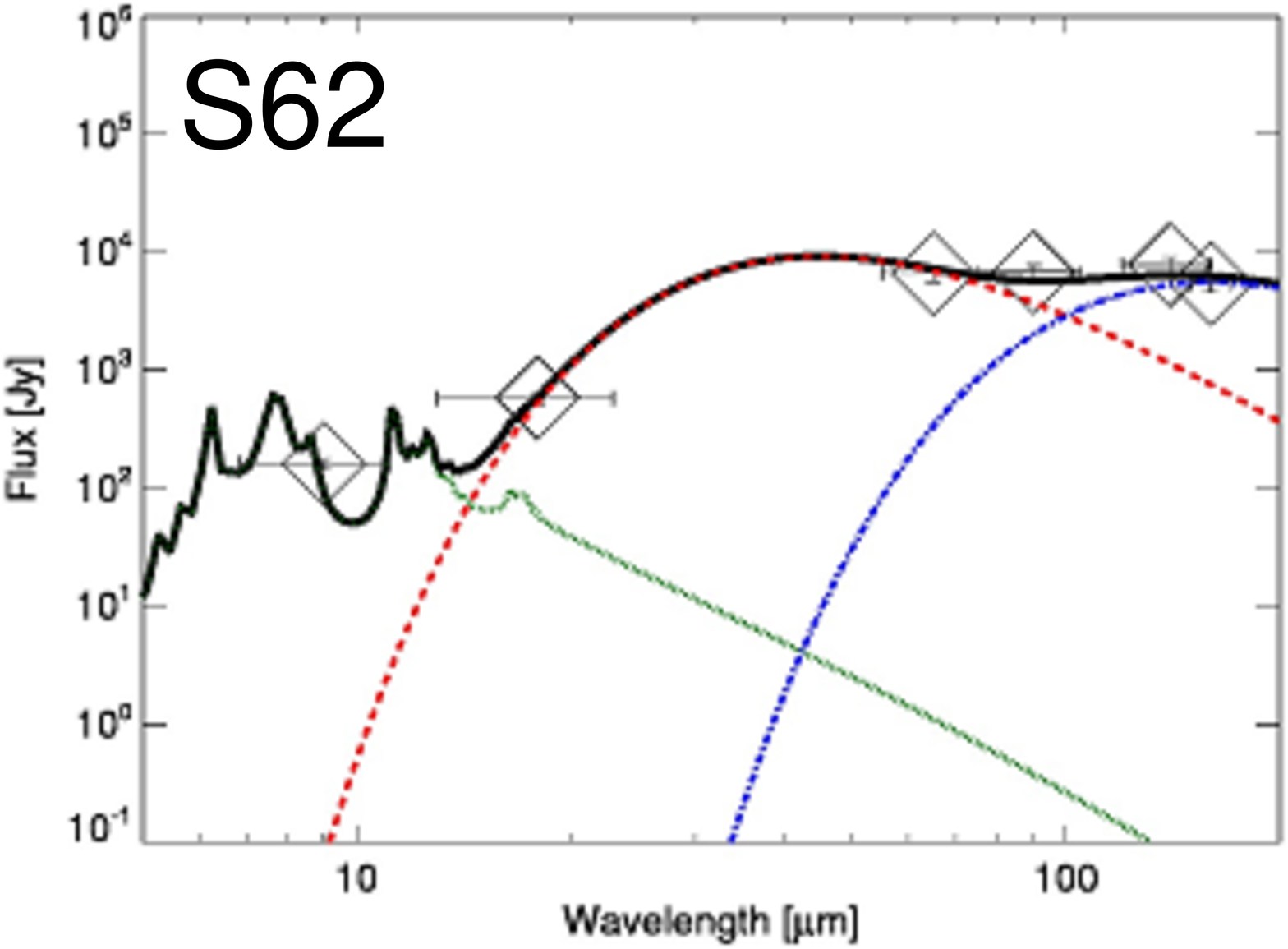}}}%
\mbox{\raisebox{6mm}{\rotatebox{90}{\small{DEC (J2000)}}}}%
\mbox{\raisebox{0mm}{\includegraphics[width=40mm]{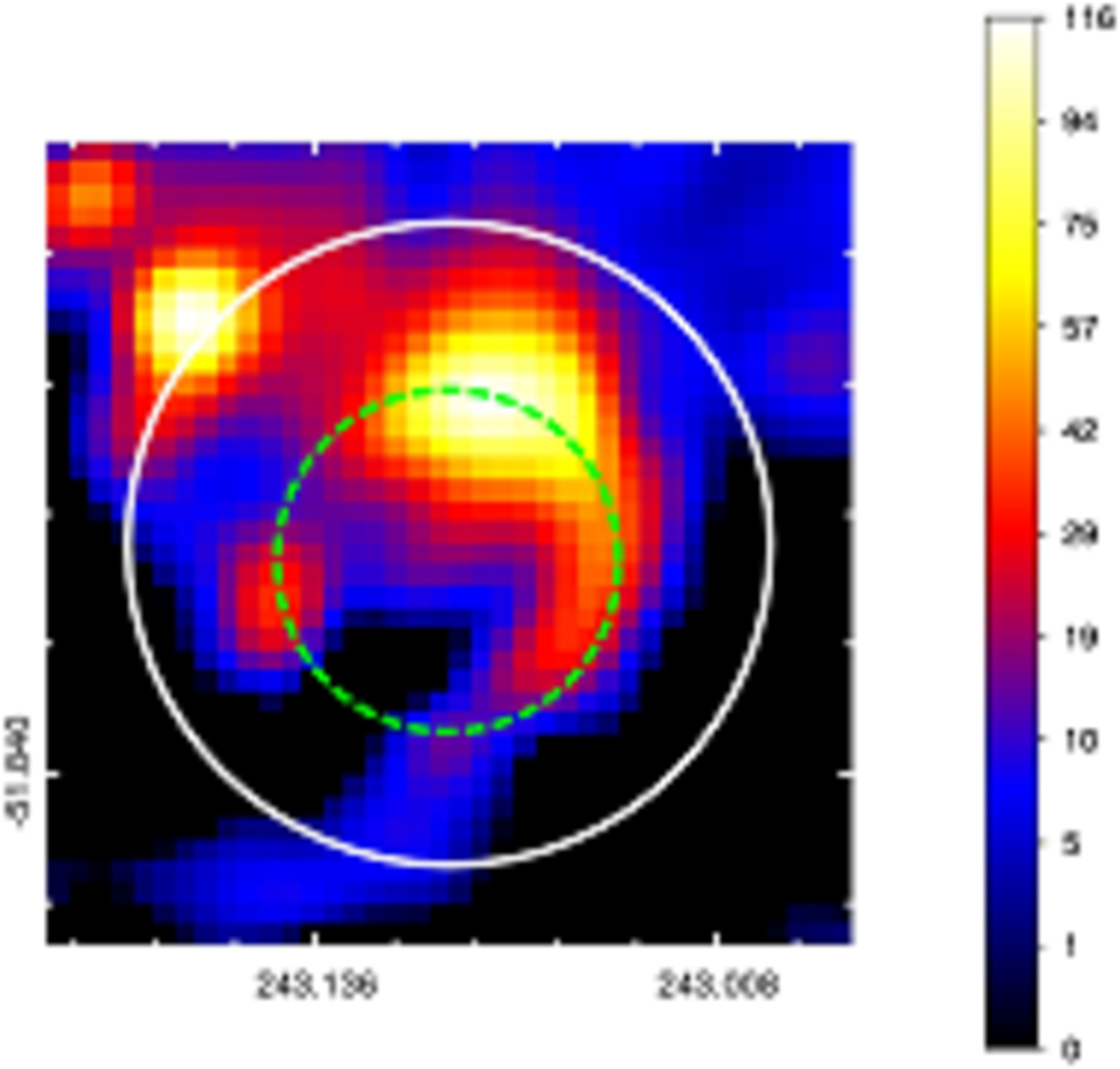}}}%
\mbox{\raisebox{0mm}{\includegraphics[width=40mm]{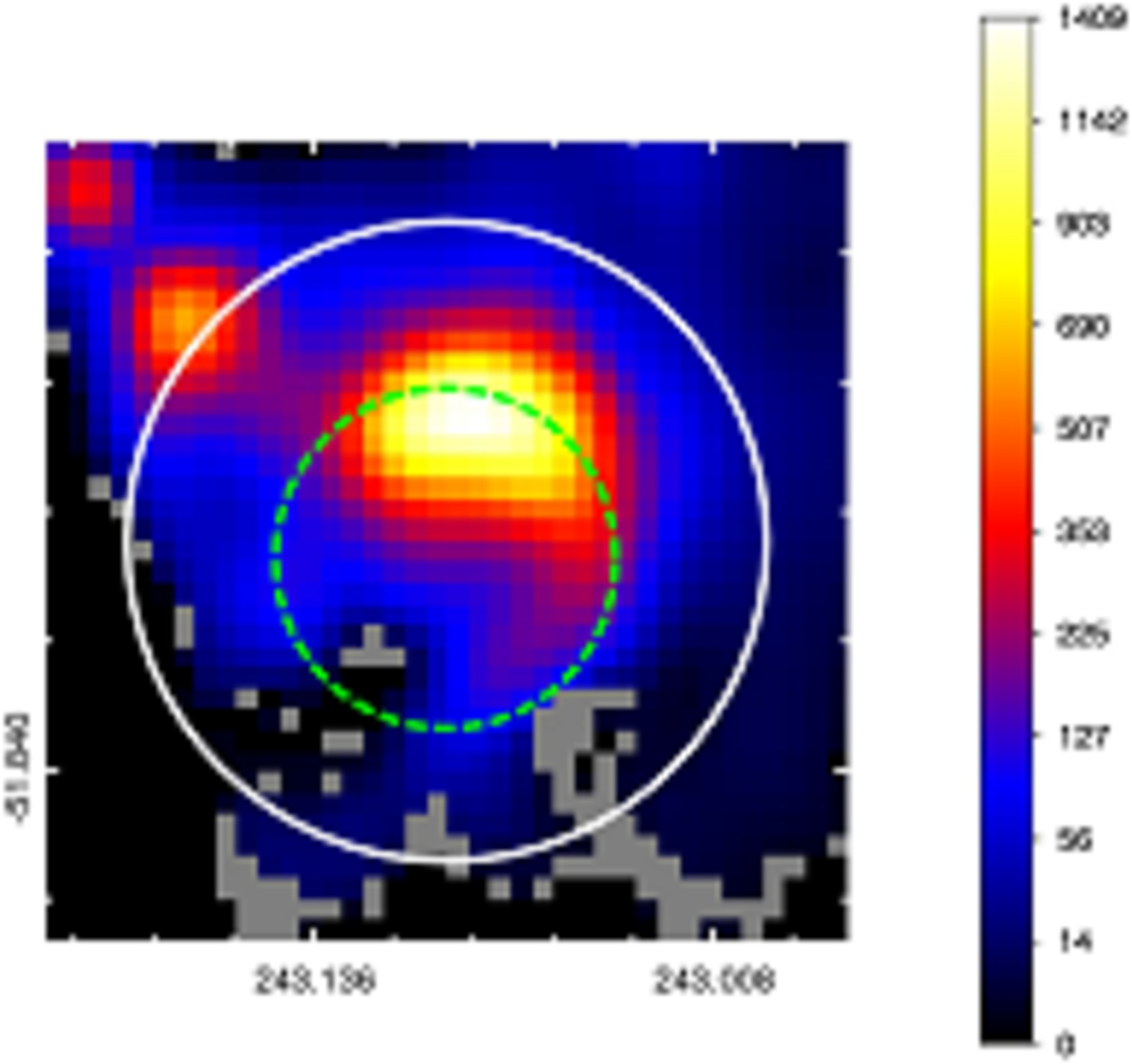}}}%
\mbox{\raisebox{0mm}{\includegraphics[width=40mm]{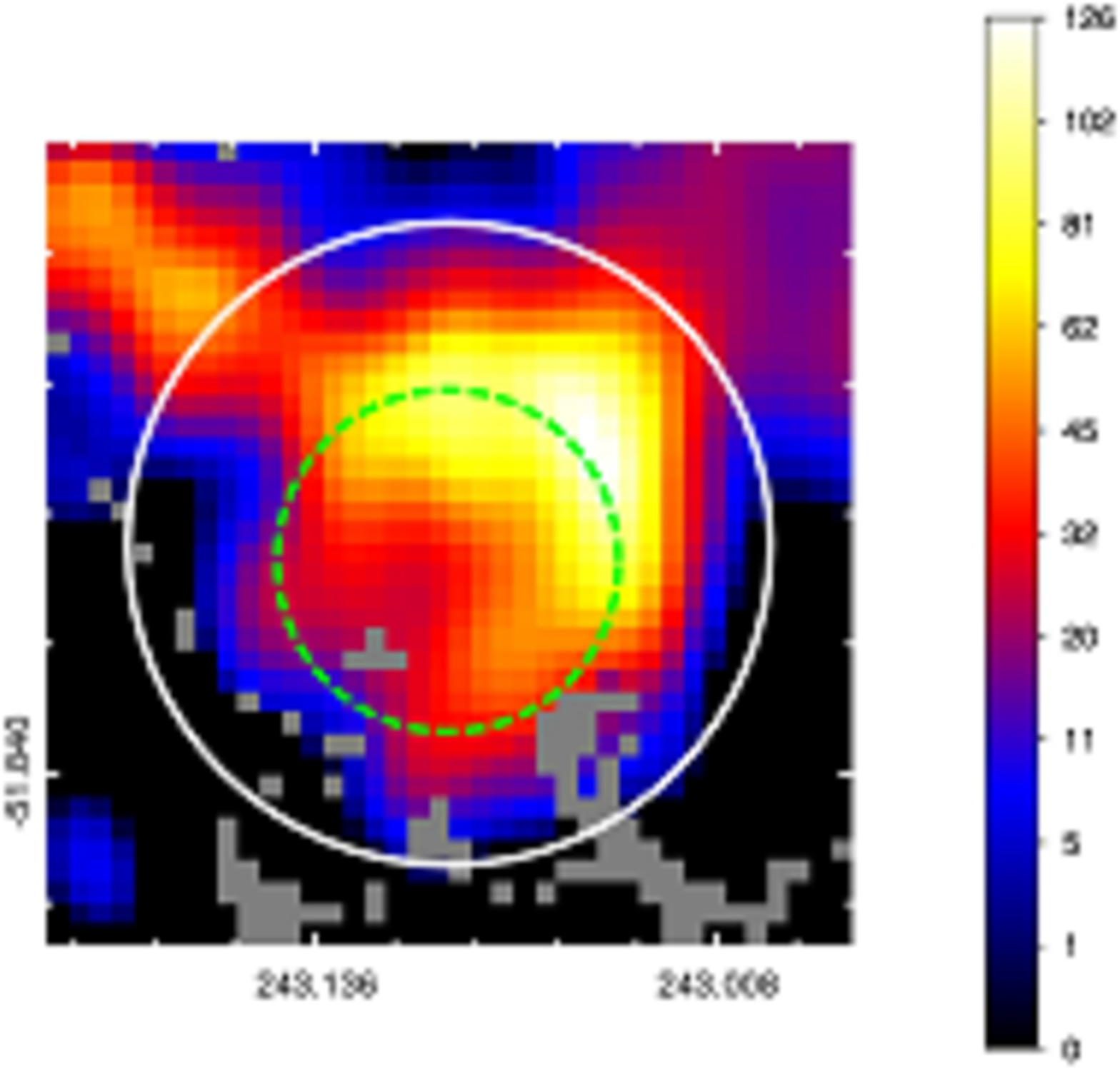}}}%
}
\subfigure{
\mbox{\raisebox{0mm}{\includegraphics[width=40mm]{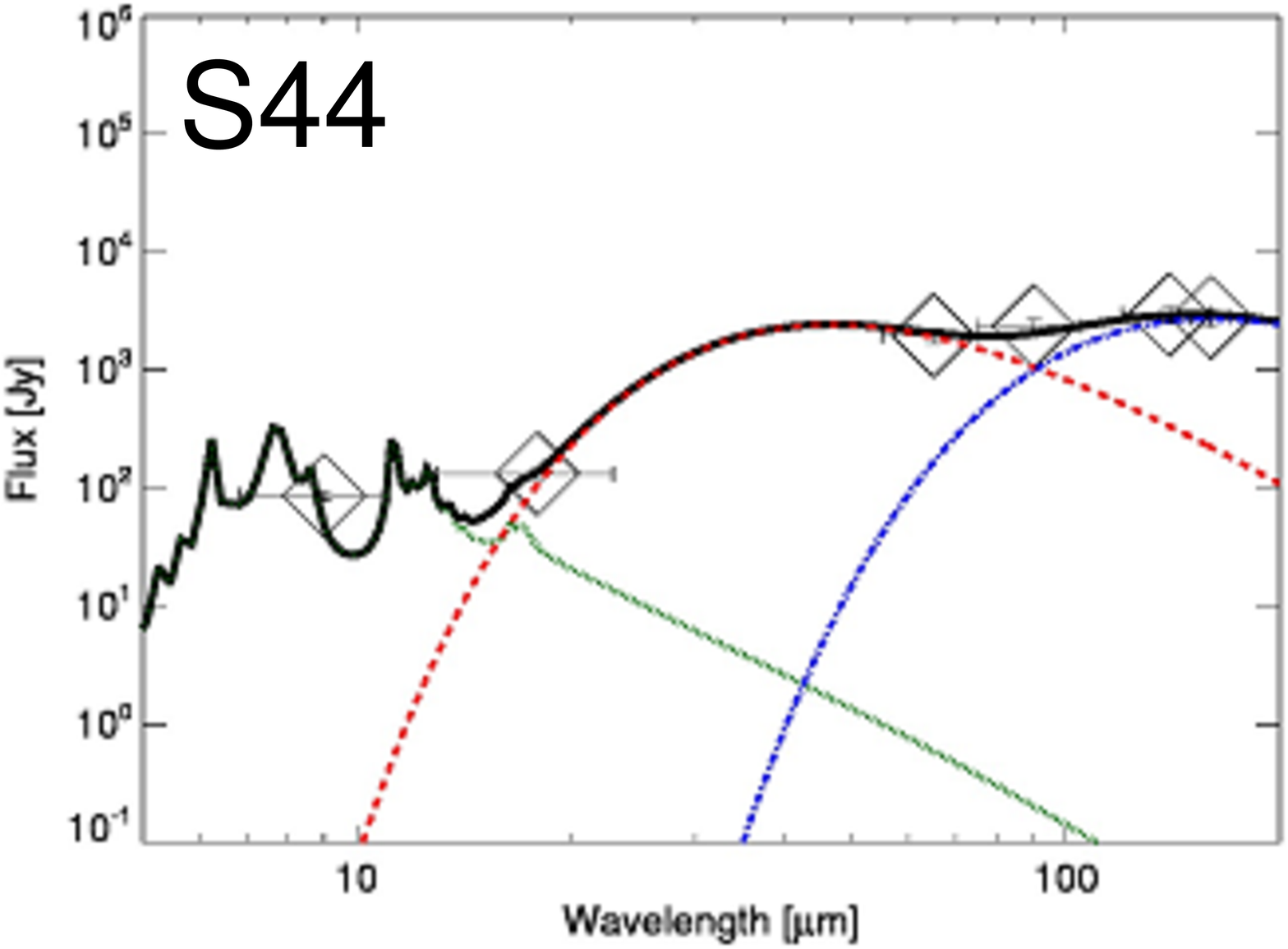}}}%
\mbox{\raisebox{6mm}{\rotatebox{90}{\small{DEC (J2000)}}}}%
\mbox{\raisebox{0mm}{\includegraphics[width=40mm]{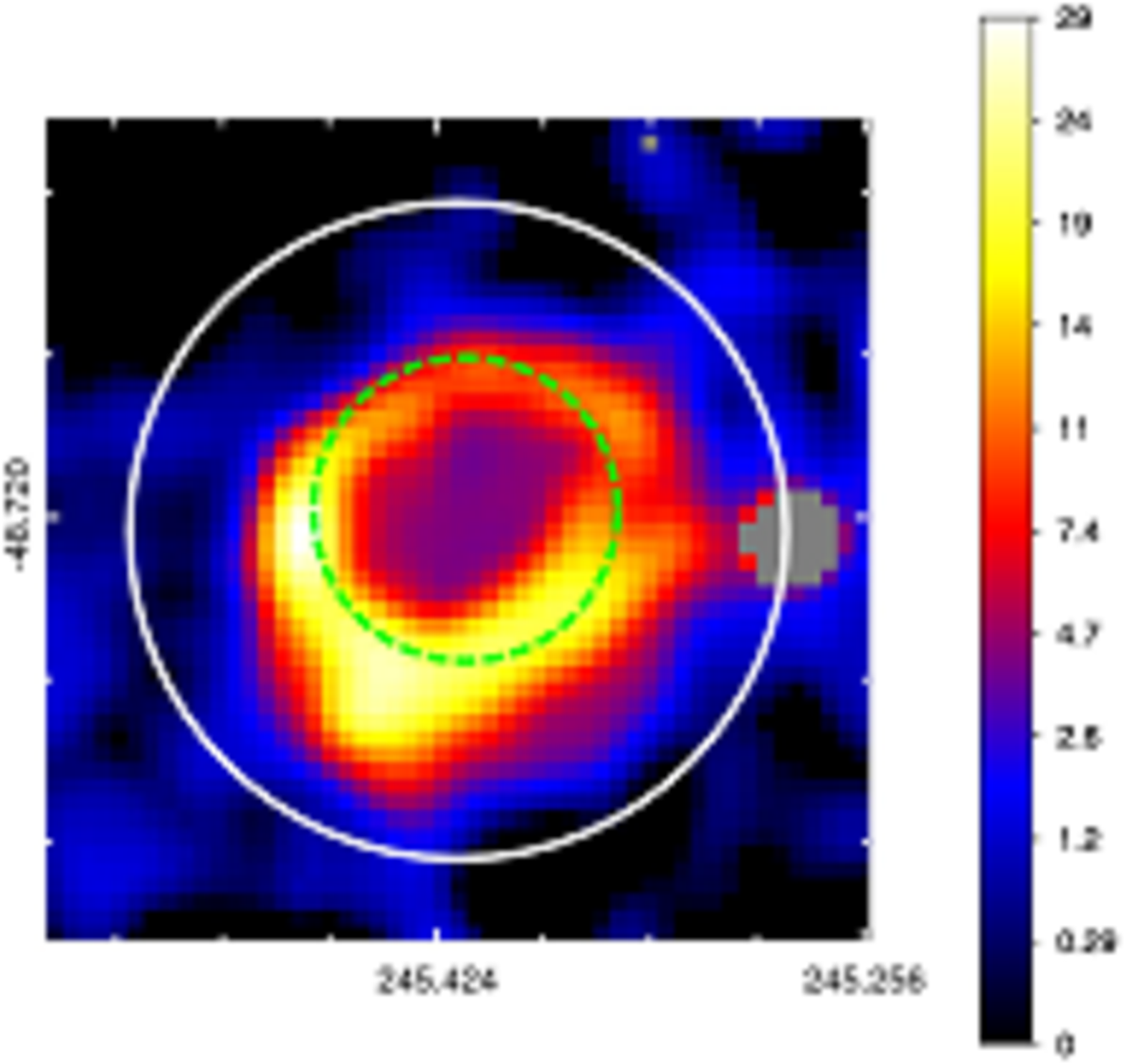}}}%
\mbox{\raisebox{0mm}{\includegraphics[width=40mm]{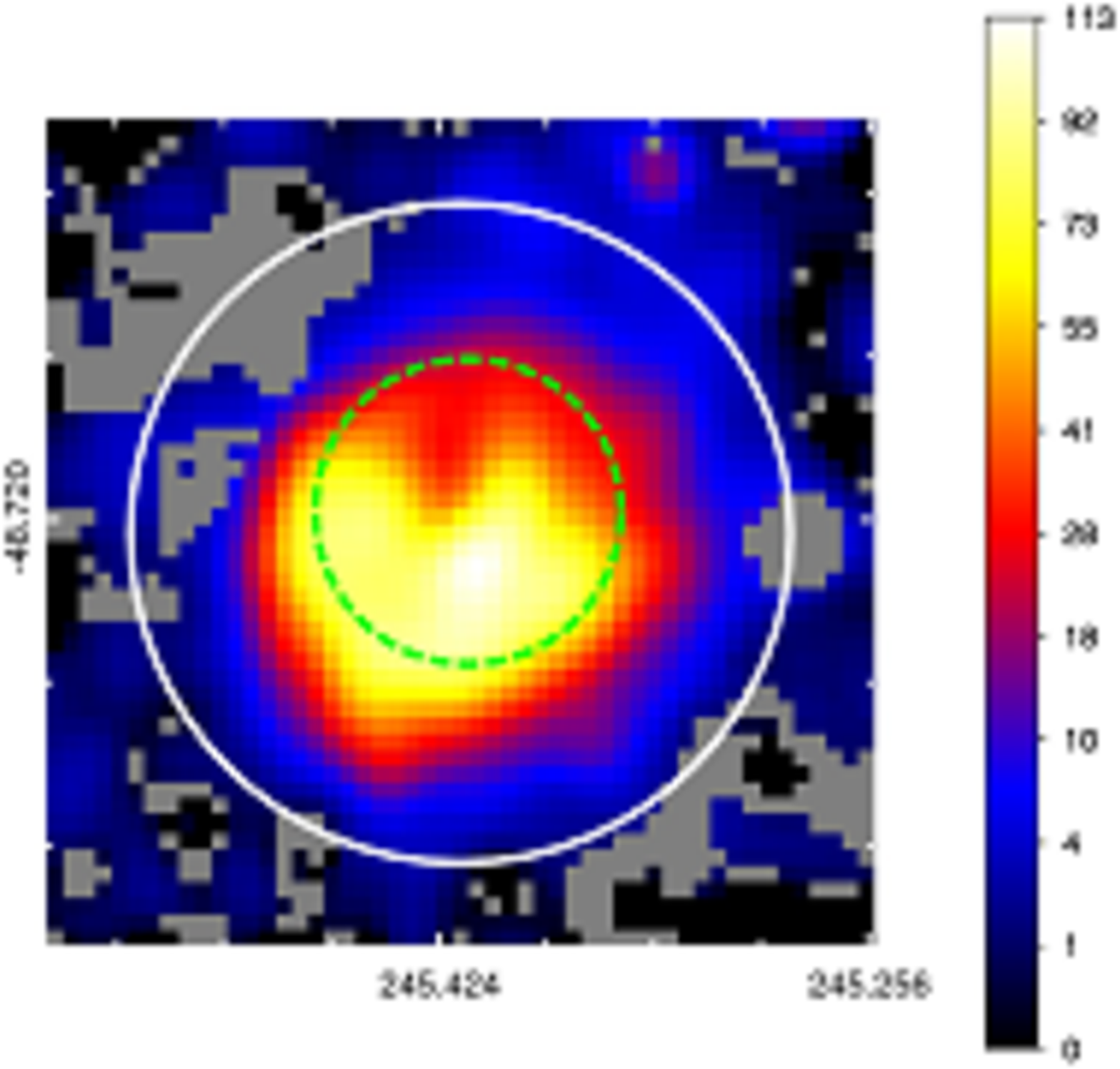}}}%
\mbox{\raisebox{0mm}{\includegraphics[width=40mm]{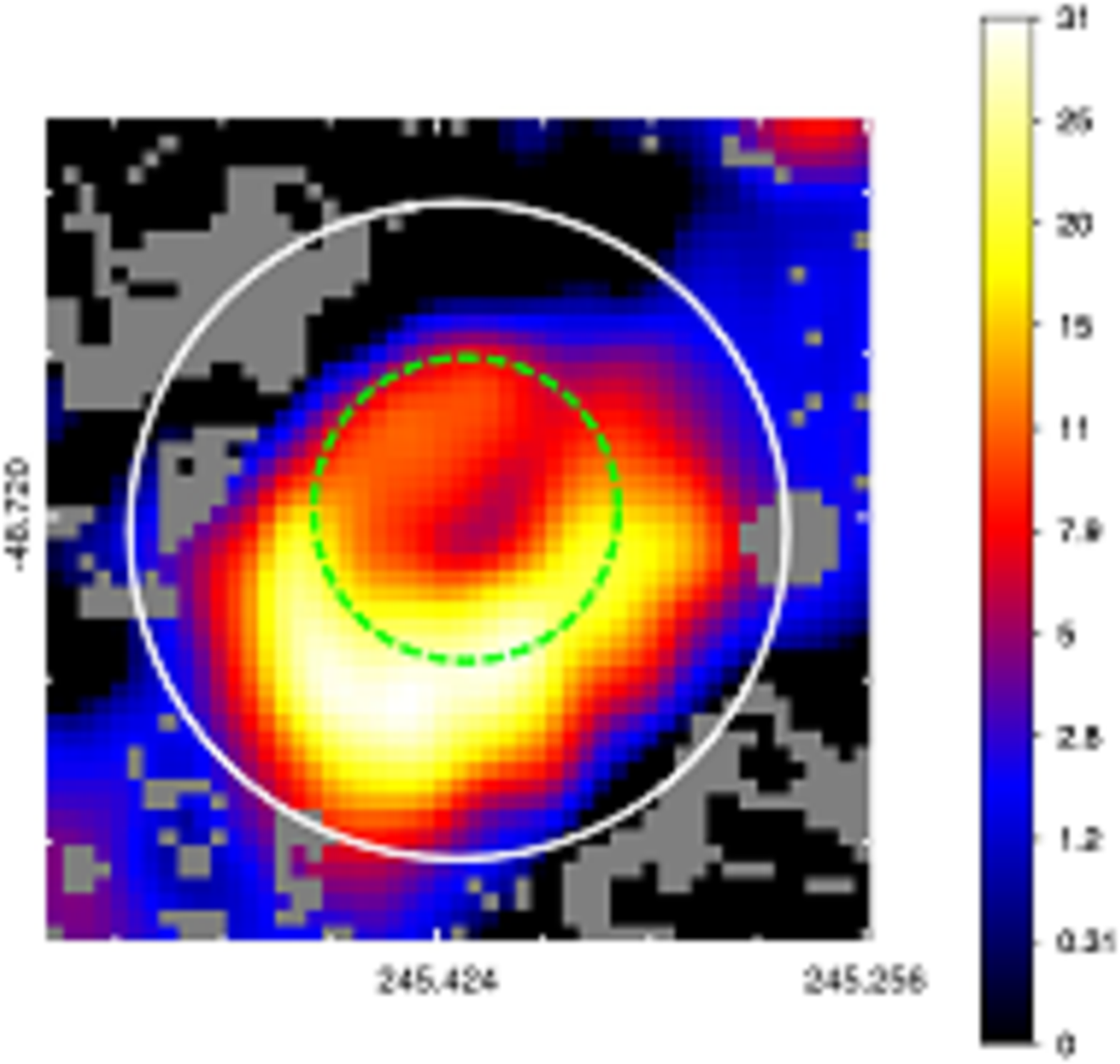}}}%
}
\subfigure{
\mbox{\raisebox{0mm}{\includegraphics[width=40mm]{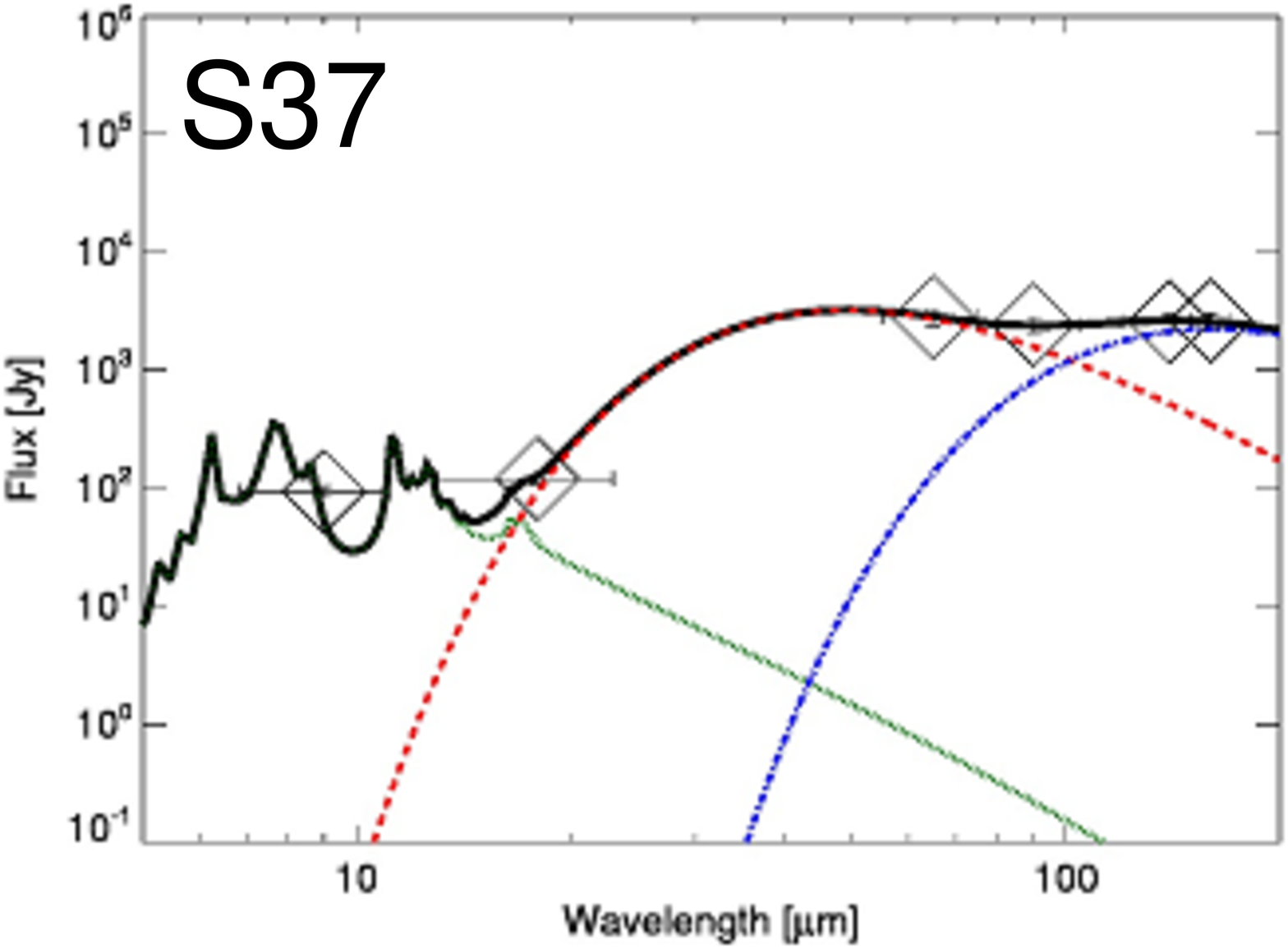}}}%
\mbox{\raisebox{6mm}{\rotatebox{90}{\small{DEC (J2000)}}}}%
\mbox{\raisebox{0mm}{\includegraphics[width=40mm]{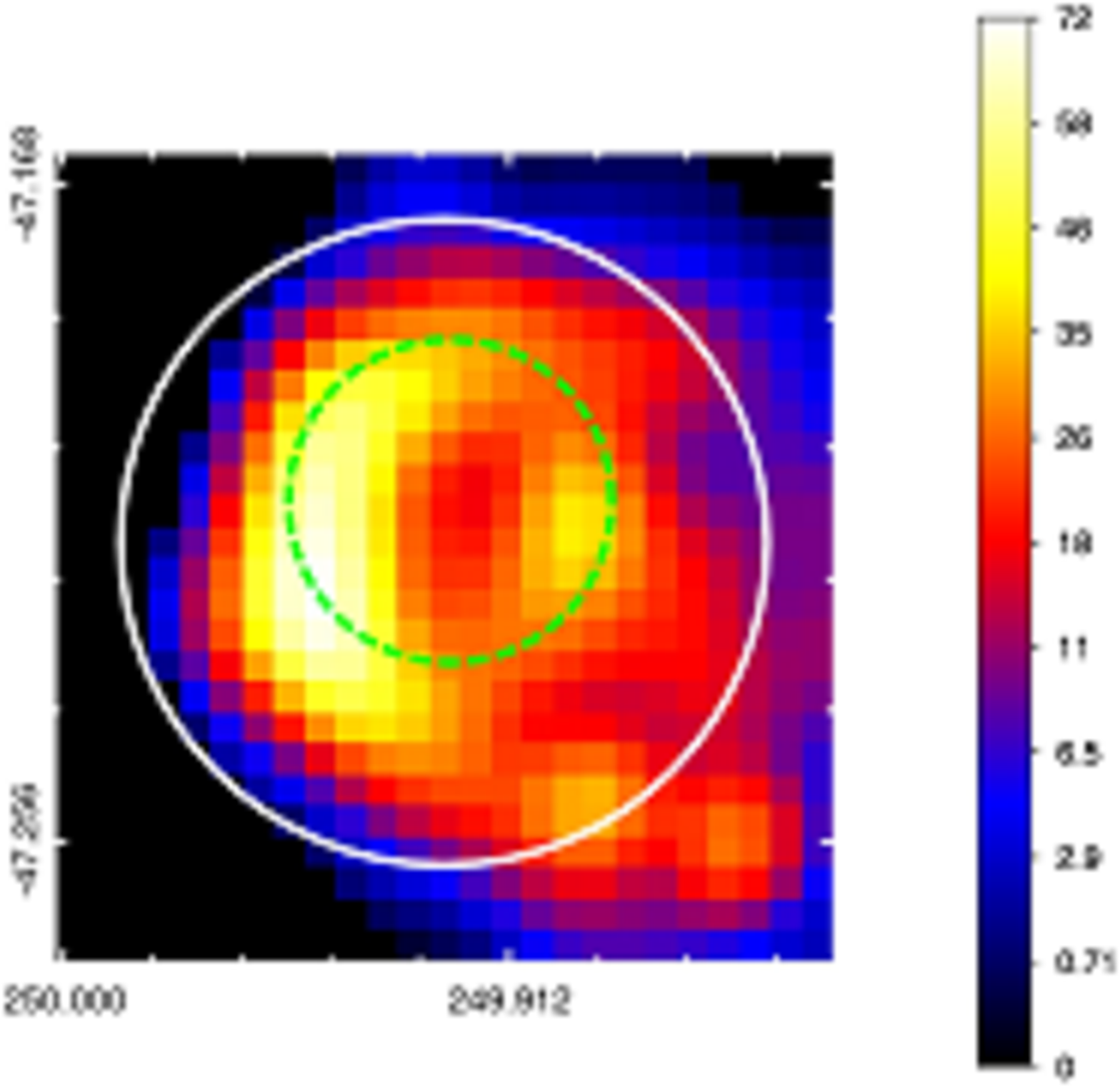}}}%
\mbox{\raisebox{0mm}{\includegraphics[width=40mm]{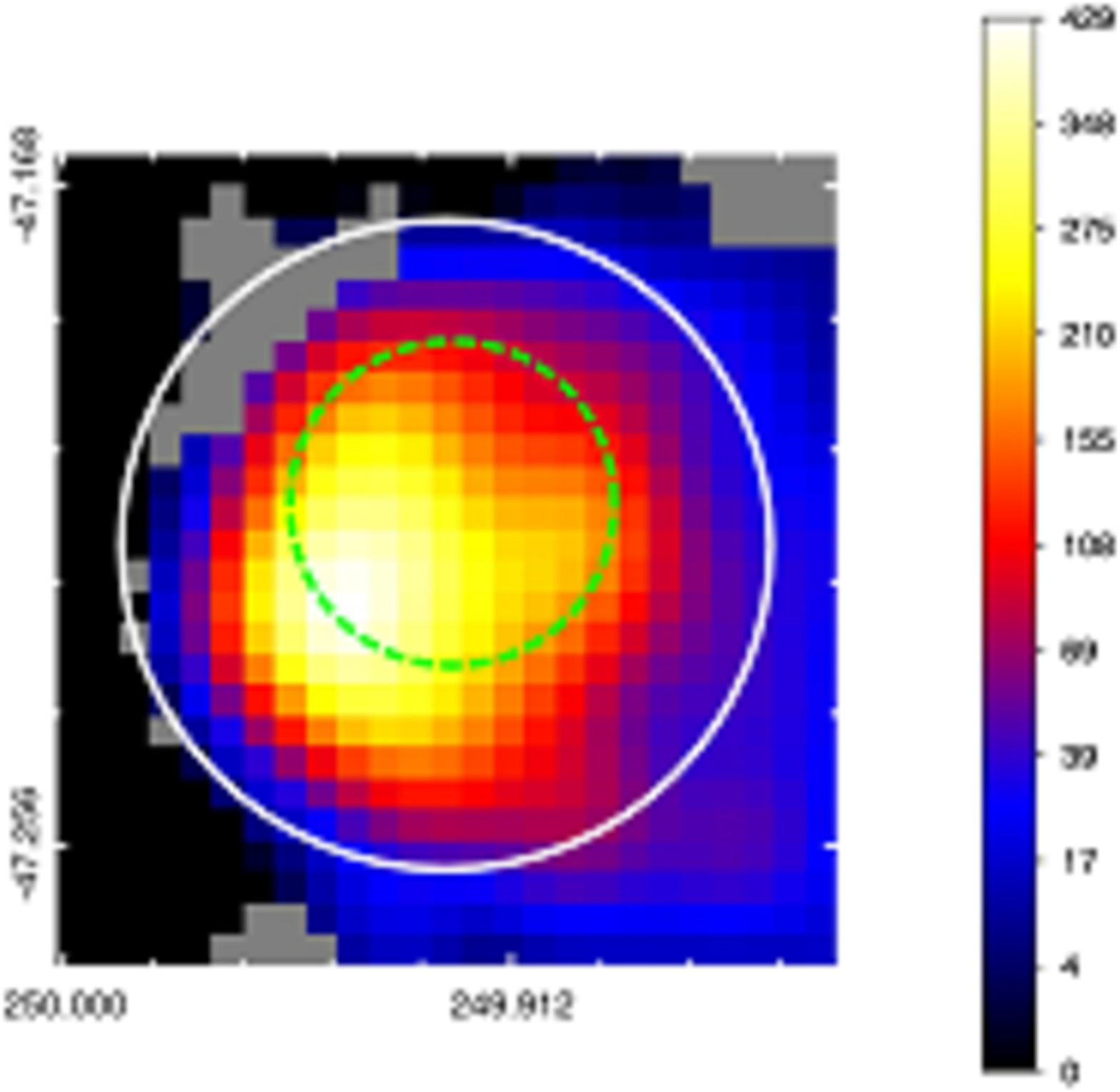}}}%
\mbox{\raisebox{0mm}{\includegraphics[width=40mm]{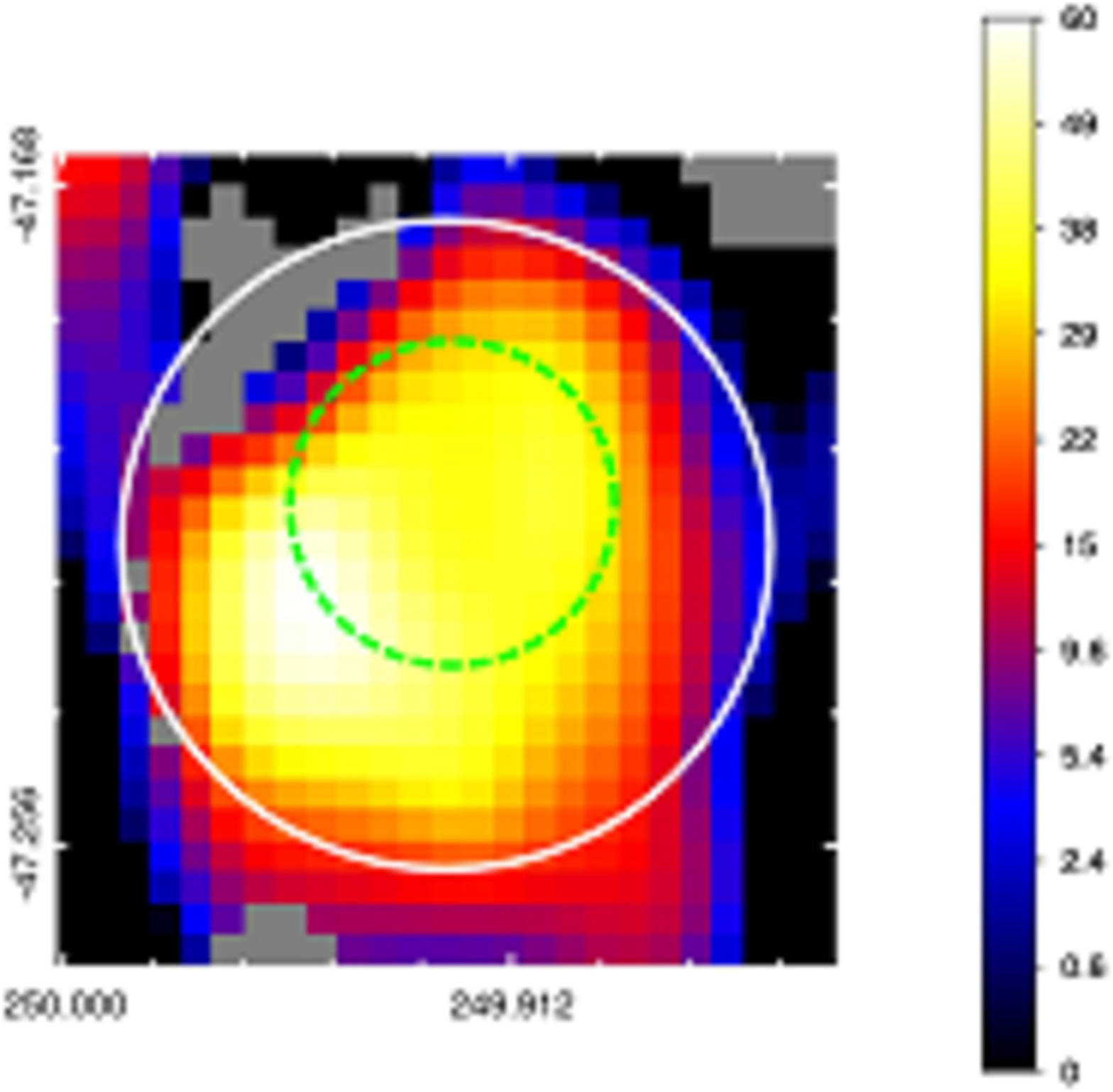}}}%
}
\subfigure{
\mbox{\raisebox{0mm}{\includegraphics[width=40mm]{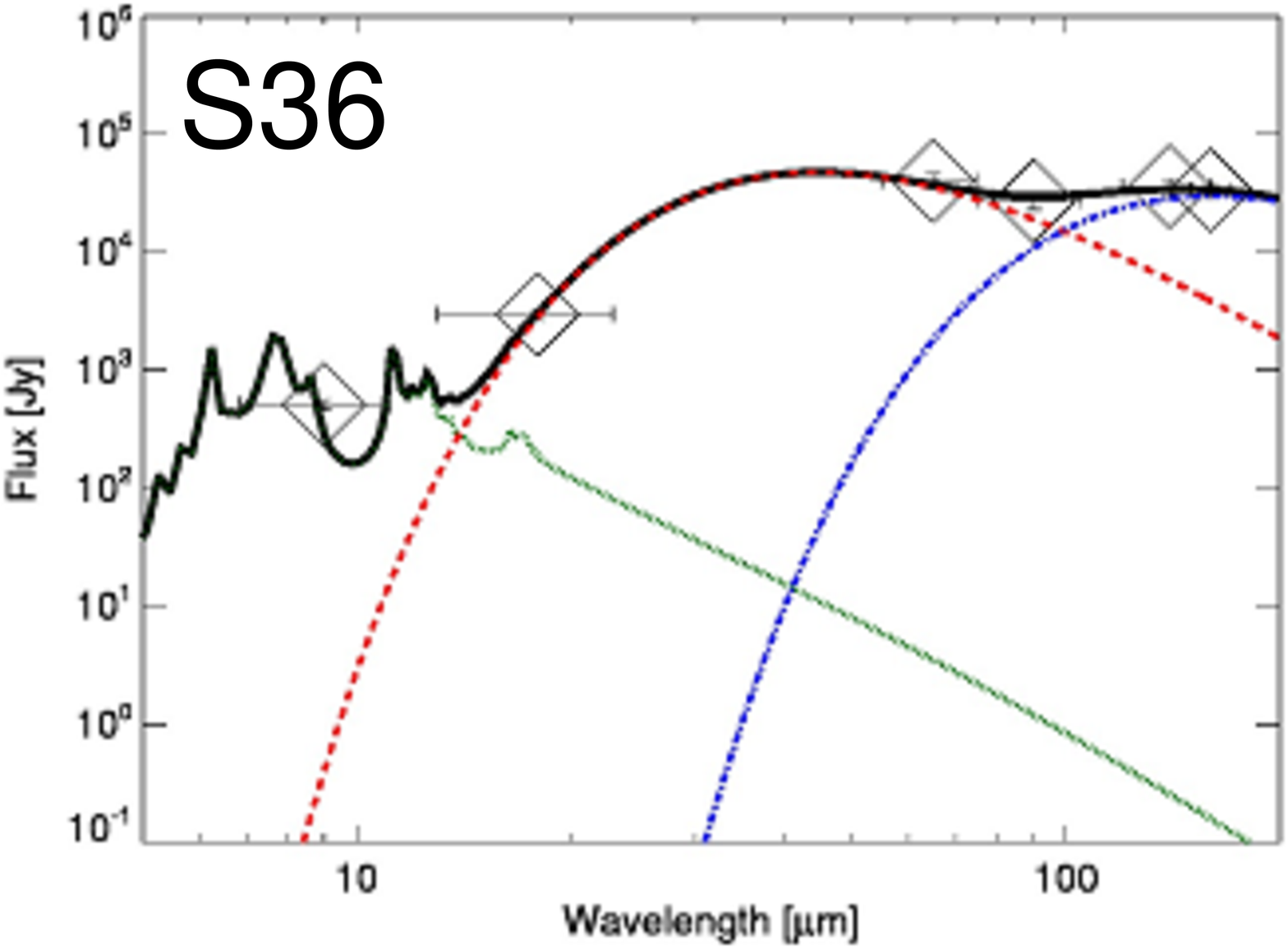}}}%
\mbox{\raisebox{6mm}{\rotatebox{90}{\small{DEC (J2000)}}}}%
\mbox{\raisebox{0mm}{\includegraphics[width=40mm]{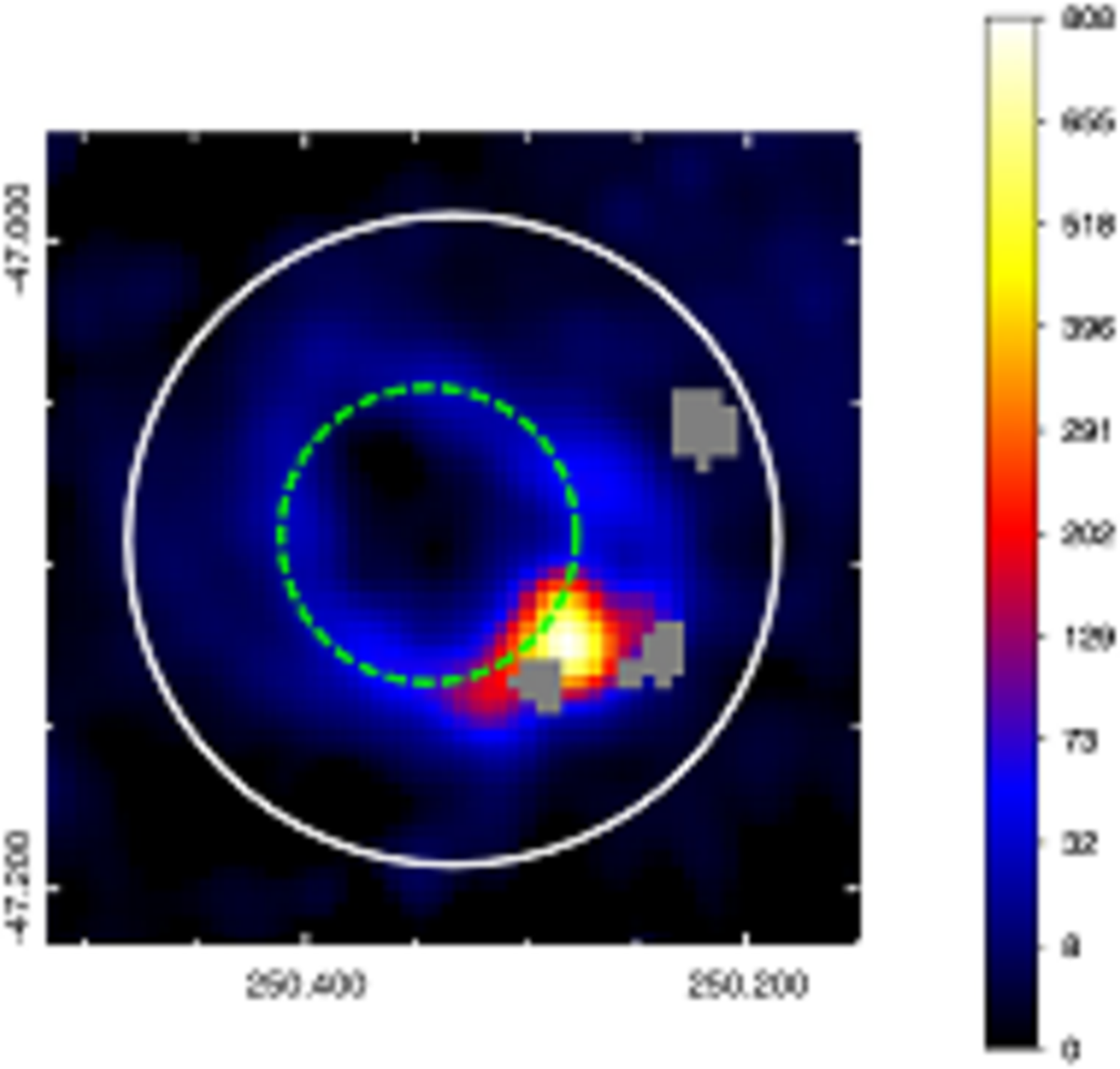}}}%
\mbox{\raisebox{0mm}{\includegraphics[width=40mm]{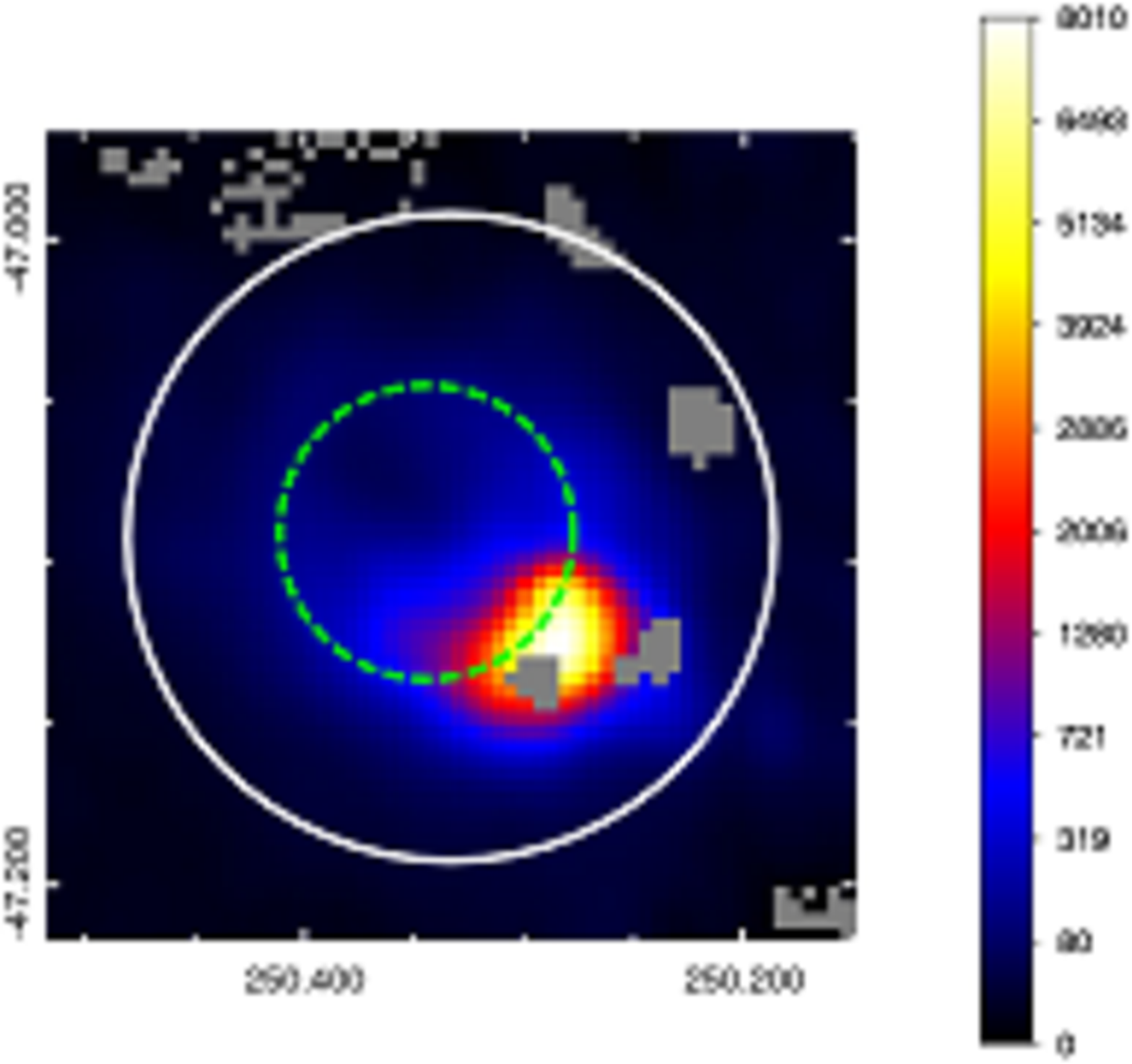}}}%
\mbox{\raisebox{0mm}{\includegraphics[width=40mm]{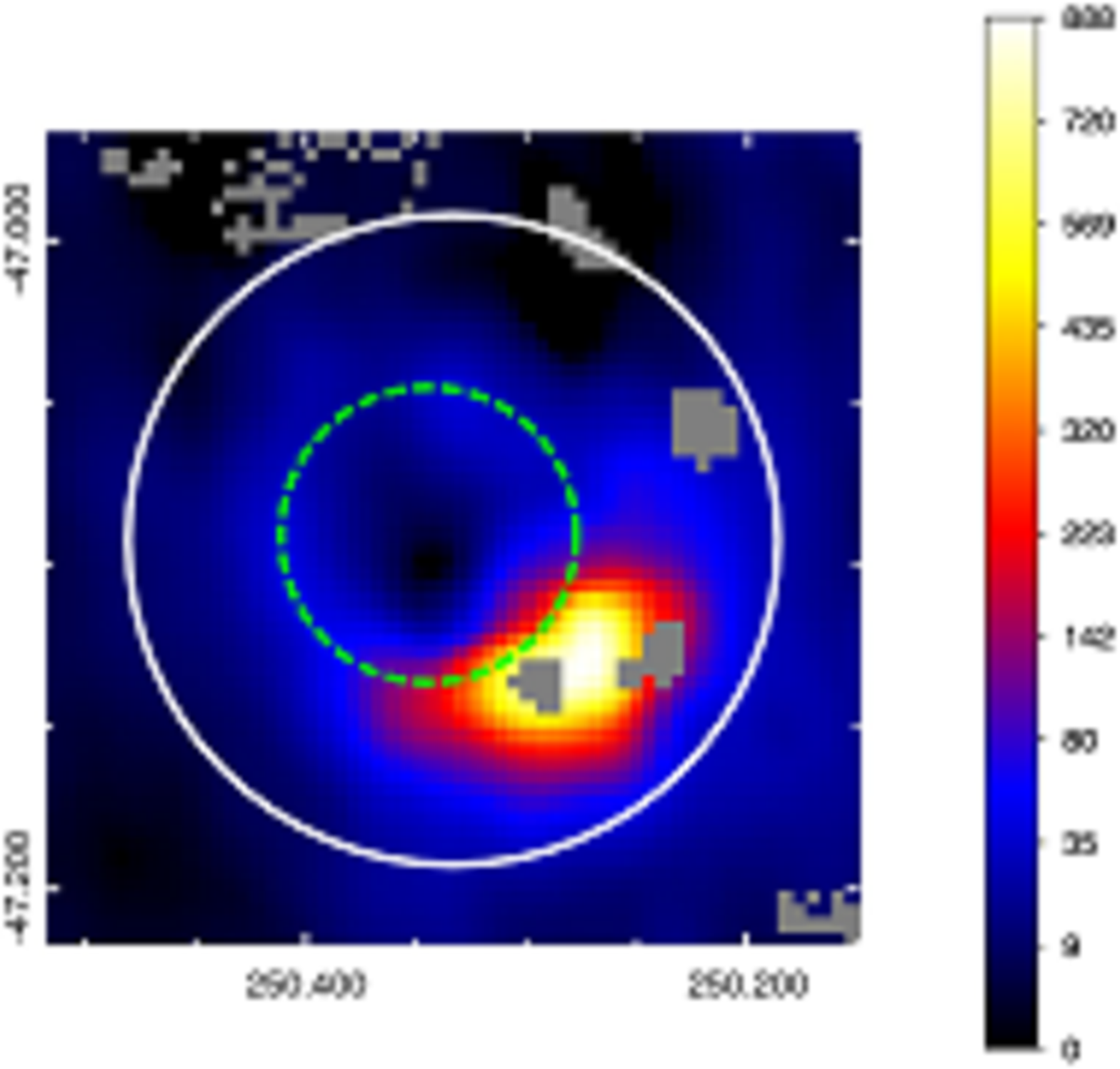}}}%
}
\caption{Continued.} \label{fig:Metfig2:g}
\end{figure*}
\addtocounter{figure}{-1}
\begin{figure*}[ht]
\addtocounter{subfigure}{1}
\centering
\subfigure{
\makebox[180mm][l]{\raisebox{0mm}[0mm][0mm]{ \hspace{20mm} \small{SED}} \hspace{27.5mm} \small{$I_{\rm{PAH}}$} \hspace{29.5mm} \small{$I_{\rm{warm}}$} \hspace{29.5mm} \small{$I_{\rm{cold}}$}}%
}
\subfigure{
\makebox[180mm][l]{\raisebox{0mm}{\hspace{52mm} \small{RA (J2000)} \hspace{20mm} \small{RA (J2000)} \hspace{20mm} \small{RA (J2000)}}}
}
\subfigure{
\mbox{\raisebox{0mm}{\includegraphics[width=40mm]{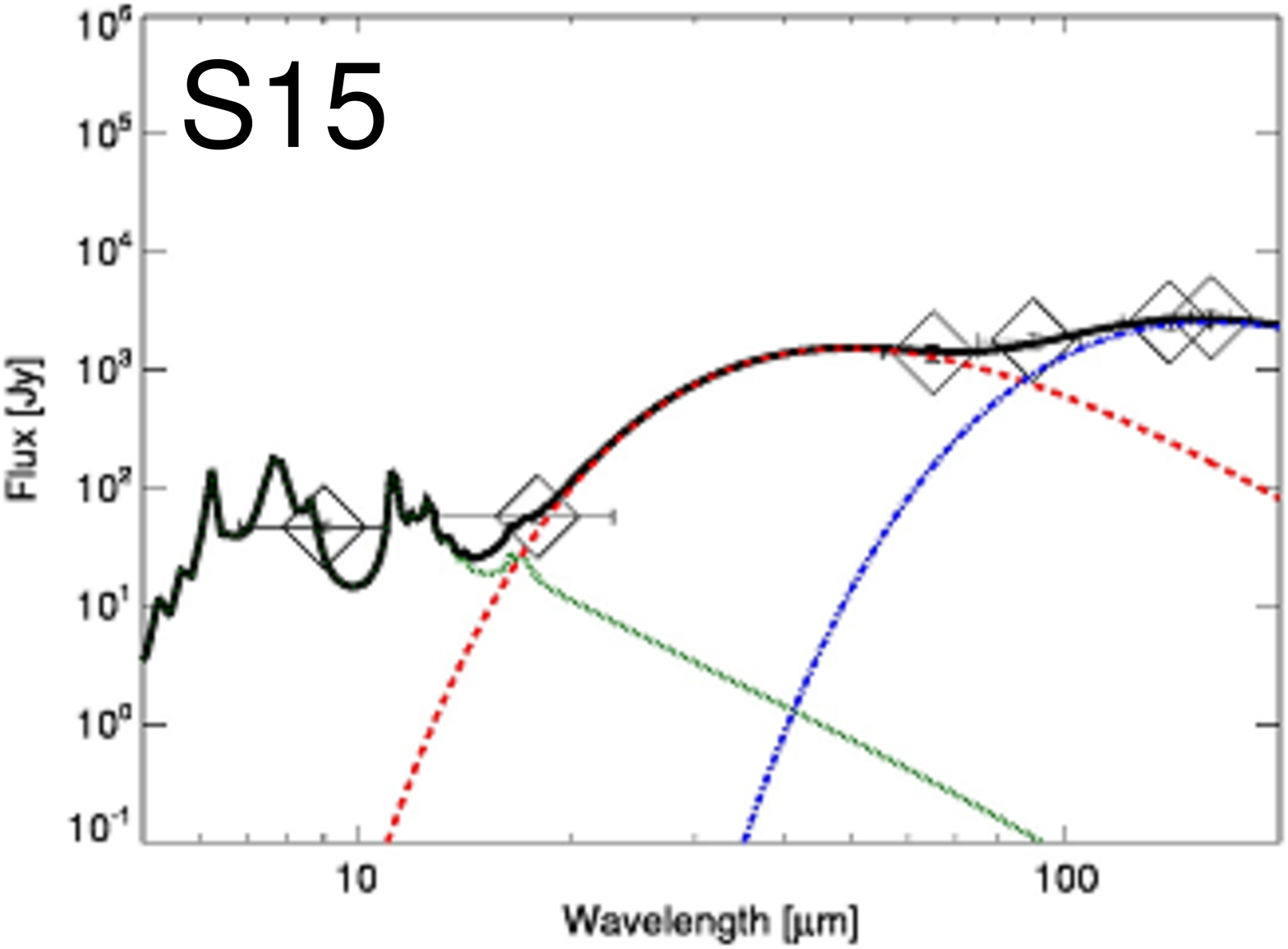}}}%
\mbox{\raisebox{6mm}{\rotatebox{90}{\small{DEC (J2000)}}}}%
\mbox{\raisebox{0mm}{\includegraphics[width=40mm]{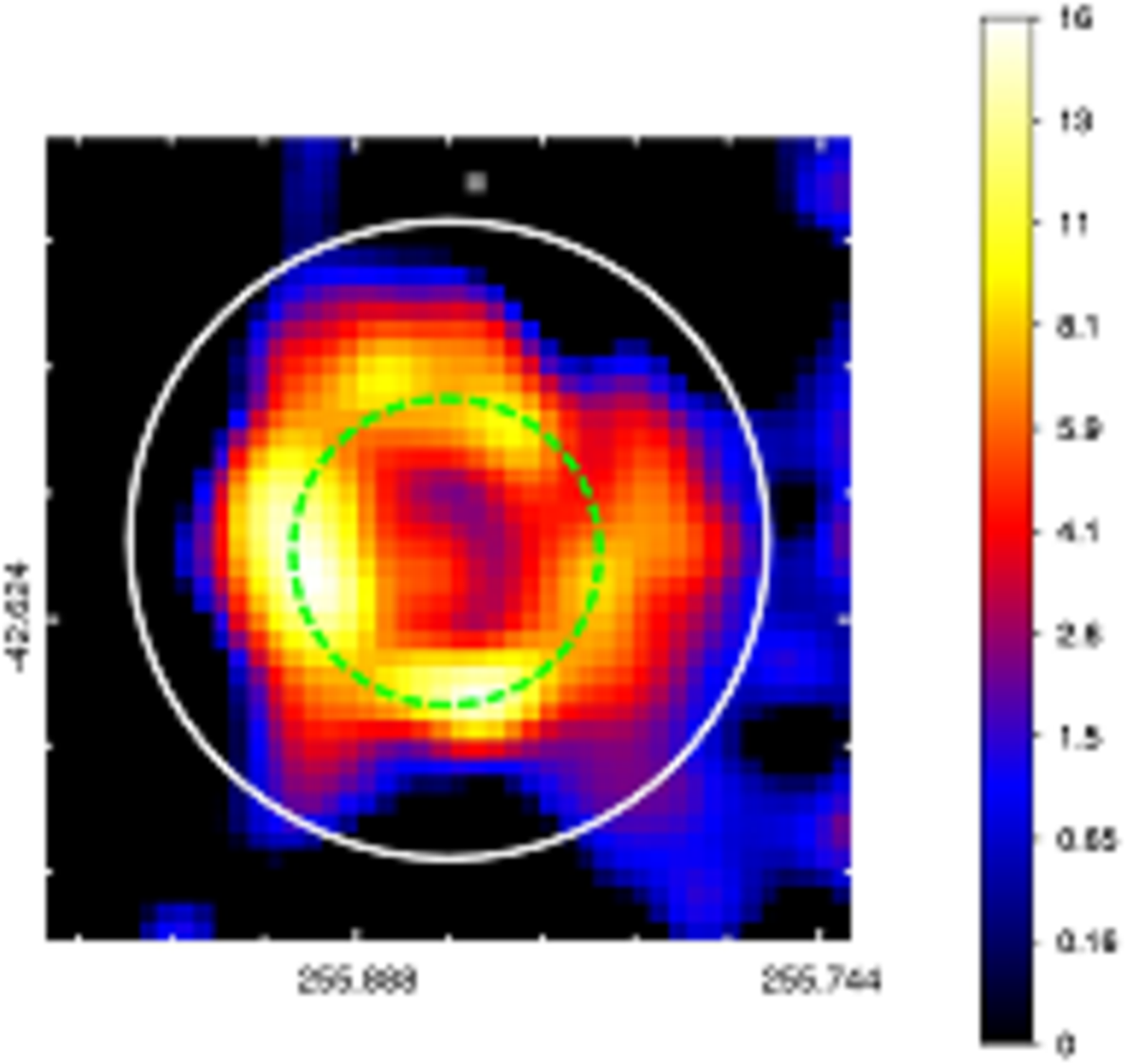}}}%
\mbox{\raisebox{0mm}{\includegraphics[width=40mm]{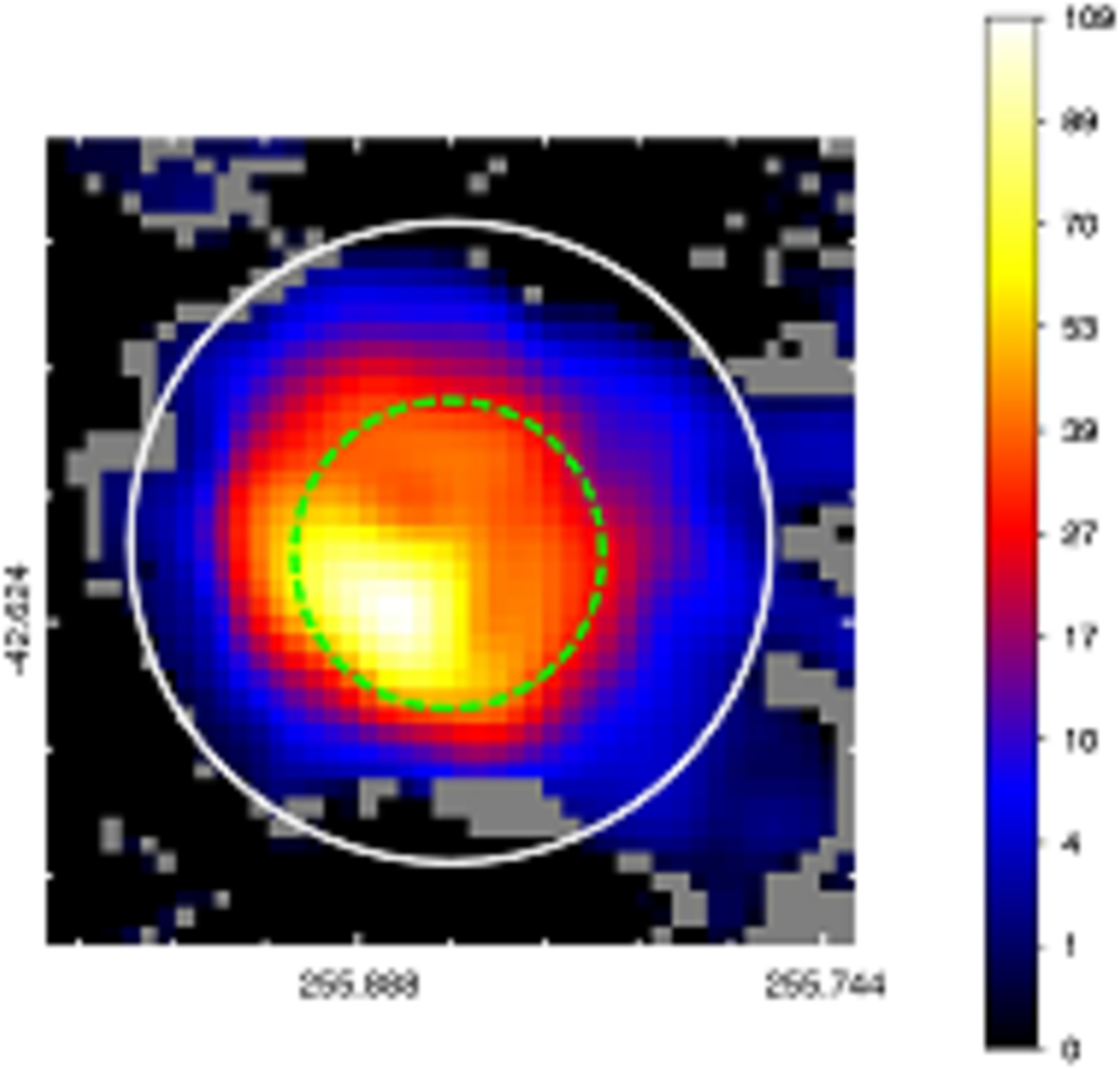}}}%
\mbox{\raisebox{0mm}{\includegraphics[width=40mm]{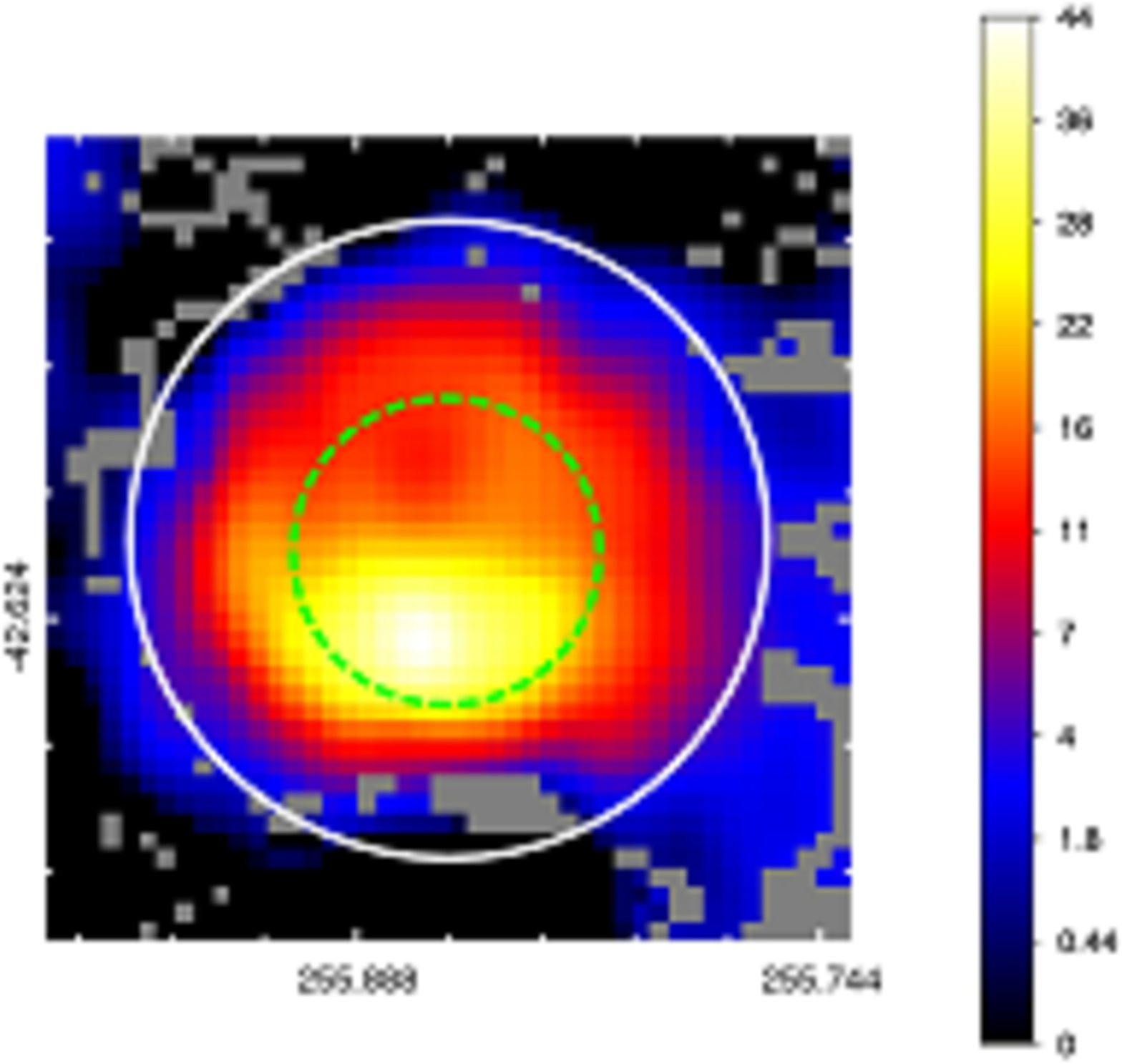}}}%
}
\subfigure{
\mbox{\raisebox{0mm}{\includegraphics[width=40mm]{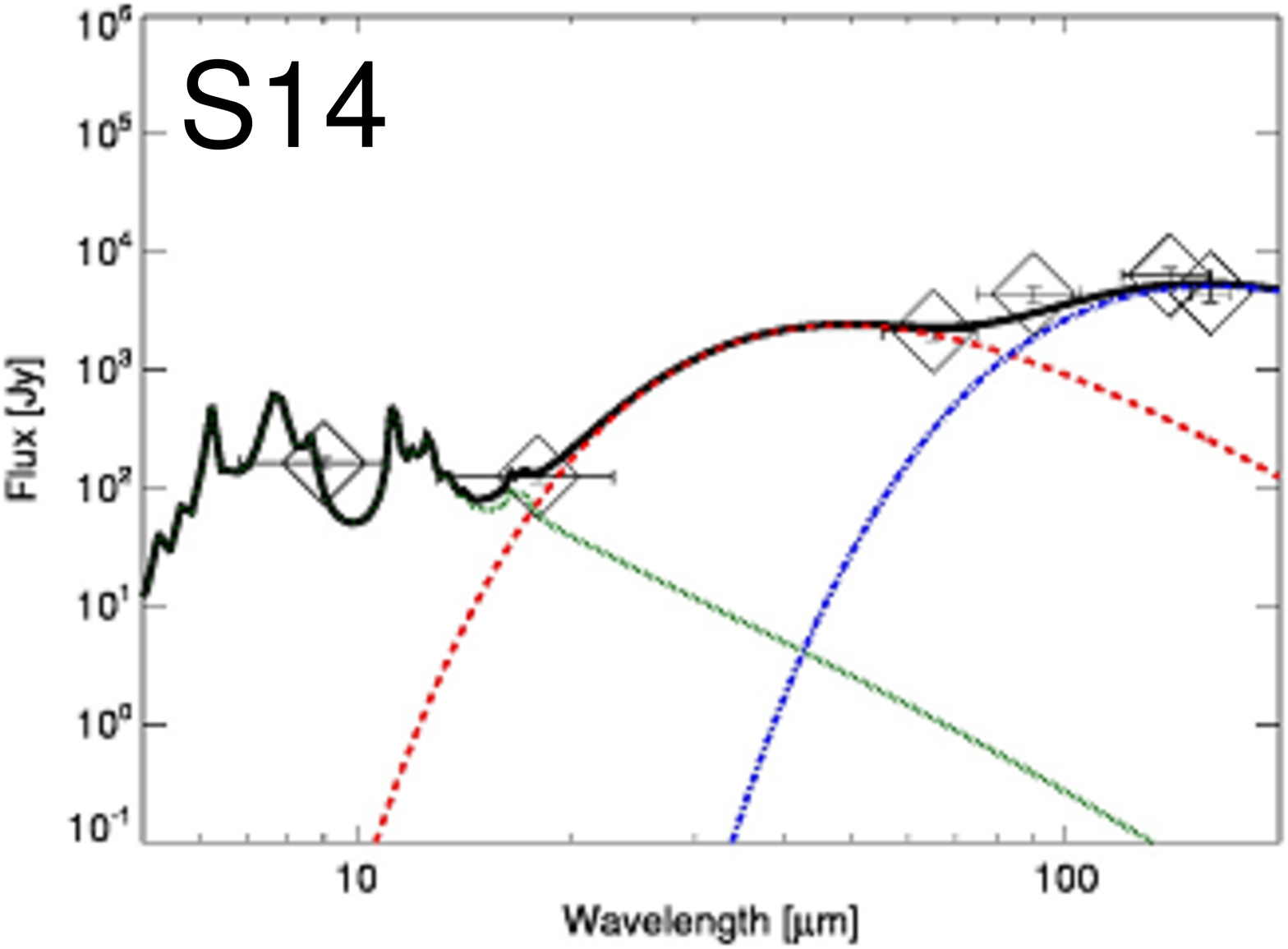}}}%
\mbox{\raisebox{6mm}{\rotatebox{90}{\small{DEC (J2000)}}}}%
\mbox{\raisebox{0mm}{\includegraphics[width=40mm]{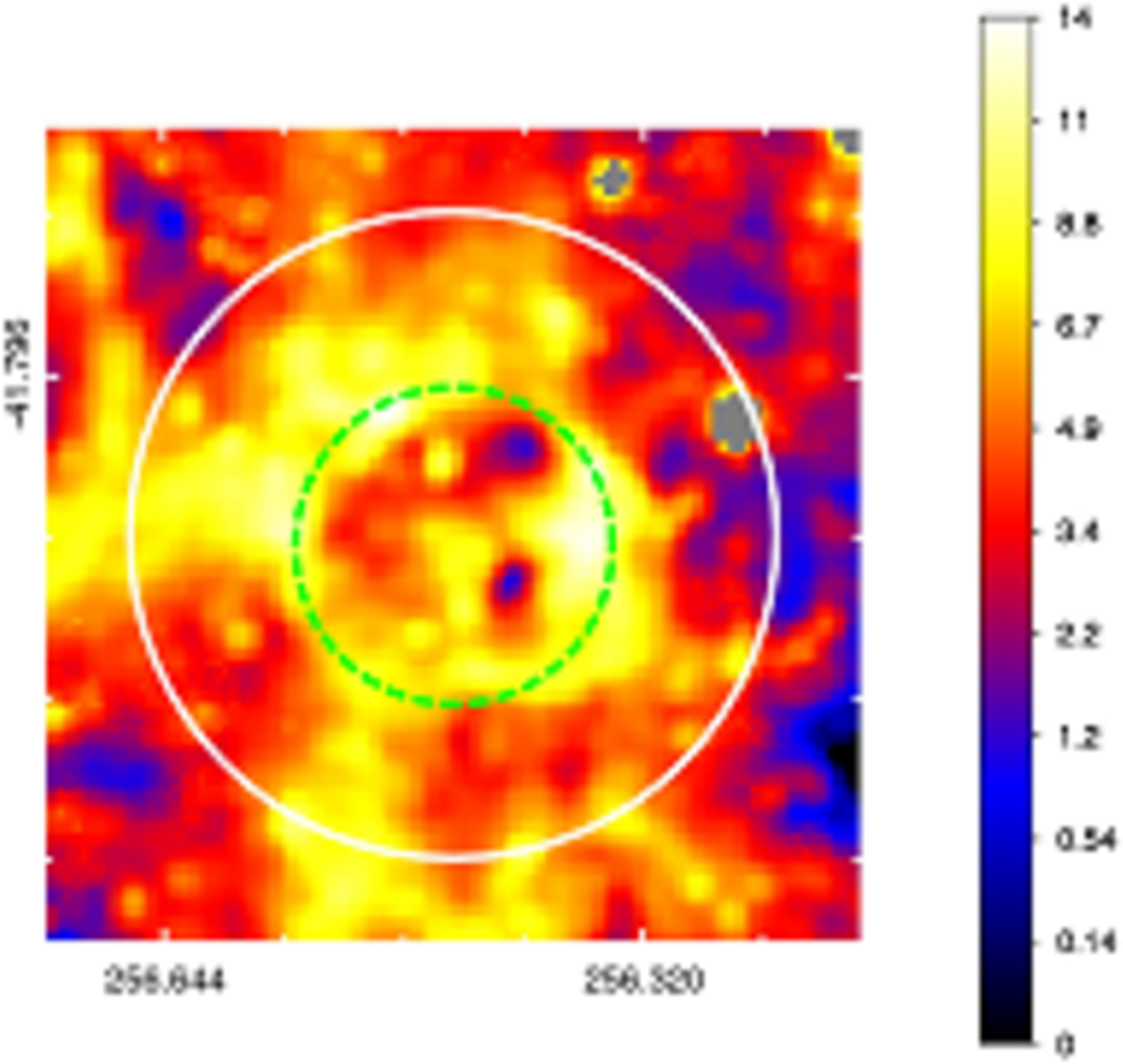}}}%
\mbox{\raisebox{0mm}{\includegraphics[width=40mm]{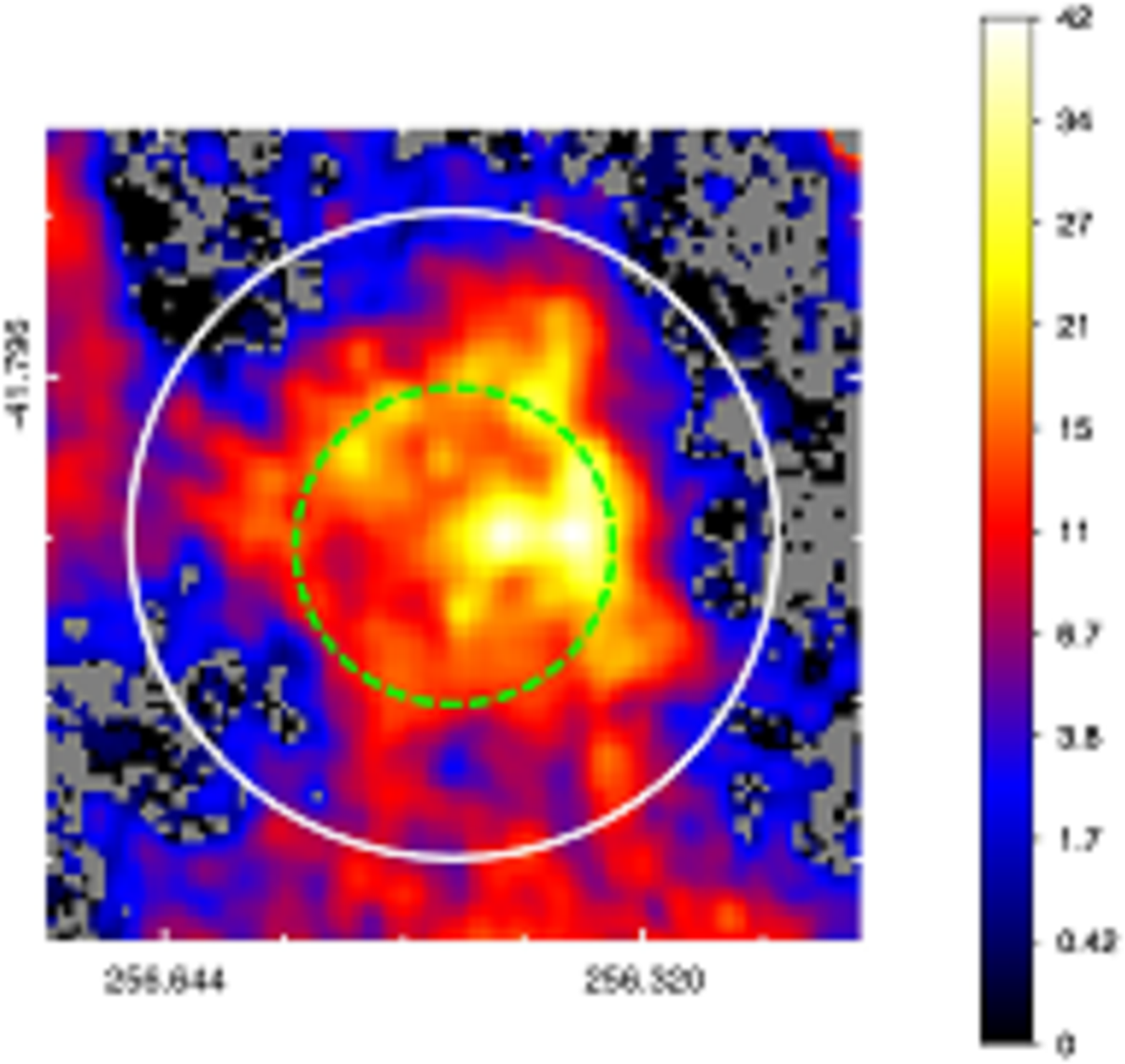}}}%
\mbox{\raisebox{0mm}{\includegraphics[width=40mm]{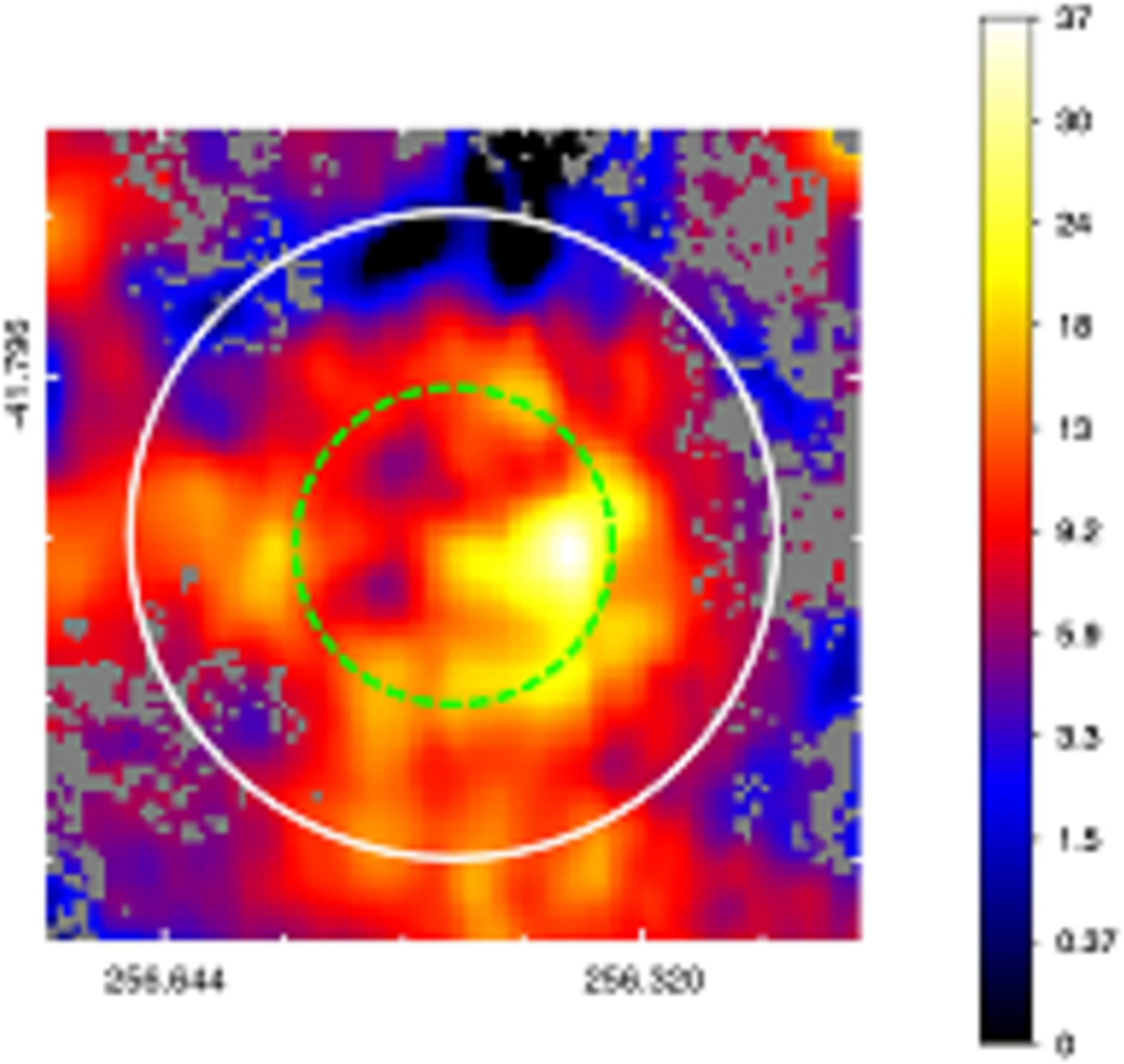}}}%
}
\subfigure{
\mbox{\raisebox{0mm}{\includegraphics[width=40mm]{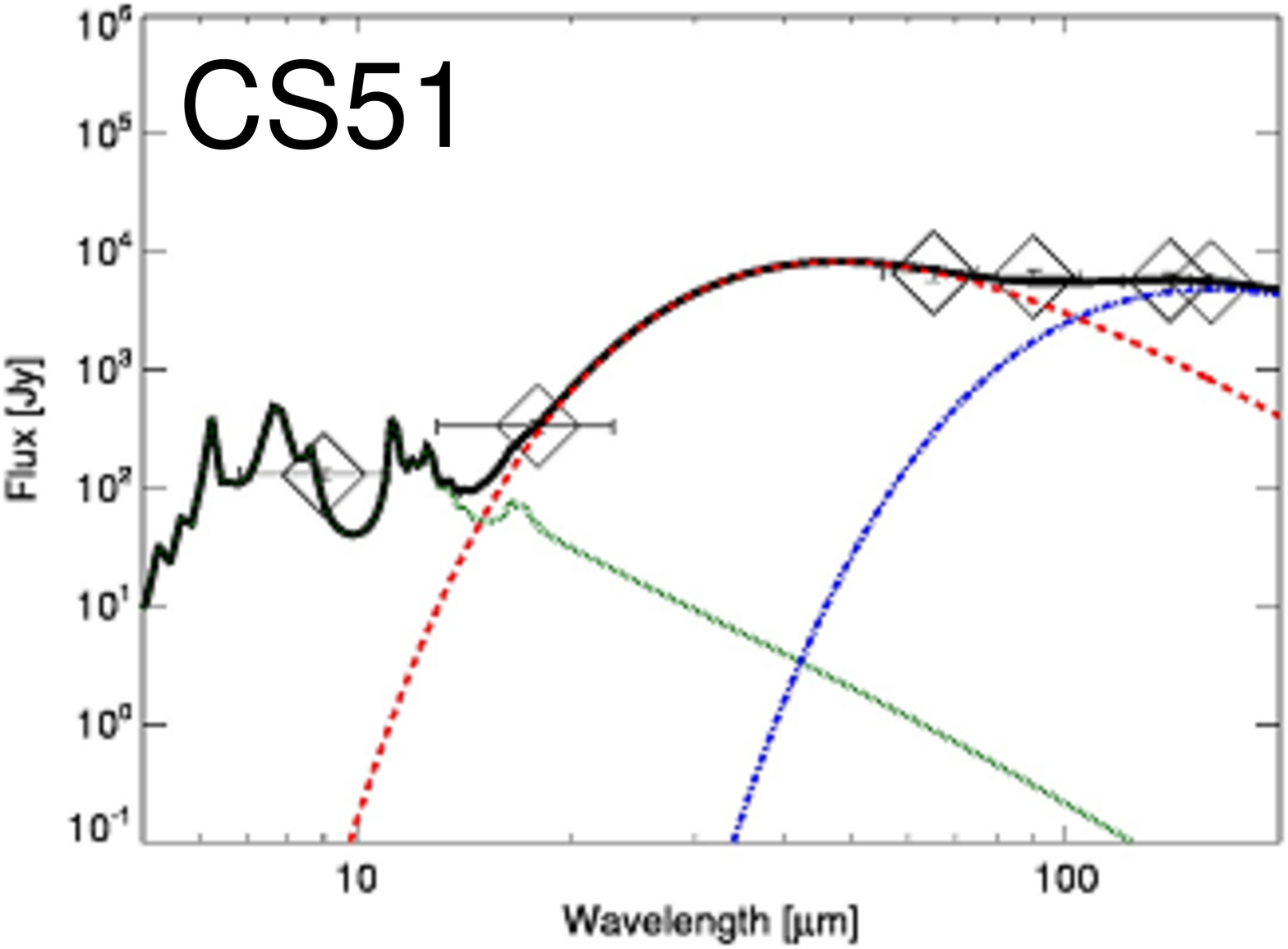}}}%
\mbox{\raisebox{6mm}{\rotatebox{90}{\small{DEC (J2000)}}}}%
\mbox{\raisebox{0mm}{\includegraphics[width=40mm]{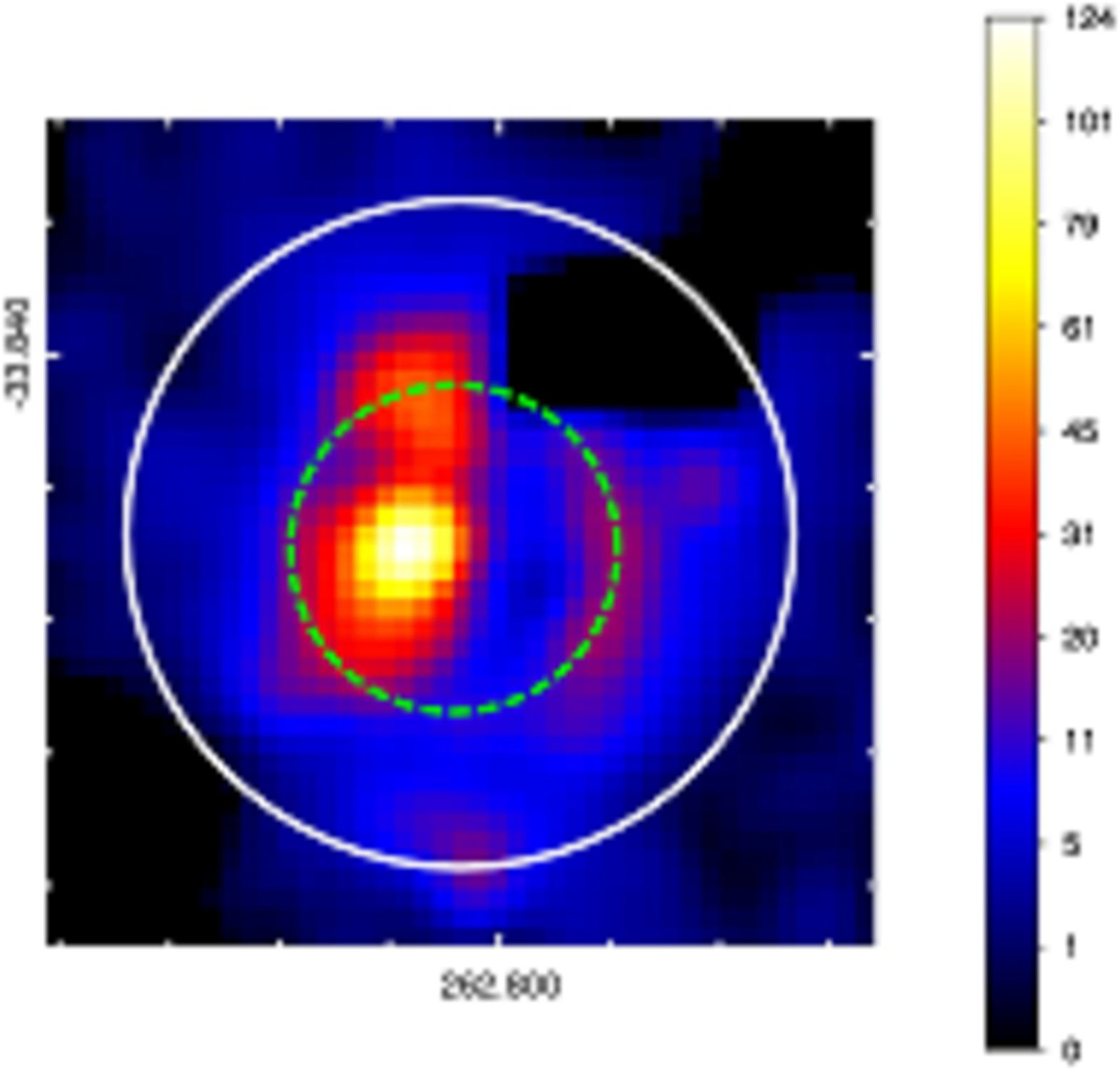}}}%
\mbox{\raisebox{0mm}{\includegraphics[width=40mm]{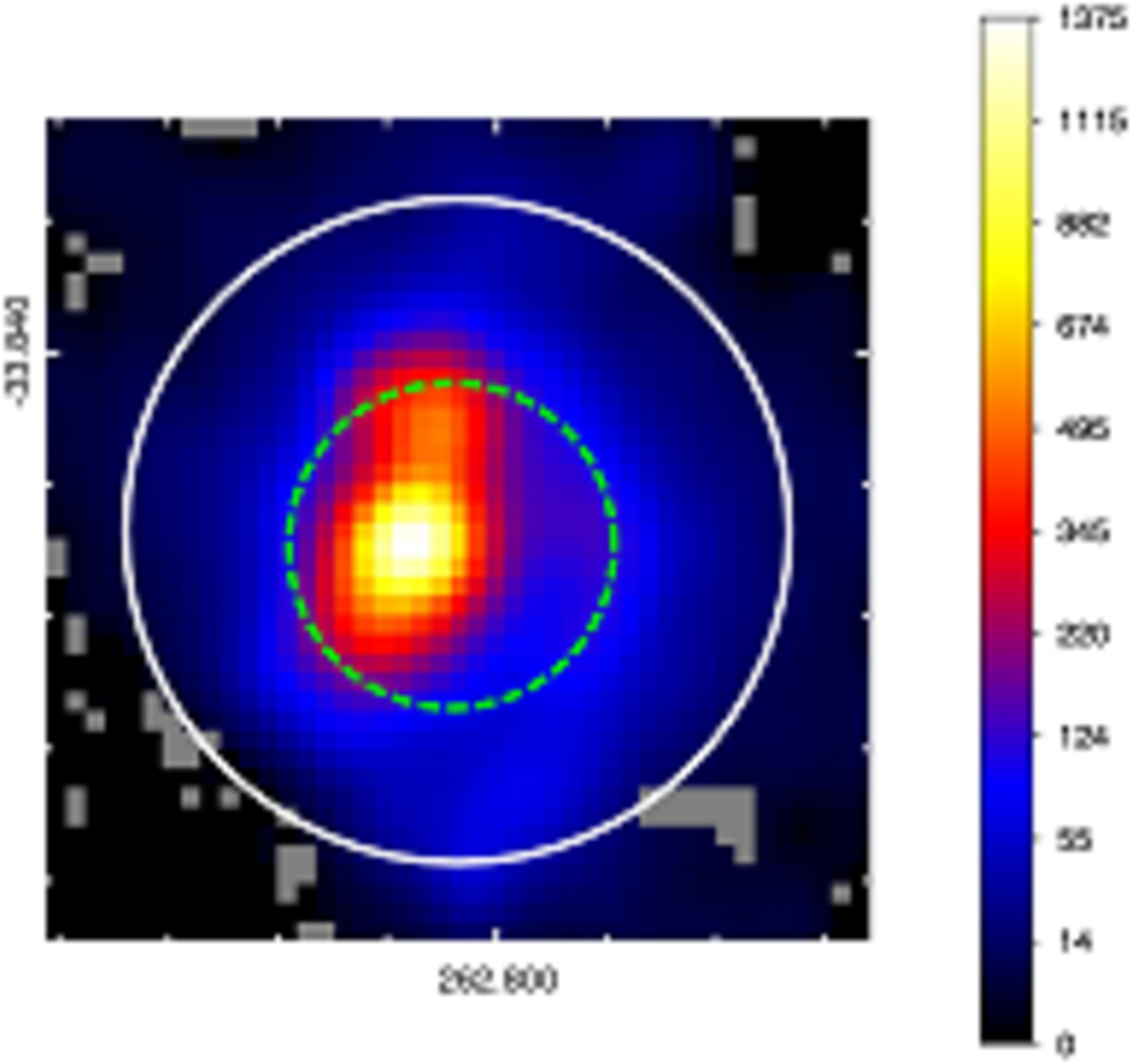}}}%
\mbox{\raisebox{0mm}{\includegraphics[width=40mm]{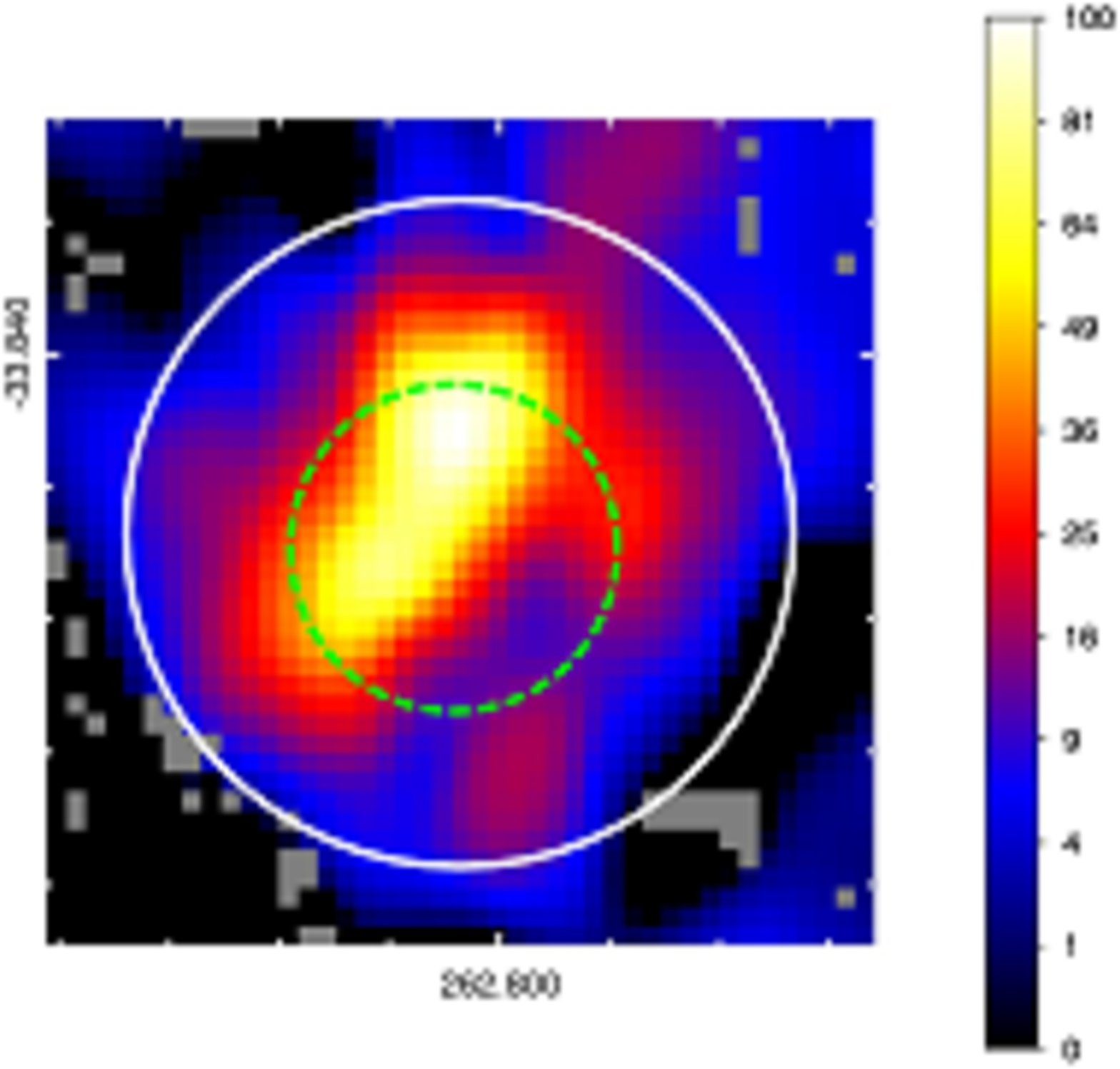}}}%
}
\subfigure{
\mbox{\raisebox{0mm}{\includegraphics[width=40mm]{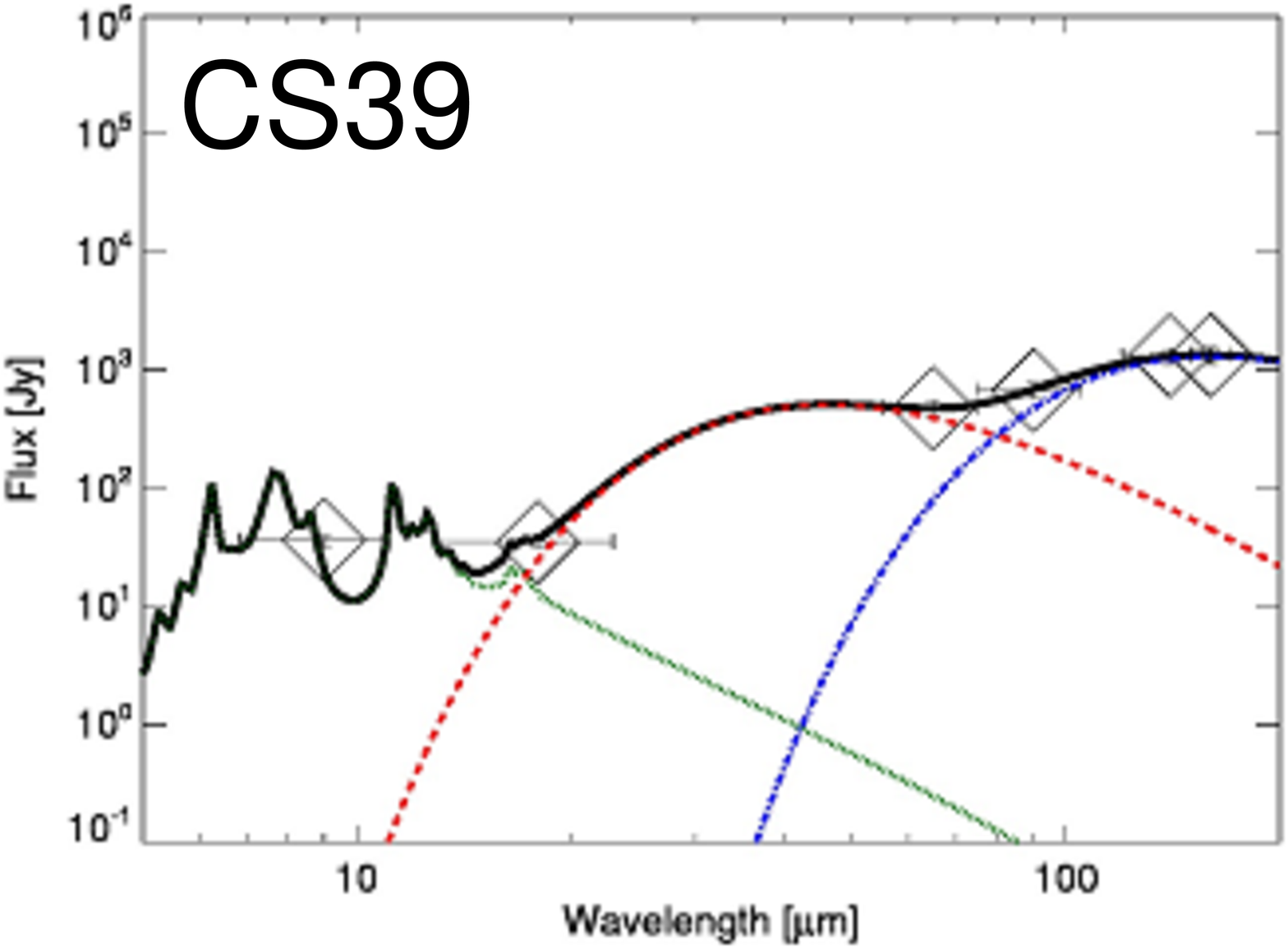}}}%
\mbox{\raisebox{6mm}{\rotatebox{90}{\small{DEC (J2000)}}}}%
\mbox{\raisebox{0mm}{\includegraphics[width=40mm]{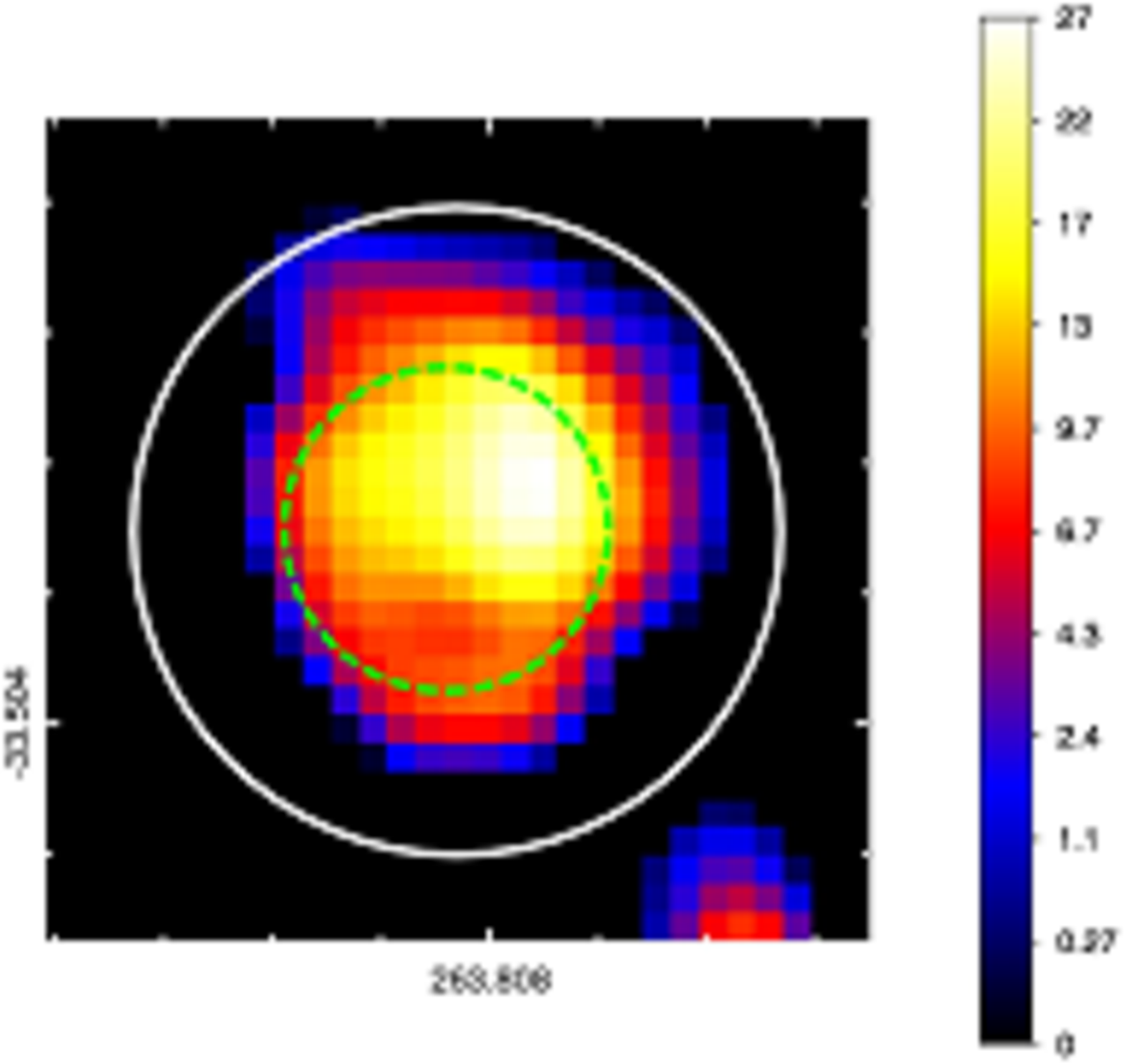}}}%
\mbox{\raisebox{0mm}{\includegraphics[width=40mm]{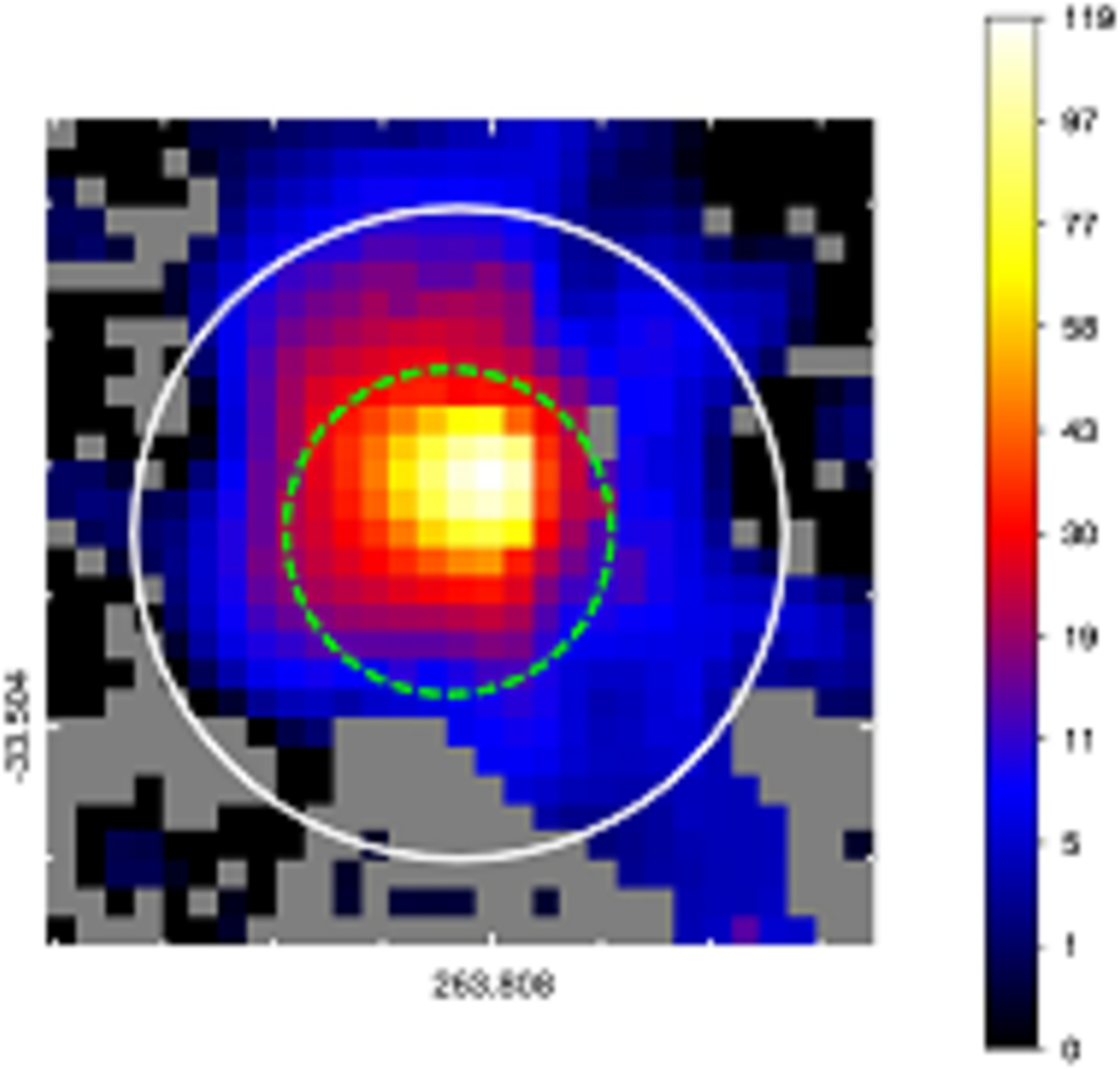}}}%
\mbox{\raisebox{0mm}{\includegraphics[width=40mm]{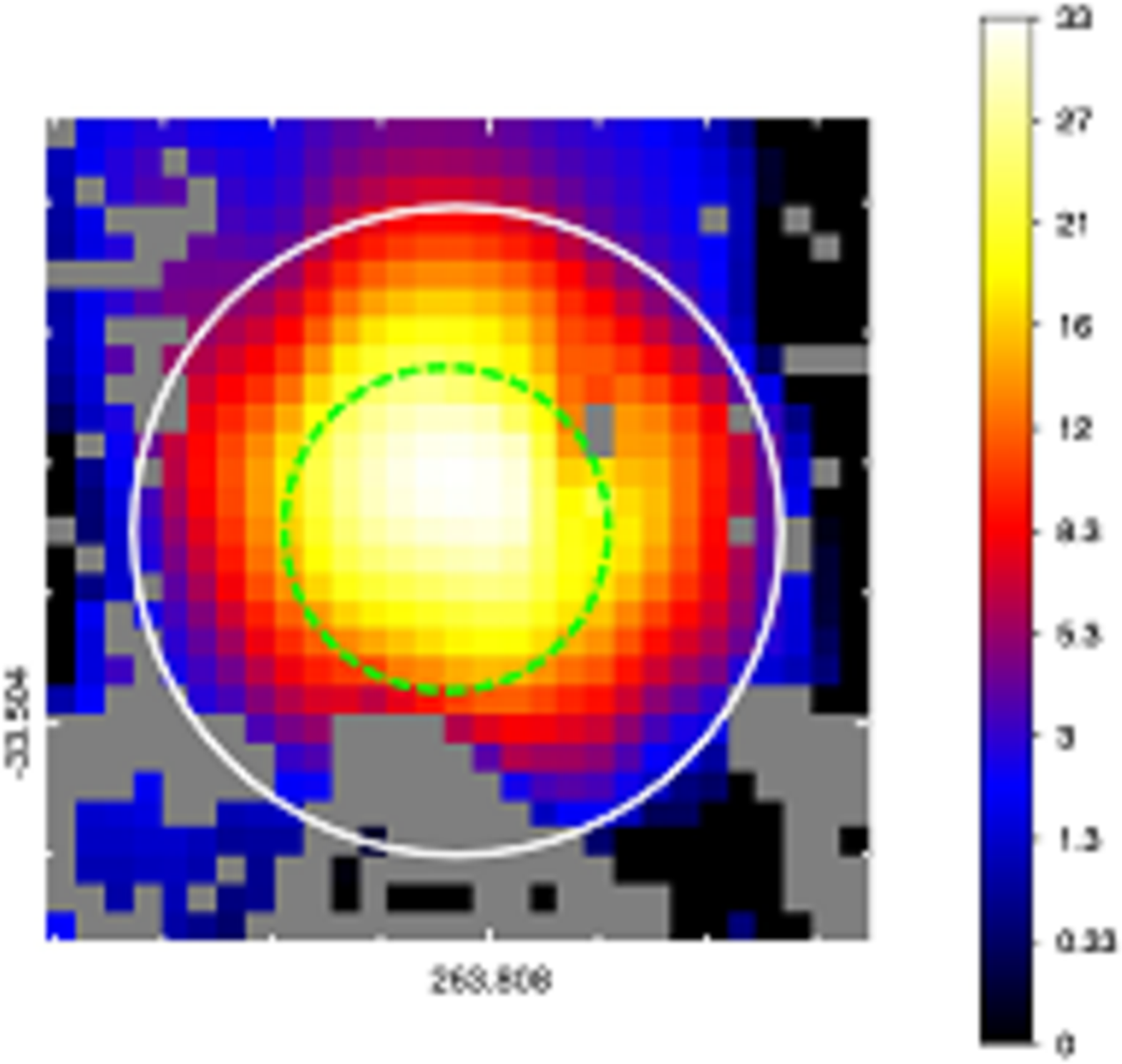}}}%
}
\subfigure{
\mbox{\raisebox{0mm}{\includegraphics[width=40mm]{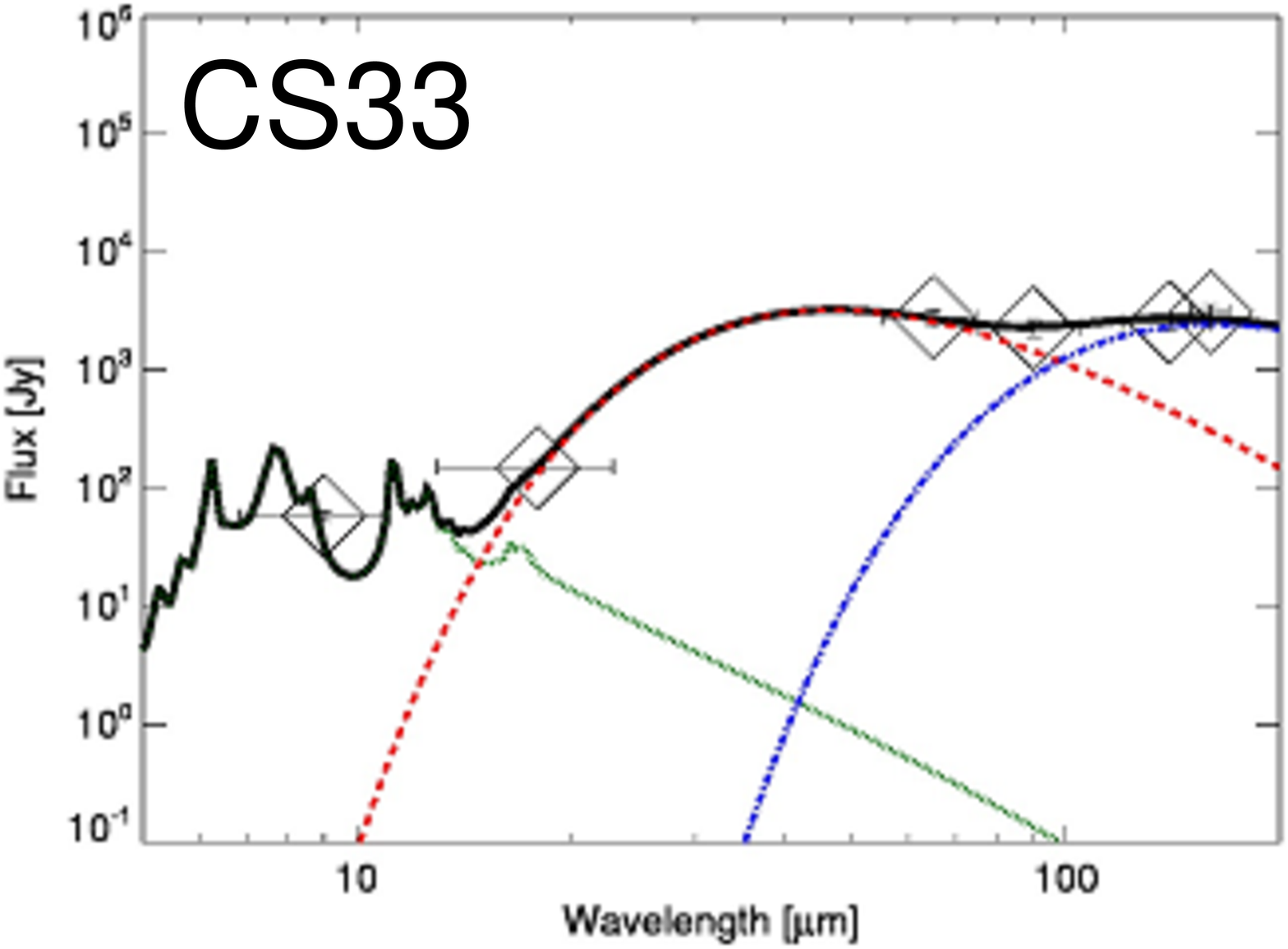}}}%
\mbox{\raisebox{6mm}{\rotatebox{90}{\small{DEC (J2000)}}}}%
\mbox{\raisebox{0mm}{\includegraphics[width=40mm]{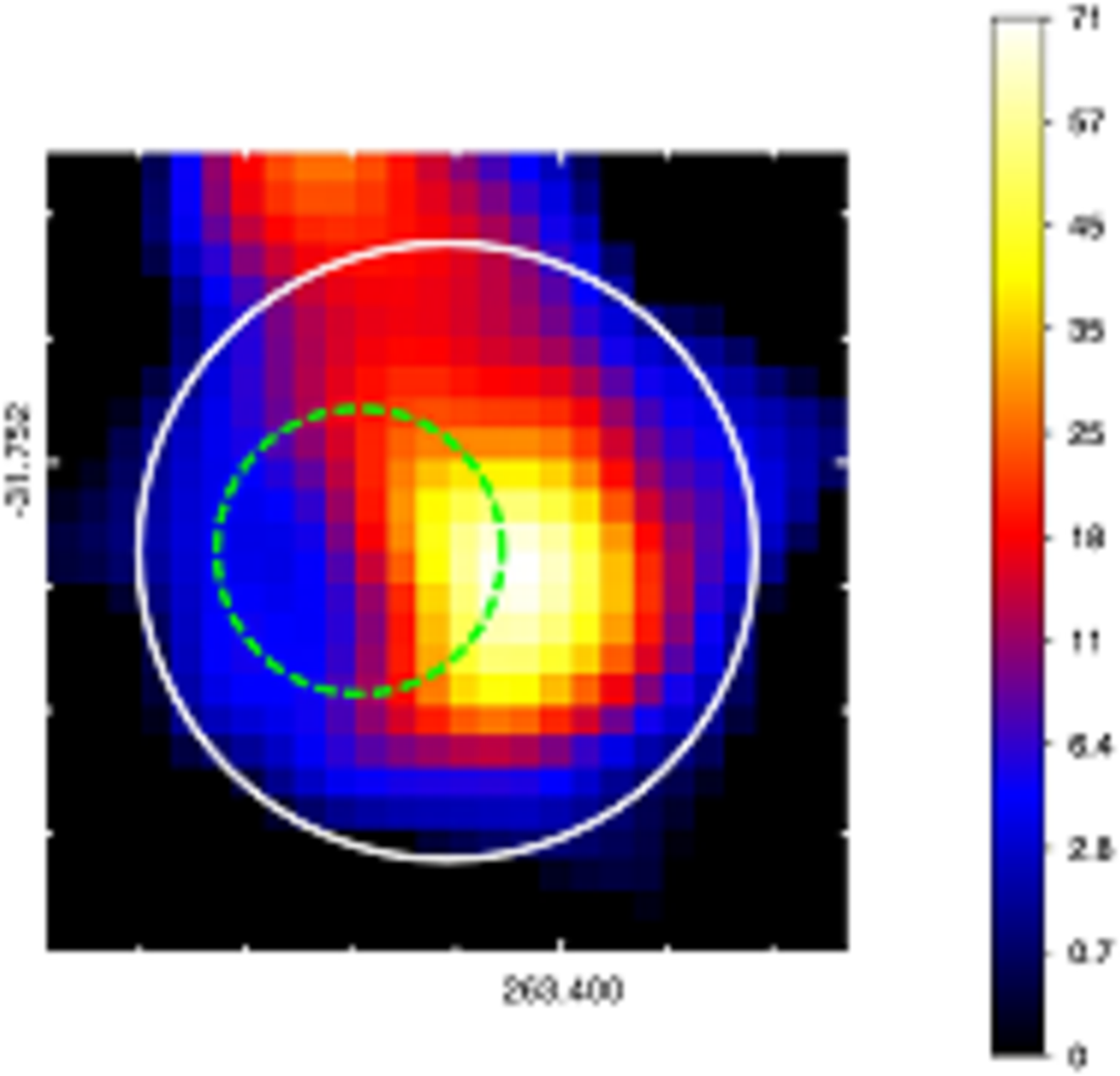}}}%
\mbox{\raisebox{0mm}{\includegraphics[width=40mm]{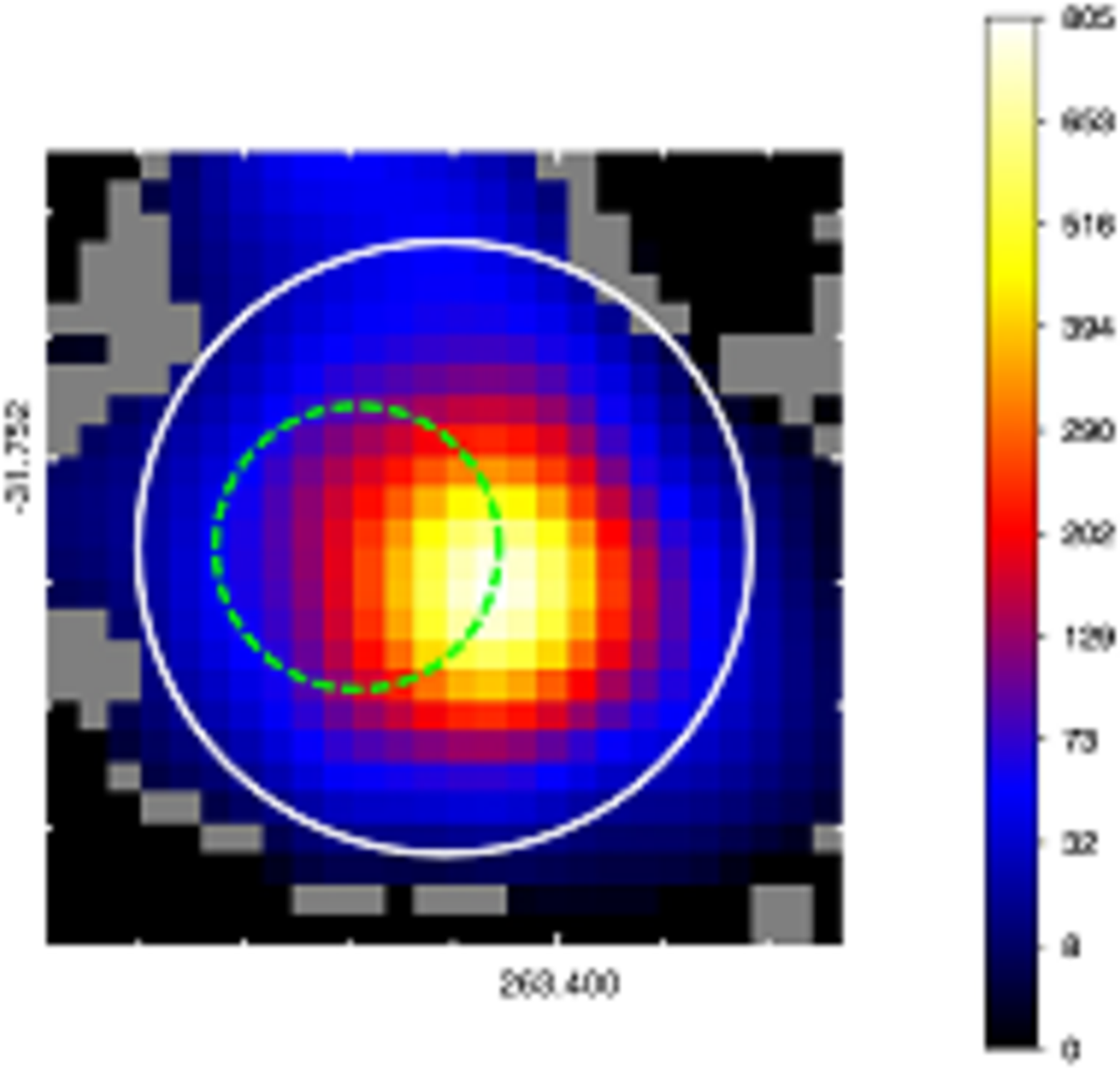}}}%
\mbox{\raisebox{0mm}{\includegraphics[width=40mm]{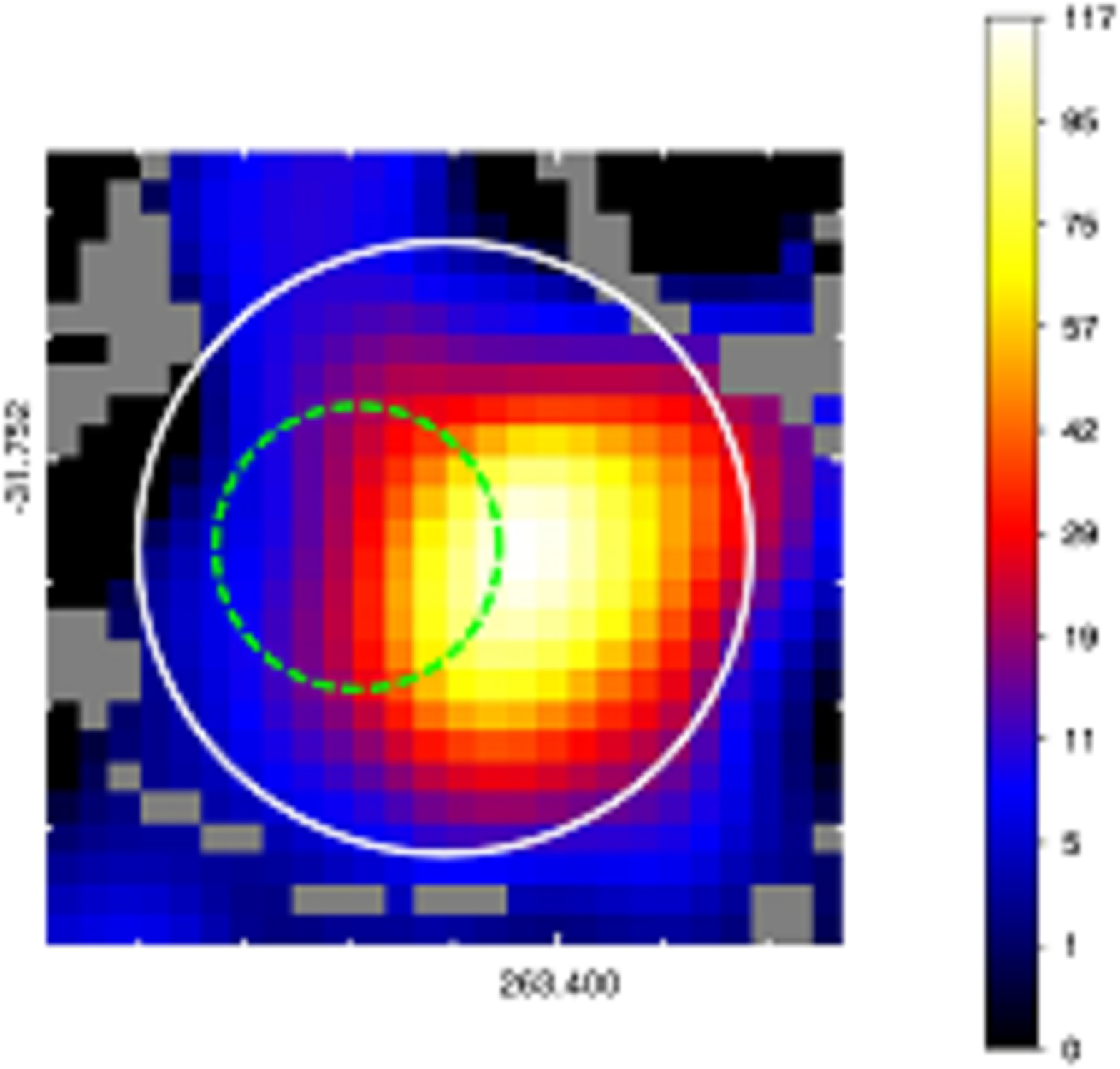}}}%
}
\caption{Continued.} \label{fig:Metfig2:h}
\end{figure*}

\clearpage

\begin{figure}[ht] %
\begin{tabular}{cc} %
\centering %
\begin{minipage}[t]{0.48\hsize} %
\begin{center} %
\includegraphics[clip, width=85mm]{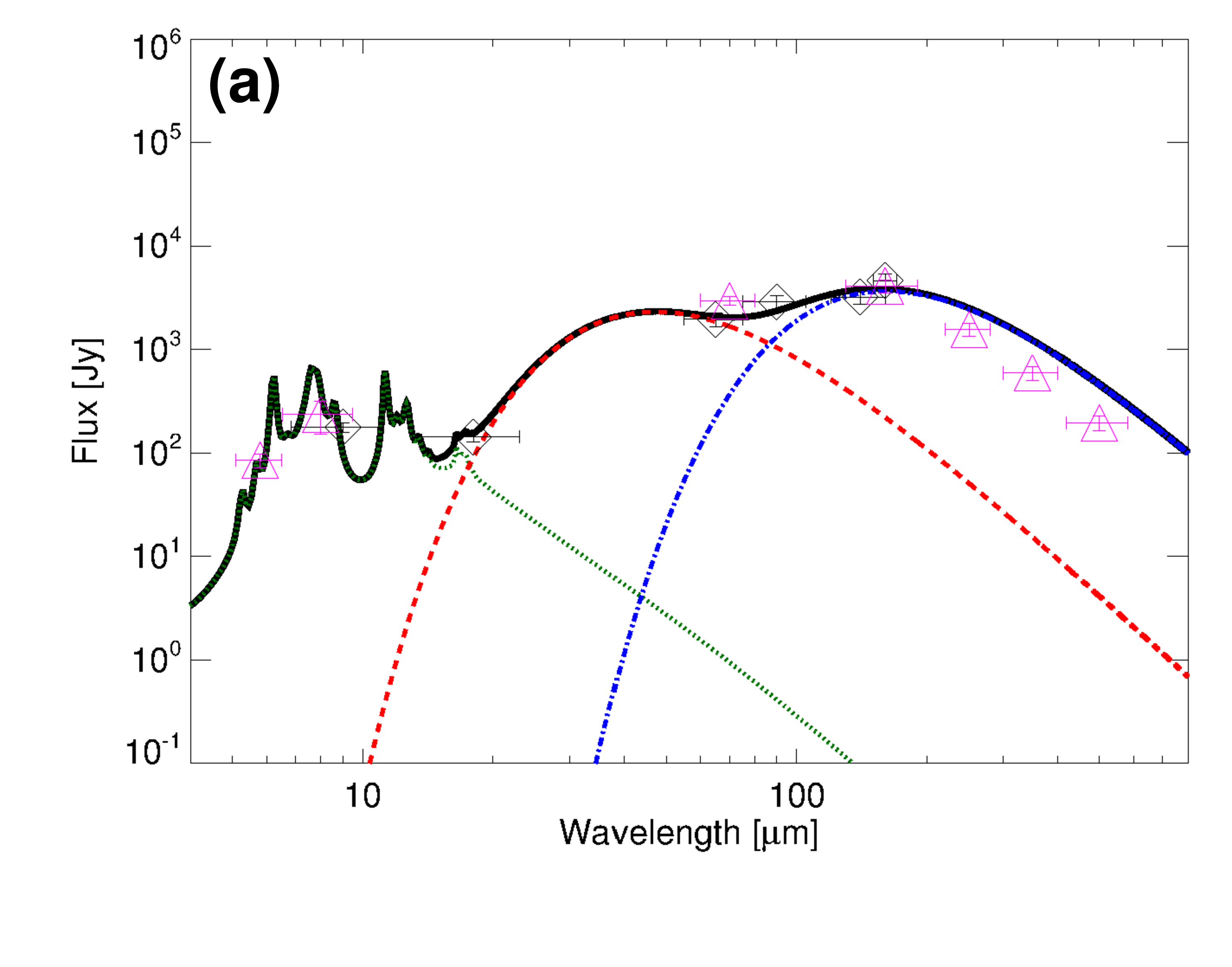} %
\end{center} %
\end{minipage} %
\begin{minipage}[t]{0.48\hsize} %
\begin{center} %
\includegraphics[clip, width=85mm]{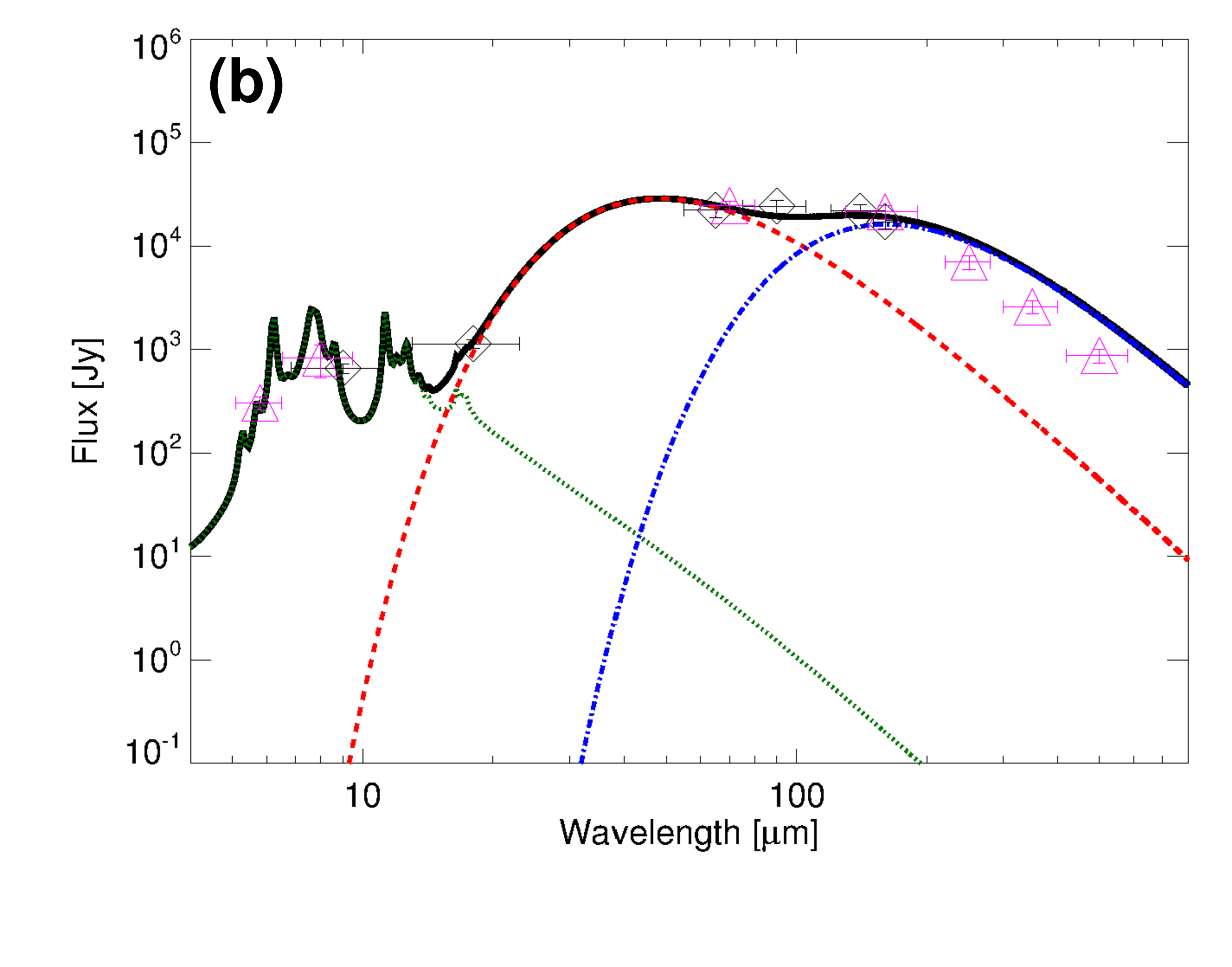} %
\end{center} %
\end{minipage} %
\end{tabular} %
\begin{flushleft} %
\caption{Global SEDs of two typical examples of (a) closed bubble S137 and (b) broken bubble S145. Diamonds represent the AKARI photometric data points, while triangles represent the Spitzer and the Herschel photometric data points. The black solid curve indicates the result of fitting to the AKARI data alone. The green dotted, the red dashed and the blue dash-dotted represent the PAH, warm and cold dust components, respectively. \label{fig:SEDwO}} %
\end{flushleft} %
\end{figure} %

\clearpage

\fontsize{11pt}{11pt}\selectfont
\begin{longtable}[ht]{ccccccccccc}
\caption{Summary of the central positions and radii of the Galactic IR bubbles estimated from the circular fitting, and the morphological types based on our quantitative classification. The central positions and radii given in \citet{Churchwell2006} and (2007), and distances to the bubbles are also listed.}\label{tab:Mettab1} \\
\hline
Name & ${\it l_{\rm Ch}}$ [deg] & ${\it b_{\rm Ch}}$ [deg] & ${\it R_{\rm Ch}}$ [\timeform{'}] & $D$ [kpc] & $D$ ref. & $l$ [deg] & $b$ [deg] & $R$ [\timeform{'}] & Type$^{**}$ \\
\endfirsthead 
\multicolumn{10}{l}{\small Continued.} \\
\hline
Name & ${\it l_{\rm Ch}}$ [deg] & ${\it b_{\rm Ch}}$ [deg] & ${\it R_{\rm Ch}}$ [\timeform{'}] & $D$ [kpc] & $D$ ref. & $l$ [deg] & $b$ [deg] & $R$ [\timeform{'}] & Type \\
\hline
\endhead
\hline
\endfoot
\hline
\multicolumn{10}{l}{\small \parbox[ht]{160mm}{\strut References: Bea10: \citet{Beaumont2010}; Cap14: \citet{Cappa2014}; Chu06: \citet{Churchwell2006}; Chu07: \citet{Churchwell2007}; Deh09: \citet{Deharveng2009}; Deh10: \citet{Deharveng2010}; Dew15: \citet{Dewangan2015}; Gen12: \citet{Gennaro2012}; Hou14: \citet{Hou2014}; Pav12: \citet{Pavel2012}; Rah10: \citet{Rahman2010}; Rod12: \citet{Rodriguez-Esnard2012}; Wat09: \citet{Watson2009}; Wat10: \citet{Watson2010}; Zha13: \citet{Zhang2013}. $^{*}$: The nearer kinematic distance is adopted. $^{**}$: C is a closed type, while B is a broken type. The others are unclassified. \strut}} \\
\endlastfoot
\hline
CN15	&	0.562	&	-0.843	&	0.71	&	1.2${\pm}$0.5		&	Gen12	&	0.564	&	-0.842	&	0.85	&	B	\\
CN17	&	0.593	&	-0.846	&	0.84	&	1.2${\pm}$0.5		&	Gen12	&	0.586	&	-0.840	&	0.75	&		\\
CN24	&	1.163	&	-0.097	&	1.92	&	11.2${\pm}$1.9		&	Hou14	&	1.150	&	-0.091	&	1.42	&	B	\\
CN39	&	3.320	&	-0.029	&	0.34	&	13.8${\pm}$2.3		&	Hou14	&	3.320	&	-0.029	&	0.25	&		\\
CN43	&	3.651	&	-0.117	&	0.77	&	14.5${\pm}$2.5		&	Hou14	&	3.650	&	-0.119	&	0.74	&	C	\\
CN58	&	4.386	&	0.073	&	0.86	&	14.2${\pm}$1.9		&	Hou14	&	4.386	&	0.073	&	0.81	&	C	\\
CN63	&	4.563	&	-0.120	&	1.77	&	12.0${\pm}$0.7		&	Hou14	&	4.563	&	-0.126	&	1.87	&		\\
CN64	&	5.191	&	-0.282	&	0.72	&	15.4${\pm}$2.3		&	Hou14	&	5.191	&	-0.282	&	0.71	&	B	\\
\end{longtable}%

\clearpage

\begin{landscape}{
\fontsize{9pt}{9pt}\selectfont
\begin{longtable}[ht]{cccccccc}
\caption{\small Summary of the flux densities obtained by the circular ($< 2R$) aperture photometry.}\label{tab:Mettab2} \\
\hline
Name & Flux 9 \mic [Jy] & Flux 18 \mic [Jy] & Flux 65 \mic [Jy] & Flux 90 \mic [Jy] & Flux 140 \mic [Jy] & Flux 160 \mic [Jy] & Notes \\
\endfirsthead
\multicolumn{8}{l}{\small Continued.} \\ 
\hline
Name & Flux 9 \mic [Jy] & Flux 18 \mic [Jy] & Flux 65 \mic [Jy] & Flux 90 \mic [Jy] & Flux 140 \mic [Jy] & Flux 160 \mic [Jy] & Notes \\
\hline
\endhead
\hline
\endfoot
\hline
\multicolumn{8}{l}{\small \parbox[ht]{210mm}{\strut sc: The background region to estimate the background level is modified from that automatically selected in the aperture photometry to avoid source contaminations. \strut}} \\ 
\endlastfoot
\hline
CN15	&$(	1.15	 \pm 	0.01	) \times 10^{	2	}$ & $(	3.69	 \pm 	0.03	) \times 10^{	2	}$ & $(	8.7	 \pm 	0.2	) \times 10^{	3	}$ & $(	4.6	 \pm 	0.1	) \times 10^{	3	}$ & $(	5.9	 \pm 	0.2	) \times 10^{	3	}$ & $(	7.8	 \pm 	0.3	) \times 10^{	3	}$ &		\\
CN17	&$(	4.94	 \pm 	0.01	) \times 10^{	1	}$ & $(	1.099	 \pm 	0.001	) \times 10^{	2	}$ & $(	1.926	 \pm 	0.005	) \times 10^{	3	}$ & $(	1.85	 \pm 	0.01	) \times 10^{	3	}$ & $(	2.04	 \pm 	0.02	) \times 10^{	3	}$ & $(	1.78	 \pm 	0.02	) \times 10^{	3	}$ &	sc	\\
CN24	&$(	1.311	 \pm 	0.007	) \times 10^{	2	}$ & $(	5.77	 \pm 	0.01	) \times 10^{	2	}$ & $(	7.07	 \pm 	0.06	) \times 10^{	3	}$ & $(	4.32	 \pm 	0.06	) \times 10^{	3	}$ & $(	6.0	 \pm 	0.2	) \times 10^{	3	}$ & $(	7.0	 \pm 	0.3	) \times 10^{	3	}$ &		\\
CN39	&$(	2.94	 \pm 	0.03	) \times 10^{	0	}$ & $(	5.98	 \pm 	0.06	) \times 10^{	0	}$ & $(	7.2	 \pm 	0.4	) \times 10^{	1	}$ & $(	5.6	 \pm 	0.3	) \times 10^{	1	}$ & $(	3.9	 \pm 	0.2	) \times 10^{	1	}$ & $(	5.0	 \pm 	0.3	) \times 10^{	1	}$ &	sc	\\
CN43	&$(	5.2	 \pm 	0.2	) \times 10^{	0	}$ & $(	1.22	 \pm 	0.02	) \times 10^{	1	}$ & $(	4.2	 \pm 	0.1	) \times 10^{	2	}$ & $(	4.4	 \pm 	0.1	) \times 10^{	2	}$ & $(	4.9	 \pm 	0.2	) \times 10^{	2	}$ & $(	7.2	 \pm 	0.2	) \times 10^{	2	}$ &		\\
CN58	&$(	1.800	 \pm 	0.008	) \times 10^{	1	}$ & $(	2.939	 \pm 	0.007	) \times 10^{	1	}$ & $(	8.15	 \pm 	0.03	) \times 10^{	2	}$ & $(	9.64	 \pm 	0.04	) \times 10^{	2	}$ & $(	1.63	 \pm 	0.02	) \times 10^{	3	}$ & $(	1.32	 \pm 	0.01	) \times 10^{	3	}$ &	sc	\\
CN63	&$(	4.22	 \pm 	0.02	) \times 10^{	1	}$ & $(	4.96	 \pm 	0.03	) \times 10^{	1	}$ & $(	1.23	 \pm 	0.02	) \times 10^{	3	}$ & $(	1.37	 \pm 	0.02	) \times 10^{	3	}$ & $(	2.66	 \pm 	0.02	) \times 10^{	3	}$ & $(	2.03	 \pm 	0.03	) \times 10^{	3	}$ &	sc	\\
CN64	&$(	6.01	 \pm 	0.07	) \times 10^{	0	}$ & $(	1.974	 \pm 	0.007	) \times 10^{	1	}$ & $(	3.31	 \pm 	0.04	) \times 10^{	2	}$ & $(	3.71	 \pm 	0.04	) \times 10^{	2	}$ & $(	3.4	 \pm 	0.1	) \times 10^{	2	}$ & $(	3.58	 \pm 	0.08	) \times 10^{	2	}$ &	sc	\\
\end{longtable}%
}
\end{landscape}%

\clearpage

\normalsize
\begin{longtable}[ht]{ccccc}
\caption{Summary of the luminosities obtained by the SED fitting.}\label{tab:Luminosity} \\ 
\hline
Name & log($L_{\rm{TIR}}$/$L_{\solar}$) & log($L_{\rm{PAH}}$/$L_{\solar}$) & log($L_{\rm{warm}}$/$L_{\solar}$) & log($L_{\rm{cold}}$/$L_{\solar}$) \\
\endfirsthead
\multicolumn{5}{l}{\small Continued.} \\
\hline
Name & log($L_{\rm{TIR}}$/$L_{\solar}$) & log($L_{\rm{PAH}}$/$L_{\solar}$) & log($L_{\rm{warm}}$/$L_{\solar}$) & log($L_{\rm{cold}}$/$L_{\solar}$) \\
\hline
\endhead
\hline
\endfoot
\hline
\endlastfoot
\hline
CN15	&	4.54	$\pm$	0.31	&	3.37	$\pm$	0.36	&	4.43	$\pm$	0.40	&	3.72	$\pm$	0.38	\\
CN17	&	4.04	$\pm$	0.29	&	3.02	$\pm$	0.36	&	3.92	$\pm$	0.38	&	3.23	$\pm$	0.37	\\
CN24	&	6.50	$\pm$	0.17	&	5.37	$\pm$	0.15	&	6.39	$\pm$	0.22	&	5.67	$\pm$	0.18	\\
CN39	&	4.55	$\pm$	0.16	&	3.76	$\pm$	0.15	&	4.44	$\pm$	0.21	&	3.23	$\pm$	0.18	\\
CN43	&	5.47	$\pm$	0.16	&	4.20	$\pm$	0.16	&	5.32	$\pm$	0.22	&	4.86	$\pm$	0.18	\\
CN58	&	5.82	$\pm$	0.13	&	4.77	$\pm$	0.12	&	5.61	$\pm$	0.20	&	5.30	$\pm$	0.15	\\
CN63	&	5.87	$\pm$	0.10	&	4.95	$\pm$	0.07	&	5.64	$\pm$	0.17	&	5.31	$\pm$	0.11	\\
CN64	&	5.50	$\pm$	0.16	&	4.31	$\pm$	0.14	&	5.36	$\pm$	0.20	&	4.85	$\pm$	0.17	\\
\end{longtable}%

\newpage

\section{Results}

\subsection{Total luminosities of Galactic IR bubbles}

\begin{figure}[ht]
\begin{minipage}[t]{0.98\hsize}
\begin{center}
\includegraphics[clip, width=140mm]{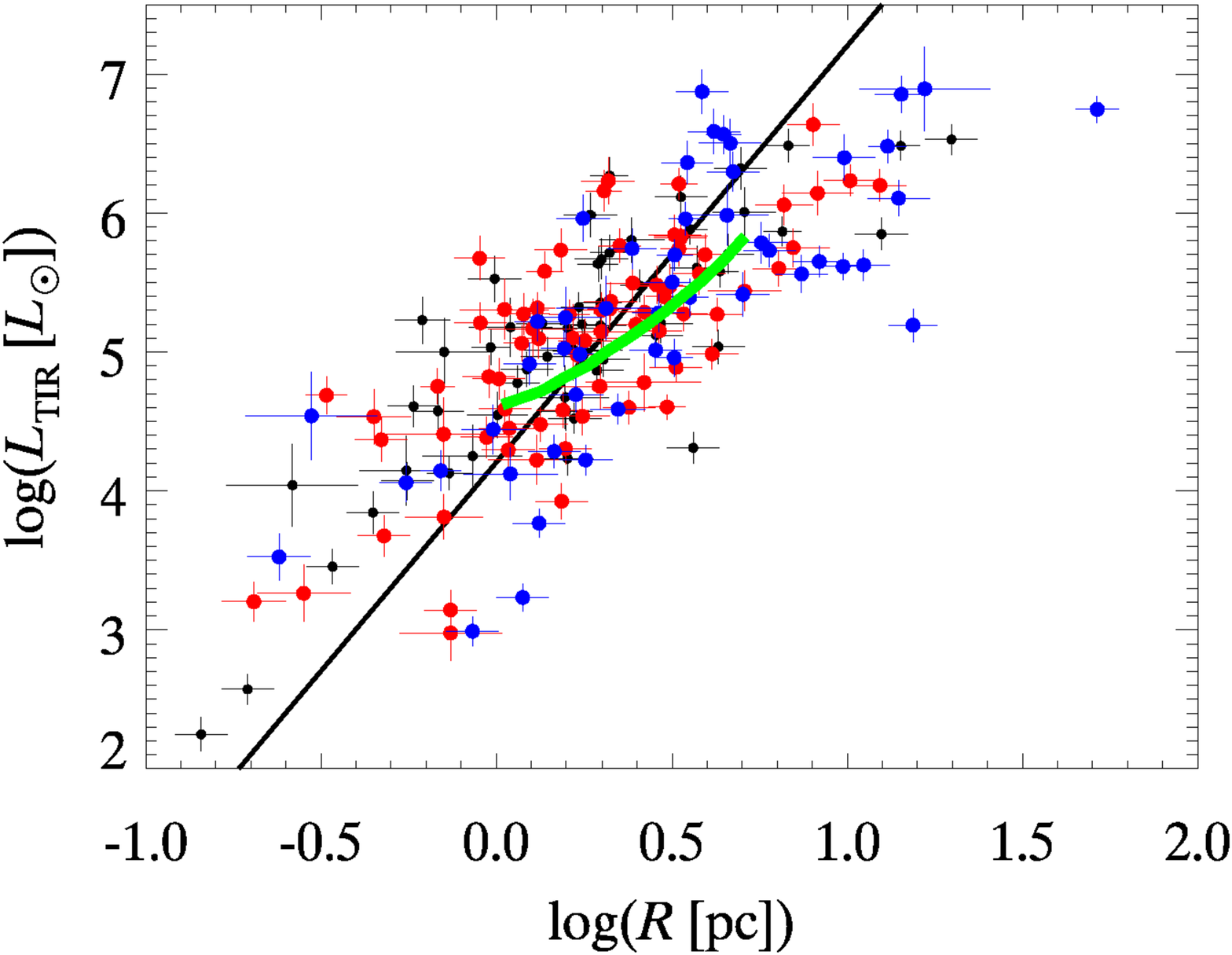}
\end{center}
\end{minipage}\hfill %
\caption{Total IR luminosity ($L_{\rm{TIR}}$) versus radius ($R$) for all the sample on a logarithmic scale. The red, the blue and the black circles represent the closed, the broken and the unclassified bubbles, respectively. The black line indicates the result of fitting to the data points of all the bubbles with $L_{\rm{TIR}} = a R^{3}$. The green curve indicates the relation calculated for O-type stars (\citealt{Martins2005}) in HII regions with the density of  $100~{\rm{cm}}^{-3}$ and temperature of $10^4$ K.  \label{fig:Resfig1}}
\end{figure}

Figure \ref{fig:Resfig1} shows the relation between $L_{\rm{TIR}}$ and the bubble radius, $R$ (pc), calculated from $R$ (arcmin) and the distance for all the 180 bubbles (i.e., 74 closed, 49 broken and 57 unclassified). The errors of $L_{\rm{TIR}}$ were estimated from those of the SED fitting and the distance uncertainties, while those of $R$ were estimated from the systematic uncertainty of 11\% in the circular fitting and the distance uncertainties. The $L_{\rm{TIR}}$ values of the bubbles range widely from $1 \times 10^2$ to $1 \times 10^7 L_{\solar}$, which correspond to the bolometric luminosity, $L_{\rm{bol}}$, of a single B-type star to many O-type stars; for example, $L_{\rm{bol}}$ of a O9$-$O3 star ranges from $5 \times 10^4$ to $6 \times 10^5 L_{\solar}$ (\citealt{Martins2005}). The figure shows that $L_{\rm{TIR}}$ and $R$ are tightly correlated with each other, where the correlation coefficient $r_{\rm c}$ is 0.78. $L_{\rm{TIR}}$ of all the bubbles follow a power-law relation with an index of $\sim$3. 

Several previous works suggested that a shell structure of a bubble is created by expansion of HII regions (e.g., \citealt{Hosokawa2005}; \citeyear{Hosokawa2006a}; \citealt{Deharveng2010}). The ionization front of an HII region, which delineates the inner edge of a shell, is described as a function of the total number of ionizing photons, $Q$, like,
\begin{equation}
\label{eq:stromgren}
Q = \frac{4}{3}\pi{R_{\rm{s}}}^3{n_{\rm{e}}}{n_{\rm{p}}}\alpha_{\rm{B}}(T_{\rm{e}}),
\end{equation}
where $R_{\rm{s}}$, $n_{\rm{e}}$, $n_{\rm{p}}$, $T_{\rm{e}}$ and $\alpha_{\rm{B}}$ are the Str$\rm{\ddot{o}}$mgren radius, electron density, proton density, electron temperature and $``$case-B$"$ recombination coefficient (\citealt{Stromgren1939}; \citealt{Osterbrock1989}). We calculated the relation between $L_{\rm{bol}}$ and $R_{\rm{s}}$ for O-type stars from $Q$ and $L_{\rm{bol}}$ given in \citet{Martins2005}, assuming $n_{\rm{e}} = n_{\rm{p}} = 100~{\rm{cm}}^{-3}$ and $T_{\rm{e}} = 10^4$ K as typical values for HII regions. In figure \ref{fig:Resfig1}, we plot the $L_{\rm{bol}}$-$R_{\rm{s}}$ relation thus calculated on the assumption that $R$ $\sim$ ${R_{\rm{s}}}$ and $ L_{\rm{TIR}} \sim L_{\rm{bol}}$. As can be seen in the figure, the $L_{\rm{TIR}}$-$R$ relation observed for the sample bubbles, as a whole, is consistent with the calculated relation. 

\begin{figure}[ht]
\begin{center}
\includegraphics[clip, width=100mm]{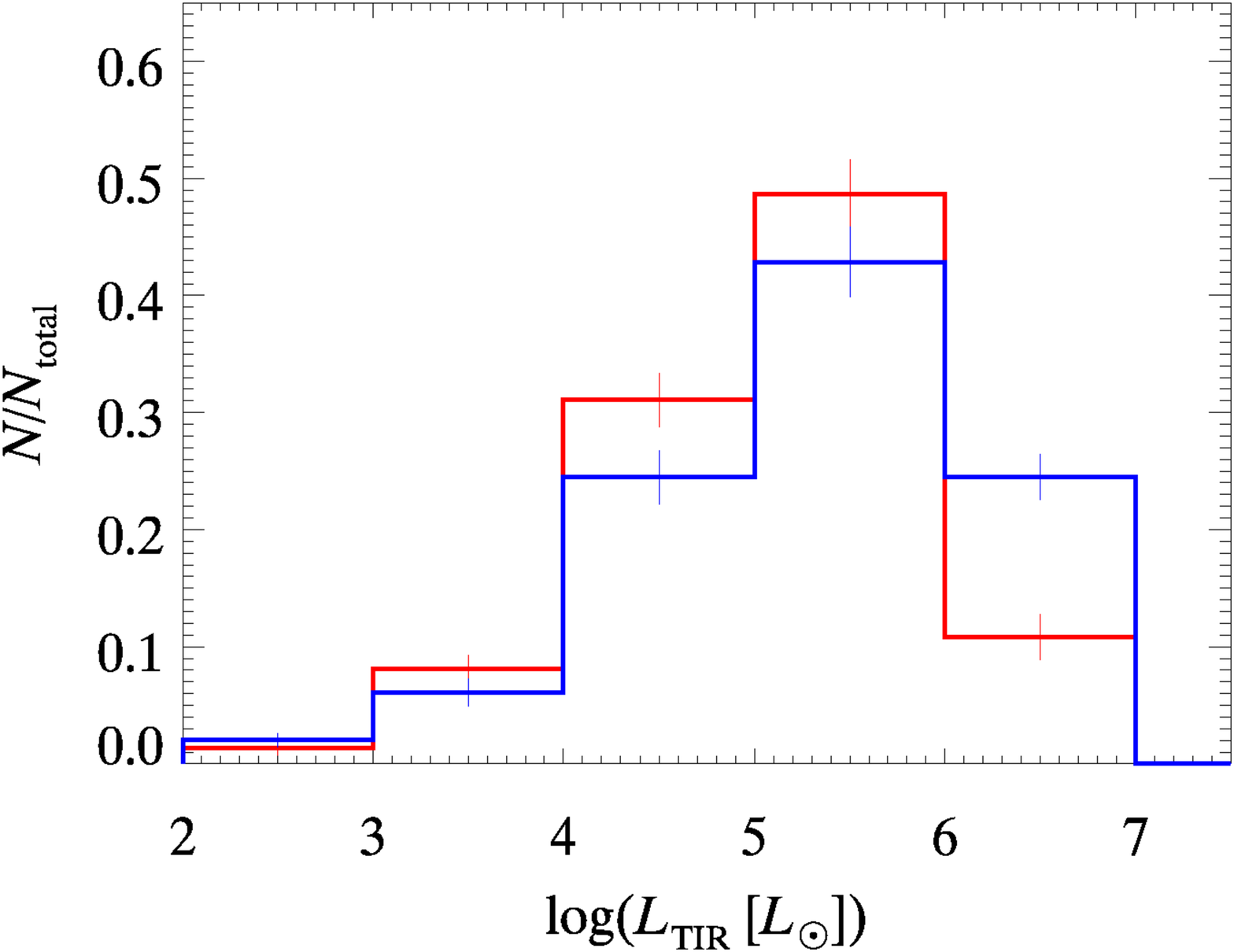}
\end{center}
\caption{Histograms of $L_{\rm{TIR}}$. The histograms in the red and the blue line correspond to the closed bubbles and the broken bubbles, respectively. The vertical axis corresponds to the fraction of the number of the bubbles to the total for each type. Poisson errors are attached as bars. \label{fig:histLTIR}}
\end{figure}

\begin{figure}[ht]
\begin{center}
\includegraphics[width=140mm]{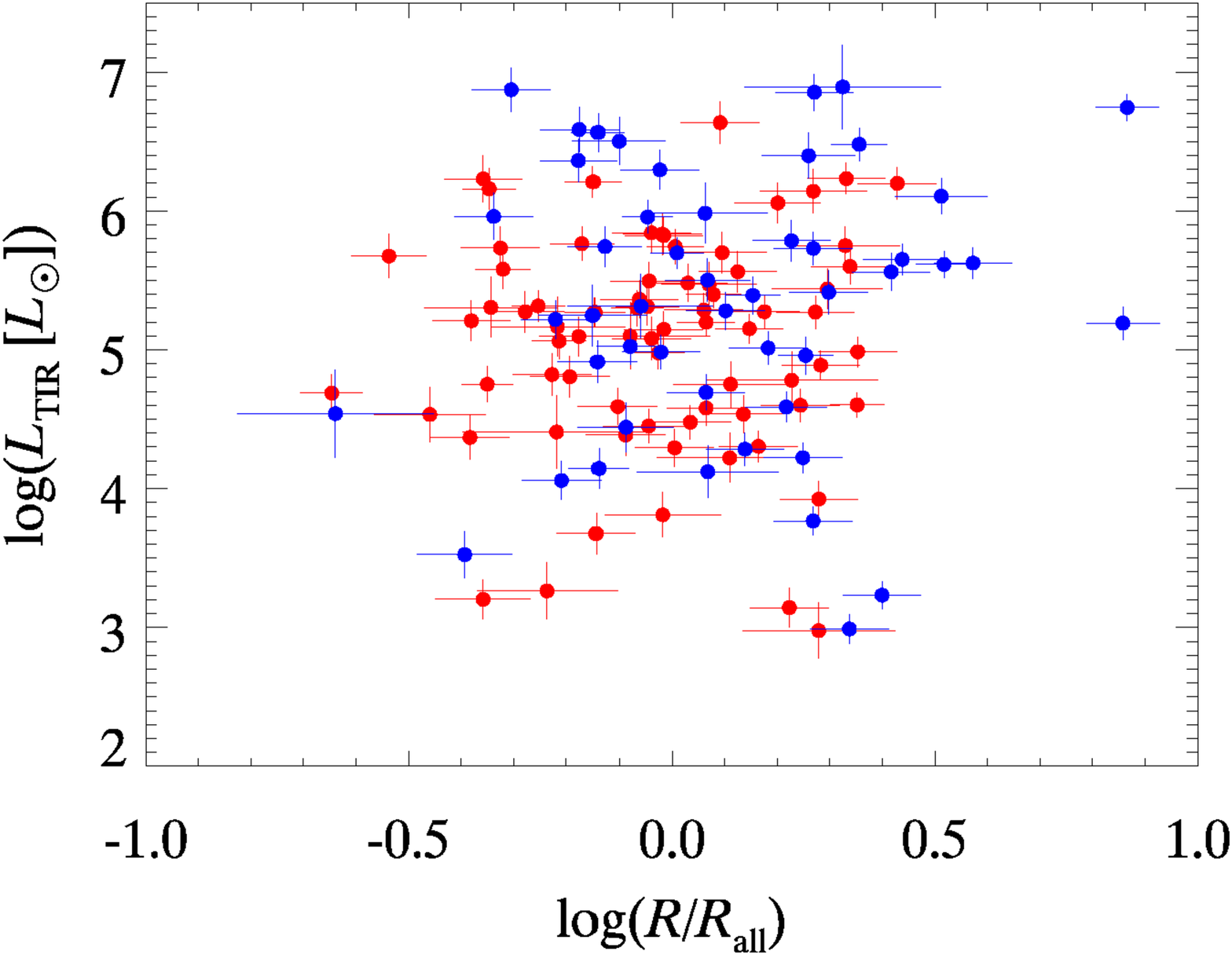}
\end{center}
\caption{$L_{\rm{TIR}}$ plotted against $R/{R_{\rm{all}}}$ for the closed and broken bubbles on a logarithmic scale, where ${R_{\rm{all}}}$ is the radius of the bubbles as a function of $L_{\rm{TIR}}$, determined by fitting to the data points of all the bubbles with $L_{\rm{TIR}} = a R^{3}$ (figure \ref{fig:Resfig1}). The symbols are the same as those in figure \ref{fig:Resfig1}. \label{fig:LRdev}}
\end{figure}

In figure \ref{fig:Resfig1}, there is an overall similarity in the distributions in the $L_{\rm{TIR}}$-$R$ plots between the closed and the broken bubbles, despite their large differences in the morphology. The $L_{\rm{TIR}}$ distribution of the broken bubbles seems to extend towards higher luminosities, as compared with that of the closed bubbles. This tendency is confirmed in the histograms of $L_{\rm{TIR}}$ in figure \ref{fig:histLTIR}; in the highest luminosity bin, the fractional number of the broken bubbles is significantly larger than that of the closed bubbles. In figure \ref{fig:LRdev}, we show the deviations of the data points with respect to the power-law relation along $R$. Here ${R_{\rm{all}}}$ is the radius of the bubbles as a function of $L_{\rm{TIR}}$, determined by fitting to the data points of all the bubbles with $L_{\rm{TIR}} = a R^{3}$ (figure \ref{fig:Resfig1}). In the figure, the broken bubbles are systematically shifted toward larger $R/{R_{\rm{all}}}$, relative to the closed bubbles, especially those with high $L_{\rm{TIR}}$ ($>10^{5}L_{\solar}$). A Kolmogorov-Smirnov (K-S) test shows that the distributions of the two types are statistically different with a 85\% confidence level (K-S value $D \sqrt{nm/(n+m)}$ of 1.15, where $D$ is a K-S statistic, and $m$ and $n$ are the numbers of the closed and the broken samples, respectively).

\clearpage

\subsection{Ratios of PAH to total luminosities of Galactic IR bubbles}

\begin{figure}[ht]
\begin{tabular}{cc}
\centering
\begin{minipage}{0.48\hsize}
\centering
\includegraphics[width=80mm]{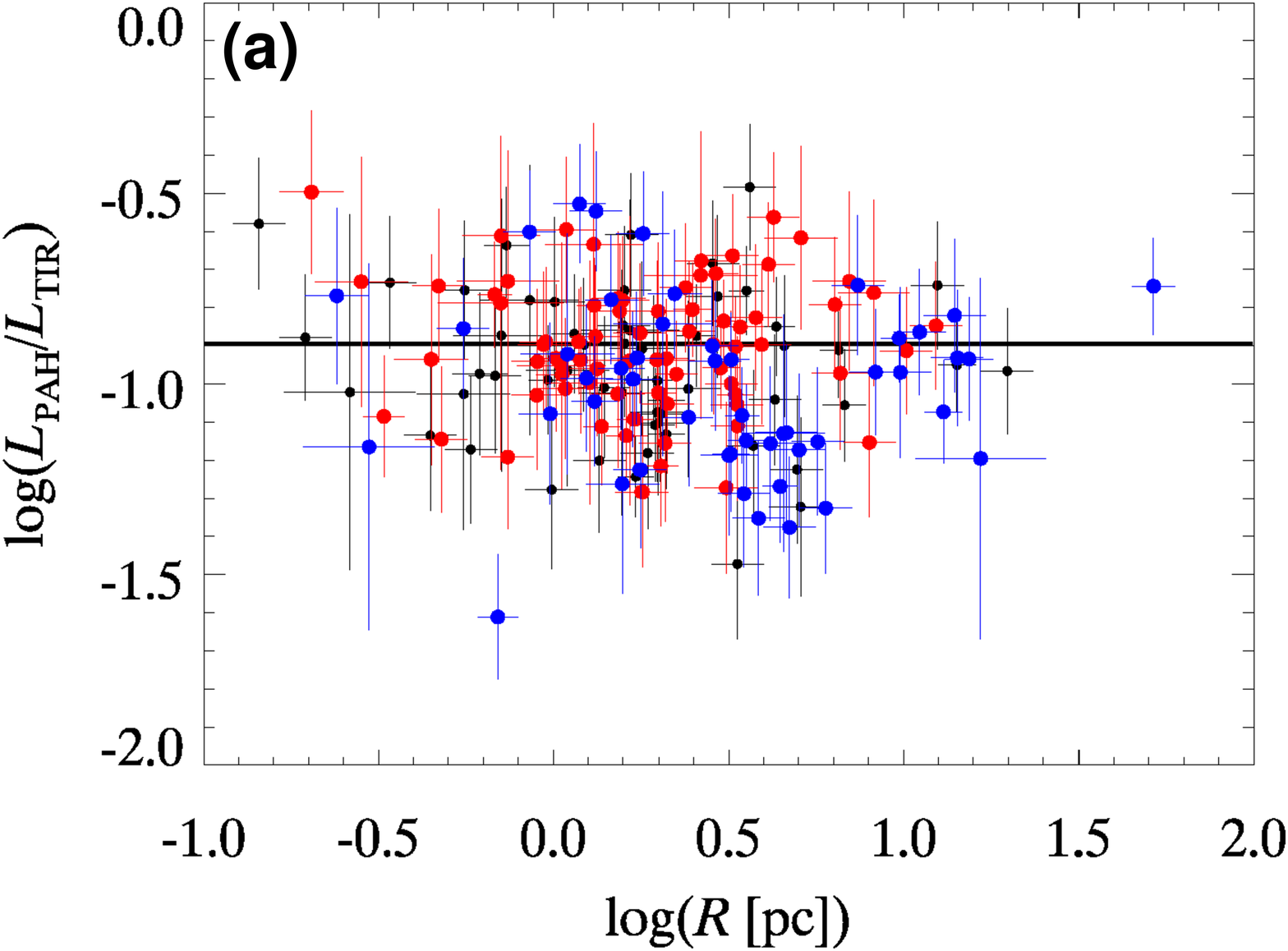}
\end{minipage}\hfill %
\begin{minipage}{0.48\hsize}
\centering
\includegraphics[width=80mm]{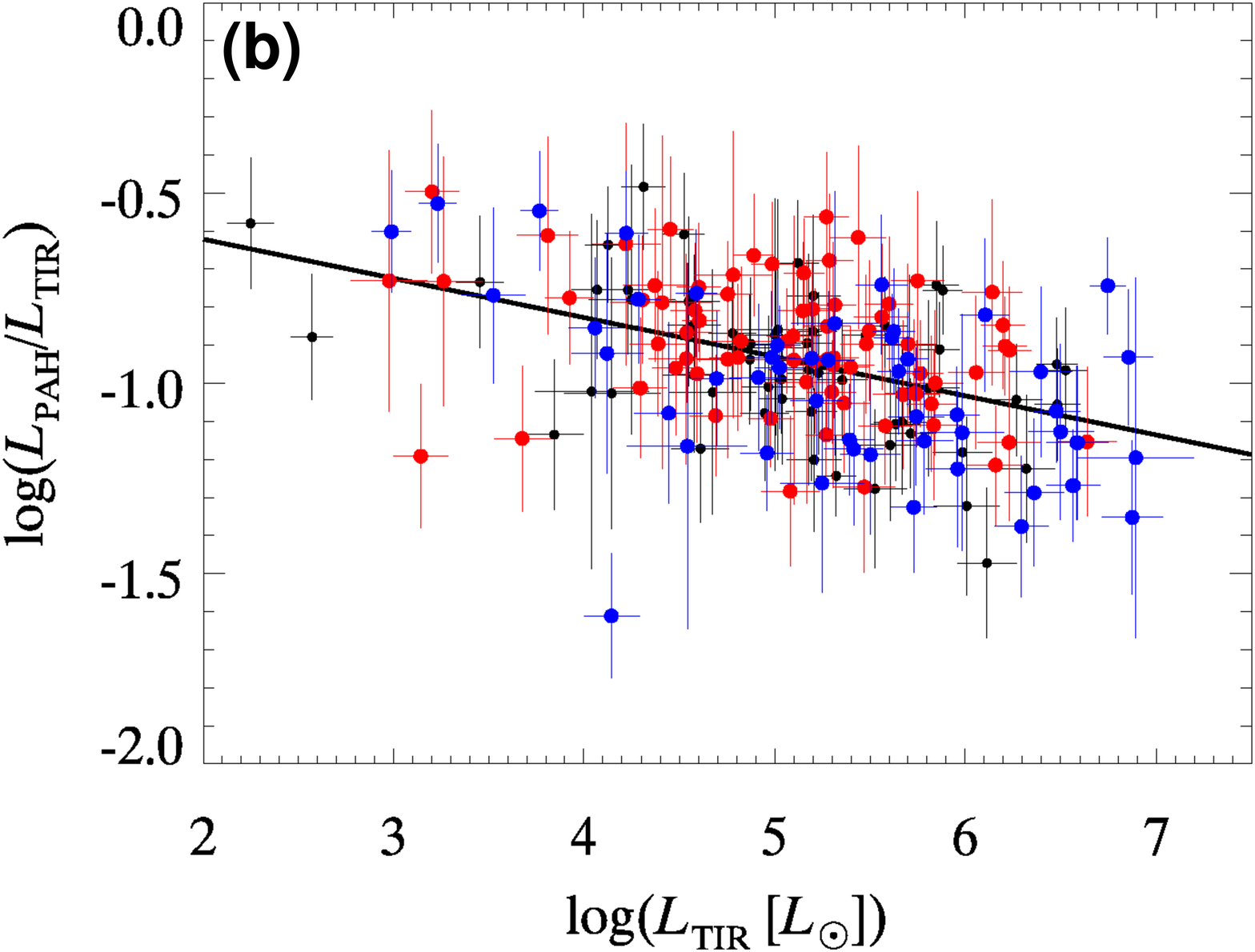}
\end{minipage}\hfill %
\end{tabular}
\caption{Ratio of the PAH to the total IR luminosity ($L_{\rm{PAH}}/L_{\rm{TIR}}$) plotted against (a) $R$ and (b) $L_{\rm{TIR}}$ for all the bubbles. The red, blue and black circles represent the closed, broken and unclassified bubbles, respectively. The line in panel (a) is the constant line fitted to the data points of the closed bubbles, while the line in panel (b) is the fitted line to the data points of all the bubbles. \label{fig:PAHplot}}
\end{figure}

Figure \ref{fig:PAHplot}a shows the relation between the ratios of the PAH to total IR luminosities ($L_{\rm{PAH}}/L_{\rm{TIR}}$) and $R$ for all the bubbles. $L_{\rm{PAH}}/L_{\rm{TIR}}$ and $R$ are not significantly correlated with each other ($r_{\rm c}=-0.11$). The result of the constant line fitting shows that $L_{\rm{PAH}}/L_{\rm{TIR}}$ is statistically constant with $R$ for the closed bubbles (${\chi}^2/{\nu}=1.19$), while it is not for the broken bubbles (${\chi}^2/{\nu}=1.81$). A K-S test shows that the distributions of the two types are statistically different with a 90\% confidence level (K-S value of 1.56). In particular, the figure reveals that the broken bubbles at radii larger than $\sim$2.5 pc (${\rm{log}}R\sim0.4$) have systematically lower $L_{\rm{PAH}}/L_{\rm{TIR}}$ ratios.


Figure \ref{fig:PAHplot}b shows the plot of $L_{\rm{PAH}}/L_{\rm{TIR}}$ against $L_{\rm{TIR}}$ for all the bubbles. In contrast to the plot of $L_{\rm{PAH}}/L_{\rm{TIR}}$ versus $R$ in figure \ref{fig:PAHplot}a, $L_{\rm{PAH}}/L_{\rm{TIR}}$ is significantly correlated with $L_{\rm{TIR}}$ ($r_{\rm c}=-0.45$). The negative correlation of $L_{\rm{PAH}}/L_{\rm{TIR}}$ with $L_{\rm{TIR}}$ can be interpreted in a straightforward manner; the intense UV radiation indicated by high $L_{\rm{TIR}}$ would accelerate photodissociation of PAHs exposed to the UV (i.e., \citealt{Smith2007}; \citealt{Bendo2008}; \citealt{Tielens2008}). On the other hand, no significant correlation of $L_{\rm{PAH}}/L_{\rm{TIR}}$ with $R$ may be explained by considering that $R$ does not strongly depend on $L_{\rm{TIR}}$ ($R \propto L^{\frac{1}{3}}$) and also that $R$ depends on the gas density for a given $L_{\rm{TIR}}$ ($R \propto n^{-{\frac{2}{3}}}$; eq. \ref{eq:stromgren}). As can be seen in figure \ref{fig:Resfig1}, $R$ scatters in a range of $\sim$1.5 orders of magnitude for a given $L_{\rm{TIR}}$. This corresponds to a variation in the gas density by one order of magnitude, which will be caused by the initial conditions of the ambient ISM and/or the later expansion of an HII region. The lower $L_{\rm{PAH}}/L_{\rm{TIR}}$ observed for the broken bubbles at larger radii (figure \ref{fig:PAHplot}a), and thus at higher $L_{\rm{TIR}}$ from figure \ref{fig:Resfig1}, seems to strengthen the negative correlation for the broken bubbles in figure \ref{fig:PAHplot}b.

\begin{figure}[ht]
\centering
\subfigure{
\makebox[180mm][l]{\raisebox{0mm}{\hspace{12mm} \small{RA (J2000)}} \hspace{17mm} \small{RA (J2000)} \hspace{17mm} \small{RA (J2000)} \hspace{17.5mm} \small{RA (J2000)}}%
}

\subfigure{
\centering
\makebox[5mm][l]{\raisebox{0mm}{\rotatebox{90}{\small{\hspace{5.5mm} DEC (J2000)} \hspace{10.5mm} \small{DEC (J2000)} \hspace{10.5mm} \small{DEC (J2000)} \hspace{10.5mm} \small{DEC (J2000)} \hspace{10.5mm} \small{DEC (J2000)}}}}%
\includegraphics[width=160mm]{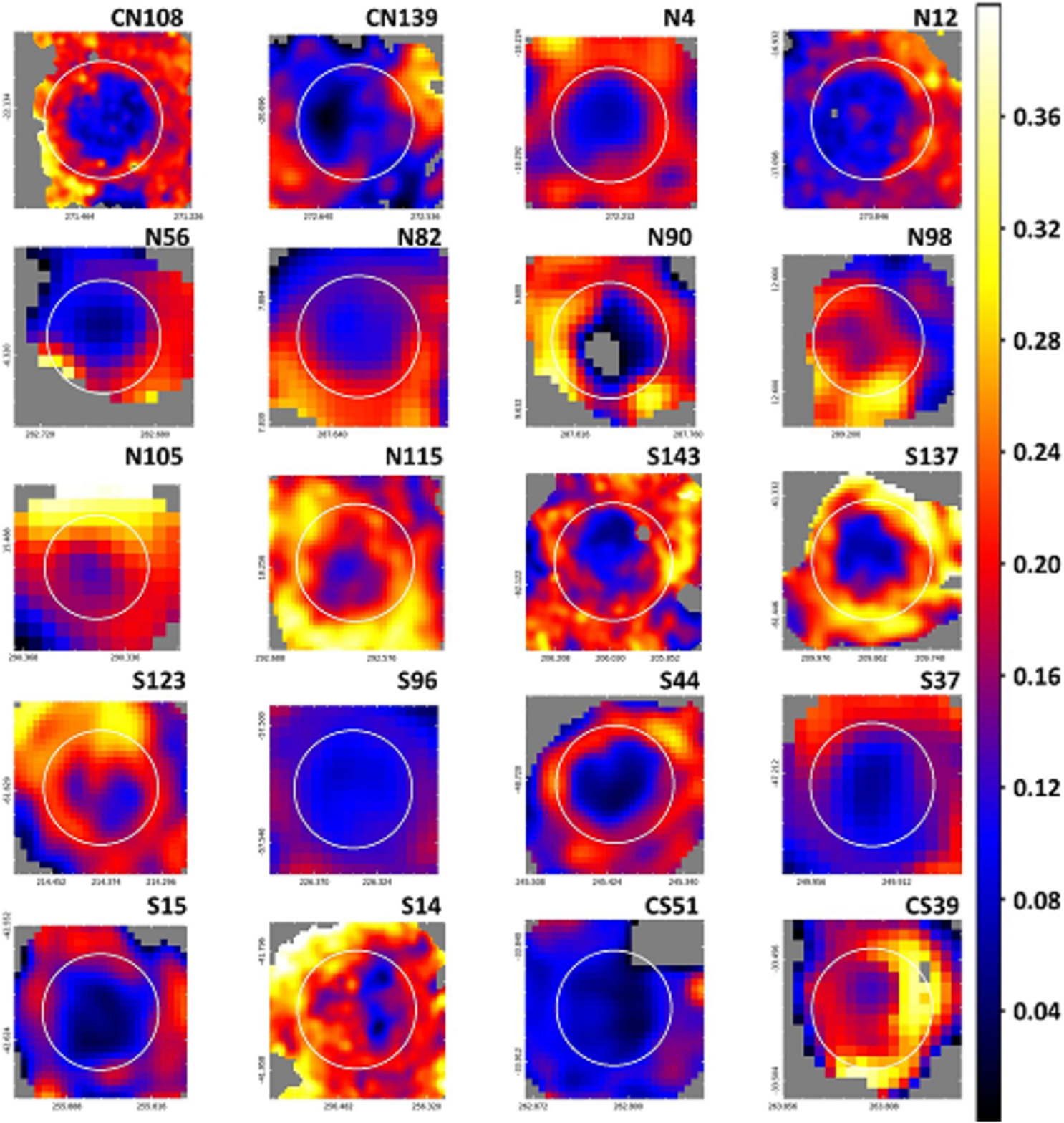}
}
\raggedright
\caption{$I_{\rm{PAH}}/I_{\rm{TIR}}$ ratio maps for the closed bubbles shown in figures \ref{fig:Introfig1} and \ref{fig:Metfig2}. The white circles are the best-fit circles determined by the fitting to the Spitzer 8 \mic images of the bubbles. The color levels are common among the bubbles. \label{fig:PAHspC}} %
\end{figure}

\begin{figure}[ht]
\centering
\subfigure{
\makebox[180mm][l]{\raisebox{0mm}{\hspace{12mm} \small{RA (J2000)}} \hspace{17.5mm} \small{RA (J2000)} \hspace{17.5mm} \small{RA (J2000)} \hspace{17.5mm} \small{RA (J2000)}}%
}

\subfigure{
\centering
\makebox[5mm][l]{\raisebox{0mm}{\rotatebox{90}{\small{\hspace{5.5mm} DEC (J2000)} \hspace{10.5mm} \small{DEC (J2000)} \hspace{10.5mm} \small{DEC (J2000)} \hspace{10.5mm} \small{DEC (J2000)} \hspace{10.5mm} \small{DEC (J2000)}}}}%
\includegraphics[width=160mm]{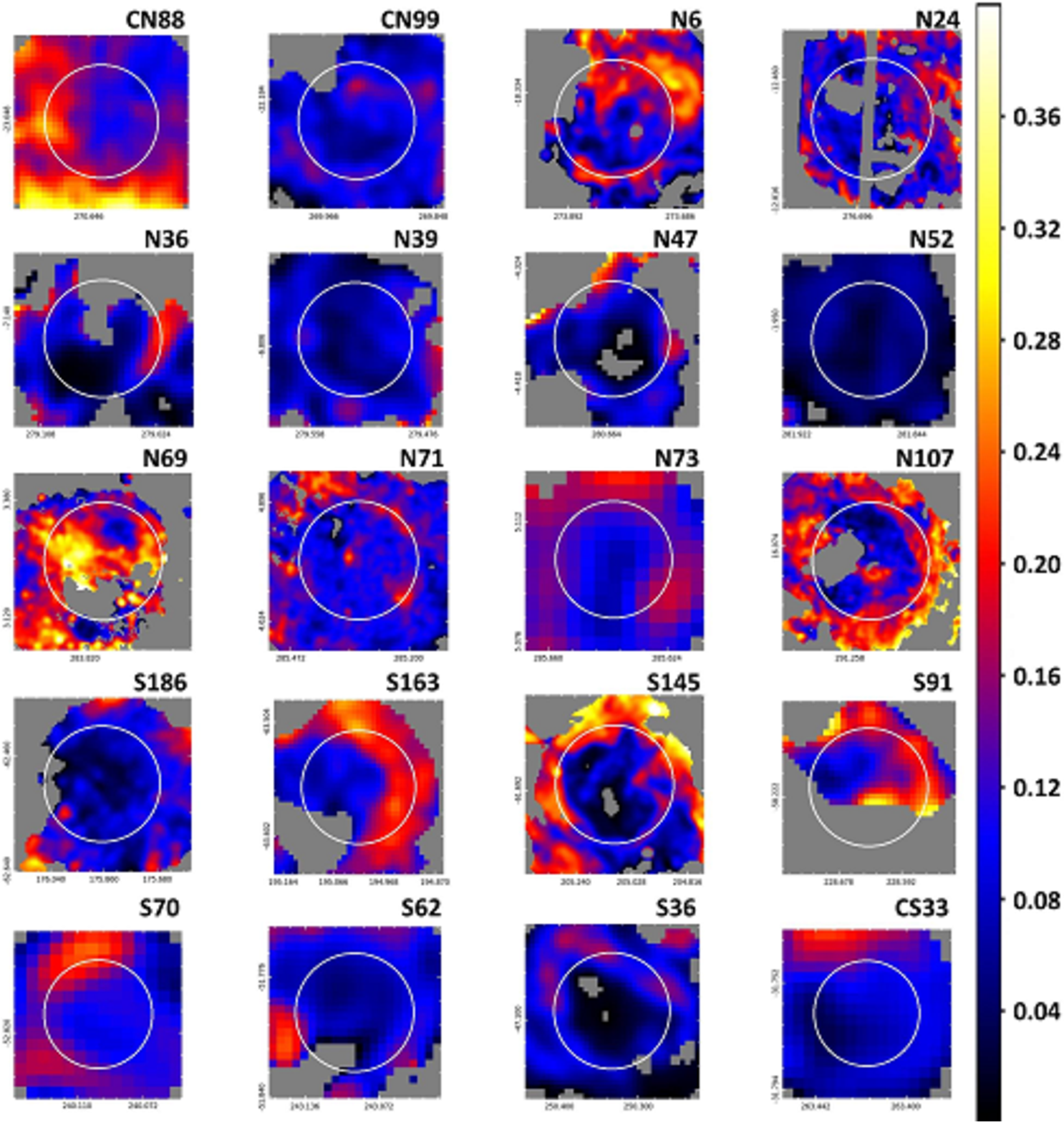}
}
\caption{Same as figure \ref{fig:PAHspC}, but for the broken bubbles shown in figures \ref{fig:Introfig1} and \ref{fig:Metfig2}.} \label{fig:PAHspB}
\end{figure}

Figures \ref{fig:PAHspC} and \ref{fig:PAHspB} show the $I_{\rm{PAH}}/I_{\rm{TIR}}$ ratio maps of 20 closed bubbles and 20 broken bubbles, respectively, which are the same sample as in figures \ref{fig:Introfig1} and \ref{fig:Metfig2}. We masked the pixels where $I_{\rm{TIR}}$ has surface brightness lower than 3 sigma of the background fluctuation above the averaged brightness at (2$-$4)${\times}R$, in order to exclude statistically insignificant pixels. Comparing figure \ref{fig:PAHspB} with figure \ref{fig:PAHspC}, we find that the whole areas of the broken bubbles tend to have lower $I_{\rm{PAH}}/I_{\rm{TIR}}$ ratios than the closed bubbles. The figures also show that $I_{\rm{PAH}}/I_{\rm{TIR}}$ ratios become lower inside the shells for both types.

\begin{figure}[ht]
\begin{center}
\includegraphics[width=85mm]{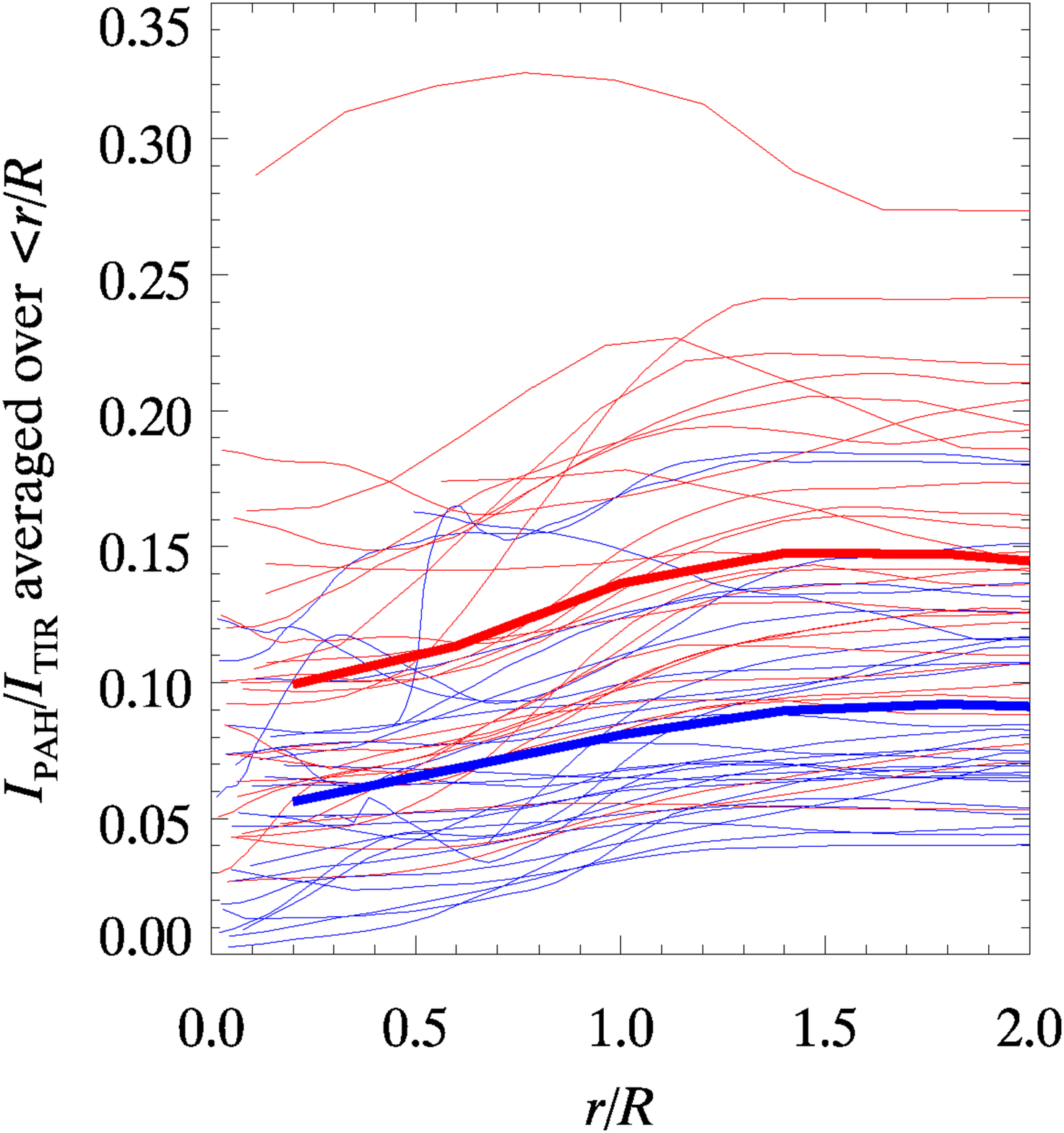}
\end{center}
\caption{Radial profiles of $I_{\rm{PAH}}/I_{\rm{TIR}}$ for the Galactic IR bubbles with R $>$ 2.5 pc. The vertical axis shows the $I_{\rm{PAH}}/I_{\rm{TIR}}$ values averaged over all the pixels within $r$. The horizontal axis shows the radial positions normalized to $R$ of each bubble. The red and the blue curves indicate the closed and broken bubbles, respectively, while the bold curves are their averages. \label{fig:Discfig2}}
\end{figure}

In order to quantify the difference in the $I_{\rm{PAH}}/I_{\rm{TIR}}$ distribution between relatively large closed and broken bubbles, we calculated the radial profiles of $I_{\rm{PAH}}/I_{\rm{TIR}}$ in figure \ref{fig:Discfig2}, where the $I_{\rm{PAH}}/I_{\rm{TIR}}$ values averaged over all the pixels within a certain radial position $r$ are plotted against $r/R$. Here we selected the Galactic IR bubbles with $R$ $>$ 2.5 pc (${\rm{log}}R > 0.4$). Almost every bubble in the figure shows that $I_{\rm{PAH}}/I_{\rm{TIR}}$ starts to increase at $r \sim R$ from the inner to outer direction, where PDRs are likely to dominate; this is reasonable since the PAH emission traces PDRs. The radial profiles of $I_{\rm{PAH}}/I_{\rm{TIR}}$ show a systematic difference in the absolute value between the closed and broken bubbles, although the profiles themselves are notably similar. This indicates that PAHs in the broken bubbles are deficient in wide areas including the outside of the bubbles. 

\clearpage

\subsection{Ratios of warm dust to cold dust luminosities of Galactic IR bubbles}

\begin{figure}[ht]
\begin{tabular}{cc}
\centering
\begin{minipage}{0.48\hsize}
\centering
\includegraphics[width=80mm]{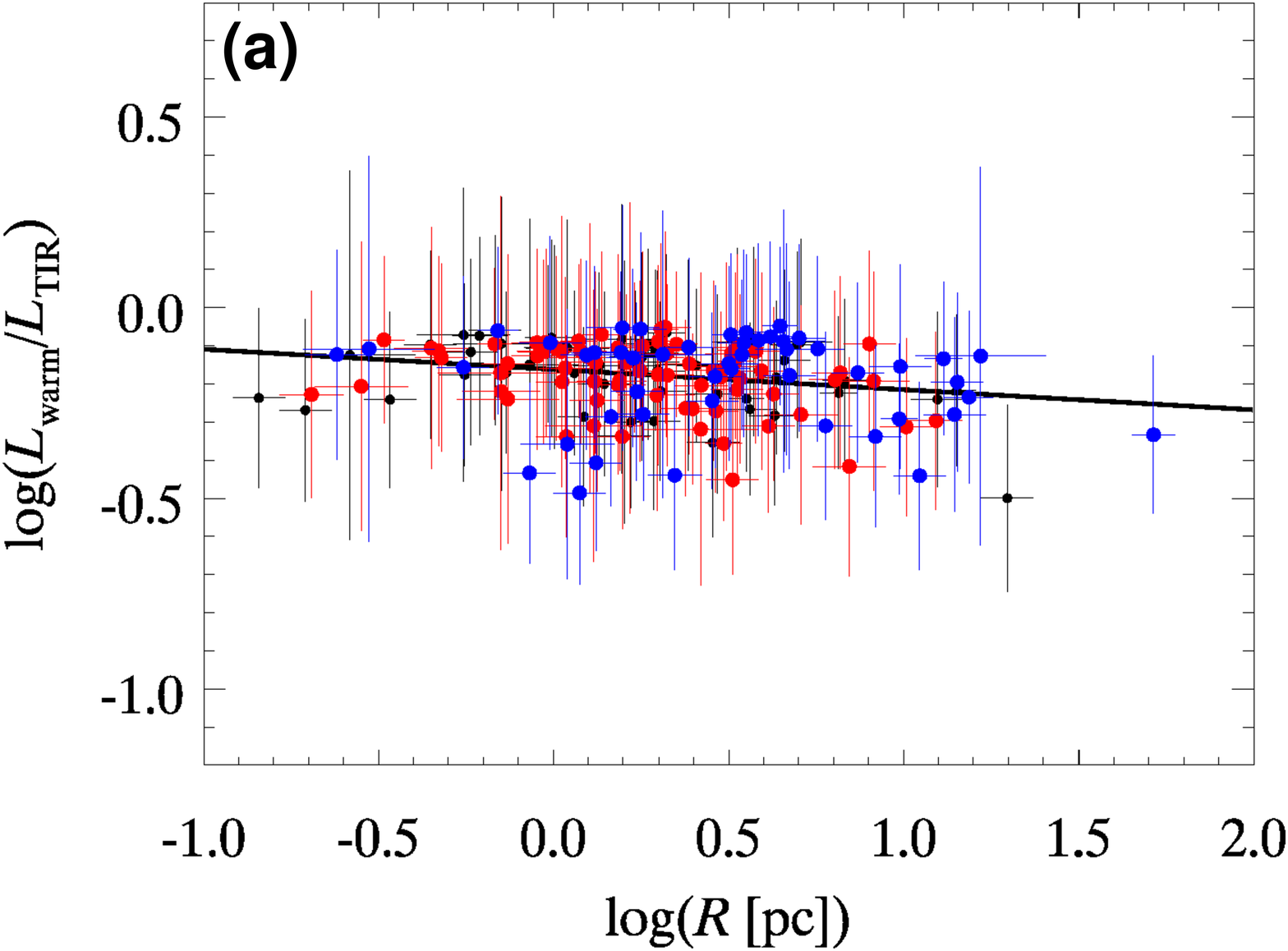}
\end{minipage}\hfill %
\begin{minipage}{0.48\hsize}
\centering
\includegraphics[width=80mm]{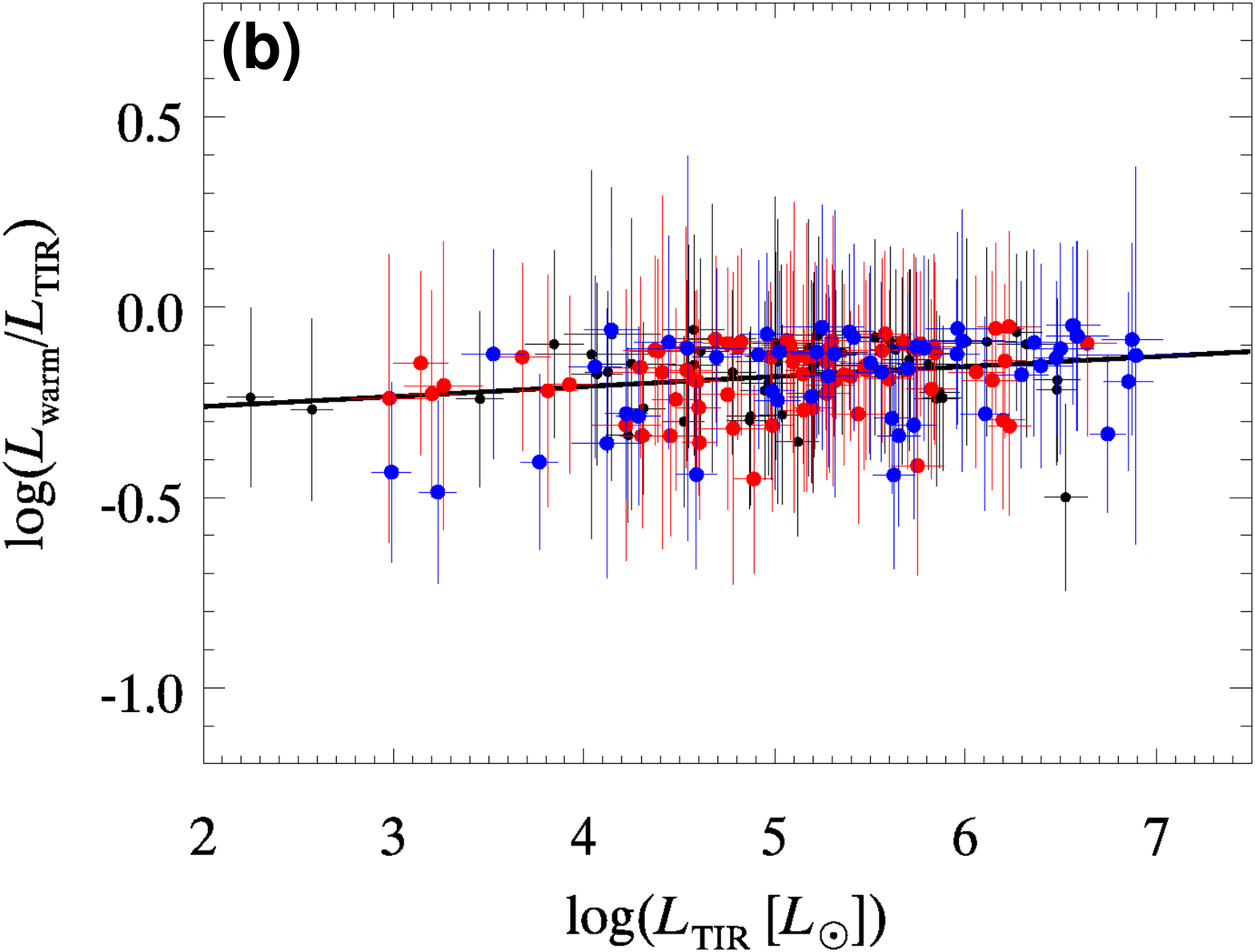}
\end{minipage}\hfill %
\end{tabular}
\begin{tabular}{cc}
\centering
\begin{minipage}{0.48\hsize}
\centering
\includegraphics[width=80mm]{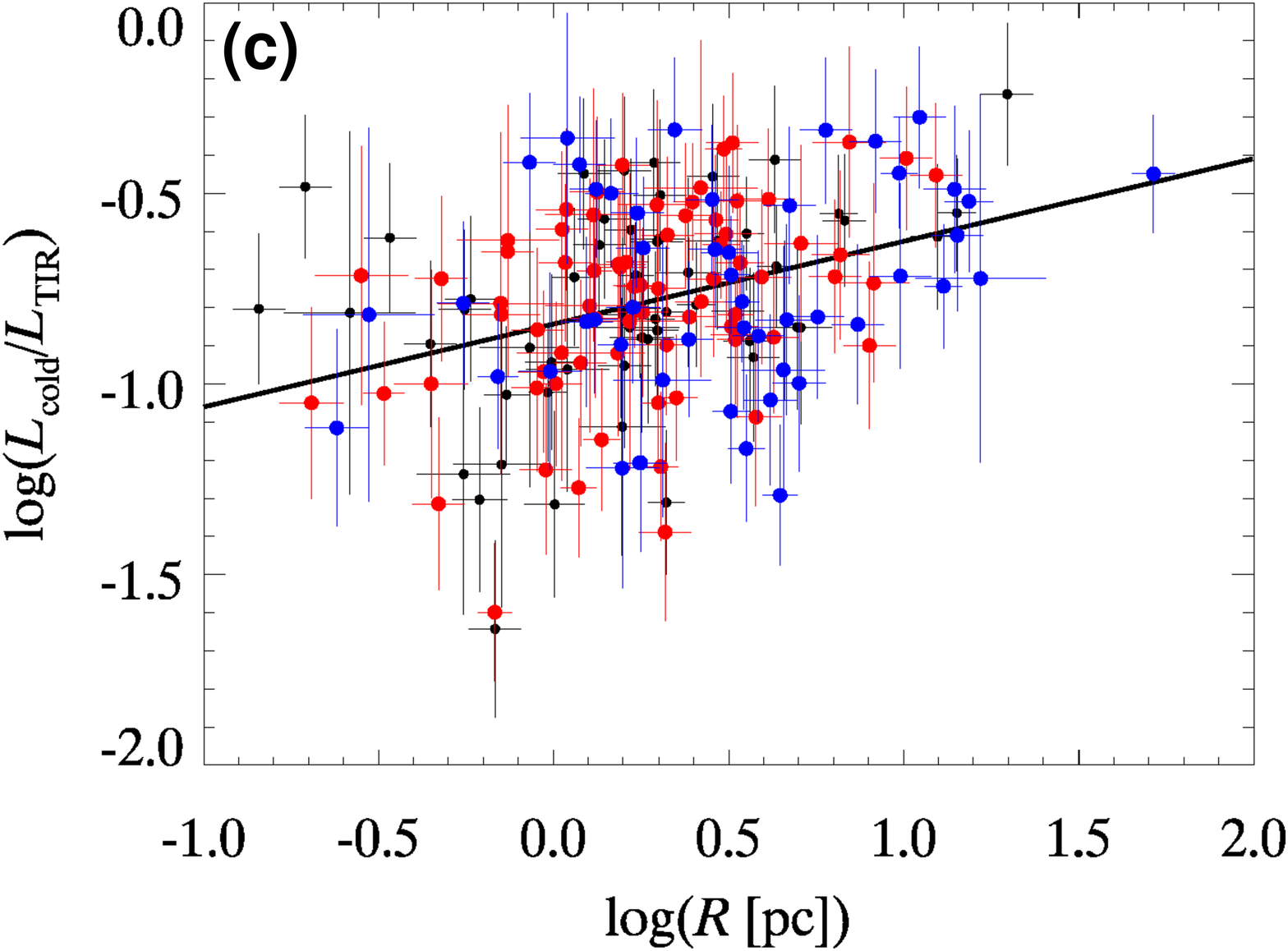}
\end{minipage}\hfill %
\begin{minipage}{0.48\hsize}
\centering
\includegraphics[width=80mm]{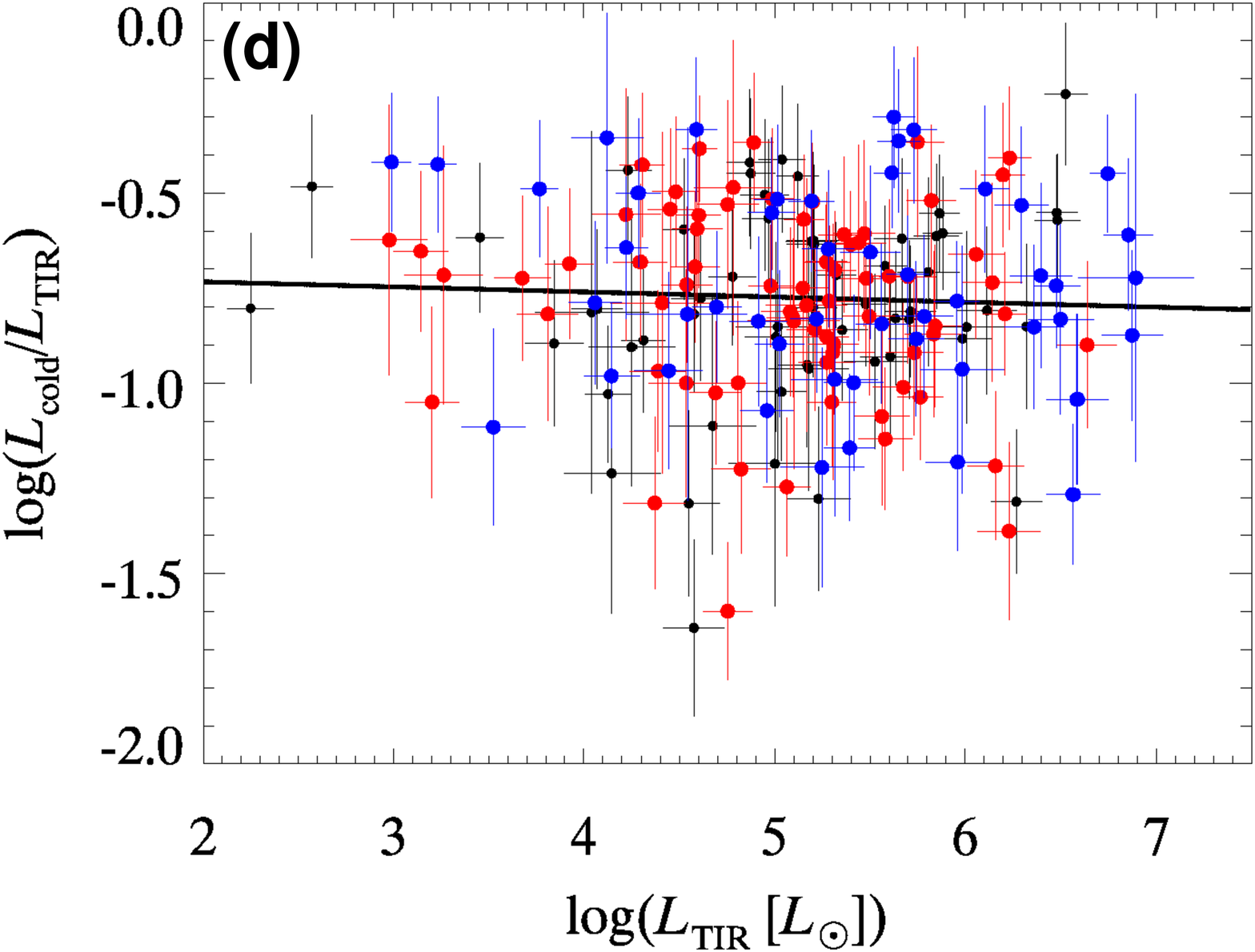}
\end{minipage}\hfill %
\end{tabular}
\begin{tabular}{cc}
\centering
\begin{minipage}{0.48\hsize}
\centering
\includegraphics[width=80mm]{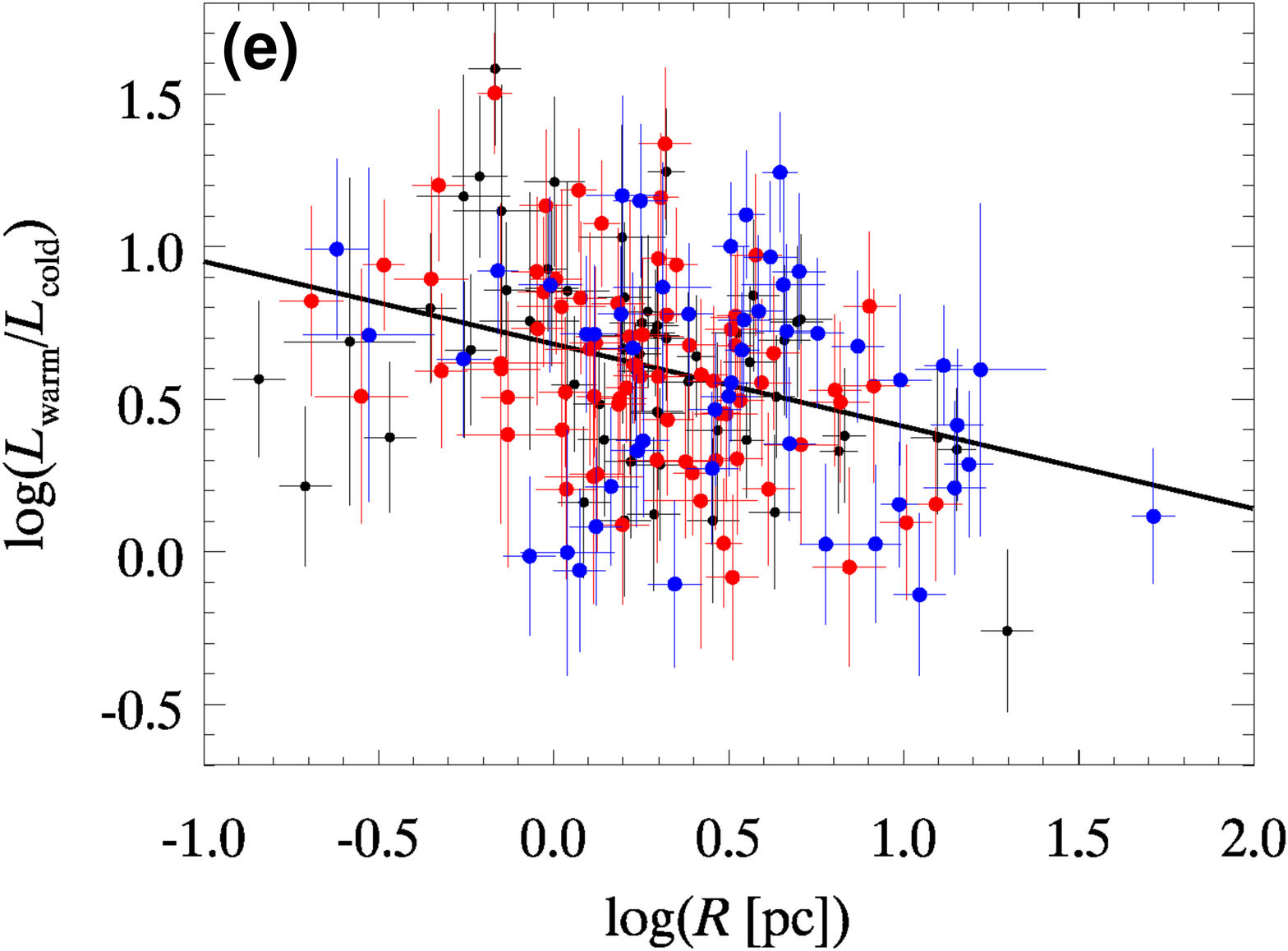}
\end{minipage}\hfill %
\begin{minipage}{0.48\hsize}
\centering
\includegraphics[width=80mm]{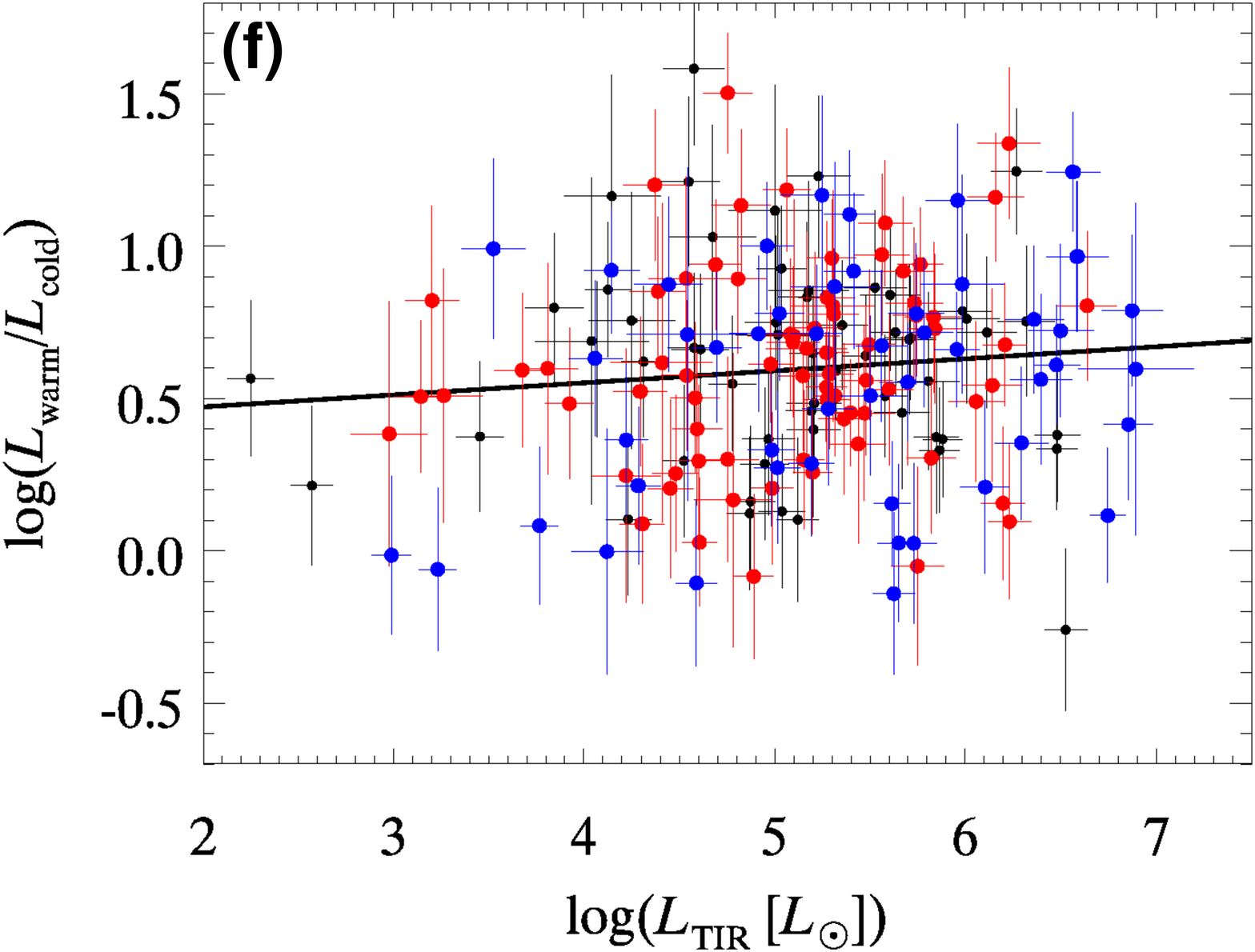}
\end{minipage}\hfill %
\end{tabular}
\caption{Ratios of (a) the warm dust to the total IR luminosity ($L_{\rm{warm}}/L_{\rm{TIR}}$), (c) the cold dust to the total IR luminosity ($L_{\rm{cold}}/L_{\rm{TIR}}$), and (e) the warm dust to the cold dust luminosity ($L_{\rm{warm}}/L_{\rm{cold}}$) plotted against $R$, and the ratios of (b) $L_{\rm{warm}}/L_{\rm{TIR}}$, (d) $L_{\rm{cold}}/L_{\rm{TIR}}$, and (f) $L_{\rm{warm}}/L_{\rm{cold}}$ plotted against $L_{\rm{TIR}}$. The symbols are the same as those in figure \ref{fig:PAHplot}. The line fitted to the data points is shown in each panel. \label{fig:warmcoldplot}}
\end{figure}

\begin{figure}[ht]
\centering
\subfigure{
\makebox[180mm][l]{\raisebox{0mm}{\hspace{12mm} \small{RA (J2000)}} \hspace{17.5mm} \small{RA (J2000)} \hspace{17.5mm} \small{RA (J2000)} \hspace{18mm} \small{RA (J2000)}}%
}

\subfigure{
\centering
\makebox[5mm][l]{\raisebox{0mm}{\rotatebox{90}{\small{\hspace{5.5mm} DEC (J2000)} \hspace{10.5mm} \small{DEC (J2000)} \hspace{10.5mm} \small{DEC (J2000)} \hspace{10.5mm} \small{DEC (J2000)} \hspace{10.5mm} \small{DEC (J2000)}}}}%
\includegraphics[width=160mm]{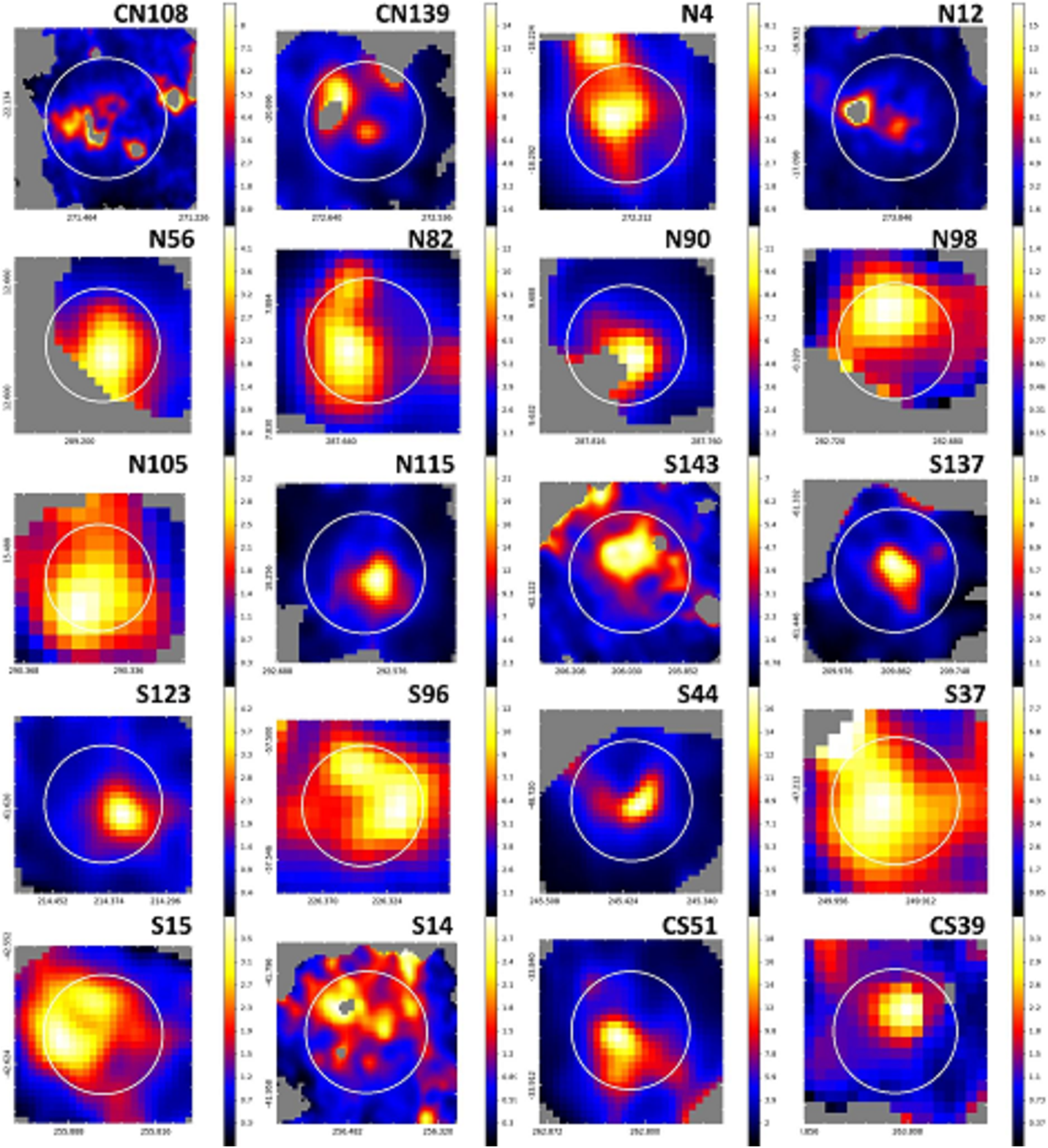}
}
\caption{Same as figure \ref{fig:PAHspC}, but $I_{\rm{warm}}/I_{\rm{cold}}$ ratio maps for the closed bubbles. The color levels are given individually. \label{fig:WoCspC}}
\end{figure}

\begin{figure}[ht]
\centering
\subfigure{
\makebox[180mm][l]{\raisebox{0mm}{\hspace{12mm} \small{RA (J2000)}} \hspace{17.5mm} \small{RA (J2000)} \hspace{17.5mm} \small{RA (J2000)} \hspace{18mm} \small{RA (J2000)}}%
}

\subfigure{
\centering
\makebox[5mm][l]{\raisebox{0mm}{\rotatebox{90}{\small{\hspace{5.5mm} DEC (J2000)} \hspace{10.5mm} \small{DEC (J2000)} \hspace{10.5mm} \small{DEC (J2000)} \hspace{10.5mm} \small{DEC (J2000)} \hspace{10.5mm} \small{DEC (J2000)}}}}%
\includegraphics[width=160mm]{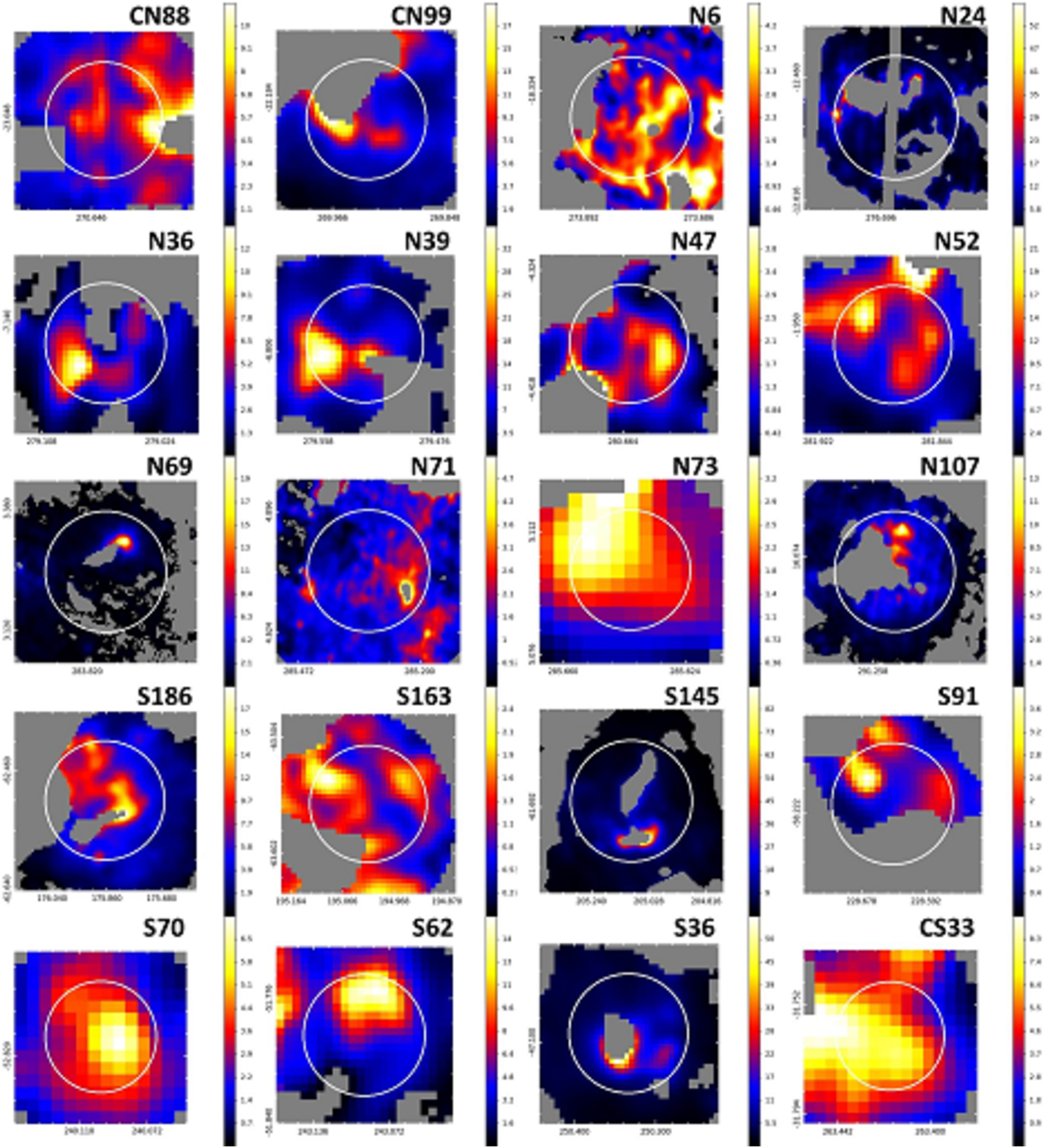}
}
\caption{Same as figure \ref{fig:WoCspC}, but for the broken bubbles.} \label{fig:WoCspB}
\end{figure}

The left panels of figure \ref{fig:warmcoldplot} show the luminosity ratios, $L_{\rm{warm}}/L_{\rm{TIR}}$, $L_{\rm{cold}}/L_{\rm{TIR}}$ and $L_{\rm{warm}}/L_{\rm{cold}}$ plotted against $R$, while the right panels of figure \ref{fig:warmcoldplot} show those against $L_{\rm{TIR}}$. From the top panels, we confirm that $L_{\rm{warm}}$ occupies a large fraction of $L_{\rm{TIR}}$ in most cases. In the middle and bottom panels, the luminosity ratios show some correlations with $R$, but not with $L_{\rm{TIR}}$. In particular, $L_{\rm{warm}}/L_{\rm{cold}}$ and $R$ are significantly correlated with each other ($r_{\rm c}=-0.34$). By interpreting $L_{\rm{warm}}$ as emission of dust in HII regions, the negative correlation of $L_{\rm{warm}}/L_{\rm{cold}}$ with $R$ may be reasonable; the dust more dilutes and less attenuates stellar UV photons as HII regions expand. On the other hand, no significant correlation of $L_{\rm{warm}}/L_{\rm{cold}}$ and $L_{\rm{TIR}}$ may be explained by considering that $L_{\rm{warm}}$ increases with $L_{\rm{TIR}}$ while $L_{\rm{cold}}$ does not necessarily because the latter depends on the conditions of the ambient ISM in parental clouds. Unlike $L_{\rm{PAH}}/L_{\rm{TIR}}$, there is no appreciable systematic difference between the closed and broken bubbles. Yet the brightness ratio maps of $I_{\rm{warm}}/I_{\rm{cold}}$ in figures \ref{fig:WoCspC} and \ref{fig:WoCspB}, which are created in the same manner as the $I_{\rm{PAH}}/I_{\rm{TIR}}$ maps in figures \ref{fig:PAHspC} and \ref{fig:PAHspB}, suggest a systematic difference between the closed and broken bubbles. The figures show that the $I_{\rm{warm}}/I_{\rm{cold}}$ distributions of the closed bubbles peak near the centers, while those of the broken bubbles tend to peak off the centers.

\begin{figure}[ht]
\begin{center}
\includegraphics[width=85mm]{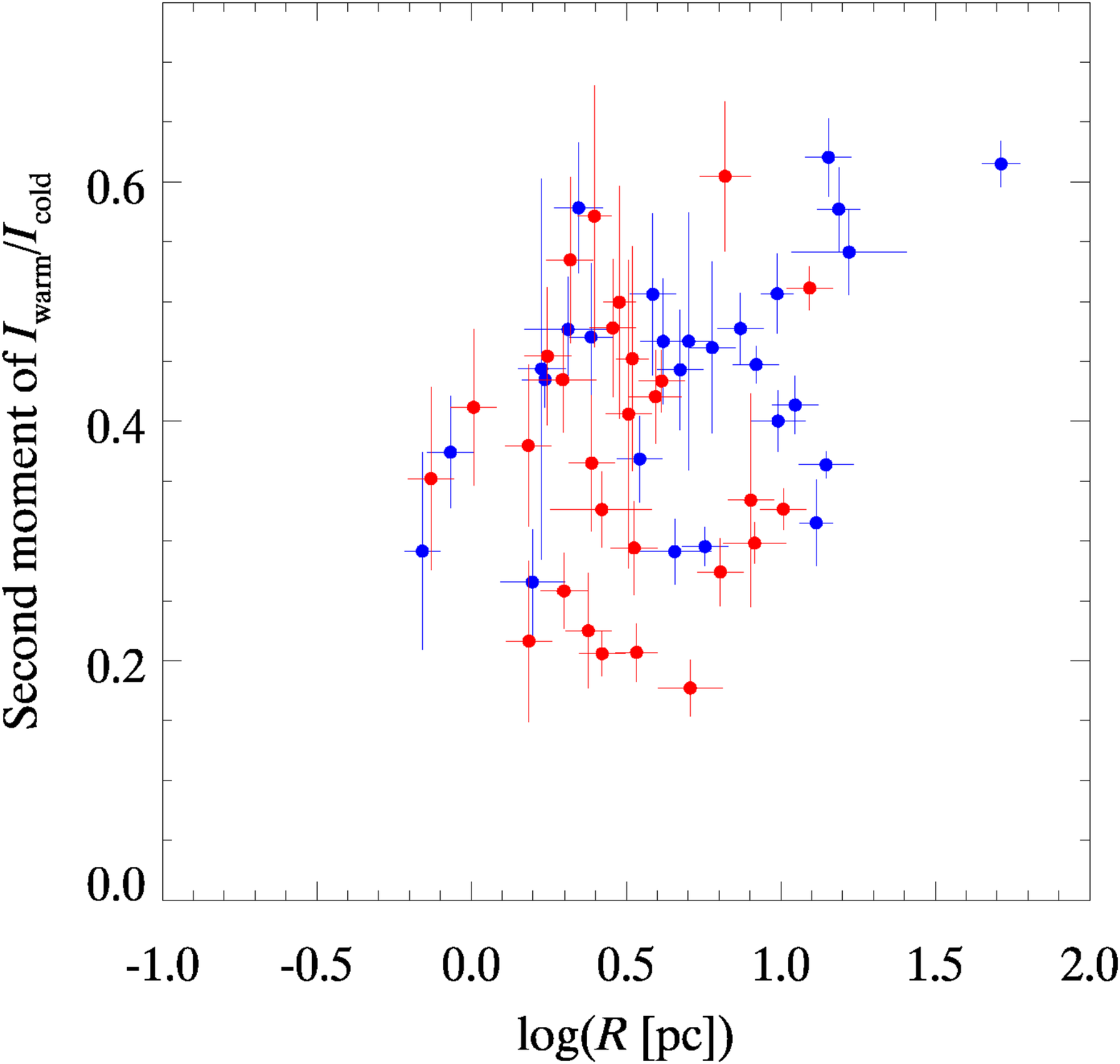}
\end{center}
\caption{Second moment of the $I_{\rm{warm}}/I_{\rm{cold}}$ distribution with respect to the center for the Galactic IR bubbles with the angular radii larger than \timeform{90''}. The symbols are the same as those in figure \ref{fig:Resfig1}. \label{fig:Discfig3}}
\end{figure}

In order to quantify the difference in the $I_{\rm{warm}}/I_{\rm{cold}}$ distribution between the closed and broken bubbles, we calculated the second moment, $I$, with respect to the center for each bubble, as shown in figure \ref{fig:Discfig3}, using the following equation:
\begin{equation}
\centering
\label{eq:SMoment}
I = \frac{\sum_{i}^{n} {r_i}^2 I_{\rm warm}(x_i,y_i)/ I_{\rm cold}(x_i,y_i)}{R^2 \sum_{i}^{n} I_{\rm warm}(x_i,y_i)/ I_{\rm cold}(x_i,y_i)},
\end{equation}
where $(x_i,y_i)$ and $r_i$ are the position of a pixel and the distance from the center to the pixel (i.e., $r_i = \sqrt{(x_i-l)^2+(y_i-b)^2}$), respectively. In the calculation, we used the pixels with $r_i < R$ and higher brightness which occupy 30\% of the total pixels at $r_i < R$. Since we require angular sizes large enough to spatially resolve the peak positions, we selected bubbles possessing angular radii larger than \timeform{90''}. As can be clearly seen in the figure, the broken bubbles tend to have larger $I$ values than the closed bubbles, indicating that the positions of the heating sources associated with the broken bubbles are substantially offset from the bubble centers. Indeed a K-S test shows that the distributions of the two types are statistically different with a 85\% confidence level (K-S value of 1.14). The averaged value of $I{\sim}0.5$ for the broken bubbles corresponds to the peak offset of ${\sim}0.7R$ from the bubble center, assuming the peak size of \timeform{1.5'} in FWHM for a bubble of \timeform{3'} in radius.

\clearpage

\section{Discussion}

Throughout the previous section, we show an overall similarity in the IR properties between the closed and broken bubbles. In particular, both types show a tight correlation of $L_{\rm{TIR}} \propto R^{3}$ as expected from the conventional picture of the Str$\rm{\ddot{o}}$mgren sphere. The fractional luminosity of PAH emission ($L_{\rm{PAH}}/L_{\rm{TIR}}$) in the shells is relatively constant at $\sim$0.1, a value typical of star-forming regions (e.g., \citealt{Smith2007}; \citealt{Kaneda2013}). The color temperature of the dust emission peaks near the bubble centers, which reflects the presence of heating sources (i.e., massive stars) in the centers. Hence, in order to interpret the results for the majority of the Spitzer Galactic IR bubbles, there is no strong reason for considering pictures other than the Str$\rm{\ddot{o}}$mgren sphere as the origin of the bubbles. However bubbles with relatively large radii tend to exhibit some differences between the closed and broken types; the large broken bubbles show relatively high $L_{\rm{TIR}}$ (figure \ref{fig:LRdev}), low $I_{\rm{PAH}}/I_{\rm{TIR}}$ (figure \ref{fig:Discfig2}) and large $I_{\rm{warm}}/I_{\rm{cold}}$ peak offsets (figure \ref{fig:Discfig3}). We below discuss the implications of the differences for a massive star formation scenario.

The low $I_{\rm{PAH}}/I_{\rm{TIR}}$ and the large $I_{\rm{warm}}/I_{\rm{cold}}$ peak offsets can be explained by the formation of massive young stars at the edges of bubbles as a consequence of triggered star-formation such as $``$globule squeezing$"$ and $``$collect and collapse$"$ (e.g., \citealt{Deharveng2010}; \citealt{Zavagno2010}; \citealt{Dirienzo2012}; \citealt{Anderson2015}). Such second-generation massive young stars emit intense UV photons from inside the bubble shells, which may cause efficient destruction of PAHs in the shells. They will also contribute to increasing the total IR luminosities. However the enlargement of $R$ may be difficult to be explained by triggered star-formation, because HII regions created by the second-generation young stars cannot substantially expand the original bubble structures. It may also be difficult to explain the systematic differences between broken and closed bubbles.

On the other hand, according to the $``$cloud-cloud collision$"$ scenario, their bubble structures may be formed by parental clouds deformed as a result of collision, which will cause the relative increase in $R$. Broken bubble structures will be favorably observed, unless a head-on collision between clouds is viewed face-on. Their high $L_{\rm{TIR}}$ may suggest the efficient formation of massive stars through the collision process. The relative decrease in $I_{\rm{PAH}}/I_{\rm{TIR}}$ and the offset of heating sources can be explained by formation of massive stars on the boundary of the collided clouds.

Nonetheless we cannot distinguish the massive star-formation scenarios based on the AKARI results alone. For further discussions, we need additional observational data of the molecular clouds, ionizing gases and young star clusters associated with the bubbles. Our AKARI results, particularly on the large broken bubbles, will provide important observational clues to revealing the origin of the massive star-formation associated with the bubbles.

\clearpage

\section{Summary}
We have investigated the mid- and far-IR properties of the Spitzer Galactic IR bubbles using the AKARI all-sky survey data at wavelengths of 9, 18, 65, 90, 140 and 160 \micron. For the 180 bubbles with known distances, we categorized them into 74 closed, 49 broken and 57 unclassified bubbles with our data analysis method. We obtained the spatial distributions and IR luminosities of the PAH, warm and cold dust components by decomposing AKARI 6-band local SEDs. As a result, the 180 sample bubbles show a wide range of the total IR luminosities corresponding to the bolometric luminosities of a single B-type star to many O-type stars. They exhibit a tight correlation between the radius and the total IR luminosity, following a power-law relation with an index of $\sim$3 as expected from the conventional picture of the Str$\rm{\ddot{o}}$mgren sphere. Their fractional luminosities of the PAH emission decrease with the total IR luminosity, which is interpreted by selective destruction of PAHs exposed to the intense UV radiation. We also investigated other relations between the radius, luminosities and luminosity ratios, and found that there are overall similarities among the bubbles regardless of their morphological types. The exceptions are large broken bubbles; they indicate larger radii, higher total IR luminosities, lower fractional luminosities of the PAH emission, and dust heating sources located nearer to the shells. One possibility to explain the different properties is the $``$cloud-cloud collision$"$ scenario for massive star-formation. Their bubble structures may have been formed by collided clouds, which caused the relative increase in the radius and total IR luminosity, PAH destruction, and formation of stars on the collision boundary.

\begin{ack}
This research is based on observations with AKARI, a JAXA project with the participation of ESA. The authors thank all the members of the AKARI project, particularly the all-sky survey data reduction team. This research has made use of the NASA/ IPAC Infrared Science Archive, which is operated by the Jet Propulsion Laboratory, California Institute of Technology, under contract with the National Aeronautics and Space Administration. This research was supported by JSPS KAKENHI Grant Number 25247020.
\end{ack}

\end{document}